\documentclass[11pt,twoside,a4paper]{cernrep} 
\usepackage{rep_common}
\usepackage{units}
\usepackage{pdflscape}
\usepackage{afterpage}
\usepackage{lineno}
\usepackage{diagbox}
\usepackage{hyperref}
\usepackage{placeins}
\usepackage{enumerate}
\usepackage[utf8]{inputenc}

\def\bibfiles{\main/bib/chapter,\main/introduction/bib/section,\main/schedule/bib/section,\main/hf/bib/section,\main/quarkonia/bib/section,\main/thermalradiation/bib/section,\main/flow/bib/section,\main/smallsystems/bib/section,\main/lightflavour/bib/section,\main/jets/bib/section,\main/smallx/bib/section,\main/beyond/bib/section,\main/helhc/bib/section,\main/accelerator/bib/section,\main/smallAexec/bib/section}
\providecommand{\biblio}{\bibliographystyle{report}\clearpage\bibliography{\bibfiles}}  
 
\begin{document}
\newcommand{\main}{.}


\newcommand{\pt}{\pT}
\providecommand{\sqrtsnn}{\sqrtsNN}
\newcommand{\Raa}{\RAA}
\providecommand{\raa}{\ensuremath{R_\text{AA}}\Xspace}

\DeclareRobustCommand{\Pepem}{\HepParticle{\Pe}{}{+}\HepParticle{\Pe}{}{-}\Xspace} 
\DeclareRobustCommand{\PGmpGmm}{\HepParticle{\PGm}{}{+}\HepParticle{\PGm}{}{-}\Xspace} 
\renewcommand{\pT}{\ensuremath{p\sb{\scriptstyle\mathrm{T}}}\Xspace}

\newcommand{\sqrts}{\ensuremath{\sqrt{s}}\Xspace}
\newcommand{\sqrtsNN}{\ensuremath{\sqrt{\sNN}}\Xspace}
\newcommand{\Npart}{\ensuremath{N_{\rm part}}\Xspace}
\newcommand{\Ncoll}{\ensuremath{N_{\rm coll}}\Xspace}
\newcommand{\RpPb}{\ensuremath{R_{\rm pPb}}\Xspace}
\newcommand{\RAA}{\ensuremath{R_{\rm AA}}\Xspace}
\newcommand{\RpA}{\ensuremath{R_{\rm pA}}\Xspace}
\newcommand{\TAA}{\ensuremath{T_{\rm AA}}\Xspace}
\newcommand{\RCP}{\ensuremath{R_{\rm CP}}\Xspace}
\newcommand{\vtwo}{\ensuremath{v_{\rm 2}}\Xspace}
\newcommand{\vone}{\ensuremath{v_{\rm 1}}\Xspace}
\newcommand{\vthree}{\ensuremath{v_{\rm 3}}\Xspace}
\newcommand{\vfour}{\ensuremath{v_{\rm 4}}\Xspace}
\newcommand{\vfive}{\ensuremath{v_{\rm 5}}\Xspace}
\newcommand{\vsix}{\ensuremath{v_{\rm 6}}\Xspace}
\newcommand{\vseven}{\ensuremath{v_{\rm 7}}\Xspace}
\newcommand{\vn}{\ensuremath{v_{\rm n}}\Xspace}

\newcommand{\nch}         {\ensuremath{N_{\mathrm {ch}}\xspace}}
\newcommand{\meannch}     {\ensuremath{\langle N_{\mathrm {ch}} \rangle\Xspace}}
\newcommand{\meanpT}      {\ensuremath{\langle\pT\rangle}}
\newcommand{\dNdeta}      {\mathrm{d}N_\mathrm{ch}/\mathrm{d}\eta}
\newcommand{\dNdy}        {\mathrm{d}N_\mathrm{ch}/\mathrm{d}y}
\newcommand{\kT}          {\ensuremath{k\sb{\scriptstyle\mathrm{T}}}}
\newcommand{\ptt}         {\ensuremath{p_{\mathrm{T, trig}}}}
\newcommand{\pta}         {\ensuremath{p_{\mathrm{T, assoc}}}}

\newcommand{\pp}          {$\mathrm{pp}$\Xspace}
\newcommand{\pPb}         {$\mathrm{p}$--$\mathrm{Pb}$\Xspace}
\newcommand{\Pbp}         {$\mathrm{Pb}$--$\mathrm{p}$\Xspace}
\newcommand{\pO}          {$\mathrm{p}$--$\mathrm{O}$\Xspace}
\newcommand{\Op}          {$\mathrm{O}$--$\mathrm{p}$\Xspace}
\newcommand{\OO}          {$\mathrm{O}$--$\mathrm{O}$\Xspace}
\newcommand{\pA}          {$\mathrm{p}$--$\mathrm{A}$\Xspace}
\newcommand{\AOnA}        {$\mathrm{A}$--$\mathrm{A}$\Xspace}
\newcommand{\PbPb}        {$\mathrm{Pb}$--$\mathrm{Pb}$\Xspace}
\newcommand{\ArAr}        {$\mathrm{Ar}$--$\mathrm{Ar}$\Xspace}
\newcommand{\XeXe}        {$\mathrm{Xe}$--$\mathrm{Xe}$\Xspace}
\newcommand{\KrKr}        {$\mathrm{Kr}$--$\mathrm{Kr}$\Xspace}
\newcommand{\AuAu}        {$\mathrm{Au}$--$\mathrm{Au}$\Xspace}
\newcommand{\CuCu}        {$\mathrm{Cu}$--$\mathrm{Cu}$\Xspace}
\newcommand{\pAu}         {$\mathrm{p}$--$\mathrm{Au}$\Xspace}
\newcommand{\dAu}         {$\mathrm{d}$--$\mathrm{Au}$\Xspace}

\newcommand{\sigmaBFPP}   {\ensuremath{\sigma_\mathrm{BFPP}}\Xspace}

\newcommand{\lsim}        {\,{\buildrel < \over {_\sim}}\,}
\newcommand{\gsim}        {\,{\buildrel > \over {_\sim}}\,}
\newcommand{\co}[1]       {\relax}
\newcommand{\nl}          {\newline}
\newcommand{\el}          {\\\hline\\[-0.4cm]}

\newcommand{\gmom}{\ensuremath{\mathrm{GeV}\kern-0.05em/\kern-0.02em c}}
\newcommand{\antip}{\ensuremath{\overline{\mathrm{p}}}}
\newcommand{\antid}{\ensuremath{\overline{\mathrm{d}}}}
\newcommand{\tritium}{\ensuremath{{}^{3}\mathrm{H}}}
\newcommand{\antitritium}{\ensuremath{{}^{3}\overline{\mathrm{\mathrm{He}}}}}
\newcommand{\hethree}{\ensuremath{{}^{3}\mathrm{He}}}
\newcommand{\hefour}{\ensuremath{{}^{4}\mathrm{He}}}
\newcommand{\antihethree}{\ensuremath{{}^{3}\overline{\mathrm{He}}}}
\newcommand{\antihefour}{\ensuremath{{}^{4}\overline{\mathrm{He}}}}
\newcommand{\hyp}        {\ensuremath{^{3}_{\Lambda}\mathrm{H}}}
\newcommand{\antihyp}{\ensuremath{^{3}_{\bar{\Lambda}}\overline{\mathrm{H}}}}
\newcommand{\hypfour}    {\ensuremath{^{4}_{\Lambda}\mathrm{H}}}
\newcommand{\antihypfour}{\ensuremath{^{4}_{\bar{\Lambda}}\overline{\mathrm{H}}}}
\newcommand{\hyphefour}    {\ensuremath{^{4}_{\Lambda}\mathrm{He}}}
\newcommand{\antihehypfour}{\ensuremath{^{4}_{\bar{\Lambda}}\overline{\mathrm{He}}}}
\newcommand{\sigmahyp}     {\ensuremath{^{3}_{\Sigma^{0}}\mathrm{H}}}
\newcommand{\antisigmahyp} {\ensuremath{^{3}_{\bar{\Sigma}^{0}}\overline{\mathrm{H}}}}

\newcommand{\sla}{\slash \hspace{-0.2cm}}
\newcommand{\slam}{\slash \hspace{-0.25cm}}
\newcommand{\no}{\nonumber}
\def\lsim{\mathrel{\rlap{\lower4pt\hbox{\hskip1pt$\sim$}}
    \raise1pt\hbox{$<$}}}         
\def\gsim{\mathrel{\rlap{\lower4pt\hbox{\hskip1pt$\sim$}}
    \raise1pt\hbox{$>$}}}         

\newcommand{\Anucl}{$\mathrm{A}$\Xspace}
\newcommand{\nbInv}{$\mathrm{nb}^{-1}$\Xspace}
\newcommand{\pbInv}{$\mathrm{pb}^{-1}$\Xspace}
\newcommand{\isospin}{$I$\Xspace}
\newcommand{\spinJ}{$J$\Xspace}
\newcommand{\BA}{$B_{\mathrm{A}}$\Xspace}
\newcommand{\significance}{$\frac{\mathrm{S}}{\sqrt{\mathrm{S}+\mathrm{B}}}$\Xspace}
\newcommand{\Tchem}{\ensuremath{T_{\mathrm{chem}}}\Xspace}

\def\Bs{{\overline{B}}_s}
\def\R{\mathcal{R}}
\newcommand{\gev}{\mathrm{GeV}}
\newcommand{\tev}{\mathrm{TeV}}
\newcommand{\mev}{\mathrm{MeV}}
\newcommand{\e}{\epsilon}
\newcommand{\tce}{\frac{t_{\rm cool}(\e)}{t_{\rm esc}(\e)}}
\newcommand{\tcer}{\frac{t_{\rm c}(\R)}{t_{\rm esc}(\R)}}
\def\Xe{X_{\rm esc}}
\def\X{X_{\rm esc}}
\def\te{t_{\rm esc}}
\def\tc{t_{\rm cool}}
\def\nb{n_{\rm B}}
\def\nc{n_{\rm C}}
\def\ni{n_{i}}
\def\rism{\rho_{\rm ISM}}
\def\nism{n_{\rm ISM}}
\def\x{(\R,\vec r,t)}
\def\xo{(\R,\vec r_\odot,t_\odot)}
\def\ap{\overline{\rm p}}
\def\ad{\overline{\rm d}}
\def\ep{e^+}
\def\Qep{Q_{e^+}}
\def\epm{$e^\pm$\ }
\def\ah{\overline{\rm ^3He}}
\def\at{\overline{\rm t}}
\def\s{$(*)$}
\newcommand{\dd}{\text{d}}
\newcommand{\gaga}{\gamma\gamma}
\newcommand{\Rp}{\mathcal{R}^\prime}
\newcommand{\Lp}{L^{\prime}}

\newcommand{\Dsc}{\ensuremath{D_{\rm s}}\Xspace}
\newcommand{\twopiTDsc}{\ensuremath{2 \pi T D_{\rm s}}\Xspace}
\newcommand{\ToverTc}{\ensuremath{T/T_{\rm c}}\Xspace}
\newcommand{\Tc}{\ensuremath{T_{\rm c}}\Xspace}
\newcommand{\chisquared}{\ensuremath{\chi^{\rm 2}}\Xspace}

\newcommand{\RunsThreeFour}{Runs~3 \& 4\Xspace}

\newcommand{\ttbar}{\ensuremath{t\overline{t}}\Xspace}



\newcommand{\qty}[2]{\ensuremath{#1\,\mathrm{#2}}}  
\newcommand{\enum}[2]{\ensuremath{#1\times10^{#2}}} 
\newcommand{\NQTY}[2]{\mbox{$[#1/{\rm #2}]$}}     
\newcommand{\UQTY}[2]{\ensuremath{#1/\mathrm{#2}}}  
\newcommand{\eqty}[3]{\qty{\enum{#1}{#2}}{#3}}  
\newcommand{\invnb}{\mathrm{nb}^{-1}}
\newcommand{\invpb}{\mathrm{pb}^{-1}}

\newcommand{\elumi}[2]{\qty{\enum{#1}{#2}}{cm^{-2}s^{-1}}}
\newcommand{\murad}[1]{\qty{#1}{\mu rad}}
\newcommand{\intlumimub}[1]{\qty{#1}{\mu b^{-1}}}

\newcommand{\sNN}{\ensuremath{s_{\mbox{\tiny NN}}}}
\newcommand{\yNN}{\ensuremath{y_{\mbox{\tiny NN}}}}
\newcommand{\bstar}{\ensuremath{\beta^{*}}}
\newcommand{\emittn}{\ensuremath{\varepsilon_n}}
\newcommand{\LAA}{\ensuremath{L_\text{AA}}}
\newcommand{\LpA}{\ensuremath{L_{pA}}}
\newcommand{\Lpp}{\ensuremath{L_{pp}}}
\newcommand{\Lpeak}{\ensuremath{\hat{L}}}
\newcommand{\LNN}{\ensuremath{ L_{\text{NN}}}}

\newcommand{\isotope}[3]{\ensuremath{^{#1}\mathrm{#2}^{#3}}}

\newcommand{\speciesheader}{ &
\isotope{16}{O}{8+}&
\isotope{40}{Ar}{18+}&
\isotope{40}{Ca}{20+}&
\isotope{78}{Kr}{36+}&
\isotope{129}{Xe}{54+}&
\isotope{208}{Pb}{82+}
}

\newcommand{\bfunc}{$\beta$-function}
\newcommand{\bstarval}[1]{$\bstar = #1\,\mbox{m}$}
\newcommand{\betarel}{\ensuremath{\beta_\text{rel}}}
\newcommand{\emittnx}{\ensuremath{\epsilon_{n,x}}}
\newcommand{\emittny}{\ensuremath{\epsilon_{n,y}}}
\newcommand{\emittnxy}{\ensuremath{\epsilon_{n,xy}}}
\newcommand{\emitts}{\ensuremath{\epsilon_s}}
\newcommand{\sigs}{\ensuremath{\sigma_s}}
\newcommand{\sigp}{\ensuremath{\sigma_p}}
\newcommand{\kb}{\ensuremath{k_b}}
\newcommand{\frev}{\ensuremath{f_0}}
\newcommand{\Nb}{\ensuremath{N_b}}
\newcommand{\Eb}{\ensuremath{E_b}}
\newcommand{\emittval}[1]{\ensuremath{\emittn=\qty{#1}{\mu m\,rad}}}
\newcommand{\Nbval}[2]{\ensuremath{\Nb=\enum{#1}{#2}}}
\newcommand{\taul}{\ensuremath{\tau_l}}
\newcommand{\taulval}[1]{\ensuremath{\taul=\qty{#1}{ns}}}
\newcommand{\sigzval}[1]{\ensuremath{\sigz=\qty{#1}{cm}}}
\newcommand{\etev}[1]{\ensuremath{\Eb=\qty{#1}{TeV}}}
\newcommand{\VRF}{\ensuremath{V_{\mathrm{RF}}}}
\newcommand{\lumival}[2]{\ensuremath{L=\qty{#1\times 10^{#2}}{cm^{-2} s^{-1}}}}

\newcommand{\aibsx}{\ensuremath{\alpha_{\mathrm{IBS},x}}}
\newcommand{\aibsy}{\ensuremath{\alpha_{\mathrm{IBS},y}}}
\newcommand{\aibsxy}{\ensuremath{\alpha_{\mathrm{IBS},x,y}}}
\newcommand{\aradd}{\ensuremath{\alpha_{\mathrm{rad}}}}
\newcommand{\aradds}{\ensuremath{\alpha_{\mathrm{rad},s}}}
\newcommand{\araddx}{\ensuremath{\alpha_{\mathrm{rad},x}}}
\newcommand{\araddy}{\ensuremath{\alpha_{\mathrm{rad},y}}}
\newcommand{\araddxy}{\ensuremath{\alpha_{\mathrm{rad},x,y}}}
\newcommand{\Z}{\ensuremath{Z_\text{ion}}}
\newcommand{\A}{\ensuremath{A_\text{ion}}}
\newcommand{\Circ}{\ensuremath{C_\text{ring}}}
\newcommand{\lumi}{\ensuremath{\mathcal{L}}}
\newcommand{\Lb}{\ensuremath{\mathcal{L}_b}}
\newcommand{\Lint}{\ensuremath{L_{\text{int}}}}
\newcommand{\Lbint}{\ensuremath{L_{b,\text{int}}}}
\newcommand{\Lbpeak}{\ensuremath{\mathcal{L}_{b,\text{peak}}}}


\newcommand{\mlna}{\langle \ln\!A \rangle}
\newcommand{\nmu}{N_\mu}
\newcommand{\lnnmu}{\ln\!\nmu}
\newcommand{\xmax}{X_\text{max}}
\newcommand{\nmult}{N_\text{mult}}
\newcommand{\tocite}{{\bf REF}}
\newcommand{\si}[1]{\ensuremath{\text{#1}}}
\newcommand{\SI}[2]{\ensuremath{#1\,\si{#2}}}

\title{{\normalfont\bfseries\boldmath\huge
\begin{center}
Future physics opportunities for high-density QCD\\ at the LHC with heavy-ion and proton beams\\
\begin{normalsize} 
  \href{http://lpcc.web.cern.ch/hlhe-lhc-physics-workshop}{Report from Working Group 5 on the Physics of the HL-LHC, and Perspectives at the HE-LHC}
\end{normalsize}
\end{center}\vspace*{0.2cm}
}}

\author{Editors: \\
Z.~Citron$^{3}$,
A.~Dainese$^{24}$,
J.F.~Grosse-Oetringhaus$^{6}$,
J.M.~Jowett$^{6}$,
Y.-J.~Lee$^{53}$,
U.A.~Wiedemann$^{6}$,
M.~Winn$^{33,43}$
\\ \vspace*{4mm} 
Chapter coordinators: \\ 
A.~Andronic$^{52}$,
F.~Bellini$^{6}$,
E.~Bruna$^{26}$
% ...
\\
$^*$ deceased
}
\institute{
\clearpage
\footnotesize
$^{1}$~AGH~University~of~Science~and~Technology,~Krak\'ow,~Poland,
$^{2}$~Academy~of~Sciences,~Prague,~Czech~Republic,
$^{3}$~Ben-Gurion~University~of~the~Negev,~Beersheba,~Israel,
$^{4}$~Brookhaven~National~Laboratory,~Upton,~USA,
$^{5}$~Brown~University,~Rhode~Island,~USA,
$^{6}$~CERN,~Geneva,~Switzerland,
$^{7}$~CFTP,~Lisbon,~Portugal,
$^{8}$~CNRS/IN2P3,~Universit\'e~Paris-Sud,~Universit\'e~Paris-Saclay,~Orsay,~France,
$^{9}$~CPHT,~CNRS,~\'Ecole~polytechnique,~Universit\'e~Paris-Saclay,~Palaiseau,~France,
$^{10}$~Center~for~Nuclear~Study,~Graduate~School~of~Science,~The~University~of~Tokyo,~Japan,
$^{11}$~Charles~University,~Prague,~Czech~Republic,
$^{12}$~Chonnam~National~University,~Gwangju,~China,
$^{13}$~Columbia~University,~New~York~City,~USA,
$^{14}$~Czech~Technical~University~in~Prague,~Czech~Republic,
$^{15}$~DESY,~Hamburg,~Germany,
$^{16}$~Duke~University,~Durham,~USA,
$^{17}$~EMMI$/$GSI~Helmholtzzentrum~f{\"u}r~Schwerionenforschung~GmbH,~Darmstadt,~Germany,
$^{18}$~Excellence~Cluster~Universe,~Technical~University~Munich,~Germany,
$^{19}$~Heidelberg~University,~Germany,
$^{20}$~Helsinki~Institute~of~Physics,~Finland,
$^{21}$~IFJ~PAN,~PL-31342~Krak\'ow,~Poland,
$^{22}$~INFN~-~Sezione~di~Bologna,~Italy,
$^{23}$~INFN~-~Sezione~di~Firenze,~Italy,
$^{24}$~INFN~-~Sezione~di~Padova,~Italy,
$^{25}$~INFN~-~Sezione~di~Roma,~Roma,~Italy,
$^{26}$~INFN~-~Sezione~di~Torino,~Italy,
$^{27}$~INFN~-~Sezione~di~Trieste,~Italy,
$^{28}$~INFN-LNF,~Frascati,~Italy,
$^{29}$~INFN-LNS,~Catania,~Italy,
$^{30}$~IPN~Orsay,~CNRS/IN2P3,~Universit\'e~Paris-Sud,~Universit\'e~Paris-Saclay,~Orsay,~France,
$^{31}$~IPN-Lyon,~CNRS/IN2P3,~Universit\'e~de~Lyon,~Lyon,~France,
$^{32}$~IPhT,~CEA~Saclay,~CNRS,~Universit\'e~Paris-Saclay,~Saclay,~France,
$^{33}$~IRFU/DPhN,~CEA~Saclay,~Universit\'e~Paris-Saclay,~Saclay,~France,
$^{34}$~IRFU/DPhP,~CEA~Saclay,~Universit\'e~Paris-Saclay,~Saclay,~France,
$^{35}$~IST~Lisbon,~Portugal,
$^{36}$~Inha~University,~Incheon,~Korea,
$^{37}$~Institute~of~Space~Science,~Bucharest,~Romania,
$^{38}$~Instituto~Galego~de~Fisica~de~Altas~Enerxias~(IGFAE),~Universidade~de~Santiago~de~Compostela,~Spain,
$^{39}$~Johann-Wolfgang-Goethe~Universit\"{a}t,~Frankfurt,~Germany,
$^{40}$~Karlsruhe~Institute~of~Technology,~Germany,
$^{41}$~Kent~State~University,~USA,
$^{42}$~Korea~University,~Seoul,~Korea,
$^{43}$~LAL,~CNRS/IN2P3,~Universit\'e~Paris-Sud,~Universit\'e~Paris-Saclay,~Orsay,~France,
$^{44}$~LIP,~Lisbon,~Portugal,
$^{45}$~LLR,~CNRS/IN2P3,~\'Ecole~polytechnique,~Universit\'e~Paris-Saclay,~Palaiseau,~France,
$^{46}$~LPSC~Grenoble,~CNRS/IN2P3,~Grenoble~INP,~Universit\'e~Grenoble~Alpes,~Grenoble,~France,
$^{47}$~Lawrence~Berkeley~National~Laboratory,~Berkeley,~USA,
$^{48}$~Lawrence~Livermore~National~Laboratory,~Livermore,~USA,
$^{49}$~Los~Alamos~National~Laboratory,~Los~Alamos,~USA,
$^{50}$~Lund~University,~Sweden,
$^{51}$~MPI~for~Nuclear~Physics,~Heidelberg,~Germany,
$^{52}$~M\"{u}nster~University,~Germany,
$^{53}$~Massachusetts~Institute~of~Technology,~Cambridge,~USA,
$^{54}$~Michigan~State~University,~East~Lansing,~USA,
$^{55}$~NNRC,~Baku,~Azerbaijan,
$^{56}$~Nanjing~University~of~Science~and~Technology,~China,
$^{57}$~National~Research~Centre~Kurchatov~Institute,~Moscow,~Russia,
$^{58}$~National~Research~Nuclear~University~MEPhI,~Moscow,~Russia,
$^{59}$~Niels~Bohr~Institute,~Copenhagen,~Denmark,
$^{60}$~Nikhef,~Amsterdam,~The~Netherlands,
$^{61}$~Oak~Ridge~National~Laboratory,~Oak~Ridge,~USA,
$^{62}$~Ohio~State~University,~Columbus,~USA,
$^{63}$~Pennsylvania~State~University,~University~Park,~USA,
$^{64}$~Petersburg~Nuclear~Physics~Institute,~Gatchina,~Russia,
$^{65}$~Politecnico~di~Bari~and~INFN~-~Sezione~di~Bari,~Italy,
$^{66}$~Politecnico~di~Torino~and~INFN~-~Sezione~di~Torino,~Italy,
$^{67}$~Purdue~University,~West~Lafayette,~USA,
$^{68}$~RIKEN~BNL~Research~Center,~Upton,~USA,
$^{69}$~Rice~University,~Houston,~USA,
$^{70}$~Rutgers~University,~New~Brunswick,~USA,
$^{71}$~SUBATECH,~CNRS/IN2P3,~IMT~Atlantique,~Universit{\'e}~de~Nantes,~France,
$^{72}$~Southern~Methodist~University,~Dallas,~USA,
$^{73}$~Stavanger~University,~Norway,
$^{74}$~Stefan~Meyer~Institute~Vienna,~Austria,~Austrian~Academy~of~Sciences,
$^{75}$~Stony~Brook~University,~USA,
$^{76}$~Technical~University~Munich,~Germany,
$^{77}$~Texas~A\&M~University,~College~Station,~USA,
$^{78}$~Tsinghua~University,~Beijing,~China,
$^{79}$~Universidade~de~Sao~Paulo,~Brazil,
$^{80}$~Universit\"{a}t~T\"{u}bingen,~Germany,
$^{81}$~Universit\'{e}~catholique~de~Louvain,~Louvain-la-Neuve,~Belgium,
$^{82}$~Universit\`a~di~Catania~and~INFN-LNS,~Catania,~Italy,
$^{83}$~University~of~Bergen,~Norway,
$^{84}$~University~of~Colorado~Boulder,~USA,
$^{85}$~University~of~Ferrara~and~INFN~-~Sezione~di~Ferrara,~Italy,
$^{86}$~University~of~Foggia~and~INFN~-~Sezione~di~Bari,~Italy,
$^{87}$~University~of~Gie{\ss}en,~Germany,
$^{88}$~University~of~Houston,~USA,
$^{89}$~University~of~Illinois,~Urbana-Champaign,~USA,
$^{90}$~University~of~Jyvaskyla,~Finland,
$^{91}$~University~of~Kansas,~Lawrence,~USA,
$^{92}$~University~of~Louisville,~USA,
$^{93}$~University~of~Massachusetts,~Amherst,~USA,
$^{94}$~University~of~Padova~and~INFN~-~Sezione~di~Padova,~Italy,
$^{95}$~University~of~Torino~and~INFN~-~Sezione~di~Torino,~Italy,
$^{96}$~University~of~Trieste~and~INFN~-~Sezione~di~Trieste,~Italy,
$^{97}$~University~of~Tsukuba,~Japan,
$^{98}$~University~of~Wroclaw,~Poland,
$^{99}$~Utrecht~University,~The~Netherlands,
$^{100}$~VECC~Calcutta,~India,
$^{101}$~Vanderbilt~University,~Nashville,~USA,
$^{102}$~Wayne~State~University,~Detroit,~USA,
$^{103}$~Weizmann~Institute~of~Science,~Rehovot,~Israel
}

\begin{titlepage}

\vspace*{-1.8cm}

\noindent
\begin{tabular*}{\linewidth}{lc@{\extracolsep{\fill}}r@{\extracolsep{0pt}}}
\vspace*{-1.2cm}\mbox{\!\!\!\includegraphics[width=.14\textwidth]{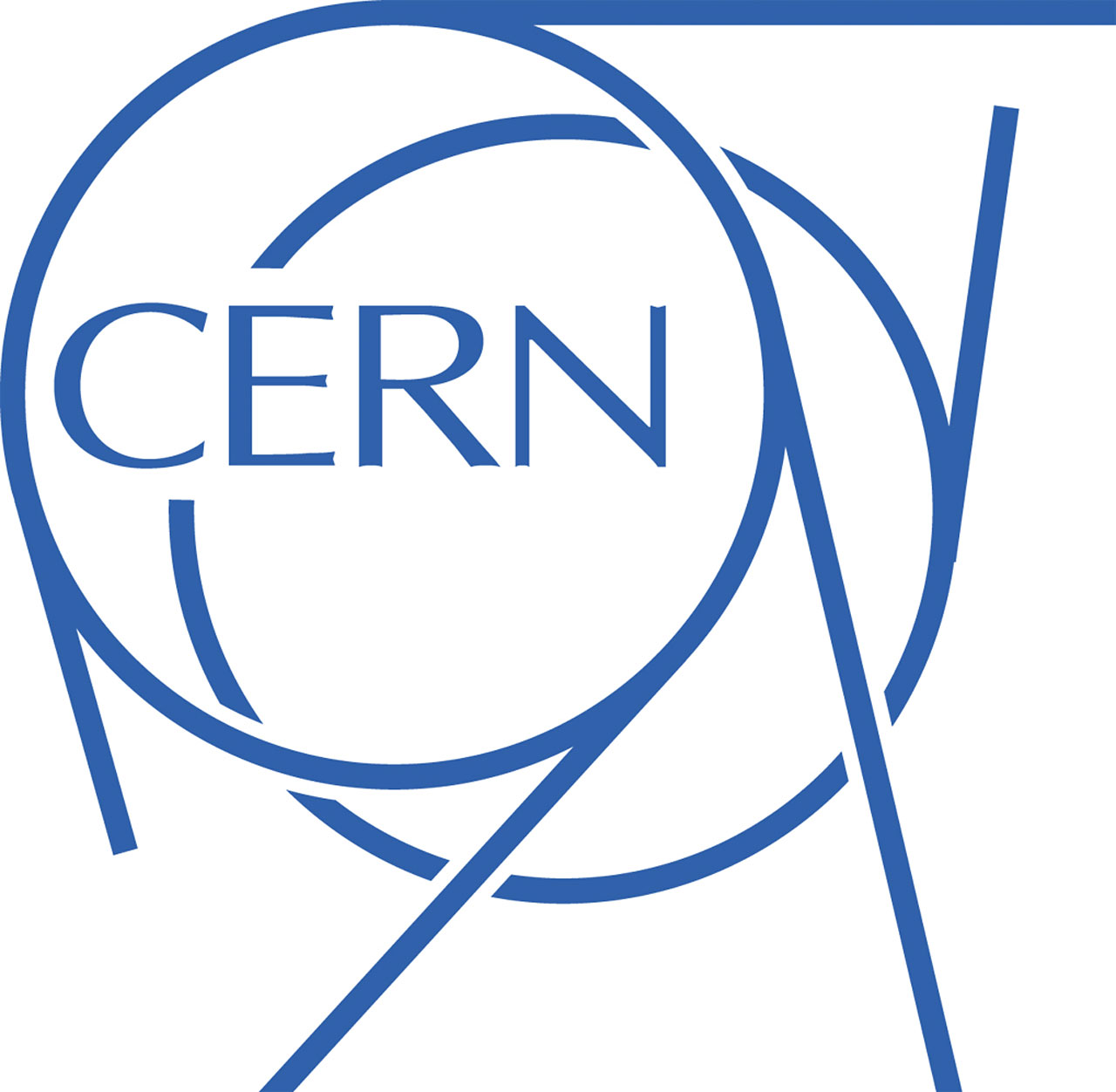}} & & \\
 & & CERN-LPCC-2018-07 \\  
 & & \today \\ 
 & & \\
\hline
\end{tabular*}

\vspace*{0.3cm}
%
%
\maketitle

\vspace*{2.0cm}
\begin{center}
    \begin{abstract}
  \noindent
  
\end{abstract}
\end{center}
The future opportunities for high-density QCD studies with ion and proton beams at the LHC are presented. Four major scientific goals are identified: the characterisation of the macroscopic long wavelength Quark--Gluon Plasma~(QGP) properties with unprecedented precision, the investigation of the microscopic parton dynamics underlying QGP properties, the development of a unified picture of particle production and QCD dynamics from small (\pp) to large (nucleus--nucleus) systems, the exploration of parton densities in nuclei in a broad ($x$, $Q^2$) kinematic range and the search for the possible onset of parton saturation. In order to address these scientific goals, high-luminosity \PbPb and \pPb programmes are considered as priorities for Runs~3 and~4, complemented by high-multiplicity studies in pp collisions and a  short run with oxygen ions. High-luminosity runs with intermediate-mass nuclei, for example $\mathrm{Ar}$ or $\mathrm{Kr}$, are considered as an appealing case for extending the heavy-ion programme at the LHC beyond Run~4. The potential of the High-Energy LHC to probe QCD matter with newly-available observables, at twice larger center-of-mass energies than the LHC, is investigated.

\vspace*{2.0cm}
\vspace{\fill}

\end{titlepage}

\setcounter{tocdepth}{2}
{ 
\baselineskip=14pt
\tableofcontents
}
\clearpage


\section{Introduction}

Experiments with heavy-ion collisions at the LHC create and diagnose strongly-interacting matter under the most extreme conditions of high density and temperature accessible in the laboratory. 
Under these conditions, QCD calculations on the lattice predict that matter undergoes a phase transition to a Quark--Gluon Plasma (QGP) in which colour charges are deconfined and chiral symmetry is restored.
Aside of its intrinsic interest, this line of research is central to our understanding of the Early Universe and the evolution of ultra-dense stars. In practice, the main focus of experimentation with nuclear beams at the LHC is on learning how collective phenomena and macroscopic properties, involving many degrees of freedom, emerge under extreme conditions from the microscopic laws of strong-interaction physics. In doing so, proton--nucleus (\pA) and nucleus--nucleus (\AOnA) collision experiments at the LHC aim at applying and extending the Standard Model of particle physics to matter properties that govern dynamically evolving systems of finite size.

Recent experiments with nuclear beams at collider energies have identified opportunities to further strengthen the connection between the rich phenomenology of ultra-dense matter, and its understanding in terms of the fundamental laws of strong-interaction physics. Two broad classes of phenomena may be highlighted in this context.

First, the observation of flow-like phenomena in essentially all measured soft particle spectra and particle correlations lends strong support to understanding bulk properties of heavy-ion collisions in terms of viscous fluid dynamics. The fluid-dynamic evolution is solely based on combining conservation laws with thermodynamic transport theories that are calculable from first principles in quantum field theory. Hence, this provides an experimentally accessible inroad to constraining QCD matter properties via soft flow, correlation and fluctuation measurements. As further explained in this document, this motivates future improved measurements of flow and transport phenomena, including in particular measurements of soft heavy flavour and electromagnetic radiation.  It also motivates improved experimental control over the system-size dependence of flow phenomena to better constrain under which conditions and in which kinematic regime ultra-relativistic \pA and \AOnA collisions show fluid dynamic behaviour and where this picture fails.

Second, the observation of quantitatively-large quenching phenomena in essentially all measured hard hadronic observables in \AOnA collisions has established the feasibility of testing the produced QCD matter with a broad set of probes whose production rates are controlled with good precision with \pp reference measurements and perturbative QCD calculations. Hard quarks and gluons are known to interact with the medium and they thus {\it probe} medium properties. As detailed in this document, important physics opportunities are related to analysing hard probes in \pA and \AOnA collisions with the greater precision and kinematic reach accessible in future LHC runs. For instance, the identification of (Rutherford-type) large-angle jet-medium scattering could constrain the quasi-particle nature of the fluid-like medium. This is of central importance since it critically tests the working hypothesis that the matter produced in \AOnA collisions is a fluid with a ratio of shear viscosity over entropy density $\eta/s$ close to the theoretical minimum value. Such a fluid would be void of quasi-particles, while QCD is definitive in predicting that a microscope with sufficiently high resolution will reveal partonic (quasi-)particle structure. Identifying the scale at which inner structure (such as quasi-particles and related non-vanishing mean free paths) arises would provide a microscopic understanding of how fluid-like behaviour arises in nucleus--nucleus collisions. Therefore, probing the inner workings of the QGP by resolving its properties at shorter and shorter length scales is one of the main motivations for future experimentation with hard probes. Also at intermediate transverse momentum, this document identifies important questions that will be accessible experimentally. For instance, more differential studies of quarkonium bound-state dissociation as a function of transverse momentum, rapidity and system size are expected to yield further insights into the mechanisms of colour deconfinement and recombination. Moreover, there is the general question of how fluid-like phenomena at low momentum scales transition to quenching phenomena at intermediate and high momentum scales.

Capitalizing on previous discoveries at RHIC, the LHC experimental programme with \PbPb and \pPb collisions has significantly advanced the state of the art in both the soft- and the hard-physics sector. In the soft-physics sector, differential measurements at the LHC have in particular allowed for precise $\pT$, particle species and rapidity-dependent measurements of all higher flow harmonics $\vn$, their mode--mode couplings and the resulting reaction-plane correlations. This flow systematics extends to charmed flavoured hadrons and possibly even to beauty ones. It is at the basis of constraining QCD transport properties today. In the hard-physics sector, the wider kinematic reach of LHC has given qualitatively novel access to quarkonium suppression and jet-quenching phenomenology, including precision measurements of bottomonium, novel observations of charmonium enhancement, and a rich phenomenology of calorimetrically defined jets and jet substructure in nuclear collisions.

LHC experiments have also led to surprises that pose significant novel challenges for the understanding of \pA and \AOnA collisions. Most notable in this context is the discovery that flow-like phenomena are not limited to nucleus--nucleus collisions but they persist with significant magnitude in \pA collisions and in high-multiplicity \pp collisions at the LHC and some of their signatures have been observed even in minimum-bias \pp collisions. However, \pp collisions are typically expected to show vanishingly small re-interaction rates between produced final-state particles. In contrast, the perfect-fluid paradigm, that underlies the successful phenomenology of flow-like phenomena in nucleus--nucleus collisions implicitly, assumes that re-interaction rates are so large that even the notions of quasi-particle and mean-free path become meaningless. Does the persistence of flow-like phenomena in \pA and \pp collisions indicate, in contrast to previous belief, that the perfect-fluid paradigm applies to these smaller collision systems? Or, if the prefect-fluid paradigm is not applicable to \pp and \pA collisions, is it conceivable that significant corrections to a fluid-dynamic picture of vanishing mean free path persist also in the larger \AOnA collision systems? The LHC discoveries in \pp and \pA collisions that give rise to these questions provide arguably the strongest motivation for a future programme of detailed experimentation that aims at constraining microscopic structures and length-scales in the produced QGP matter and that is expected to clarify in this way the microscopic mechanisms underlying the apparent fluid-dynamic behaviour of \pp, \pA and \AOnA collisions.   

Historically, experimental heavy-ion programmes have always addressed a very diverse set of phenomenological opportunities. Some of the proposed experimental measurements have always reached out to other areas of science and could be clearly related to fundamental open question such as the origin of mass in the Universe, QCD deconfinement, or the determination of thermodynamic transport properties (that led in the past to unforeseen connections between string theory and the thermodynamics of quantum field theories). Other parts of the experimental programme were originally not related to a working hypothesis based on an open fundamental question, but they sometimes revealed themselves a posteriori as elements of crucial insight. This can be said for instance about the LHC \pA programme, that was not part of the original LHC design, that was first conceived mainly as a set of benchmark measurements for establishing the cold nuclear matter baseline for interpretation of heavy-ion data, and that has resulted in one of the most surprising discoveries made in the LHC nuclear-beams programme. We therefore emphasize that heavy-ion physics at the LHC, in the future as well as in the past, is likely to have multiple ways of reaching out and contributing to physics at large. At the time of writing this report, questions about the origin of collectivity in small \pp and \pA collision systems, and their implications for the interpretation of collective phenomena observed in \AOnA collisions, are arguably identified as the most pressing conceptual issue in the scientific debate. As outlined so far, they are clearly related to further experimentation with soft processes, and to research on the internal structure of QGP matter utilizing hard processes.  However, future experimentation at the LHC is not limited to this set of questions. From improved constraints on nuclear parton distribution functions that may inform us about the physics reach of future electron--ion facilities, to improved measurements of anti-nuclei, to ultra-peripheral collisions of electromagnetic Weizs\"acker-Williams photons at unprecedented field strength, to the search for qualitatively novel signatures of ultra-strong QED magnetic fields, the LHC nuclear beams programme can provide new insight in a much broader range of subject areas. 

As detailed in this report, the HL/HE-LHC physics working group 5 has identified future physics opportunities for high-density QCD with ions and proton beams that can be grouped broadly into the following four goals that are coming now within experimental reach:

\begin{enumerate}
\item Characterizing the macroscopic long-wavelength QGP properties with unprecedented precision. 
\item Accessing the microscopic parton dynamics underlying QGP properties.
\item Developing a unified picture of particle production from small (\pp) to larger (\pA and \AOnA) systems.
\item Probing parton densities in nuclei in a broad $(x,\,Q^2)$ kinematic range and searching for the possible onset of parton saturation.
\end{enumerate}
 
In the following, we summarize how the four general goals are addressed by the measurements discussed in this report. As the density and temperature attained in hadronic collisions changes mildly but distinctly with the centre-of-mass energy, the physics opportunities listed below can be further enhanced by a combined interpretation of future measurements at the LHC and at RHIC. In particular, new or upgraded experiments at RHIC aim at providing measurements of highly-improved quality and precision in the sectors of hard probes (jets and jet correlations, heavy flavour, quarkonia) and electromagnetic observables.

\subsection{Macroscopic QGP properties}
At sufficiently long wavelength, essentially all forms of matter can be described by fluid dynamics. The observation of flow-like behaviour in nucleus--nucleus collisions and in smaller collision systems demonstrates that this universal long-wavelength limit of hot and dense QCD matter can be accessed experimentally at the LHC. This provides an experimental inroad to fundamental questions about QCD thermodynamics and hydrodynamics since i) the QGP properties entering a fluid dynamic description are calculable from first principles in quantum field theory, and ii) hydrodynamic long wavelength properties depend on the effective physical degrees of freedom in the plasma and they are thus sensitive to the microscopic dynamics that governs their interactions. The following properties of the QCD matter produced in TeV-scale collisions are accessible via future measurements at the LHC

\begin{enumerate}
\item {\it Temperature}\\
Within the programme of determining the QCD equation of state, QCD lattice simulations at finite temperature have established since long a precise relation between the QCD energy density and pressure that determine the fluid dynamic expansion, and the temperature of QCD matter.  While energy density and pressure can be constrained experimentally by many measurements, an independent determination of temperature is of great value for testing the idea of local equilibration in heavy-ion collisions or for establishing deviations thereof. Future LHC experiments will constrain the temperature and its time evolution with unprecedented precision, in particular via thermal radiation of real and virtual (dileptons) photons (Chapter~\ref{chapter:electromagnetic_radiation}).
\item {\it QCD phase transition at $\mu_B \simeq 0$}\\ Collisions at the LHC realize systems of close-to-zero baryo-chemical potential. QCD calculations on the lattice predict in this regime a smooth cross-over transition from a hot partonic plasma to a cold hadron gas. Fluctuation measures of conserved charges are sensitive to the characteristics of the phase transition. In future LHC experiments, they are accessible with unprecedented precision and completeness (Chapter~\ref{sec:lf}). Future measurements of low-mass dileptons will also be, for the first time at the TeV-scale, sensitive to in-medium modifications of the $\rho$-spectral function (Chapter~\ref{chapter:electromagnetic_radiation}). This provides unique access to the transition between phases with restored and broken chiral symmetry that is predicted by lattice QCD.   
\item {\it Viscosity and further QCD transport coefficients}\\ Existing flow measurements provide tight upper bounds on the value of $\eta/s$ and they have been a cornerstone in supporting the perfect-fluid paradigm. In the future, measuring higher-order cross-correlations of flow coefficients will significantly extend this line of research.  In addition, we note that the value of $\eta/s$ can be related to the existence and size of the mean free path (isotropization length scale), which in turn results from the existence of (quasi-)particle-like excitations in the produced matter. This motivates increasing the precision on $\eta/s$ with the aim of establishing the tightest lower bound on this quantity (Chapter~\ref{sec:flow}), and it motivates detailed studies of flow in smaller systems with the idea of identifying the scale at which the system size becomes comparable to the mean free path (Chapter~\ref{chapter:smallsystems}). Also, measurements of soft dileptons will give access to the electric conductivity of the strongly interacting medium (Chapter~\ref{chapter:electromagnetic_radiation}), and flow and correlation measurements may help to constrain the bulk viscous corrections (Chapter~\ref{sec:flow}).
\item {\it Heavy-quark transport coefficients}\\ Heavy quarks provide unique tools for testing collective phenomena in nuclear collisions. As they are produced in initial hard-scattering processes and flavour is conserved throughout the collective dynamical evolution, they are the best experimental proxy to the idea of putting coloured test charges of well-defined mass into the medium and testing how they participate in the evolution. Of particular interest are precision measurements of transverse-momentum anisotropies $\vn$ and nuclear-modification factors of open heavy-flavoured mesons that are known to constrain e.g.\,the heavy-quark diffusion coefficient $2\pi T D_s$ and its dependence on the temperature $T$, that can be compared to first-principle calculations of QCD on the lattice (Chapter~\ref{sec:HI_HF}).
\item {\it Searching for transport phenomena related to the presence of strong electrodynamic fields.}\\ Heavy-ion collisions produce the largest electromagnetic field of any system accessible to laboratory experiments. The field is largest in the early phase of the collision, thus the early-produced heavy quarks are expected to be the most sensitive to its strength (Chapter~\ref{sec:HI_HF}).
As the maximal field strengths are estimated to be of the order of the pion mass ($e\, B^2 \sim m_\pi^2$), effects are also likely to be present for light-flavour charged hadrons (Chapter~\ref{sec:flow}). Other measurements of interest include transport coefficients such as the electric conductivity with which the plasma responds to an electromagnetic field, and that are calculable within QCD. In addition, as a consequence of the chiral anomaly, QCD coupled to QED gives rise to {\it anomalous hydrodynamics} that displays various qualitatively novel phenomena, including for instance a component of the electromagnetic currents that flows parallel to the magnetic field. The existence of these anomalous phenomena follows from first principles in field theory and thermodynamics, but the size of potential experimental signatures is model-dependent.  Beyond determining conventional QED transport phenomena, LHC allows to search with increased precision for these intriguing signatures of anomalous fluid dynamics (Chapter~\ref{sec:flow}). 
\end{enumerate}

\subsection{Accessing the inner workings of hot QCD matter}
Previous experiments at the LHC and at lower centre-of-mass energy have established that the QCD matter produced in nucleus--nucleus collisions is subject to strong collective evolution. However, the nature of the effective constituents of that matter, and its characteristic inner length scales (such as screening lengths or mean scattering times, if any) are not yet understood. The scale dependence of QCD implies that one must be able to resolve partonic constituents of hot QCD matter at sufficiently high resolution scale. This motivates the use of high-momentum transfer processes (hard probes) to study the inner workings of hot QCD matter. In addition, the current status of phenomenological modelling does not exclude the existence of a sizeable mean free path of hot QCD matter (which is assumed in transport model simulations of heavy-ion collisions, but which is not assumed in almost perfect fluid dynamic models). This motivates to learn about the inner workings at hot QCD matter also from particle production at intermediate transverse momentum (such as heavy quark transport at intermediate and low $\pT$). Here, we highlight the following opportunities for further experimentation.
\begin{enumerate}
\item {\it Constraining with jet quenching the colour field strength of the medium}\\ In general, the fragments of jets produced in nucleus--nucleus collisions are medium-modified due to  interactions with the hot QCD matter. These jet quenching effects depend on the inner structure of that matter. In particular, the average medium-induced colour field strength experienced by the escaping jet can be quantified e.g.\,with the quenching parameter $\hat q$, which measures the average exchanged transverse momentum squared per unit path length. Experiments at the LHC will provide improved constraints on this field strength measurement (Chapter~\ref{sec:HI_Jets}).  Qualitatively novel opportunities for testing the time evolution of the medium opacity to hard partons could arise if boosted tops could be studied in nuclear matter. A run at the LHC with lighter (than $\mathrm{Pb}$) nuclei, like e.g.\,$^{40}\mathrm{Ar}$, would provide sufficient luminosity to this end, as well as largely enhanced kinematic and statistics reach for $\gamma$--jet and Z--jet recoil measurements (Chapter~\ref{sec:smallAsum}). The opportunities for boosted top measurements in \PbPb and lighter nuclei collisions at the HE-LHC are also discussed (Chapter~\ref{sec:HELHC}).
\item {\it Investigating the quasi-particle structure of QCD matter with jet and heavy-quark measurements}\\ 
Hard quarks and gluons with different energies can be used to investigate the constituents of the QCD matter at various resolution scales. On one side of the scale, high-energy quarks and gluons (leading to high-energy jets) address the smallest spatial scale.
While $\hat q$  characterizes the effects of jet--medium interactions in the coherent regime in which individual constituents in the medium are not resolved, Z/$\gamma$--jet correlations and modern jet substructure measurements on the high-statistics samples of future LHC Runs are expected to access a regime of Rutherford-type large angle jet--medium scattering, in which the detection of recoil or of large angle deflections gives insight into the microscopic structure of the produced matter (Chapter~\ref{sec:HI_Jets}). On the other side
of the scale, the scattering of low-momentum heavy quarks, characterized using for example the diffusion coefficient,
addresses the nature of the QCD constituents at long-wavelength scale (Chapter~\ref{sec:HI_HF}).
\item {\it Testing colour screening with bottomonium production}\\The family of bottomonium bound states gives access to a set of well-defined length scales that are embedded in the hot and dense QCD matter. As a consequence, the dissociation of the various bound states sensitively depends on the interplay of color screening and the coupling to the strongly interacting medium through dissociation reactions. This opens a unique window on the in-medium modifications of the fundamental QCD force that binds bottomonium. Increasingly-tight bound states are expected to melt with increasing temperature, providing a laboratory for in-medium spectroscopy. The increased precision of future measurements will allow to test for additional physics mechanisms in the production of bottomonium, such as regeneration processes that may affect the yield of \PGUP{2S} and \PGUP{3S}. Also, the higher rates of future experiments will allow one to cross-correlate measurements of bottomonium suppression with other manifestations of collectivity, such as elliptic flow (Chapter~\ref{sec:quarkonia}).
\item {\it Testing colour screening and regeneration dynamics with charmonium production}\\ In close similarity to bottomonium, the medium-modification of charmonium bound states is sensitive to colour screening and it is subject to the same QCD dissociation dynamics. However, since charm quarks and anti-quarks are produced abundantly in nucleus--nucleus collisions at the TeV scale, c and $\mathrm{\bar{c}}$ produced in different hard processes can form bound states. Indications of this qualitatively novel bound-state formation process are accessible at low and intermediate transverse momentum, and they motivate high-precision measurements of nuclear modification factor $\RAA$ and elliptic flow $\vtwo$. In addition, future open heavy-flavour measurements reaching down to zero $\pT$ with high precision will help to determine the total charm cross section which is a central input for the precise quantification of regeneration (Chapters~\ref{sec:HI_HF} and \ref{sec:quarkonia}). 
\item {\it Formation of hadrons and light nuclei from a dense partonic system}\\ The question of how a collective partonic system of many degrees of freedom evolves into the hadronic phase and produces colour singlet hadrons and light nuclei is essential for a complete dynamical understanding of nucleus--nucleus collisions. Recent LHC measurements in proton--proton and proton--nucleus collisions suggest that also in small-system hadronic collisions the hadronization process may be modified with respect to elementary \Pepem collisions. 
Future measurements at the LHC will enable comparative and multi-differential studies of these modifications with unprecedented precision, for both the heavy-flavour sector ($\PDps$, $\PBs$, charm and beauty baryons, see Chapter~\ref{sec:HI_HF}) and for light nuclei and hyper-nuclei (Chapter~\ref{sec:lf}). For open heavy flavour, these measurements are also crucial to disentangle the role of mass-dependent radial flow and of recombination, as well as to constrain the parameters of hadronization in the models that are used to estimate QGP properties like the heavy-quark diffusion coefficients. For light nuclei, precise measurements of nuclei and hyper-nuclei with mass numbers 3 and 4 as well as possible observation of exotic baryonic states will address the question whether their production is dominated by coalescence of protons, neutrons and $\Lambda$ baryons or by statistical hadronization of a partonic system. These measurements, in addition to that of high-momentum deuteron production, also have important astrophysical implications (dense compact stars as well as dark matter searches in the Cosmos).
\end{enumerate}

\subsection{Developing a unified picture of QCD collectivity across system size}
As discussed at the beginning of this chapter, recent LHC discoveries of signatures of collectivity in proton--nucleus and in proton--proton collisions question common beliefs. For the smallest collision systems, these measurements indicate that more physics effects are at work in multi-particle production than traditionally assumed in the modelling of proton--proton collisions. For the larger collision systems (proton--nucleus and nucleus--nucleus collisions), they question whether the origin of signatures of collectivity is solely (perfect) fluid dynamical, given that these signatures persist in proton--proton collisions. This raises important qualitative questions like: What is the smallest length scale on which QCD displays fluid-dynamic behaviour? Is there a non-vanishing characteristic mean free path for the production of soft and intermediate $\pT$ hadrons, and if so, is it smaller or larger than the proton diameter? What are the novel physics concepts with which underlying event simulations in proton--proton collisions need to be supplemented (e.g. in multi-purpose event generators) to account for the totality of observed phenomena?  While some of these questions sound technical, it needs to be emphasized that the size of dissipative properties of QCD matter, such as its shear viscosity, are quantitatively related to the presence or absence of intrinsic length scales such as a mean free path. Any systematic experimental variation of the system size therefore relates directly to a search for intrinsic length scales that determine the dissipative properties of hot QCD matter. Within the present report, we identify in particular the following future opportunities for an improved understanding of the system-size dependence of collective phenomena (Chapter~\ref{chapter:smallsystems}):

\begin{enumerate}
\item {\it Flow measurements in \pp and \pA systems: Onset and higher-order correlations}\\ While flow signals have been established in smaller collision systems in recent years, their detailed characterization lacks behind the state of the art achieved in nucleus--nucleus collisions. Future measurements will allow for characterizing higher-order cumulants in largely non-flow suppressed multi-particle correlations, and test whether there is a system-size dependence in the characteristic correlations between different flow harmonics \vn, or the characteristic reaction plane correlations. 
\item {\it Flow of heavy flavour and quarkonium in smaller systems}\\  Is there a minimal system size needed to transport heavy quarks within a common flow field? Given that the local hard production of heavy flavour is expected to be independent of any collective direction, precision measurements of heavy-flavour flow in \pp and \pA collisions will provide decisive tests of heavy quark transport, thus experimentally addressing the question of how QCD flow field build up efficiently and on short length and time scales.
\item  {\it Strangeness production as a function of system size}\\ One of the recent surprising LHC discoveries that is not accounted for in traditional models of minimum-bias \pp collisions but that may be accounted for in a thermal picture is the smooth increase of strangeness with event multiplicity across system size. We discuss in detail how future measurement at the LHC, such as the study of strange D-mesons or baryons, can extend the systematics underlying this observation (Chapters~\ref{sec:HI_HF} and Section~\ref{sect:smallsystems_strangeness}).
\item {\it Searching for the onset/existence of energy-loss effects in small systems}\\ All dynamical models of collectivity involve final-state interactions. This implies the existence of jet--medium final-state interactions and, a fortiori, the existence of parton energy-loss effects. The latter have not been identified experimentally, yet. In this report, we discuss novel opportunities to test for their existence, including tests in future \pPb collisions, as well as opportunities specific to \OO collisions.
\item {\it Searching for the onset/existence of thermal radiation in small systems}\\ If the collectivity observed in smaller collision systems is due to final-state isotropization and equilibration phenomena, it must be accompanied by thermal radiation. The search for the corresponding conceptually clean electromagnetic signatures, such as thermal dilepton and photon production in \pPb collisions, is an important part in developing a unified  picture of QCD collectivity. This report discusses the experimental opportunities in light of experimental upgrades.   
\end{enumerate}

\subsection{Nuclear parton densities and search for non-linear QCD evolution}
Future experiments at the LHC offer a variety of opportunities for precision measurements with nuclear beams. Here, we highlight three opportunities that are clearly related to the main physics challenges of the heavy-ion programme (Chapter~\ref{sec:smallx}):

\begin{enumerate}
\item {\it Precise determination of nuclear PDFs at high $Q^2$}\\ As high-momentum transfer processes have a short space-time scale, they are not affected by the long-wavelength particle excitations of the QCD matter in nuclear collisions. This implies that the primary production rates of hard processes in nucleus--nucleus collisions are determined perturbatively, whereas their medium modifications arise from traversing a dense QCD matter of considerable spatial extent and considerable colour-field strength. For  a dynamical understanding of jet quenching, control over primary-production rates is indispensable and this necessitates the knowledge of nuclear parton distribution functions. Global nPDF-fits that reflect the current state of the art of nuclear parton distributions could be improved at the LHC in the near future at high $Q^2$ and $x\sim 10^{-3}$--$10^{-2}$ in particular with high-precision W, Z and dijets measurements in \pPb and \PbPb collisions (Sections~\ref{sec:pPbex} and~\ref{sec:nPDFs}). High-luminosity \ArAr collisions would enable for the first time using top quarks to constrain nuclear PDFs at very high $Q^2$ and large $x\sim 10^{-2}$--$10^{-1}$ and would contribute to constraining experimentally the nuclear mass number $A$ dependence of nuclear PDFs. 
\item {\it Constraining nuclear PDFs at low $Q^2$}\\ Drell-Yan and photon measurements (Section~\ref{sec:pPbex}) in \pPb collisions could provide significantly improved constraints on the nuclear parton distribution functions at low $Q^2$ and low $x\sim 10^{-5}$--$10^{-3}$, where nuclear effects are larger. In addition, although so far global PDFs do not use measurements from ultra-peripheral collisions (UPC), it is thought that quarkonia and dijet production in UPC can constrain nuclear PDFs in the future. This report identifies opportunities for the corresponding measurements (Section~\ref{sec:UPCqcd}). 
\item {\it Access to non-linear QCD evolution at small-x}\\ The scale-dependence of parton distribution functions is known to obey linear QCD evolution equations within a logarithmically wide range in $Q^2$ and $\ln x$. However, where partonic density in the incoming hadronic wave function are not perturbatively small, qualitatively novel non-linear density effects are expected to affect the QCD evolution. For $Q^2$ smaller than a characteristic saturation scale $Q^2_s(x)$, these effects are dominant, and as the saturation scale $Q^2_s(x)$ increases with decreasing $x$, one expects on general grounds at sufficiently small $\ln x$ a qualitatively novel saturation regime in which non-linear QCD evolution occurs at perturbatively large $Q^2$. Future measurements at LHC will provide novel test for these saturation effects with previously unexplored measurements (Section~\ref{sec:pPbex}). The larger centre-of-mass energy of the HE-LHC would extend the small-$x$ coverage by an additional factor two. Measurements of relevance include in \pPb collisions dilepton and photon production at small-$x$ and forward measurements of dihadron and dijet correlations. The perspectives for such measurements are discussed in Chapter~\ref{sec:HELHC}. 
\end{enumerate}

\subsection{Physics performance studies by ALICE, ATLAS, CMS and LHCb}

The performance results presented in this report were obtained by experimental groups within the four Collaborations
and they are described in more detail in these documents: ALICE~\cite{ALICE-PUBLIC-2019-001,Abelevetal:2014cna,Abelevetal:2014dna,CERN-LHCC-2015-001}, ATLAS~\cite{ATL-PHYS-PUB-2018-019,ATL-PHYS-PUB-2018-020,ATL-PHYS-PUB-2018-039}, CMS~\cite{CMS-PAS-FTR-17-002,CMS-PAS-FTR-18-024,CMS-PAS-FTR-18-025,CMS-PAS-FTR-18-026,CMS-PAS-FTR-18-027}, LHCb~\cite{LHCb-CONF-2018-005}. Two types of results are included: {\it simulation} studies used full or fast simulations of the detector setups for Run 3 and/or Run 4; {\it projection} studies are based on existing measurements where their uncertainties have been reduced as expected with the future detectors and data samples. The impact of the detector upgrades on the various observables is discussed in the corresponding chapters and in more detail in the referenced documents by the Collaborations. The integrated luminosities used for the physics studies are summarised in Chapter~\ref{sec:schedule}.

\clearpage

\section{Heavy-ion performance of LHC, HL-LHC and HE-LHC}

{ \small
\noindent \textbf{Coordinator}: John M. Jowett (CERN)

\noindent \textbf{Contributors}: 
R.~Bruce (CERN), 
M.~Schaumann (CERN),  
M.A.~Jebramcik (Johann-Wolfgang-Goethe Universit\"{a}t, Frankfurt \& CERN) 
}

\subsection{Heavy-ion performance of LHC in Runs~1 and 2}

The 2018 \PbPb run of the LHC brought Run~2 to an end and launched the hardware upgrades to the collider,
and to the ALICE experiment, that should allow the full  
``HL-LHC'' heavy-ion performance to be delivered from 2021 onward.
Beyond \pp collisions, the 2004 LHC Design Report~\cite{lhc1}, specified only    
\PbPb collisions   with a peak \PbPb luminosity of \lumival{1}{27}.   
Now, much  of the upgraded performance is already in hand.
Not only has that peak \PbPb luminosity goal been exceeded by a factor of more than 6, but the \pPb collision mode---an upgrade beyond the initial design whose   feasibility was widely doubted---has yielded similarly high luminosity
in multiple operating conditions 
(see Ref.~\cite{Jowett:2018yqk} and references therein). 
Table~\ref{tab:LHCparamsPbPb} summarises the main parameters of the runs to date.
Additionally, in 2017, the LHC has collided beams of Xe nuclei \cite{Schaumann:2018qat},
providing many   new results and 
demonstrating the potential for colliding  lighter species.
The goal for 2018 was to complete the accumulation
of an integrated \PbPb\
luminosity of \qty{1}{nb^{-1}} to
each of the ALICE, ATLAS and CMS experiments and this was substantially exceeded. 
The LHCb experiment also received over \qty{0.26}{nb^{-1}}. 
	
\begin{table*}[t]
\caption{Representative simplified beam parameters 
at the start of the highest luminosity   physics   fills, 
in conditions that lasted for $>\qty{5}{days}$,   
in each annual \PbPb\  run~(Ref.~\cite{Jowett:2018yqk} and references therein).
The original design values for \PbPb~\cite{lhc1} 
  collisions 
and future upgrade \PbPb goals are also shown 
(in this column the integrated luminosity goal is to be attained over the 4~\PbPb\  
runs in the
{10-year periods} before and after 2020).
Peak  luminosities are averages for ATLAS and CMS 
(ALICE being levelled).
The smaller luminosities delivered to LHCb from 2013--2018  are not shown.
Emittance and bunch length are RMS values. 
The series of runs with $\sqrt{\sNN}=\qty{5.02}{TeV}$ also included \pp\ reference runs, not shown here.
Design and record achieved nucleon-pair luminosities are \protect\fbox{boxed}, and some key parameters related to \pPb\ parameters in Table~\ref{tab:LHCparamspPb} are 
set in \textcolor{red}{red type}, for easy comparison. 
The upgrade peak luminosity is reduced by a factor $\simeq3$ from its potential value by levelling.
\label{tab:LHCparamsPbPb}}	
\centering
    {
{\renewcommand{\arraystretch}{1.4}
\newcolumntype{L}[1]{>{\raggedright\let\newline\\\arraybackslash\hspace{0pt}}m{#1}}
\begin{tabular}{L{6cm}|c|cccc|c@{}}
\hline                   
  Quantity                        &  {design} & \multicolumn{4}{c|}{achieved} & upgrade \\
  \hline
Year                                & (2004)   & 2010 & 2011  & 2015  & 2018 & $\ge$2021 \\
Weeks in physics & -  & 4 & 3.5  & 2.5  & 3.5 &-\\
Fill no. (best)  & & 1541 & 2351  & 4720  &7473& -\\
Beam energy \qty{E[Z}{TeV]}  &  { 7 } & \multicolumn{2}{c}{3.5}   & 6.37   & 6.37 & 7 \\
Pb beam energy \qty{E[A}{ TeV]}  & {2.76} & \multicolumn{2}{c}{ 1.38 }   &   2.51   &2.51 & 2.76 \\
\raggedright Collision energy \qty{\sqrt{\sNN}}{[TeV]}  & 5.52 & \multicolumn{2}{c}{2.51}  & \textbf{5.02}&\textcolor{red}{\textbf{5.02}} & 5.52  \\
\hline
Bunch intensity \qty{N_b}{[10^8]}       &  {0.7} & 1.22 &1.07 & 2.0  &2.2 &1.8 \\
No.\ of bunches $k_b$             & 592 &137& 338     & 518   &733 &1232 \\
Pb norm. emittance \qty{\epsilon_N}{[\mu m]}  &  { 1.5 }  & 2.& 2.0   & 2.1   &2.0&   1.65   \\
Pb bunch length \qty{\sigma_{z}}{m}   &  {0.08} & \multicolumn{4}{c|}{0.07--0.1} &  0.08 \\
\qty{\bstar}{[m]}&        {0.5}& 3.5 & 1.0   & 0.8    &0.5 &0.5\\
Pb stored energy   MJ/beam    & 3.8   &0.65   & 1.9   & 8.6    &13.3&  21 \\
\hline
Luminosity \qty{\LAA}{[10^{27}cm^{-2}s^{-1}]}   & 1  &0.03  & 0.5 & 3.6   &\textcolor{red}{6.1}& 7   \\
NN luminosity \qty{ \LNN}{[10^{30} {cm}^{-2} {s}^{-1}]}   &\fbox{43}      & 1.3  & 22.  & 156   &\fbox{264}& \fbox{303}  \\
Integrated luminosity/experiment [\qty{}{\mu b^{-1}}]& 1000   &  9  & 160   & 433,585   &\textcolor{red}{900,1800} & $10^4$ \\
Int. NN lumi./expt. [\qty{}{pb^{-1}}]& 43    &  0.38  & 6.7   & 19,25.3 &39,80 & \enum{4.3}{5} \\
 \hline
\end{tabular}
}

    }
 
\end{table*} 

\begin{table*}[t]
\caption{Representative simplified beam parameters 
at the start of the highest luminosity physics   fills, 
in conditions that lasted for $>\qty{5}{days}$,   
in the one-month   p--Pb runs~
(Ref.~\cite{Jowett:2018yqk} and references therein).
The very short pilot run in 2012 is not shown.
The original ``design'' values for  \pPb~\cite{Salgado:2011wc} collisions 
 are also shown 
(in this column the integrated luminosity goal  was supposed to be obtained over a few runs. 
Peak  luminosities are averages for ATLAS and CMS 
(ALICE being levelled).
The smaller luminosities delivered to LHCb from 2013--2016 and 
in the minimum-bias part of the run in 2016 are not shown.
Emittance and bunch length are RMS values. 
Single bunch parameters for these \pPb\ or \Pbp\ runs 
are generally those of the $\mathrm{Pb}$ beam.
Design and record achieved nucleon-pair luminosities are \protect\fbox{boxed}, and some key  parameters related to \pPb\ parameters in Table~\ref{tab:LHCparamsPbPb} are 
set in \textcolor{red}{red type}, for easy comparison. 
\label{tab:LHCparamspPb}}	
\centering
    { 
        
{\renewcommand{\arraystretch}{1.4}
\newcolumntype{L}[1]{>{\raggedright\let\newline\\\arraybackslash\hspace{0pt}}m{#1}}
\footnotesize
\begin{tabular}{L{6cm}|c|cc}
\hline                   
  Quantity                        & {``design''} & \multicolumn{2}{c}{achieved}  \\
  \hline
Year                            &(2011)    &2012--13  &2016   \\
Weeks in physics & -  &  3  & 1, 2  \\
Fill no. (best)  &   & 3544  &5562 \\
Beam energy \qty{E[Z}{TeV]}  &  { 7 } &   4    & 4,6.5  \\
Pb beam energy \qty{E[A}{ TeV]}  &  {2.76} &       2.51 & 1.58,2.56   \\
\raggedright Collision energy \qty{\sqrt{\sNN}}{[TeV]}  & 5.52  & \textcolor{red}{\textbf{5.02}}  & \textcolor{red}{\textbf{5.02},8.16}  \\
\hline
Bunch intensity \qty{N_b}{[10^8]}       &  {0.7}    & 1.2   &2.1  \\
No. \ of bunches $k_b$             & 592  & 358   & 540  \\
Pb norm. emittance \qty{\epsilon_N}{[\mu m]}  &  { 1.5 }    & 2.   & 1.6   \\
Pb bunch length \qty{\sigma_{z}}{m}   &  {0.08} & \multicolumn{2}{c|}{0.07--0.1}  \\
\qty{\bstar}{[m]}&        {0.5}  & 0.8   & 10, 0.6 \\
Pb stored energy   MJ/beam    & 3.8     &  2.77    & 9.7   \\
\hline
Luminosity \qty{\LAA}{[10^{27}cm^{-2}s^{-1}]}   & 150   & \textcolor{red}{116} &\textcolor{red}{850}    \\
NN luminosity \qty{ \LNN}{[10^{30} {cm}^{-2} {s}^{-1}]}   &\fbox{43}       & 24  & \fbox{177}  \\
Integrated luminosity/experiment [\qty{}{\mu b^{-1}}]  & \textcolor{red}{$10^5$}   & \textcolor{red}{32000}   &\textcolor{red}{\enum{1.9}{5}}  \\
Int. NN lumi./expt. [\qty{}{pb^{-1}}]  & 21   & 6.7  & 40  \\
 \hline
\end{tabular}
}

    }
 
\end{table*}

\subsection{\PbPb\ luminosity in Run~3 and Run 4 (HL-LHC)}  \label{sec:HLPbPb} \label{sec:LHCpresentbaseline}

The High Luminosity LHC (HL-LHC) is an upgrade of the LHC  to achieve instantaneous \pp\ luminosities a factor of five larger than the LHC nominal value.  
Its operational phase is scheduled to start in LHC Run~4,  
in the second half of the 2020s, 
for the \pp physics programmes described in the other chapters of this report.   
The HL-LHC project also includes  
hardware upgrades  of the present LHC that will  allow the LHC to 
operate with   potential peak
\PbPb\ luminosities an order of magnitude larger than the nominal~\cite{lhc1}.  
These upgrades will be completed during Long Shutdown 2 and can already be exploited 
in Run 3, starting in 2021. 
Upgrades to the heavy-ion injector chain, in the framework of the 
LHC Injectors Upgrade project will increase the 
total stored intensity of heavy-ion beams
and will also be completed for Run~3.   
Finally,  the   ALICE experiment will   be upgraded to accept higher peak luminosity. 

The heavy-ion performance of the LHC will be similar in Run~3 and in Run~4. Therefore, 
the two Runs are discussed together in this report in terms of their contribution to the HL-LHC 
heavy-ion physics programme. 
To achieve the performance parameters given in the last column of Table~\ref{tab:LHCparamsPbPb} 
a detailed specification of the requirements on the beams at LHC injection 
has been given~\cite{HLLHCPbPbspec} and later updated in Ref.~\cite{cham2017:jowett}. 
In a typical one-month \PbPb run, this will yield an integrated luminosity of \qty{~3.1}{nb^{-1}}.  
The necessary single-bunch intensities have already been attained but an implementation of 
slip-stacking in the SPS  will be required to obtain a basic bunch spacing of \qty{50}{ns} and 
store over 1200 $\mathrm{Pb}$ bunches in each LHC ring.  
The necessary upgrades of the SPS RF system will be implemented during LS2 and it is planned to commission 
this  new mode of operation  in 2021.

\subsubsection{Secondary beams from the IPs}

Ultra-peripheral electromagnetic interactions of Pb nuclei lead to copious 
lepton-pair production. Most of this is innocuous except for the (single) bound-free pair production (BFPP1):
\begin{equation}
^{208}\mathrm{Pb}^{82+} + ^{208}\mathrm{Pb}^{82+} \longrightarrow ^{208}\mathrm{Pb}^{82+} + ^{208}\mathrm{Pb}^{81+} +  e^+,
\end{equation}
in which the electron is bound to one nucleus.
As extensively discussed in e.g. Refs.~\cite{Klein:2000ba,cham2003,PhysRevSTAB.12.071002} and elsewhere, the modified nuclei emerge from the collision point, as a narrow secondary beam with modified magnetic rigidity, following a dispersive trajectory that impacts on the beam screen in a superconducting magnet in the dispersion suppressor (DS) downstream.  
These secondary beams emerge in both directions from every interaction point (IP) where ions collide. 
Each carries a power of  
 \begin{equation}
 \mathrm{P_{BFPP}} = L \sigmaBFPP\Eb,
 \end{equation}
 where $L$ is the luminosity and  $\sigmaBFPP \simeq \qty{276}{b}$  is  the cross-section at the 2015/18 run energy of $\Eb=\qty{6.37Z}{TeV}$.
These losses carry much greater power than the luminosity debris
(generated by the nuclear collision cross-section of \qty{8}{b})
and can quench magnets and directly limit luminosity. 
With a peak luminosity of \lumival{6.1}{27} each secondary beam carries $\mathrm{P_{BFPP}} \lesssim \qty{120}{W}$, which is more than enough to quench an LHC dipole as demonstrated in 2015~\cite{ipac2016:BahamondeCastro:2207361}.

To reduce the risk of quenching these magnets, orbit bumps were implemented around the impact locations in IP1 and IP5 in order to move the losses out of the dipole and into the adjacent connection cryostat (``missing dipole'' in the DS) that does not contain a superconducting magnet coil and therefore is less likely to quench. This technique was first used in 2015.   
It was almost fully proved in 2018 when the ATLAS and CMS \PbPb 
luminosities were sustained at  values very close to the nominal levelling values for Runs 3 and 4. Beam-loss monitor thresholds were set, based on the measured quench level in 2015 and it was clear that there was sufficient margin for still higher luminosity. 
In IP2, the method of orbit bumps alone is not applicable with present optics and layout. It is therefore foreseen to install an additional collimator in the connection cryostat on the outgoing beam on each side of IP2. In combination with this, orbit bumps will then be deployed to steer the BFPP beams onto the collimators. The installation will take place in LS2 in order to allow the HL-LHC design luminosity for ALICE (corresponding to a hadronic event rate of \qty{50}{kHz}) in subsequent runs.

\subsubsection{Collimation and intensity limit}

While the LHC stores unprecedented beam energies, superconducting magnets are needed to bend and focus these beams, most of which are operated at 1.9~K. A loss of a tiny fraction of the beam is enough to induce a magnet quench, and it is therefore vital to avoid any uncontrolled beam losses. To safely intercept losses and provide protection of the magnets, the LHC uses a multi-stage collimation system~\cite{assmann05chamonix,guillaume-thesis,assmann06,chiara-thesis}. During the first two runs of the LHC, this system has shown a very good performance with proton beams~\cite{bruce14-PRSTAB-sixtr,valentino15-evian,mirarchi16-evian,bruce17-NIM-beta40cm} and ion beams~\cite{pascal-thesis,hermes16-nim}. 

LHC collimation is much less efficient with heavy-ion beams than with protons, since ions have a high cross section for undergoing nuclear fragmentation inside the primary collimators~\cite{epac2004:Braun:IonCollimation}. The angular offsets of the out-scattered fragments are frequently not large enough to reach the secondary collimators in the straight collimation insertion (IR7).  At the same time these fragments have a magnetic rigidity different from the main beam, so that they risk being lost where the dispersion starts to rise in the first few dipoles of the DS. This was the most critical beam loss location during the $\mathrm{Pb}$ ion runs in Run~1 and Run~2, with a local cleaning inefficiency of about a factor~100 worse than for protons~\cite{pascal-thesis,hermes16-nim}. Therefore, even though the total stored beam energy is about a factor 10 lower with $\mathrm{Pb}$ ions than with protons, collimation of heavy ions is critical. Still, ion collimation has worked well in the LHC and did not introduce operational bottlenecks so far.
However, extrapolations of the losses in the DS from a 2015 experimental tests to Run~3 and HL-LHC show that, if nothing is done, the total stored $\mathrm{Pb}$ beam energy is limited to around 10~MJ, if drops of the instantaneous beam lifetime down to 12~minutes are assumed~\cite{hermes16-ion-quench-test}.  
At the same time, it is foreseen to increase the stored $\mathrm{Pb}$ beam energy to about 24~MJ. To alleviate this limitation and safely intercept the losses, it is planned to install additional collimators, called TCLDs, in the dispersion suppressors~\cite{bruce14ipac-DS-coll,lechner14ipac-DS-coll,hl-lhc-tech-design}.  
On the other hand, the LHC was successfully operated with $\mathrm{Pb}$ beams containing over 13.5 MJ each in 2018 thanks to good control of beam lifetimes. 
In order to make space for the TCLDs, standard 8.3~T LHC dipoles will be replaced by an assembly consisting of two shorter higher-field 11~T dipoles with the TCLD in between~\cite{hl-lhc-tech-design}. The solution that gives the best simulated $\mathrm{Pb}$ cleaning efficiency uses two TCLDs per side of IR7. However, this is not possible within tight constraints of long shutdown~2 and the HL-LHC project, and the baseline is therefore to install one TCLD per side. If this turns out to be a real limitation, it could be considered at a later stage to install a second TCLD. 
As an alternative and complementary alleviation method, it is under study whether crystal collimation could help in reducing the losses in the DS. In this collimation scheme, bent crystals are used instead of the standard LHC primary collimators~\cite{daniele-thesis}. Incoming beam particles follow the curvature of the crystal planes, the so-called channelling, and exit with a significant angular kick. They can then be efficiently steered onto an absorber. Nuclear interactions inside the the channels of well-aligned crystals are significantly suppressed. Initial experiments using an LHC test installation~\cite{Mirarchi2017_crystals} have shown very promising results with $\mathrm{Xe}$ and proton beams~\cite{scandale16}.  Channelling has very recently also been observed with $\mathrm{Pb}$ beams in 2018 and potential improvements of the collimation system are presently being assessed experimentally. Studies with $\mathrm{Pb}$ beams are not yet conclusive but it is hoped that this will be further clarified by analysis of data taken during the 2018 $\mathrm{Pb}$ ion run. 

Collimation of lighter ion species has not yet been studied in detail, although some first simulations are presented in Ref.~\cite{pascal-thesis}. Results for $\mathrm{Ar}$ and $\mathrm{Xe}$ beams show that the amount of expected losses in the DS is similar to $\mathrm{Pb}$ but the longitudinal loss distribution changes. The fractional change in magnetic rigidity for every lost nucleon in the collimators is larger  for light ions, and it is hence expected that out-scattered fragments have larger effective energy deviations and are lost more upstream. It is thus likely that the  TCLD should help significantly also for lighter ions, although comparative studies on intensity limits for different ion species still remain to be done.

\subsection{Proton-lead operation in Run~3 and HL-LHC}

Within colliding nuclei,
with charges $Z_1$, $Z_2$
and nucleon numbers ${A_1}$, ${A_2}$,
in  rings with magnetic field set for protons of momentum $p_p$\footnote{Conditions imposed by the two-in-one magnet design of the LHC.},
the  colliding nucleon pairs will have an average centre-of-mass energy
\begin{equation}\label{eq:sNN}
\sqrt{\sNN}  \approx 2c\,{p_p}\sqrt{\frac{{{Z_1}{Z_2}}}{{{A_1}{A_2}}}}
\approx 2c\,{p_p}
\begin{cases}
1 & \text{\pp}\\
0.628 & \text{\pPb}\\
0.394 & \text{\PbPb}
\end{cases}
\end{equation}
and a central rapidity shift in the direction of the $(Z_1,A_1)$ beam   
\begin{equation} 
\yNN \approx \frac{1}{2}\log \left( \frac{{{Z_1}{A_2}}}{{{A_1}{Z_2}}}\right)
\approx
\begin{cases}
0 & \text{\pp, \PbPb}\\
0.465 & \text{\pPb} \\
-0.465 & \text{\Pbp}
\end{cases}. 
\end{equation} 
We present parameters for operation at the nominal LHC momentum
\(p_p c=\qty{7}{TeV} \)
extrapolating from the experience of the last \pPb run in 2016.

The injection and ramp of protons and lead ions with equal magnetic rigidity leads to moving long-range beam-beam encounters in the four interaction regions of the LHC. 
These beam-beam encounters were one of the reasons why the feasibility of \pPb\ operation in the LHC was initially questioned. This effect has been proven small in the LHC and calculations have this will remain true 
for the HL-LHC era despite larger bunch numbers and higher proton bunch intensities.
The dynamic range of the interlock strip-line BPMs, common for the lead and proton beam, limited the proton intensity to $N_b<\enum{5}{10}$ protons per bunch during Run 1. 
Gating the stripline BPM read-out appropriately removed this constraint a few days before the end of the 2016 run. 
The higher proton intensity of 
$\Nb=\enum{2.8}{10}$ 
protons per bunch resulted in increased luminosities at the IPs but also led to the substantial deposition of collision debris from the Pb~beam in the dispersion suppressors at ATLAS and CMS risking a beam dump \cite{ipac2017:Jowett:2289686}. 
The   collision debris collimators (TCLs), which could have intercepted emerging fragments from the IPs, were not commissioned at tighter settings 
for the 2016 \pPb\ run. 
Appropriate  TCL settings are expected to neutralise these fragments and should allow for higher peak luminosities in the future.

\begin{figure}[!t]
    \centering
    \includegraphics[scale=.8]{\main/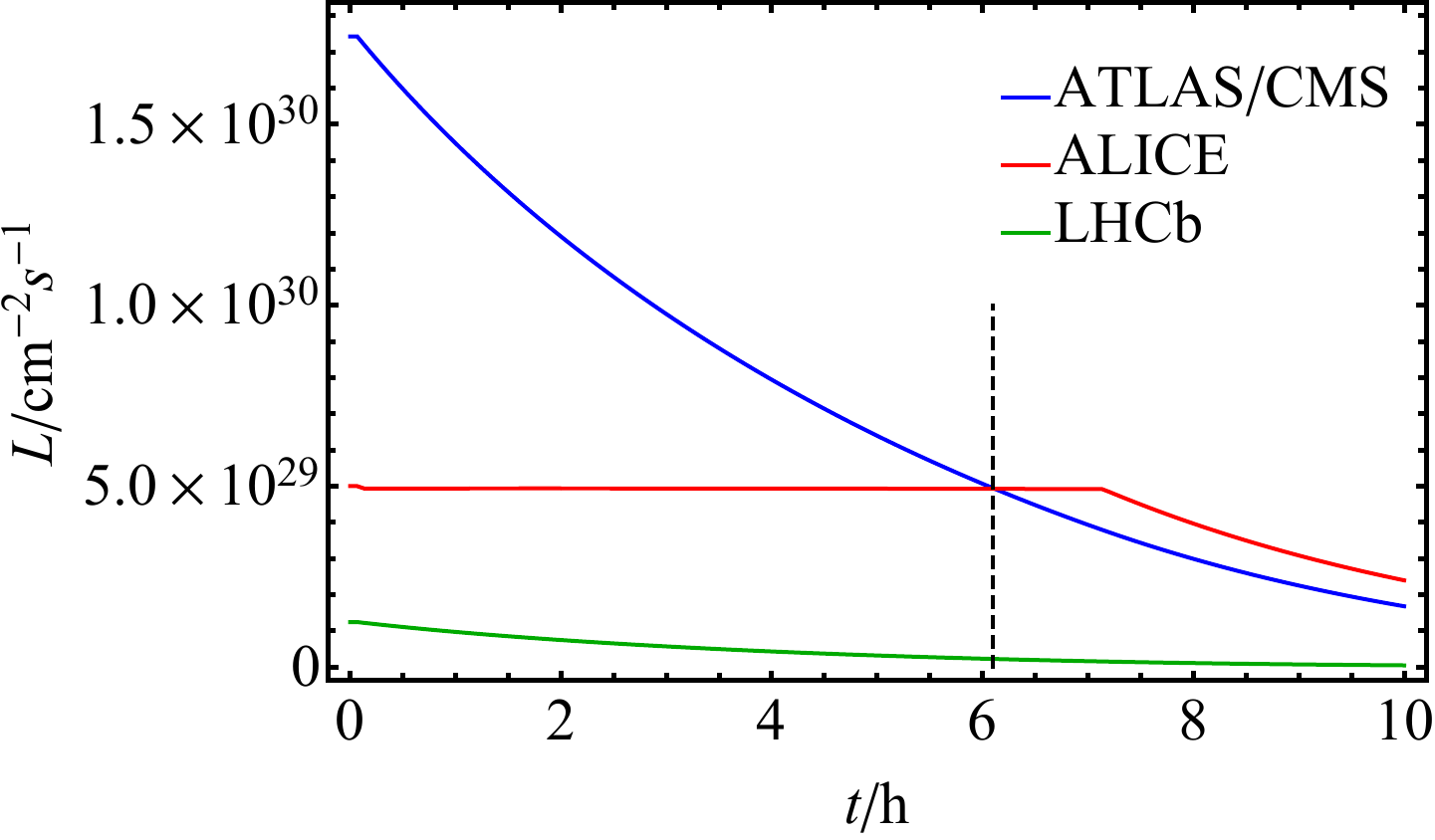} 
\caption{Evolution of the instantaneous luminosity in the LHC experiments during a p-Pb fill. 
At around \qty{6.1}{h} (dashed line), the interlock BPM threshold is  reached for some bunches, 
limiting the fill length.} 
\label{fig:p_pb_lumi_evo}
\end{figure}

A potential \pPb\ run during Run 3 and beyond will greatly benefit from the longitudinal slip stacking in the SPS and the small $\bstar=\qty{0.5}{m}$ in three experiments. 
The proton intensity cannot be pushed to values much larger than the maximum achieved in 2016 as bunches colliding in multiple IPs and especially in ATLAS and CMS will approach the interlock BPM threshold of \enum{2}{9} charges per bunch too quickly if the luminosities of ATLAS and CMS are not levelled.  
This would lead to an undesirable early beam dumps while ALICE is still levelled. 
In order to predict the potential performance of a future \pPb\ run, the expected \PbPb\ filling pattern \cite{cham2017:jowett} is used providing 1136 collisions in ATLAS/CMS, 1120 collisions in ALICE and 81 collisions in LHCb. 
This approximation is made since the proton injection should be flexible enough to reproduce most of the respective Pb pattern. This calculations assumes $N_b=3\times10^{10}$ protons per bunch and ALICE being levelled to the instantaneous luminosity of $L_\mathrm{AA}=\qty{5\times 10^{29}}{cm^{-2}s^{-1}}$. 
$\LAA=\elumi{5}{29}$
ATLAS and CMS are not luminosity levelled in this scenario since the loss of integrated luminosity for ATLAS and CMS outweighs the marginal gain for ALICE. A simulation of the beam evolution based on ordinary differential equations including intra-beam scattering and radiation damping leads to a luminosity evolution in the different IPs as displayed in Fig.~\ref{fig:p_pb_lumi_evo}.

At around $\qty{6.1}{h}$, the bunch intensity of the bunches colliding in ATLAS, ALICE and CMS drop below the interlock BPM threshold ultimately limiting the fill length, even though the potential levelling time for ALICE has not been reached.  Detailed engineering of the filling scheme might avoid this.    
Key results from the beam evolution study are listed in Tab.~\ref{tab:p_Pb_table}. The expected peak luminosity in ATLAS and CMS is at around $L_\mathrm{AA}=\qty{17.4\times 10^{29}}{cm^{-2}s^{-1}}$, i.e., roughly a factor $2$ larger than in 2016. The integrated luminosity in ATLAS and CMS are expected to approach $\qty{0.7}{pb^{-1}}$ outperforming the 2016 integrated luminosity by a rough factor $3.5$. Since the nominal \mbox{HL-LHC} normalised proton emittance of $\epsilon_N=\qty{2.5}{\mu m}$ is assumed, the actual performance may exceed these predictions since normalised proton emittances in the range of $\epsilon_N=\qty{1.3}{\mu m}$ have already been achieved.
\begin{table}[htb]
    \centering
        \caption{Key parameters and results of the p--Pb beam evolution calculation. A turn-around time, i.e., the time between Beam Dump and Stable Beams, of \qty{3}{h} and an operational efficiency factor of $0.5$ is assumed. The final result was scaled down by additional $\qty{5}{\%}$ to take potential deviations of the proton filling pattern into account. 
    The time the first bunches need to reach the interlock BPM threshold is used as the fill time.}
\begin{tabular}{l|ccc}
  \hline
Species                                     &\multicolumn{1}{c}{p} & \multicolumn{2}{c}{Pb} \\
\hline
Beam energy \qty{E[Z}{TeV]}                 & \multicolumn{3}{c}{7} \\
Collision energy \qty{\sqrt{\sNN}}{[TeV]}   & \multicolumn{3}{c}{8.78}  \\
Bunch intensity \qty{N_b}{[10^8]}           &\multicolumn{1}{c}{300} &\multicolumn{2}{c}{1.8} \\
No,\ of bunches $k_b$                       &\multicolumn{1}{c}{1232} &\multicolumn{2}{c}{1232} \\
Norm. emittance \qty{\epsilon_N}{[\mu m]}   &\multicolumn{1}{c}{2.5} &\multicolumn{2}{c}{1.65}  \\
Bunch length \qty{\sigma_{z}}{[m]}          &\multicolumn{1}{c}{0.09} &\multicolumn{2}{c}{0.08} \\
Fill time \qty{t_\mathrm{fill}}{[h]}         & \multicolumn{3}{c}{6.1} \\
\hline
IP                                          & ATLAS/CMS & ALICE & LHCb \\
\hline
\qty{\bstar}{[m]}                           & 0.5 &  0.5 &  0.5\\
Colliding bunches  $k_c$                    & 1136 & 1120 &  81 \\
Luminosity \qty{\LAA}{[10^{29}cm^{-2}s^{-1}]}               & 17.4 & 5.0 & 1.2  \\
NN luminosity \qty{ \LNN}{[10^{31} {cm}^{-2} {s}^{-1}]}     & 36.3 & 10.4 & 2.6  \\
$\int _{\text{month}}L_{\text{AA}}\text{ dt}$ [nb${}^{-1}$] & 674 & 328 & 41.1   \\
$\int _{\text{month}}L_{\text{NN}}\text{ dt}$ [pb${}^{-1}$] & 140 & 68 & 8.5   \\
 \hline
\end{tabular}
    \label{tab:p_Pb_table}
\end{table} 

\subsection{Colliding lighter nuclei at HL-LHC} \label{sec:lightions}

The bunch intensity limits in the injectors depend largely on the ion charge which changes at the various stripping stages which must be optimised for space-charge limits, intra-beam scattering, efficiency of electron-cooling,   beam losses on residual gas and other effects in the ion source, Linac4,  LEIR, the PS and SPS.  
Given the uncertainties, the deliverable intensity for other species can only be determined after sufficient time spend commissioning and empirically optimising the many parameters and operating modes of the whole injector chain. 
To simplify present considerations, we postulate a simple form relating the number of ions per bunch, \Nb, to the well-established value
(\Nb(82,208)=\enum{1.9}{8}) for Pb beams  
\begin{equation}
{\Nb}(Z,A)   = {\Nb}(82,208){\left( {\frac{Z}{{82}}} \right)^{ - p}} 
\label{eq:pscaling}
\end{equation}
Fitting such an expression to the limited information~\cite{Manglunki:2016qzl} from the few species used for SPS fixed-target in recent years (since the commissioning of the present ECR ion source and LEIR) yields a value of the fit parameter \(p=1.9\).    
Beam quality requirements for fixed-target beams are, of course, less stringent than for injection into the collider.
Fitting to the first commissioning of Xe beams for the LHC~\cite{Schaumann:2018qat}, on the other hand, gives a much less optimistic \(p=0.75\).   
Although this was the only occasion where any other species than Pb was delivered to the collider, only the simplest version of the injection scheme was used and it is clear that, 
given time, significantly higher intensities could be achieved.  
We consider that  \(1.5\le p\le 1.9\) corresponds to a representative range of possibilities that could be 
realised in fully-prepared future operation.

In addition, we make a number of simplifying assumptions to allow a simplified, yet meaningful, comparison between species 
\begin{itemize}
    \item  The     \emph{geometric} transverse beam emittances at the start of collisions will be equal to those of Pb beams~\cite{HLLHCPbPbspec}.
    This is justified, at least at the level of the LHC, since the scaling of intra-beam scattering with \(\Nb\), \(Z\) and \(A\), 
    given by  the parameter 
    $f_{\text{IBS}}\text{/(m Hz)}$  
    is generally smaller than for Pb as long as 
    $p\lsim1.9$.  A similar scaling should hold in the injectors such as the SPS where intra-beam scattering may also blow up the emittances.  
    This ignores possible space-charge limits in the injectors which
    should also be considered once the appropriate stripping schemes and charge states have been defined. 
    
    \item Same filling scheme and number of bunches.
    \item No luminosity-levelling in any experiment.
     \item Fill length optimised for intensity evolution dominated by luminosity burn-off. 
     \item Equal operational efficiency of 50\%.     Following conventional practice for HL-LHC, the integrated luminosity for a 1-month run is estimated assuming back-to-back ideal fills of optimal length and a turn-around time of \qty{2.5}{h} between the end  of one fill and the resumption of ``Stable Beams''  for collisions in the next.  The operational efficiency factor is then applied as a way of taking into account the time needed for commissioning, intensity ramp-up, faults and other non-availability of LHC and injector systems. 
     
\end{itemize} 
The parameters are estimated using analytical approximations unlike the more elaborate simulations used in Section~\ref{sec:HLPbPb}.  
Together with the assumption that there is no luminosity levelling, these lead to a higher estimate of integrated luminosity in a one-month run.  
Nevertheless they can be used as a guide to the relative gain factors in integrated nucleon-nucleon luminosity by changing from Pb to a lighter nucleus. 

\begin{table}
\centering
   \caption{Parameters and performance for a range of light nuclei with a moderately optimistic value of the scaling parameter \(p=1.5\) in \eqref{eq:pscaling}. }
   
   {\small
	{\renewcommand{\arraystretch}{1.2}
\begin{tabular}{*{1}{l}{*{6}{|c}}}
\speciesheader\\
\hline
$\gamma$                                                                       &  \(3760.\) & \(3390.\) & \(3760.\) & \(3470.\) & \(3150.\) & \(2960.\) \\
$\sqrt{s_{\text{NN}}}\text{/TeV}$                                              &  \(7.\) & \(6.3\) & \(7.\) & \(6.46\) & \(5.86\) & \(5.52\) \\
$\sigma _{\text{had}}\text{/b}$                                                &  \(1.41\) & \(2.6\) & \(2.6\) & \(4.06\) & \(5.67\) & \(7.8\) \\
$\sigma _{\text{BFPP}}\text{/b}$                                               &  \(2.36\times 10^{-5}\) & \(0.00688\) & \(0.0144\) & \(0.88\) & \(15.\) & \(280.\) \\
$\sigma _{\text{EMD}}\text{/b}$                                                &  \(0.0738\) & \(1.24\) & \(1.57\) & \(12.2\) & \(51.8\) & \(220.\) \\
$\sigma _{\text{tot}}\text{/b}$                                                &  \(1.48\) & \(3.85\) & \(4.18\) & \(17.1\) & \(72.5\) & \(508.\) \\
$N_b$                                                                          &  \(6.24\times 10^9\) & \(1.85\times 10^9\) & \(1.58\times 10^9\) & \(6.53\times 10^8\) & \(3.56\times 10^8\) & \(1.9\times 10^8\) \\
$\epsilon _{\text{xn}}\text{/$\mu $m}$                                         &  \(2.\) & \(1.8\) & \(2.\) & \(1.85\) & \(1.67\) & \(1.58\) \\
$f_{\text{IBS}}\text{/(m Hz)}$                                                 &  \(0.0662\) & \(0.0894\) & \(0.105\) & \(0.13\) & \(0.144\) & \(0.167\) \\
$W_b\text{/MJ}$                                                                &  \(68.9\) & \(45.9\) & \(43.6\) & \(32.5\) & \(26.5\) & \(21.5\) \\
$L_{\text{AA0}}/\text{cm}^{-2}s^{-1}$                                          &  \(1.46\times 10^{31}\) & \(1.29\times 10^{30}\) & \(9.38\times 10^{29}\) & \(1.61\times 10^{29}\) & \(4.76\times 10^{28}\) & \(1.36\times 10^{28}\) \\
$L_{\text{NN0}}/\text{cm}^{-2}s^{-1}$                                          &  \(3.75\times 10^{33}\) & \(2.06\times 10^{33}\) & \(1.5\times 10^{33}\) & \(9.79\times 10^{32}\) & \(7.93\times 10^{32}\) & \(5.88\times 10^{32}\) \\
$P_{\text{BFPP}}\text{/W}$                                                     &  \(0.0031\) & \(0.179\) & \(0.303\) & \(5.72\) & \(43.4\) & \(350.\) \\
$P_{\text{EMD1}}\text{/W}$                                                     &  \(4.98\) & \(16.5\) & \(16.9\) & \(40.5\) & \(76.7\) & \(141.\) \\
$\tau _{\text{L0}}\text{/h}$                                                   &  \(16.4\) & \(21.3\) & \(23.\) & \(13.5\) & \(5.87\) & \(1.57\) \\
$T_{\text{opt}}\text{/h}$                                                      &  \(9.04\) & \(10.3\) & \(10.7\) & \(8.23\) & \(5.42\) & \(2.8\) \\
\qty{\left\langle L_{\text{AA}}\right\rangle}{cm^{-2}s^{-1}}             &  \(8.99\times 10^{30}\) & \(8.34\times 10^{29}\) & \(6.17\times 10^{29}\) & \(9.46\times 10^{28}\) & \(2.23\times 10^{28}\) & \(3.8\times 10^{27}\) \\
\qty{\left\langle L_{\text{NN}}\right\rangle}{cm^{-2}s^{-1}}       &  \(2.3\times 10^{33}\) & \(1.33\times 10^{33}\) & \(9.87\times 10^{32}\) & \(5.76\times 10^{32}\) & \(3.71\times 10^{32}\) & \(1.64\times 10^{32}\) \\
$\int _{\text{month}}L_{\text{AA}}\text{ dt/}\text{nb}^{-1}$                   &  \(1.17\times 10^4\) & \(1080.\) & \(799.\) & \(123.\) & \(28.9\) & \(4.92\) \\
$\int _{\text{month}}L_{\text{NN}}\text{ dt/}\text{pb}^{-1}$                   &  \(2980.\) & \(1730.\) & \(1280.\) & \(746.\) & \(481.\) & \(213.\) \\
$R_{\text{had}}\text{/kHz}$                                                    &  \(2.07\times 10^4\) & \(3340.\) & \(2440.\) & \(653.\) & \(270.\) & \(106.\) \\
$\mu$                                                                          &  \(1.64\) & \(0.266\) & \(0.194\) & \(0.0518\) & \(0.0215\) & \(0.00842\) \\
\end{tabular}
}
	}

  \label{tab:species1}
\end{table}

\begin{table}
\centering
   \caption{Parameters and performance for a range of light nuclei with an optimistic value of the scaling parameter \(p=1.9\) in \eqref{eq:pscaling}. }
   {\small
	{\renewcommand{\arraystretch}{1.2}
\begin{tabular}{*{1}{l}{*{6}{|c}}}
\speciesheader\\
\hline
$\gamma$                                                            &    \(3760.\) & \(3390.\) & \(3760.\) & \(3470.\) & \(3150.\) & \(2960.\) \\
$\sqrt{s_{\text{NN}}}\text{/TeV}$                                   &    \(7.\) & \(6.3\) & \(7.\) & \(6.46\) & \(5.86\) & \(5.52\) \\
$\sigma _{\text{had}}\text{/b}$                                     &    \(1.41\) & \(2.6\) & \(2.6\) & \(4.06\) & \(5.67\) & \(7.8\) \\
$\sigma _{\text{BFPP}}\text{/b}$                                    &    \(2.36\times 10^{-5}\) & \(0.00688\) & \(0.0144\) & \(0.88\) & \(15.\)& \(280.\) \\
$\sigma _{\text{EMD}}\text{/b}$                                     &    \(0.0738\) & \(1.24\) & \(1.57\) & \(12.2\) & \(51.8\) & \(220.\) \\
$\sigma _{\text{tot}}\text{/b}$                                     &    \(1.48\) & \(3.85\) & \(4.18\) & \(17.1\) & \(72.5\) & \(508.\) \\
$N_b$                                                               &    \(1.58\times 10^{10}\) & \(3.39\times 10^9\) & \(2.77\times 10^9\) & \(9.08\times 10^8\) & \(4.2\times 10^8\) & \(1.9\times 10^8\) \\
$\epsilon _{\text{xn}}\text{/$\mu $m}$                              &    \(2.\) & \(1.8\) & \(2.\) & \(1.85\) & \(1.67\) & \(1.58\) \\
$f_{\text{IBS}}\text{/(m Hz)}$                                      &    \(0.168\) & \(0.164\) & \(0.184\) & \(0.18\) & \(0.17\) & \(0.167\) \\
$W_b\text{/MJ}$                                                     &    \(175.\) & \(84.3\) & \(76.6\) & \(45.2\) & \(31.4\) & \(21.5\) \\
$L_{\text{AA0}}/\text{cm}^{-2}s^{-1}$                               &    \(9.43\times 10^{31}\) & \(4.33\times 10^{30}\) & \(2.9\times 10^{30}\) & \(3.11\times 10^{29}\) & \(6.66\times 10^{28}\) & \(1.36\times 10^{28}\) \\
$L_{\text{NN0}}/\text{cm}^{-2}s^{-1}$                               &    \(2.41\times 10^{34}\) & \(6.93\times 10^{33}\) & \(4.64\times 10^{33}\) & \(1.89\times 10^{33}\) & \(1.11\times 10^{33}\) & \(5.88\times 10^{32}\) \\
$P_{\text{BFPP}}\text{/W}$                                          &    \(0.0199\) & \(0.601\) & \(0.935\) & \(11.\) & \(60.6\) & \(350.\) \\
$P_{\text{EMD1}}\text{/W}$                                          &    \(32.\) & \(55.6\) & \(52.2\) & \(78.3\) & \(107.\) & \(141.\) \\
$\tau _{\text{L0}}\text{/h}$                                        &    \(6.45\) & \(11.6\) & \(13.1\) & \(9.74\) & \(4.96\) & \(1.57\) \\
$T_{\text{opt}}\text{/h}$                                           &    \(5.68\) & \(7.62\) & \(8.08\) & \(6.98\) & \(4.98\) & \(2.8\) \\
\qty{\left\langle L_{\text{AA}}\right\rangle}{cm^{-2}s^{-1}}        &    \(4.54\times 10^{31}\) & \(2.45\times 10^{30}\) & \(1.69\times 10^{30}\) & \(1.68\times 10^{29}\) & \(2.95\times 10^{28}\) & \(3.8\times 10^{27}\) \\
\qty{\left\langle L_{\text{NN}}\right\rangle}{cm^{-2}s^{-1}}        &    \(1.16\times 10^{34}\) & \(3.93\times 10^{33}\) & \(2.71\times 10^{33}\) & \(1.02\times 10^{33}\) & \(4.91\times 10^{32}\) & \(1.64\times 10^{32}\) \\
$\int _{\text{month}}L_{\text{AA}}\text{ dt/}\text{nb}^{-1}$        &    \(5.89\times 10^4\) & \(3180.\) & \(2190.\) & \(218.\) & \(38.2\) & \(4.92\) \\
$\int _{\text{month}}L_{\text{NN}}\text{ dt/}\text{pb}^{-1}$        &    \(1.51\times 10^4\) & \(5090.\) & \(3510.\) & \(1330.\) & \(636.\) &\(213.\) \\
$R_{\text{had}}\text{/kHz}$                                         &    \(1.33\times 10^5\) & \(1.12\times 10^4\) & \(7540.\) & \(1260.\) & \(378.\) & \(106.\) \\
$\mu$                                                               &    \(10.6\) & \(0.893\) & \(0.598\) & \(0.1\) & \(0.03\) & \(0.00842\) \\
\end{tabular}
}

	}
 \label{tab:species2}
\end{table} 
 

\subsection{Short run for \OO\ and \pO}
\label{sec:pOrun} 

As discussed in Section~\ref{sec:pOcosmic}, a  short \pO\ collision run is of interest for cosmic-ray physics.  If O beams were available from the injectors, this could be combined with a  short, low-luminosity, \OO\ run, which would be of  value for the main high-density QCD programme.  
Limiting the beams to low-intensities would allow a rapid set-up in LHC on the successful 
model of the 2012 \pPb\ run 
which was later re-used in the 
2017 \XeXe\ run\cite{Schaumann:2018qat}.  
Each of those runs took  about 16~h of LHC operation time, 
including set-up and physics data-taking but a combination \OO/\pO run could take a few days.

Because oxygen is used as the carrier gas in the CERN heavy-ion source, 
the idea has been mooted that 
it may be possible to switch from Pb to
\isotope{16}{O}{8+}  beams for the LHC, and back,  
somewhat more rapidly than other species. 
Commissioning of the O beam in the injectors for single-bunch injection into the LHC
would need to be scheduled, 
in parallel with \pp\ operation, and use of the injectors for other programmes,   
in the period preceding the \OO/\pO\ run.    
The possibilities are under study and include either inserting the run at
the end of one of the annual Pb--Pb runs or  scheduling it  
earlier in the year in order to provide time for 
the source to be switched back to Pb operation
afterwards. 


\subsection{Heavy-ion performance of HE-LHC} 

Heavy-ion operation of HE-LHC awaits a fully detailed study.   
First results were presented in~\cite{JowettHELHC}. 
Since the HE-LHC would occupy the same tunnel as the LHC, one can, 
for the moment, assume the same injected beams as HL-LHC. 
Future  possible upgrades to the injectors might improve this.
The total integrated luminosity obtainable per fill, summed over all experiments, 
is bounded by the total intensity divided by the burn-off cross-section and 
will therefore be similar to the HL-LHC.  
The same can be said for  time taken to inject a fill. 
Only a  modest increase  in integrated luminosity, given by somewhat shorter times spent in collision and, most likely, 
a reduction in the number of experiments, can be envisaged.  
For purposes of this report, we estimate that the integrated luminosity obtained by each
of two experiments in a one-month run will be of order \qty{6}{nb^{-1}}. 
However, the BFPP power with \PbPb collisions will be very high and this could be a strong
argument for running with somewhat  lighter species. 
In that case, one can expect the luminosity to scale similarly to  HL-LHC in Sect.~\ref{sec:lightions}.

\clearpage

\clearpage
\section{Light flavour sector: (anti-)(hyper-)nuclei and fluctuations of conserved charges}
\label{sec:lf}

{ \small
\noindent \textbf{Coordinator}: Francesca Bellini (CERN)

\noindent \textbf{Contributors}: 
M.~Arslandok (Heidelberg University), 
N.~K. Behera (Inha University), 
R.~Bellwied (Houston University), 
K.~Blum (CERN and Weizmann Institute of Science), 
A.~Borissov (M\"{u}nster University), 
P.~Braun-Munzinger (EMMI$/$GSI Helmholtzzentrum f{\"u}r Schwerionenforschung GmbH), 
B.~Doenigus (Frankfurt University), 
L.~Fabbietti (TU Munich), 
S.~Floerchinger (Heidelberg University), 
A.P.~Kalweit (CERN), 
R.~Lea (University and INFN, Trieste), 
A.~Mastroserio (Foggia University and INFN, Bari),
A.~Ohlson (Heidelberg University), 
V.~Okorokov (National Research Nuclear University MEPhI, Moscow), 
S.~Piano (INFN Trieste), 
M.~Puccio (University and INFN, Torino), 
K.~Redlich (University of Wroclaw and EMMI$/$GSI Helmholtzzentrum f{\"u}r Schwerionenforschung GmbH), 
A.~Rustamov (NNRC Baku, GSI Helmholtzzentrum f{\"u}r Schwerionenforschung GmbH and Heidelberg University), 
J.~Stachel (Heidelberg University), 
A.~Timmins (University of Houston),
S.~Trogolo (University and INFN, Torino).
}

\subsection{Introduction}
The analysis of the data collected at the LHC during Run 1 and Run 2 has consolidated our understanding of a standard model for the production of light-flavour hadrons (containing u, d and s quarks) in heavy-ion collisions:
particle chemistry (described by integrated particle yields) is well described by the thermal-statistical model \cite{Andronic:2017, Acharya:2017bso} and kinetic equilibrium (reflected in the \pT-dependence of particle production) is well described by a common radial expansion governed by hydrodynamics \cite{Abelev:2013vea, Acharya:2018zuq}.
While the physics of light-flavour particles is often perceived as not statistics hungry, the unprecedented large integrated luminosities expected in Run 3 and Run 4 at the LHC offer a unique physics potential. 
Despite containing only u, d and s valence quarks, light \mbox{(anti-)(hyper-)nuclei} are very rarely produced because of their composite nature and very large mass. Their study will enormously profit from the significant increase in luminosity for heavy-ion collisions expected in the years 2021 until 2029.
The same holds true for the study of event-by-event fluctuations of the produced particles, which is closely linked to the production of light \mbox{(anti-)(hyper-)nuclei} in the scenario of a common chemical freeze-out determining light-flavour hadrons and (hyper-)nuclei abundances. 
If, as indicated by the recent experimental findings \cite{Acharya:2017bso}, the thermal-statistical approach is the correct model to describe (anti-)(hyper-)nuclei production, the chemical freeze-out temperature is most precisely determined by measurements of light \mbox{(anti-)(hyper-)nuclei} as they are not subject to feed-down corrections from strong decays \cite{Andronic:2017}. This is the same temperature at which event-by-event fluctuations of conserved quantities are compared to lattice QCD (lQCD).
The physics of light \mbox{(anti-)(hyper-)nuclei} and exotic multi-quark states together with the related observables that will become experimentally accessible in \PbPb collisions at the LHC Runs 3 and 4 are discussed in Sec.~\ref{sec:nuclei}. 
In Sec.~\ref{sec:fluctuations}, measurements of fluctuations of particle production and conserved charges are discussed as they give experimental access to fundamental properties of the QCD phase transition at $\mu_{\mathrm{B}}$ and allow for direct comparison with lQCD calculations. 
\\\noindent In small collision systems (\pp, \pPb), measurements of light-flavour hadrons provide fundamental input to the study of particle production mechanisms and collectivity across systems, as discussed in Ch.~\ref{chapter:smallsystems}. At the same time, the physics programme with \pp and \pPb collisions in Runs 3 and 4 will open the possibility for system-size dependent studies of (anti-)nuclei production and for precision measurements of the hyperon-nucleon potentials. 
The physics case for these measurements in small colliding systems is motivated in this chapter in Sec.~\ref{sec:astro}, as well as the implications of the findings at the LHC for astrophysics and searches for dark matter in space-based experiments.

\subsection{(Anti-)(hyper-)nuclei production} 
\label{sec:nuclei}
\subsubsection{Testing thermal production and nucleon coalescence models}
\label{sec:producionmodels}

The production of light (hyper-)nuclei and their anti-matter counterparts is modeled within the scenarios of thermal-statistical hadronisation and nucleon coalescence.
In the thermal-statistical approach~\cite{Andronic:2010qu,Andronic:2017}, particles are produced from a fireball in thermal equilibrium with temperatures of the order of \Tchem~$\approx$~156~MeV that are near the temperature of the QCD phase transition boundary, as predicted by lQCD calculations \cite{Bazavov:2014pvz,Bellwied:2013cta}.
The yields of the produced objects depend on the chemical freeze-out temperature \Tchem (when inelastic collisions cease) and the mass $m$ of the object, and approximately scale as d$N$/d$y \propto \exp(-m/\Tchem)$.  
Thermal-statistical models have been successful in describing light-flavour particle production across a wide range of energies in nucleus-nucleus collisions \cite{Andronic:2017, Acharya:2017bso}.  
Due to their large mass, light (anti-)(hyper-)nuclei are particularly sensitive to \Tchem and since they are not affected by feed-down from higher mass states \cite{Andronic:2017}, the measurement of their production constitutes a precision test for the thermal model.  

In the coalescence scenario, composite objects are formed at kinetic freeze-out by coalescence of nucleons that are close in configuration and momentum space \cite{Butler:1963, Kapusta:1980, Bergstrom:1979gpv, Sato:1981ez, Nagle:1996vp, Scheibl:1998tk}. Calculations of the coalescence probability based on a density matrix approach \cite{Scheibl:1998tk} require the knowledge of the nucleus wave function and identify the volume of the particle source as the homogeneity volume that can be extracted via Hanbury-Brown--Twiss interferometry \cite{Wiedemann:1999qn}. 
The size of the (hyper-)nucleus is identified with the size parameter of its wave-function, which is related to the (measurable) rms of the charge distribution by simple relations \cite{Sato:1981ez, Bellini:2018epz}. 

\begin{figure}[!t]
\begin{center}
\includegraphics[width=\textwidth]{\main/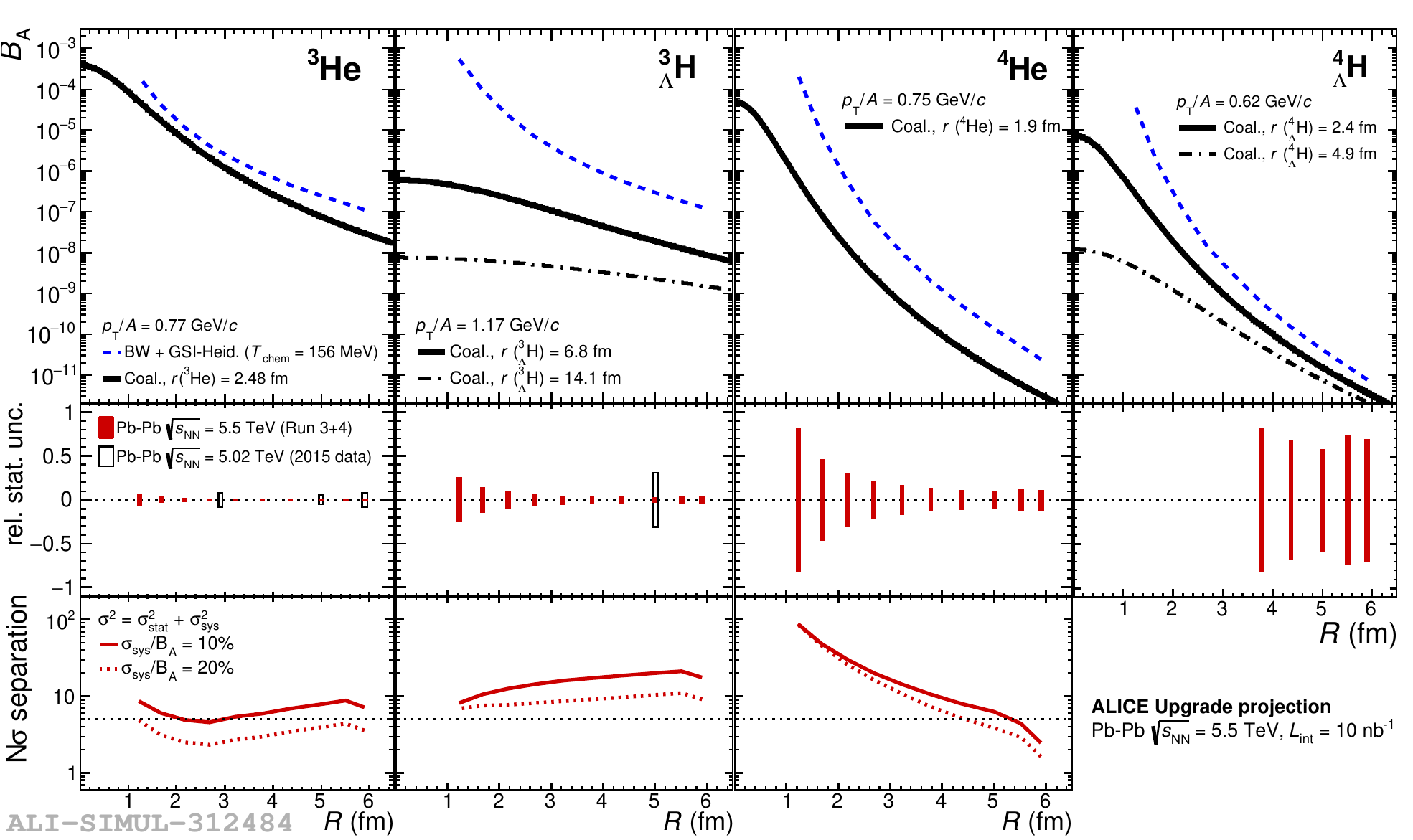}
\end{center}
\caption{
Top: comparison of predictions for the coalescence parameters for (hyper-)nuclei with \Anucl = 3, 4 from the Blast-Wave + GSI-Heidelberg thermal statistical model and nucleon coalescence as a function of the radius ($R$) of the particle emitting source. For each (hyper-)nucleus, the radius $r$ considered by the coalescence model is reported in the legend. For \hyp~(\hypfour), two values of the radius are considered: the lower value represents the average separation of the three (four) constituents, whereas the larger $r$ corresponds to the average separation between the $\Lambda$ and the deuteron (triton) core. See \cite{Bellini:2018epz} for full details on the models. Middle: projection of the relative statistical uncertainty achievable with a minimum bias \PbPb integrated luminosity of \Lint = 10 \nbInv and the upgraded ALICE detector (in red) compared to the relative statistical uncertainty of the Run 2 measurements (in black). Bottom: significance in the discrimination between the two models, assuming 10$\%$ and 20$\%$ systematic uncertainty in addition to the statistical uncertainty expected with \Lint = 10 \nbInv. For \hyp, the coalescence prediction considered is  for $r = 6.8$ fm (corresponding to the black continuous lines in the top panel). Figure from Ref.~\cite{ALICE-PUBLIC-2019-001}.}
\label{fig:BAmodels}
\end{figure}

While there are several theory groups working on the calculation of the expected coalescence \cite{Scheibl:1998tk, Cho:2017dcy, Zhang:2018euf, Bazak:2018hgl, Zhao:2018lyf} and thermal production rates \cite{Andronic:2010qu, Wheaton:2004qb, Petran:2013dva}, predictions reported in Fig.~\ref{fig:BAmodels} rely on the study presented in~\cite{Bellini:2018epz}, which contrasts the two production scenarios. 
In order to distinguish them, a measurement of the coalescence parameter for (anti-)(hyper-)nuclei that differ by mass, spin and size 
as a function of source volume (or source radius) is proposed. 
The coalescence parameter \BA is defined as 
\begin{equation}
E_{A}\frac{\mathrm{d}^{3}N_{A}}{\mathrm{d}p_{A}^{3}}=B_{A}{\left(E_{\mathrm{p,n}}\frac{\mathrm{d}^{3}N_{\mathrm{p,n}}}{\mathrm{d}p_{\mathrm{p,n}}^{3}}\right)^{A}}\left\vert_{\vec{p}_{\mathrm{p}}=\vec{p}_{\mathrm{n}}=\frac{\vec{p}_{A}}{A}} \right.,
\label{eq:BA}
\end{equation}
\noindent where $p_{\mathrm{p,n}}$ are the momenta of the proton and neutron and $E_{p,n}$ their energies. In the coalescence model (black curves in top panels of Fig.~\ref{fig:BAmodels}), the coalescence parameter is determined analytically. The thermal model predicts \pT-independent particle yields at a given \Tchem, therefore a Blast-Wave (BW) model is used in \cite{Bellini:2018epz} to describe the \pT-dependence of (hyper-)nuclei and nucleon production. With the \pT~spectra of (hyper-)nuclei and protons obtained in this way, Eq.~\ref{eq:BA} is used to extract \BA (dashed blue curve in top panels of Fig.~\ref{fig:BAmodels}). 
Similarly, the coalescence parameter is obtained experimentally from Eq.~\ref{eq:BA} using the measured (hyper-)nucleus and proton \pT~distributions as input.
It is considered that for BW, little energy dependence of the fit parameters is observed in \PbPb collisions from \sqrtsNN = 2.76 to 5.02 TeV.
The thermal model yields only depend on temperature and no collision energy dependence of the temperature is expected in the LHC energy range.
The size of the source can be sampled by means of multiplicity- and centrality-differential measurements.

The particle with the strongest sensitivity to the production mechanism appears to be the hypertriton (a p$\Lambda$n bound state) with its large charge rms radius of about 10 fm, for which the coalescence and the thermal model predictions differ by up to three orders of magnitude as a function of the source radius. 
While the hypertriton seems to be largely suppressed with respect to \hethree~($\mathrm{pnn}$), the \hypfour\ ($\mathrm{pp}\Lambda\mathrm{n}$) is predicted to have only a slightly lower coalescence probability with respect to \hefour~($\mathrm{ppnn}$). 
Moreover, for small $R$, i.e. in small systems as those formed in \pp and \pPb collisions, \hyp~is predicted by coalescence to be suppressed by about a factor of 100 with respect to \hethree. 
These considerations motivate systematic multi-differential measurements of \Anucl = 3 and \Anucl = 4 nuclei and hyper-nuclei as a function of multiplicity and from small (\pp, \pPb) to large systems (\PbPb) to test the validity of the coalescence picture as opposed to thermal production.

With an integrated luminosity \Lint~=~10~\nbInv in \PbPb collisions in Runs 3 and 4, \BA for \hethree, \hyp~and \hefour~can be measured in ALICE in up to ten centrality classes with a statistical precision lower than 5$\%$, 10$\%$ and 20$\%$, respectively.
The projected relative statistical uncertainties on \BA ($\sigma_{\mathrm{stat}}$/\BA) for (hyper-)nuclei with \Anucl~>~2 are reported in the central row of panels of Fig.~\ref{fig:BAmodels}.
These uncertainties have been estimated by scaling the significance of the nuclei and hyper-nuclei spectra measurements in \PbPb collisions at \sqrtsNN~=~5.02~TeV \cite{ALICE-PUBLIC-2017-006, Trogolo:2017oii} to the expected integrated luminosity of Runs 3 and 4 and assuming thermal production for the states with \Anucl~=~4. The uncertainties on the proton spectra are negligible already in the existing measurements. 
 
The experimental discrimination power between the models has been extracted as $(B_{\mathrm{A}}^{therm} - B_{\mathrm{A}}^{coal})/\sigma$, where $\sigma = \sqrt{\sigma_{\mathrm{stat}}^{2} + \sigma_{\mathrm{sys}}^{2}}$, and is reported in the lower panels of Fig. \ref{fig:BAmodels}.
Relative systematic uncertainties $\sigma_{\mathrm{sys}}$/\BA = 10$\%$ and 20$\%$ have been considered, to be compared with a typical 15$\%$ uncertainty of the Run 1 and 2 measurements. 
Measurements of \hyp~allow for a 10$\sigma$ discrimination between models, even in a pessimistic scenario in all centralities. The discrimination power rises above the 10$\sigma$ level for \hefour~in semi-central and peripheral collisions.  

\subsubsection{Light (anti-)(hyper-)nuclei observables in Runs 3 and 4}
\label{sec:hyper}

\begin{figure}
\begin{center}
\includegraphics[width=0.49\textwidth]{\main/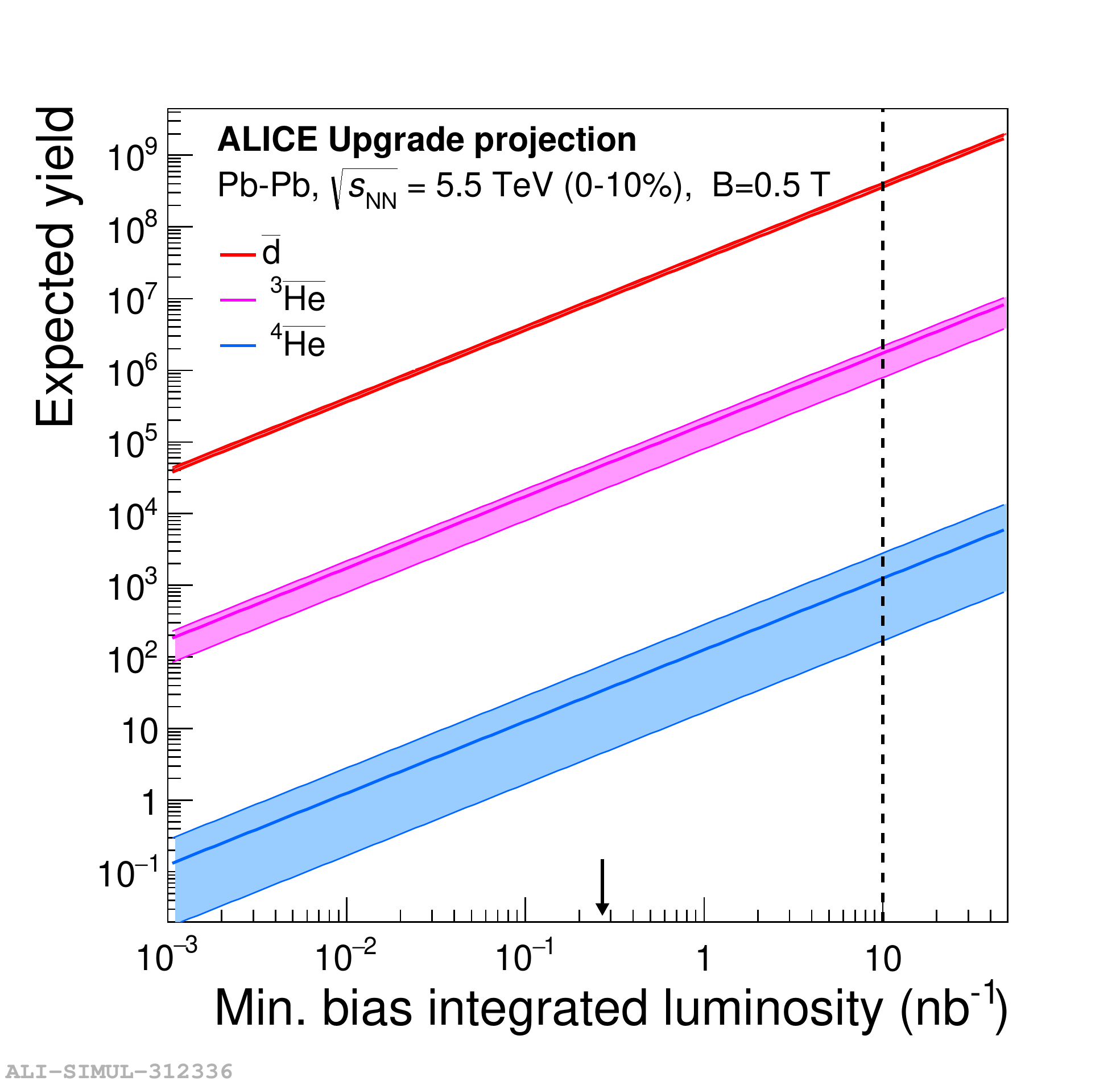}
\includegraphics[width=0.49\textwidth]{\main/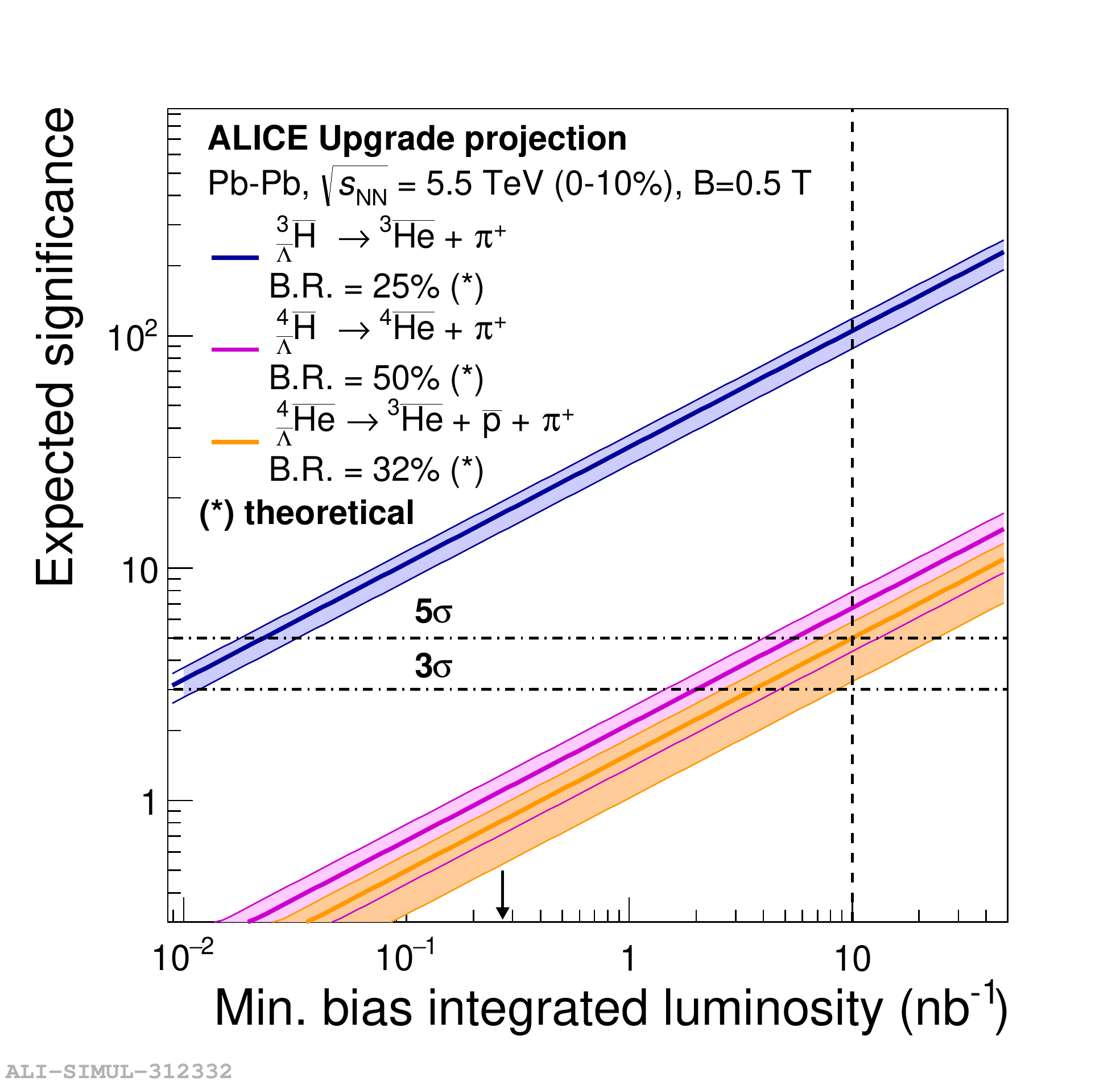}
\end{center}
\caption{Left: (raw) yield of anti-nuclei in the $2 <\pT< 10~\gmom$ interval, detectable in 0--10$\%$ central \PbPb collisions with the ALICE detector as a function of the minimum bias luminosity. 
Right: Projected significance of anti-hyper-nuclei measurements in central \PbPb collisions in Runs 3 and 4 with ALICE as a function of the integrated minimum-bias luminosity. In both panels, the arrow represents the minimum bias \PbPb luminosity anticipated for the end of Run 2. The dashed vertical line marks the projections with \Lint~=~10~\nbInv. The bands represents the uncertainty on model prediction for the yield (see text for details).
Figures from Ref.~\cite{ALICE-PUBLIC-2019-001}.
}
\label{fig:yieldrun34}
\end{figure}

Measurements of (anti-)(hyper-)nuclei and exotic QCD bound states require large event samples collected with a minimum-bias trigger, as well as high tracking precision for the separation of secondary vertices and charged-hadron (light nucleus) identification. The upgraded ALICE detector after LS2 \cite{Abelevetal:2014dna, ALICE:2014qrd,Antonioli:1603472,Buncic:2015ari} fulfills these requirements, developing further the potential already explored in Runs 1 and 2.
The yields of (hyper-)nuclei (d,  \hethree, \hefour, \hyp, \hypfour, \hyphefour) and their anti-particles in \PbPb collisions at the LHC in Runs 3 and 4 have been estimated for measurements with ALICE. 
The detectable yield and significance for (anti-)(hyper-)nuclei have been estimated for 0--10\% central \PbPb collisions considering the acceptance and detection efficiency in the nominal magnetic field of the ALICE detector (B~=~0.5~T).
These projections are reported for anti-particles in Fig.~\ref{fig:yieldrun34} as a function of the minimum-bias integrated luminosity. The detectable particle and anti-particle yields are equivalent in the considered \pT\ range. 
All projections have been extracted in the $2 <\pT< 10~\gmom$ range, where the lower limit is given by the \pT\ down to which nuclei with \Anucl~$\geq$~3 can be reconstructed without ambiguity in ALICE. 
In a scenario in which ALICE will take data with a central-barrel low-field configuration (B~=~0.2~T), it will be possible to extend the low-\pT\ limit for \mbox{(anti-)nuclei} identification down to 1~\gmom, increasing the expected number of detectable light \mbox{(anti-)(hyper-)nuclei} (by about 20$\%$ for \hethree). 
In Fig.~\ref{fig:yieldrun34}, the bands indicate the uncertainty on the yield (significance) associated with different model predictions:
the central line is obtained assuming statistical-thermal production with \Tchem~=~156~MeV~\cite{Andronic:2010qu}, the upper line is the yield (significance) assuming thermal production at \Tchem~=~158~MeV, and the lower one using for the yields the expectation from coalescence (see Sec.~\ref{sec:producionmodels}). 
The arrow represents the recorded luminosity at the end of the LHC Run 2. 
It has to be noted that for this study, the geometry of the ALICE Inner Tracking System (ITS) in Run 2 has been considered. 
The new geometry and acceptance of the upgraded ITS system \cite{Abelevetal:2014dna} are expected to increase the detection efficiency by up to 20\%.

The expected yield per unit of rapidity at mid-rapidity are reported for \antid, \antihethree\ and \antihefour\ in left panel of Fig.~\ref{fig:yieldrun34}.
With \Lint~=~10~\nbInv recorded with the nominal magnetic field, a measurement of the elliptic flow ($v_2$) of \hethree~and \tritium~(and anti-nuclei) in \PbPb collisions will become feasible with ALICE with a statistical precision better than 5$\%$ in the 2--$10~\gmom$ transverse momentum range in at least eight centrality intervals. 
Elliptic flow measurements for anti-nuclei provide a powerful independent test of coalescence scenarios as already demonstrated with deuterons \cite{Acharya:2017dmc} and might provide an indirect assessment of the neutron flow comparing the \hethree\ and \tritium\ results. 
In addition, the large data sample that will be collected for light anti-nuclei will lead to the first precise measurements of the mass of light anti-nuclei with \Anucl~=~3, by means of the Time-Of-Flight detector~\cite{Adam:2015pna}. 
This measurement will make it possible to test Charge Symmetry Breaking (CSB) in the anti-nuclei sector due to the differences in the up and down quark masses and due to electromagnetic effects \cite{Miller:1990ky}. 
The differences in \Anucl~=~3 systems are extensions of the neutron-proton difference. Although the mass difference for the lightest ``mirror pair'' with \Anucl~=~3 (i.e.\,\tritium,\hethree), is well known (at the level of O(eV)~\cite{Audi:2002rp}), no measurement has been performed in the anti-matter sector and will be accessible with \Lint~=~10~\nbInv. 

In the right panel of Fig.~\ref{fig:yieldrun34}, the expected significance of anti-hyper-nuclei measurements in central \PbPb collisions is reported as a function of the minimum bias integrated luminosity. 
For each species, the decay channels with the minimum number of charged particles in the final state and with the highest detection efficiency in ALICE have been considered for this study, as reported in the legend of Fig.~\ref{fig:yieldrun34}. 
The study of other decay channels, e.g. the three body decay of \hyp\ that has larger theoretical branching ratio with respect to the 2-body decay \cite{ Phys.Rev.C57}, but lower detection efficiency in ALICE, will be also carried out profiting from the large integrated luminosity.
Considering the thermal model predictions at \Tchem~=~156~MeV, the expected significance of \antihyp, \antihypfour\ and \antihehypfour\ at \Lint~=~10~\nbInv is 100, 7 and 5, respectively. 
The collected sample will enable very precise measurements of the production of \hyp\ and \antihyp\ and the first ever measurement of their elliptic flow as a function of \pT.
The discovery of \antihypfour\ and \antihehypfour\ will be in reach with \Lint~=~10~\nbInv at the end of Run 4.

\subsubsection{The hypertriton lifetime}
The experimental measurement of the $\Lambda$ separation energy of the \hyp, B$_{\Lambda}$ = 0.13 $\pm$ 0.05 (stat.) $\pm$ 0.04 (syst.) MeV \cite{davis20053}, led to the hypothesis that the lifetime of the hypertriton is equal to or only slightly below the free $\Lambda$ lifetime $\tau$($\Lambda$) = 263.2 $\pm$ 0.2 ps \cite{pdg:2018}.
Three different experimental techniques have been used to tackle this question: photographic emulsion, He bubble chambers, and counter experiments. The average for the emulsion experiments is 203$^{+40}_{-31}$ ps \cite{agnello:2016}, for the He bubble chambers is 193$^{+15}_{-13}$ ps \cite{agnello:2016}, and for the combination of both visualizing techniques is 193$^{+14}_{-13}$ ps \cite{agnello:2016}. The most recent results, 181 $^{+54}_{-39} \pm$ 33 ps and 142 $^{+24}_{-21} \pm$ 29 ps, have been obtained with the counter technique in heavy-ion collisions by the ALICE \cite{PhysLettB.754.360} and STAR \cite{PhysRevC.97.054909} experiments, respectively. This technique is currently the one with the highest precision (14-16$\%$) and the weighted average of heavy-ion experiments results is 185$^{+28}_{-23}$ ps \cite{agnello:2016}.
However, the few existing theoretical calculations point in the direction of the hypothesis mentioned at the beginning of this section. 
The first theoretical determination of $\tau$(\hyp) (by Dalitz and Rayet, \cite{NuovCim.A46}) ranged from 239.3 to 255.5 ps. More recent calculations from Congleton \cite{jphysg.18.339} and Kamada \cite{Phys.Rev.C57} estimated values of 232 ps and 256 ps, respectively. The deviation of the experimental results from the theoretical calculations and the free $\Lambda$ lifetime, by more than 2$\sigma$, is known as the ``hypertriton lifetime puzzle''.
 
With the expected integrated \PbPb luminosity at the end of the LHC Run 4, the statistical uncertainty on the lifetime will be reduced down to 1$\%$. 
In parallel, a reduction of the systematic uncertainty ($\sim$ 10$\%$ in the most recent ALICE measurements), will be achieved with the upgraded ALICE ITS that will allow for a reduction of the uncertainty due to tracking and material budget. 
To improve even further down the control on the systematic uncertainty, a better understanding of the corrections for the absorption in the material will be crucial. 

\subsubsection{$\Sigma$-hypernuclei}
In addition to measurements of $\Lambda$-hypernuclei, also the search for $\Sigma$-hypernuclei is to be considered with the luminosities forseen for the LHC Runs 3 and 4. 
Theory calculations for the $\Sigma$NN system suggest the presence of a near-threshold narrow ($\sim$2 MeV wide) quasi-bound state in the \isospin~=~1 and \spinJ~=~1/2 configuration, where the possible isospin and spin states are 0, 1, 2 and 1/2, 3/2, respectively ~\cite{cite:SigmaTriton-theory}. 
Among $\Sigma$-hypernuclei, only the $_{\Sigma^0}^{4}\mathrm{He}$ bound state has been observed so far, using the ${^4}\mathrm{He}(K^-,\pi^-)$ reaction \cite{cite:SigmaTriton-data}.
When the $\Sigma^0$ hyperon is bound inside a nucleus, the electromagnetic decay is dominated by the conversion reaction $\Sigma^{0}\mathrm{N}\to\Lambda\mathrm{N}$, thus the partial width of electromagnetic decay is expected to be reduced substantially. 
However, for the \isospin~=~2 state the conversion reaction is not allowed and the electromagnetic decay becomes prominent.
Experimental searches for  $\Sigma$-hypertriton bound states will also profit from the \PbPb data-taking programme of the LHC Runs 3 and 4 to exploit the strong decay
$\sigmahyp~(\antisigmahyp)~\rightarrow~\Lambda~(\overline{\Lambda})~+~\mathrm{d}$
and the decay $\sigmahyp~(\antisigmahyp)~\rightarrow~\hyp~(\antihyp)~+~\gamma$
following a similar strategy to the detected electromagnetic decay of 
$\Sigma^{0}~(\overline{\Sigma}^{0})~\rightarrow~\Lambda~(\overline{\Lambda})~+~\gamma$~\cite{Borissov:2015ura}.
The signal of hypertriton can be reconstructed in ALICE as discussed in Sec.~\ref{sec:hyper}. The soft photon can be identified by exploiting the conversion into electron pairs in the detector material of the ITS and Time Projection Chamber ($X/X_0~\approx~9~\%$ considering the upgraded ITS and the TPC together), covering the pseudorapidity range $\vert\eta\vert~<~0.8$, over the full azimuth ($\Delta\varphi~=~2\pi$)~\cite{Abelev:2014ffa}.
Alternatively, the photon can be detected in the PHOS calorimeter, but with limited acceptance of $\Delta\varphi~=~100^{\mathrm{o}}$ and $\vert\eta\vert<~0.12$~\cite{Abelev:2014ffa}.
The search for $\Sigma$-hypernuclei via electromagnetic decay will be carried out in ALICE profiting from the expected detector performance and detection of about $10^5$ hypertriton candidates (see Fig.~\ref{fig:yieldrun34}) in 0--10\% central \PbPb collisions at \Lint~=~10~\nbInv.

\subsubsection{Exotic QCD bound states}
At LHC energies, potential QCD bound states that have more complex structures such as pentaquarks, tetraquarks, hadron molecules or dibaryon states could be produced.
In particular, the possibility to detect and measure f$_{0}$(980), N(1875), N$\Xi$, N$\Omega$ and N$\Lambda_c$ in heavy-ion collisions with the unprecedented statistics of the LHC Runs 3 and 4 programme has been investigated. The advanced capabilities of the ALICE experiment in terms of hadron identification, including topological reconstruction of weak decays, are particularly suited for these studies.

The per-event yields of these states (d$N$/d$y$)$_{\mathrm{th}}$) predicted by quark- and hadron-coalescence models \cite{Cho:2017dcy} and the statistical-thermal model \cite{Andronic:2017} are reported in Tab.~\ref{tab:yield}.

\begin{table}[!b]
\centering
\caption{Properties and yields of exotic states in 0--10\% central \PbPb collisions at $\sqrt{s_{\mathrm{NN}}} = 5.5$~TeV. Theoretical predictions of yields per event, (d$N$/d$y$)$_{\mathrm{th}}$, are given in three different scenarios: quark- and hadron-coalescence \cite{Cho:2017dcy}, and thermal model \cite{Andronic:2010qu}. $\mathrm{S_{raw}}$ represents the total detectable yield at the \PbPb luminosity of 10~\nbInv, taking into account the branching ratios (B.R.) of the decay channels considered and assuming the ALICE detector performance as in Run 2 \cite{Jacazio:2017dvy, Acharya:2018ckj}. For f$_0(980)$, a $\mathrm{K\overline{K}}$ state and a decay into $\mathrm{K\overline{K}}$ with B.R. = 10$^{-3}$ is assumed for hadron coalescence~$^{\dagger}$. A tetraquark state is assumed for quark coalescence and a decay into $\pi\pi$. The same decay channel is assumed for the thermal production case. Masses are from \cite{pdg:2017}.}
\label{tab:yield}
\resizebox{\textwidth}{!}{
\begin{tabular}{|c|c|c|c|c|c|c|}
\hline
 & Model & f$_{0}$(980)& N(1875)& N$\Xi$& N$\Omega$  & N$\Lambda_{c}$ \\
\hline
\hline
Structure & & $\mathrm{qq\bar{q}\bar{q}}$ or $\mathrm{K}\overline{\mathrm{K}}$ & hadron molecule & dibaryon & dibaryon  & dibaryon \\
 \hline
 &  q-coal. & 5.4 $\times$ 10$^{-2}$ &  - & - & 1.8 $\times$ 10$^{-3}$ & 1.5 $\times$ 10$^{-3}$ \\
 
$\big(\frac{\mathrm{d}N}{\mathrm{d}y}\big)_{\mathrm{th}}$ & h-coal. & 3.2 $^{\dagger}$       &  - & - & 1.6 $\times$ 10$^{-3}$ & 5   $\times$ 10$^{-3}$ \\

 & thermal &  10  & 3 $\times$ 10$^{-1}$ & 8.7 $\times$  10$^{-3}$ & 5.7 $\times$  10$^{-3}$ & 4 $\times$  10$^{-3}$ \\
\hline \hline
 Decay channel &
 & $\pi\pi$ / $\mathrm{K}\overline{\mathrm{K}}$ 
 & $\Sigma^{\ast}(\rightarrow \Lambda \pi)$K    
 & $\Xi\rightarrow \Lambda \pi$ & $\Omega \rightarrow \Lambda$K  
 & $\Lambda_{c}\rightarrow \pi$Kp + $\Lambda_{c}\rightarrow $K$^0_{\mathrm{S}}$p
\\

 B.R. ($\%$) & & dominant / seen$^{\dagger}$ & unknown (87) & 99.9 & 67.8 & 6.2 + 1.58 \\
Mass (MeV/$c^{2}$) & & 990 & 1850 -- 1920 & - & - & - \\
Width (MeV/$c^{2}$) & & 10 -- 100 & 120 -- 250 & - & - & - \\
\hline
\hline
					 & q-coal. & 1.8 $\times$ 10$^{8}$ & - & - & 6.2 $\times$ 10$^{4}$ & 1.5 $\times$ 10$^{4}$\\
$\mathrm{S}_{\mathrm{raw}}$ & h-coal. & 6.4 $\times$ 10$^{6}$ $^{\dagger}$ & - & - & 5.5 $\times$ 10$^{4}$ & 5.1 $\times$ 10$^{4}$ \\
  					& thermal & 3.6 $\times$ 10$^{10}$ & 5.5 $\times$ 10$^{7}$& 6.7 $\times$ 10$^5$ & 1.9 $\times$ 10$^{5}$ & 4.1 $\times$ 10$^{4}$\\

\hline
               & q-coal. & 130-3.5 & - & - & - & - \\
\significance & h-coal. & - & - & - & - & - \\
 			   & thermal & 2600-70 & 520-360 & - & - & - \\

\hline
\end{tabular}
}
\end{table}

The total number of signals (S$_{\mathrm{raw}}$) detectable in ALICE with a minimum bias \PbPb integrated luminosity of 10~\nbInv have been estimated assuming the same detector performance as in Run 2 \cite{Jacazio:2017dvy, Acharya:2018ckj}. 
The significance is defined as \significance, where S and B are the integrals of the signal and background distributions, respectively, in a $\pm~3~\sigma$ window centered at the nominal mass from \cite{pdg:2018}. 
$\sigma$ is $\Gamma/2.35$, where $\Gamma$ is the resonance width taken from \cite{pdg:2018}. 
The significance for f$_0$(980) and N(1835) was extracted assuming a combinatorial background in the invariant mass range under study. 
Such combinatorics was computed based on particle species that can populate the invariant mass distribution, 
making use of the corresponding momentum distribution as measured in ALICE in Run 2 (e.g. individual primary charged pions paired as candidates for the f$_{0}$(980)$~\rightarrow~\pi^{+}+\pi^{-}$ channel)
and assuming a uniform distribution in $\varphi$ and $\eta$ \footnote{An additional factor is introduced if the decay particle is reconstructed via invariant mass, since the candidate may belong also to the background.}.
The resulting significance is reported in the last row of Tab.~\ref{tab:yield} for the most pessimistic scenario, in which production occurs via quark coalescence, and the most optimistic scenario, corresponding to thermal production.

Measurements of f$_{0}$(980) and N(1875) will be feasible in Runs 3 and 4 and will shed light on the highly-debated nature of the states (hadrons or hadronic molecules). 
In particular, the N(1875) can be considered a molecular bound state and at the same time the strange partner of the recently discovered pentaquark P$_{c}$ \cite{He:2017aps}. 
Because the structure of exotic states is related to the fundamental properties of Quantum Chromodynamics (QCD), their observation can provide new insights on the properties of QCD at finite temperature and density, for instance that tetraquark condensation may lead to a second chiral phase transition \cite{Cho:2017dcy}.
Several possible states have been studied and predictions are available on the expected yields at LHC energies \cite{Cho:2017dcy}. Among the possible dibaryon bound states, the N$\Omega$, N$\Xi$ and N$\Lambda_c$ look promising in terms of detection feasibility. Their detection and baryon-baryon correlations will be useful for hyperon correlation studies, providing new insights into the baryon-baryon attractive potential as well as upper limits on the formation of such bound states in central heavy-ion collisions.

Very recently, a study was reported on the production of exotic charmonia, specifically  the possible tetraquark state X(3872),  in Pb-Pb and Kr-Kr collisions at LHC collision energies \cite{Andronic:2019wva}. For  X(3872) production in Pb-Pb collisions the authors predict, using the statistical hadronization model for charm, a transverse momentum distribution similar in shape to that for $J/\psi$ mesons with a strong enhancement at low transverse momenta and a production yield of about 1\% relative to that for $J/\psi$. It would be interesting and feasible to test this prediction in LHC Run 3 and especially Run 4.

\subsubsection{Implications of anti-nuclei measurements for cosmic-ray physics and dark-matter searches}
\label{sec:astro}

The HL-LHC physics program with \pp and \pPb collisions will allow for precision measurements of anti-nuclei production and related observables that have implications for cosmic-ray physics and dark-matter searches.
Cosmic-ray (CR) anti-nuclei \antip, \antid, and \antihethree~have long been considered as probes of new physics, such as dark matter annihilation~\cite{Donato:1999gy, Baer:2005tw, Donato:2008yx, Brauninger:2009pe, Kadastik:2009ts, Cui:2010ud, Dal:2012my, Ibarra:2012cc, Fornengo:2013osa, Carlson:2014ssa, Aramaki:2015pii,Korsmeier:2017xzj}. Detecting these particles is one of the main goals of various CR experiments (e.g. AMS-02 ~\cite{Giovacchini:2007dwa, kounineHebar}, GAPS~\cite{vonDoetinchem:2015zva, Aramaki:2015laa}, BESS-Polar \cite{Abe:2011nx}). 

The galaxy produces CR anti-nuclei as secondaries, due to collisions of CR protons and helium with interstellar matter. Information from accelerator experiments is essential for the theoretical description of the background constituted by these secondary anti-nuclei.
The flux of secondary anti-nuclei can be calculated with only minor sensitivity to the details of CR astrophysics. The point is to use secondary-to-secondary flux ratios, where astrophysical uncertainties largely cancel. 
Secondary \antip, \antid, and \antihethree~are all formed dominantly by the same set of reactions. 
%
Using this basic fact, an explicit prediction for the locally observable flux of secondary \antid, relative to the flux of secondary \antip~can be derived~\cite{Ginzburg:1990sk,Katz:2009yd,Blum:2017qnn}:
\begin{eqnarray}
\label{eq:ad2ap}\frac{J_{\ad}(\R)}{J_{\ap}(\R)}&=&\frac{\int \rmd\epsilon\,J_{\mathrm{p}}(\epsilon)\,\frac{\rmd\sigma_{\mathrm{pp}\to\ad}(\epsilon,\epsilon_{\ad})}{\rmd\epsilon_{\ad}}}{\int \rmd\epsilon\,J_{\mathrm{p}}(\epsilon)\,\frac{\rmd\sigma_{\mathrm{pp}\to\ap}(\epsilon,\epsilon_{\ap})}{\rmd\epsilon_{\ap}}+\left(\sigma_{\ad}(\epsilon_{\ad})-\sigma_{\ap}(\epsilon_{\ap})\right)J_{\ap}(\R)}.
\end{eqnarray}
\noindent 
Here $J_{\ad}(\R)$ is the predicted \antid~ flux, given at magnetic rigidity $\R = p/Z$, where $p$ is the momentum and $Z$ is the electric charge. $J_{\ap}(\R)$ is the (already well-measured~\cite{Aguilar:2016kjl}) \antip~ flux at the same rigidity, and  
$J_{\mathrm{p}}(\epsilon)$ is the proton flux~\cite{Aguilar:2015ooa} at energy $\epsilon$. 
$\frac{\rmd\sigma_{\mathrm{pp}\to\bar x}(\epsilon,\epsilon_{\bar x})}{\rmd\epsilon_{\bar x}}$ and $\sigma_{\bar x}(\epsilon_{\bar x})$ are the inclusive production and inelastic cross sections, respectively, with $x={\rm d},\,{\rm p}$. 
%
%
The particle energy $\epsilon_{\bar x}$ for a nucleus with mass number \Anucl is evaluated at $\R$: $\epsilon_{\ap}=\sqrt{\R^{2}+\mathrm{A}^2m_{\mathrm{p}}^{2}}$. 
To describe \antihethree~ we use an analogous expression to Eq.~(\ref{eq:ad2ap}), adding the production of $\at$ which decays to \antihethree. More details, including the relation of the differential cross section appearing in Eq.~(\ref{eq:ad2ap}) to the Lorentz-invariant differential cross section measurable at the LHC, can be found in~\cite{Blum:2017qnn}.
%

%

The cross section for producing an anti-nucleus can be parameterized in terms of the anti-proton cross section, using the coalescence factor \BA: 
$(\epsilon_\mathrm{A}\rmd\sigma/\rmd^3p)_{\mathrm{pp}\to \mathrm{A}} = B_{\mathrm{A}}/\sigma_{\mathrm{pp}}^{\mathrm{A}-1} [(\epsilon_{\antip} \rmd\sigma/\rmd^3p)_{\mathrm{pp}\to \antip}]^\mathrm{A}$, 
where $\sigma_{\mathrm{pp}}$ is the total inelastic \pp cross section. 
Here, for simplicity, threshold effects are omitted~\cite{Duperray:2002pj,Duperray:2003tv,Blum:2017qnn}. 
%
Using Eq.~(\ref{eq:ad2ap}), and plugging in the coalescence factors experimentally obtained at the LHC~\cite{Acharya:2017fvb}, the predicted flux ratios can be obtained. 
Secondary CR production is dominated by the low \pT\ region. As a result, the impact on the CR flux, due to \pT-dependent \BA, can be factored out to good approximation, allowing us to derive simple approximate formulae 
\footnote{Note that the rigidity $\mathcal{R}=100$~GV refers to the CR experiment rest frame, which is boosted w.r.t. the proton-proton collision centre of mass frame. In the proton-proton collision centre of mass frame, the anti-nuclei are formed close to threshold.}~\cite{Blum:2017qnn}:
\begin{eqnarray}\label{eq:J2}\frac{J_{\overline{\rm d}}\left(\R\right)}{J_{\overline{\rm p}}\left(\R\right)}\huge|_{\R=100{\rm GV}}&\approx&4\times10^{-4}\,\left(\frac{B_2}{1.5\times10^{-2}~\rm GeV^2}\right),\\
%
%
\label{eq:J3}\frac{J_{\antihethree}\left(\R\right)}{J_{\antip}\left(\R\right)}\huge|_{\R=100{\rm GV}}&\approx&2\times10^{-7}\,\left(\frac{B_3}{1.5\times10^{-4}~\rm GeV^4}\right),\end{eqnarray}
where, for CR studies, the $B_2$ and $B_3$ values should be read from the average yield in the range $\pT/\mathrm{A}=(0-0.5)~\gmom$ in the accelerator analysis. 
The precision requirements (O(10$\%$)) on the flux ratio determination for the astrophysical applications discussed here will be matched by measuring $B_{2}$ and $B_{3}$ in the lowest \pT\ bin with a relative systematic uncertainty of the order of 10$\%$ \cite{Acharya:2017fvb}. The latter largely dominates the statistical uncertainty that is expected to be of O(0.1$\%$)  already with \Lint~=~6~\pbInv in \pp collisions at $\sqrt{s}~=~5.5$~TeV. 
Moreover, a first measurement of $B_{4}$ in \pp collisions will be achievable in the same sample. The statistical precision on $B_{4}$ can be lowered to the 10$\%$ level if a luminosity of 200~\pbInv in \pp collisions at $\sqrt{s}~=~14$~TeV can be inspected with a dedicated trigger for (anti-)nuclei.

It is important to note that the \BA measurement~\cite{Acharya:2017fvb} performed by ALICE during the LHC Run 1 was confined to mid-rapidity, $|y|<0.5$. Possible $y$ dependence of the coalescence factor \BA at $y=\mathcal{O}(1)$, or variation of the \antip\ differential cross section with respect to current parameterisations~\cite{Donato:2017ywo}, would affect the prediction in Eqs.~(\ref{eq:J2}-\ref{eq:J3}). It would be an important task of future LHC measurements to test these effects. Similarly important, albeit -- possibly -- beyond the reach of the LHC, would be to study the low $\sqrt{s}=\mathcal{O}(10)$~GeV behaviour of \BA~\cite{Blum:2017qnn}.

\subsubsection{Implications of anti-nuclei measurements and hyperon-nucleon correlations for neutron star physics}

Although the neutron star crust is composed of neutrons, within the innermost core hyperons could be present \cite{RMP-88-035004-2016}.
Whether or not this scenario holds true depends on the two- and three-body hyperon nucleon interactions (YN and YNN) that are still only rather scarcely constrained experimentally.\\
At present the mass range for observed neutron stars is about (0.9 --  3.0)$M_{\bigodot}$ within errors \cite{ARNPS-62-485-2012}, where $M_{\bigodot}$ stands for the solar mass. 
The equation of state (EoS) of neutron stars is constrained by the mass-radius relationship, in particular, the maximum mass ($M_{\scriptsize{\mbox{max}}}$). An EoS with "conventional" (N$+\pi$) degrees of freedom provides $M_{\scriptsize{\mbox{max}}}$ invariably above 2$M_{\bigodot}$ \cite{HyperfineInteract-233-131-2015,Nature-467-1081-2010,AstrophysJ-832-167-2016}. 
However, adding the $\Lambda$ hyperon in the hadronic basis softens the EoS and, as a consequence, significantly reduces $M_{\scriptsize{\mbox{max}}}$. 
The solution of this so-called "hyperon puzzle" is non-trivial, and is presently the subject of very active research.

Thanks to the large yields of free hyperons and  exotic (anti-)hyper-nuclei that can be produced in collider experiments and the excellent particle identification capabilities of the ALICE experiment, the upcoming experimental campaigns in Runs 3 and 4 at the LHC offer a unique opportunity to quantitatively characterise hypermatter under controlled (laboratory) conditions and infer on the equation of state of compact objects as neutron stars.

One of the crucial  element to solve the ``hyperon puzzle'' is the determination of the $\Lambda$NN three-body forces. 
Calculations show that with a parameterization of these forces compatible with the hyper-nuclear binding energies, the admixture of $\Lambda$'s in neutron star matter gets strongly reduced such that the pressure to support a 2$M_{\bigodot}$ neutron star can be maintained \cite{Lonardoni:2014bwa}. 
The observation of neutron-rich $\Lambda$ hyper-nuclei like \hypfour\ etc. at colliders could be very promising for studying the effects of the
three-body $\Lambda$NN forces in dense strongly interacting matter since a precise knowledge of light neutron-rich hyper-nuclei energy level structure could imply far-reaching consequences for  dense strange stellar matter properties.

Another promising way to contribute to the understanding of the hyperon puzzle is to pin down the hyperon-nucleon two-body interaction
for hyperons such as $\Sigma^{-}$ and $\Xi^-$. These hyperons can also be produced within neutron rich matter 
(n + n $\rightarrow \Sigma^- +$ p, $\Lambda$ + n $\rightarrow \Xi^- +$ p) depending on their interaction with the surrounding neutrons.
Some models assume a repulsive p$\Sigma$ interaction and postulate that $\Sigma^{-}$ can appear in neutron rich matter only starting from baryon densities $\rho \simeq 4\rho_{0}$ (where $\rho_{0}$ is the density of standard nuclear matter) \cite{PRC-93-035808-2016}.
For $\Xi$, no reliable experimental information about the interaction is available.
Recent studies \cite{Acharya:2018gyz} showed that the femtoscopy technique applied to \pp and \pPb collisions at LHC energies are particularly suited to study the final state interaction between nucleons and strange baryons (e.g.: $\Lambda$--p) and between two strange baryons (e.g.: $\Lambda$--$\Lambda$). 
Indeed, small colliding systems such as \pp and \pPb lead to hadron sources of rather small dimensions, of the order of 1 fm, in the range where the strong interaction is mostly evident.
Also, the production mechanism of hadrons in minimum bias \pp and \pPb collisions is not affected by the intermediate creation of a QGP and its time-dependent evolution as in \PbPb collisions at LHC energies. This allows for a more precise study of the hadron-hadron interactions.

\begin{figure}
\centering
\includegraphics[width=.5\textwidth]{\main/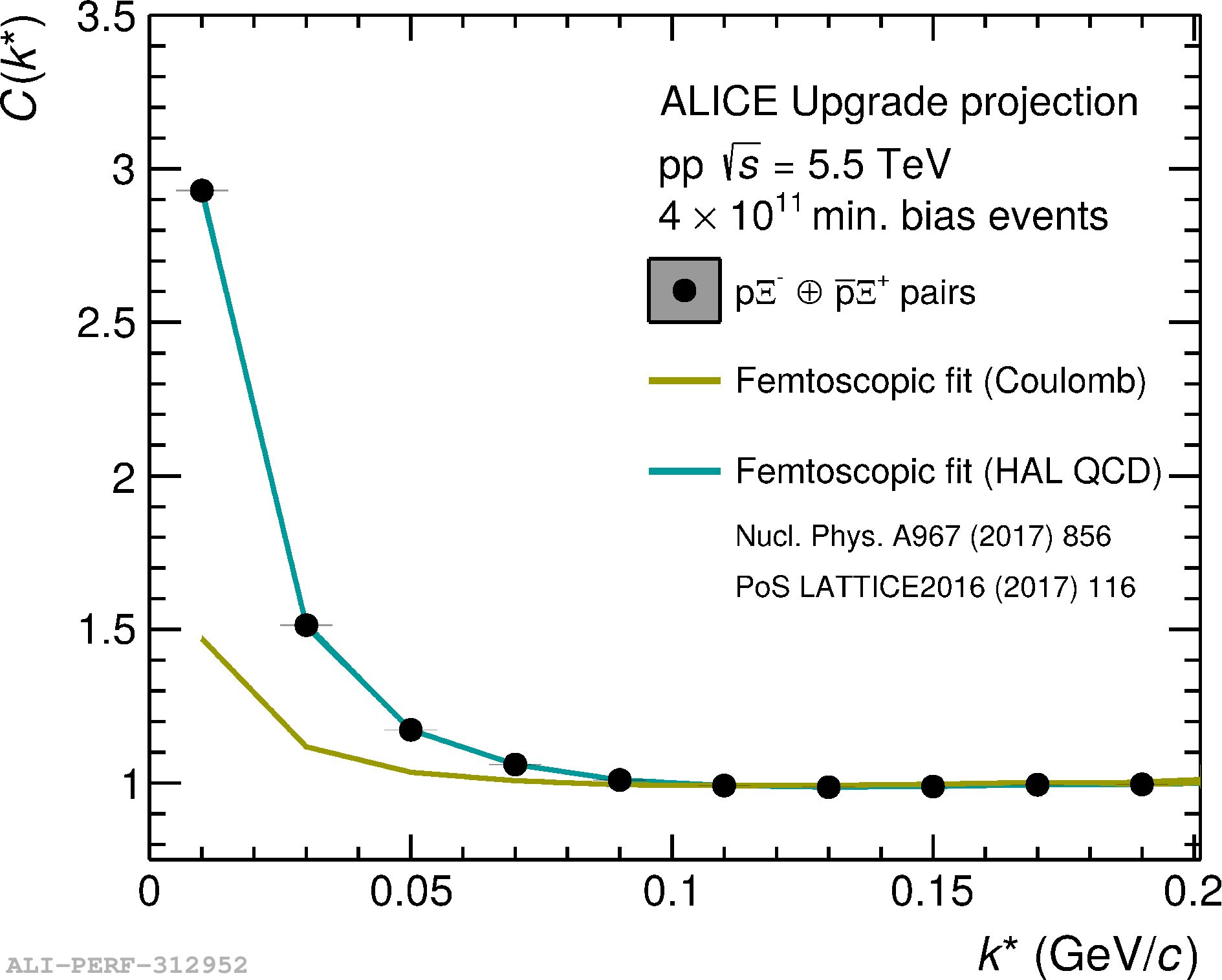}
\caption{Expected p$\Xi^-$ + $\bar{\mathrm{p}}\Xi^+$ correlation for \pp collisions at $\sqrt{s} = 5.5$ TeV and $4\times 10^{11}$ minimum bias events, corresponding to \Lint~=~6~\pbInv. Only statistical errors have been estimated. Figure from Ref.~\cite{ALICE-PUBLIC-2019-001}.}
\label{pXiProj}
\end{figure}

Among the quantitative results obtained by ALICE in Run 2 at the LHC is the first observation of the attractive p$\Xi^-$ interaction. 
Figure \ref{pXiProj}  
shows the expected p$\Xi$ correlation for the Run 3 \pp sample as a function of the relative momentum $k*$. 
The projection is obtained on the basis of the current prediction by the HAL-QCD lQCD group \cite{Sasaki:2017ysy,Hatsuda:2017uxk} that is in agreement with the Run 2 results.
The clear deviation from the Coulomb-only correlation function shows the effect of the strong attractive interaction 
and the expected statistics will allow for a quantitative determination of the scattering parameters and the test of different hadronic models \cite{Haidenbauer:2018jvl,Rijken:2006kg}.
The investigation will also be extended to the $\Sigma^0$ hyperon, since in Run 3 and 4 we expect a total of 500,000 p$\Sigma^0$ pairs to be used to study the femtoscopy correlation.

In summary, massive neutron stars with $M \sim 2M_{\bigodot}$ are very intriguing recent observations in relativistic astrophysics. 
An improved account of the two-body YN interaction, the three-body $\Lambda$NN forces, and the contribution of multi-strange hyperons in the EoS is crucially important for more realistic description of compact astrophysical objects, in particular neutron and hybrid stars. The measurement of hyper-nuclei and hyperon correlations with the HL-LHC project are suggested as a promising tool for astrophysical applications.

\subsection{Fluctuations of conserved charges} 
\label{sec:fluctuations}
\subsubsection{Physics introduction and observables}

In the phase diagram of strongly interacting matter at zero net baryon density, the presence of a chiral phase transition between hadronic matter and a QGP has been conjectured \cite{Pisarski:1983ms}, and arguments have been presented~\cite{Ejiri:2009ac,Ding:2018auz} in lQCD that the transition, for vanishing light quark masses, is of second order and belongs to the O(4) universality class. Due to the small but finite physical quark masses, in lQCD a rapid crossover is found \cite{Aoki:2006we,Aoki:2009sc,Borsanyi:2010bp,Bazavov:2011nk,Bhattacharya:2014ara} which, however, exhibits pseudo-critical features due to the smallness of the u- and d-quark masses and the proximity of the crossover region to the O(4) line \cite{Ejiri:2009ac,Ding:2013lfa}. 

In general, fluctuations can be linked to critical behaviour associated with a phase transition, and it has been pointed out that fluctuations of conserved charges in heavy-ion collisions can provide an experimental observable to test for critical behaviour in the phase diagram of strongly interacting matter \cite{Ejiri:2005wq,Bazavov:2017dus,Friman:2011pf,Bazavov:2012jq}. These fluctuations can be related to susceptibilities, specifically to the derivatives of the pressure with respect to the chemical potentials corresponding to the conserved charges. Here, the relevant `charges' are baryon number $B$, strangeness $S$, and electrical charge $Q$, and the corresponding chemical potentials are $\mu_B$, $\mu_S$, and $\mu_Q$. The susceptibilities are defined (see e.g. \cite{Bazavov:2012jq,Bellwied:2015lba}) in terms of dimensionless normalized chemical potentials \(\hat{\mu}_X\equiv \mu_X/T \) \: as

\begin{equation}
\chi_{ijk}^{BQS}(T) = \left.
\frac{\partial P(T,\hat{\mu})/T^4}{\partial\hat{\mu}_B^i \partial\hat{\mu}_Q^j \partial\hat{\mu}_S^k}\right|_{\hat{\mu}=0} \; .
\label{suscept}
\end{equation} 

\noindent The generalized susceptibilities can be computed in lQCD at vanishing chemical potential, exactly the conditions probed by experiments at the LHC. Within the Grand Canonical Ensemble (GCE), these generalized susceptibilities can be related to experimental measurements of the fluctuations of particle multiplicities, such as the net number of baryons. For instance, a measurement of higher moments or cumulants of net baryon number in relativistic nuclear collisions can be directly related \cite{Karsch:2010ck,Skokov:2012ds,Karsch:2012wm,Karsch:2017mvg,Borsanyi:2013hza,Borsanyi:2014ewa} to theoretical predictions from lQCD or from more phenomenological models of the chiral phase transition \cite{Almasi:2017bhq,Parotto:2018pwx} to shed light on the possible critical behaviour near the QCD phase boundary. 
For a distribution of the net baryon number, $\Delta N_B = N_B - N_{\overline{B}}$, with moments defined as 

\begin{equation}
\mu_i = \langle (\Delta N_B - \langle \Delta N_B \rangle )^i \rangle ,
\end{equation}
the cumulants $\kappa_i$ can be directly linked to the generalized susceptibilities such as

\begin{equation}
\kappa_2 = \mu_2 = VT^3 \chi_2^B
\end{equation}
\begin{equation}
\kappa_3 = \mu_3 = VT^3 \chi_3^B  
\end{equation}
\begin{equation}
\kappa_4 = \mu_4 - 3\mu_2^2 = VT^3 \chi_4^B.
\end{equation}

\begin{figure}[ht]
\begin{center}
\includegraphics[width=0.45\textwidth]{\main/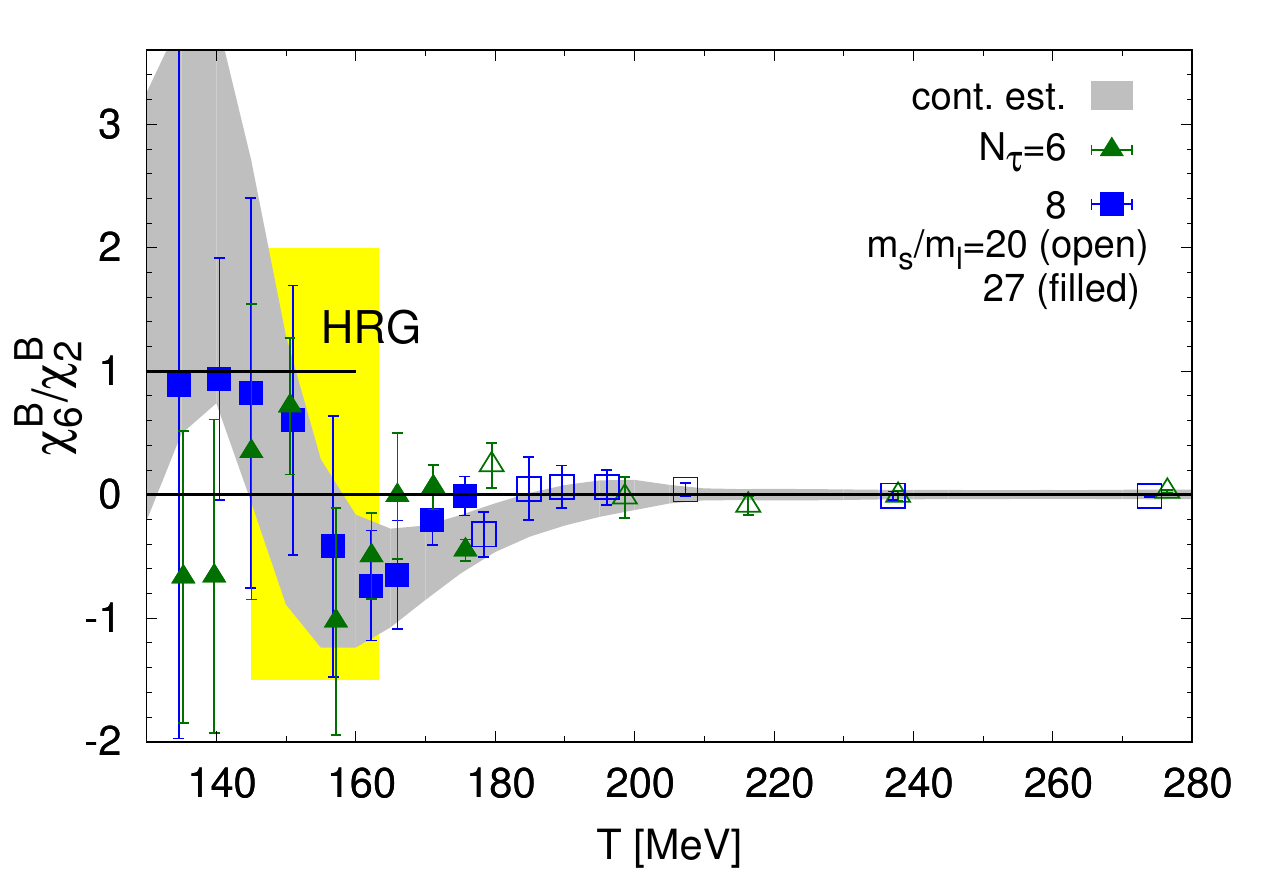}
\includegraphics[width=0.46\textwidth]{\main/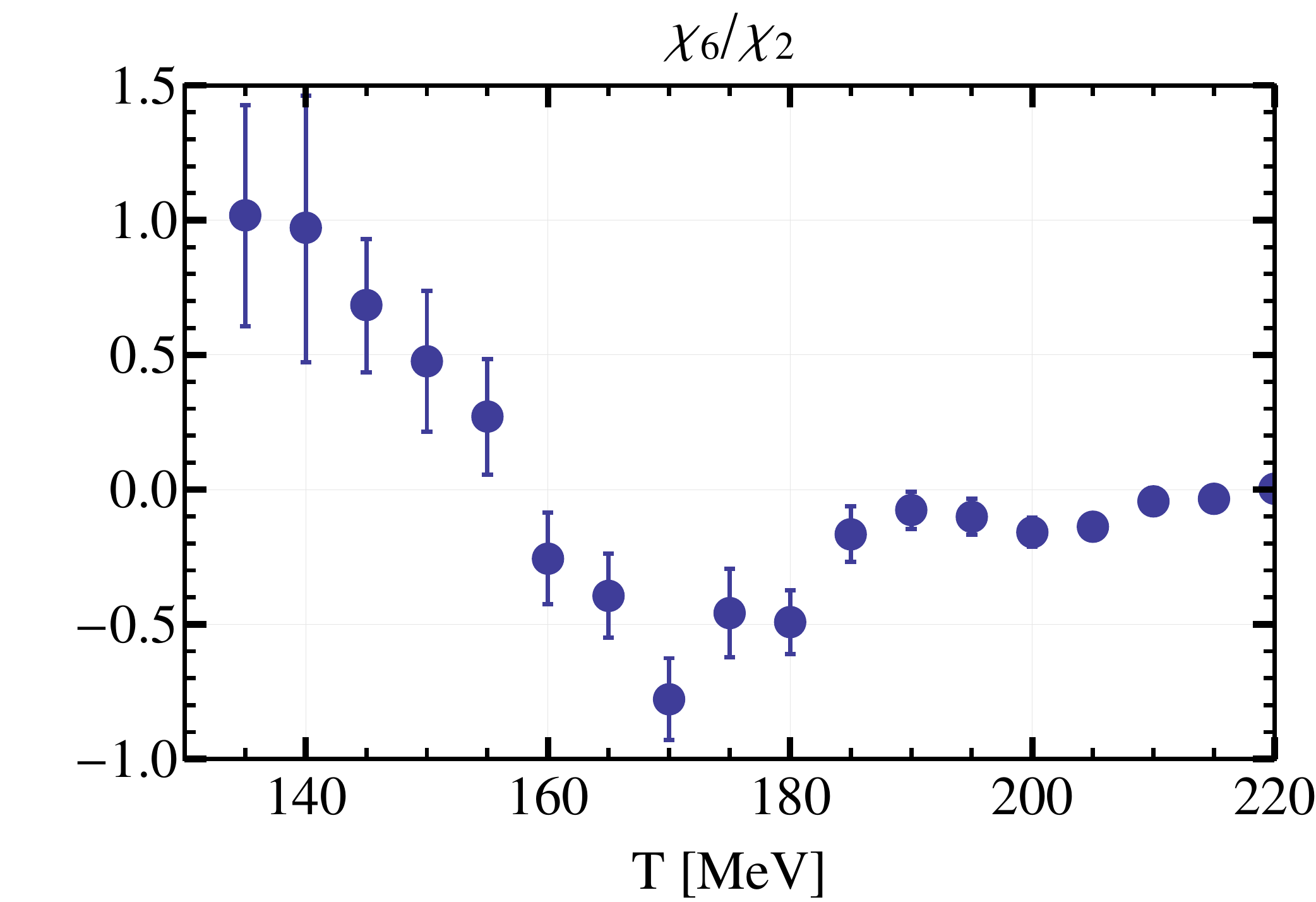}
\end{center}
\caption{Ratio of sixth to second-order baryon number susceptibilities from lQCD. The left-hand figure is from \cite{Bazavov:2017dus}. The right-hand figure is calculated from recent lQCD data on  sixth and  second  order susceptibilities from \cite{Borsanyi:2018grb}. }  
\label{fig:chi62}
\end{figure}

In the O(4) universality class, a singular contribution to the pressure shows up for higher order  moments.
More specifically, at vanishing chemical potential, all odd susceptibilities of the net baryon number vanish.  
In addition, in the O(4) universality class, the second- and fourth-order susceptibilities remain finite  at  the  phase  transition  temperature  at $\mu_B~=~0$  in  the  chiral  limit,
implying  that  only  sixth-  and  higher-order  susceptibilities  diverge \cite{Ejiri:2005wq,Friman:2011pf}. Thus, for  physical  quark  masses and at $\mu_B~=~0$, only higher order cumulants $\kappa_n$ with $n\geq 6$ can exhibit O(4) criticality,  whereas at finite $\mu_B$ this is already the case for $\kappa_n$ with $n\geq 3$.  

Sensitivity to chiral criticality due to the vicinity of the O(4) line at $\mu_B~=~0$ is borne out in phenomenological models as is shown in \cite{Friman:2011pf,Almasi:2017bhq}, and in lQCD predictions \cite{Bazavov:2017dus,Borsanyi:2018grb}, by strong deviations of $\chi_6^B/\chi_2^B$ from unity as shown in Fig.~\ref{fig:chi62}.

We note that a convenient baseline for the cumulants of multiplicity distributions and fluctuations of produced particles in relativistic nuclear collisions can be obtained in the framework of the hadron re\-so\-nance gas (HRG) \cite{Allton:2005gk,Karsch:2010ck,BraunMunzinger:2011ta,Borsanyi:2018grb,Luo:2017faz}. In this model, uncorrelated Poissonian fluctuations of baryon and anti-baryon multiplicities result in a Skellam distribution for the net baryon number, in which the higher moments and cumulants can all be related to the first moments in the following way \cite{BraunMunzinger:2011ta,BraunMunzinger:2011dn, Braun-Munzinger:2018yru}:
\begin{equation}
\label{lbaseline}
\kappa_n(N_B - N_{\overline{B}}) = \langle (N_B +(-1)^n N_{\overline{B}}) \rangle
\end{equation}
For zero net baryon number then all odd cumulants vanish and all even cumulants are identical.

Measuring such cumulants with precision poses a formidable experimental challenge due to the requirement of very large data sets ($> 10^9$ events of a particular event or centrality class) with superb control of systematic uncertainties. As a first physics case to consider along this line, the impact of measuring the distribution of net protons as a proxy for net baryons needs to be studied further. We note that, at LHC energy and low transverse momentum, particle production near mid-rapidity takes place mostly in gluonic processes, implying that isospin asymmetries, as in the colliding nuclei, are absent. As a consequence, the production yields of protons and neutrons should be very close. For light nuclei this isospin symmetry has been checked experimentally, albeit with significant uncertainties. 
In addition, non-critical contributions to the cumulants from volume fluctuations and global baryon number conservation \cite{Skokov:2012ds,Braun-Munzinger:2016yjz, Braun-Munzinger:2018yru} need to be evaluated and the data corrected accordingly. Furthermore, in particular for comparison to lQCD predictions, care needs to be taken to keep experimental cuts such as in \pT\ to a minimum insofar as such cuts cannot be introduced in lQCD \cite{Karsch:2015zna,Alba:2015iva}.
     
Two-particle correlations with net baryons can also be used to explore transport properties of the hydrodynamic evolution. The baryon diffusion constant $D$ is a fundamental transport property of the quark-gluon plasma, similar to shear viscosity $\eta$ or bulk viscosity $\zeta$. It characterizes the mobility of baryon number, and is predicted to be finite at the LHC despite the fact that $\mu_{B}\sim0$. A two-particle correlation function as been proposed~\cite{Floerchinger:2015efa}, which explores correlations of net-baryon fluctuations as a function of separations in azimuthal angle and rapidity, and can provide experimental constraints on the diffusion coefficient $D$. As $\mu_{B}\sim0$ at the LHC, such an analysis has yet to be carried out in Run 1 and 2 data since it is statistically challenging, and will be greatly aided by the increase by about a factor 100 in the \PbPb integrated luminosity foreseen for Runs 3 and 4.

\subsubsection{State of the art experimental measurements and present limitations}

Net proton fluctuations measured by the ALICE experiment and in the STAR beam energy scan program provide interesting and stimulating results. The measurements at STAR~\cite{Adamczyk:2013dal} complement the corresponding measurements from ALICE, which will make it possible to pin down the global structure of the phase diagram of strongly interacting matter in a wide range of temperatures and net-baryon densities. However, before drawing firm conclusions by confronting theoretical calculations with  data, non-dynamical contributions stemming from unavoidable fluctuations of  participant nucleons and  overall baryon number conservation have to be subtracted from the experimental measurements. Both of these non-dynamical contributions, which exist neither in lQCD nor in the HRG model, lead to deviations from the baseline as defined in Eq.~\ref{lbaseline}. Indeed, the acceptance dependence of the second-order cumulants of net-protons measured by ALICE~\cite{Rustamov:2017lio} exhibits deviations from the non-critical (Skellam) baseline. However, these deviations were explained by global baryon number conservation~\cite{Rustamov:2017lio, Braun-Munzinger:2016yjz, Braun-Munzinger:2018yru}, which, in accordance with the experimental findings, decreases the amount of fluctuations with the increasing acceptance. This is the first experimental verification of the lQCD predictions for the second-order cumulants of net-baryon distributions. This also serves as a strong support of the HRG model, in that experimental measurements of the second cumulants of net-protons do not show any evidence of criticality and actually coincide with the second cumulants of the Skellam distribution.  In order to probe critical phenomena, higher cumulants beyond the second order have to be addressed. 

\begin{figure}[ht]
\begin{center}
\includegraphics[width=0.6\textwidth]{\main/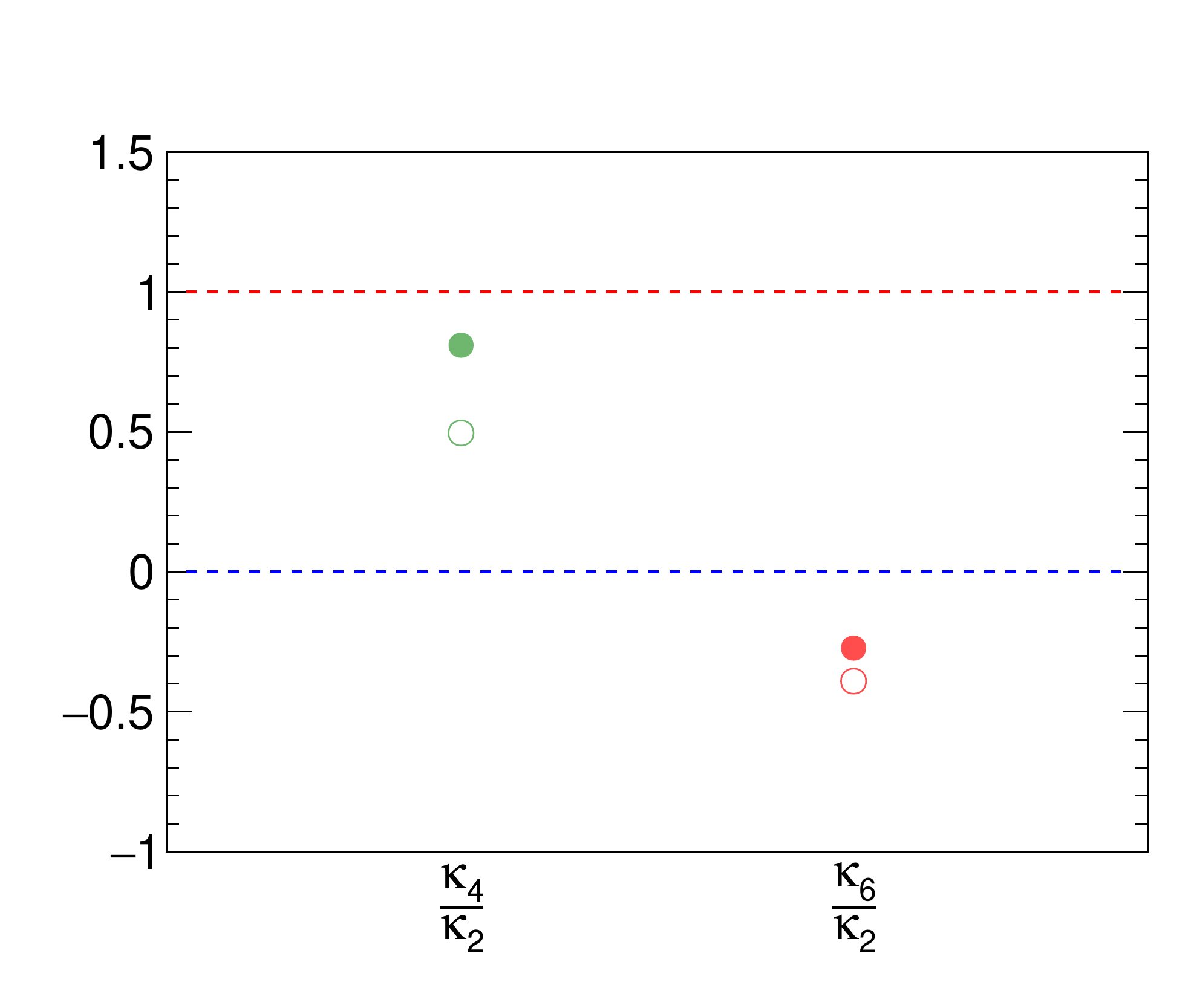}
\end{center}
\caption{$\kappa_{4}/\kappa_{2}$ and $\kappa_{6}/\kappa_{2}$  as calculated within PQM~\cite{Almasi:2017bhq} model (open symbols).  After taking into account contributions from participant nucleon fluctuations and global baryon number conservation~\cite{Braun-Munzinger:2016yjz, Braun-Munzinger:2018yru}, the deviations from unity decrease (closed symbols).} 
\label{netpALICE_STAR}
\end{figure}

As mentioned in the previous section, even at vanishing net-baryon densities, lQCD and other theoretical calculations such as Polyakov-loop extended Quark- Meson model (PQM)~\cite{Almasi:2017bhq} predict critical fluctuations encoded in the deviations of net-baryon $\kappa_{4}/\kappa_{2}$ and $\kappa_{6}/\kappa_{2}$ from unity. Moreover, at the pseudo critical temperature of about 156 MeV the magnitudes of $\kappa_{4}/\kappa_{2}$ and $\kappa_{6}/\kappa_{2}$  are predicted in Ref.~\cite{Almasi:2017bhq} to be 0.5 and -0.39, respectively. Similar values of $\kappa_{6}/\kappa_{2}$ are quoted in different lQCD calculations as presented in Fig~\ref{fig:chi62}.  These numbers, shown in  Fig.~\ref{netpALICE_STAR}, do not take into account experimental artefacts such as global net-baryon number conservation and unavoidable fluctuations of participating nucleons from event to event. Also shown are the values of of $\kappa_{4}/\kappa_{2}$ and $\kappa_{6}/\kappa_{2}$ after accounting for these non-dynamical effects using the procedure in Refs.~\cite{Braun-Munzinger:2016yjz, Braun-Munzinger:2018yru, Bzdak:2012an}. Even after accounting for participant fluctuations and global baryon number conservation we observe deviations in $\kappa_{4}/\kappa_{2}$ and $\kappa_{6}/\kappa_{2}$ from unity, although they are somewhat reduced.   This motivates our experimental program  of measuring higher moments of net-proton distributions at the LHC energies. Also, fluctuation measurements  are underway in the strange baryon sector to approach measurements of net baryon number fluctuations. All this will be greatly helped by the anticipated dramatic increase in statistics in Runs 3 and 4.

\subsubsection{Projections for HL-LHC}

As discussed above, precise studies of the higher-order cumulants of particle multiplicity distributions are needed to verify theoretical predictions. 
In this section we estimate the statistics needed to address these measurements with the ALICE experiment. For this purpose two distinct Monte Carlo simulations have been developed. 
In the first approach, following recent developments in~\cite{Bzdak:2018axe}, the probability distribution function of net-protons is approximated  by a superposition of two Gaussian distributions which has four free parameters. 
Using the experimentally measured second cumulant of net-protons for 0-5 \% most central \PbPb collisions~\cite{Rustamov:2017lio} and the $\kappa_{4}/\kappa_{2}$ and $\kappa_{6}/\kappa_{2}$ ratios from the PQM model~\cite{Almasi:2017bhq}, absolute values for $\kappa_{4}$ and $\kappa_{6}$ were obtained first. 
These values were adjusted to account for fluctuations from participant nucleons in 0-5$\%$ most central \PbPb collisions and global baryon number conservation~\cite{Braun-Munzinger:2016yjz, Braun-Munzinger:2018yru}. Finally, the free parameters of the double Gaussian distribution were fixed using the expected values of $\kappa_{1}$ , $\kappa_{2}$ , $\kappa_{4}$ and $\kappa_{6}$ , where $\kappa_{1}$ equals zero by definition. 
The event-by-event net-proton number was sampled from the double Gaussian function  thus generating the net-proton distribution for a given number of events. 
In the second approach the probability distribution functions of protons and anti-protons are calculated separately by exploiting the Pearson curve method \cite{Behera:2017xwg}. This approach also needs four measurements as inputs, which are taken as the first four cumulants of the proton and anti-proton multiplicities measured by ALICE~\cite{Behera:2018wqk}.
The net-proton distribution for a given number of events is constructed by sampling the obtained proton and anti-proton probability distribution functions. 
In each approach, the resulting statistical uncertainties are obtained using the subsample method. 

\begin{figure}[ht]
\begin{center}$
\begin{array}{cc}
\includegraphics[width=0.45\textwidth]{\main/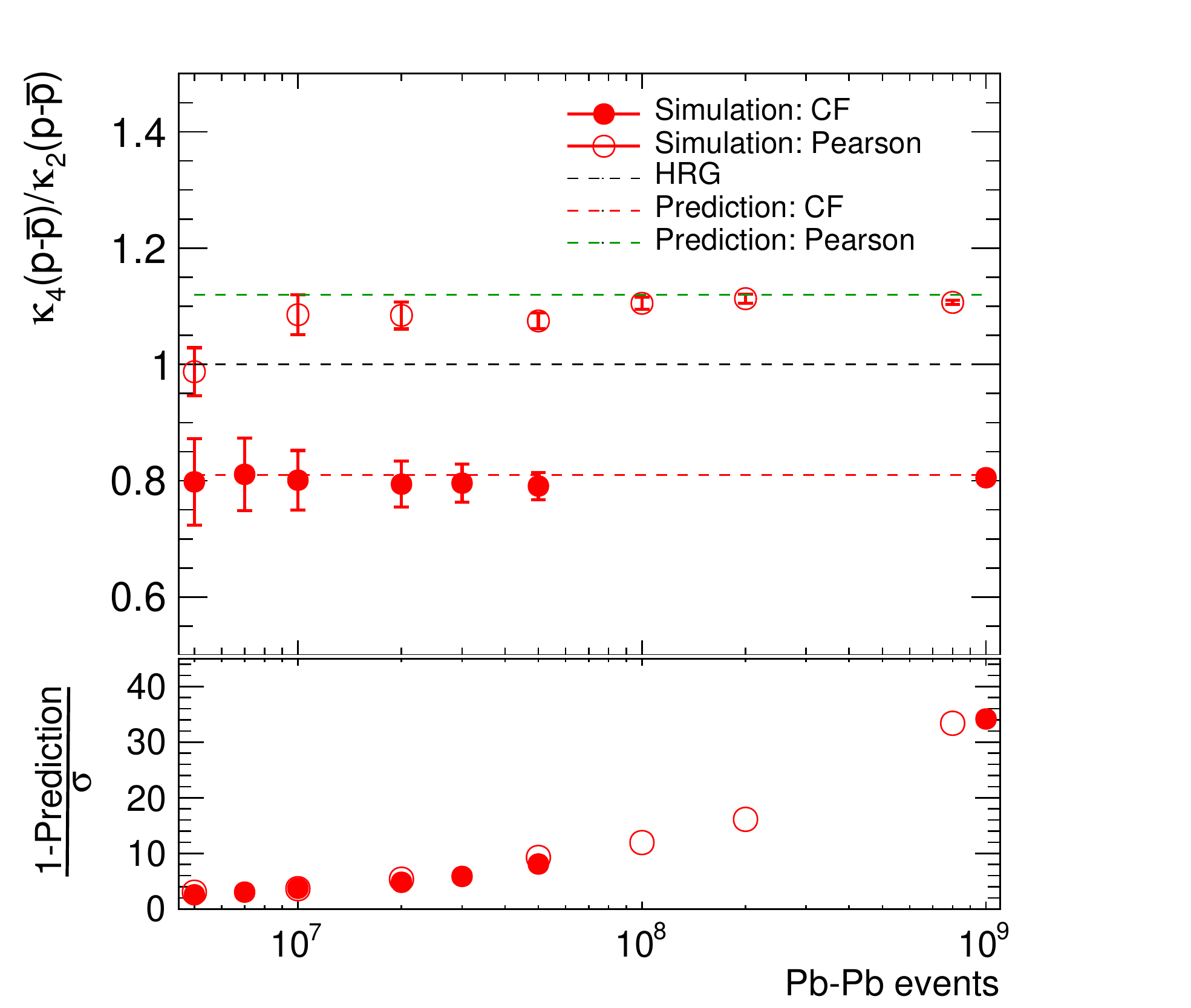} &
\includegraphics[width=0.45\textwidth]{\main/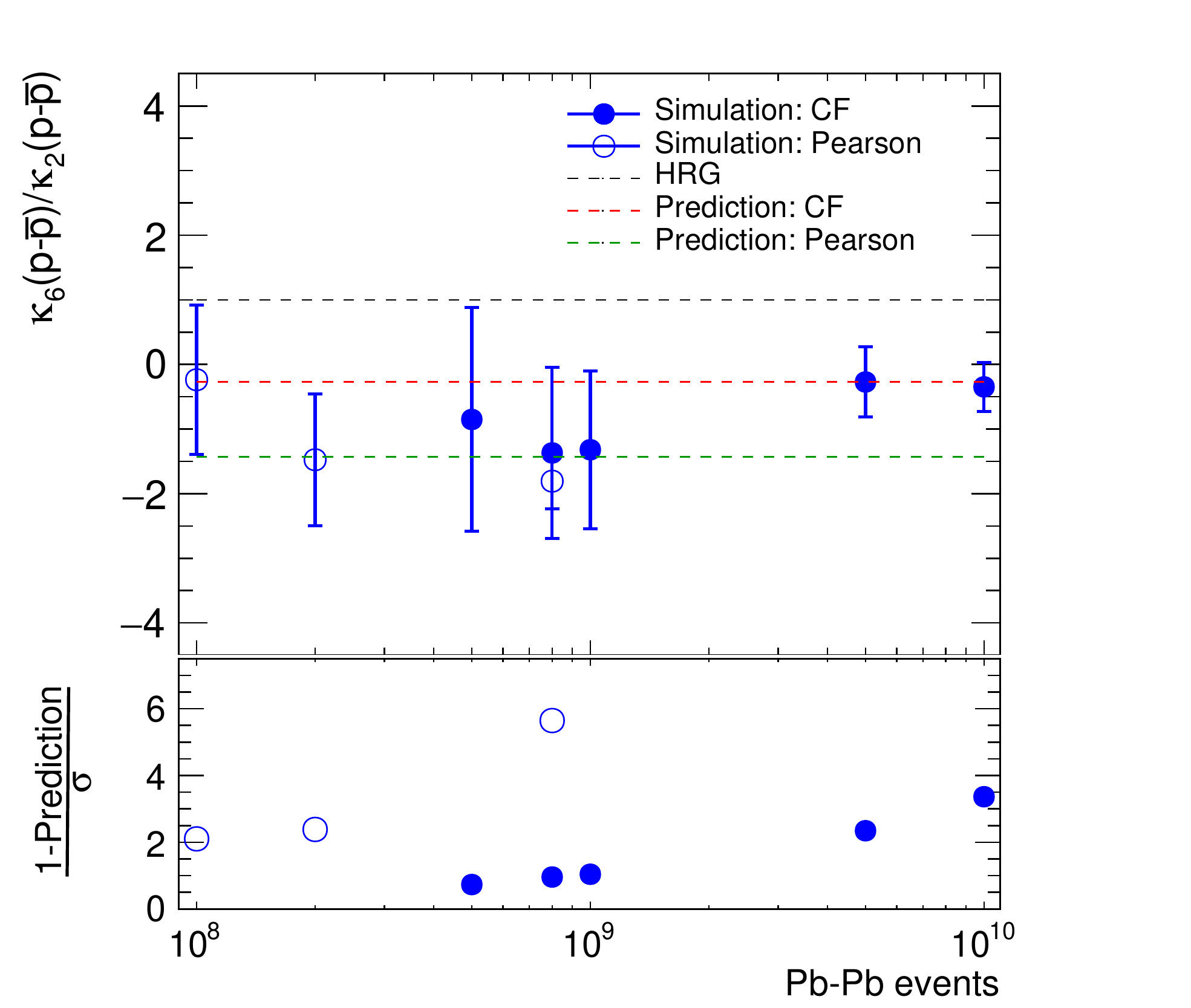}
\end{array}$
\end{center}
\caption{Simulated values of  $\kappa_{4}/\kappa_{2}$ (left panel) and $\kappa_{6}/\kappa_{2}$ (right panel) as functions of the generated number of events. Full symbols represent results obtained with the double Gaussian approach adjusted to reproduce critical fluctuations (CF) predicted in the PQM model~\cite{Almasi:2017bhq}. Open symbols are obtained with the Pearson Curve Method\cite{Behera:2017xwg}.}
\label{fig:c4c2toymc}
\end{figure}

The obtained results for $\kappa_{4}/\kappa_{2}$ and $\kappa_{6}/\kappa_{2}$ and their corresponding statistical uncertainties are shown in Fig.~\ref{fig:c4c2toymc} as a function of the simulated event statistics. The dashed red lines correspond to the input values predicted by PQM calculations of critical fluctuations (CF) and assuming a double Gaussian net-proton distribution, while the green dashed lines come from the Pearson curve method based on the lower-order cumulants measured by ALICE. As expected, with increasing statistics both $\kappa_{4}/\kappa_{2}$ and $\kappa_{6}/\kappa_{2}$ approach their nominal values. The statistics necessary to measure these cumulants are presented in the bottom panels of Fig.~\ref{fig:c4c2toymc}, where the deviations of the expected values from unity are quantified in units of the magnitudes of the statistical uncertainty ($\sigma$). As seen from the left panel of Fig.~\ref{fig:c4c2toymc}, for $\kappa_{4}/\kappa_{2}$ already 10 million events are sufficient to distinguish the expected critical fluctuations signal from unity with a statistical significance of $4\sigma$. Similar conclusions are obtained with the Pearson curve method.  
Several times this amount of data has already been recorded by ALICE, and the expected statistics in Runs 3 and 4 will make it possible to measure  $\kappa_{4}/\kappa_{2}$ with unprecedented precision.  

For $\kappa_{6}/\kappa_{2}$, however, significantly larger event sample is needed. As seen from the right panel of Fig.~\ref{fig:c4c2toymc}, more than 5 billion 0-5 $\%$ central events generated with the double Gaussian approach are needed in order to observe statistically significant deviations from unity in favor of the critical values indicated with the red dashed line. 
These would correspond to a minimum bias \PbPb integrated luminosity of 12.5 \nbInv in Runs 3 and 4.
Results obtained with the Pearson curve method indicate that more than 200 million 0-5 $\%$ central events (corresponding to a minimum bias \PbPb integrated luminosity of 0.5 \nbInv) would be sufficient in order to claim a significant deviation from unity in favour of the corresponding expected value. 
This difference in the estimation of the required statistics for $\kappa_{6}/\kappa_{2}$ comes mainly from the different baseline values of -1.43 and -0.27 used in the Pearson and double-Gaussian methods, respectively. 
In addition, the value of $\kappa_{2}$ used in the Pearson method is about two times smaller than measured in the experiment and used in the double Gaussian method. 
Track reconstruction and particle identification efficiency in the fiducial acceptance in $\eta$ and \pT~efficiencies, which would increase the required number of events for a given statistical precision, are not included in the study presented here because they depend on the details of the analysis.
Considering that these efficiencies are expected to range from 60$\%$ to 80$\%$, we conclude that the \PbPb integrated luminosity of 13 \nbInv foreseen in Runs 3 and 4 (see Ch.~\ref{sec:schedule}) will be sufficient to probe the critical phenomena contained in $\kappa_{6}/\kappa_{2}$.
      
\clearpage

\section{Flow and Correlations}
\label{sec:flow}

{ \small
\noindent \textbf{Coordinators}: Soumya Mohapatra (Nevis Labs, Columbia University)

\noindent \textbf{Contributors}: 
A.F.~Dobrin (CERN), 
S.~Floerchinger (Heidelberg University), 
P.~Huo (Stony Brook University), 
M.~Konyushikhin (Wayne State University),
W.~Li (Rice University), 
J.~Noronha-Hostler (Rutgers University), 
V.~Okorokov (National Research Nuclear University MEPhI), 
B.~Schenke (Brookhaven National Laboratory), 
M.~Sievert (Rutgers University), 
A.~Trzupek (IFJ PAN, Krak\'ow),
U.A.~Wiedemann (CERN)
}

\subsection{Introduction}

It is particularly interesting to study the macroscopic properties of 
  the QGP fluid because - at least conceptually - they are fully fixed 
  by the microscopic properties of a renormalizable, fundamental quantum 
  field theory, namely QCD. 
One key question is how macroscopic properties of QGP arise from its 
  microscopic interactions.
Many theoretical methods ranging from perturbative to non-perturbative 
  techniques are being developed to understand this in detail and one can 
  expect that the insights gained here will be valuable for many related 
  problems in fields ranging from condensed matter theory to cosmology in 
  the future.  
Many different fronts of research are being explored at the moment. 
This ranges from conceptual questions on how to consistently formulate 
  relativistic fluid dynamics or how to solve quantum field theory in 
  non-equilibrium situations to very concrete practical questions about 
  the thermodynamic and transport properties (such as viscosities or 
  conductivities) of the QGP. 
The description of the initial state -- prior to QGP formation --
  and the mechanism by which the products of the collision rapidly thermalize
  are also under investigation.
Besides the role of strong interactions, also electromagnetic interactions 
  and in particular the role of magnetic fields
  are being explored. 
Other fronts of research concern the role of quantum anomalies, chirality 
  and vorticity or the dependence of collective behavior on system size 
  (nucleus-nucleus versus proton-nucleus and proton-proton collisions), 
  on centrality and collision energy, the initial state directly after 
  the collision, or various types of fluctuations. 
These challenges are discussed in more detail in this section.

\subsection{Review of current status of theory on bulk and flow observables}
\subsubsection{QCD Equation of State}
The QCD equation of state, accessible in high-energy collisions (and in 
  the region around mid-rapidity) is one that has vanishing baryon chemical 
  potential. 
It has been established for some time that it features 
  a crossover transition to a chirally symmetric quark gluon plasma~\cite{Aoki:2006we}. 
Most recent lattice calculations~\cite{Steinbrecher:2018phh} have determined 
  the cross-over temperature to be $T_c\simeq 156.5 \pm 1.5\,{\rm MeV}$. 
Recent efforts are also exploring the equation of state at finite $\mu_B$, 
  which at LHC would have relevance mainly at very forward rapidities. 
Here, because of the fermion sign problem in lattice QCD, methods like 
  Taylor expansion~\cite{Kaczmarek:2011zz,Endrodi:2011gv,Bazavov:2015zja,Bonati:2018nut} 
  or imaginary chemical potentials~\cite{Cea:2014xva,Bonati:2014kpa,Bonati:2015bha,Bellwied:2015rza,Cea:2015cya} 
  have to be used. 
To employ lattice QCD based equations of state in hydrodynamic calculations, 
  they need to be matched to a hadron resonance gas model at low temperatures. 
Various equations of state~\cite{Huovinen:2009yb, Borsanyi:2013bia, Bazavov:2014pvz}, 
  using different lattice data and different matching conditions have been used 
  in simulations. 
A comparison of some of them can be found in~\cite{Moreland:2015dvc}, where 
  the sensitivity of observables to the choice of equation of state 
  was studied. 
For a recent theoretical proposal on how to gain experimental sensitivity on the equation of state, see ref.~\cite{Monnai:2017cbv}.
Currently available data are consistent with the lattice QCD equation of state, 
  however with an $\sim 50\,\%$ error. 
To reduce the uncertainty, measurements of particle spectra over a wide 
  range of colliding energies using a single detector with good particle 
  identification, especially at low transverse momentum, would be needed.
Another possibility to constrain the equation of state from experimental data 
  would be to extend state of the art Bayesian techniques~\cite{Moreland:2018gsh} 
  to include free parameters describing the equation of state and fit them 
  along with other free parameters such as shear and bulk viscosities.

\subsubsection{Shear and bulk viscosities of hot nuclear matter\label{sec:macro}}
Ideal fluid dynamics has been very successful in describing a variety of 
  bulk observables in heavy ion collisions~\cite{Kolb:2003dz,Huovinen:2003fa,Hirano:2002ds}, 
  indicating early on that the shear and bulk viscosities of the produced 
  matter cannot be large. 
Calculations in the strong coupling limit using gauge gravity duality have 
  found a value of $\eta/s=1/4\pi$ for an $N = 4$ Supersymmetric Yang-Mills 
  quantum system~\cite{PhysRevLett.87.081601,PhysRevLett.94.111601}. 
This value was significantly smaller than the $\eta/s$ obtained in 
  perturbative QCD calculations, which were however, beset by 
  significant uncertainties, mainly resulting from uncertainties in the 
  relevant scales~\cite{Arnold:2003zc}. 
Recently, such perturbative calculations have been extended to include 
  next-to-leading order corrections and a significant reduction compared 
  to the leading order result was found~\cite{Ghiglieri:2018dgf}: 
At temperatures of the order of the QCD transition the $\eta/s$ obtained using NLO corrections
  is smaller by a factor of $5$ compared to the LO result, and reaches 
  values of approximately $2/4\pi$. 
Extractions of transport coefficients from lattice QCD calculations~\cite{Nakamura:2004sy,Meyer:2007ic,Pasztor:2018yae} 
  are extraordinarily hard, because a numerically difficult analytic 
  continuation from imaginary to real times is necessary.

There are also several theoretical indications that bulk viscosity could play an important role in the QGP to hadron gas transition region (see~\cite{Ryu:2017qzn} and references therein). 
Perturbative calculations have shown that the simple estimate 
$\zeta\approx 15 \eta(1/3-c_s^2)^2$~\cite{Horsley:1985dz} is parametrically 
correct for QCD~\cite{Arnold:2006fz}, where $(1/3-c_s^2)$ is the deviation 
  from conformal symmetry. 
Lattice calculations using the Kubo formula yield large values of 
  $\zeta/s$ ($\sim1$) around $T_c$~\cite{Karsch:2007jc,Meyer:2007dy} with 
  large uncertainties~\cite{Kharzeev:2007wb}.
The calculations also show a fast drop of $\zeta/s$  with increasing $T$. 
Parametrizations of the bulk viscosity over entropy density's temperature 
  dependence were performed in~\cite{Denicol:2009am} with input from~\cite{Karsch:2007jc} 
  for the QGP phase and~\cite{NoronhaHostler:2008ju} for the hadronic phase. 
Similar to the case of shear viscosity, bulk viscosity over entropy density 
ratios have been determined also in holographic models that are supplemented with non-conformal features~\cite{Buchel:2007mf,Finazzo:2014cna}. In this approach, the $\zeta/s$ reaches a much lower peak value $\sim 0.05$ at temperatures slightly above $T_c$~\cite{Finazzo:2014cna}. It remains to be understood to what extent this provides semi-quantitative guidance for the value in  QCD.   


Apart from theoretical calculations of the shear and bulk viscous properties, one may also constrain them by means of fluid dynamic simulations and comparison to experimental data~\cite{Gale:2013da,Heinz:2013th}. 
This method suffers mainly from uncertainties in the initial state (see also Section~\ref{sec:initialstate}) and has an uncertainty of approximately a factor of $3$ (for $\eta/s$) at this point. 
Some of the latest constraints come from simulations using the IP-Glasma initial 
  state~\cite{Schenke:2012wb,Schenke:2012fw}, the EKRT model~\cite{Niemi:2015qia} 
  and Bayesian analyses employing the Trento initial state model~\cite{Moreland:2018gsh}. 
In terms of the Viscous corrections to the distribution function at freeze-out,  
  the low-\pt\ range of the spectrum is more sensitive to the bulk viscosity 
  than to the shear viscosity~\cite{Bozek:2009dw,Paquet:2015lta}. 
Consequently, the uncertainties resulting from bulk viscous corrections are 
  typically larger than for shear when studying \pt\ integrated observables. 
Precise measurements of the low-\pt\ spectra and mean-\pt\ in different 
  collision systems  will help in lowering the current uncertainties, 
  specifically in the extraction of $\zeta/s$.
In order to disentangle features of the initial state and medium properties, 
  it might be useful to study additional collision systems such as \ArAr\ or 
  \OO\ and to perform a more detailed global analysis including refined data 
  on harmonic flow coefficients for identified particles that become available 
  during Run~3 and 4 (see Section~\ref{sec:identified_particle_vn}).


\subsubsection{Initial conditions} \label{sec:initialstate}
Modelling the exact geometry and initial conditions for a fluid dynamic 
  description of heavy ion collisions is not a simple task, because it 
  involves non-perturbative physics.
The available descriptions for the initial state thus range from very 
  simplistic models that assign deposited energy densities based on the 
  wounded nucleons or binary collisions determined in a Monte-Carlo 
  Glauber prescription, to classical effective theories of QCD that are 
  valid in the high energy limit. 
The major ingredient that needs to be provided by an initial state model 
  is the geometry of the interaction region in the plane transverse 
  to the beam. 
It is entirely dominated by the positions of incoming nucleons whose 
  fluctuations also play an important role. 

Initial conditions for hydrodynamic simulations have to provide, in principle, 
  all components of the energy momentum tensor as a function of spatial position 
  (and initial conditions for other conserved charges, if considered). 
This includes, apart from the always included energy density distribution, 
  the initial flow as well as initial viscous corrections. 
Initial flow is included in many recently developed models, that either assume 
  free streaming~\cite{Moreland:2018gsh}, including Yang-Mills evolution, which 
  is close to free streaming~\cite{Gale:2012rq} or an initial flow distribution 
  motivated by strong coupling calculations~\cite{Weller:2017tsr}. 
Initial viscous corrections are often set to zero. 
Only a few works~\cite{Mantysaari:2017cni,Schenke:2018fci,Moreland:2018gsh} 
  include the full viscous stress tensor provided by the initial state description.

Since the initial state models that provide the entire $T^{\mu\nu}$ all switch 
  from essentially a freely streaming system to strongly interacting hydrodynamics 
  at a fixed time $\tau$, that transition is somewhat abrupt and unphysical. 
To improve over this situation an intermediate step using effective kinetic theory 
  has been introduced~\cite{Kurkela:2018wud,Kurkela:2018vqr}. 
This procedure allows for a somewhat smoother matching but has yet to be used in 
  full fledged hydrodynamic simulations. A first study that matches full kinetic theory to full viscous fluid dynamics indicates that the intermediate kinetic transport formulation becomes more important with decreasing system size~\cite{Kurkela:2018qeb}.

As already discussed in Section~\ref{sec:macro}, the choice of initial state has 
  a significant effect on the extraction of transport coefficients. 
A more compact initial state and the presence of initial flow lead to a larger 
  transverse flow, which requires a larger bulk viscosity to compensate for it 
  and produce agreement with experimental data~\cite{Schenke:2018fci}. 
In addition the initial flow also affects the value of the extracted 
  shear viscosity.
Also, the models' eccentricities will affect the final momentum anisotropies, 
  influencing the extracted shear viscosity to entropy density ratio. 
Two possible attempts to solve this problem have been pursued: 
  the first is to perform a combined Bayesian analysis of all parameters~\cite{Moreland:2018gsh}, 
  including those of the initial state, to find the best fit for all transport 
  coefficients along with the initial state description.
The second is to constrain an initial state description as well as possible 
  using data from experiments other than heavy ion collisions e.g. $e$--$p$ 
  scattering data, which will hopefully be extended to $e$--A in a future 
  electron ion collider facility.
As mentioned above, at the moment the two approaches lead to some similar 
  features of the initial state (product of thickness functions, presence 
  of subnucleon structure), but also show discrepancies (size of the nucleon 
  and sub-nucleon scales along with the size of the extracted bulk viscosity). 
In the near future, flow measurements in light ion collisions such as 
  \ArAr, \OO\ etc. can also provide independent experimental constraints 
  on initial conditions (see Section~\ref{sec:system_size_dependence}). 
Similarly, new flow observables can also help constrain the initial conditions.
In particular the measurement of flow fluctuations has provided constraints
  on initial geometry models, in both the approaches discussed above.
Analogously, the more recent studies of longitudinal flow 
  fluctuations~\cite{CMS-HIN-14-012, CMS-HIN-15-008, HION-2016-04}
  and their extensions in Run~3 and 4 (see Section~\ref{sec:longitudinal_flow_fluctuations}) can provide additional constraints.

\subsubsection{Response functions}
\label{sec:ResponseFunctions}
In a fluid dynamic description of heavy ion collisions, one can understand 
  the azimuthal harmonic flow coefficients \vn\ as a response to deviations 
  of the initial state from an azimuthally isotropic form. 
Mathematically, one can formulate this in terms of response functions that 
  describe how the solution of the fluid dynamic evolution equations,  
  as well as resulting experimental observables such as azimuthal particle 
  distributions, get modified when the initial values of the fluid fields 
  are changed~\cite{Teaney:2010vd,Floerchinger:2013rya,Floerchinger:2014fta}. 
In the simplest implementation,  linear response functions describe the 
  linear response of flow coefficients to eccentricities $v_n \sim\epsilon_n$,  
  while the quadratic response functions describe terms like $v_n \sim \epsilon_a\epsilon_b$ 
  where symmetry reasons imply $|n|=|a\pm b|$~\cite{Teaney:2012ke,Floerchinger:2013tya}. 
Response functions can not only be used to study deviations from azimuthal 
  rotation symmetry but also for deviations from (approximate) Bjorken boost 
  invariance,  vanishing baryon number density, for electric fields and so on. 
Quite generally, response functions carry interesting information about fluid 
  properties such as thermodynamic and transport properties. 
  Where the response functions are known, one can infer properties of the 
  initial state by reverse engineering.
Experimentally, one can constrain properties of response functions indirectly 
  via measurements of various particle correlation functions. 
It is particularly interesting to compare situations with strong deviations 
  from a symmetry (such as peripheral collisions for the case of azimuthal 
  rotation invariance) to situations with small deviations (such as central collisions) 
  in order to differentiate between linear and non-linear response. 
For existing experimental work in this direction see~\cite{HION-2012-03,HION-2014-03,ALICE:2016kpq} 
  and for an example of a recent further going theoretical proposal see~\cite{Giacalone:2017uqx}.

Detailed comparison of flow observables between experiment and theoretical 
  calculations, especially regarding the dependence on external parameters 
  like system size and collision energy as well as differential information 
  such as on centrality, or particle identification will be helpful to make 
  further progress in constraining response functions. Improvements in 
  particle identification, transverse momentum and longitudinal coverage 
  in Run~3 and 4 will be particularly valuable to this end.

\subsection{Experimental constraints from Run~3 and 4}

Since measurements of flow and correlations provide arguably the most direct 
  manifestations of collectivity, they play naturally a central role in the 
  scientific programme of exploring finite temperature QCD via collectivity. 
At the HL-LHC, much more stringent tests of the collective dynamics in 
  nucleus--nucleus collisions will be possible.
These will constrain QGP medium properties and initial conditions, 
  as discussed in the previous section, in great detail. 
In the following, these newly arising opportunities are illustrated with a 
  set of physics performance studies exploiting: 1) high-statistics 
  particle-identified flow measurements, 2) system-size dependence of flow, 
  and 3) longitudinal flow fluctuations.

\FloatBarrier
\subsubsection{Identified particle \vn\label{sec:identified_particle_vn}}

\begin{figure}[!htb]
\begin{center}
\includegraphics[width=0.495\textwidth]{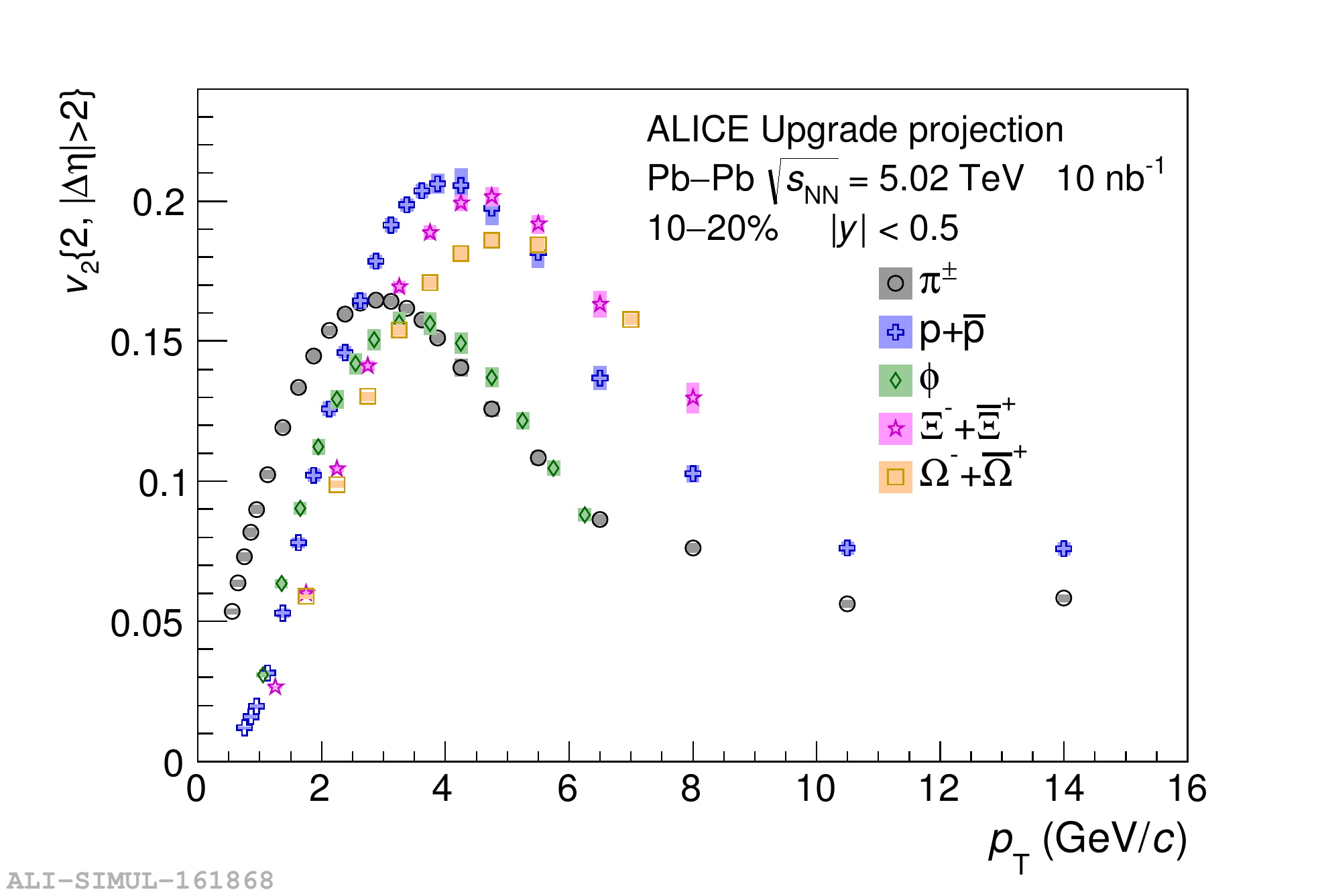}
\includegraphics[width=0.495\textwidth]{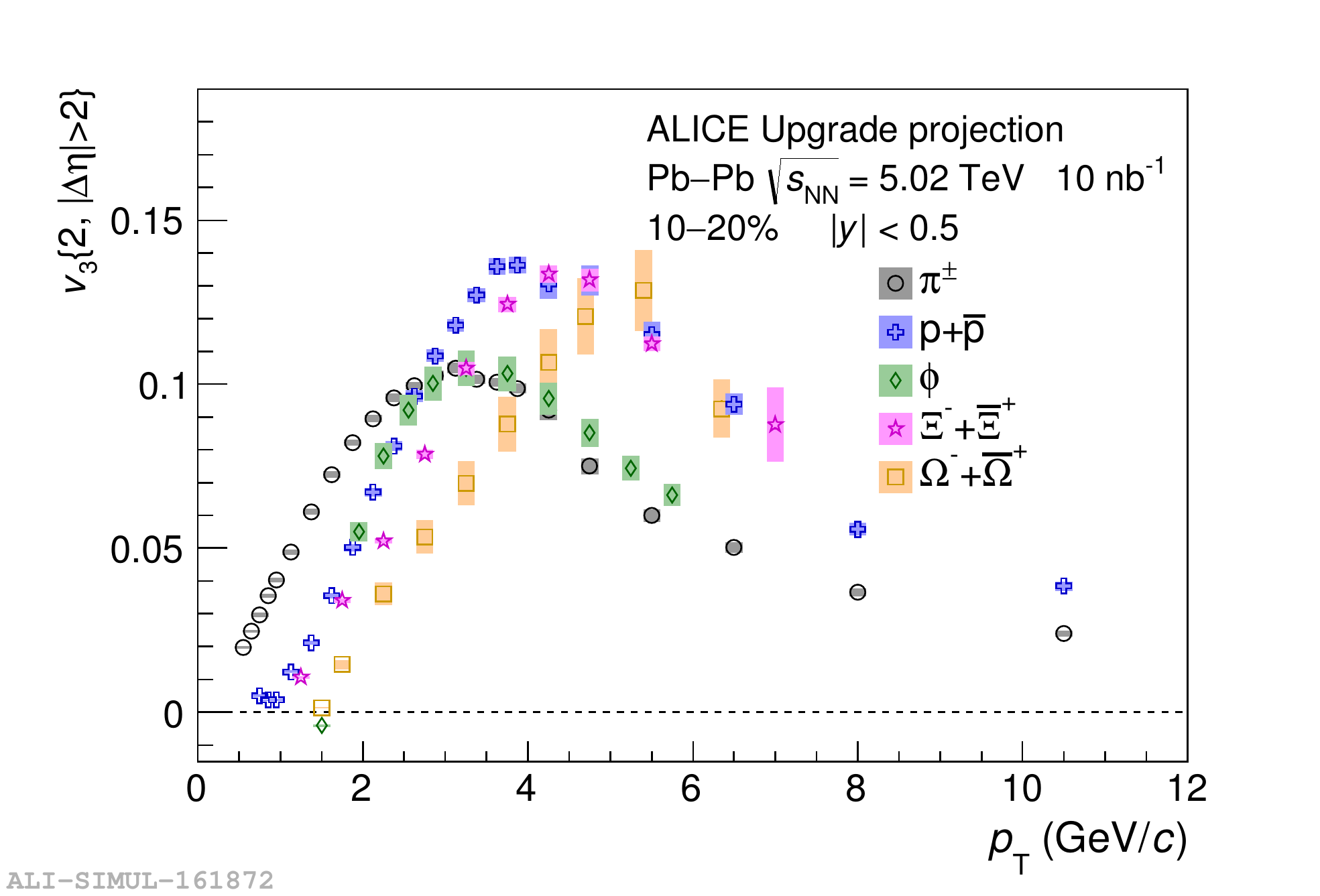}
\caption{
ALICE projections for $v_2$ (left) and $v_3$ (right) of $\pi^\pm$, 
  $\mathrm{p}$+$\overline{\mathrm{p}}$, $\Xi$+$\overline{\Xi}$, 
  $\Omega$+$\overline{\Omega}$, and the $\phi$-meson 
  in the 10--20\% centrality interval
  for an integrated luminosity of 10~nb$^{-1}$. 
Error bars (shaded boxes) represent the projected statistical 
  (systematic) uncertainties. Figures from Ref.~\cite{ALICE-PUBLIC-2019-001}.}
\label{fig:alice_vn}
\end{center}
\end{figure}

In a fluid dynamic picture of A--A collisions, hadrons decouple from the fluid 
  at late times, when the density of the system is sufficiently low and the 
  mean-free path is sufficiently large so that a fluid dynamic description 
  ceases to apply. 
Therefore, it is a direct consequence of this late time-scale, that the 
  PID-dependence of \vn\ measurements tests a limitation of fluid 
  dynamic behavior. 
Do all hadron species emerge from the same common flow field? Or can one see at 
  the higher accuracy of future \vn\ measurements particle specific 
  differences in the decoupling which are related to the differences in 
  hadronic cross sections? 
Quantitative questions of this kind will allow one to better constrain how 
  significant the hadronic stage of A--A collisions contribute to \vn,  
  and since the viscous properties of QCD change significantly in the hadronic 
  stage, this is of direct relevance for extracting information about viscous 
  transport coefficients with higher precision. 
The same class of improved PID-sensitive \vn\ measurements is also important 
  for testing different dynamical pictures of hadronization. 
In particular since fragmentation is expected to be the dominant hadronization 
  mechanism at high \pT\ while recombination becomes relevant at lower \pT, 
  extending these PID measurements with precision over the largest possible 
  transverse momentum range will be of interest. 
To this end, figure~\ref{fig:alice_vn} shows projections from the ALICE collaboration for 
  the $v_2$ and $v_3$ of several light-flavor species, that are expected for 
  an integrated luminosity of 10~nb$^{-1}$ expected in Run~3 and 4.
The projected statistical uncertainties are typically negligible over the 
  entire \pt\ range and in most cases the systematic uncertainties are 
  quite small as well.
These measurements will be much more precise compared to those in Run~1 and 2, and will allow 
  for the recombination/fragmentation descriptions to be examined with unprecedented accuracy.
Similar projections for heavy-flavor particles are discussed further in Chapter~\ref{sec:HI_HF}
  together with their physics implications. 

\FloatBarrier
\subsubsection{System size dependence\label{sec:system_size_dependence}}
\label{sec:flow_sizedep}

\begin{figure}[!ht]
\begin{center}
\includegraphics[width=0.8\textwidth]{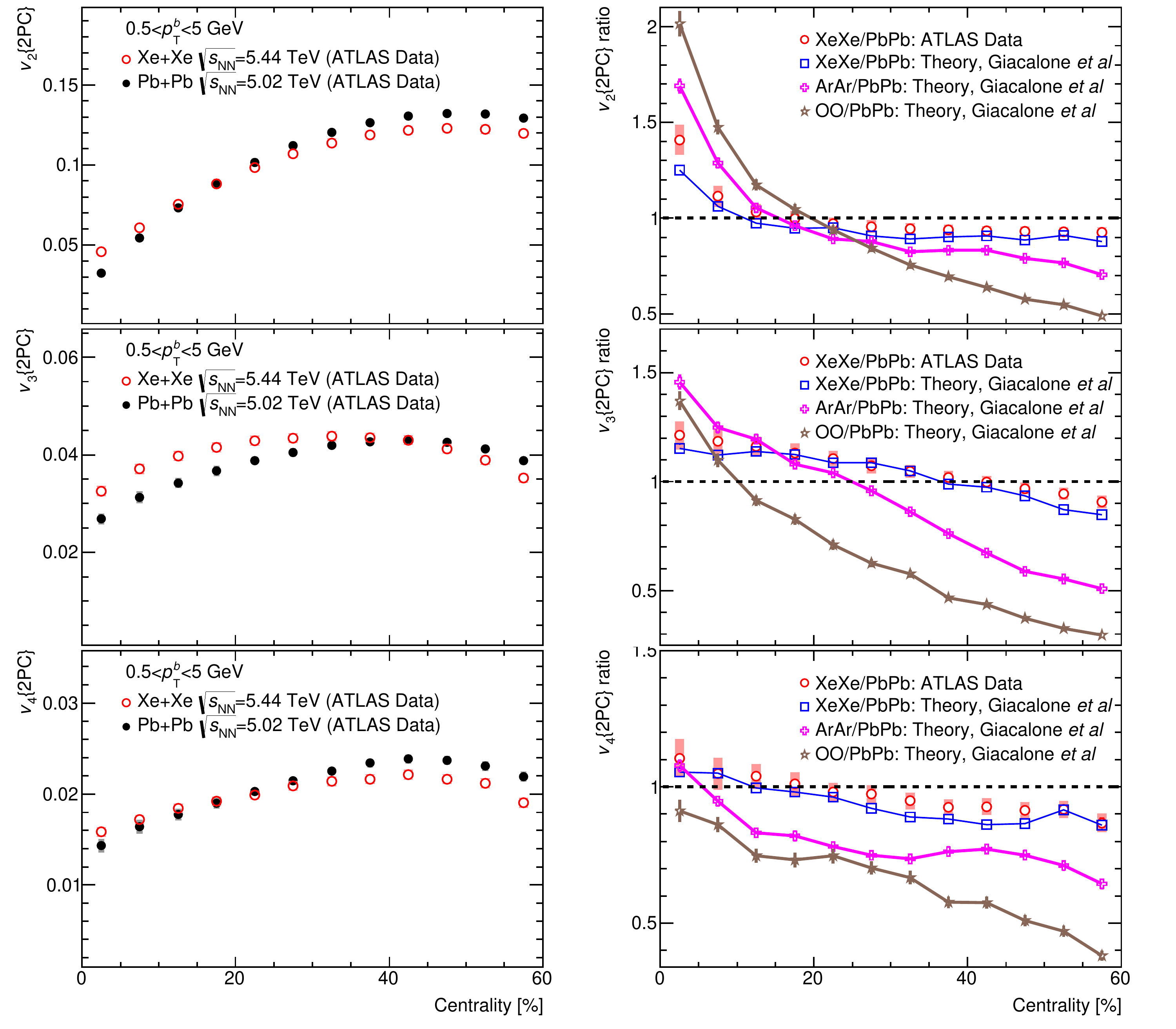}
\caption{
Left panels: comparisons of the centrality dependence of the \vn\ measured 
  in \PbPb\ collisions at \snn=5.02~TeV to \XeXe\ measurements. 
The plots are for the 0.5--5~\UGeVc\ \pT\ interval. 
From top to bottom each row corresponds to a different harmonic order $n$.
The ratios are compared to theoretical predictions from Ref.~\cite{Giacalone:2017dud}.
Also shown are theory calculations of the ratios for \ArAr\ and \OO.
ATLAS Data taken from Ref.~\cite{ATLAS-CONF-2018-011}.
}
\label{fig:atlas_xexe_vn}
\end{center}
\end{figure}

In \PbPb\ collisions, previous \vn\ measurements at the LHC have accessed the 
  system size dependence of flow via the centrality dependence. 
However, this centrality dependence is biased by a strong variation of 
  the spatial eccentricity of the nuclear overlap. 
While modelling allows one to control this eccentricity dependence to some 
  extent, studying smaller collision systems at highest centrality (i.e. impact parameter close to zero) remains conceptually the cleanest way of establishing the system size dependence of flow. 
It also provides a way of disentangling the event-averaged spatial eccentricity 
  from the event-by-event eccentricity fluctuations, and can thus contribute 
  significantly to constraining the initial condition from which collective 
  behaviour emerges. 
These are important motivations for studying soft multi-particle production 
  and its transverse asymmetries in the collision of lighter nuclei.

Figure~\ref{fig:atlas_xexe_vn} shows ATLAS comparisons of the \vn\
  in \XeXe\ and \PbPb\ collisions as a function of centrality (left panels)
  and their ratios (right panels).
Also shown for comparison in the right panels are theoretical predictions
  for the ratios from Ref.~\cite{Giacalone:2017dud}.
It is seen that in most central collisions the ratio \vn(Xe--Xe)/\vn(\PbPb) 
  is larger than unity for $n=2$ and 3.
With decreasing centrality the ratios for all harmonics show a decreasing trend.
These trends can be explained as follows:
  \XeXe\ being a smaller system than \PbPb, the effect of fluctuations
  is more important.
The fluctuations increase the initial eccentricities of the collision
  geometry and therefore enhance the \vn.
However, because \XeXe\ is a smaller collision system the viscous
  effects (which suppress the \vn) are larger, and play a bigger role
  with decreasing centrality and increasing harmonic order.
In most central events, the effect of the increased fluctuations wins
  for $v_2$.
But with increasing harmonic order and/or decreasing centrality, 
  eventually the viscous effects reduce the \vn compared to \PbPb.
These observations indicate the ability of such cross-system 
  \vn\ measurements to be very sensitive to initial conditions of 
  the heavy ion collision as well as the transport coefficients
  of the QGP. 
The measured ratios for the \vn(Xe--Xe)/\vn(\PbPb) are qualitatively reproduced 
  by the theory predictions from Ref.~\cite{Giacalone:2017dud}.
In order to illustrate the sensitivity of such models to the collision species, 
  predictions are also presented for \ArAr\ and \OO\ collisions. 
The predictions show considerably larger variation of the centrality dependence
  of the \vn\ going from \XeXe\ to \ArAr\ to \OO, as compared to
  the variation going from \PbPb\ to \XeXe.
Given such strong trends in the theory predictions, performing \vn\ measurements 
  in light ion species such as \ArAr\ and \OO\ in Run~3 can provide strong 
  constrains on the theoretical models.

Furthermore, there has been much work in studying long-range correlations
  observed in \pA, d--A, $^{3}\mathrm{He}$--A, and more recently in 
  \pp\ collisions (see Chapter~\ref{chapter:smallsystems}).
Measuring flow in medium and light ions would allow for a continuous study 
  of how collective phenomena vary from large (\PbPb) to small 
  (\pA\ and \pp) systems.

\FloatBarrier
\subsubsection{Longitudinal flow fluctuations~\label{sec:longitudinal_flow_fluctuations}}

The characterization of how the longitudinal scales at which the symmetry 
  planes $\Psi_n$ associated to \vn\ decorrelate, and how this relates to the 
  variation of the signal strength \vn\ with rapidity is still far from the 
  state of the art reached in \vn\ measurements at mid-rapidity. 
However, any deviation from the simple picture of a rapidity-independent 
  (Bjorken-like) longitudinal dynamics directly impacts our understanding of 
  the time evolution of matter in all rapidity windows, including the 
  well studied mid-rapidity one. 
Multiple recent measurements at the LHC indicate the presence of 
  considerable longitudinal dynamics.
These include measurements from CMS of event-plane decorrelation in \pPb\ and \PbPb\ 
  collisions~\cite{CMS-HIN-14-012,CMS-HIN-15-008} and from ATLAS on flow-decorrelations~\cite{HION-2016-04}
  and forward-backward multiplicity fluctuations~\cite{HION-2015-13}.
It is therefore important that experiments at the HL-LHC will parallel higher 
  precision measurements at mid-rapidity with improved experimental control 
  over the longitudinal evolution.

\begin{figure*}[!htb]
\begin{center}
\includegraphics[width=0.8\textwidth]{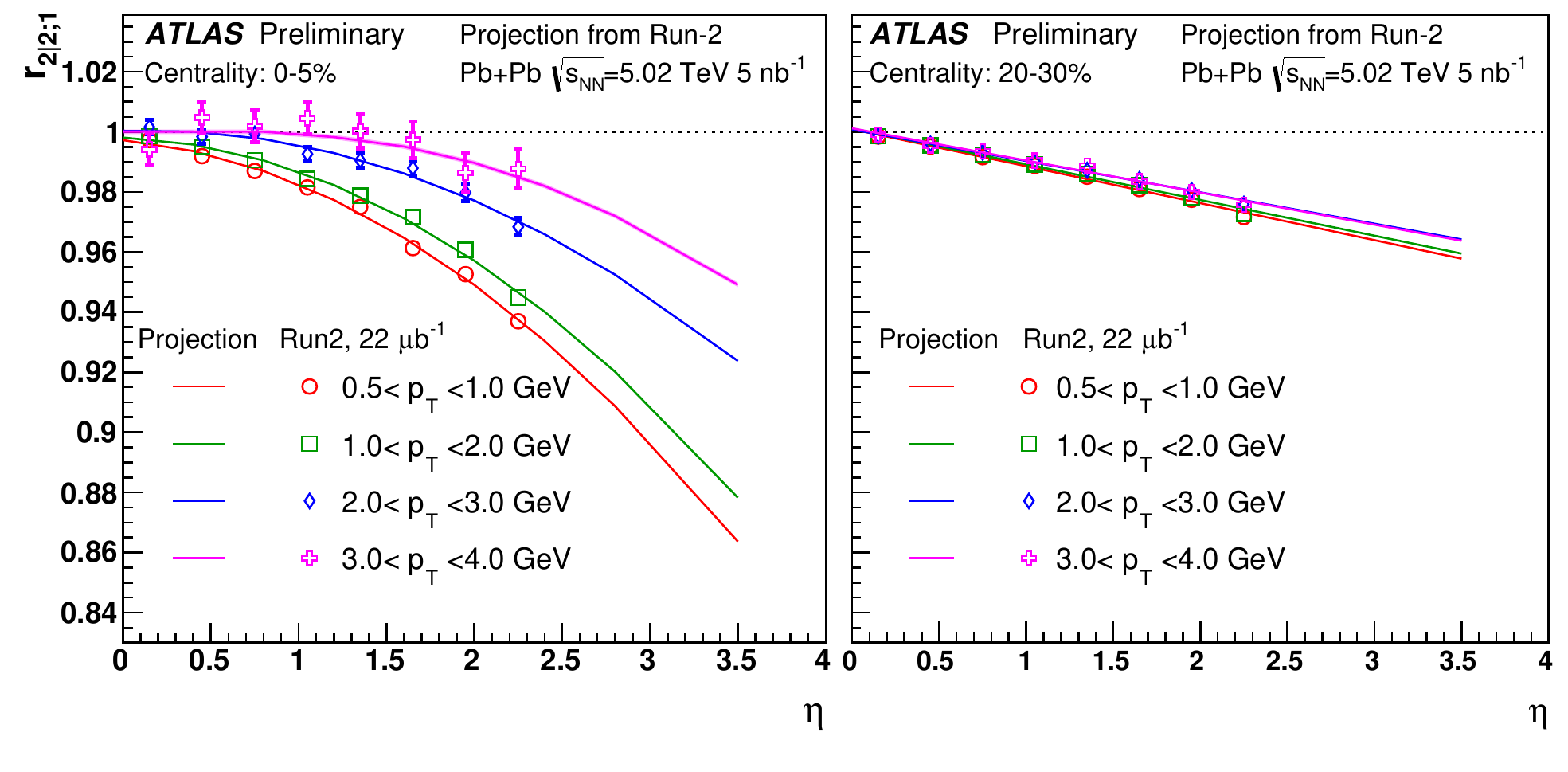}
\caption{
ATLAS projections of the flow-decorrelation observable $r_{2|2;1}$ 
  as a function of $\eta$ (lines).
The markers indicate the present measurements from Ref.~\cite{HION-2016-04}.
The left and right panels show projections for the 0--5\% and 20--30\%
  centrality intervals, respectively.
The width of the projection bands indicates the expected statistical uncertainty.
}
\label{fig:atlas_r221}
\end{center}
\end{figure*}

In the ATLAS measurements in Ref.~\cite{HION-2016-04}, the flow decorrelation 
  is quantified by constructing a correlator $r_{n|n;1}$ defined as:
\begin{equation}
r_{n|n;1}(\eta)=%
  \frac{\langle \mathbf{v}_n(-\eta)\mathbf{v}_n^*(\eta_{\mathrm{ref}})\rangle}%
  {\langle \mathbf{v}_n(\eta)\mathbf{v}_n^*(\eta_{\mathrm{ref}})\rangle},
\end{equation}
where $\mathbf{v}_n$ is the normalized flow vector, and $\eta_{\mathrm{ref}}$ is 
  the reference pseudo-rapidity~\cite{HION-2016-04}.
The correlator,  $r_{n|n;1}$,  measures the relative difference between flow
  $v_ne^{in\psin}$ at $\eta$ and $-\eta$.
If flow were boost-invariant, then $r_{n|n;1}$ would equal unity.
However any  difference in the $\eta$ dependence of the flow magnitude
  \vn\ and the event plane angle \psin\ will lead to $r_{n|n;1}$
  become smaller than unity.
The ATLAS measurements of $r_{2|2;1}$ over the 0--2.5 $\eta$ range 
  are shown in Figure~\ref{fig:atlas_r221} by the markers. 
It is observed that the $r_{2|2;1}$ is significantly smaller than unity 
  in central collisions, which indicates stronger flow decorrelation.
For a given centrality, $r_{2|2;1}$ decreases faster at low \pt\ 
  that at higher-\pt.
In the 20--30\% mid-central $r_{2|2;1}$ decreases linearly with $\eta$, 
  however in the 0--5\% most central collisions there are indications
  that the decorrelations are possibly quadratic.

Repeating this measurement in Run~4 will lead to significant improvements 
  due to increased luminosity and especially due to increased tracking 
  acceptance in $\eta$ to $\pm4$ units.
Figure~\ref{fig:atlas_r221} also shows the ATLAS projections for 
  $r_{2|2;1}$ made for Run~4, indicated as dashed lines.
The ATLAS tracking acceptance in Run~4 will extend the $\eta$ range 
  to $\pm$4 units, but the projected measurement is made
  to $\pm$3.5 units, in order to leave a gap between the ITk and the 
  region of the forward calorimeter in which the reference measurement 
  is made ($4.4<|\eta|<4.9$).
The projections are made by fitting the existing data with a linear 
  function for the 20--30\% centrality range and with a quadratic 
  function for the 0--5\% centrality range.
It is seen that with the increased $\eta$ acceptance the non-linearity 
  in the flow decorrelation can be studied in much more detail.

The longitudinal flow-decorrelation observables are sensitive to the 
  event by event fluctuations of the initial energy density profile 
  in the longitudinal direction. 
Thus precise measurement of these decorrelations should give
  a better understanding of the initial conditions along the 
  longitudinal direction and in the development of full three-dimensional 
  viscous hydrodynamic models.
These would in turn result in a more accurate estimation of $\eta/s$.

\FloatBarrier
\subsection{Vorticity and polarization}

An interesting open question for relativistic fluids is to what extent the 
  spin degrees of freedom thermalize locally and to what extend spin 
  polarization results as a consequence of the fluid motion. 
Intuitively, one might expect that spin aligns locally with the rotational 
  motion of the fluid as measured by vorticity, corresponding to the curl 
  of the fluid velocity.

The relativistic generalization of the non-relativistic fluid vorticity is 
  not unambiguous, however. 
The vorticity of a fluid in local equilibrium is characterized by the so-called 
  thermal vorticity tensor, corresponding to $\omega_{\mu\nu} = \frac{1}{2} 
  (\nabla_\nu \beta_\mu - \nabla_\mu \beta_\nu)$ where $\beta_\mu=u_\mu / T$ 
  is the ratio of fluid velocity to temperature~\cite{Becattini:2013fla}. 
This thermal vorticity includes contributions from global rotational motion, 
  local fluid acceleration, and temperature gradients. 
It has been argued that this thermal vorticity should lead to local 
  spin polarization. 
If this holds at chemical freeze-out, one should be able to find traces of the 
  thermal vorticity in the spin polarization of particles and resonances, 
  such as $\Lambda$ ($\overline{\Lambda}$) particles.

Spin polarization is in this picture closely tied to angular momentum of the 
  expanding fireball. 
For non-central events, the angular momentum of the produced matter 
  is perpendicular to the event plane. 
Via the spin-vorticity coupling mechanism, this leads to a global polarization 
  in the transverse plane aligning with the global angular momentum (also known 
  as the ``transverse polarization''). 
This global transverse polarization has recently been observed in the measurement 
  of $\Lambda$ spin polarization at RHIC~\cite{STAR:2017ckg}. 
For this global effect following global angular momentum, one expects a 
  decreasing magnitude with increasing collision energy and the effect is 
  expected to be relatively small at LHC energies. 

\begin{figure*}[!htb]
\begin{center}
\includegraphics[width=0.8\textwidth]{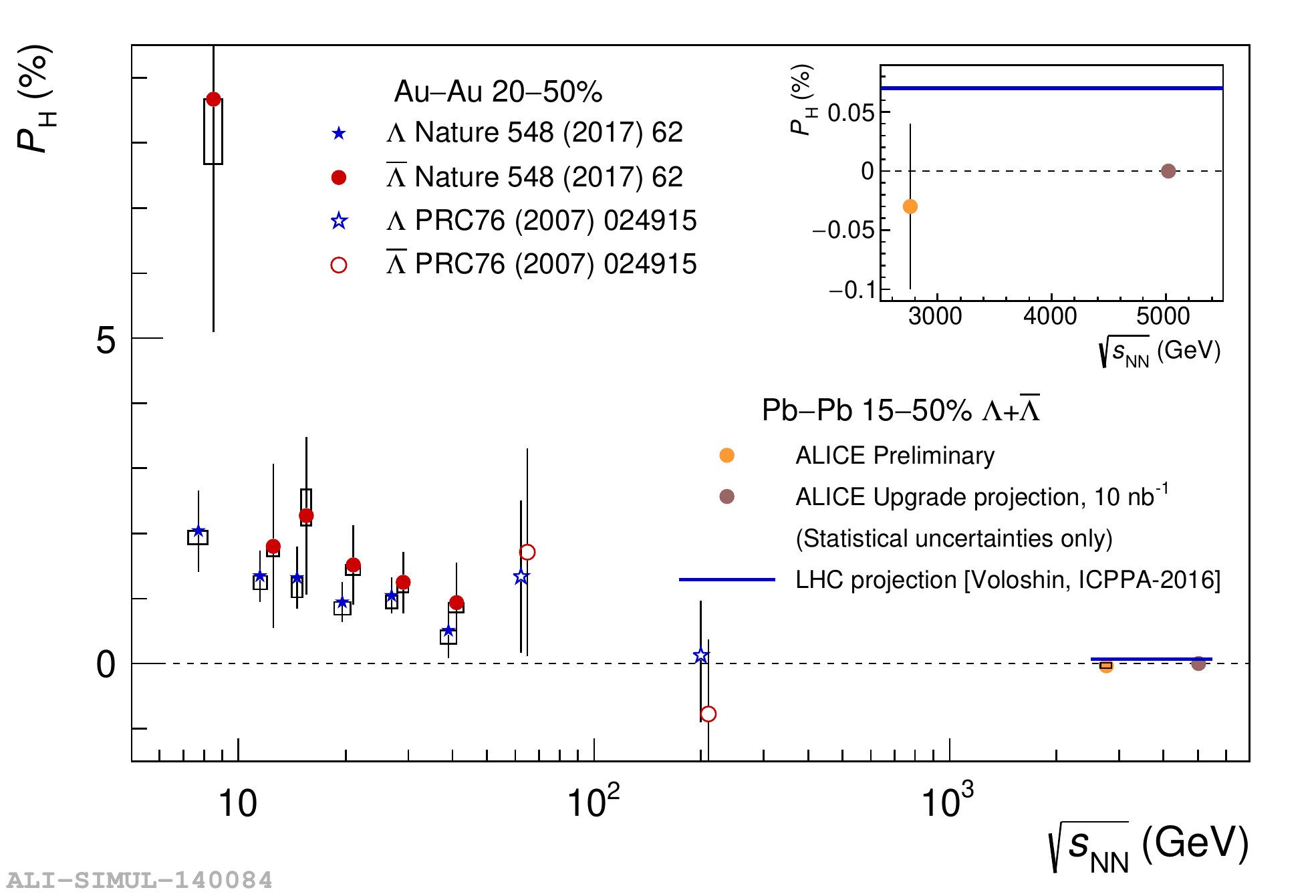}
\caption{
ALICE projections for the Global hyperon polarization in \PbPb\ 
  collisions at $\snn=2.76$~TeV for an integrated luminosity of 
  10~nb$^{-1}$ (blush symbol), together with the present measurements (orange symbol) 
  compared to analogous measurements at various collision energies from the STAR 
  collaboration~\cite{STAR:2017ckg, Abelev:2007zk} (blue and red symbols). 
The blue line indicates the prediction for the maximum value at the 
  LHC~\cite{Voloshin:ICPPA2}. 
The inlay plot shows a zoomed in version of the plot around the ALICE 
  measurement and Run~3 and 4 projection, together with the prediction 
  for the maximum value at the LHC.
The points for $\overline{\Lambda}$ are slightly shifted along the horizontal 
  axis for visibility.  
Error bars (open boxes) represent the statistical (systematic) uncertainties. Figure from Ref.~\cite{ALICE-PUBLIC-2019-001}.}
\label{fig:alice_lambda}
\end{center}
\end{figure*}

Figure~\ref{fig:alice_lambda} shows the energy dependence of the global 
  transverse polarization of $\Lambda$ and $\overline{\Lambda}$ for
  semi-central heavy ion collisions. 
The RHIC results show the decrease of polarization with increasing \snn. 
The preliminary ALICE data point at $\snn=2.76$~TeV is consistent with zero
  within 1$\sigma$ statistical/systematic uncertainties.
However it is also consistent with the predicted maximum value (blue line)
  within $\sim$2$\sigma$ statistical/systematic uncertainties.
But the ALICE upgrade projection at twice large collision energy,
  (assuming zero signal) shows that the polarization in Run~3 
  and 4 can be measured with very high precision. 
Therefore the study of global polarization of $\Lambda$ and $\overline{\Lambda}$ 
  within HL--LHC project allows the unambiguous conclusion with regard of the 
  values of this physics quantity in the TeV-energy domain. 

In addition to the transverse polarization, an azimuthal-dependent, 
  longitudinal polarization (in the direction of the beam pipe) has 
  also been predicted and recently observed at RHIC~\cite{Niida:2018hfw}. 
This is mainly a consequence of an azimuthal dependence of local acceleration 
  and temperature gradient (e.g., the elliptic flow), which could lead to an 
  elliptic modulation of longitudinal spin polarization in 
  non-central collisions. 
Unlike the global transverse polarization, this longitudinal polarization 
  effect has a much weaker dependence on collision energy from RHIC to 
  the LHC~\cite{Karpenko:2017dui}, mainly because the anisotropic flow has a 
  weak collision energy dependence. 
With increased data sample and upgraded detectors covering a wider rapidity 
  range in the HL--LHC era, there will be exciting opportunities for precision
  study of the $\Lambda$ polarization and to map out the dependence on variables 
  such as azimuthal angle, rapidity, transverse momentum.

\FloatBarrier
\subsection{Chiral Magnetic Effect}

An important property of the strong interaction which is potentially observable in heavy-ion collisions is parity violation. Although it is allowed by 
quantum chromodynamics (QCD), global parity violation is not observed in strong interaction. However, QCD predicts the existence of topologically 
non-trivial configurations of the gluonic field, instantons and sphalerons, which might be responsible for local parity violation in microscopic QCD 
domains at finite temperature \cite{Lee:1973iz, Lee:1974ma, Morley:1983wr, Kharzeev:1998kz}. The $P$- and $CP$-odd interactions between quarks 
and such fields with non-zero topological charge~\cite{Chern:1974ft} change the quark chirality, breaking parity symmetry by creating an imbalance 
between the number of left- and right-handed quarks. Furthermore, an extremely strong magnetic field is expected to be produced in heavy-ion collisions 
\cite{Deng:2012pc, Gursoy:2014aka} (of the order of $10^{19}$ Gauss at the LHC) because the charges of initial ions add coherently. This strong 
magnetic field aligns the spins of the positively (negatively) charged quarks in the direction parallel (anti-parallel) to magnetic field orientation. Moreover, 
right-handed (left-handed) quarks have their direction of momentum parallel (anti-parallel) to the spin orientation. The spin alignment coupled with the 
local imbalance between the number of left- and right-handed quarks leads to the development of a quark current. The current moves the positively 
charged quarks along its direction and the negatively charged quarks in the opposite direction. This implies a charge separation along the direction 
of the magnetic field, which is on average perpendicular to the reaction plane, 
a phenomenon called Chiral Magnetic Effect (CME) \cite{Kharzeev:2004ey, Kharzeev:2007tn, Kharzeev:2007jp, Fukushima:2008xe}. 

The sign of the topological charge can give rise to a positive or negative current in the magnetic field direction with equal probability. Therefore, the charge separation averaged 
over many events is zero. This makes the observation of the CME experimentally difficult and possible only via azimuthal particle correlations, which 
introduces a large flow related background into the measurements.

\begin{figure}[!ht]
\begin{center}
\includegraphics[width=0.6\textwidth]{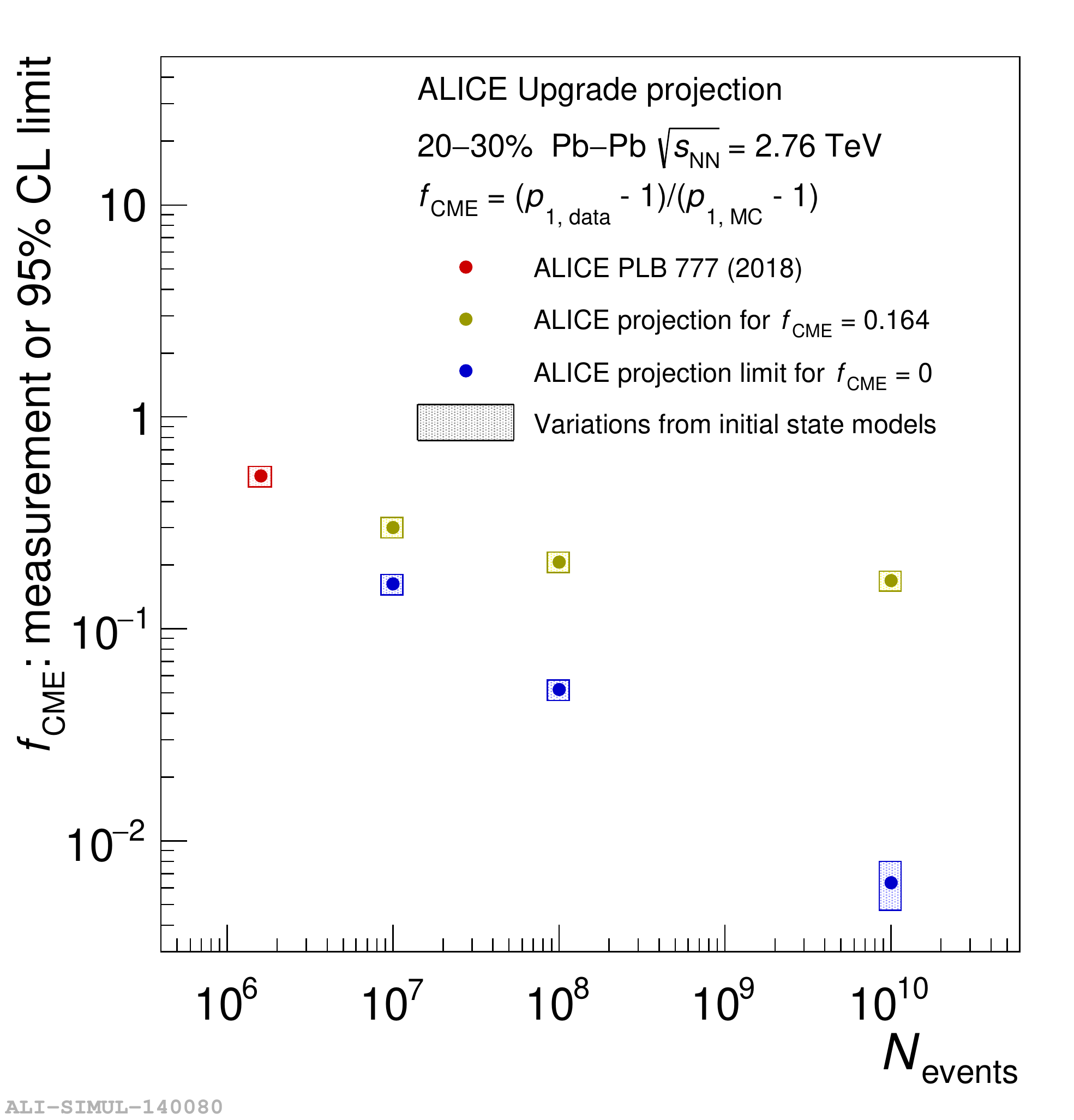}
\caption{
ALICE projections for the upper limit on the CME fraction at 95\% confidence 
  level as a function of the number of events in the 20--30\% 
  centrality interval.
The right-most projection point corresponds to an integrated luminosity 
  of 10~nb$^{-1}$.
The result reported by the ALICE collaboration~\cite{Acharya:2017fau} 
  is shown together with expectations for $f_{\rm CME}=0.164$ (current estimate) 
  and $f_{\rm CME}=0$ (null hypothesis). 
The shaded boxes denote variations from various initial 
  state models (see text for details). Figure from Ref.~\cite{ALICE-PUBLIC-2019-001}. 
}
\label{fig:alice_fcme}
\end{center}
\end{figure}

The three-particle correlator $\gamma_{\alpha\beta} = \langle \cos(\varphi_{\alpha} + \varphi_{\beta} - 2\Psi_{\rm 2}) \rangle$ \cite{Voloshin:2004vk}, 
where $\varphi_{\alpha}$ is the azimuthal angle of the particle of charge $\alpha$ and $\Psi_{\rm 2}$ is the second harmonic symmetry plane 
angle, was proposed to measure charge-dependent azimuthal correlations. This correlator eliminates correlations independent of symmetry 
plane orientation, suppressing background contributions at the level of $\sim v_2$. However, the interpretation of the experimental results is 
complicated by the remaining background (e.g. local charge conservation (LCC) coupled with elliptic flow~\cite{Schlichting:2010qia, Pratt:2010zn}). 
Recent observation of similar charge-dependent azimuthal correlations in \pPb\ (where the CME is not expected) and 
\PbPb\ collisions~\cite{Khachatryan:2016got} indicates 
the $\gamma_{\alpha\beta}$ correlator be dominated, if not all, by the background effect.
The ALICE \cite{Acharya:2017fau} and CMS \cite{Sirunyan:2017quh} collaborations have used the Event Shape Engineering (ESE) technique 
\cite{Schukraft:2012ah} to estimate the CME fraction to the charge dependence of $\gamma_{\alpha\beta}$, $f_{\rm CME}$, in \PbPb\ collisions. 
ALICE extracted 
$f_{\rm CME}$ by relating measurements of the charge dependence of $\gamma_{\alpha\beta}$ from the ESE analysis to CME signal expectations from 
various initial state model calculations including a magnetic field. It has been assumed that the CME signal is proportional to 
$\langle |B|^2 \cos(2(\Psi_{\rm B} - \Psi_2)) \rangle$, where $|B|$ and $\Psi_{\rm B}$ are the magnitude and direction of the magnetic field, respectively. Within current experimental uncertainties, the CME signal contribution to the $\gamma_{\alpha\beta}$ correlator is consistent with zero.

Figure~\ref{fig:alice_fcme} shows the upper limit on $f_{\rm CME}$ at 95\% confidence level for the 20--30\% centrality interval reported by the 
ALICE collaboration together with expectations for $f_{\rm CME}=0.164$ (current estimate) and $f_{\rm CME}=0$ as a function of the number of events. The shaded 
boxes denote variations due to different estimates of the magnetic field from the investigated models. The ALICE upgrade projection indicates that 
stringent constraints for the CME contribution to the charge dependence of $\gamma_{\alpha\beta}$ can be achieved at a level of less than 1\% 
with the expected HL--LHC statistics. 

One key ingredient needed for the observation of the CME is the strong magnetic field in the QGP medium. It is important to establish direct
evidence for this field and determine its strength, which will help significantly constrain theoretical predictions on the magnitude of the CME signal. Measurement of the pseudorapidity-odd component of directed 
flow, $v_1^{\rm odd}$, separately for positive and negative charged particles has been proposed as a probe to 
the magnetic field~\cite{Gursoy:2014aka}. Any difference will indicate the presence of induced electromagnetic 
currents and will allow to estimate the magnitude of the effect. It will also provide information on the electric conductivity of the QGP medium.

\begin{figure}[!htb]
\begin{center}
\includegraphics[width=0.6\textwidth]{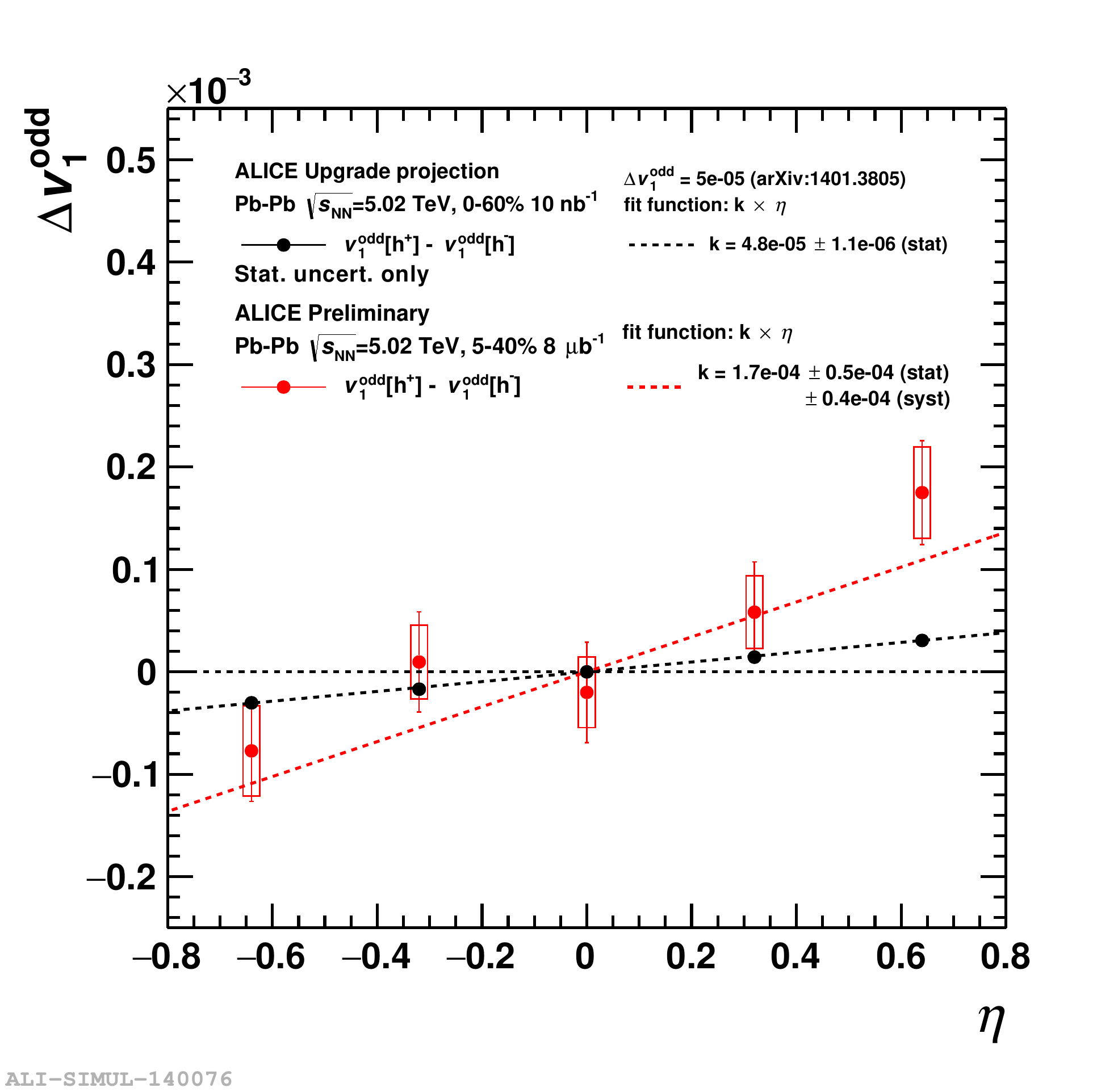}
\caption{
Charge difference of $v_1^{\rm odd}$ as a function of pseudorapidity measured 
  by the ALICE collaboration in \PbPb\ collisions at $\snn=5.02$~TeV~\cite{Margutti:2017lup} 
  (red symbols) and the projection for a $5 \times 10^{-5}$ difference~\cite{Gursoy:2014aka} 
  from 10~nb$^{-1}$ (black symbols) together with linear fits (dashed lines). 
Error bars (open boxes) represent the statistical (systematic) uncertainties. Figure from Ref.~\cite{ALICE-PUBLIC-2019-001}.}
\label{fig:alice_delta_v1}
\end{center}
\end{figure}

Figure~\ref{fig:alice_delta_v1} shows the charge difference of $v_1^{\rm odd}$, 
$\Delta v_1^{\rm odd}=v_1^{\rm odd +} - v_1^{\rm odd -}$, as a function of pseudorapidity measured by the ALICE collaboration in Pb--Pb collisions 
at $\snn=5.02$ TeV \cite{Margutti:2017lup} together with a linear fit. A hint of a charge-dependent difference is observed and quantified 
by the slope $k$ with a total significance of 2.6 $\sigma$. 
This difference, which differs both in magnitude and sign compared to 
  predictions for $\pi^{\pm}$ at $\snn=2.76$ TeV and similar 
  $\langle p_{\rm T} \rangle$~\cite{Gursoy:2014aka}, needs to be 
  measured with better precision.
This will be achieved with the large data sample expected at the HL--LHC
  which will be sensitive to a difference as small as $5 \times 10^{-5}$ 
  (about three times smaller that the current measurement), as reported by 
  the ALICE upgrade projection in Fig.~\ref{fig:alice_delta_v1}.
Furthermore, similar measurement can also be performed
in the heavy flavor sector, e.g., for $\rm D^{0}$ and $\rm \overline{D^{0}}$ meson directed flow~\cite{Das:2016cwd}, this is further discussed in Chapter~\ref{sec:HI_HF}. 
Heavy flavor quarks have the advantage of being produced at a very early stage, 
and thus potentially have a better sensitivity to the magnetic field at its evolution at early time.

\subsection{Summary}
The measurements of inclusive hadron \vn\ by traditional methods such as 
  two-particle correlations, event-plane/scalar-product methods, 
  multi-particle cumulants etc. have been performed with high precision 
  by the ALICE, ATLAS and CMS experiments at the LHC.
These inclusive hadron \vn\ measurements are not statistically limited 
  across most of the centrality-\pt\ phase space and further improvement 
  in the measurements is not a high priority for Run~3 and 4.
However, in the case of identified hadrons the increased statistics 
  will lead to further improvement in the \vn\ measurements.
This is true for both light hadrons such as pions, protons, $\phi$-mesons 
  as shown in Figure~\ref{fig:alice_vn}, as well as for heavy-flavor 
  particles such as $\rm D^0$, $\rm D^{\pm}$, J/$\psi$, $\Upsilon$ which are 
  discussed in Chapters~\ref{sec:HI_HF} and \ref{sec:quarkonia}, respectively.
Significant improvements are expected in measurements of 
  longitudinal flow fluctuations, which have only been briefly investigated
  in Run~1 and 2.
These are largely driven by the increases $\eta$ acceptance of the
  ATLAS and CMS tracking detectors in Run~4, the acceptance is planned 
  to reach $\pm$5 units.
The study of longitudinal flow fluctuations will allow comparisons to predictions 
  of 3+1D hydro models.
Flow measurements in light ions such as \ArAr\ and \OO, will lead 
  to stronger constraints on theoretical models describing different 
  stages of a heavy ion collision 
  -- initial conditions, equation of state, transport coefficients etc.
This is difficult presently, as flow observables are dependent 
  on all of these, so it becomes difficult to constrain any one of these 
  without full knowledge of the others.
Flow measurements across a variety of colliding species will provide independent 
  data that will improve our understanding of the different 
  stages of heavy ion collisions.
Further physics motivations for colliding light ions are discussed in Chapter~\ref{sec:smallAsum}.

Other observables related to collective phenomena where current 
  measurements are statistics limited and are expected to improve 
  considerably are related to effects of vorticity and magnetic fields.
The current measurements of $\Lambda$ polarization from ALICE are statistics limited
  and consistent with both the null hypothesis as well as with the theoretically
  predicated value.
The ALICE projections for $\Lambda$ polarization in Run~3 and 4 show that 
  the measurements will have significantly smaller statistical uncertainties 
  and will differentiate between the null and predicted values. 
ALICE and CMS have measured the fraction of the three-particle correlator
  $\gamma_{\alpha\beta}$ that arises from CME effects: $f_{\mathrm{CME}}$.
The measured $f_{\mathrm{CME}}$ by ALICE is consistent with zero but due to 
  large uncertainties its upper limit at 95\% CL can be as large as $\sim$0.5.
ALICE projections for Run~3 and 4 show that the $f_{\mathrm{CME}}$ can 
  be determined with a precision of better than 1\%.


\clearpage

\clearpage
\section{Open heavy flavour}
\label{sec:HI_HF}

{ \small
\noindent 
\textbf{Coordinators}: Elena Bruna (INFN Torino) and Gian Michele Innocenti (Massachusetts Institute of Technology, CERN)

\noindent \textbf{Contributors}: 
J.~Aichelin (SUBATECH, CNRS/IN2P3, IMT Atlantique, Universit{\'e} de Nantes),
S.A.~Bass (Duke University),
C.~Bedda (Utrecht University),
A.~Beraudo (INFN Torino),
G.E.~Bruno (Politecnico di Bari and INFN), 
Z.~Citron (Ben-Gurion University of the Negev),
Z.~Conesa~del~Valle (CNRS/IN2P3, Universit\'e Paris-Sud, Universit\'e Paris-Saclay, Orsay),
A.~Dainese (INFN Padova),
A.~Dubla (GSI Helmholtzzentrum f{\"u}r Schwerionenforschung GmbH),
F.~Fionda (University of Bergen),
P.B.~Gossiaux (SUBATECH, CNRS/IN2P3, IMT Atlantique, Universit{\'e} de Nantes),
V.~Greco (INFN Catania, Universit\`a di Catania, INFN-LNS),
F.~Grosa (Politecnico di Torino and INFN),
Y.-J.~Lee (Massachusetts Institute of Technology),
J.~Margutti (Utrecht University),
V. Minissale (INFN-LNS, Catania),
A.~Mischke (Utrecht University),
S.~Mohapatra (Columbia University, New York),
C.C.~Peng (Purdue University),
M.~Peters (Massachusetts Institute of Technology),
S.~Plumari (Universit\`a di Catania),
F.~Prino (INFN Torino),
A.~Rossi (Padova University and INFN),
J.~Sun (Tsinghua University, Beijing),
C.~Terrevoli (Padova University and INFN),
B.~Trzeciak (Utrecht University),
A.~Uras (Universit{\'e} de Lyon, CNRS/IN2P3, IPN-Lyon),
L.~Van Doremalen (Utrecht University, Utrecht),
I.~Vitev (Los Alamos National Laboratory),
J.~Wang (Massachusetts Institute of Technology),
T.-W.~Wang (Massachusetts Institute of Technology),
M.~Winn (LAL, now DPhN, CEA/IRFU),
R.~Xiao (Purdue University),
Y.~Xu (Duke University)
}

\vspace{3mm}
\begin{center}
{\em This chapter is dedicated to the memory of our colleague Andre Mischke.}
\end{center}


\subsection{Perspectives for heavy-flavour observables in LHC Run 3 and 4}
Charm and beauty quarks are produced in hard scattering processes occurring in the early stage of heavy-ion collisions. They subsequently traverse the QGP medium and interact with its constituents through inelastic (gluon radiation) and elastic (or collisional) processes.
These interactions may lead to the thermalisation of low-momentum heavy quarks, which would thus take part in the expansion and hadronisation of the medium.
For these reasons, heavy-flavour hadrons provide information on all stages of the system evolution and they uniquely probe the quark-mass dependence of the QGP inner workings (see Refs.~\cite{Andronic:2015wma,Prino:2016cni,Rapp:2018qla} for recent reviews).

Many experimental observations from RHIC and LHC showed evidence that charm and beauty quarks interact strongly with the QGP and that beauty quarks lose less energy at low transverse momentum compared to charm quarks~\cite{Adam:2015nna,Khachatryan:2016ypw}. 
While data are becoming more and more precise to start imposing constraints on theoretical calculations, there are still several unresolved questions: What are the microscopic mechanisms that drive heavy-flavour interaction and diffusion in the QGP, and what are their implications for the QCD matter structure? What is the relative relevance of collisional and radiative processes? Can the same QCD process can describe both the heavy-quark interaction with the strongly coupled plasma and the mechanisms of hadronisation?

The Run 3 and 4 of the LHC will open a new precision era for heavy-flavour measurements in heavy-ion collisions that will address the above questions.
With the upgrades of the machine and of the tracking detectors, the higher accumulated statistics and higher precision will make it possible to quantify the properties of the QGP with heavy-flavour probes. This high-precision era will also make new and more differential observables accessible for the first time.
The key measurements that are expected to have a strong impact on the characterisation of the QGP with heavy-flavour observables are discussed in this chapter and summarized below.
\begin{itemize}
\item Nuclear modification factor and flow harmonics: these measurements for particles with charm and beauty in the large kinematic range covered by combining the different LHC experiments will put the strongest constraints on the transport coefficients of the QGP, clarifying the microscopic mechanisms governing the interactions of heavy quarks with the medium, and quantifying their degree of thermalisation.
\item Strange D and B mesons, charm and beauty baryons: currently limited by statistics, these measurements will help to quantify not only the degree of thermalisation of heavy quarks, but also the contribution of recombination with lighter quarks to the hadronisation process. They are also sensitive to the mass scaling of the hyrodynamical flow in the heavy-flavour sector.
\item Heavy-flavour correlations and jet observables: they will provide new insights on the parton mass effects in parton showers, on the redistribution of the radiated energy, and on the role of collisional and radiative energy loss.

\end{itemize}

\subsection{Impact of detector upgrades on heavy-flavour measurements}
The upgrades of the four large LHC experiments during LS2 and LS3 will strongly enhance their performance for open heavy-flavour measurements. The detector improvements that will have the largest impact are the new silicon trackers, with higher granularity and precision, as well as extended pseudo-rapidity coverage. Brief descriptions of these improvements are reported in the following.

\begin{itemize}
\item ALICE. The new Inner Tracking System~\cite{Abelev:1625842}, which will be installed during LS2, is composed of seven layers of pixel detectors with an intrinsic spatial precision of about $5\times 5~$\Uum$^2$ and a material thickness of 0.3\% of the radiation length in the innermost layers. The track pointing resolution will be improved by a factor 3 in the direction transverse to the beam line and by a factor 5 in the longitudinal direction, down to values of about \unit[20]{$\Uum$} for tracks with $\pT=1~$\UGeVc. A Muon Forward Tracker~\cite{CERN-LHCC-2015-001}, composed of 5 disks of pixel detectors with the same spatial resolution as the Inner Tracking System, will instrument the region $2.5<\eta<3.6$, in front of the muon spectrometer, enabling the separation from the primary vertex of single and dimuons from D, B and \PJGy decays. The upgraded TPC with GEM-based readout chambers, together with readout upgrades of several other detectors and with a new Online-Offline computing system, will enable the full recording of \PbPb interactions with a minimum-bias trigger at a rate of 50~kHz, which is 50-fold larger than for the present apparatus.    
\item ATLAS.  The Inner Tracker (ITk)~\cite{ATL-PHYS-PUB-2016-025} will be an all-silicon tracker composed of pixels and strips installed during LS3 for ATLAS phase II.  The ITk will provide charged-particle tracking acceptance for $|\eta|<4$.  The performance of the ITk in \PbPb collisions is expected to be comparable to pp collisions.  The High Granularity Timing Detector~\cite{Collaboration:2623663} has been proposed to complement the spatial information of the ITk with timing information.  These detectors will improve jet reconstruction capabilities, and in particular tagging of heavy-flavour jets, as well as all studies using charged particles.
\item CMS. The following upgrades scheduled for LS3 will largely enhance the performance for heavy-flavour measurements, in particular in the low-momentum region~\cite{CMS:FTR-18-012}. The upgraded inner tracker will cover a large acceptance up to $|\eta|<4$~\cite{Contardo:2020886}. The improved L1 and DAQ rate (up to 60~GB/s) will allow more sophisticated triggers and to record a larger number of minimum-bias triggered events. In addition, the proposed MIP Timing Detector~\cite{Collaboration:2272264} with a radius of 1.16~m and a time resolution of $\approx 30$~ps could provide, in conjunction with other detectors, proton, pion and kaon separation in the interval $0.7<\pT < 2$~\UGeVc~in 
$|\eta|<1.5$. 
\item  LHCb. The experiment is preparing to run at five times larger instantaneous luminosities in pp collisions, processing the full event rate with a software trigger and preserving or exceeding the present performance.
All tracking detectors will be upgraded during LS2~\cite{Collaboration:1624070,Collaboration:1647400}. 
Most notably for heavy-flavour observables, the active area of the upgraded Vertex Locator, the pixel detector replacing the present silicon strip detector, will move as close as 5.1~mm to the nominal beam spot.
The larger granularity for the majority of phase space will improve the performance in \PbPb collisions whereas proton-induced reactions will result in average in lower detector occupancies than the standard pp running.  
\end{itemize}

\subsection{Nuclear modification factor and collective flow}



The standard observable used to study the medium effects on heavy-flavour meson production is the nuclear modification factor ($\RAA$), defined as the ratio of the \PbPb yield to the pp cross-section scaled by the nuclear overlap function. In the view of the pQCD-based models, heavy quarks interact with the medium constituents via radiative and collisional processes. While radiative interactions only lead to energy loss, collisional ones can also result in an increase of the heavy-quark momentum. The dead-cone effect~\cite{DOKSHITZER2001199} is expected to reduce small-angle gluon radiation of heavy quarks when compared to both gluons and light quarks. At low $\pt$, the production rate of heavy-flavour mesons in heavy-ion collisions is sensitive to the elastic energy loss of the heavy quark in medium, the nuclear shadowing effect in the initial state, and the recombination of the heavy quark with light quarks at the hadronization stage. At high $\pt$, the nuclear modification factor is sensitive to the medium-induced radiative energy loss of heavy quarks. Precise measurements of the $\RAA$ thus provide insights on the momentum dependence of heavy quark energy loss, and provide important tests of QCD predictions, in particular for the expected flavour and mass dependence of the energy loss processes.

Another interesting observable is the azimuthal anisotropy of open heavy flavour production, which can be characterized by the Fourier coefficients $v_n$ in the azimuthal angle ($\varphi$) distribution of the heavy-flavour hadron yield with respect to the reaction plane in non-central \PbPb collisions. 
At low $\pt$, the $\vtwo$ measurements can provide important insights into the mechanisms of interaction of heavy quarks with the medium and on their strength (as discussed in more details in Sec.~\ref{sec:HFDs}). Heavy quarks are indeed expected to acquire a positive $\vtwo$ mostly as a consequence of their interaction with the light quarks of the medium.
Measurements of elliptic flow of heavy hadrons are also sensitive to hadronisation processes. In particular, they can be used to study the relevance of heavy-flavour recombination (see Sect.~\ref{sec:HFhadro1}) in which heavy-quarks can acquire additional $\vtwo$ by combining with light quarks at the hadronisation stage. 
At high $\pt$, $\vtwo$ of heavy-flavour hadrons is sensitive to the path-length dependence of heavy quark energy loss. The simultaneous description of $\RAA$ and $\vtwo$ for heavy-flavour hadrons is still challenging for most of the theoretical calculations, because it entails accurate modelling of the initial heavy-quark production and its modification in nuclei, of the medium and its expansion, of the various quark-medium interaction mechanisms and of the possible modification of hadronisation processes.


\subsubsection{Experimental performance of the ALICE, ATLAS and CMS experiments}
\label{sec:HFRAAv2}

\begin{figure}[ht]
  \begin{center}
    \includegraphics[width=0.43\textwidth]{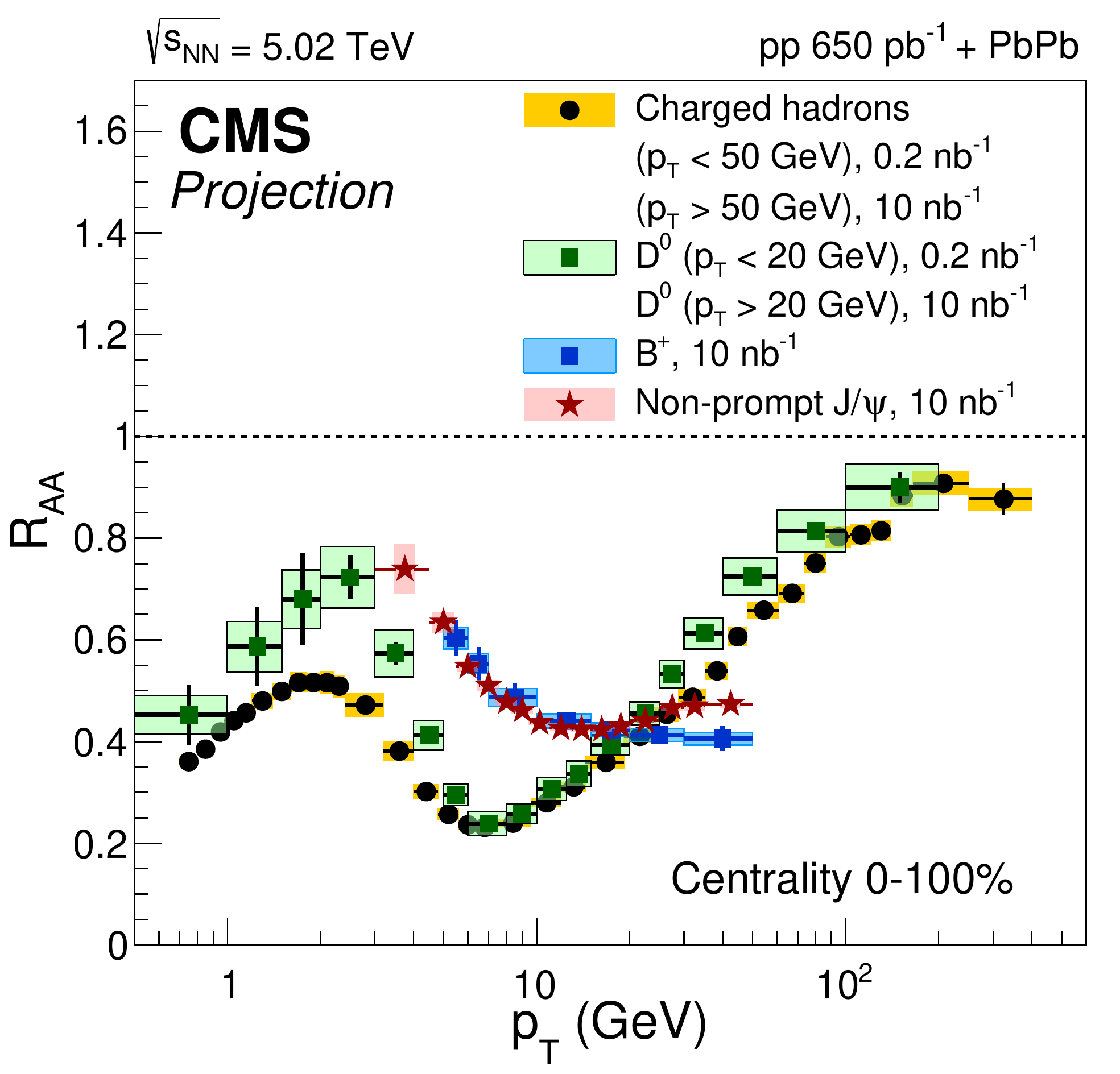}
   \includegraphics[width=0.54\textwidth]{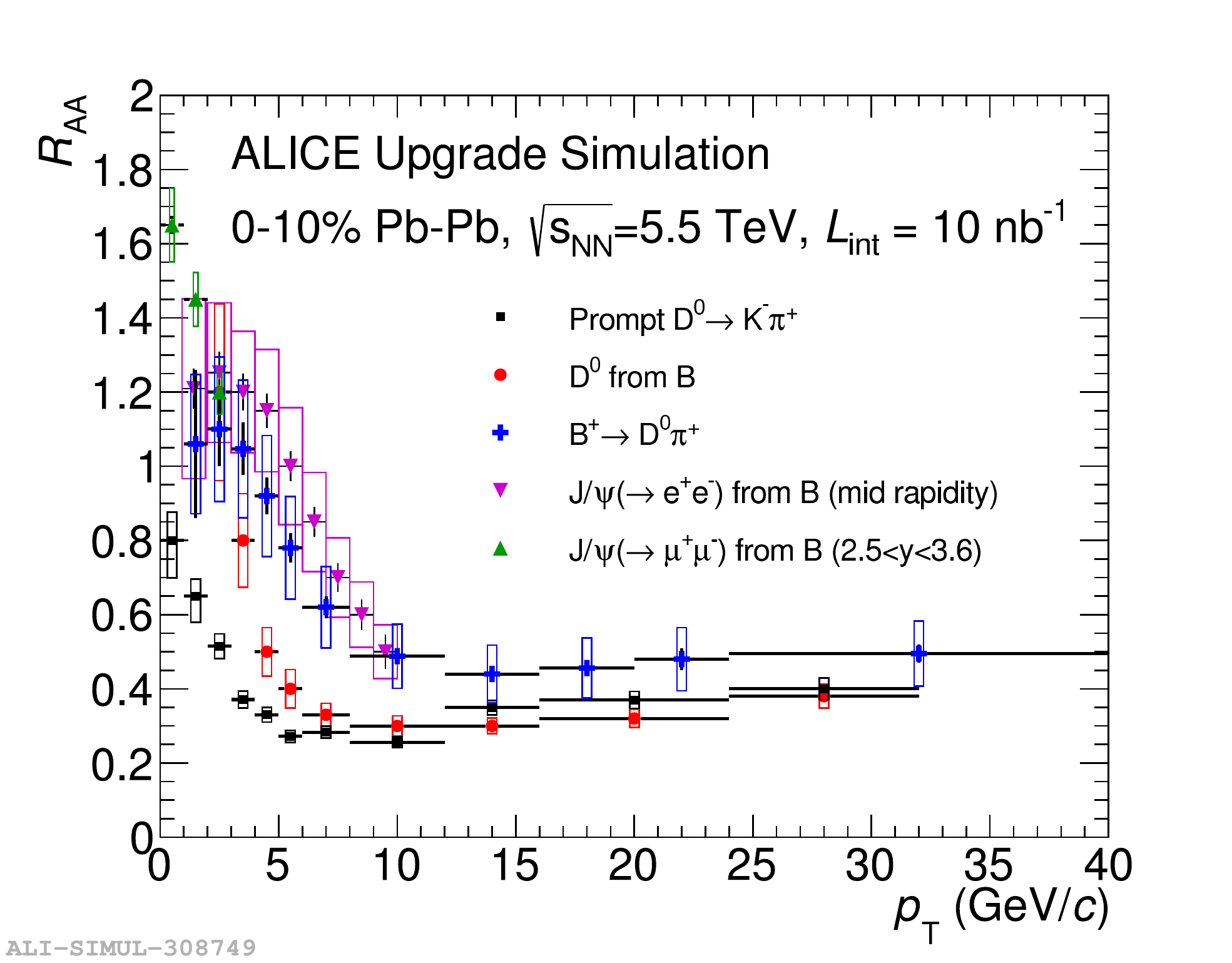}
    \caption{Nuclear modification factors of charged particles, $\PDzero$, $\PBp$ and non-prompt \PJGy in CMS~\cite{CMS-PAS-FTR-17-002} (left). $\RAA$ of $\PDzero$, non-prompt \PJGy and non-prompt $\PDzero$ in ALICE in central \PbPb collisions for $\Lint=10~\invnb$~\cite{Abelev:1625842,ALICE-PUBLIC-2019-001} (right).}
    \label{fig:RAAv2.RAA}
  \end{center}
\end{figure}

Figure~\ref{fig:RAAv2.RAA} shows the projected performance for the $\RAA$ of several heavy-flavour hadrons or decay channels with $\Lint=10~\invnb$. The left panel presents the projection of charged particles, $\PDzero$, $\PBp$ and non-prompt \PJGy from b-hadron decay which can be measured by CMS. The right panel shows the ALICE simulation results for $\PDzero$, non-prompt \PJGy, non-prompt $\PDzero$, $\PBp$ ($\rightarrow \PDzero \PGpp$) 
---in addition, the $\rm B^0\rightarrow D^{*+}\pi^-$ reconstruction was studied by ALICE and it provides an alternative channel for the study of the beauty meson $\RAA$ with a significance of larger than $5\,\sigma$ at $\pt > 3$~GeV/$c$. With the high luminosity and the Inner Tracking System Upgrade in ALICE, the $\RAA$ of light hadrons, charm hadrons and beauty hadrons can be clearly separated in a wide kinematic range.

\begin{figure}[t]
  \begin{center}
    \includegraphics[width=0.43\textwidth]{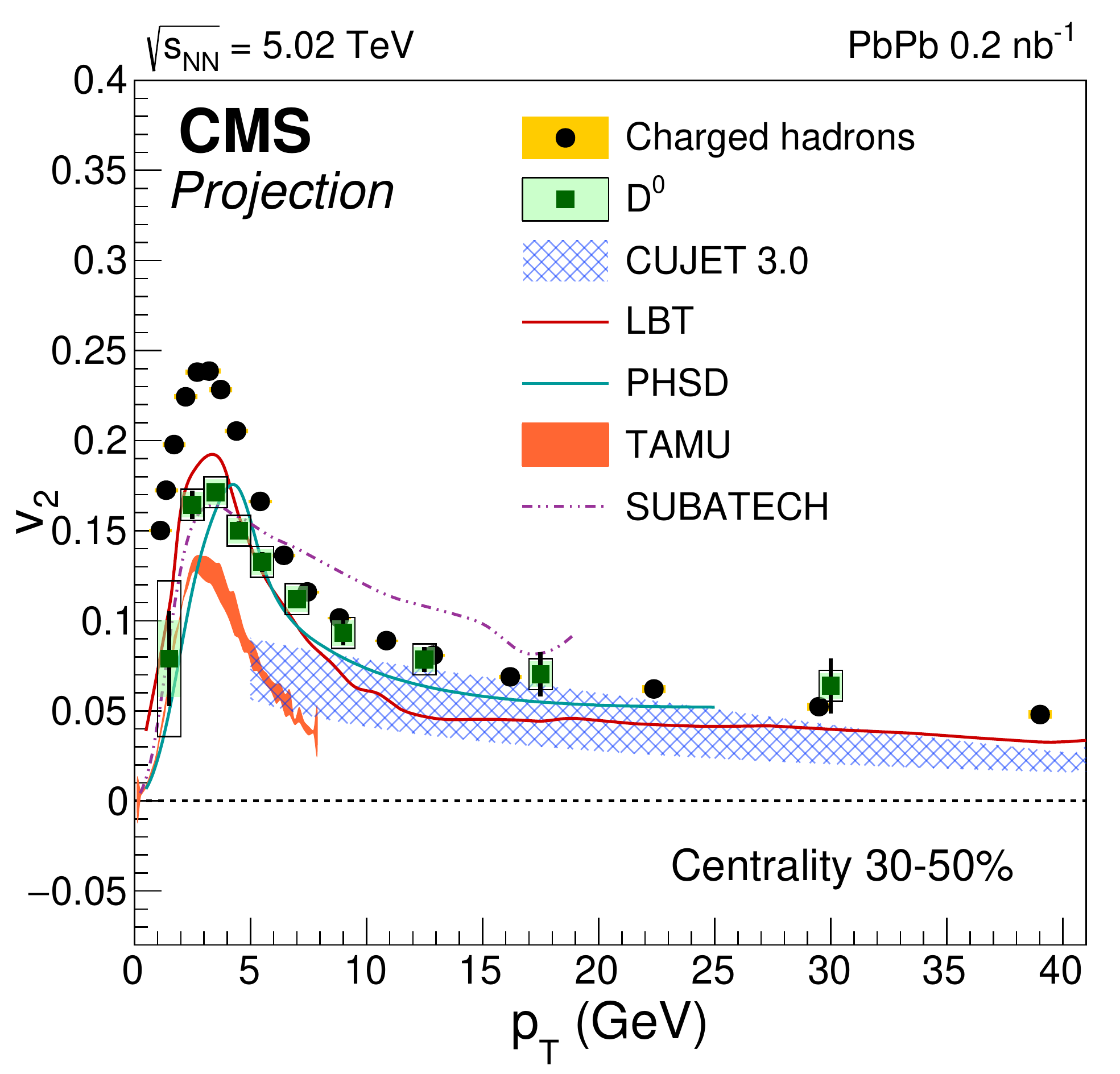}
   \includegraphics[width=0.54\textwidth]{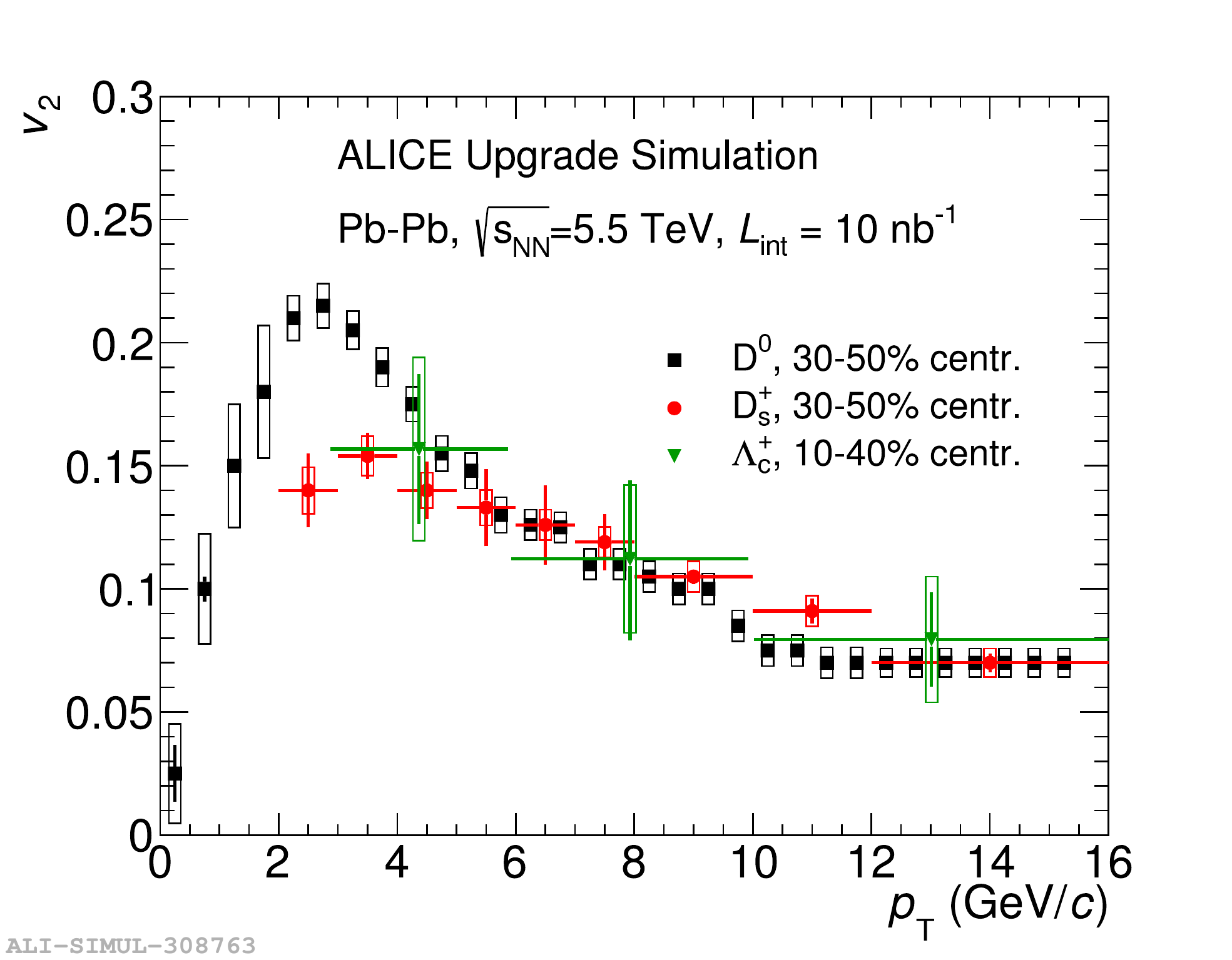}
    \caption{$\vtwo$ of charged particles and $\PDzero$ in CMS~\cite{CMS-PAS-FTR-17-002} (left), charm hadrons ($\PDzero$, $\PDs$, \PGLc) in ALICE (right) in \PbPb collisions with $\Lint=10~\invnb$~\cite{Abelev:1625842,ALICE-PUBLIC-2019-001}.}
    \label{fig:RAAv2.v2charm}
  \end{center}
\end{figure}

\begin{figure}[t]
  \begin{center}
    \includegraphics[width=0.49\textwidth]{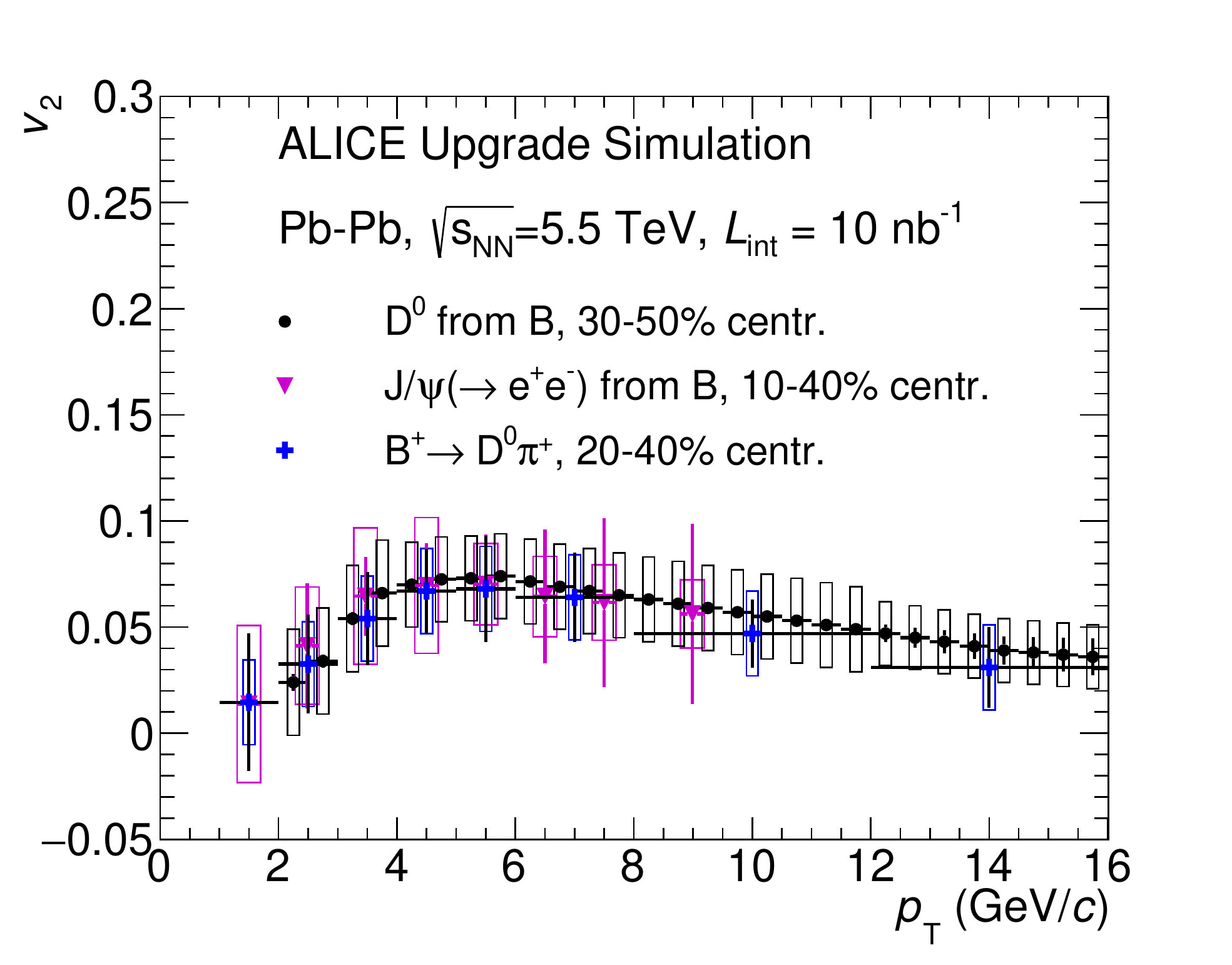}
    \includegraphics[width=0.49\textwidth]{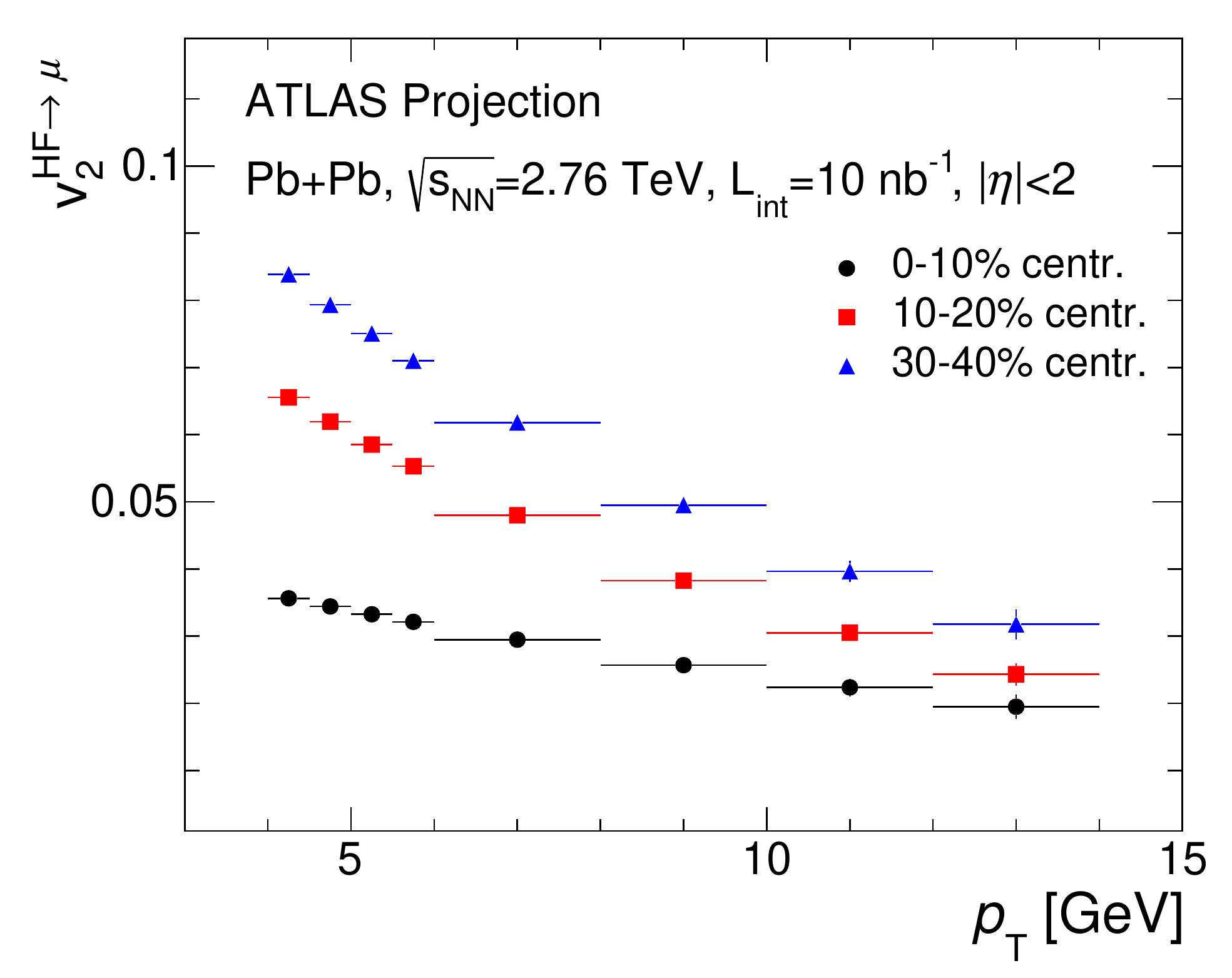}
    \caption{$\vtwo$ of non-prompt $\PDzero$, non-prompt \PJGy and $\PBp$ ($\rightarrow \PDzero$) in ALICE (left)~\cite{Abelev:1625842,ALICE-PUBLIC-2019-001} and of heavy-flavour decay muons in ATLAS (right)~\cite{ATL-PHYS-PUB-2018-020} in \PbPb collisions for $\Lint=10~\invnb$.}
    \label{fig:RAAv2.v2beauty}
  \end{center}
\end{figure}

Figure~\ref{fig:RAAv2.v2charm} shows the projected performance for $\vtwo$ of charm hadrons with $\Lint=10~\invnb$. The left panel shows the projection for $\PDzero$ in CMS, with the charged particle $\vtwo$ also shown for comparison. The right panel presents the projection for $\PDzero$, $\PDs$ and \PGLc in ALICE~\cite{Abelev:1625842}. Precise measurements of charm hadron $\vtwo$ will allow the study of the thermalization of heavy quarks and the wide kinematic range allows to get insights on different process, as coalescence hadronization and energy loss. Figure~\ref{fig:RAAv2.v2beauty} (left) shows the projected performance for $\vtwo$ of $\rm B^+$ mesons, non-prompt $\PDzero$ and non-prompt \PJGy. These will be the first precise measurements of B meson elliptic flow at the LHC.
Heavy-flavour flow will also be measured with high precision using decay electrons and muons. As an example the projection for the measurement with muons by ATLAS is shown in Fig.~\ref{fig:RAAv2.v2beauty} (right)~\cite{ATL-PHYS-PUB-2018-020}.

\subsubsection{Constraining the heavy-quark diffusion coefficient $\twopiTDsc$}
\label{sec:HFDs}

Many theoretical efforts have been recently undertaken to understand the properties of the QGP medium and the interaction between heavy quarks and the medium constituents, see Refs.~\cite{Andronic:2015wma,Prino:2016cni,Rapp:2018qla} for recent reviews. 
Although the interaction mechanism can widely vary among different theoretical models, the reduction to a few transport coefficients allows one to compare these models and evaluate different microscopic pictures. 
Most of the present theoretical models explain the interactions of heavy quarks as dominated by collisional (elastic) processes in the low transverse momentum region (up to about 5--10~GeV/$c$) and by radiative energy loss (inelastic process with gluon radiation off the heavy quark) at higher $\pt$.

The extraction of the heavy-quark spatial diffusion coefficient, which is one of the main QGP properties regulating the strength of collisional processes, is considered here to illustrate the impact of the high-precision measurements in Run 3 and Run 4. In particular, the $\pt$-dependence of this coefficient provides important constraints on the weakening of the interaction strength with increasing $\pt$. For illustration, the extracted values of $D_s$ is considered at a fixed $\pt$ value. The heavy-quark spatial diffusion coefficient $D_s$ in the QGP is related to the relaxation (equilibration) time of heavy quarks $\tau_{\rm Q} = \frac{m_{\rm Q}}{T}D_s$, where $m_{\rm Q}$ is the quark mass and $T$ is the medium temperature~\cite{Moore:2004tg}.


\begin{figure}[t]
	\begin{center}
		\includegraphics[width=0.54\textwidth]{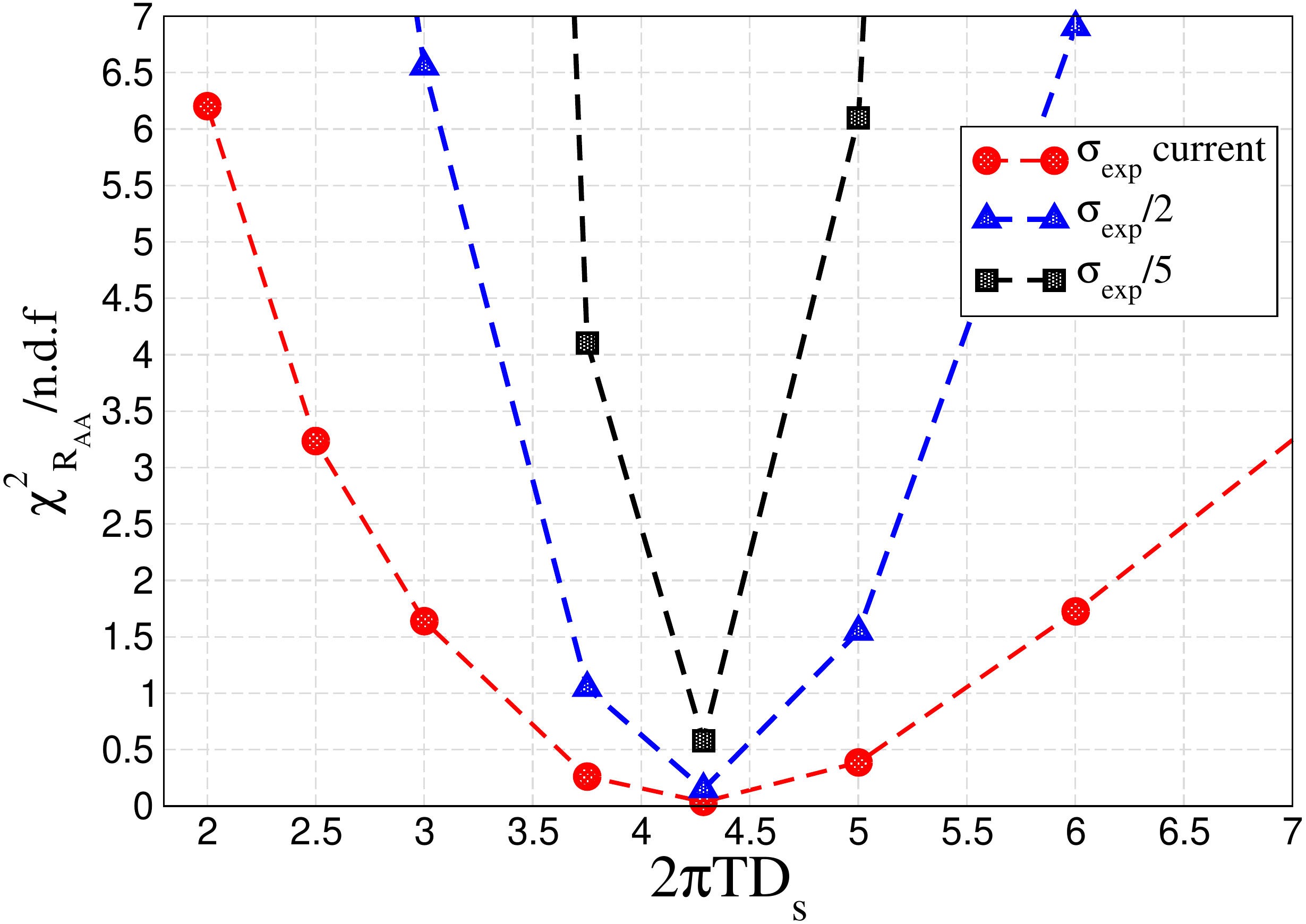}
		\includegraphics[width=0.45\textwidth]{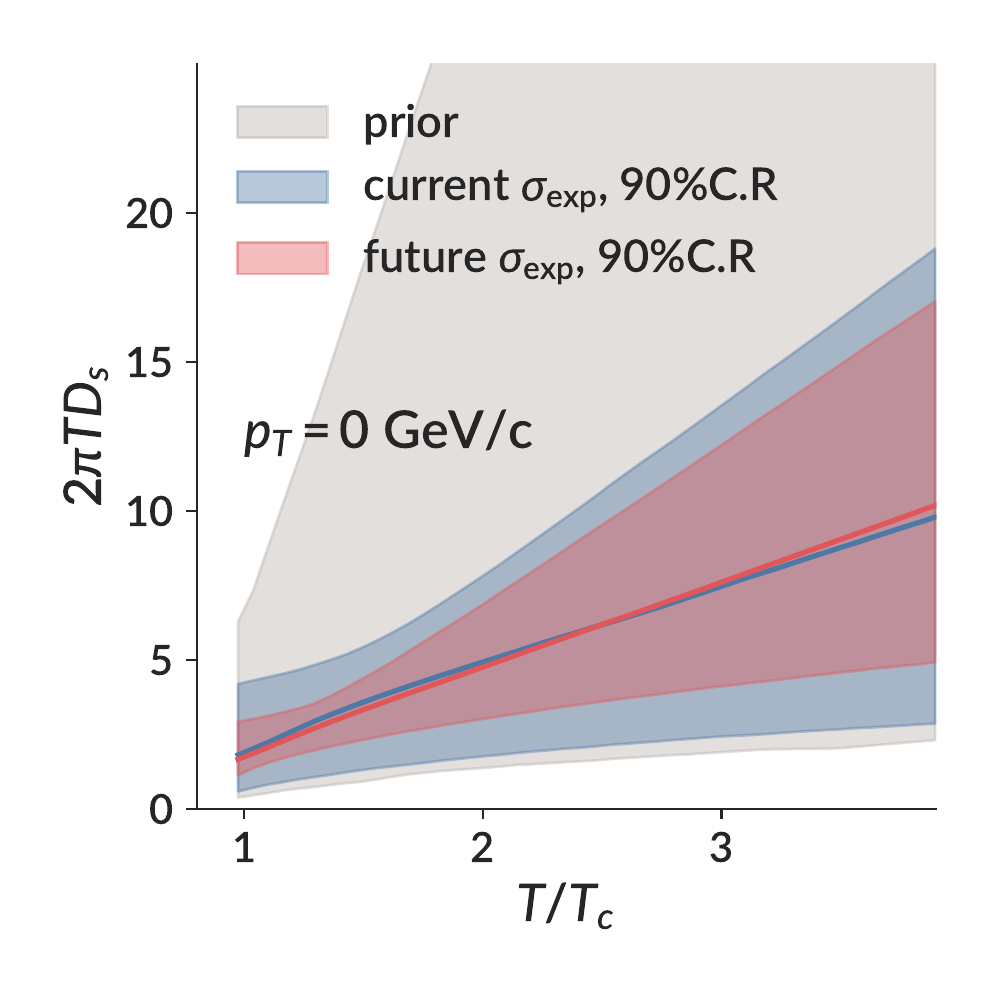}
		\caption{Left: Normalized $\chisquared$ as a function of spatial diffusion coefficient ($\twopiTDsc$) for different experimental precision levels from the Catania Fokker-Plank transport model~\cite{Das:2015ana,Das:2013kea}. Right: Coefficient range (90\% credibility region) for $\twopiTDsc$ as a function of $\ToverTc$ for different experimental precision levels, estimated by a model-to-data Bayesian analysis using a modified Langevin framework~\cite{PhysRevC.97.014907}.}
		\label{fig:RAAv2.Dstheory}
	\end{center}
\end{figure}

Figure~\ref{fig:RAAv2.Dstheory} shows the constraining power of future experimental measurements of  $\RAA$ and $v_2$ on the heavy quark diffusion coefficient ($\twopiTDsc$) using two different transport models: Catania model with Fokker-Plank equation~\cite{Das:2015ana,Das:2013kea} on the left and a modified Langevin framework on the right~\cite{PhysRevC.97.014907}. The left figure presents a normalized $\chisquared$ as a function of spatial diffusion coefficient by comparing the model calculation~\cite{Das:2015ana,Das:2013kea} of D-meson $\RAA$ in Pb--Pb collisions at 5.02~TeV in a single centrality class (0--10\%). 
The cases of the present experimental uncertainties (2015 Pb--Pb sample) and of these uncertainties reduced factors of two or five are considered for illustration. In the projections for $\Lint=10~\invnb$ shown in the previous section, the D-meson $\RAA$ uncertainties are reduced by a factor between two and five, depending on $\pt$, with respect to the present measurements.
Considering the $\twopiTDsc$ range with $\chisquared_{\RAA}/{\rm n.d.f}<1.5$ (corresponding to 85\% confidence), it is found that by reducing the present experimental uncertainty by a factor two or five, the uncertainty on the estimation $\twopiTDsc$ would be also reduced almost to 50$\%$ or 20\%, respectively. The right panel presents the diffusion coefficient as a function of temperature, which is estimated using a Bayesian calibration on D-meson $\RAA$ and $\vtwo$ in \PbPb collisions at 5.02~TeV for different centralities. The $\twopiTDsc$ shows a positive temperature dependence with the minimum value around $\Tc$. Such behaviour is consistent with the Bayesian estimation for shear viscosity $\eta/s$. The potential improvement with Run 3 and Run 4 measurements is estimated using the $\RAA$ and $v_2$ projections by ALICE and CMS shown in the previous section and it is shown by the red band. With these future experimental measurements, the diffusion coefficient around $\Tc$ could be constrained with an uncertainty of about 30--50\% of the present one. 




\subsubsection{D-meson analyses with Event Shape Engineering}

\begin{figure}[t]
  \begin{center}
    \includegraphics[width=0.7\textwidth]{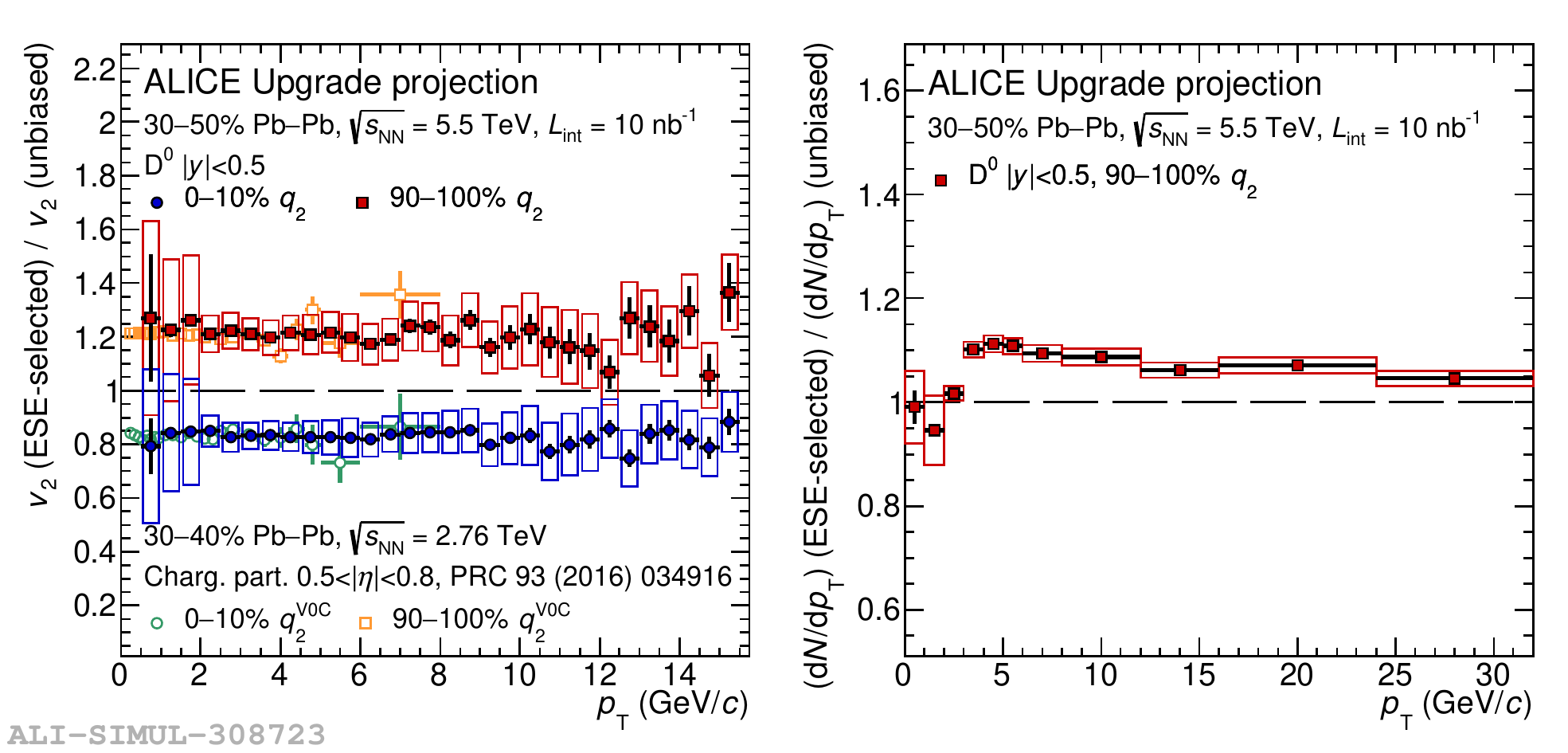}
 
    \caption{Left: projection of the expected ratio of $\PDzero$-meson $\vtwo$ in the 10\% events with larger (smaller) $q_2$ with respect to the unbiased one as a function of $\pt$ for the 30--50\% centrality class. The modification of the $\PDzero$-meson $\vtwo$ was assumed to be equal to that measured for the charged particles in Pb--Pb collisions at $\sqrtsNN=2.76~{\UTeV}$ in the the 30--40\% centrality class (superimposed for comparison)~\cite{Adam:2015eta}. Right: projection of the expected ratio of $\PDzero$-meson $\pt$-differential yield in the 10\% events with larger $q_2$ with respect to the unbiased one, estimated considering the prediction provided by the POWLANG model~\cite{Beraudo:2018bxb}. Figure from Ref.~\cite{ALICE-PUBLIC-2019-001}.}
    \label{fig:ESE}
  \end{center}
\end{figure}

Further insight into the dynamics of heavy quarks in the medium can be
obtained from measurements of the yield and elliptic flow of heavy-flavour
particles with the Event Shape Engineering (ESE) 
technique~\cite{Schukraft:2012ah}.
This technique consists of selecting events with the same centrality but 
different magnitude of the average bulk elliptic flow and therefore initial-state geometry eccentricity.
The analyses with ESE will allow us to investigate the correlation between 
the flow coefficients of heavy-flavour hadrons and soft hadrons, to study the interplay between elliptic and radial flow, and to further constrain the path-length dependence of the energy loss suffered by the heavy quarks in the QGP. A first analysis was published by ALICE using 2015 Pb--Pb data~\cite{Acharya:2018bxo}.
In the left-hand panel of Fig.~\ref{fig:ESE}, the prospects for the measurement of the ${\PDzero}$-meson $\vtwo$ with the ESE technique in the 30--50\% centrality class with $\Lint=10~\invnb$ are shown.
In particular, the ratio between the $\vtwo$ of D mesons in the 10\% of the events with larger (smaller) elliptic flow of the bulk (quantified through the magnitude of the so-called reduced flow vector $q_2$) and the $\vtwo$ in all the collisions in the considered centrality class is reported as a function of $\pt$.
It is compared to the current measurement of the same ratio for charged 
particles, which is dominated by light-flavour hadrons.
The expected statistical uncertainties for the $\PDzero$-meson $\vtwo$ in the 0--10\% of events with larger (smaller) $q_2$ are of the order of about 1--2\% in the interval $1 < \pt < 8~\UGeVc$. This will allow us to resolve a possible difference of a few percent in the response of the $\vtwo$ to the ESE selection between the $\PDzero$ mesons and the light hadrons, providing new insight on the coupling of the charm quark with the medium constituents and on its degree of thermalisation. 
The performance for the measurements of $\PDzero$-meson 
$\pt$-differential yield in event-shape classes is displayed in the right-hand panel of Fig.~\ref{fig:ESE}.
The expected performance will provide a sensitivity of a few percent for the modification of 
the D-meson $\pt$ spectra in events with small (large) initial 
geometrical anisotropy.
This will open the way for precise studies on the interplay between the initial geometrical anisotropy (the collective flow of the bulk) and the heavy-flavour radial flow and energy loss.

\subsection{Studies of heavy-quark hadronisation}
\subsubsection{Hadronisation mechanisms for c and b quarks}
\label{sec:HFhadro1}


The hadronisation mechanisms belong to the non-perturbative domain of QCD and a
first-principle description of these processes is still missing, for both light and heavy flavours.
However, from the study of charm dynamics in nucleus--nucleus collisions in the last decade
there is a general consensus that the details of hadronisation have a large effect on both the heavy-flavour observables 
$\RAA$ and $\vtwo$~\cite{Dong:2018ntm,Rapp:2018qla,Scardina:2017ipo}.
This section introduces 
the two main microscopic hadronisation mechanisms for the production of heavy-flavour hadrons: fragmentation and coalescence (also denoted as recombination).
Fragmentation is one of the most common approaches for the calculation of inclusive hadron production and it is appropriate
for high-momentum partons emerging
from initial hard processes, where high-momentum quarks fragment directly and independently into high-momentum hadrons. Independent fragmentation has also been widely applied at low momentum in $\rm e^+e^-$, ep and pp collisions.
On the other hand, coalescence is expected to dominate in the low-momentum regime in nucleus--nucleus collisions, where partons are abundant and heavy quarks
can hadronise by recombination with light quarks~\cite{PhysRevC.68.044901,GRECO2004202,Fries:2008hs}. Recent measurements at the LHC indicate that fragmentation may not be sufficient to 
describe charm quark hadronisation at low momentum in pp and p--Pb collisions, at least
for what concerns baryon production~\cite{Acharya:2017kfy,Acharya:2017lwf}.

The hadron momentum spectra produced by heavy-quark fragmentation are given by:
\begin{equation}
\frac{{\rm d}N_{\rm had}}{{\rm d}^{2}\pT\,{\rm d} y}=\sum \int {\rm d} z \frac{{\rm d} N_{\rm fragm}}{{\rm d}^{2}\pT\, {\rm d} y} \frac{D_{\rm had/Q}(z,Q^{2})}{z^{2}} 
\label{Eq:frag}
\end{equation}
where $z=p_{\rm had}/p_{\rm Q}$ is the fraction of quark momentum carried by the hadron, $Q^2=(p_{\rm had}/2 z)^2$ 
is the momentum scale for the fragmentation process and $D_{\rm had/Q}$ is the fragmentation function. 

In the basic coalescence model developed in Refs.~\cite{Greco:2003mm,Greco:2003vf,Fries:2003kq,Fries:2003vb,Oh:2009zj,Minissale:2015zwa,Plumari:2017ntm} and used here for illustration,
the spectrum of heavy-flavour hadrons formed by coalescence of heavy-light quarks can be written as
\begin{equation}
\label{eq-coal}
\frac{{\rm d}^{2}N_{\rm had}}{{\rm d}\pT^{2}}
= g_{\rm had} \int \prod^{n}_{i=1} \frac{{\rm d}^{3}p_{i}}{(2\pi)^{3}E_{i}} p_{i} 
\cdot {\rm d}\sigma_{i}  \cdot f_{q_i}(x_{i}, p_{i})  
f_{\rm had}(x_{1}...x_{n}, p_{1}...p_{n})\, 
\delta^{(2)} \left(p_{\rm T,had}-\sum^{n}_{i=1} p_{{\rm T},i}\right)
\end{equation}
where ${\rm d}\sigma_{i}$ denotes an element of a space-like hypersurface and $f_{q_i}$ are the quark (antiquark)
phase-space distribution functions for $i$-th quark (antiquark), while $g_{\rm had}$ is the statistical factor
to form a colourless hadron from quarks and antiquarks with spin 1/2. An alternative approach proposes resonance formation in quark-antiquark scatterings as the main production channel of D mesons through coalescence. In this approach, the same QCD process is responsible for both the interaction of charm quarks with the strongly-coupled medium and for the process of D-meson formation through recombination~\cite{RAVAGLI2007126,PhysRevC.86.014903}.

Hadronisation via coalescence leads to a modification of the relative abundance of the
various heavy-flavour hadron species produced. 
The most striking effect is an enhancement of the baryon-to-meson ratios for heavy-flavour hadrons.
First studies of these ratios~\cite{Oh:2009zj,Plumari:2017ntm} indicated a significant change in the relative abundances of the heavy-flavour hadron species, and in particular a ratio of $\PGLc/\PDzero$ close to unity, which
is nearly an order of magnitude larger than what predicted by the fragmentation
process implemented in the PYTHIA event generator. 
This value is also much larger than the
prediction of the statistical hadronisation model~\cite{Andronic:2007zu}, in which the hadronisation occurs by recombination of an equilibrated system of quarks and the hadron abundances are mainly determined by their masses.
In addition, hadronisation via coalescence in a strangeness-rich Quark-Gluon Plasma is predicted to lead to a large enhancement in the production of strange heavy-flavour hadrons, like $\PDs$ and $\rm B_s$~\cite{Kuznetsova:2006hx,He:2014cla,Song:2015ykw}.
Hints of an enhancement of the $\PDs/\PDzero$ and $\PGLc/\PDzero$ ratios in nucleus--nucleus collisions have recently been reported by ALICE at the LHC~\cite{Acharya:2018hre,Peng:2018pwj} and by STAR at RHIC~\cite{Radhakrishnan:2018xxx}.

The hadronisation by coalescence plus fragmentation also affects the $\pT$ distribution  of heavy-flavour hadrons. For D mesons the effect can be roughly seen as a
shift in $\pT$ of about 1.0--1.5~GeV/$c$ in the region of $\pT$ of 1.5--6~\UGeVc, resulting in a
significant enhancement of $\RAA(\pT)$. The degree of these enhancements for D mesons, however, is
significantly different  among the different implementations of the hadronisation
by coalescence and it is nearly unknown for heavy-flavour baryons, like \PGLb and \PGLc.

Finally, the coalescence process leads to a significant enhancement of the $\vtwo(\pT)$ in the
intermediate $\pT$ region that, for D mesons, amounts to about 20--40\% depending on the
specific modelling. The \PGLc is expected to acquire instead a much 
larger enhancement of at least a factor of two for the $\vtwo$. Therefore, a combined measurement of the
$\rm \PGLc/D$ and $\rm \Lambda_b/B$ ratios and of the baryon elliptic flow would impose strong constraints
on the hadronisation mechanism and lead to a better determination of the diffusion transport coefficient of heavy quarks (see Section~\ref{sec:HFhadro3}).

\subsubsection{Measurement performance studies (ALICE and CMS)}
\label{sec:HFhadro2}

Simulation studies for the measurement of the $\PDs$, \PGLc and \PGLb
production were carried out by the ALICE Collaboration~\cite{Abelev:1625842} and are updated in the present document. A projection of the performance for the $\PBs$ meson by the CMS Collaboration is also reported~\cite{CMS-PAS-FTR-17-002}. The $\PDs$, \PGLc and \PGLb ($\to\Lambda_c\pi$) reconstruction strongly benefits from the improved track spatial resolution of the ALICE Inner Tracking System Upgrade, because they have small mean proper decay lengths (e.g.\,about 60~$\mu$m for \PGLc)
and large combinatorial backgrounds. The \PGLc, \PGLb and $\PBs$, in particular, require very large integrated luminosities, because the decay branching ratios are very small (e.g.\,about $3\times 10^{-4}$ for $\rm \Lambda_b\to\Lambda_c(\to pK\pi)\pi$) and the combinatorial background is very large, in case for the \PGLc, which has a small separation from the interaction vertex and lower invariant-mass than the b-hadrons. In~\cite{CMS-PAS-FTR-17-002} it has been shown that the statistical uncertainty in the lowest accessible $\pt$ intervals for these hadrons would increase above 20--30\% with integrated luminosity significantly lower than 10~nb$^{-1}$.

\begin{figure}[!t]
\centering
\includegraphics[width=0.55\textwidth]{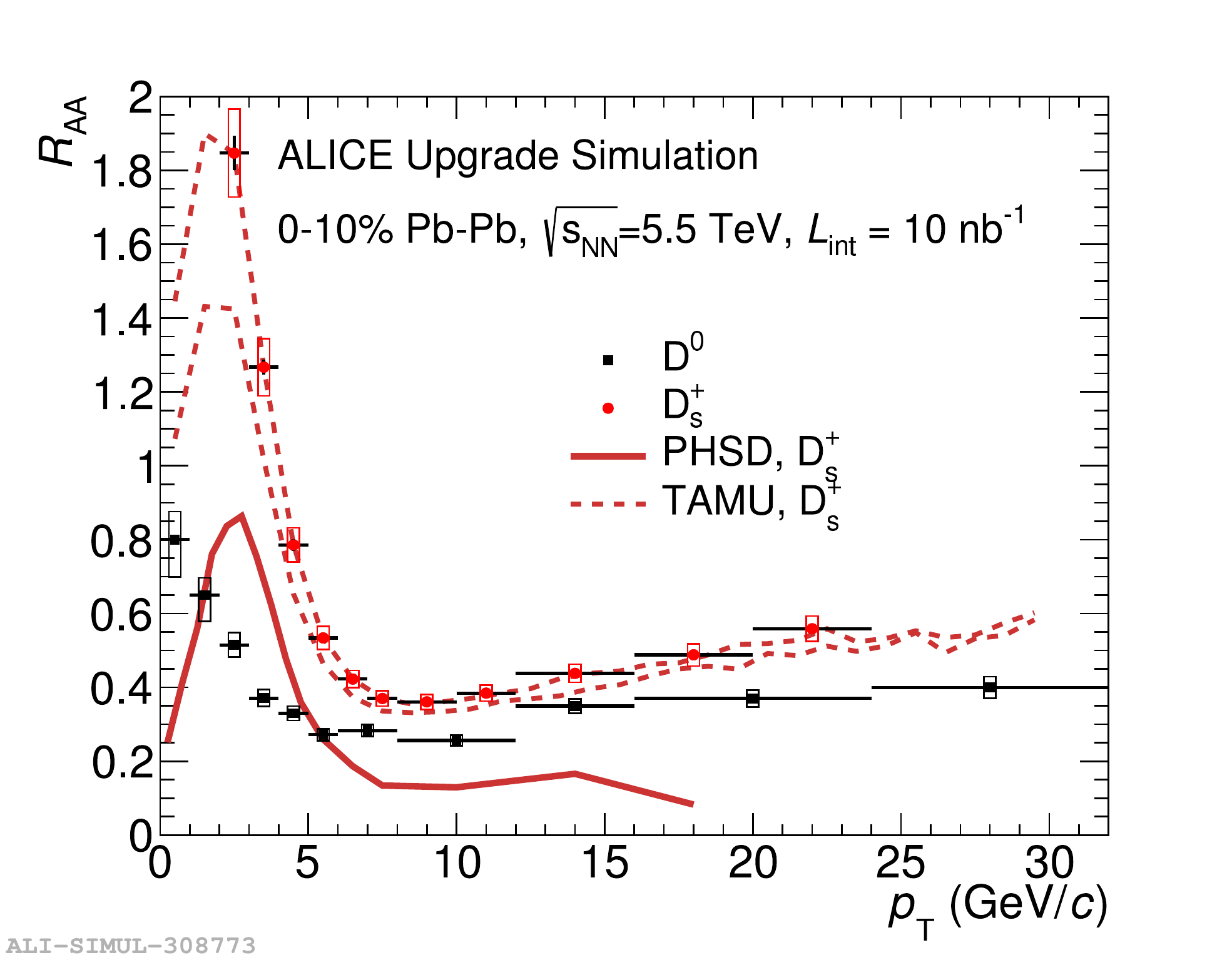}
\includegraphics[width=0.44\textwidth]{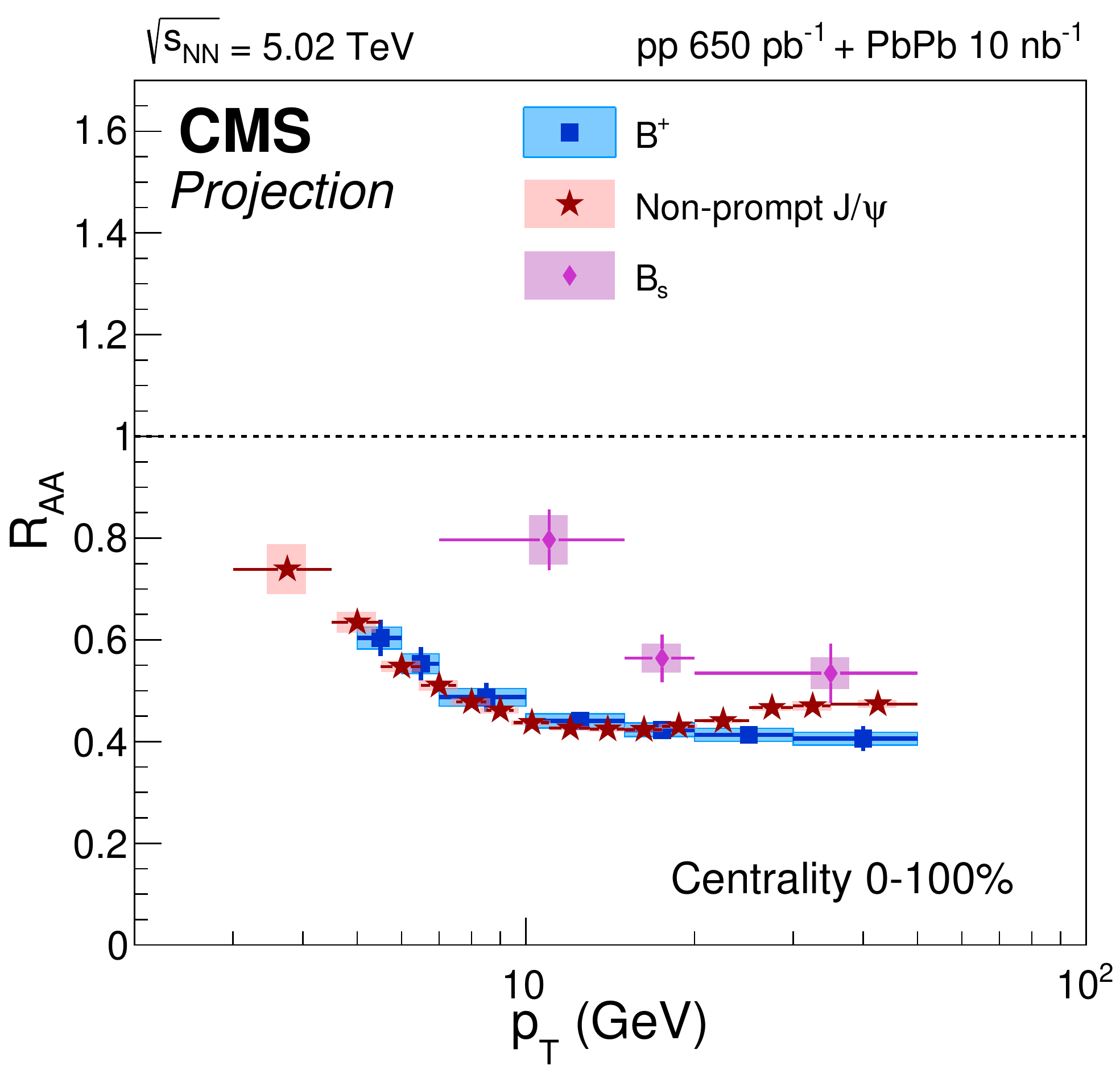}
\caption{Measurement performance projections for the nuclear modification factor  $\RAA$ of $\PDs$ (left) and $\PBs$ (right) mesons in \PbPb collisions ($\Lint=10~\invnb$). The ALICE study for $\PDs$ is based on full simulation~\cite{Abelev:1625842,ALICE-PUBLIC-2019-001}. The CMS projection is based on scaling of uncertainties from existing measurements~\cite{CMS-PAS-FTR-17-002,CMS-PAS-FTR-18-024}.}
\label{fig:HFDsBs}
\end {figure}

Figure~\ref{fig:HFDsBs} shows the performance for the $\RAA$ of $\PDs$ (left) and $\PBs$ (right), compared with the corresponding non-strange mesons. The predicted difference between the $\rm D^0$ and $\PDs$ $\RAA$ will be measured very precisely and the difference in the beauty sector could be observed with a significance of about 3\,$\sigma$.
The measurements were studied only for $p_{\rm T}>2$ and 8~GeV$/c$, respectively, but an extension to lower $\pt$ is considered within reach.

\begin{figure}[!t]
\centering
\includegraphics[width=0.49\textwidth]{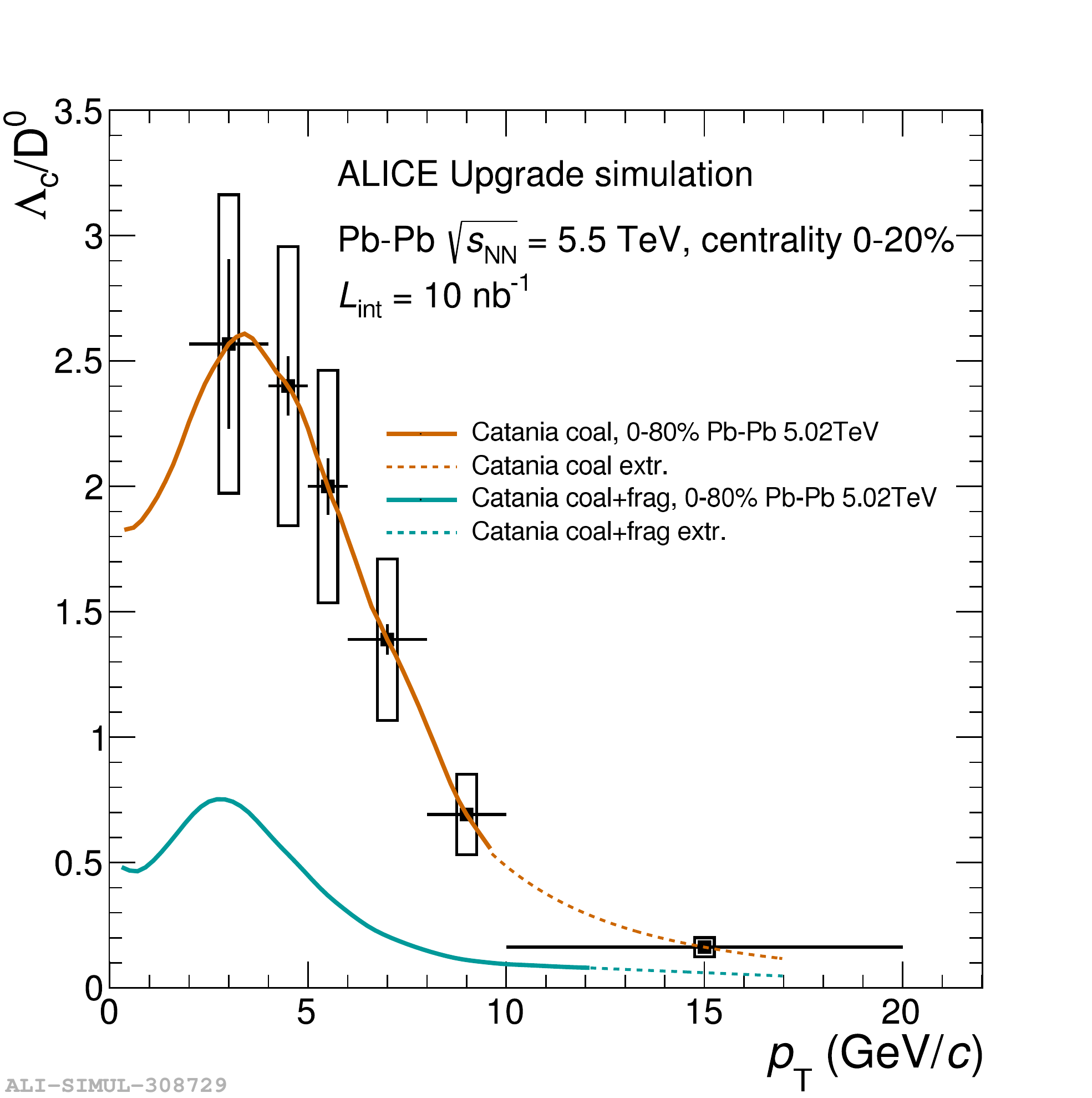}
\includegraphics[width=0.49\textwidth]{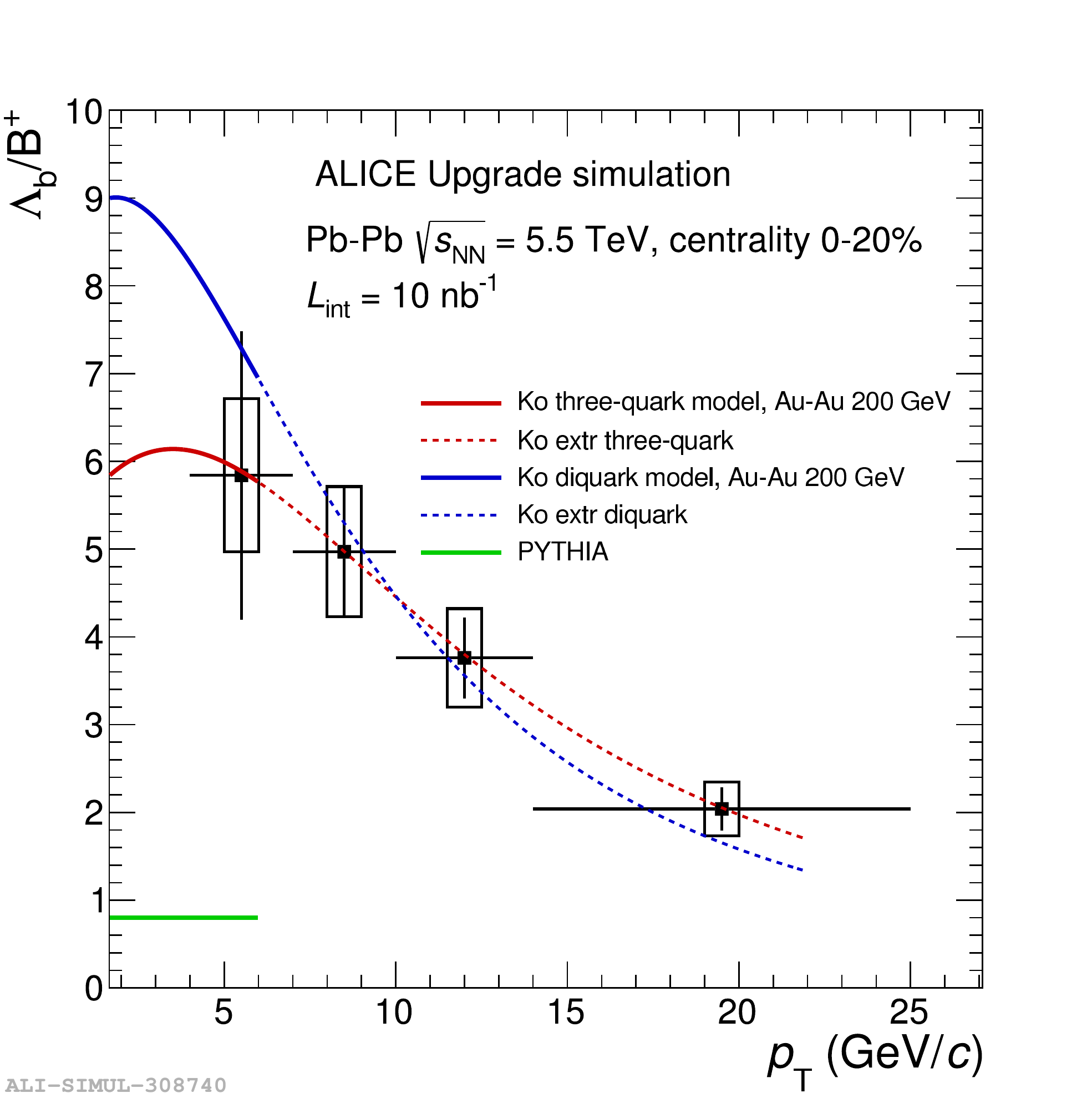}
\caption{ALICE measurement performance for the $\Lambda_c/D^0$ (left) and $\Lambda_b/B^+$ ratios in central \PbPb collisions ($\Lint=10~\invnb$), based on studies from~\cite{Abelev:1625842}. Figures from Ref.~\cite{ALICE-PUBLIC-2019-001}.}
\label{fig:HFLcLb}
\end {figure}

Figure~\ref{fig:HFLcLb} shows the performance for the charm and beauty baryon-to-meson ratios as they can be measured by ALICE with $\Lint=10~\invnb$ ~\cite{Abelev:1625842}. The measurements are compared with predictions based on various mechanisms for heavy-quark recombination in the medium~\cite{Plumari:2017ntm,Oh:2009zj}.
Figure~\ref{fig:RAAv2.v2charm} (right) shows the performance for the elliptic flow coefficient $\vtwo$ of $\rm D^0$, $\PDs$ and \PGLc 
in semi-central \PbPb collisions~\cite{Abelev:1625842}. The precision of the $\PDs$ $\vtwo$ should be sufficient to enable a significant comparison with $\rm D^0$ and with model calculations, in which the 
observable is found to be sensitive to the interactions of $\rm D$ mesons in the hadronic phase that characterises the late stages of the collision~\cite{He:2014cla}. 
Both measurements cannot be extended to the very-low-momentum region, where the separation of the heavy-flavour secondary vertex from the primary vertex is small.  
This limitation motivates studies for a further improvement of the ALICE inner tracker during LS3~\cite{ALICE:ITS3LoI}.  A more precise measurement would open the possibility to test in the charm sector some features at present only observed for the $\vtwo$ of light-flavour hadrons: the mass scaling at low $\pt$ and 
the baryon--meson grouping at high $\pt$.


\subsubsection{Impact of hadronisation models on QGP characterisation}
\label{sec:HFhadro3}


\begin{figure}[!ht]
\begin{center}
\includegraphics[width=0.49\textwidth]{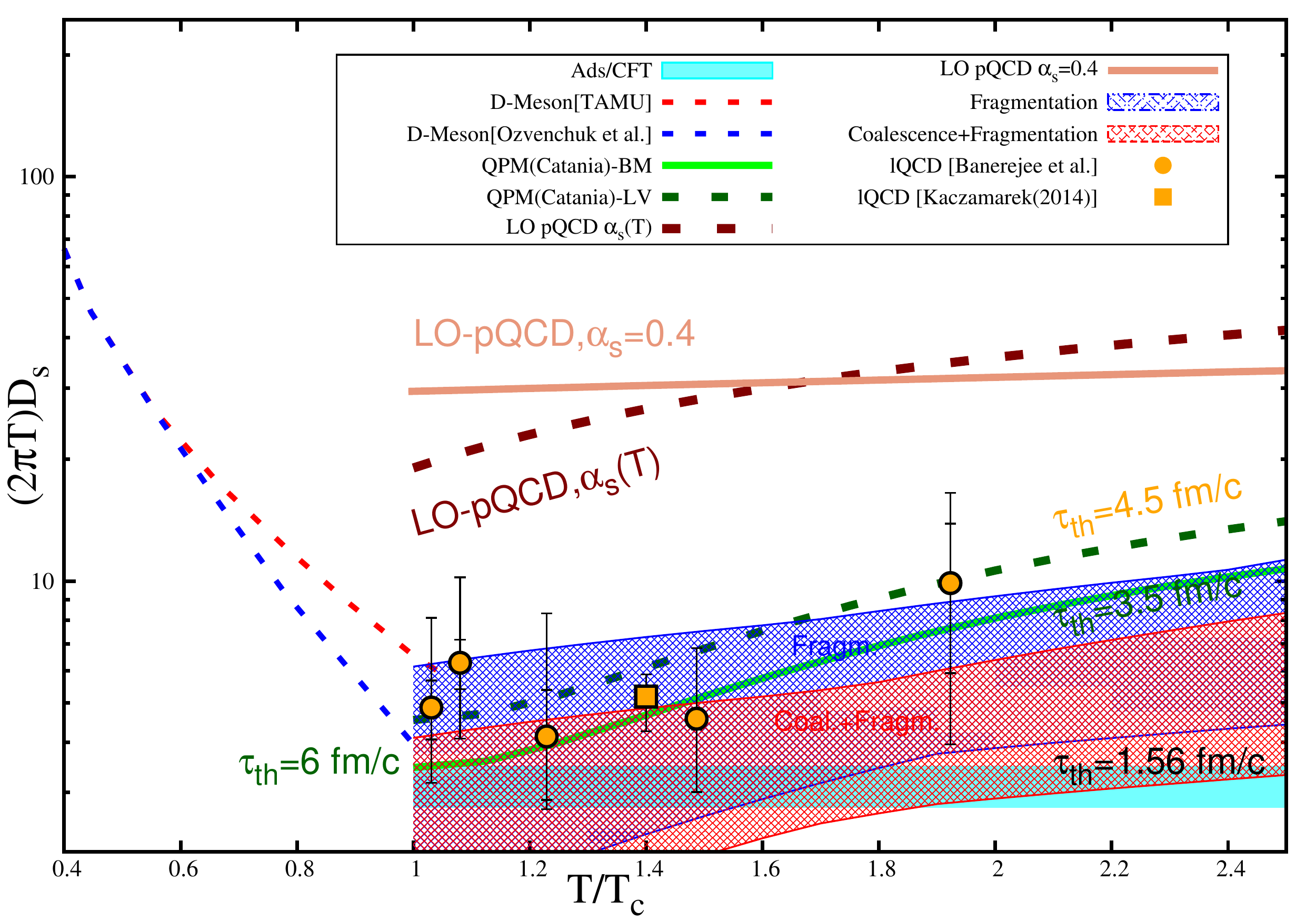}
\includegraphics[width=0.49\textwidth]{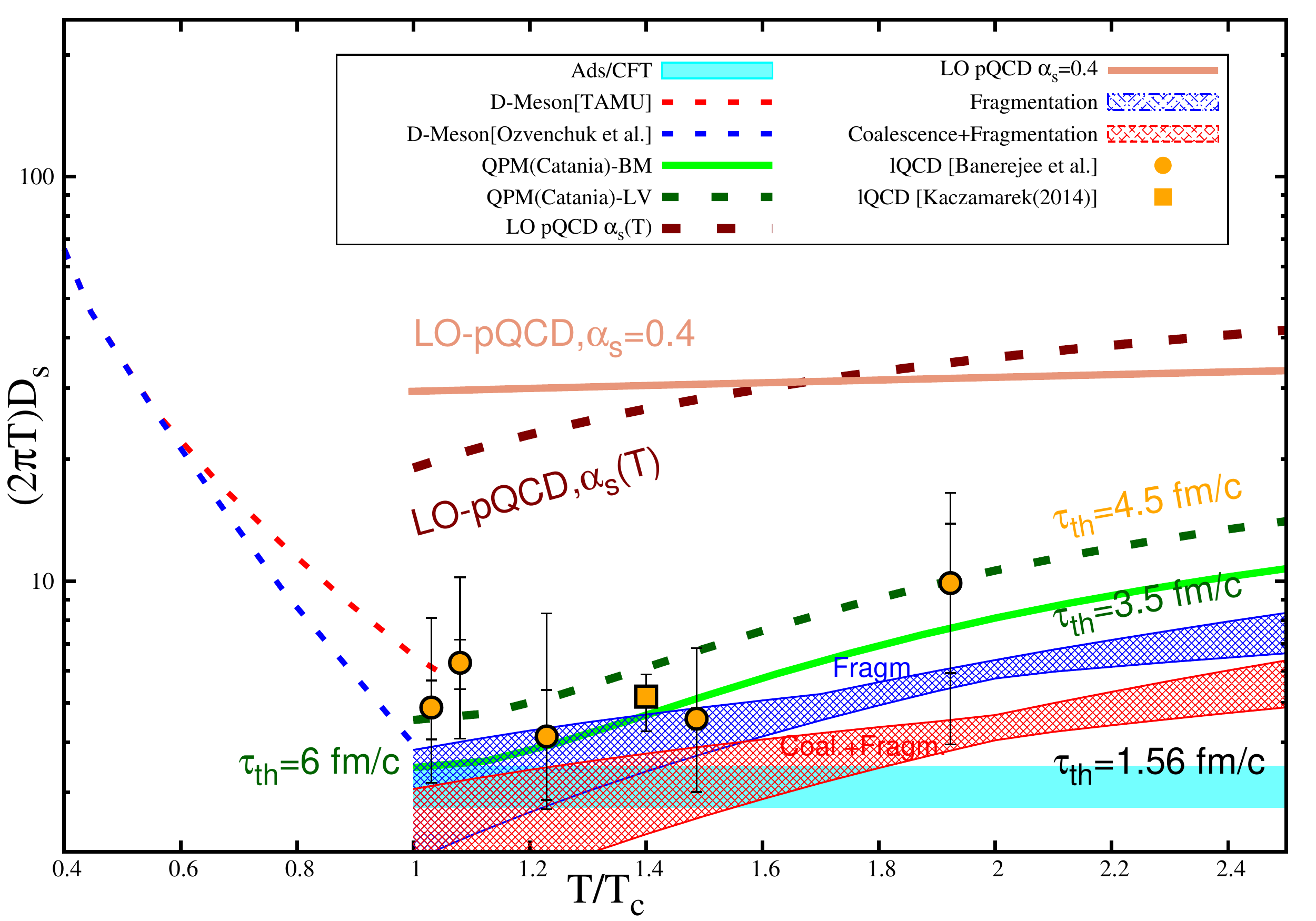}
\end{center}
\caption{Illustration of charm diffusion coefficient $2 \pi T D_s(T)$ 
estimate from D meson $R_AA$ data using different hadronisation assumptions in the Catania model~\cite{Das:2015ana,Das:2013kea} as a function of temperature. The blue band refers to fit with fragmentation only, the red band to the fit with coalescence plus fragmentation. In the left panel the $D_s$ estimation obtained with the current experimental uncertainties is shown, while in the right panel the estimation considering the case with more precise measurements as expected with $\Lint=10~\invnb$ is shown.}
\label{Fig:CHI2}
\end{figure}

The hadronisation mechanism of heavy quarks is important for the description of the measured
heavy-flavour $\RAA$ and $\vtwo$ at RHIC and LHC energies. In particular, it has been recognized that recombination
play a dominant role in describing simultaneously both D meson $\RAA$ and $\vtwo$~\cite{PhysRevC.73.034913,Gossiaux:2008jv,He:2011qa,Scardina:2017ipo}.
Moreover, as discussed in Section~\ref{sec:HFhadro1}, the charm baryon-to-meson ratio $\PGLc/\PDzero$ measured at RHIC and LHC is not consistent with a fragmentation only scenario~\cite{Oh:2009zj,Plumari:2017ntm}.
Therefore, a combined study including also the heavy baryon-to-meson ratio provides further
information to solve the ambiguity on the
recombination fraction. In the following, the sensitivity
of the QGP characterisation to the
recombination fractions described in Section~\ref{sec:HFhadro1} is illustrated.
To estimate how different model implementations of the hadronisation mechanisms can affect the extraction of the charm quark diffusion coefficient $2\pi T D_s$,
a global quantitative $\chi^2$ analysis was carried out by comparing experimental data on the D meson $\RAA$ with theoretical
results obtained using the Fokker-Planck equation under a standard bulk medium evolving
hydrodynamically with $\eta/s=0.1$~\cite{Plumari:2015cfa,Ruggieri:2013ova}.
The model developed in~\cite{Das:2015ana,Das:2013kea} with diffusion and drag
coefficients related by the fluctuation-dissipation theorem was used, considering two different
implementations of the hadronisation process: one using only fragmentation of charm quarks to D mesons (with the Peterson function)
and another one including a hybrid hadronisation by coalescence plus fragmentation~\cite{Plumari:2017ntm,Scardina:2017ipo}.
For the estimate of the temperature dependence of $2\pi T D_s$, a schematic model in which the
drag coefficient is parametrized as $\gamma=\gamma_0 (T/T_{c})^{\beta}$ was used. The two parameters $\gamma_0$ and $\beta$ were determined by minimizing the
$\chi^2/{\rm n.d.f.}$ of the model with respect to the measured D meson $\RAA$. The exercise could in principle be repeated using both $\RAA$ and $\vtwo$. 
The spatial diffusion coefficient is directly related to
the drag coefficient by $D_s=T/(M\cdot\gamma (p=0))$, where $M$ is the charm quark mass.
The left panel of Fig.~\ref{Fig:CHI2} shows the spatial diffusion coefficient
$2 \pi T D_s(T)$ estimated from the fit to the present data with $\chi^2/ {\rm n.d.f} < 2.5$. The blue band corresponds to the fragmentation 
only results and the red band to the coalescence plus fragmentation result.
The right panel shows the same calculations obtained with experimental data 
with smaller error bars on the D-meson $\RAA$ as expected with $\Lint=10~\invnb$.
The comparison between the two cases highlights the difference in the estimation 
of the $D_s$ coefficient obtained with the two different hadronisation mechanism.
Clearly, an optimal estimate of the diffusion coefficient requires an accurate description of the hadronisation mechanisms in the model.

\subsection{Heavy-flavour correlations and jets}
\subsubsection{Heavy-flavour correlations}
Although heavy quarks at the LHC energies are mostly produced in primary hard scatterings between the incoming partons in hadronic collisions, a non negligible fraction of c and b quarks are originated from processes of gluon splitting. At the leading order, $\rm c\overline c$ pairs are produced with an azimuthal opening angle of 180$^\circ$. At the next to leading order, gluon splitting and flavour excitation processes can generate $\rm c\overline c$ pairs typically at small opening angles. The role of next-to-leading order production is currently poorly understood even in proton-proton collisions \cite{LHCb-PAPER-2012-003,LorentzpaperHF} and introduces sizeable uncertainties in the models for \PbPb collisions. As a result, heavy-ion calculations which use proton-proton generators as baseline can be biased because of the different energy loss of quarks and gluons. Correlations between D and $\APD$ represent a promising way to study the relevance of these mechanisms in different kinematic ranges of the charm quarks. In \pPb collisions, the study of open heavy-flavour correlations at forward and backward rapidity provides information to test modifications of parton distribution functions in nuclei (see Sect.~\ref{sec:cgctmd}). 

\begin{figure}
\centering
\includegraphics[width=0.49\textwidth]{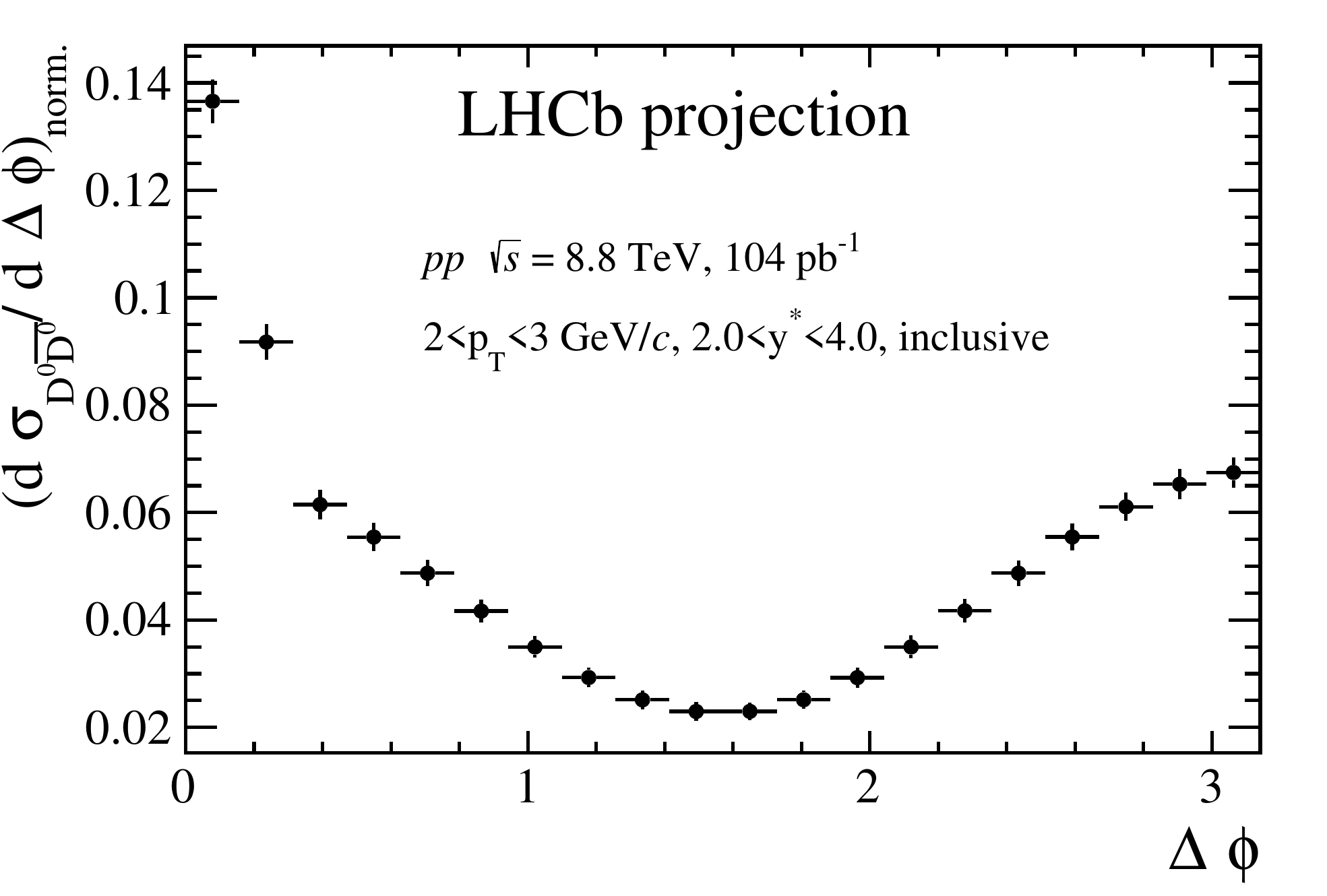}
\includegraphics[width=0.49\textwidth]{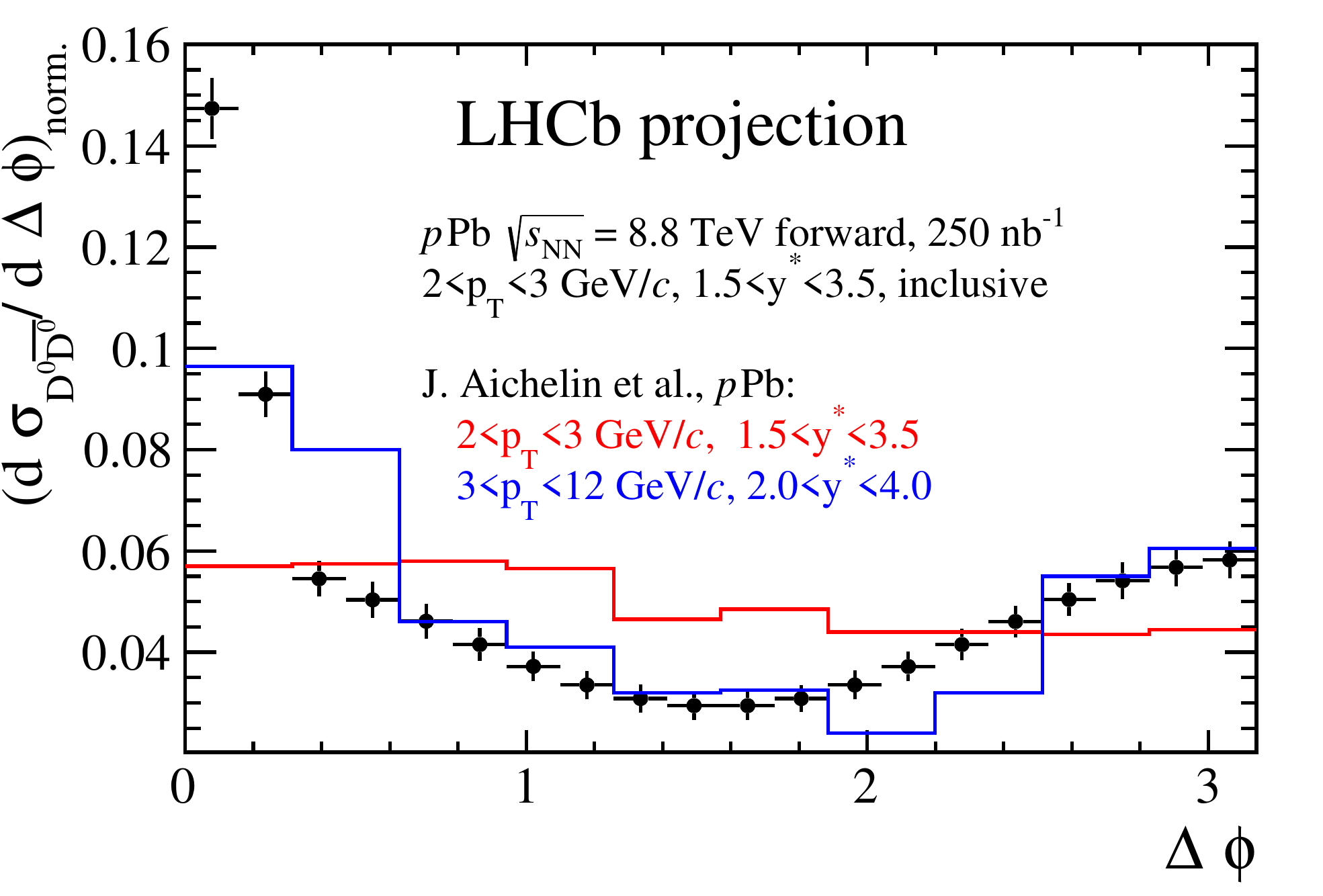}
\caption{ LHCb projection of the azimuthal $\PD \APD$ correlations in \pPb and in \pp collisions at $\sqrts=$ 8.8 $\UTeV$ for the integrated Run~3 and Run~4 luminosity. The \pPb~panel shows a calculation for two different kinematic selection in \pPb collisions within the EPOS3-HQ generator. A detailed description of the experimental estimates can be found in Ref.~\cite{LHCb-CONF-2018-005}. }
\label{fig:DD}
\end {figure}

Figure~\ref{fig:DD} shows the LHCb projections for the azimuthal ${\PD \APD}$ correlations at forward rapidity in \pp and \pPb collisions at the same collision energy per nucleon-nucleon-pair of 8.8~$\UTeV$. The figure shows only statistical uncertainties (dominant with respect to the systematic uncertainties) as expected with the Run 3 and 4 integrated luminosity. The measurement of the $\PD \APD$ correlation can be performed in intervals of D meson $\pT$, providing differential information to test theoretical models with precision. The experimental projections are compared with predictions obtained with EPOS3-HQ event generator for two different kinematic selections in case of $p$Pb collisions. In particular, the two selected \pT~ranges, 2-3~GeV/$c$ and 3-12~GeV/$c$, yield to significantly different correlation shapes in the calculation. This dynamic change of shape demonstrates the necessity to provide precise quantitative tests of the importance of different production mechanism at the partonic level, the hadronisation and potential medium effects in \pp and in \pPb~collisions across different kinematic configurations.

In nucleus-nucleus collisions, ${\PD \APD}$ correlations are sensitive observables to discriminate among different mechanisms of in-medium energy loss of heavy-quarks, like radiative and collisional processes. Such measurements, presently challenged by statistical limitations, will greatly benefit from future high-luminosity heavy-ions runs. 

\subsubsection{Heavy-flavour jet measurements}
Further insights into the mechanism of parton energy loss in the QGP can be provided by the study of reconstructed heavy-flavour jets. These measurements provide complementary information to the studies of D and B mesons since they enable to better determine the energy of the initial heavy quark and give access to the jet energy profile. By comparing the production of heavy-flavour jets with light jets one can test the expected flavour dependence of the in-medium energy loss in a wide transverse momentum range and study the different energy loss mechanism of quarks and gluons. The study of more differential observables related to the production of heavy-flavour particles in jets, like the fragmentation function and the angular correlations, provides additional constraints into the mechanisms of redistribution of the energy lost by the parton inside the medium. 

\begin{figure}
\centering
\resizebox{0.45\textwidth}{!}{  
\includegraphics{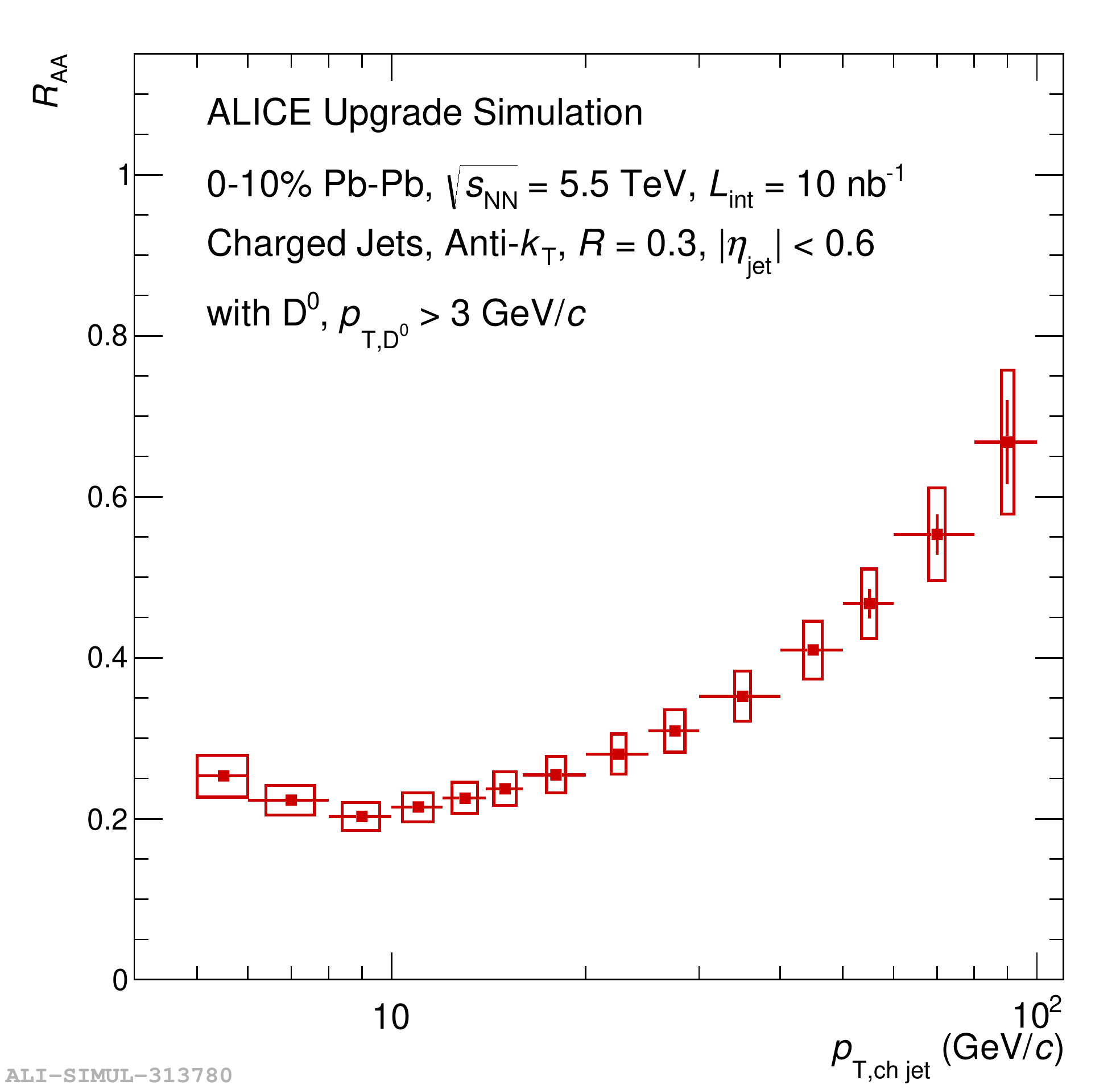}}
\resizebox{0.45\textwidth}{!}{  
\includegraphics{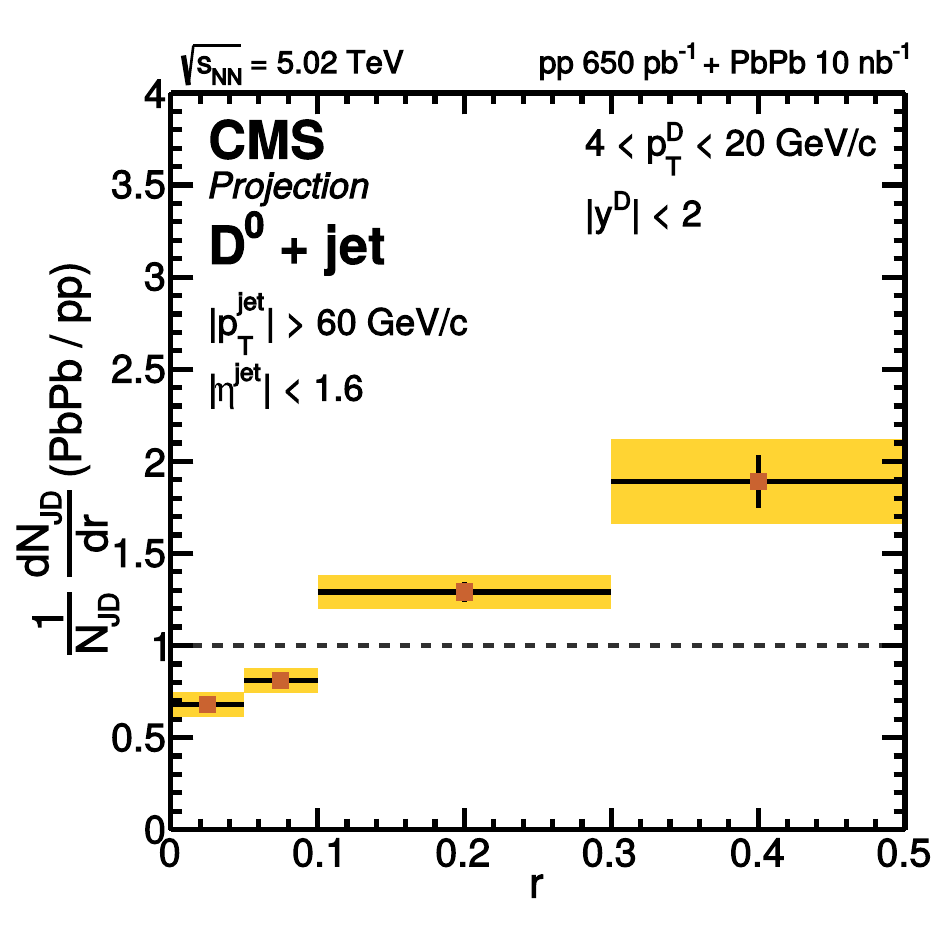}}
\caption{(Left) ALICE simulation results for the nuclear modification factor of $\PDzero$-meson tagged jets in central \PbPb collisions~\cite{ALICE-PUBLIC-2019-001}. (Right) CMS projection for the distribution of $\PD$ mesons in jets as a function of the distance from the jet axis for jets of $\pT>60$\,GeV/$c$ in \PbPb collisions divided by the distribution in pp collisions.}
\label{fig:DjetCMS}
\end {figure}

Figure~\ref{fig:DjetCMS} (left) shows the projections for the nuclear modification factor of $\PDzero$-meson tagged jets in \PbPb collisions at $\sqrtsNN=5.02~{\UTeV}$ with ALICE. The measurement will provide a precise estimation of the suppression of $\PDzero$-tagged jets down to low $\pT$, opening the way to study possible modifications of charm fragmentation. Future measurements of the fragmentation functions of charmed hadrons in \pp collisions will also help to reduce the uncertainties on the charm fragmentation mechanisms, which are currently among the main sources of uncertainties for theoretical calculations that describe heavy-quark production at the LHC. 

Figure~\ref{fig:DjetCMS} (right) shows the CMS projection for the distribution of $\PD$ mesons as a function of the distance from the jet axis for jets of $\pT>60$~$\UGeV/c$ in \PbPb collisions divided by the distribution in pp collisions.  With the precision achievable with the high-luminosity data, one would be able to measure precisely the effect of heavy-quark in-medium energy loss and the redistribution of the energy at large angle with respect to the jet axis. This is expected to provide new constraints on in-medium energy loss calculations. By comparing the ratio measured for D mesons with the one obtained for charged particles, it will be possible to assess the relevance of medium response phenomena, which can induce modification of the jet shape at large angles and are expected to be less relevant for heavy quarks due to their large masses.

Another area of research where the LHC experiments can make key contributions is the study of heavy-flavour jet substructure. New experimental observables related to the inner structure of heavy-flavour jets, like the splitting functions, can provide insights into the mass dependence of the parton shower in new kinematic regimes. A precise measurement of these observables at the LHC down to low $\pT$ would also provide a unique opportunity to further investigate the dead cone effect~\cite{DOKSHITZER2001199}, currently not well understood and constrained.

\subsection{Sensitivity to early magnetic fields and vorticity phenomena}
\begin{figure}
\centering
\includegraphics[width=0.50\textwidth]{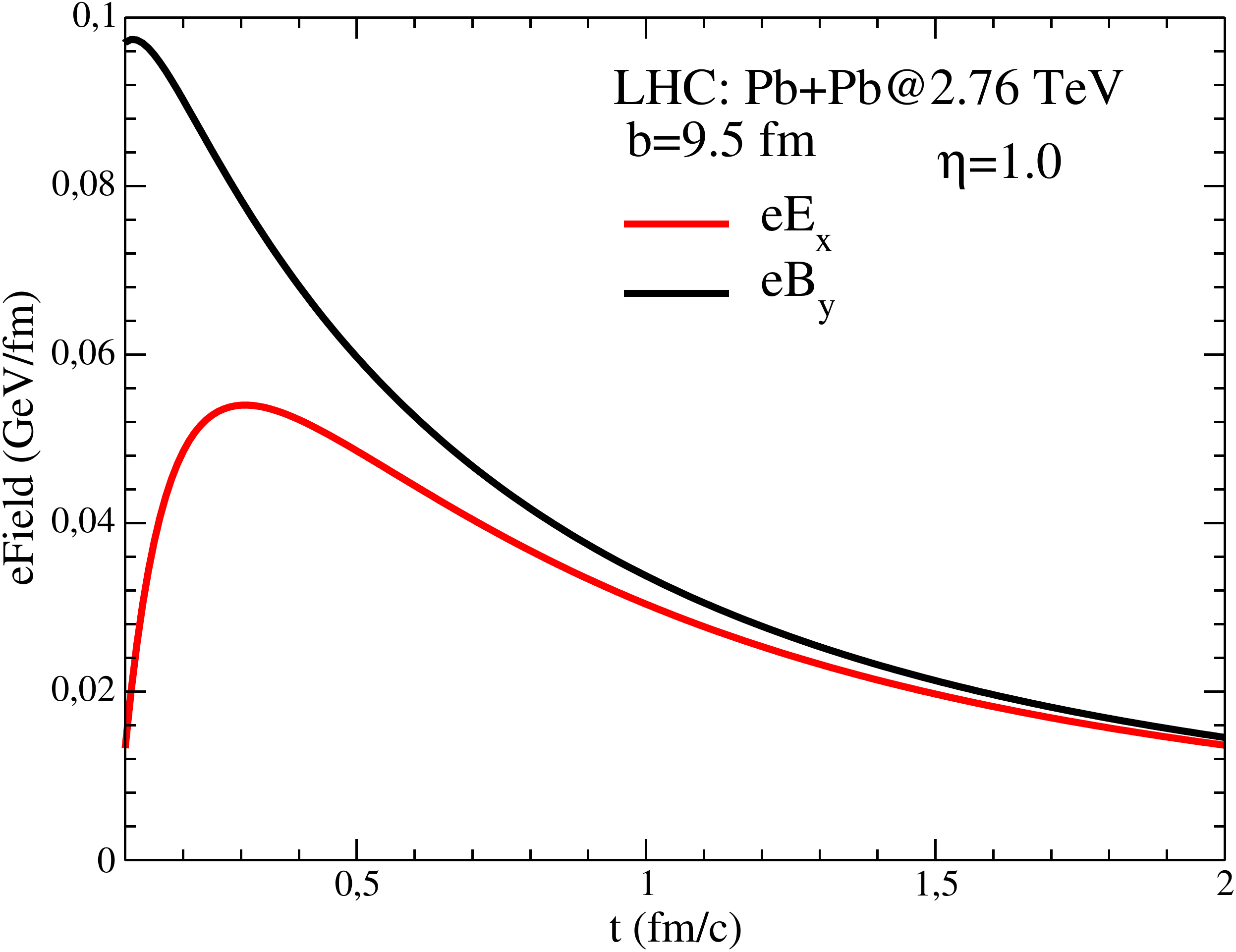}
\includegraphics[width=0.46\textwidth]{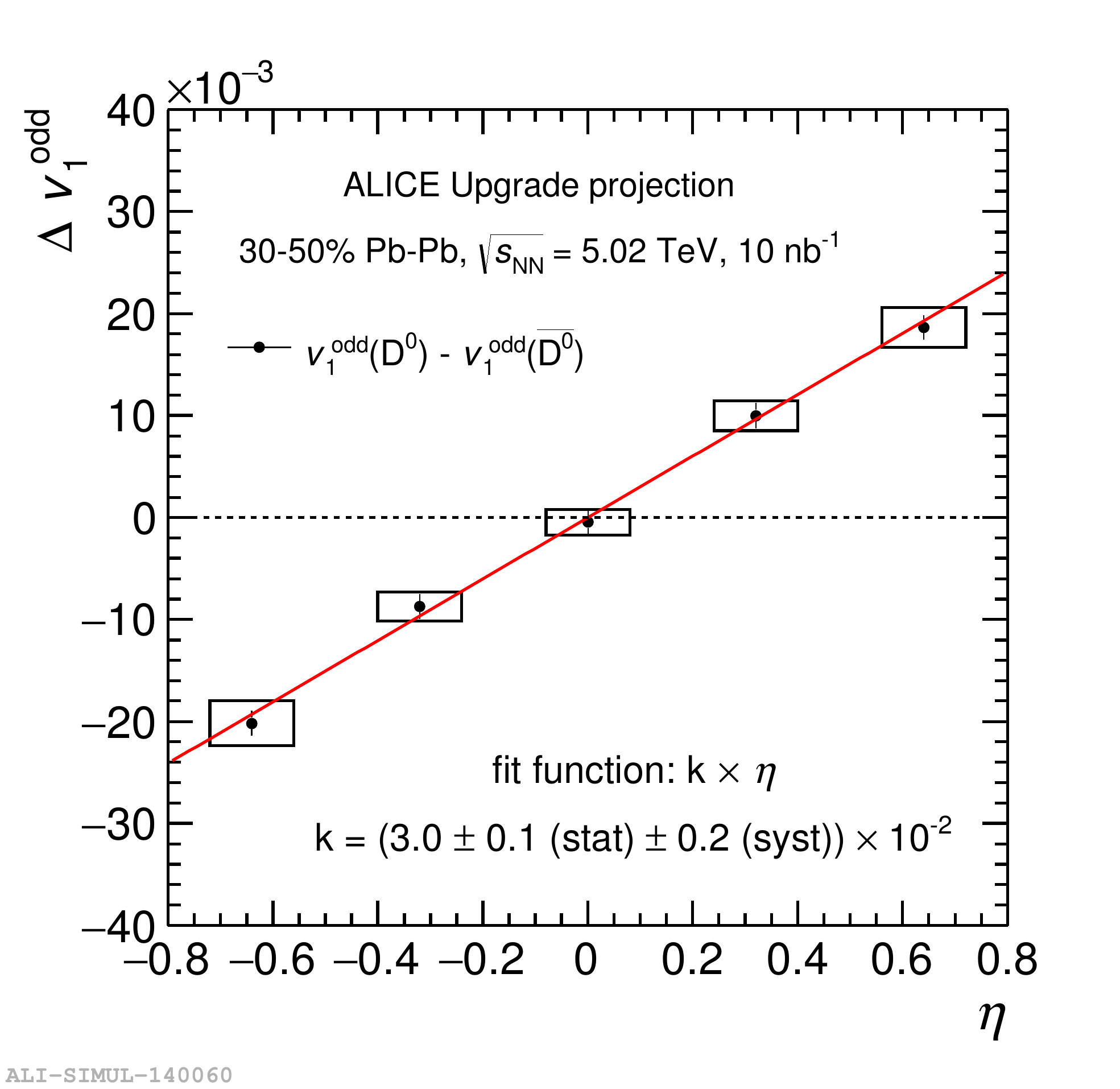}
\caption{Left: Time evolution in the forward rapidity region of the magnetic field $\rm B_{y}$ and electric field $\rm E_{x}$ in \PbPb collisions at $\sqrtsNN=$ 2.76$\UTeV$ with impact parameter $b=9.5\,\Ufm$~\cite{Das:2016cwd}. Right: ALICE projection of the difference of directed flow $\vone$ for $\PDz$ and $\APDzero$ as a function of pseudorapidity in semi-central \PbPb collisions at $\sqrtsNN=$ 5.5 $\UTeV$~\cite{ALICE-PUBLIC-2019-001}.}
\label{fig:v1}
\end {figure}

Recently, it has been recognized that very strong electric and magnetic fields are created at early times of ultra-relativistic heavy-ion collisions. In Fig.~\ref{fig:v1} (left), the time dependence of the magnetic ($\rm B_{y}$) and electric ($\rm E_{x}$) fields in \PbPb collisions at $\sqrtsNN=$ 2.76$\UTeV$ with impact parameter b=9.5$\Ufm$ is presented for a system with electric
conductivity $\rm \sigma_{el}=$0.023$\Ufm^{-1}$. The space-time evolution is obtained as solution of the Maxwell's equations as developed in \cite{Gursoy:2014aka}. The magnetic field produced in non-central heavy ion collisions is dominated by the $\rm B_{y}$ component which induces a current in the $xz$ plane while the time dependence of $\rm B_{y}$ generates an electric field which is directed in the $x$ direction. The combined effect of both fields is a current in the $xz$ plane.

The presence of early magnetic fields produced in heavy-ion collisions is expected to have an effect on the charm directed flow~\cite{Das:2016cwd,Chatterjee:2018lsx}, resulting in a $\vone$ value larger than that of lighter particles (short charm formation time and therefore sensitive to the maximum magnetic field strength), and opposite for particles with charm and anti-charm (due to the Lorentz force). Also the initial vorticity of non-central collision is expected to affect the directed flow observable.
The STAR collaboration recently presented the first measurements of the directed flow $\vone$ coefficient for mesons. This first observation of non-zero $\vone$ for charm mesons, larger than that of lighter particles, is in qualitative agreement with theoretical models including both electromagnetic and vorticity effects. The uncertainties on the difference between the $\vone$ of $\PDz$ and $\APDzero$
from STAR are still too large to draw conclusions on the effects of the early magnetic fields.
The LHC Run 3 and 4 will enable more precise measurements on the charm directed flow, which will give additional insights into the initial vorticity of the Quark Gluon Plasma and the strength of the electromagnetic fields. Figure~\ref{fig:v1} (right) shows the precision level for the difference of directed flow $\vone$ for $\PDz$ and $\APDzero$ which ALICE can measure as a function of pseudorapidity in semi-central \PbPb collisions at $\sqrtsNN=$ 5.5 $\UTeV$.

\subsection{Heavy flavour measurements in small colliding systems}

Traditionally, proton-nucleus collisions were considered just as a benchmark to investigate Cold-Nuclear-Matter (CNM) effects in the absence of a deconfined medium. As discussed in the devoted section, recent results suggest to partially reconsider such a paradigm. 
In particular, the study of soft observables in high-multiplicity \pPb (and even pp) collisions led to the observation of signals traditionally attributed to the formation of a hot deconfined medium: long-range azimuthal 2-particle correlations~\cite{Khachatryan:2010gv,CMS:2012qk,Abelev:2012ola,Aad:2012gla}, non-vanishing flow harmonics~\cite{ABELEV:2013wsa,Khachatryan:2015waa}, increasing baryon/meson ratio as a function of $\pT$~\cite{Abelev:2013haa}, enhancement of strange-hadron production~\cite{Adam:2015vsf}. Many authors interpreted these observations as indications that the same strongly-interacting medium -- with a collective hydrodynamic expansion driven by pressure gradients -- supposed to be formed in nucleus-nucleus collisions is also produced in smaller systems. If the hydrodynamic scheme provides a consistent framework describing the above observations, its strong assumptions (e.g. a mean-free-path much smaller than the system size $\lambda_{\rm mfp}\ll L$) look challenging to satisfy for such small systems. Hence, some authors looked for alternative interpretation of the data in terms, for instance, of initial-state effects (Color-Glass-Condensate~\cite{Dusling:2013qoz}), of the formation of a system with small parton-parton cross-sections~\cite{Bzdak:2014dia}, of the presence of anisotropic parton escape mechanisms~\cite{HE2016506} or of quantum-mechanical interference in the presence of multiple sources of particle production, entailing to reconsider the no-interaction baseline before looking for final-state collective effects~\cite{Blok:2017pui} .

On the other hand, no signature of jet-quenching or suppression of high-$\pT$ particle production was observed in high-multiplicity proton-nucleus collisions~\cite{Acharya:2018qsh}.  At first glance this appears in contradiction with the measurements involving soft observables. One should in any case consider that the quenching of jets due to parton energy-loss in the QGP has a strong dependence on the thickness of the crossed matter.
On the contrary, if one accepts the hydrodynamic paradigm, the smaller size of the medium with respect to the nucleus-nucleus case would lead to even larger pressure gradients and hence to a larger acceleration of the fluid, compensating its shorter lifetime.

In light of the above findings, as an independent probe, it is clearly of interest to address the study of small systems also through heavy-flavour observables. Due to their large mass c and b quarks are in fact produced in the very first instants in hard processes described by pQCD, at most affected by nuclear modifications of the Parton Distribution Functions (nPDFs) and by an initial transverse-momentum broadening acquired in CNM. 
It looks then natural to extend transport calculations, developed to describe heavy-flavour propagation through the plasma formed in nucleus-nucleus collisions, also to the case of small systems, assuming as a working hypothesis that also in this case a hot deconfined medium is formed. The theoretical modelling involves the same processes as in the heavy-ion case: the so-called CNM effects (nPDFs and initial $k_{\rm T}$-broadening) modifying the hard $Q\overline{Q}$ production, the propagation of the heavy quarks throughout the fireball and finally their hadronization in the presence of a hot medium. The only difference is that, in the case of small systems, it is mandatory to include event-by-event fluctuations in the initial conditions.

\begin{figure}[ht]
\centering
\includegraphics[width=0.5\textwidth]{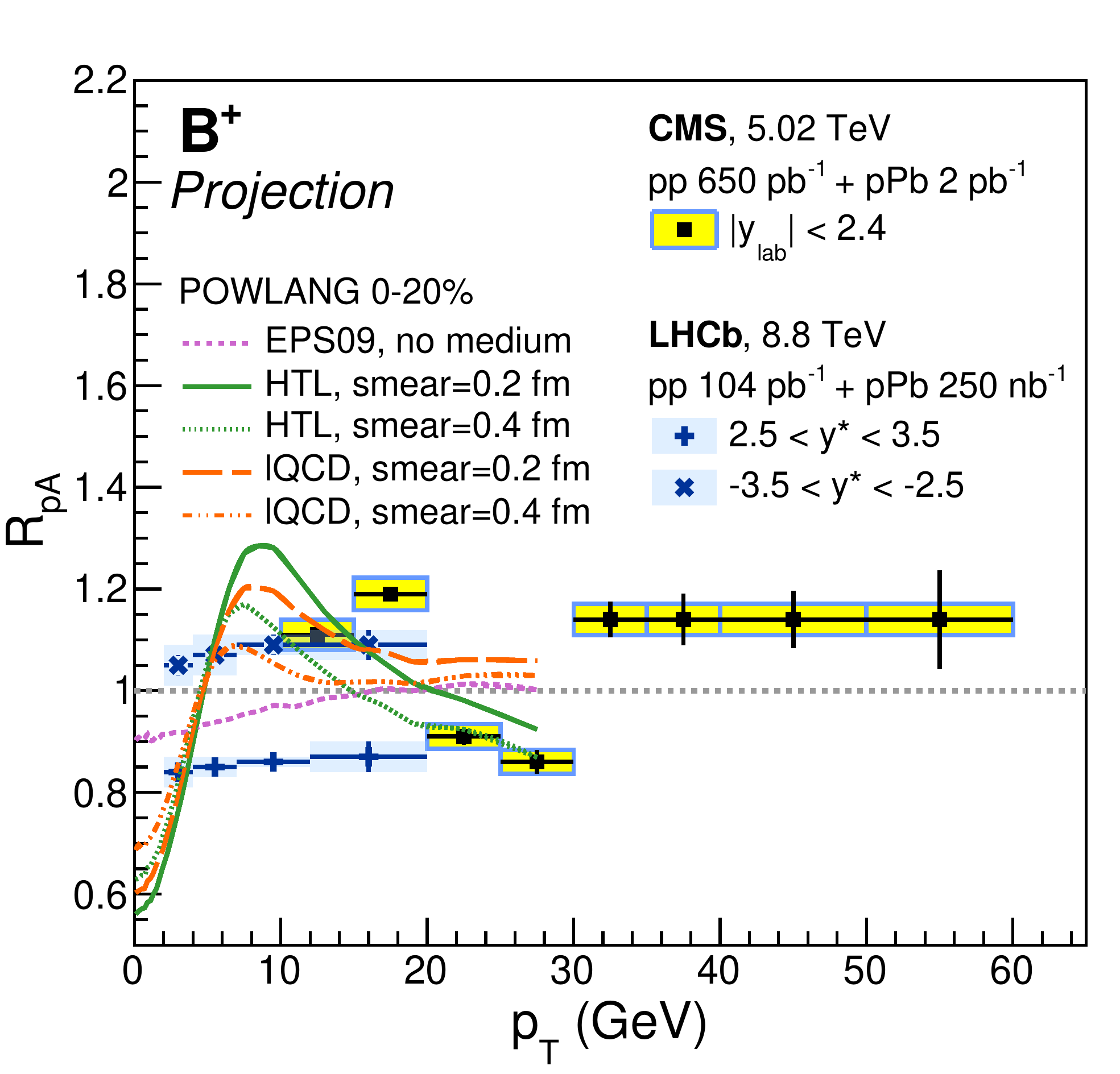}
\caption{Projection for the measurement of the nuclear modification factor of B mesons in \pPb collisions achievable by CMS~\cite{CMS-PAS-FTR-18-024} and LHCb at $\sqrtsnn=5.02~\UTeV$ and 8.8 $\UTeV$ respectively. The predictions of the POWLANG model with different choices of the transport coefficients and of the smearing of the initial condition at $\sqrtsnn=5.02~\UTeV$ are shown. Also reported, in magenta, the curve containing only Cold-Nuclear-Matter effects.}
\label{fig:POWLANG-small1}
\end{figure}
\begin{figure}[ht]
\centering
\includegraphics[width=0.4\textwidth]{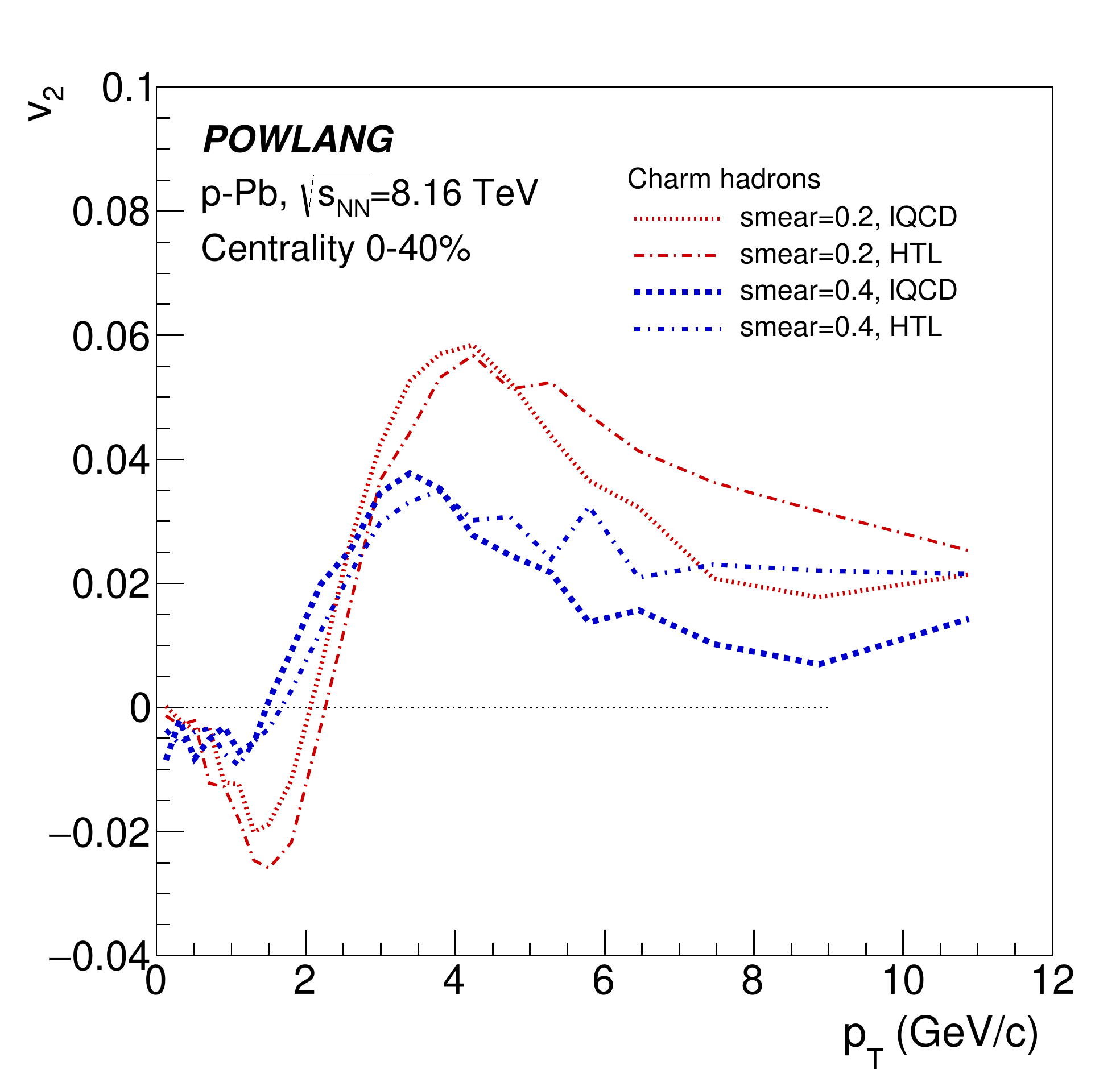}
\includegraphics[width=0.4\textwidth]{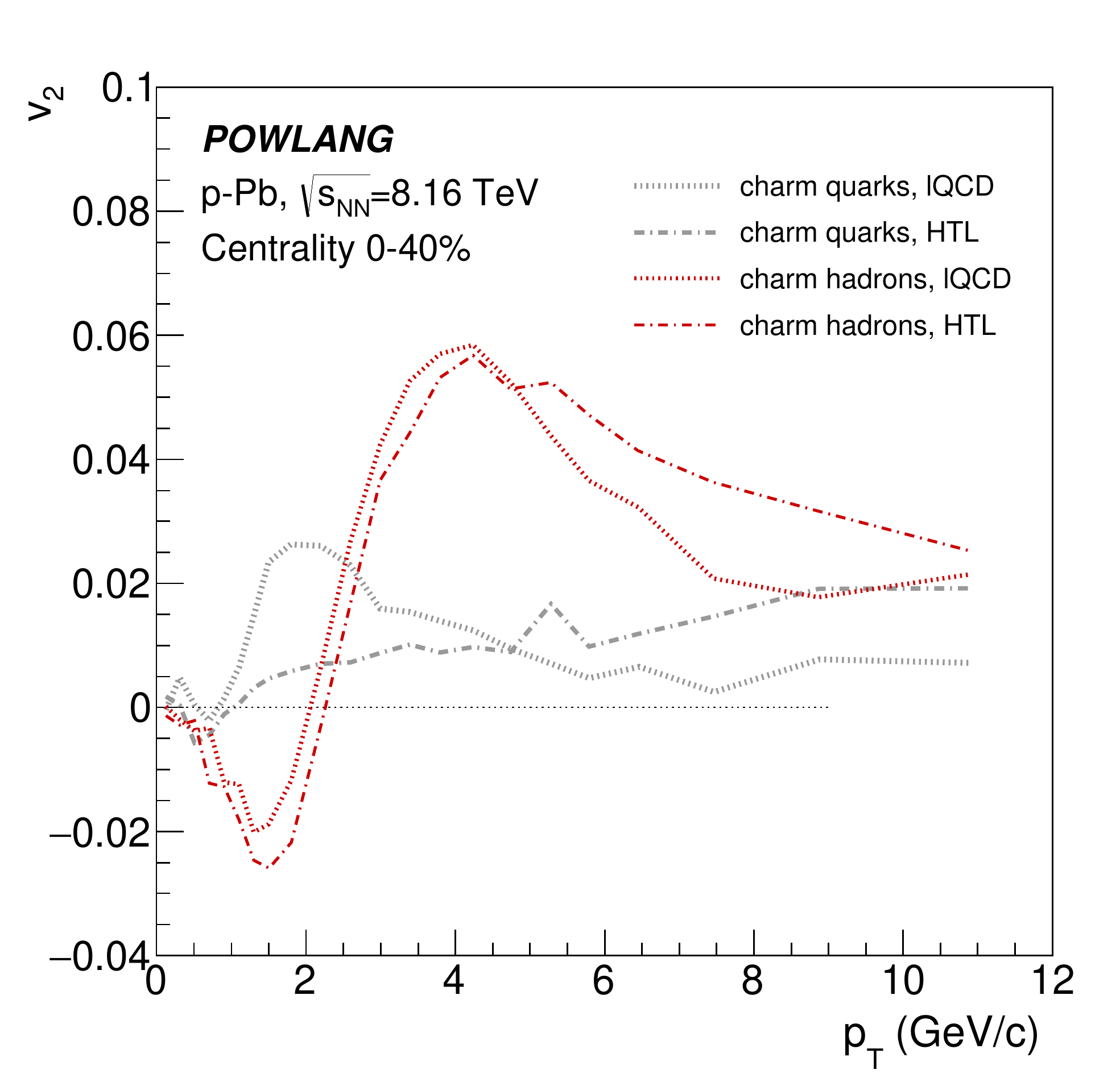}
\caption{
Left: The elliptic flow of charmed hadrons in the 0-40\% most central \pPb collisions at $\sqrtsNN=8.16$~$\UTeV$. Results of the POWLANG model with different choices of the transport coefficients and of the smearing of the initial condition are shown. 
Right: a comparison of the results at the level of charm quarks and hadrons. An important fraction of the flow is acquired at hadronization via recombination with light partons from the medium.}
\label{fig:POWLANG-small2}
\end{figure}
\begin{figure}[ht]
\centering
\includegraphics[width=0.4\textwidth]{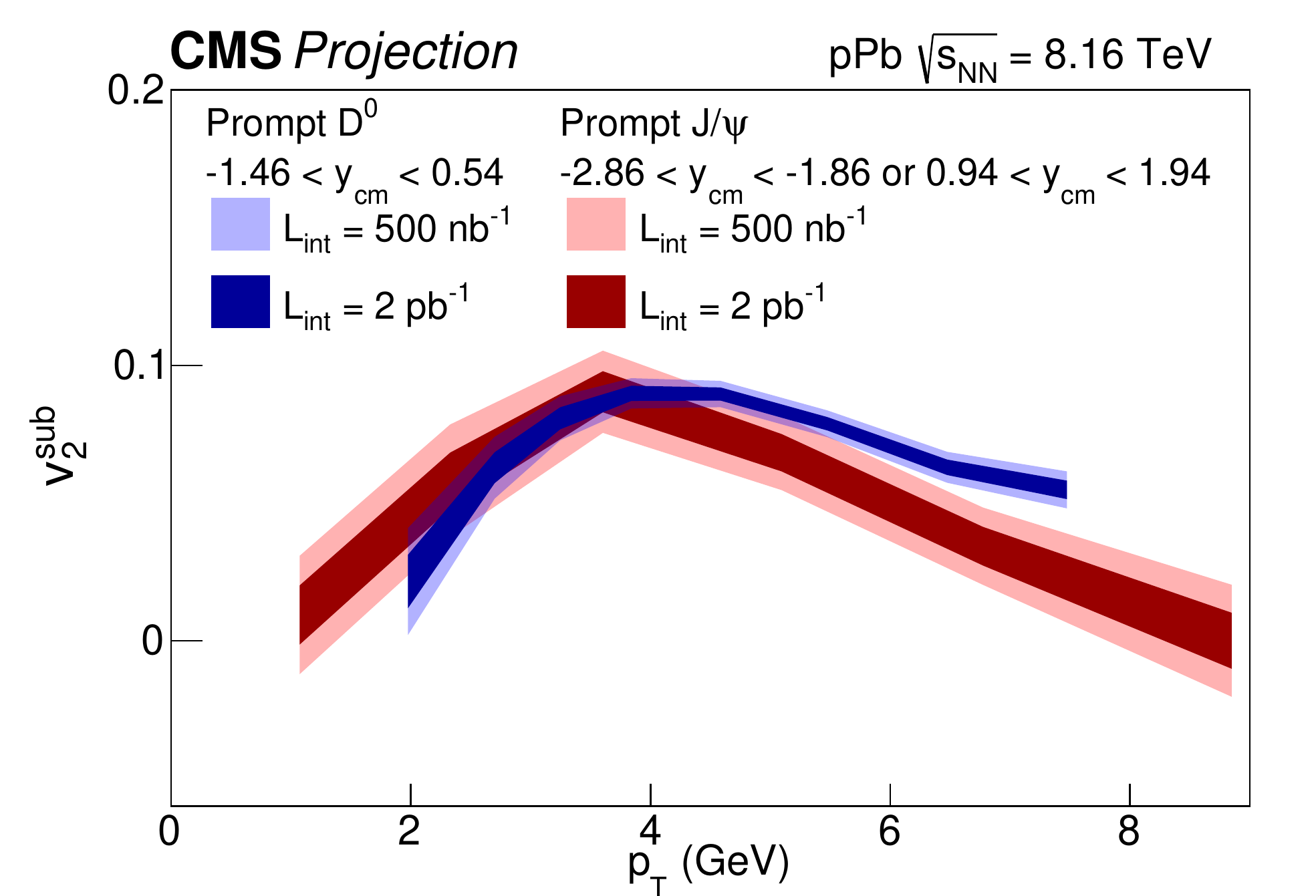}
\includegraphics[width=0.43\textwidth]{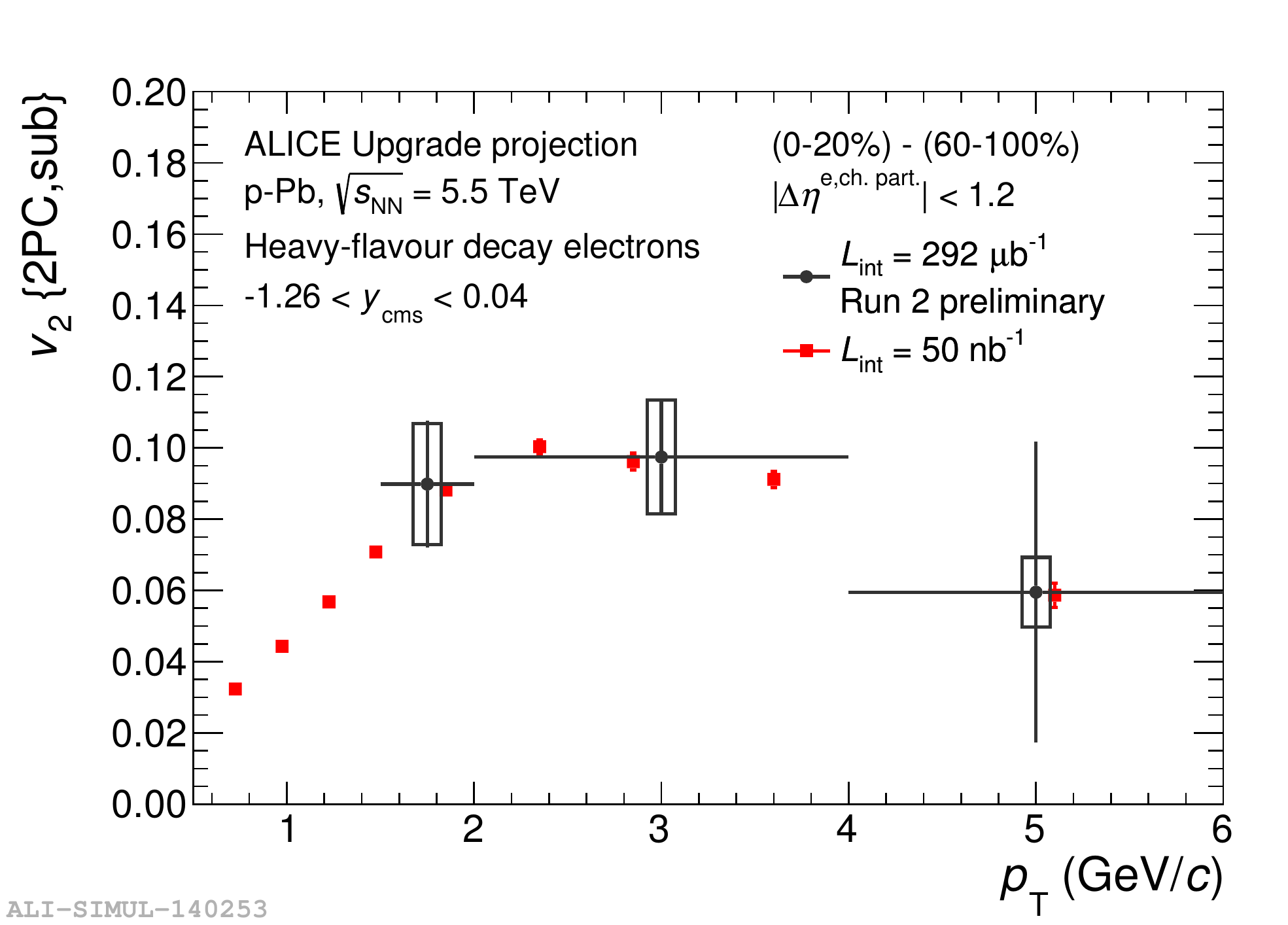}
\caption{
Left: projection of prompt D-meson and prompt J$/\psi$ elliptic flow as a function of $\pt$ in high-multiplicity \pPb collisions at $\sqrtsNN=8.16$~$\UTeV$ with CMS~\cite{CMS-PAS-FTR-18-026}.
Right: projection of elliptic flow of electrons from heavy-flavour hadron decays as a function of $\pt$ in 0--20\% central \pPb collisions at $\sqrtsNN=5.02$~$\UTeV$ with ALICE~\cite{ALICE-PUBLIC-2019-001}.}
\label{fig:elliptic-data}
\end{figure}
In order to illustrate the level of precision which future experimental measurements must reach to discriminate among different scenarios of initial and final-state effects, various sets of predictions based on the POWLANG model are reported in the following~\cite{Beraudo:2015wsd}.
In Fig.~\ref{fig:POWLANG-small1} results of transport calculations for the nuclear modification factor of beauty hadrons in \pPb collisions at $\sqrtsnn=5.02~\UTeV$ are displayed and compared to experimental projections by CMS for the B-meson $R_{\rm pPb}$ obtained from Run 2 data and projections of LHCb at lower transverse momentum for the B-meson $R_{\rm pPb}$~\cite{LHCb-CONF-2018-005} obtained using as central points the calculations based on Ref.~\cite{Kusina:2017gkz}. As can be seen, model predictions are sensitive to the different choices of the transport coefficients (see discussion in~\cite{Beraudo:2015wsd}) and of the initial conditions (each nucleon-nucleon collision is assumed to deposit some entropy with a Gaussian smearing). In order to answer the question about possible final-state hot-medium effects, experimental measurements should be extended to lower $\pT$, as planned for the different experiments (see ALICE plans to perform beauty measurements down to low $\pT$ in nucleus-nucleus collisions in Sec.~\ref{sec:HFRAAv2}). In fact, for $\pT$ larger than 10--15~GeV/$c$ the curves accounting for the transport and hadronization of the heavy quarks in a hot medium (green curves) are very close to the (magenta) one which includes only the effect of the nPDF's and of the initial $k_{\rm T}$-broadening acquired in CNM, all of them being very close to unity as the experimental projections by CMS. On the contrary, at lower $\pT$ the radial flow of the beauty hadrons, acquired in part during the propagation through the hot medium and in part at hadronization via recombination, leads to a depletion of the spectrum at low $\pT$ and to an enhancement at intermediate $\pT$ which would allow one to distinguish this scenario from the case of pure CNM effects. 

As displayed in Fig.~\ref{fig:POWLANG-small2} the possible production of a hot deconfined medium in proton-nucleus collisions would also leave its fingerprints in the azimuthal distribution of the final hadrons, leading in particular to a non-vanishing elliptic-flow coefficient. Notice how one gets a positive $v_2$ for charmed hadrons only starting from $p_T\!\approx\!2$~GeV/$c$, in agreement with recent CMS data~\cite{Sirunyan:2018toe}. Differently from the heavy-ion case, in such small systems it is important how the initial entropy deposition is modelled with varying the smearing parameter (left panel). As displayed in the right panel of Fig.~\ref{fig:POWLANG-small2}, within the framework of the model, an important role is played by hadronization via recombination with light partons from the fireball, whose collective flow enhances the azimuthal asymmetry of the charmed hadrons.

The projected precision with p--Pb integrated luminosities $\Lint\sim 2~\invpb$  (ATLAS and CMS) and $\sim 1$~pb$^{-1}$ (ALICE) has the potential to shed light on the different mechanisms behind the observed anisotropy (see Fig.~\ref{fig:elliptic-data} for D mesons and prompt J$/\psi$ with CMS~\cite{CMS-PAS-FTR-18-026} and electrons from heavy-flavour hadron decays with ALICE).


Besides the analysis of kinematic distributions (in momentum and angle) of heavy-flavour particles, further interesting information on the onset of possible medium effects may come from the study of the yields of the various charm and beauty hadrons as a function of charged-particle multiplicity going from pp to \pPb and Pb--Pb, as done in the past for the case of strangeness production. As an example, with high-multiplicity triggers in pp collisions at $\sqrts =14$~$\UTeV$ ($L_{\rm int}=200$~pb$^{-1}$), the ALICE experiment will have the potential to detect D mesons with a precision better than 10\% in a wide kinematic range up to about 10--12 times the average charged-particle multiplicity.





\clearpage

\section{Jets and parton energy loss}\label{sec:HI_Jets}

{ \small
  \noindent \textbf{Coordinators}: Marta Verweij (Vanderbilt University and RIKEN BNL Research Center)
  
  \noindent \textbf{Contributors}:
L.~Apolinario (LIP and IST Lisbon),
R.~Bi (Massachusetts Institute of Technology),
Z.~Citron (Ben-Gurion University of the Negev),
Y.~Chen (CERN),
L.~Cunqueiro Mendez (Oak Ridge National Laboratory),
D.~d'Enterria (CERN),
P.M.~Jacobs (Lawrence Berkeley National Laboratory),
F.~Krizek (Academy of Sciences, Prague),
Y.-J.~Lee (Massachusetts Institute of Technology),
M.~van Leeuwen (NIKHEF),
C.~Mcginn (Massachusetts Institute of Technology),
G.~Milhano (LIP and IST Lisbon, CERN),
D.V.~Perepelitsa (University of Colorado Boulder),
M.~P\l osko\'n (Lawrence Berkeley National Laboratory),
M.~Rybar (Columbia University),
A.M.~Sickles (U. Illinois, Urbana-Champaign),
M.~Spousta (Charles University),
K.~Tatar (Massachusetts Institute of Technology),
J.~Wang (Massachusetts Institute of Technology)
}

\subsection{Introduction}

The most direct way to measure the structure of matter is the controlled scattering of a beam of probe particles. This approach was used to discover the atomic nucleus, and quarks and gluons, and it is employed today to explore the partonic structure of nucleons and nuclei. However, the partonic phase of the QGP lives for $\sim10^{-23}$ seconds before breaking up into its hadronic remnants, so that probing it by the scattering of an externally-generated beam is impossible in practical terms. As an alternative, energetic jets arising from high-$Q^2$ processes in the same nuclear collision that generates the QGP provide internally-generated probes that may be applied for this purpose~\cite{Bjorken:1982tu,Gyulassy:1990ye,Baier:1994bd,Zakharov:2018rst,Gyulassy:1999zd,Wiedemann:2009sh}. 

Jets, observed as collimated sprays of energetic particles, were predicted by Quantum Chromo Dynamics (QCD) to form in high energy collisions. They constitute a substantial part of the background in beyond the Standard Model physics searches and were instrumental in the Higgs boson discovery. While jet evolution in vacuum is well understood, the question of how jets interact with a dense deconfined medium remains an active field of study, that is largely driven in the recent years by the unprecedented experimental capabilities of the RHIC and LHC accelerators and detectors.  Understanding from first principles how a jet evolves as a multi-partonic system, spanning a large range of scales (from  $\sim$1 GeV to $\sim$1 TeV) is crucial to quantitatively probe the Quark Gluon Plasma (QGP).  The successful description of bulk observables by viscous hydrodynamic calculations with a small viscosity to entropy ratio have led to the standard picture of a strongly coupled plasma. However, due to the property of asymptotic freedom in QCD, the produced matter is expected to behave differently at smaller and smaller distances which can only be accessed with well calibrated probes, namely, QCD jets.

High-$Q^2$ processes between the partonic constituents of colliding nucleons occur early in the collision. Further interactions of the outgoing partons with the hot and dense QCD medium produced in heavy ion collisions are expected to modify the angular and momentum distributions of final-state jet fragments relative to those in proton-proton collisions. This process, known as jet quenching, can be used to probe the properties of the QGP~\cite{Bjorken:1982tu,Gyulassy:1990ye,Baier:1994bd,Zakharov:2018rst,Gyulassy:1999zd,Wiedemann:2009sh}. Jet quenching was first observed at RHIC, BNL~\cite{Adcox:2001jp,Adler:2003qi,Adler:2002xw,Adler:2002tq,Adams:2003kv,Adams:2003im,Adams:2006yt,Arsene:2003yk,Back:2003qr,Adamczyk:2016fqm,Adamczyk:2017yhe} and then at the CERN LHC~\cite{Aamodt:2010jd,Aamodt:2011vg,Aad:2015wga,CMS:2012aa,Aad:2010bu,Chatrchyan:2012nia,Aad:2012vca,Abelev:2013kqa,Adam:2015ewa,Khachatryan:2016odn,Adam:2015doa} by studying the redistribution of energy radiated from the parton because of interactions with the QGP. More recent detailed analyses have focused on modifications to the distribution of final-state particles emitted in the parton's shower~\cite{Chatrchyan:2013kwa,Chatrchyan:2014ava,Aaboud:2017bzv,Acharya:2017goa,Acharya:2018uvf,Sirunyan:2017bsd,Sirunyan:2018gct}.

One of the main goals of the RHIC and LHC heavy ion physics programs is utilization of jets and their decay products, including high $p_{\mathrm{T}}$ hadrons formed by light and heavy quarks, to investigate the QGP properties. A milestone in this program is the extraction of the transport coefficient $\hat q $ by the JET Collaboration~\cite{Burke:2013yra}, based on inclusive hadron suppression measurements at RHIC and the LHC. However, this result has significant systematic uncertainties, due both to theoretical issues and to the limited view provided by inclusive hadron suppression measurements into the fundamental processes underlying jet quenching. A more complete picture requires measurements of reconstructed jets and their in-medium modification. 

At the LHC, the collision energy is over an order of magnitude larger than at RHIC. Jet production cross-sections are correspondingly larger, enabling the study of hard processes over a wider kinematic range. Detectors at both facilities have extensive capabilities to study fully-reconstructed jets by grouping the detected particles within a given angular region into a jet, thereby capturing a significant fraction of the parton shower. Jets are a key diagnostic of the QGP, as their interactions with this new state of matter reveal its properties. The interaction with the medium can result in a broadening of the jet profile with respect to vacuum fragmentation. In this case, for a given jet size and a fixed initial parton energy, the energy of the jet reconstructed in heavy ion collisions will be smaller than in vacuum. In the case where the gluons are radiated inside the cone, the jet is expected to have a softer fragmentation and a modified density profile compared to jets in vacuum. Jets may also scatter coherently in the medium, and measurements of jet deflection may provide a direct probe of the micro-structure of the QGP. Fully reconstructed jets provide better theoretical control than high $p_{\mathrm{T}}$ hadrons because they are less sensitive to non-perturbative physics and therefore have the potential to provide a better characterization of the QGP. 
Furthermore, major theoretical and experimental advances were made recently in understanding the evolution of parton showers in a QCD medium with the development of novel jet substructure observables.

In the following sections the expected performance using a total integrated luminosity of 10 $\mathrm{nb}^{-1}$ of Pb--Pb data, which is expected for HL-LHC, for a selection of key jet observables will be discussed. 

\subsection{Out-of-cone radiation}\label{sec:preceloss}

\begin{figure}[!ht]
\begin{center}
\includegraphics[width=.45\textwidth]{\main/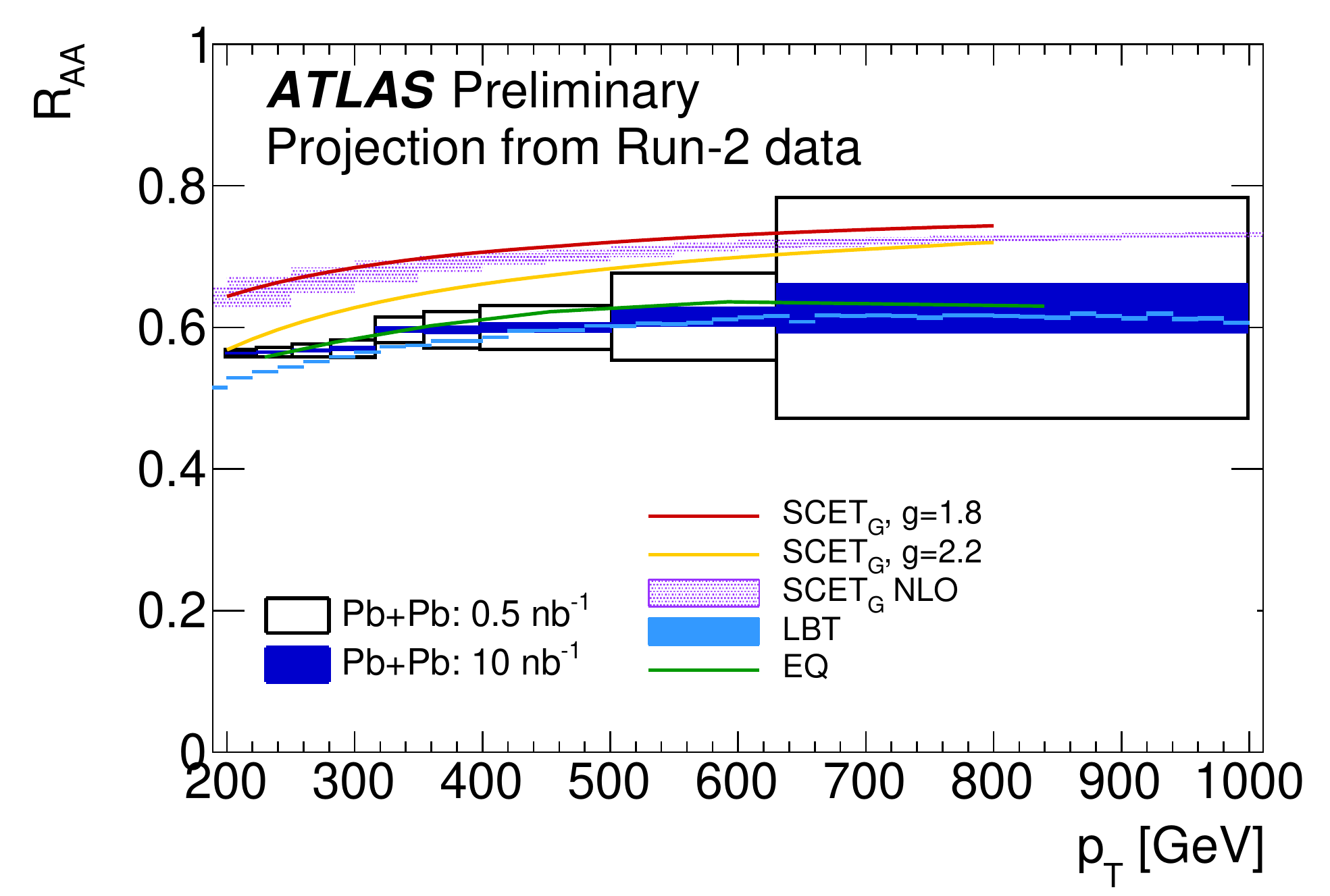}
\includegraphics[width=.45\textwidth]{\main/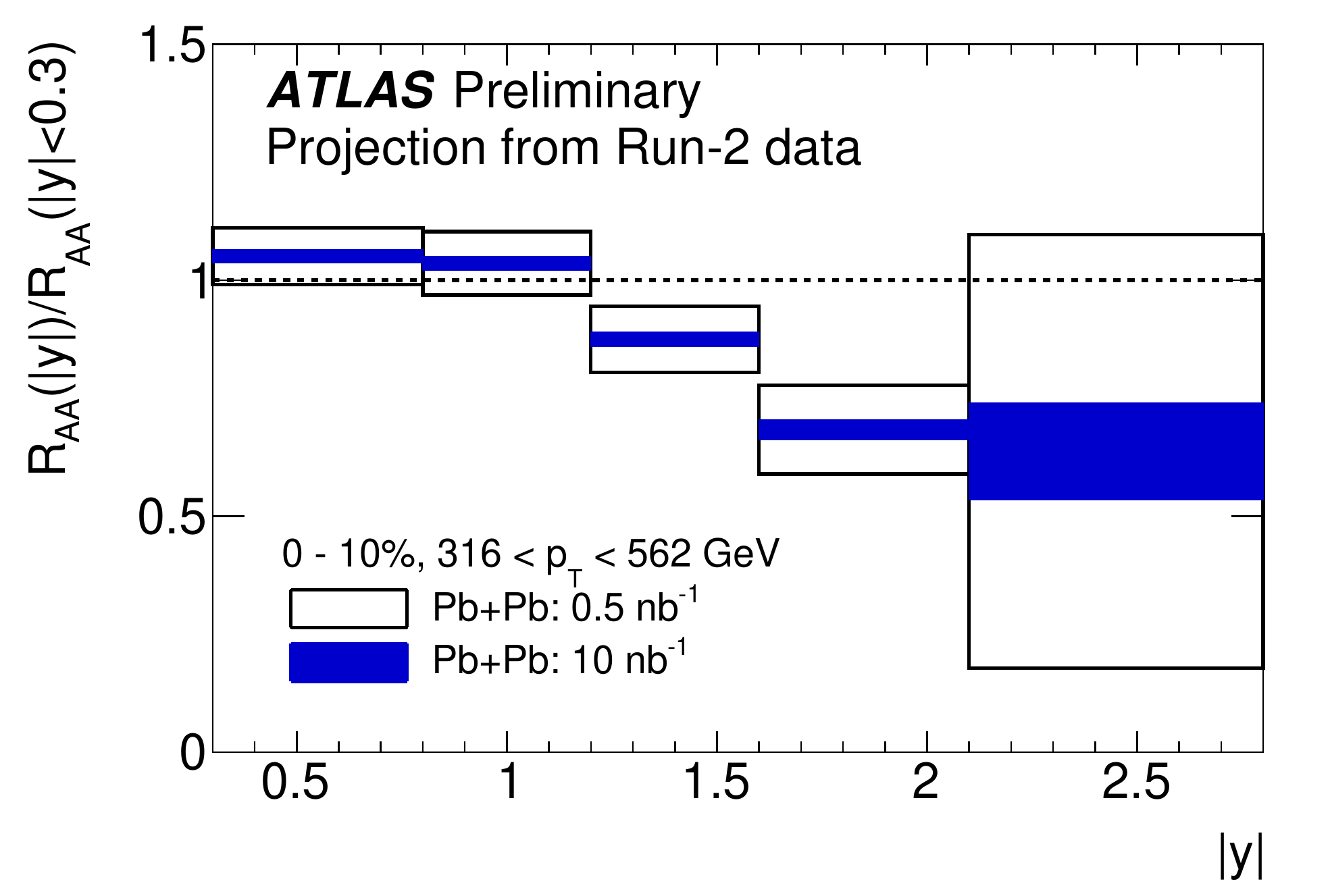}
\caption{Projection of the precision that can be reached for jet $R_{\mathrm{AA}}$ at the HL-LHC using calorimeter jets at ATLAS as function of $\pT$ (left panel) and rapidity (right panel)~\cite{ATL-PHYS-PUB-2018-019}. See text for model details.}
\label{fig:jetRAA}
\end{center}
\end{figure}

One of the classic observables to measure the out-of-cone radiation due to jet quenching is the jet nuclear modification factor $R_{\mathrm{AA}}$ defined as: 
  \begin{equation} 
  \RAA=\dfrac{ \left.\dfrac{1}{N_{\mathrm{evt}} }\dfrac{\mathrm{d}^{2} 
N_{\mathrm{jet}}}{\mathrm{d} \pT \mathrm{d} y} \right|_{\mathrm{cent}}
}
{
\left. \TAA \dfrac{\mathrm{d}^2 \sigma_{\mathrm{jet}}}{\mathrm{d} \pt \mathrm{d} y}\right|_{pp}
}\, ,
\label{eq:RAAjet}
\end{equation} 
where $N_{\mathrm{jet}}$ and $\sigma_{\mathrm{jet}}$ are the jet yield in Pb--Pb collisions and the jet production cross-section in pp collisions, 
respectively, both measured as a function of transverse momentum, \pT, and rapidity, $y$, and where $N_{\mathrm{evt}}$ is the total number of Pb--Pb collisions within a chosen centrality interval. 
Measurements of the jet $R_{\mathrm{AA}}$ at the LHC have shown a suppression of a factor of two in central collisions over a wide range of jet transverse momentum~\cite{Aad:2012vca,Abelev:2013kqa,Khachatryan:2016jfl}. Figure~\ref{fig:jetRAA} shows the current precision obtained with 0.5 $\mathrm{nb}^{-1}$ and what can be achieved at the HL-LHC with a factor of 20 more data (10 $\mathrm{nb}^{-1}$). Especially at high transverse momentum a strong reduction of the experimental uncertainties is expected, which will allow a detailed study of the momentum dependence of the out-of-cone radiation. The jet $R_{\mathrm{AA}}$ is sensitive to various physics mechanisms such as color coherence, flavor dependence of energy loss, and the medium response to the jet. Models incorporating these various physics effects can be confronted with the high precision data from HL-LHC with a goal of determining what the relative contribution of each of these phenomena is. The expected performance is compared with several recent model predictions: the Linear Boltzmann Transport model (LBT)~\cite{He:2015pra}, three calculations using Soft Collinear Effective Theory (SCET)~\cite{Chien:2015vja,Chien:2015hda,Vitev:2008rz,Kang:2017frl}, and the Effective Quenching model (EQ)~\cite{Spousta:2015fca}. The higher precision data will allow tighter constraints on or falsification of theoretical model predictions. In addition to the inclusive jet $R_{\mathrm{AA}}$ it is particularly interesting to study the mid- and forward rapidity region separately since it allows to study the interplay between flavor and spectral steepness, and the path-length dependence of jet quenching. 
The right panel of Fig.~\ref{fig:jetRAA} shows the improvement in statistical precision in the forward rapidity region. 
The statistical precision should be sufficient to quantitatively assess the rapidity dependence of the $R_\mathrm{AA}$ up to a rapidity of $|y|=2.8$. Both of these predictions indicate that HL-LHC should bring a definitive understanding of the intriguing features of the jet $R_\mathrm{AA}$ as seen in the current data.

\begin{figure}[!ht]
\begin{center}
\includegraphics[width=.45\textwidth]{\main/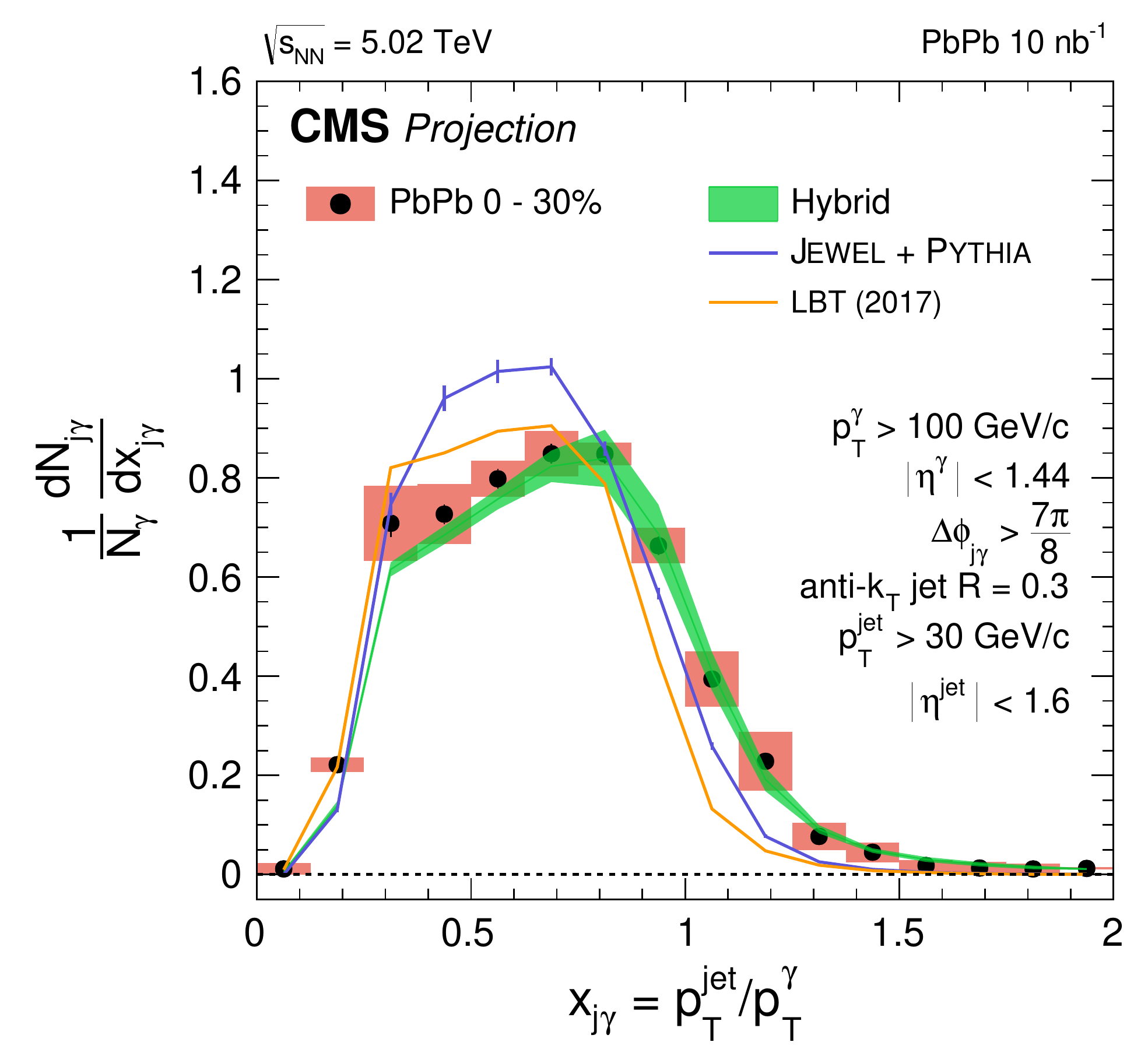}
\includegraphics[width=.45\textwidth]{\main/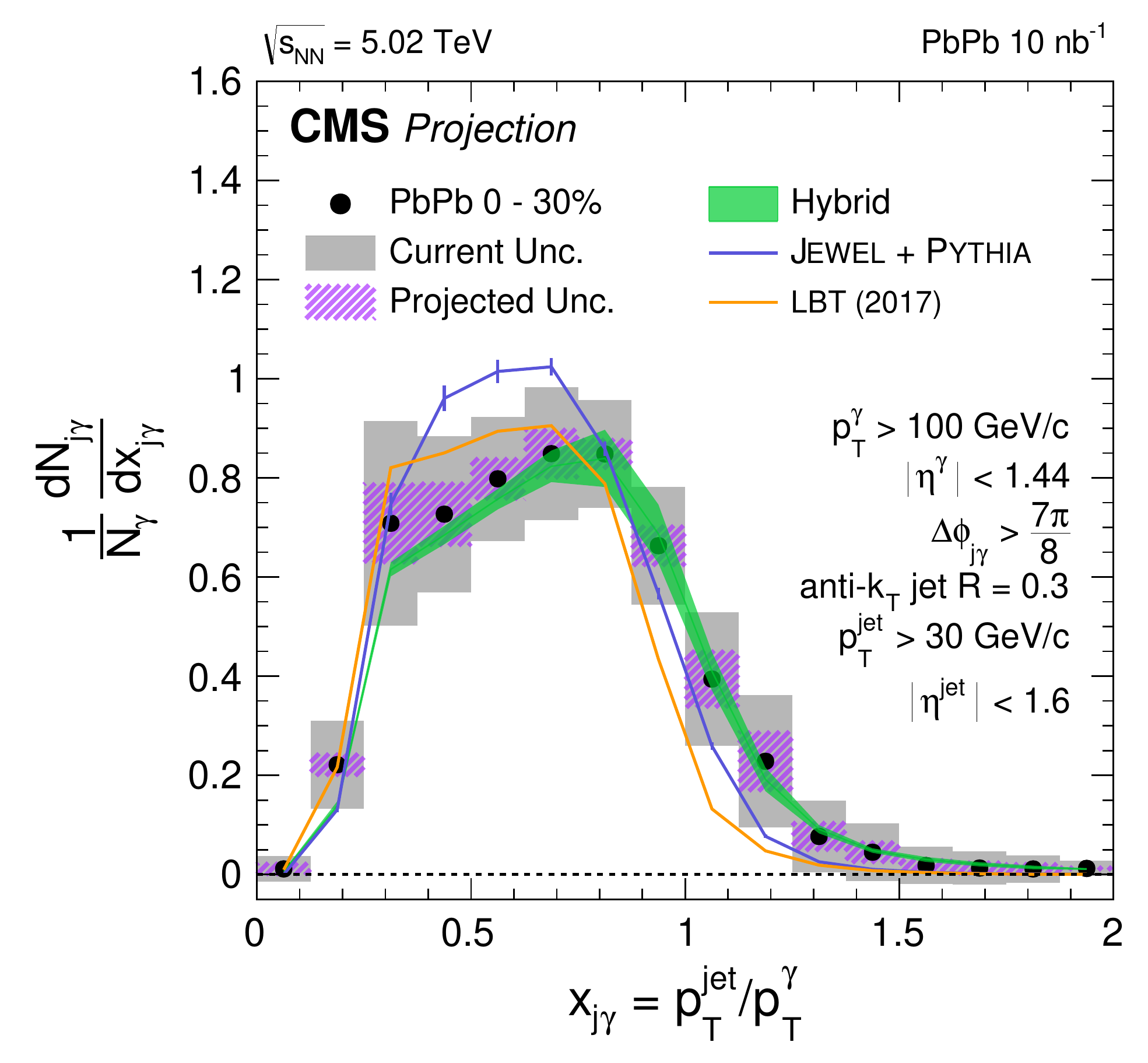}
\caption{(Left Panel) Photon-jet momentum balance $x_{j\gamma}$ distribution for isolated-photon+jets of $p_{\gamma}$ $> $ 100 GeV/c and $|\eta_{\gamma}|<1.44$, $p_{\rm jet} > $ 30 GeV/c and $|\eta_{\rm jet}| < 1.6$ in the HL-LHC data (Right Panel). Comparison between the current performance with 0.4 nb$^{-1}$ of Pb--Pb data collected in 2015 and with HL-LHC data~\cite{CMS-PAS-FTR-17-002}.}
\label{fig:photonjet}
\end{center}
\end{figure}

\begin{figure}[!ht]
\begin{center}
\includegraphics[width=.45\textwidth]{\main/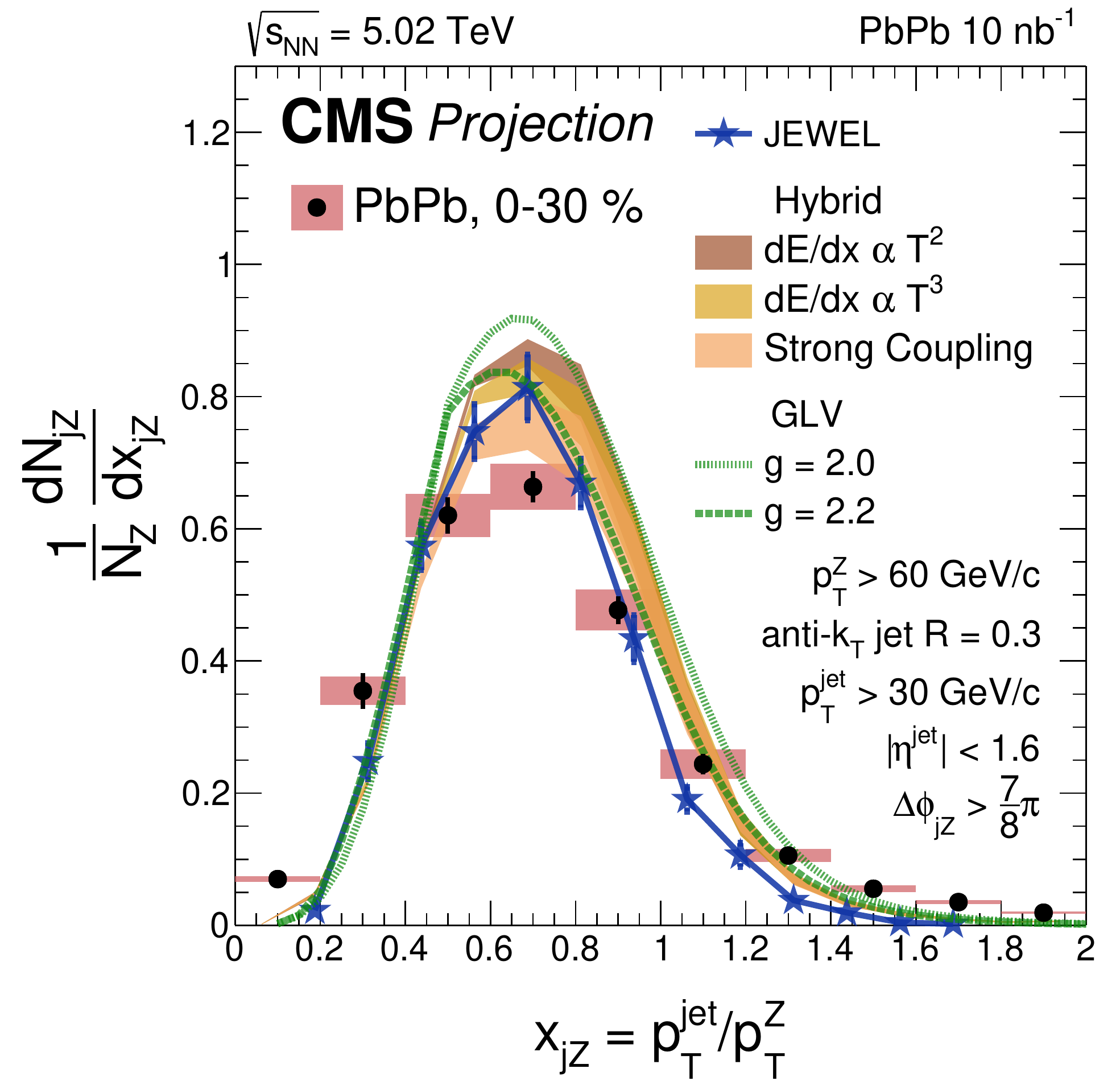}
\includegraphics[width=.45\textwidth]{\main/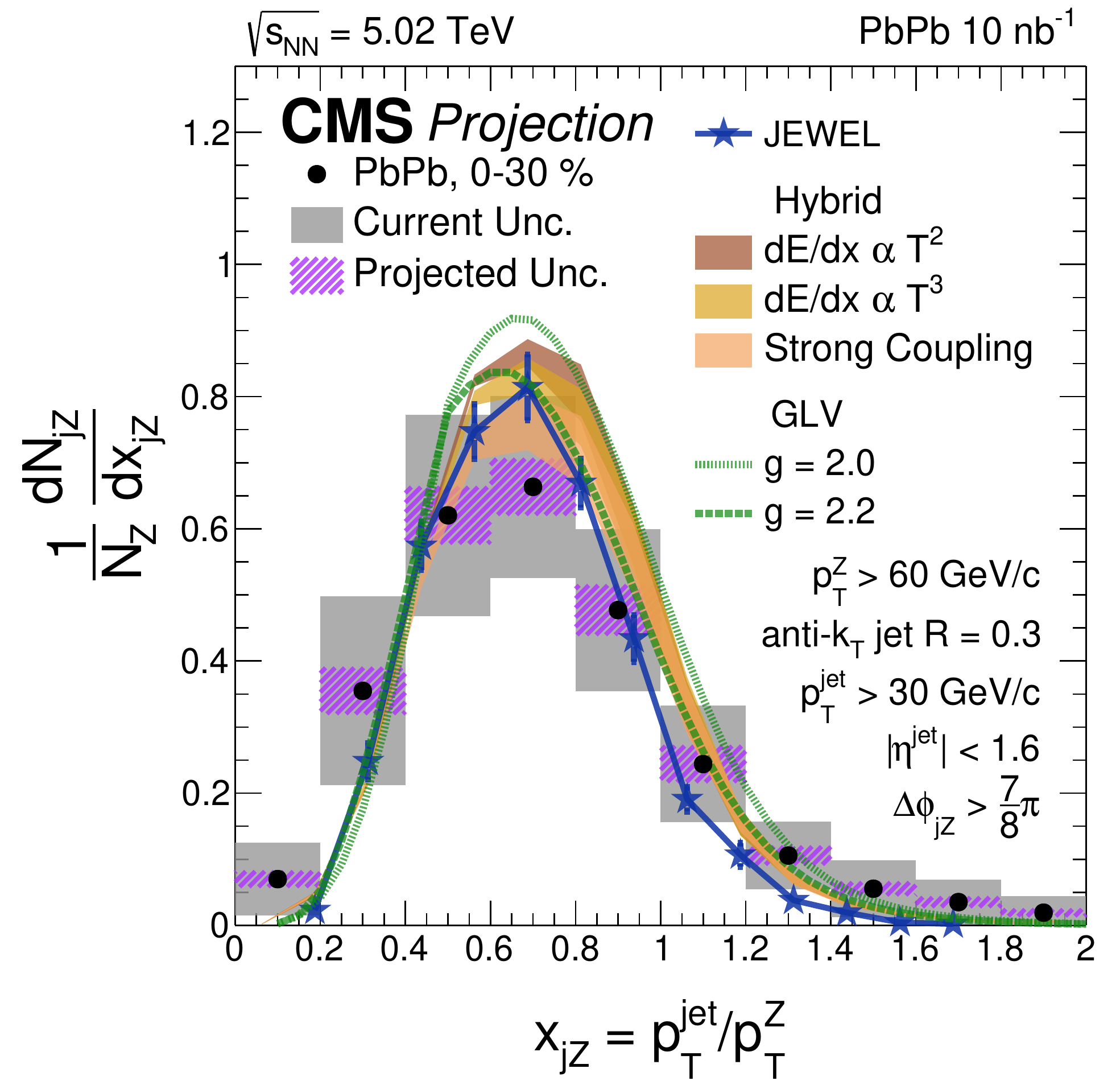}
\caption{(Left Panel) $X_{jZ}$ distribution for Z boson-jet pairs with $p_\mathrm{T}^{Z}$ $> $ 60 GeV/c, $p_{\rm jet} > $ 30 GeV/c and $|\eta_{\rm jet}| < 1.6$ in the HL-LHC data (Right Panel) Comparison between the current performance with 0.4 nb$^{-1}$ of Pb--Pb data collected in 2015 and with HL-LHC data~\cite{CMS-PAS-FTR-17-002}.}
\label{fig:Zjet}
\end{center}
\end{figure}

Parton energy loss can be studied more differentially using boson tagged jets. The bosons (photons or Z$^{0}$ bosons) escape the region of the hot dense medium unmodified. This has been confirmed through the absence of significant modification of both photon and Z$^{0}$ boson production rates in Pb--Pb collisions relative to expectations from measured cross section in pp collisions by both ATLAS and CMS collaborations~\cite{Aad:2012ew,Aad:2015lcb,Chatrchyan:2012vq,Chatrchyan:2014csa}. However, the partons recoiling from the boson is modified in heavy ion collisions due to interactions with the QCD medium. Furthermore, jets produced opposite to the isolated boson
are more likely to originate from quarks, while dijet and hadron+jet correlations usually involve significant gluon contributions. 
Comparing Z+jet and $\gamma$+jet observables to dijets~\cite{Chatrchyan:2011sx,Chatrchyan:2012nia} (or hadron+jets~\cite{Adam:2015doa}) allows to explore the difference between energy loss for quark and gluon initiated jets. Figures~\ref{fig:photonjet} and~\ref{fig:Zjet} show the expected performance at the HL-LHC for the transverse momentum balance between the jet and the boson. The central values are based on the smoothed data from the previous CMS publications~\cite{Sirunyan:2017jic,Sirunyan:2017qhf}. The systematic uncertainties are reduced by a factor of two with respect to the results with the 2015 Pb--Pb data due to improvements on the jet energy scale and jet energy resolution uncertainties available with the larger data sample at the HL-LHC. The collected number of $\gamma$+jet events will also be sufficient to study the path length dependence of jet quenching by performing measurements as a function of angle with the reaction plane for the first time. In addition to the smaller uncertainties due to the enhanced statistics at the HL-LHC, it will also be possible to utilize higher momentum photons and Z$^{0}$ bosons allowing the measurement of larger jet energy losses. The LHC experiments also envision extending the jet momentum reach to lower transverse momentum in certain analyses, allowing to recover those heavily quenched jets which are currently not selected for such measurements due to limitation arising from the fluctuating background. A distinct effect due to large backgrounds is that of limited jet energy resolution, which can be improved by using more sophisticated techniques for the background correction as was recently shown in Ref.~\cite{Haake:2018hqn}.




\subsection{Jet deflection}

Angular deflection of the jet relative to its initial direction due to momentum transfer with the medium can be used as a direct probe of the QGP. 
Jet deflection can be measured by coincidence observables, in which an axis is determined by a hard reference object, and the deflection of the jet recoiling from hard object is measured relative to that axis. Such scattering measurements, carried out over a wide range in energy and resolution scale, can be used to explore the microscopic structure of the QGP. Modification of the rate of rare, large-angle jets with respect to the hard reference object in nuclear collisions compared to the production rate in vacuum may arise from the scattering off quasi-particles (quarks and gluons or composite objects) of the QGP, thereby probing their nature~\cite{DEramo:2012uzl}. In addition, the recoil jet distribution at small recoil angles relative to the reference axis (the axis of the hard object selected at the opposite hemisphere) may be modified by soft multiple scattering in the QGP, which can be used to extract the jet transport parameter $\hat{q}$ by comparison to models~\cite{Chen:2016vem}.

Measurements of the angular distribution of jets relative to a reference axis have been reported for either dijet, photon-jet, Z$^{0}$-jet and hadron-jet coincidences, at RHIC~\cite{Adamczyk:2017yhe} and LHC ~\cite{Adam:2015doa,Sirunyan:2017jic,Sirunyan:2017qhf,Chatrchyan:2012nia,Aaboud:2017eww}. 
These current measurements exhibit no significant evidence of in-medium modification of angular distributions, both at small and large angles to the reference axis. While they impose constraints on the magnitude of in-medium scattering effects, their statistical precision is limited. Measurements during HL-LHC will either discover in-medium modification to the recoil jet angular distributions, or improve these constraints substantially.


There is an intensive ongoing effort to develop analysis tools and calculable approaches that discriminate the various contributions to in-medium jet deflection and shower modification, in both experiment and theory~\cite{Andrews:2018jcm} (see Sec.~\ref{sec:preceloss} and~\ref{sec:jetsub}). In this section we focus solely on jet-centroid deflection measurements, without consideration of the effects of shower broadening or other shower modification (see Sec.~\ref{sec:jetsub}). Measurements of both classes of jet quenching observable must ultimately be interpretable in a single consistent picture, but such an approach is beyond current experimental and theoretical capabilities. 

The most significant background to the measurement of medium-induced jet deflection is broadening of the angular difference between the two leading jets due to well-established vacuum QCD effects, in particular Sudakov radiation, which is radiation outside the jet cone that generates a broad peak in the recoil jet angular distribution relative to a reference axis (for example a high momentum hadron)~\cite{Chen:2016vem,Mueller:2016xoc}. 


Low-energy jet measurements are expected to experience larger deflection for a given momentum transfer between the jet and medium~\cite{DEramo:2018eoy,Gyulassy:2018qhr} and are therefore more likely to show large angle deflection. 
A recent calculation, that includes the effects of vacuum Sudakov radiation and jet-medium interactions based on the few-hard (GLV) or multiple-soft (BDMPS) scattering approaches to jet quenching, finds that the acoplanarity distributions for these different jet quenching pictures differ by a few percent in the range $20<\ptjet<40$ \gevc~\cite{Gyulassy:2018qhr}. This sets the precision required for the observation of medium-induced jet deflection during HL-LHC. Additional theoretical considerations of in-medium \pT-broadening can be found in~\cite{Zakharov:2018rst,Ghiglieri:2018ltw}.

In light of such considerations, it is necessary to utilize analysis techniques that can attain few percent precision in the measurement of recoil jet angular distributions for low \ptjet\ and large jet radius $R$, over the large and complex uncorrelated backgrounds in central Pb--Pb collisions at the LHC. This precision is achievable using the statistical approach to jet background correction~\cite{Adam:2015doa,Adamczyk:2017yhe,Sirunyan:2017jic,Sirunyan:2017qhf}, in which the discrimination of correlated and uncorrelated recoil jet yield is carried out in a fully data-driven way, at the level of ensemble-averaged distributions. The statistical correction approach has been used to measure the azimuthal distribution for charged jets with $R=0.5$ and $40<\ptjetch<60$ \gevc\ recoiling from a high-\pt\ hadron in central Pb--Pb collisions at the LHC~\cite{Adam:2015doa}, and for charged jets with $R=0.5$ and $\ptjetch\sim10$ \gevc\ in central \AuAu\ collisions at RHIC~\cite{Adamczyk:2017yhe}, as well as for photon-jet and Z-jet correlations~\cite{Sirunyan:2017jic,Sirunyan:2017qhf}. We expect that reaching as low as $\ptjet=10$ \gevc\ is likewise achievable at the LHC, with good systematic precision. 

The required experimental approach is therefore in hand, and we explore here the statistical precision achievable using it for such measurements during HL-LHC.
We utilize the \jewel\ event generator~\cite{Zapp:2013vla} for these projections, which incorporates medium-induced interactions of partons propagating in the QGP. Calculations are carried out for central Pb--Pb collisions at $\sqrtsnn=5.02$ TeV with integrated luminosity of 10 nb$^{-1}$ , and pp collisions at $\sqrts=5.02$ TeV with integrated luminosity of 6 pb$^{-1}$. 
The \jewel\ calculations for central Pb--Pb collisions are carried out with the ``Recoil off" configuration in which the partons from the medium response are neglected. 

\begin{figure}[tbh!]
\centering
\includegraphics[width=0.6\textwidth]{\main/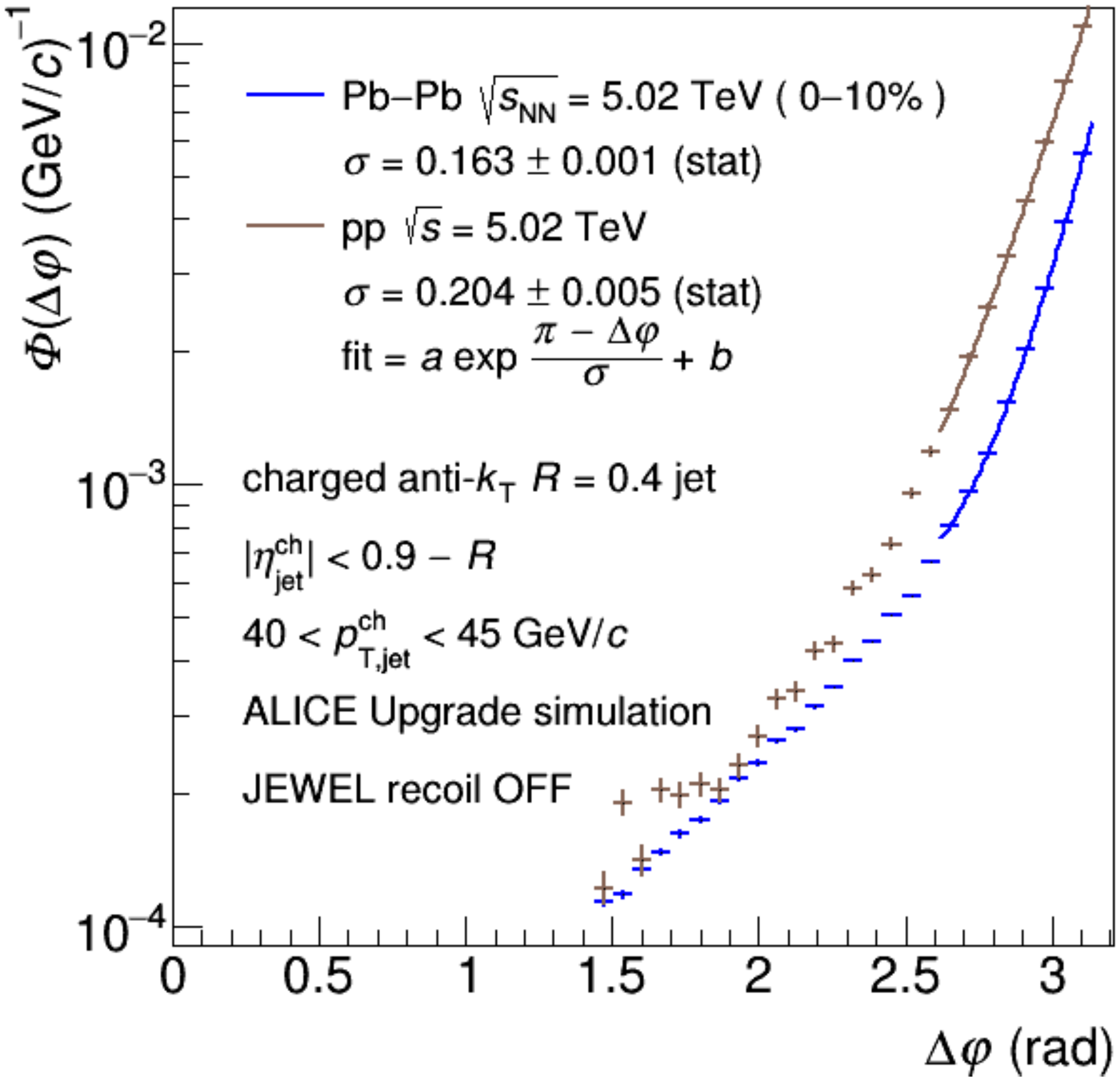}
\caption{\jewel\ simulation of the angular distribution of charged jet yield in the ALICE acceptance for  $40<\ptjetch<45$ \gevc\ and $R=0.4$ recoiling from a high-\pt\ reference hadron ($20<\ptt<50$ \gevc), for central Pb--Pb collisions at $\sqrtsnn=5.02$ TeV with 10 nb$^{-1}$ int. luminosity, and pp collisions at $\sqrts=5.02$ TeV with 6 pb$^{-1}$ int. luminosity. The recoil jet azimuthal angle \Dphi\ is defined with respect to the reference axis. The observable shown is $\Phi(\Dphi)$ which incorporates statistical suppression of uncorrelated background. Figure from Ref.~\cite{ALICE-PUBLIC-2019-001}.}
\label{fig:JetDeflectionPhi}
\end{figure}

Figure~\ref{fig:JetDeflectionPhi} shows the recoil jet azimuthal angle, \Dphi, defined with respect to the reference axis~\cite{Adam:2015doa} as simulated by the \jewel\ event generator. The background-corrected azimuthal distribution of recoil jets recoiling from a high-\pT\ hadron, with the statistics expected by ALICE for central Pb--Pb and pp collisions during the HL-LHC phase is shown. The distribution for central Pb--Pb collisions exhibits an overall yield suppression, corresponding to jet quenching, but also a slight narrowing of the main peak at $\Dphi\sim\pi$ and an enhancement at large deflection angle. The narrowing is characterized by extracting the width of the \Dphi distribution which is $0.204 \pm 0.005$ in the pp simulation and $0.163 \pm 0.001$ for the Pb--Pb simulation with \jewel\. 
In order to quantify the difference at large recoil jet deflection angle between pp and central Pb--Pb collisions, we integrate the $\Phi(\Dphi)$ from $\pi/2$ to a threshold angle \dphithresh~\cite{Adam:2015doa},

\begin{equation}
\Sigma(\dphithresh) = 
\int_{\pi/2}^{\pi-\dphithresh}
\Phi(\Delta\varphi)\mathrm{d}\Delta\varphi.
\label{eq:Sigma}
\end{equation}

\begin{figure}[tbh!]
\centering
\includegraphics[height=0.6\textwidth]{\main/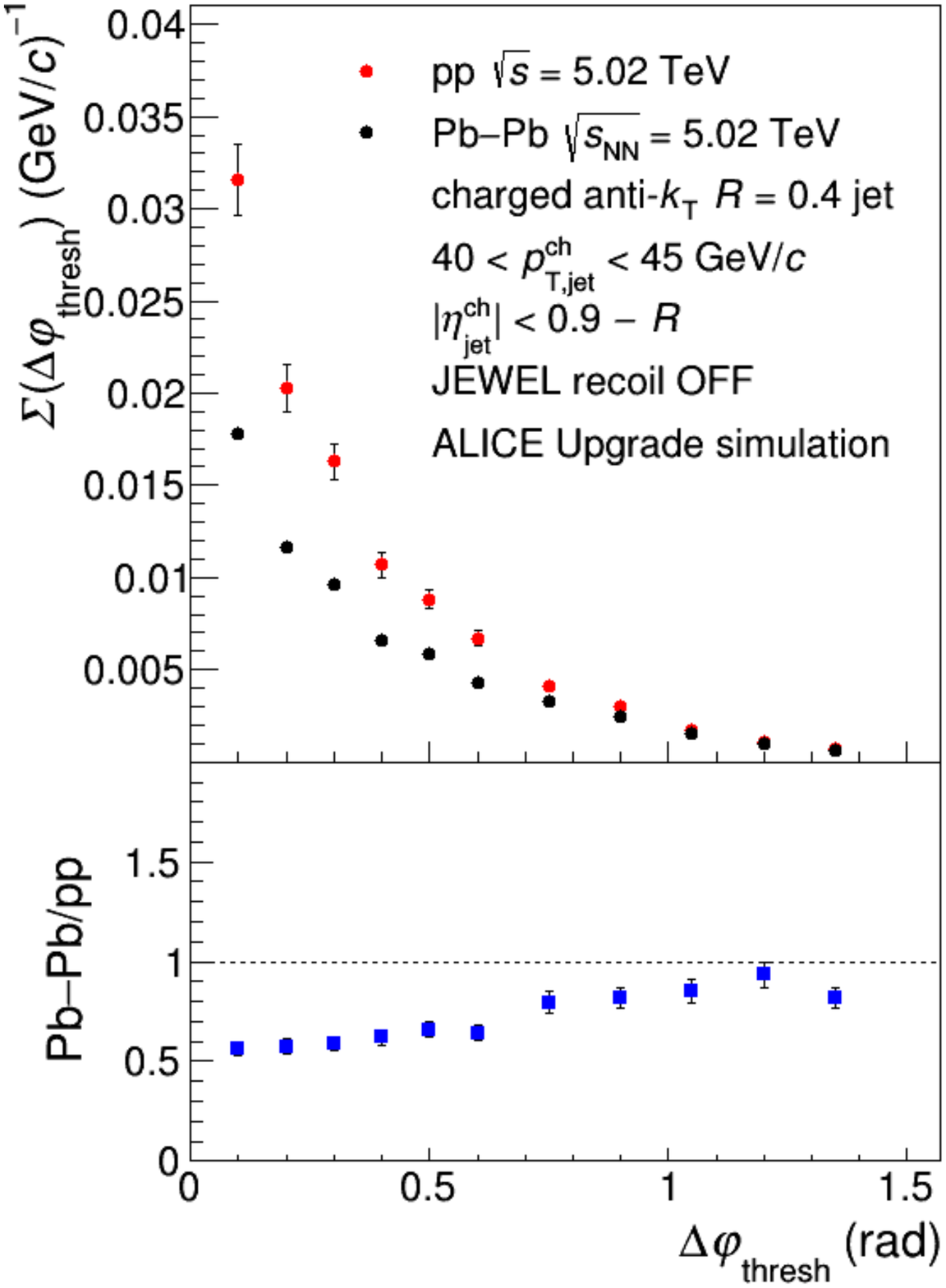}
\caption{Cumulative large-angle yield $\Sigma(\dphithresh)$ (Eq.~\ref{eq:Sigma}) vs. \dphithresh\ for the pp and central Pb--Pb distributions $\Phi(\Dphi)$ in Fig.~\ref{fig:JetDeflectionPhi}. See text for details.
Figure from Ref.~\cite{ALICE-PUBLIC-2019-001}.
}
\label{fig:JetDeflectionSigma}
\end{figure}

\noindent
Figure~\ref{fig:JetDeflectionSigma} shows $\Sigma(\dphithresh)$ for the $\Phi(\Dphi)$ distributions in Fig.~\ref{fig:JetDeflectionPhi}, together with their ratio. In this calculation, the value of $\Sigma$ at $\dphithresh=0$ is around 0.5, which is the yield suppression averaged over the full recoil hemisphere. The ratio grows to $\Sigma\sim1$ at $\dphithresh=1.2$, indicating a factor two enhancement in large-angle yield relative to the hemisphere average. The statistics of the measurement are clearly sufficient to measure the effect predicted by this calculation.
However, the calculation in~\cite{Gyulassy:2018qhr} predicts a difference of only a few percent in these distributions for GLV-like and BDMPS-like in-medium scattering, which is more difficult to discriminate. The statistical error in the ratio in Fig.~\ref{fig:JetDeflectionSigma} is around 5\% at $\dphithresh\sim1$, due predominantly to the statistical precision of the pp distribution. 


\subsection{Jet internal structure}\label{sec:jetsub}
The first measurements of jet quenching through full jet reconstruction at the LHC revolutionized our understanding of parton energy loss in a hot and dense medium. Nevertheless, there remains a gap in our understanding of the jet quenching mechanism that could be resolved by measuring the exact properties of the parton evolution through the medium. High statistics of collected jets during HL-LHC will provide a prime opportunity to explore the details of the internal structure of high energy jets that undergo interactions with the QGP. Observables probing the internal structure of jets can be defined using all measured hadrons in a jet or by using subjet techniques selecting only a specific region of the radiation phase space. In the following sections both approaches and their potential will be discussed.

\subsubsection{Substructure with hadrons}
\begin{figure}[!ht]
\begin{center}
\includegraphics[width=.45\textwidth]{\main/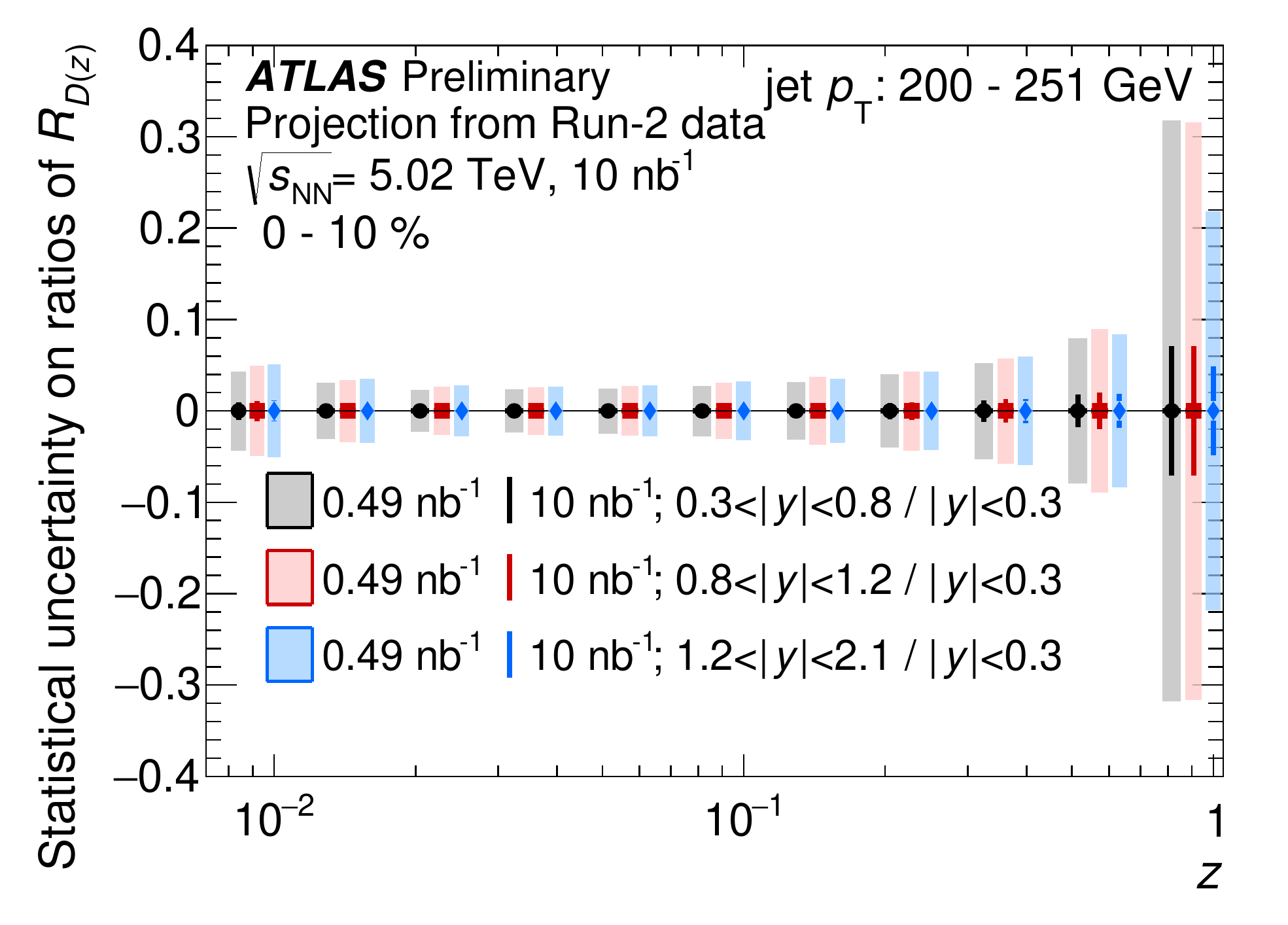}
\includegraphics[width=.45\textwidth]{\main/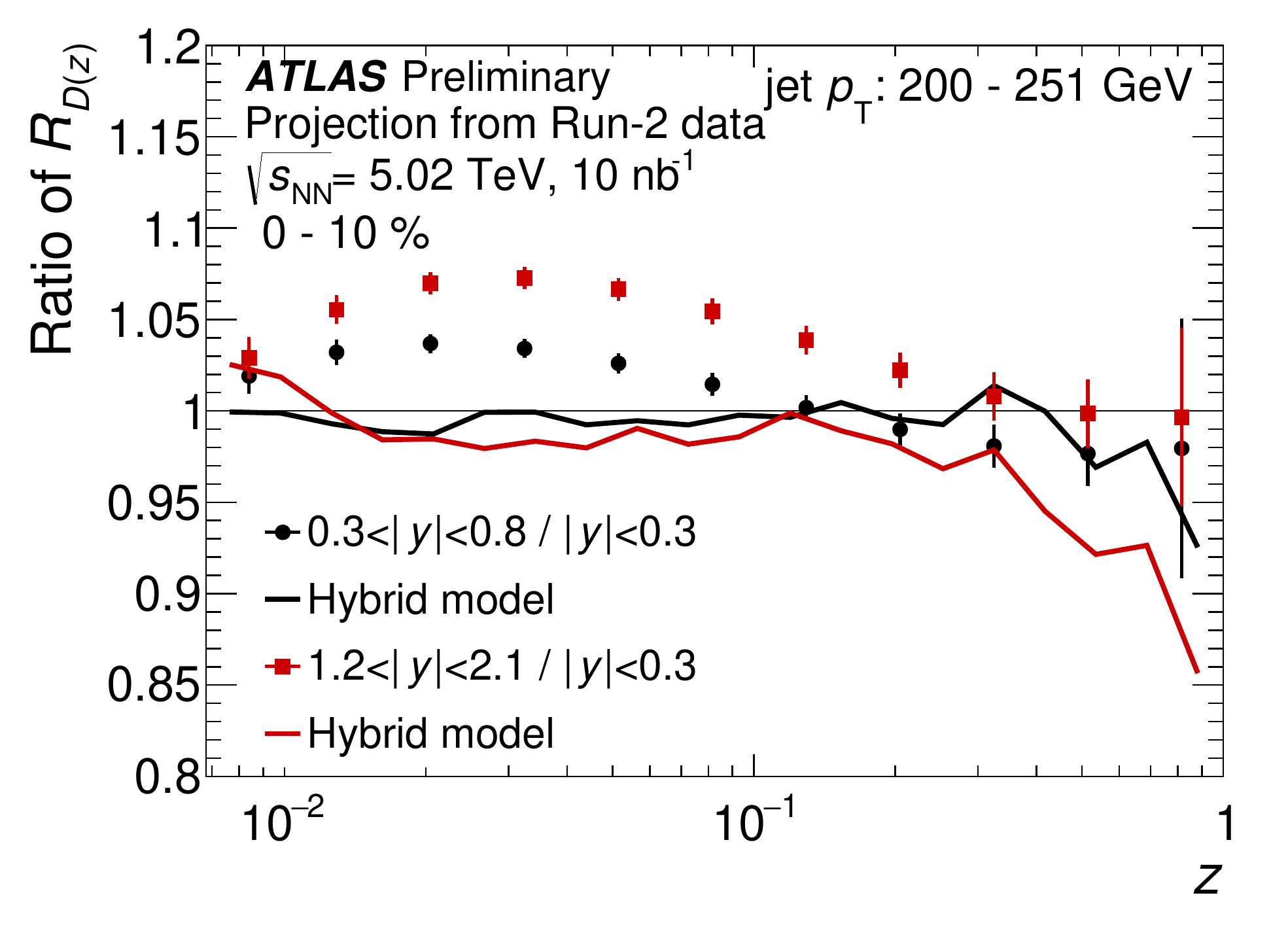}
\caption{Projection of the precision that can be reached for the modification of jet fragmentation function, $R_{D(z)}$, measured in jet \pT\ interval $200-251$ \gevc. In the left panel the statistical uncertainty on the measurement with the shaded boxes corresponding to 0.49 $\mathrm{nb}^{-1}$ while the vertical bars are for 10 $\mathrm{nb}^{-1}$. The right panel shows a comparison of $R_{D(z)}$ with a theory model (see text for more details)~\cite{ATL-PHYS-PUB-2018-019}.}
\label{fig:jetDz}
\end{center}
\end{figure}

Inclusive measurements of the longitudinal and transverse momentum distribution of hadrons in inclusive jets have been performed with high accuracy at the LHC~\cite{Aaboud:2018hpb,Sirunyan:2018jqr}. The modification due to jet quenching is studied by comparing the results in pp and Pb--Pb collisions.
For certain kinematic selections, for example at large $z$ where the leading particle in the jet is carrying a large fraction of the total jet momentum, the current experimental uncertainties are however large (see left panel of Fig.~\ref{fig:jetDz}) limiting the constraints on the jet quenching mechanism that can be extracted by comparing data to theoretical models. The expected statistical precision at HL-LHC for the ratio of fragmentation functions in Pb--Pb and pp collisions is shown in Fig.~\ref{fig:jetDz}. This precision will allow detailed characterization of the excess in yield of hard (large $z$) and soft (small $z$) fragments and the suppression in the region between these two excesses providing strong constraints to theoretical models. Measurements of the rapidity dependence of jet observables are of great interest since the fraction of quark- and gluon-initiated jets varies with rapidity. However, current measurements of the fragmentation function are statistics limited and no significant rapidity dependence is observed~\cite{Aaboud:2018hpb}. The right panel of Fig.~\ref{fig:jetDz} shows the ratio of fragmentation functions of high momentum jets for most central collisions with the expected accuracy at the HL-LHC. The projection are compared to the hybrid model~\cite{Casalderrey-Solana:2014bpa,Hulcher:2017cpt} which implements energy loss according to the strong coupling description of the radiation of low energy gluons associated with the hot QCD matter which predicts a rapidity-dependent suppression of particle yield at high $z$. 

\begin{figure}[!ht]
\begin{center}
\includegraphics[width=.485\textwidth]{\main/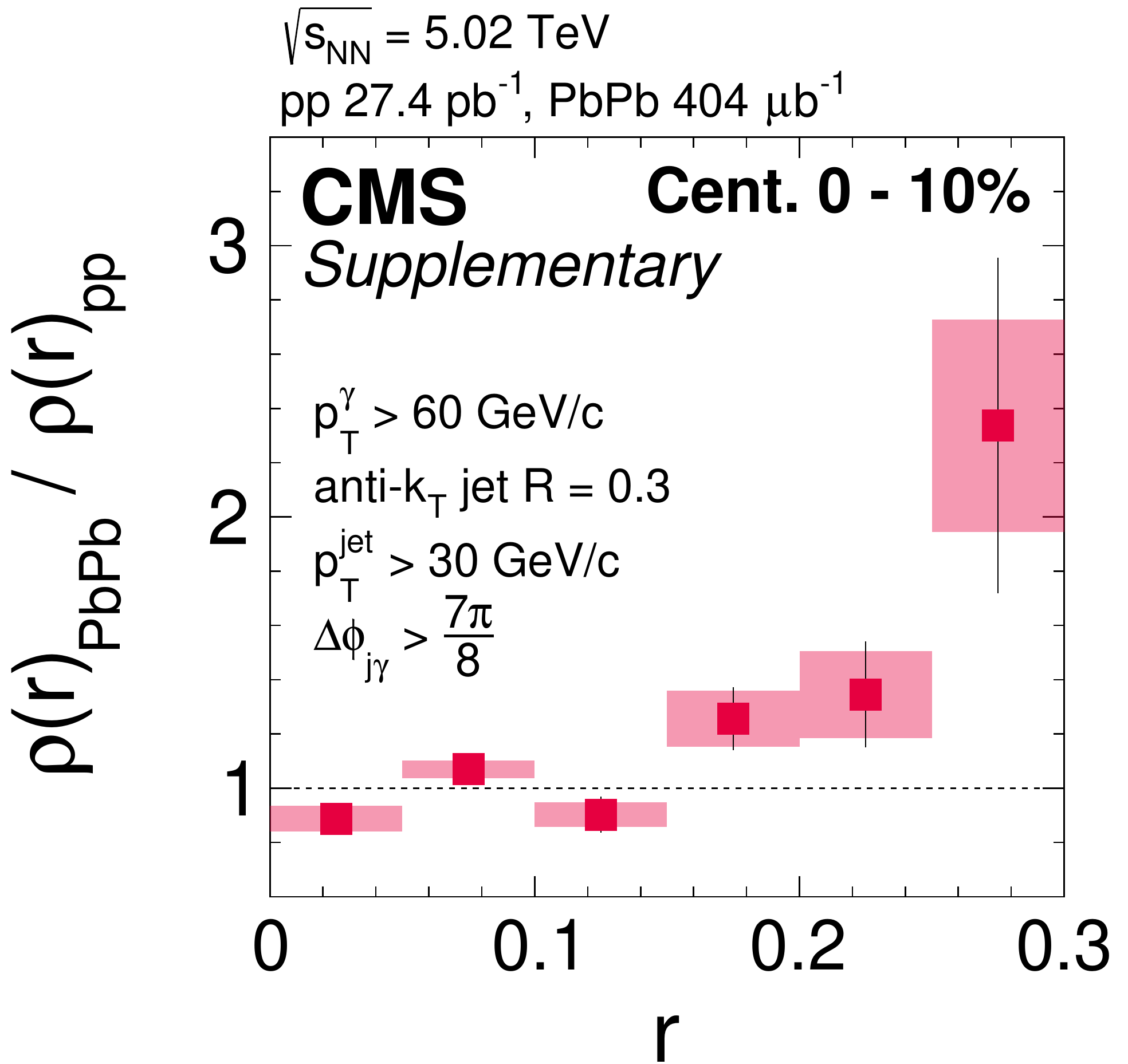}
\includegraphics[width=.45\textwidth]{\main/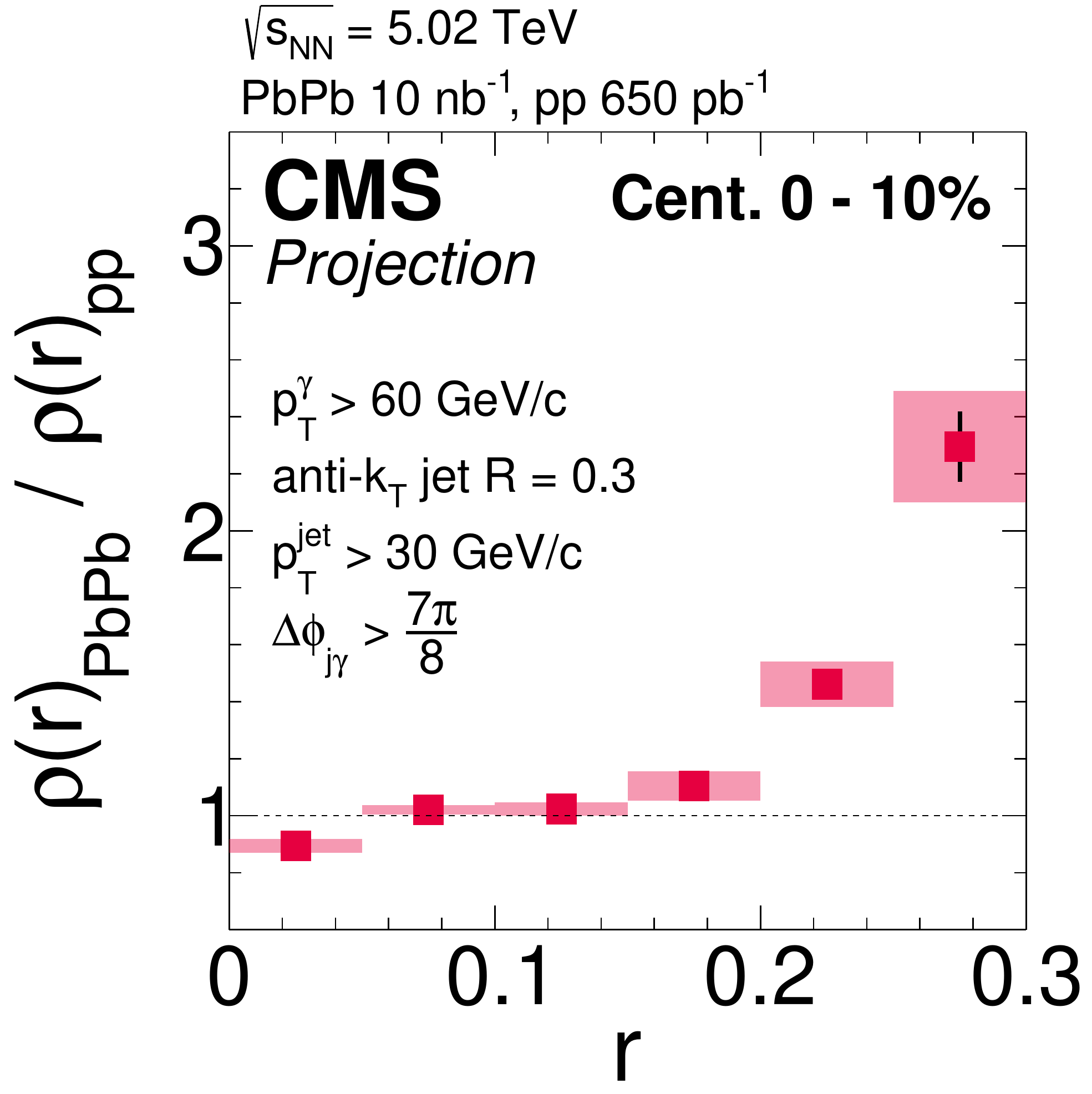}
\caption{(Left Panel:) The ratio of measured photon-tagged jet shape in Pb--Pb and pp collisions with the 2015 Pb--Pb data~\cite{Sirunyan:2018ncy}. (Right Panel:) The expected performance of the jet shape ratio in the HL-LHC data, using a third-order polynomial for smoothing the data. [REF to be added when note is public]}
\label{fig:jetshape}
\end{center}
\end{figure}

\begin{figure}[!ht]
\begin{center}
\includegraphics[width=.45\textwidth]{\main/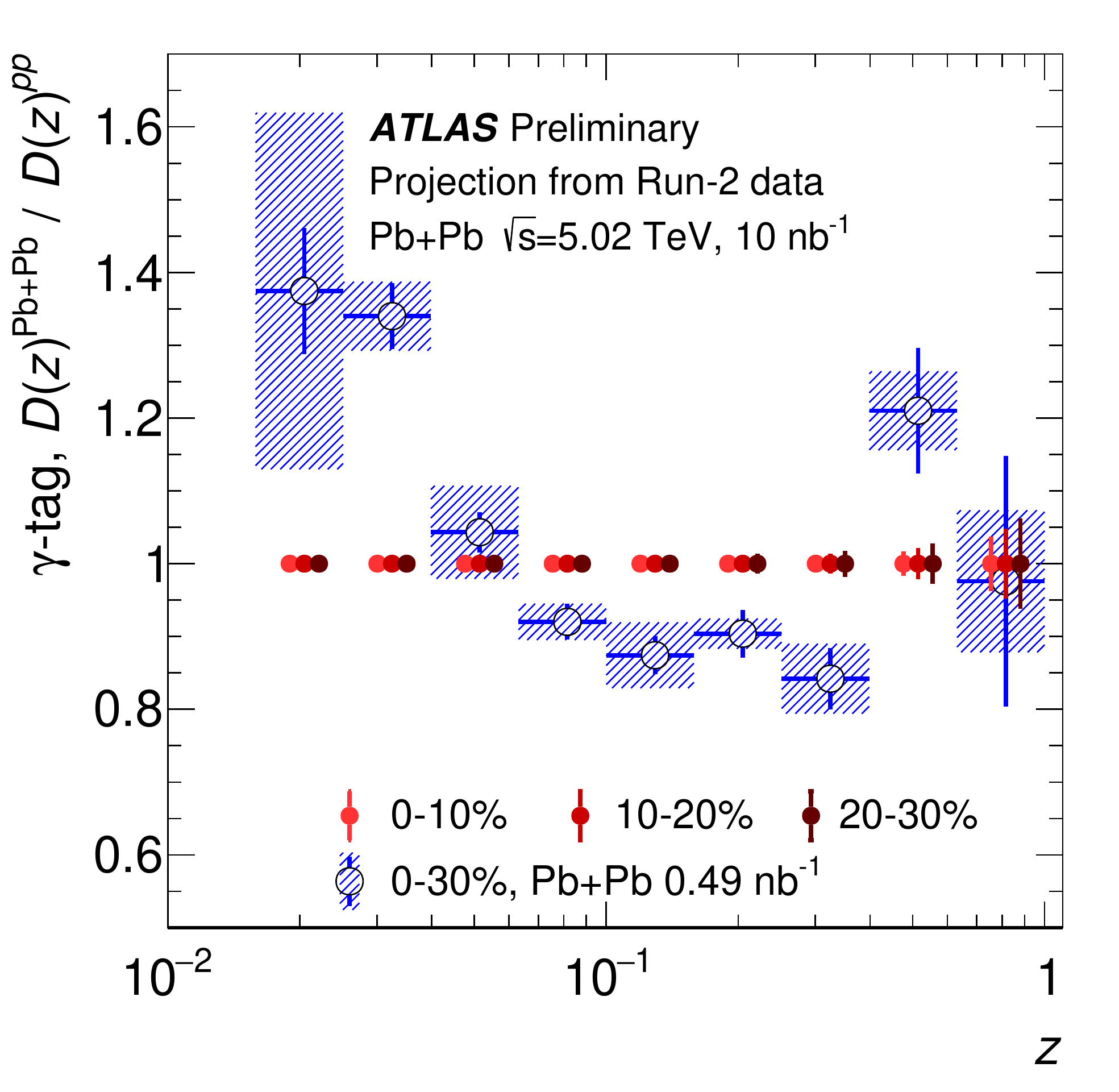}
\includegraphics[width=.45\textwidth]{\main/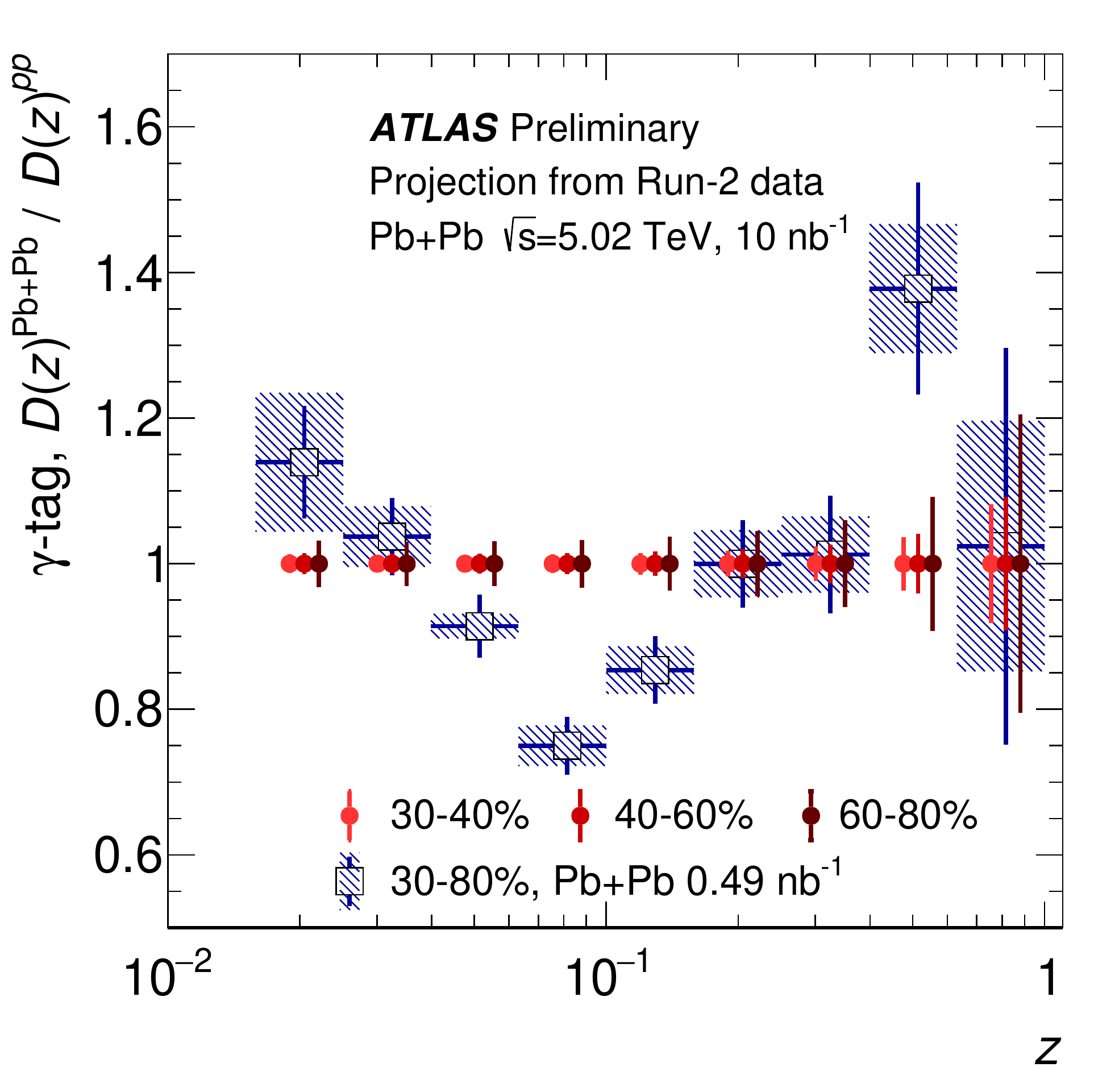}
\caption{Projection of the statistical precision that can be reached for the ratio of jet fragmentation functions in Pb--Pb and pp collisions, $R_{D(z)}$, of jets recoiling from a photon. The left panel shows the projection for the most central collisions while the right panel for the more peripheral events~\cite{ATL-PHYS-PUB-2018-019}.}
\label{fig:jetDzPh}
\end{center}
\end{figure}

When interpreting the modification of inclusive jets one has to realize that by requiring a certain jet momentum range a different sample of partons initiating the jet is selected in pp and Pb--Pb collisions. Incorporation of this effect in model calculations introduces an additional uncertainty limiting the constraints that can be put on a model. This can be overcome by using the rare process of jets recoiling from photons. The expected performance of the radial \pT\ profile in jets recoiling from a high momentum photon at HL-LHC is shown in Fig.\ref{fig:jetshape}. The central values of the extrapolated spectra are obtained by smoothing the results from~\cite{Sirunyan:2018ncy} by a third order polynomial. The systematic uncertainties shown are obtained by reducing by a factor of two those from the 2015 Pb--Pb data results, considering the expected improvements on the jet energy scale and jet energy resolution uncertainties. The results show that the photon-tagged jet shape could be measured with high precision providing insights about the modification of the jet transverse structure of quark initiated jets in the strongly interacting medium.
Figure~\ref{fig:jetDzPh} shows the expected statistical precision of the fragmentation function on photon-tagged events. The larger data sample will enable the measurement for finer centrality selections with respect to the current preliminary results~\cite{ATLAS-CONF-2017-074} allowing an exploration of the temperature and path length dependence of jet quenching. 

\subsubsection{Substructure with subjets}
\begin{figure}[!ht]
\begin{center}
\includegraphics[width=.8\textwidth]{\main/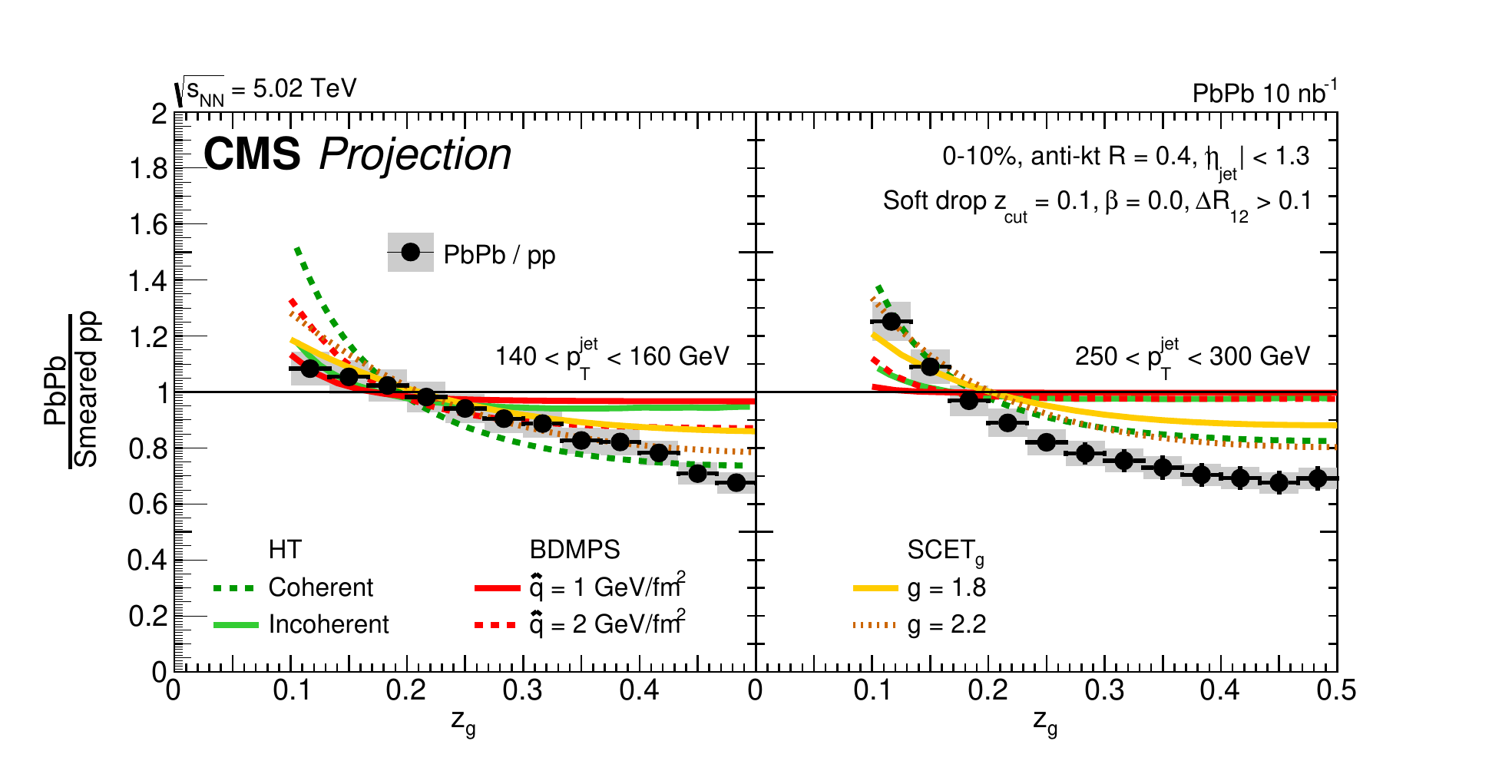}
\caption{Performance of jet splitting function measurement with HL-LHC data in Pb--Pb collisions for two different selections in jet transverse momentum.~\cite{CMS-PAS-FTR-17-002}}
\label{fig:ZG}
\end{center}
\end{figure}

\begin{figure}[!ht]
\begin{center}
\includegraphics[width=.8\textwidth]{\main/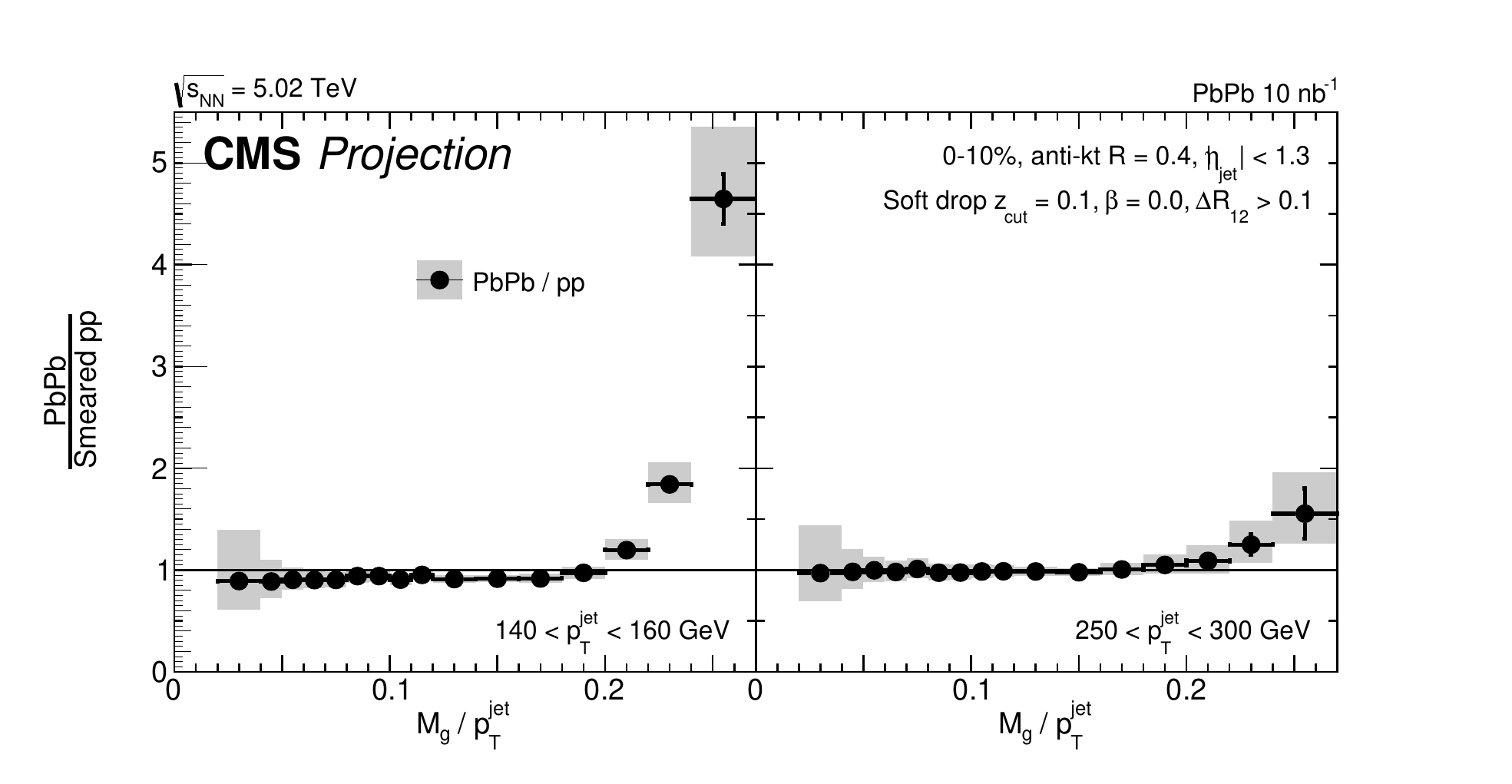}
\includegraphics[width=.8\textwidth]{\main/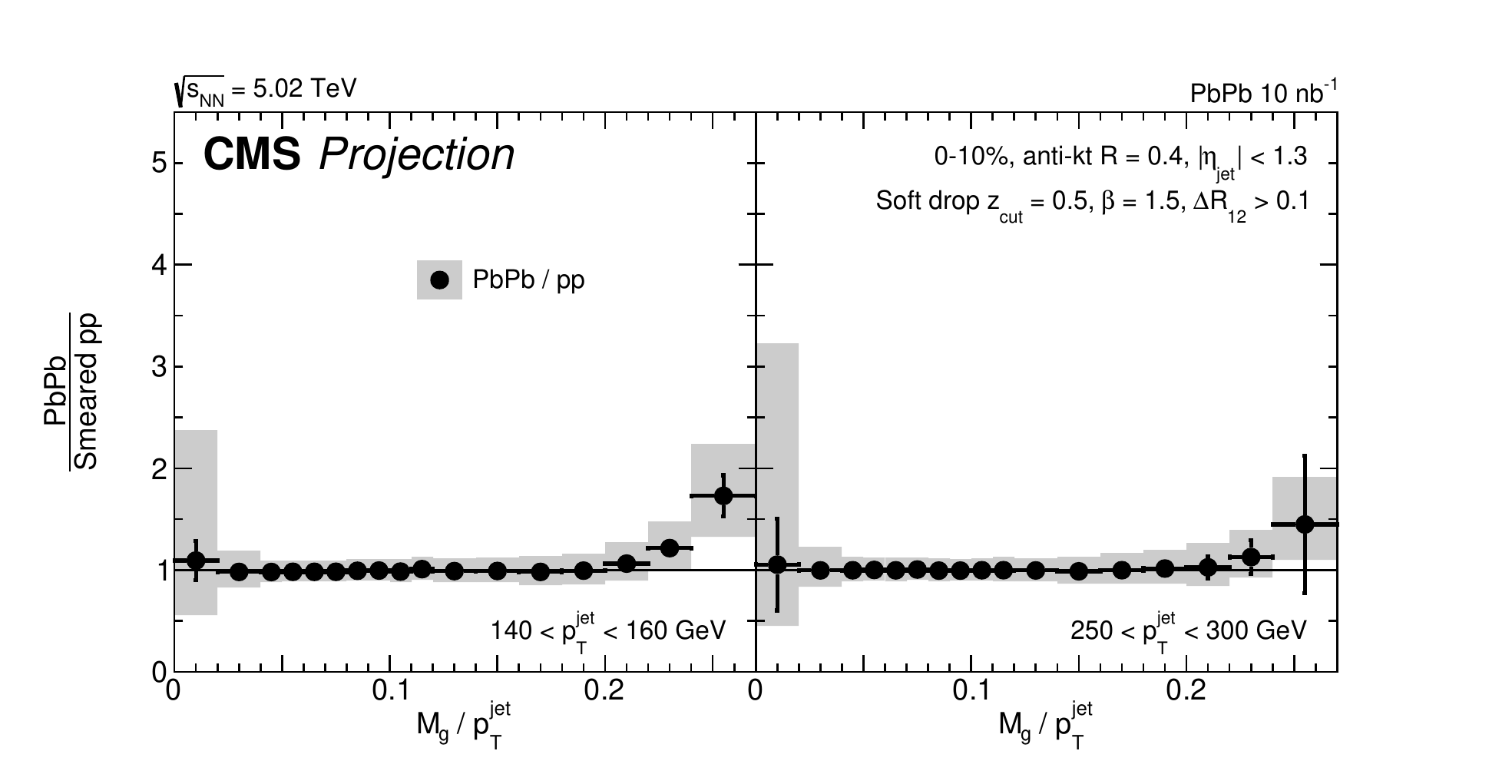}
\caption{Jet mass distribution with grooming setting $(z_{cut},\beta)=(0.1,0.0)$ (Upper panels) and $(z_{cut},\beta)=(0.5,1.5)$ (Lower panels).~\cite{CMS-PAS-FTR-17-002}}
\label{fig:Mass}
\end{center}
\end{figure}

Early hard splittings in a parton shower may result in two partons with high transverse momentum that are well separated in angle. 
Information about these leading partonic components can be obtained by removing the softer wide-angle radiation contributions. This is done through the use of a jet grooming algorithm called ``soft drop'', an extension of the modified mass drop tagger (mMDT), that attempt to split a single jet into two subjets, a process referred to as ``declustering''~\cite{Ellis:2009me,Butterworth:2008iy,Krohn:2009th,Dasgupta:2013ihk,Larkoski:2014wba}. For a parton shower in vacuum, these subjets provide access to the properties of the first splitting in the parton evolution~\cite{Altarelli:1977zs,Larkoski:2015lea}. Figure~\ref{fig:ZG} shows the expected performance for the momentum sharing fraction, $z_{\mathrm{g}}$~\cite{Larkoski:2015lea}, in the HL-LHC phase. The central values of $z_{\mathrm{g}}$ and jet mass are from previous CMS publications~\cite{Sirunyan:2017bsd,Sirunyan:2018gct}. The systematic uncertainties are reduced by a factor of two with respect to the results with 2015 Pb--Pb data due to the expected improvements on the jet energy scale and jet energy resolution uncertainties.  
While the current data is not precise enough to constrain the medium properties further, the expected luminosity at the HL-LHC will allow more detailed constraints as can be observed from the different results of the BDMPS~\cite{Mehtar-Tani:2016aco} and SCETg~\cite{Chien:2016led} calculations when the medium density ($\hat{q}$ for BDMPS and $g$ for SCETg) is varied. In addition, the expected precision will also provide the ability to distinguish different physical mechanisms and scales relevant for jet quenching as is shown for the role of coherence in Fig.~\ref{fig:ZG} in the HT theoretical calculations~\cite{Chang:2017gkt}. A measurement of the groomed jet mass with the 2015 LHC Pb--Pb data already showed that jet quenching might cause an increase of high mass jets~\cite{Sirunyan:2018gct}. Figure~\ref{fig:Mass} shows the expected performance for the groomed jet mass at HL-LHC which will allow measuring the high mass region with higher precision.

\newpage
\subsubsection{Radiation phase space with Lund diagram}

Recently, a theoretical representation of the radiation phase space within jets inspired by Lund diagrams~\cite{Andersson:1988gp} has been proposed~\cite{Andrews:2018jcm} to study medium modification of the radiation pattern. The so-called Lund jet plane~\cite{Dreyer:2018nbf} - a portrayal of the internal structure of jets - was designed to build a conceptual connection between manually constructed observables and approaches that use Machine Learning techniques to study QCD jets and/or discriminate between signal and background jets.
The diagram is constructed by mapping the available phase-space within a jet to a triangle in a two dimensional (logarithmic) plane that shows the transverse momentum and the angle of any given emission with respect to its emitter.
Such a triangular diagram, a representation of the radiation within any given jet, can be created through repeated Cambridge/Aachen declustering.

To demonstrate the potential of future measurements at the LHC we constructed Lund diagrams using the \jewel\ Monte Carlo event generator~\cite{Zapp:2013vla}.
To study the differences in the Lund diagram due to medium effects the results are compared to a vacuum reference (jets produced in \pp\ collisions). For the simulations the \jewel
 generator with the default settings is used without the optional calculation of the so-called medium response retaining the partons / scattering centres that interacted with the jet was not used (i.e. {\it recoils off} setting of the MC generator was used).
The substructure of jets was analysed by reclustering the constituents of the jet with the Cambridge/Aachen (C/A) algorithm as implemented in the \fastjet\ package~\cite{Cacciari:2011ma,Cacciari:2005hq} for two selections of jet \pt\ 80--120~$\gevc$ and 200--250~$\gevc$.

\begin{figure}[ht]
	\centering
	\includegraphics[width=0.45\textwidth,page=1]{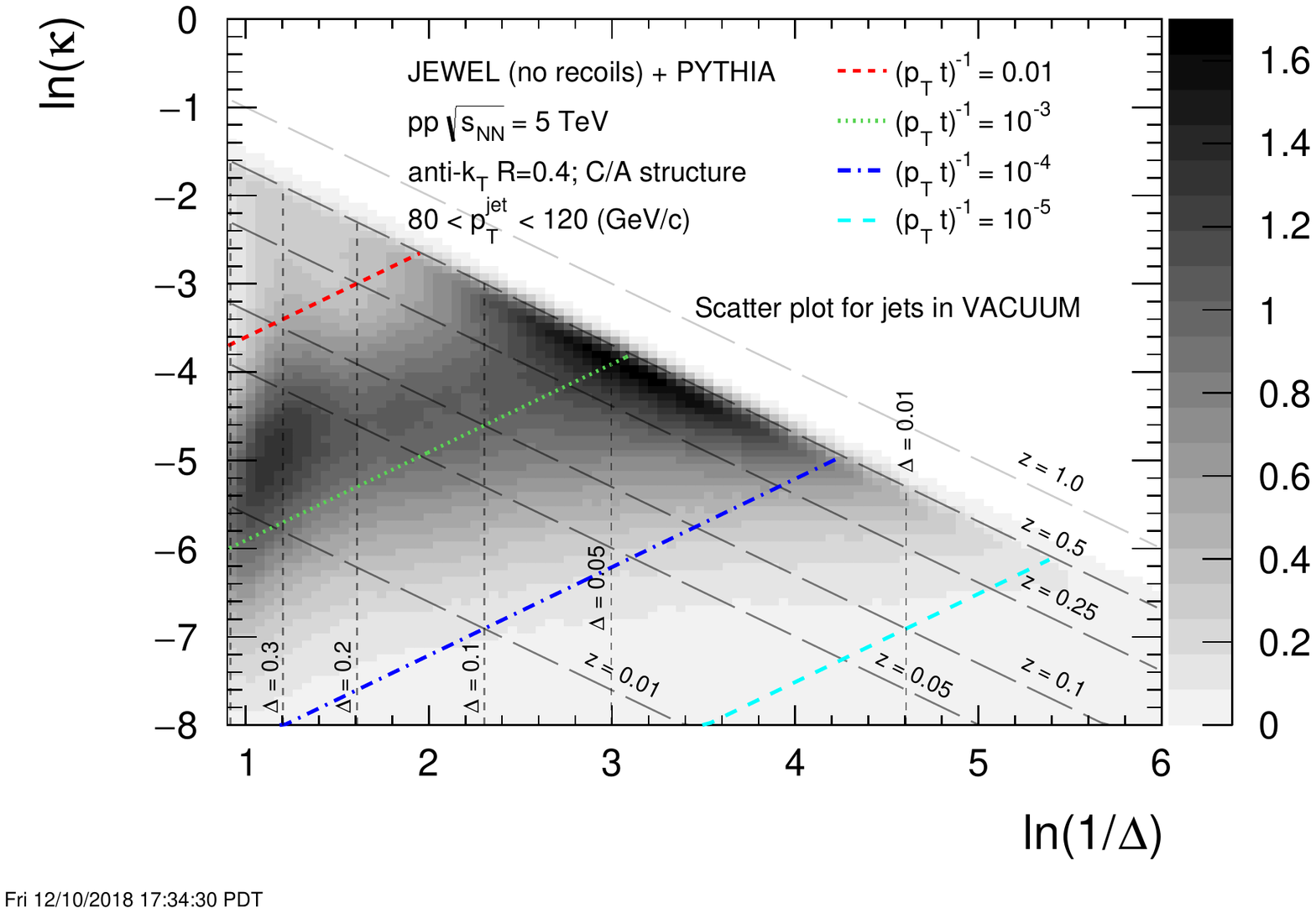}
	\includegraphics[width=0.45\textwidth,page=2]{\main/jets/figures/lund/lund_t}
	\includegraphics[width=0.45\textwidth,page=4]{\main/jets/figures/lund/lund_t}
	\includegraphics[width=0.45\textwidth,page=5]{\main/jets/figures/lund/lund_t}
	\caption{The density of points of a Lund diagram for anti-\kT\ $R=0.4$ jets for two \pt\ selections: $80 < \pt\ < 120$ \gevc\ in the upper row and $200 < \pt\ < 250$ \gevc\ in the lower row. Result of the \jewel\ Monte Carlo generator with left column: jets in \pp\ collisions; Right column: jets from Pb--Pb collisions - some with in-medium modifications. Each of the density plots shows curves of the average quantities of the densities over the other axis.}
	\label{fig:Lund_jets}
\end{figure}

The Lund diagram density can be constructed experimentally and compared to analytic predictions and parton-shower Monte-Carlo simulations, such as \jewel. For this purpose a density map of points (emissions) is defined following formulations in~\cite{Dreyer:2018nbf}: 
\begin{equation}
\bar{\rho}(\Delta, \kappa) = \frac{1}{N_{\rm jet}} \frac{\mathrm{d}n_{\rm emission}}{\mathrm{d} \ln \kappa~\mathrm{d} \ln 1/\Delta},
\end{equation}
where for two clusters $1$ and $2$ labeled such that $p_{\rm T,1} > p_{\rm T,2}$, $\Delta^{2} = (y_1 - y_2)^2 + (\varphi_1 - \varphi_2)^2$ with $\varphi$ being the azimuthal angle and $y$ the rapidity of a cluster, and $\kappa=\frac{p_{T,2}}{p_{\rm T,1} + p_{\rm T,2}}\Delta$. Figure~\ref{fig:Lund_jets} shows the density $\bar{\rho}$ from the \jewel\ simulation without (left panels) and with (right panels) medium effects.
The $z_{g}$ variable which was defined in~\cite{Larkoski:2015lea} and studied in heavy-ion collisions~\cite{Sirunyan:2017bsd} is related to the variables in the Lund plane in the following way: $z_{\mathrm{g}} = \kappa/\Delta$ from the first of the entries ($1...i...N$) in the primary declustering sequence that satisfies $z^{(i)} \geq z_{\mathrm{cut}}(\Delta^{(i)})^{\beta}$ \cite{Dreyer:2018nbf} resulting in diagonal lines with negative slope in the Lund diagram for a constant value of $z_\mathrm{g}$.
\begin{figure}[htbp]
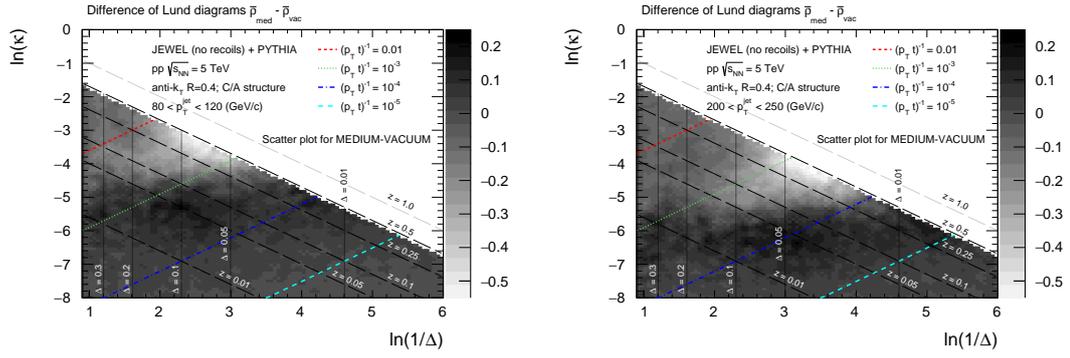

	\centering
	\includegraphics[width=0.45\textwidth,page=3]{\main/jets/figures/lund/yellow_jet_lund}
	\includegraphics[width=0.45\textwidth,page=6]{\main/jets/figures/lund/yellow_jet_lund}
	\caption{Result of the JEWEL+PYTHIA MC simulation: MEDIUM-VACUUM difference of the calculations shown in Fig.~\ref{fig:Lund_jets} for two jet \pt\ selections.}
	\label{fig:Lund_jets_vac_med}
\end{figure}

The effect of jet quenching on the Lund diagram is quantified by taking the difference between the diagram with and without medium effects as shown in Fig.~\ref{fig:Lund_jets_vac_med} for the two transverse momentum ranges considered in this study. 
The average density integrated over $\ln \kappa$ calculated for Pb--Pb (MEDIUM) case shows little deviation from the pp (VACUUM) reference. 
The most pronounced differences between VACUUM and MEDIUM calculations are visible for the region of $-3 < \ln \kappa < -3$ and large $\ln 1\Delta$ which correspond to the hard-collinear splittings (Region-A), and a band along $\ln 1/\Delta$ for small $\ln \kappa$ (Region-B): $-5 < \ln \kappa < -6$ for the lower \pt\ selection and $-5.5 < \ln \kappa < -7$ for higher \pt\ jets; that corresponds to an enhancement of soft (moderate $\ln 1/\Delta$) and hard collinear splittings (large $\ln 1/\Delta$). 
These observations are consistent with soft and hard collinear splittings being modified by the medium.

\begin{figure}[htbp]
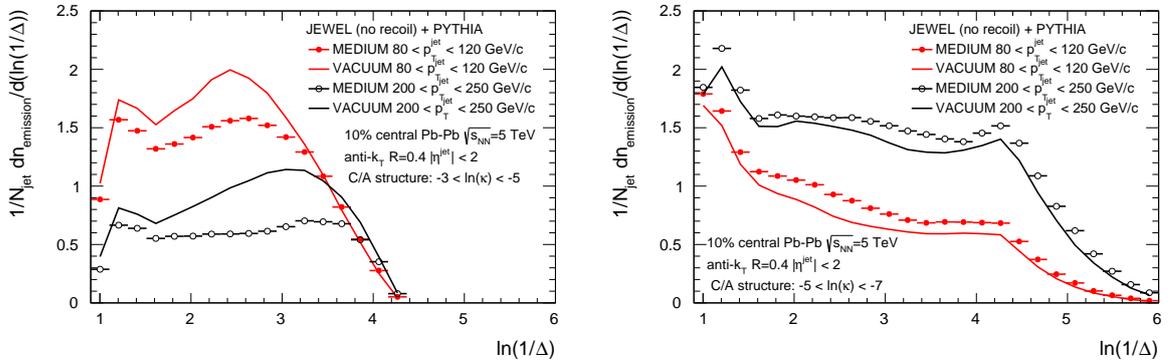

	\centering
	\includegraphics[width=0.49\textwidth,page=7]{\main/jets/figures/lund/yellow_jet_lund}
	\includegraphics[width=0.49\textwidth,page=8]{\main/jets/figures/lund/yellow_jet_lund}
	\caption{Projections of the lund diagram along the angular separation $\ln 1/\Delta$ of the splittings for the two selections of jet \pT. In-medium suppression of splittings for moderate $\ln{\kappa}$ according to \jewel\ (left). Enhancement for small $\ln{\kappa}$ (right).}
	\label{fig:Lund_projections}
\end{figure}

To illustrate the different modifications of the Lund diagram density for the two regions identified in Fig.~\ref{fig:Lund_jets_vac_med}, projections along $\ln 1/\Delta$ are shown in Fig.~\ref{fig:Lund_projections}. For Region-A we observe 30\%-40\% depletion of splittings for the MEDIUM case whereas in Region-B a moderate increase of splittings induced by the medium is visible. The depletion in Region-A is consistent a sample of more collimated jets consistent with previous measurements in heavy-ion collisions~\cite{Acharya:2018uvf,Sirunyan:2018jqr}. The increase seen in Region-B is consistent with a small in-medium enhancement of splittings with moderate dependency on the angle of the splitting but favoring the soft collinear medium-induced radiation (moderate $ln 1/\Delta$).

\begin{figure}[ht]
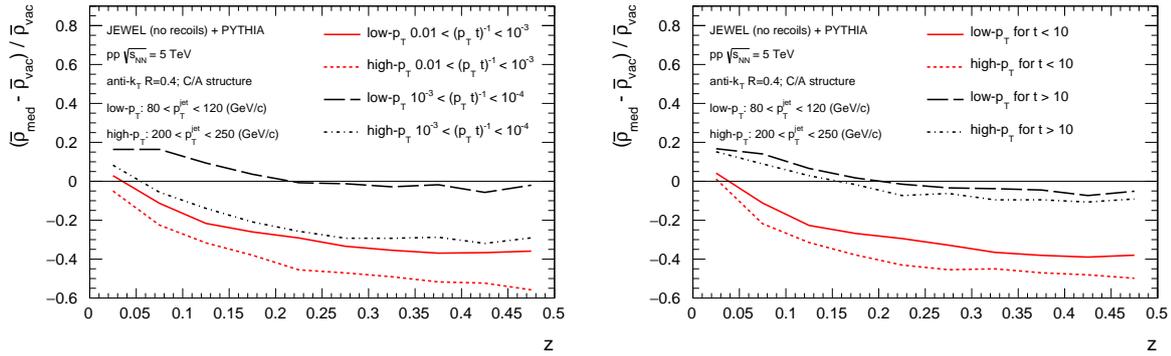

	\centering
	\includegraphics[width=0.49\textwidth,page=11]{\main/jets/figures/lund/yellow_jet_lund}
	\includegraphics[width=0.49\textwidth,page=12]{\main/jets/figures/lund/yellow_jet_lund}
	\caption{Projections of the relative difference of the Lund diagram onto momentum imbalance of the splittings for two selections of jet \pT.
	Left: selection of splittings for different $\frac{1}{p_{\rm T} t}$.
	Right: selection of splittings for different $t$.
	}
	\label{fig:Lund_projections_z}
\end{figure}

As discussed in Ref.~\cite{Andrews:2018jcm} specific regions in the Lund plane are sensitive to different type of parton splittings.
These regions can be identified by selecting the desired area using linear functions $\ln\kappa = \ln1/\Delta + \ln \frac{1}{p_{\rm T} t}$, where $t$ is related to the decoherence time (thus formation time).
Depending on the selection, different formation times are probed and splittings will occur within or outside the medium.
Several arbitrary regions selected by the diagonal lines for constant $\frac{1}{p_{\rm T} t}$ are indicated in Fig.~\ref{fig:Lund_jets}. 
To illustrate the in-medium effects and their dependence on the jet momentum, the $p_{\rm T}$ and the formation (decoherence) time the density of the splittings can be projected along the momentum imbalance $z = p_{\rm T,2}/(p_{\rm T,1} + p_{\rm T,2})$.
In the left panel of Fig.~\ref{fig:Lund_projections_z} we show the relative difference of the splitting density $\Delta_{\rm}\bar{\rho} = (\bar{\rho}_{\mathrm med} - \bar{\rho}_{\mathrm vac})/\bar{\rho}_{\mathrm vac}$ for a selection of $p_{\rm T} t$.
For small $p_{\rm T} t$ the splitting density is suppressed for the in-medium calculations whereas for large $p_{\rm T} t$ the modification is smaller. 
This is consistent with the expectation that for large formation times the medium effects should be of smaller magnitude as compared to splittings formed early.
As the suppression seen in $\Delta_{\rm}\bar{\rho}$ selected on $\pT t$ depends on the jet \pT\ we find similar suppression for high- and low-\pT\ jet selection for substantially different product of the \pT\ and formation time.
In particular, for low momentum jets the modifications of $\Delta_{\rm}\bar{\rho}$ for large $p_{T} t$ is small.
To further exploit the formula providing the approximate formation time dependence of the splittings we select two regions of the Lund diagram: ``late'' $t > 10$ and ``early'' $t < 10$.
The ``early'' region should be dominated by splittings that form within the medium, whereas the ``late'' splittings are to be dominated by the shower evolution outside the medium of length $L \sim t$.
In the right panel of Fig.~\ref{fig:Lund_projections_z} we present $\Delta_{\rm}\bar{\rho}$ for two selections of decoherence time. 
As expected, a small dependence on jet \pT\ for ``late'' and ``early'' splittings is seen - a similar suppression for ``early'' splittings independent of jet \pT\ and almost identical $\Delta_{\rm}\bar{\rho}$, with small deviations from unity, for ``late'' emmissions. 
The residual differences could be attributed to different fractions of the splittings resolved by the medium, and likely, different impact of non-perturbative effects (such as hadronization) for the two jet \pT\ selections.

\subsection{Opportunities for jet quenching studies  with light-ion collisions}
\label{sec:jetslightions}

The ability of the LHC to collide ions lighter than Pb as discussed in section \ref{sec:lightions} provides an opportunity to enhance the heavy-ion programme with a very large number of rare probes as summarized in Sect.~\ref{sec:smallAsum}.  \ArAr\ collisions are used as a test case for light ion running, although the optimal choice of ion is still under study. It is clear that due to the larger integrated luminosity obtainable for a given heavy-ion running period the number of jets produced in \ArAr\ collisions will be significantly larger than in \PbPb\ collisions.  The question which will determine the value of the of light-ion running to the study of jets and parton energy loss is to what extent jet suppression effects will be reduced in smaller systems. To address this question, projections for \ArAr\ and \XeXe\ collisions at the LHC are considered using the \jewel\ Monte Carlo event generator~\cite{Zapp:2013vla}.

The jet nuclear modification factor \RAA\ (discussed in Sect.~\ref{sec:preceloss}) in \ArAr\ collisions is here computed as the ratio of the jet transverse momentum spectrum in medium (\ArAr) over that in vacuum (pp)
\begin{equation}
    R_{\rm AA}^{\rm jet} = \frac{ (\rmd N_{\rm jets}/\rmd p_{\rm T})^{\rm med}}{(\rmd N_{\rm jets}/\rmd p_{\rm T})^{\rm vac}} \, .
\end{equation}
Table~\ref{tab:medium_pars} summarizes the parameters used in \jewel\ to calculate this quantity in 0--10\% centrality \ArAr collisions, compared with the parameterization for \PbPb and \XeXe. The $T_i^{v1}$ values were obtained assuming the same pre-factors as in \cite{Dainese:2016gch}:
\begin{equation}
    T (t) = \left( \epsilon(t) \frac{30}{47.5 \, \pi^2} \right)^{1/4} \, ,
\end{equation}
where the energy density $\epsilon(t)$ follows a Bjorken evolution:
\begin{equation}
    \epsilon(t) = \frac{1}{\pi R_{nucl}^2 t} \frac{\rmd E}{\rmd\eta} \, .
\end{equation}
The energy per unit of pseudo-rapidity is taken from centrality-dependent measurements in Pb--Pb collisions. Finally the temperature $T_i^{v1}$ is evaluated at $\tau_i$.

\begin{table}[!ht]
    \caption{Energy and medium parameter used in \jewel\ simulation of dijets and Z+jet events.}
    \centering
    \begin{tabular}{|c|c|c|c|}
        \hline
         &  \PbPb & \XeXe & \ArAr \\
        \hline
        $\sqrtsNN$ (TeV) & 5.02 & 5.80 & 6.30 \\
         $\left\langle N_{\rm part} \right\rangle$ & 353 & 210 & 66\\
         $R_{\rm nucl}$ (fm) & 6.6 & 5.4 & 3.6\\
         $\tau_i$ (fm/$c$) & 0.6 & 0.57 & 0.63 \\
         $T_i^{v1}$ (MeV) & 360 & 350 & 318 \\
         $T_i^{v2}$ (MeV) & 260 & 250 & 218 \\
         \hline
    \end{tabular}
    \label{tab:medium_pars}
\end{table}



\begin{figure}[!t]
\centering
\includegraphics[width=.45\textwidth,page=2]{\main/jets/figures/jewel/jewel_params_v3.pdf}
\includegraphics[width=.45\textwidth,page=1]{\main/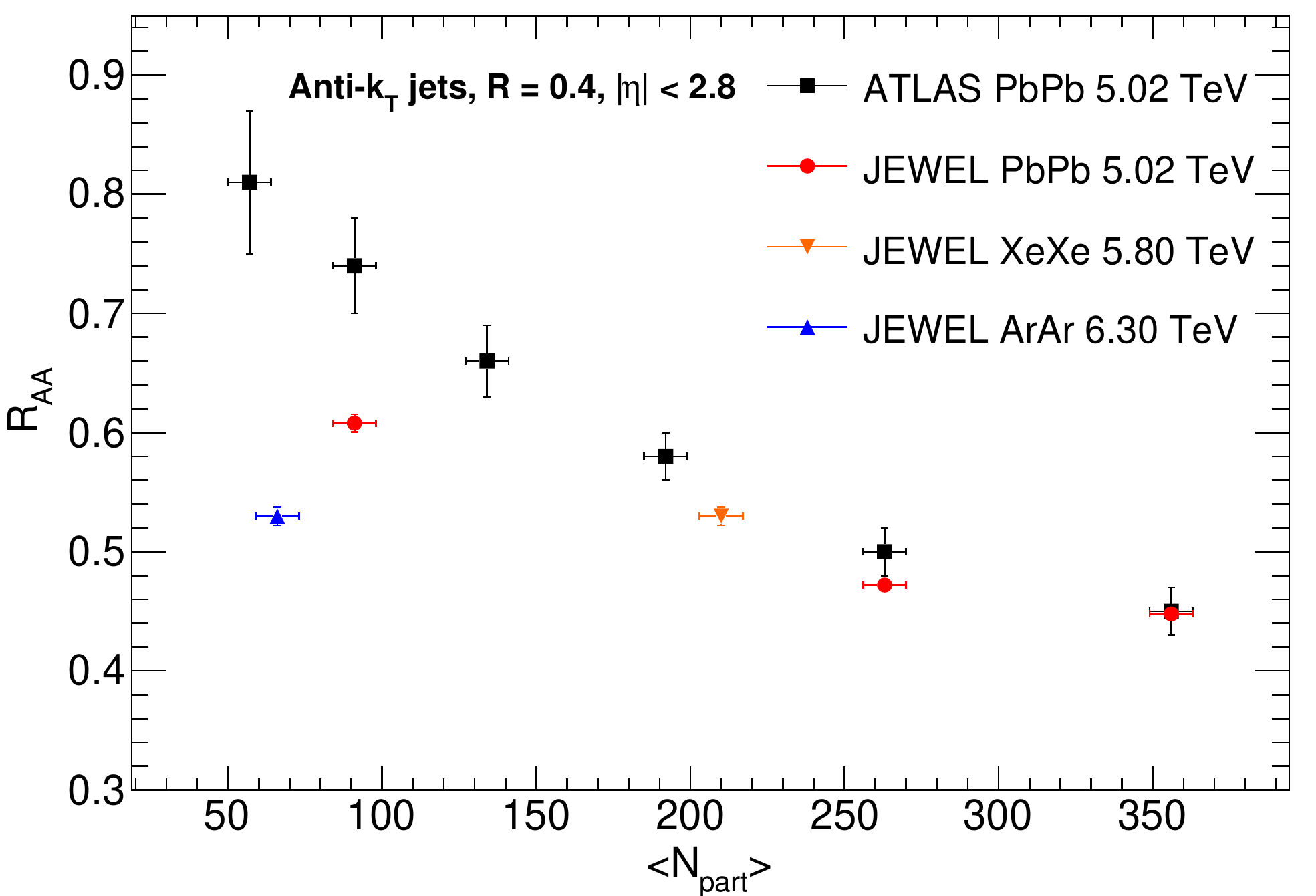}
\caption{Jet \RAA\ obtained from a \jewel\ simulations using the medium parameters listed in Table \ref{tab:medium_pars}, with the temperatures listed as $T_{i}^{v2}$. On the left, the jet \RAA\ is shown as a function of the jet $\pt$, and on the right as a function of $\left\langle N_{\rm part} \right\rangle$ for a jet $100<\pt <126$~GeV/$c$.}
\label{fig:jewel_raa_v3}
\end{figure}

With these medium parameters, \jewel\ results lie quite below the ATLAS \RAA\ results for the \PbPb\ 0--10\% centrality class \cite{Aaboud:2018twu}. \jewel was run with medium recoil effects off, although they are known to contribute to increase the jet \RAA\ by $\sim 0.1$--0.2 in the most central events \cite{KunnawalkamElayavalli:2017hxo}.  This may explain the discrepancy. Alternatively, the discrepancy can be eliminated by reducing  the temperature in \jewel.  Starting from the \PbPb\ temperature changed in order to match the ATLAS results for the \PbPb\ 0--10\% centrality class, temperatures for \XeXe\ and \ArAr\ are obtained by assuming that the energy density scales with $A^{1/3}$. Thus, for an arbitrary collision system $XX$ one has:
\begin{equation}
    \left( \frac{T_{XX}}{T_{\rm Pb-Pb}} \right )^{4} = \left( \frac{A_{XX}}{A_{\rm Pb}} \right)^{1/3} \, .
\end{equation}
This parameterisation is used to obtain the $T_i^{v2}$ values listed in Table \ref{tab:medium_pars} which are used to calculate the jet \RAA\ shown in Figure \ref{fig:jewel_raa_v3}.  The figure shows the \jewel calculations for \PbPb, \XeXe, and \ArAr\ along with ATLAS \PbPb\ measurements in centrality classes chosen to match the $\left\langle N_{\rm part} \right\rangle$ values in Table \ref{tab:medium_pars}.

\begin{figure}[!ht]
\centering
\includegraphics[width=.45\textwidth,page=1]{\main/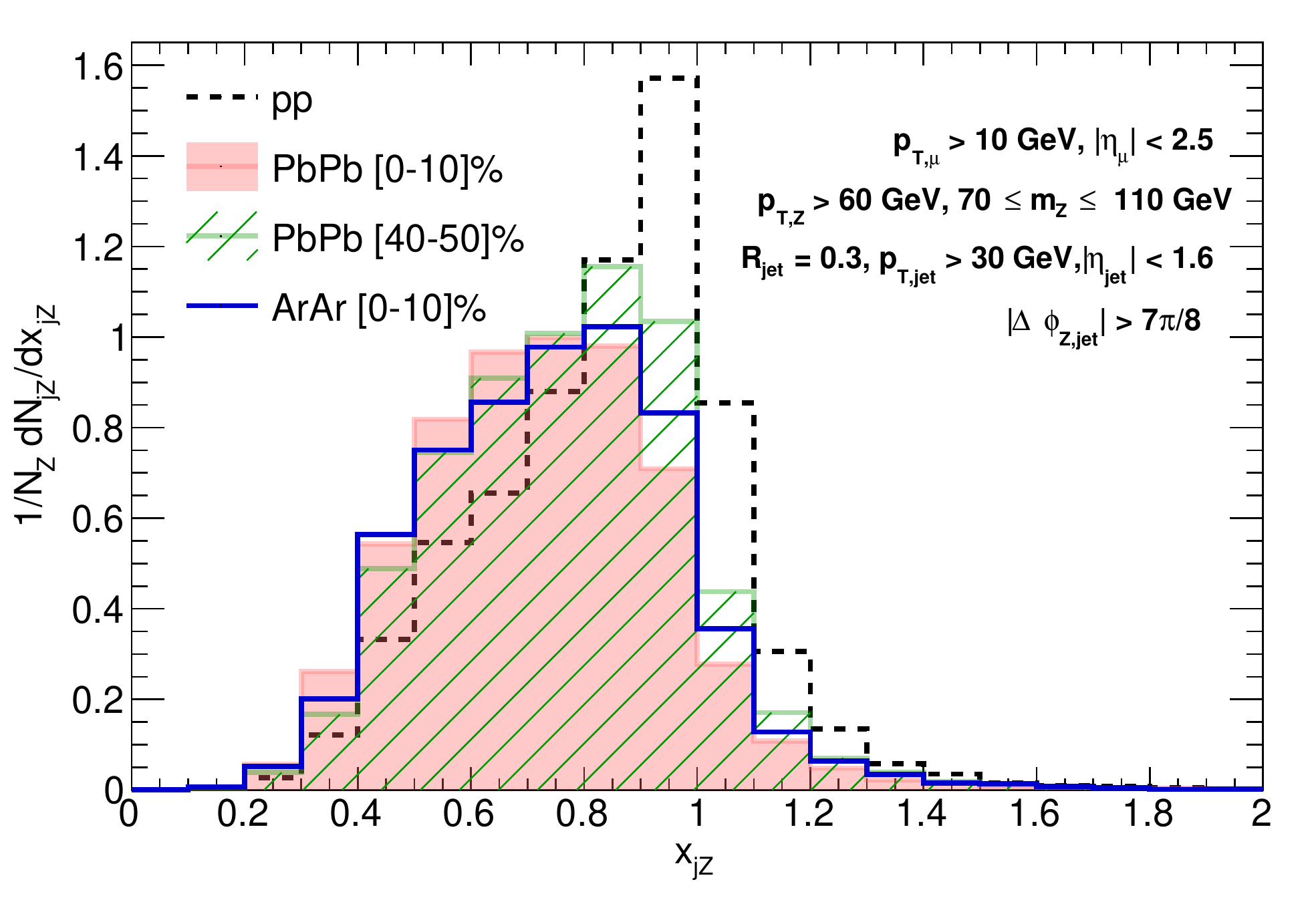}
\caption{Boson-Jet energy asymmetry, $x_{\rm jZ}$ obtained from \jewel\ simulations using the medium parameters listed in table \ref{tab:medium_pars}, with the temperatures listed as $T_{i}^{v2}$.}
\label{fig:jewel_xjz}
\end{figure}

To further investigate jet energy loss in \ArAr\ collisions, Z boson + jet events are studied within the same \jewel framework.
The importance of boson + jet events for the precision study of energy loss is discussed in Sect.~\ref{sec:preceloss}.  
Events with a Z boson decaying into $\mu^+ \mu^-$ associated with a jet were simulated with \jewel + \pythia Monte Carlo \cite{KunnawalkamElayavalli:2016ttl}. Events were selected for a reconstructed Z boson with a mass within 70--110~GeV, a minimum \pt of 10~GeV/$c$ for its decay muons, and an associated jet with a $\pt > 30$~GeV/$c$ and a $|\delta \varphi| > 7 \pi/8$ with respect to the boson momentum  direction. The resulting energy asymmetry distributions, $x_{\mathrm{jZ}}=p_\mathrm{T}^{\textrm{jet}}/p_\mathrm{T}^{\rm Z}$, normalized to the number of reconstructed Z bosons are shown in Figure \ref{fig:jewel_xjz}.
The distribution for central Ar--Ar collisions is similar to those for central and semi-central Pb--Pb collisions. The effect of jet quenching is very significant, as apparent in the comparison with the distribution in pp collisions.

Taken together, these studies suggest that \jewel does somewhat over-predict suppression in smaller systems, e.g.\,\XeXe. However, considering that central collisions are reproduced well by this model and that allowing as an upper limit the suppression measured in the \PbPb\ 40--50\% centrality class, a significant suppression is expected in central \ArAr collisions.  
The expected suppression combined with the much larger nucleon--nucleon integrated luminosity (e.g.\,8--25 larger with \ArAr\ than with \PbPb\ collisions) makes lighter ion collisions at the LHC an attractive possibility for the study of parton energy loss.


\clearpage

\section{Quarkonia}

\label{sec:quarkonia}

{ \small
\noindent \textbf{Coordinators}: Anton Andronic (M\"{u}nster University) and Emilien Chapon (CERN)

\noindent \textbf{Contributors}: 
E.~G.~Ferreiro (Instituto Galego de Fisica de Altas Enerxias (IGFAE) Universidade de Santiago de Compostela), 
J.-P.~Lansberg (Institut de Physique Nucl\'{e}aire d'Orsay), 
R.~Rapp (Texas A\&M University, College Station),
J.~Castillo Castellanos (IRFU/DPhN, CEA Saclay),
C.~Cheshkov (IPN Lyon),
J.~Martin Blanco (Laboratoire Leprince Ringuet),
J.~Park (Korea University), 
X.~Du (Texas A\&M University, College Station),
M.~Strickland (Kent State University), 
R.~Venugopalan (BNL), 
I.~Vitev (Los Alamos National Laboratory)
}

\subsection{Introduction} %
\label{sec_intro}
A key objective in high-energy heavy-ion physics is to determine the in-medium forces that give rise to the remarkable many-body features of the QGP.
In the QCD vacuum, the unravelling of the fundamental force between two static Color charges was made possible by the discovery of the charmonium and bottomonium states in the 1970's. 
Subsequent quantitative analyses of the bound-state spectra established a phenomenological potential of the Cornell type~\cite{Eichten:1979ms}, 
\begin{equation}
V(r) = -\frac{4}{3} \frac{\alpha_s}{r} + \sigma r \ ,
\label{V}
\end{equation} 
with a colour-Coulomb term due to gluon exchange dominant at short distances, and a linear term with string tension $\sigma\simeq0.9$\,GeV/fm to account for confinement at large distance. 
This potential has also been quantitatively confirmed by lattice-QCD (lQCD) calculations~\cite{Bali:2000gf,Brambilla:2004jw}. 
The corresponding effective field theory of QCD, potential non-relativistic QCD (pNRQCD), allows for the definition of the static potential~\cite{Schroder:1998vy} in a  $1/m_Q$ expansion for large heavy-quark mass, $m_Q$~\cite{Brambilla:1999xf,Brambilla:2004wf}. 
The heavy-quark (HQ) potential thus provides a well calibrated starting point to probe the QCD medium, and the in-medium spectroscopy of quarkonia is the natural tool to carry this out in heavy-ion collisions, cf.~\cite{Rapp:2008tf,BraunMunzinger:2009ih,Kluberg:2009wc,Mocsy:2013syh,Liu:2015izf} 
for recent reviews.
The string term in the HQ potential, eq.~(\ref{V}), characterises the long-range nonperturbative part of the force and is associated with the confining property of QCD. It is expected to play a critical role in the transition from hadronic to partonic degrees of freedom, and may be responsible for the remarkable transport properties of the QGP, {\it i.e.}, its strongly coupled nature, up to temperatures of 2-3 times the (pseudo-)critical temperature, \Tc~\cite{Liu:2016ysz}. 

Much like in vacuum, a systematic investigation of the in-medium force must involve the {\em spectroscopy} of different states, as they subsequently dissolve with increasing temperature. 
The complexity in describing the in-medium properties of quarkonia and their
implementation into transport calculations in heavy-ion collisions prevents their 
use as a straightforward thermometer of the medium produced in these reactions. 
On the contrary, using information on the space-time and temperature evolution in
heavy-ion collisions from other sources (\eg, hydrodynamics and electromagnetic 
radiation), on can utilize quarkonium observables to deduce their in-medium properties
and infer the fundamental interactions in QCD matter. 
In the vacuum, only the 1S ground-state bottomonia (\PGUP{1S} and \PGhb) are small enough in size to be mostly bound by the colour-Coulomb force. All excited bottomonia and all charmonia are predominantly bound by the nonperturbative string term (and/or residual mesonic forces). 
Thus, charmonia and excited bottomonia are excellent probes of the in-medium confining force, as originally envisioned for the \PJgy~\cite{Matsui:1986dk}.
However, in the cooling of the expanding fireball, quarkonia can also be ``(re)generated'' through recombination of individual heavy quarks and anti-quarks diffusing through the medium. It is important to emphasise that quarkonium formation occurs also from quarks and antiquarks from different initial pairs.
This mechanism \cite{BraunMunzinger:2000px,Thews:2000rj,Young:2008he} has turned out to be critical in understanding the rise of \PJgy production from RHIC to the LHC where (re)generation seems to constitute the major part of the yield observed in central \PbPb collisions \cite{Rapp:2017chc}.
The data is also compatible with production of \PJgy exclusively through statistical hadronisation at the crossover phase boundary \cite{Andronic:2017pug}.
Precise measurements of the $c\bar{c}$ production cross section and the extraction of the charm-quark diffusion coefficient in \RunsThreeFour will be important for making a more definite statement; these are key objectives discussed in the chapter~\ref{sec:HI_HF} on open heavy-flavor production. Information from \pt spectra and elliptic flow will help to complete the picture.

\begin{figure}[!h]
\begin{center}
\includegraphics[width=0.6\columnwidth]{\main/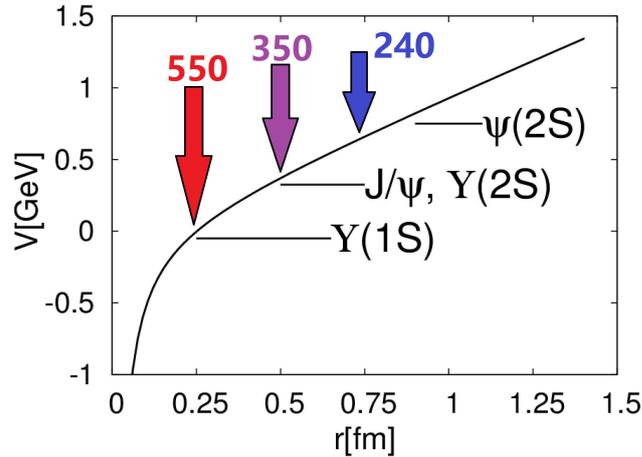}
\end{center}
\vspace{-0.5cm}
\caption{The vacuum heavy-quark potential as a function of $Q\bar Q$ separation. The horizontal lines indicate the approximate locations of the vacuum bound states while the vertical arrows indicate the minimal screening distances of the media produced at the SPS, RHIC and LHC, as deduced from approximate initial temperatures reached in these collisions extracted from data (indicated above the arrows in MeV) and from quarkonium production systematics in \PbPb and Au-Au collisions. Figure taken from Ref.~\cite{Rapp:2017chc}.}
\label{FigQ:pot}
\end{figure}

Regarding bottomonia, the current understanding suggests that (re)generation is less important for \PGUP{1S}, but possibly figures as a major component in the strongly suppressed yield of excited states~\cite{Du:2017qkv,Krouppa:2017jlg}. It is therefore of great importance to obtain additional information about the typical time at which quarkonia are produced, in particular through $\pt$ spectra and elliptic flow which contain information about the fireball collectivity imprinted on the quarkonia by the time of their decoupling. A schematic illustration of the current knowledge extracted from in-medium quarkonium spectroscopy, {\it i.e.}, their production systematics in heavy-ion collisions is shown in Fig.~\ref{FigQ:pot}.        

On the theoretical side, the basic objects are the quarkonium spectral functions which encode the information on the quarkonium binding energies, in-medium HQ masses and the (inelastic) reaction rates.  Ample constraints on the determination of the quarkonium spectral functions are available from thermal lQCD, {\it e.g.}, in terms of the heavy-quark free energy, euclidean and spatial quarkonium correlation functions, and HQ susceptibilities, and are being implemented into potential model calculations~\cite{Wong:2004zr,Mocsy:2005qw,Alberico:2006vw,Brambilla:2008cx,Riek:2010py,Burnier:2015tda,Liu:2017qah}.
In particular, the role of dissociation reactions has received increasing attention. Early calculations of gluo-dissociation~\cite{Bhanot:1979vb,Kharzeev:1994pz} or inelastic parton scattering~\cite{Grandchamp:2001pf} have been revisited and reformulated, \eg as a singlet-to-octet transition mechanism~\cite{Brambilla:2008cx} or in terms of an imaginary part of a two-body potential~\cite{Laine:2006ns}, respectively. In particular, the latter accounts for interference effects which reduce the rate relative to ``quasi-free'' dissociation~\cite{Grandchamp:2001pf} in the limit of small binding; interference effects can also be calculated diagrammatically~\cite{Park:2007zza}; they ensure that, in the limit of vanishing size, a Color-neutral $Q\bar Q$ dipole becomes ``invisible'' to the Color charges in the QGP.

The information from the spectral functions can then be utilised in heavy-ion phenomenology via transport models. The latter provide the connection between first-principles information from lQCD and experiment that greatly benefits the extraction of robust information on the in-medium QCD force and its emergent transport properties, most notably the (chemical) equilibration rates of quarkonia. 
Thus far most transport models are based on rate equations and/or semiclassical Boltzmann equations. In recent years quantum transport approaches have been developed using, \eg a Schr\"odinger-Langevin~\cite{Blaizot:2015hya,Katz:2015qja,Kajimoto:2017rel,Petreczky:2016etz} or density-matrix~\cite{Akamatsu:2014qsa,Brambilla:2016wgg} formulation. These will enable to test the classical approximation underlying the Boltzmann and rate equation treatments and ultimately quantify the corrections.  Quantum effects may be particularly relevant at high $\pt$ in connection with the in-medium formation times of quarkonia, augmented by the Lorentz time dilation in the moving frame; schematic treatments of this effect in semiclassical approaches suggest that varying formation times can leave observable differences for high-momentum charmonia and bottomonia~\cite{Song:2015bja,Hoelck:2016tqf,Du:2017qkv,Aronson:2017ymv,Krouppa:2017jlg}. 
Finally, the implementation of phase-space distributions of explicitly diffusing heavy quarks into quarkonium transport is being investigated by various groups (see, \eg Ref.~\cite{Yao:2017fuc}), which, as mentioned above, will provide valuable constraints on the magnitude and \pt dependence of (re)generation processes.  
In particular, the role of non-perturbative effects in the HQ interactions in the QGP 
  (which are believed to be essential to explain the large elliptic flow observed for 
  D-mesons~\cite{Rapp:2018qla}) needs to be accounted for; the associated large 
  scattering widths are likely to require quarkonium transport implementations beyond 
  semi-classical (or perturbative) approximations, which reiterates the need for 
  a quantum treatment of recombination processes.

The larger experimental data samples in \RunsThreeFour, combined with improved detector performance and measurement techniques, will allow one to significantly improve over the current measurements, with extended kinematic coverage (in \pT) and allowing one to reach also currently-unobserved quarkonium states, like \PGUP{3S}.
The complementarity (and overlap) of all 4 LHC experiments is crucial in this endeavour and will call for a data combination strategy, for instance for \PGU\ azimuthal anisotropy.
Quarkonia are measured in the dimuon channel in ATLAS ($|y|<2.0$), CMS ($|y|<2.4$), LHCb ($2.0<y<4.5$), and ALICE ($2.5<y<4.0$), and in the dielectron channel with ALICE ($|y|<0.9$).
We present below data projections and simulations for a selection of observables and compare to model predictions (which sometimes constitute the basis for the projections). The model uncertainties shown in this section represent the current knowledge; significant improvements are expected both in what concerns the conceptual aspects discussed above as well for the input parameters, which will be constrained by data and theory (for instance in what concerns nuclear PDFs, see also Section~\ref{sec:nPDFs}).

All four LHC experiments will benefit from a large upgrade program, during the Long Shutdown 2 (2019--2020) for ALICE and LHCb, and during Long Shutdown 3 (2024--2025) for ATLAS and CMS~\cite{CMS:FTR-18-012}.
The addition of the Muon Forward Tracker (MFT) will allow ALICE to separate the prompt charmonium from the contribution from B meson decays. In addition, the background will be reduced, yielding to better signal over background ratios. 
Regarding ATLAS and CMS, the upgraded inner tracker will extend to $|\eta|\lesssim 4.0$ after LS3, and the muon system coverage to $|\eta|\lesssim 2.7$ (3.0) for ATLAS (CMS). While the detector improvements will have a smaller impact than the
increase in data sample size, this increase in pseudorapidity coverage is appreciable in also giving an overlap with the range of ALICE and LHCb.
Better track momentum resolution is also expected from these upgraded inner trackers, with an improvement of about 30\% of the mass resolution of quarkonia for CMS~\cite{CMS-TDR-15-02}. The expected improvement in the 
relative statistical uncertainty, due to a better signal over background ratio, is in the range 10--25\%~\cite{CMS-PAS-FTR-18-024}.

\subsection{Charmonia in \PbPb collisions} %

A remarkable discovery at the LHC was that the suppression of \PJgy is significantly reduced in comparison to lower energies~\cite{Abelev:2012rv} and that this reduction is concentrated at lower \pT~\cite{Abelev:2013ila,Adam:2016rdg}, compatible with predictions of (re)generation at the phase boundary of QCD~\cite{BraunMunzinger:2000px} or throughout the deconfined phase~\cite{Thews:2000rj,Zhao:2011cv,Yan:2006ve}, via recombination of diagonal (correlated pairs) or off-diagonal c$\bar{\text{c}}$ pairs~\cite{Young:2008he}. No significant difference is however found between measurements at $\sqrtsnn = \unit[2.76]{TeV}$ and $\sqrtsnn = \unit[5.02]{TeV}$~\cite{Adam:2016rdg,Sirunyan:2017isk}.
Recently, the measurement of a significant elliptic flow coefficient $v_2$ both for D mesons~\cite{Abelev:2014ipa,Acharya:2017qps,Sirunyan:2017plt} and \PJgy~\cite{Khachatryan:2016ypw,ALICE:2013xna,Acharya:2017tgv,Aaboud:2018ttm,Acharya:2018pjd}, which was shown to be correlated to the flow of the bulk particles~\cite{Acharya:2018bxo,Acharya:2018pjd}, can be seen as another indication for the thermalisation of charm quarks in the QGP. Transport model calculations \cite{Zhou:2014kka,Du:2015wha} currently underestimate the data for $\pT\gtrsim 6\UGeVc$~\cite{Acharya:2017tgv,Khachatryan:2016ypw,Aaboud:2018ttm}.
  
\begin{figure}[h]
\begin{center}
 \includegraphics[width=0.66\textwidth]{\main/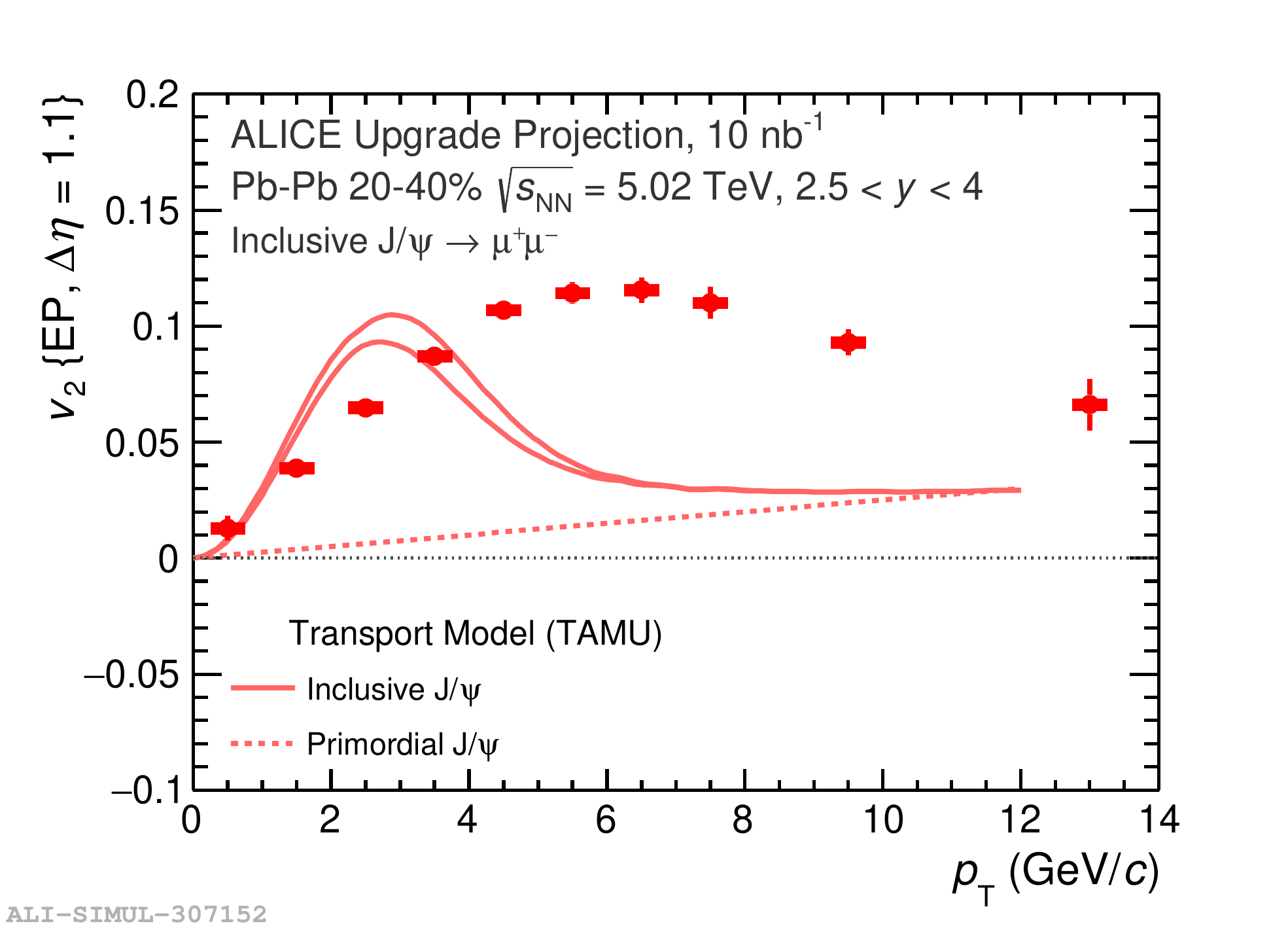}
\end{center}
 \caption{Projected measurement of elliptic flow coefficient $v_2$ as a function of \pT~ for \PJgy mesons (measured in ALICE, for $2.5<y<4$), for the centrality class 20--40\%, in comparison to model calculations \cite{Du:2015wha}. Figure from Ref.~\cite{ALICE-PUBLIC-2019-001}.}
\label{FigQ:v2pTPbPb}
\end{figure}

The projected ALICE measurement of inclusive \PJgy $v_2$ as a function of \pT\ for the centrality class 20--40\%, for $2.5<y<4$, in comparison to model calculations \cite{Du:2015wha} is shown in Fig.~\ref{FigQ:v2pTPbPb}. 
Disentangling the contributions of prompt and non-prompt \PJgy and considering (path-length dependent) energy loss seems mandatory to understand the details of the \PJgy $v_2$ pattern, which will be facilitated with the detector upgrades and higher luminosity of \RunsThreeFour.
The measurement of higher harmonics, \eg $v_3$, which are sensitive to initial state energy density fluctuations, will also become available
and  provide further insight into the charmonium production mechanisms. Precise prompt and non-prompt \PJgy $v_2$ and $v_3$ measurements
at low \pt will be reachable using the ALICE central barrel. Polarisation will be measured too~\cite{Abelevetal:2014cna}, providing further insight in the different production mechanisms involved in \PbPb collisions as compared to \pp. Under the statistical hadronisation paradigm, the prompt \PJgy yield in \PbPb collisions should be unpolarised with the 3 polarisation states equally populated.

\begin{figure}[h]
\begin{center}
 \includegraphics[width=0.48\textwidth]{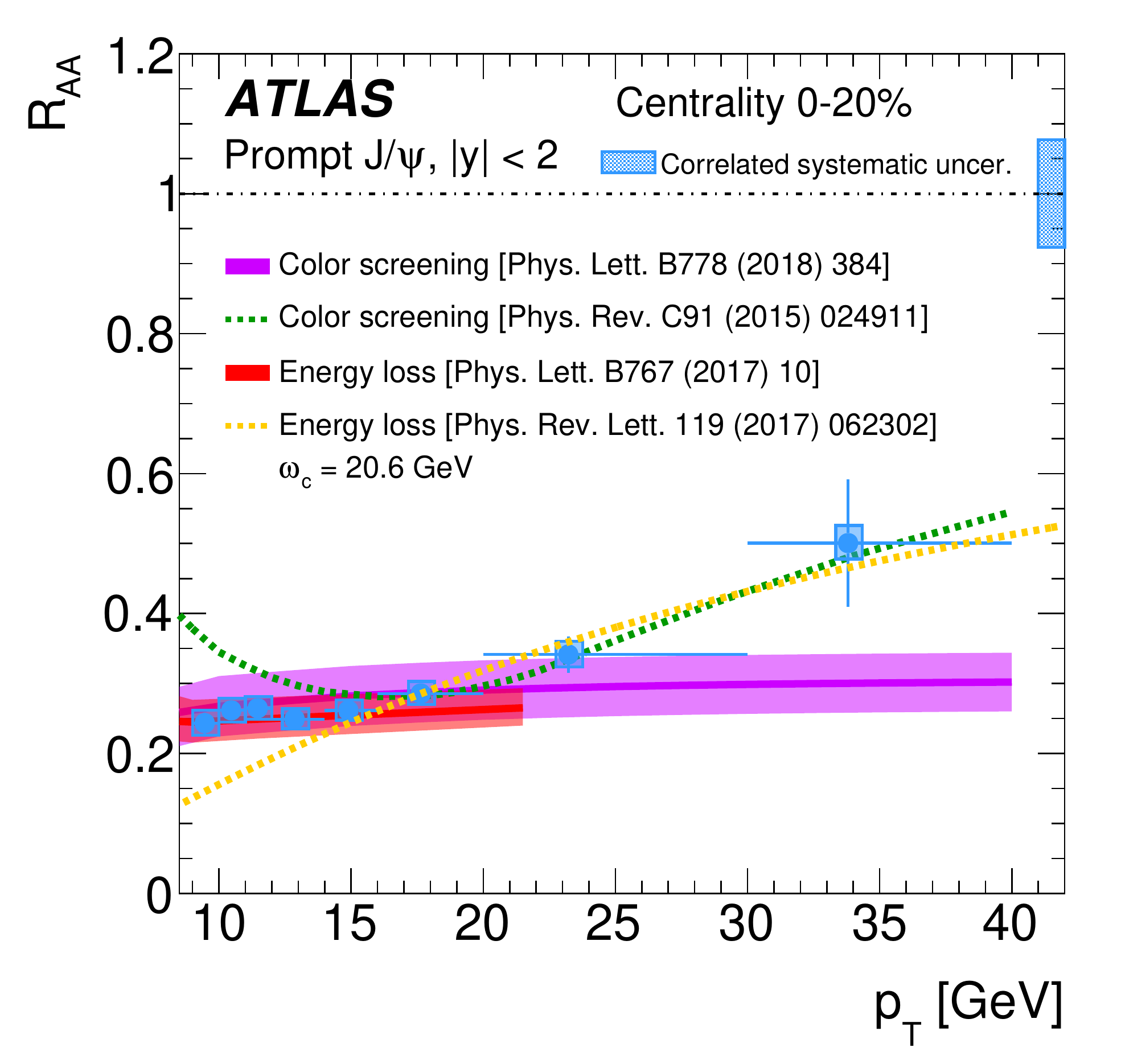}
 \includegraphics[width=0.47\textwidth]{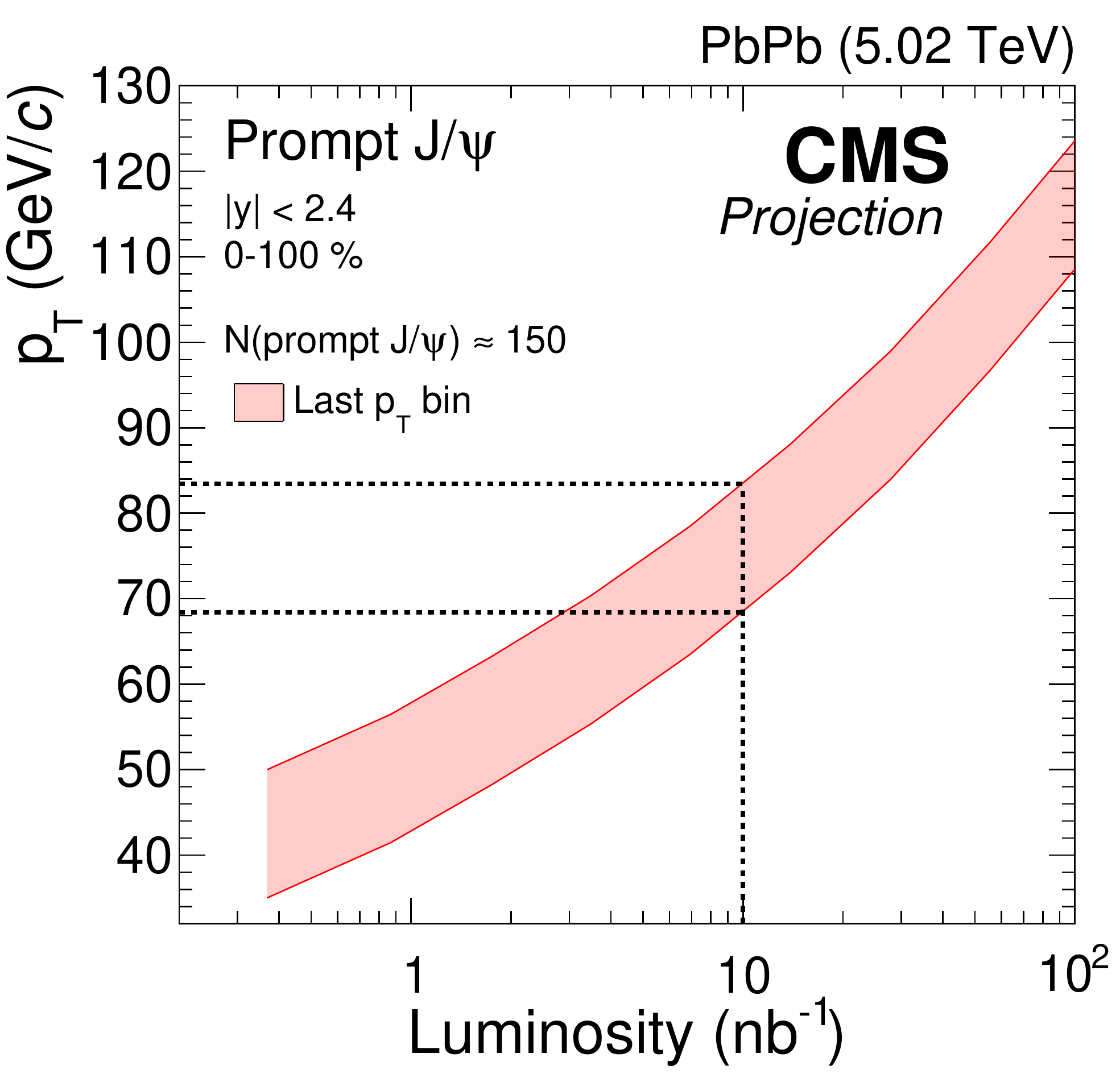}
\end{center}
\caption{Left: \RAA vs \pT~ for prompt \PJgy in central (0--20\%) collisions (ATLAS, $|y|<2$, \cite{Aaboud:2018quy}). Right: Prompt \PJgy high \pt bin boundaries as a function of luminosity with the CMS experiment~\cite{CMS-PAS-FTR-18-024}. The boundaries are chosen in such a way the number of mesons in the bin for the corresponding luminosity equals the number of mesons found in the last \pt bin of the current measurement~\cite{Sirunyan:2017isk}.
}
\label{FigQ:JpTPbPb}
\end{figure}

At high \pT, where a raising trend is currently hinted by Run~2 \raa measurements~\cite{Sirunyan:2017isk,Aaboud:2018quy}, the production mechanisms cannot currently be resolved, given the statistical limitation in the data (see Fig.~\ref{FigQ:JpTPbPb}, left). The high \pt reach of \RunsThreeFour data (illustrated in Fig.~\ref{FigQ:JpTPbPb} (right) for CMS) will allow one to conclude on the important question of whether \PJgy
formation at high \pt is determined by the Debye screening mechanism \cite{Kopeliovich:2014una,Aronson:2017ymv}, or by energy loss of the charm quark or the $\text{c}\bar{\text{c}}$ pair~\cite{Spousta:2016agr,Arleo:2017ntr}.

\begin{figure}[h]
\begin{center}
 \includegraphics[width=0.49\textwidth]{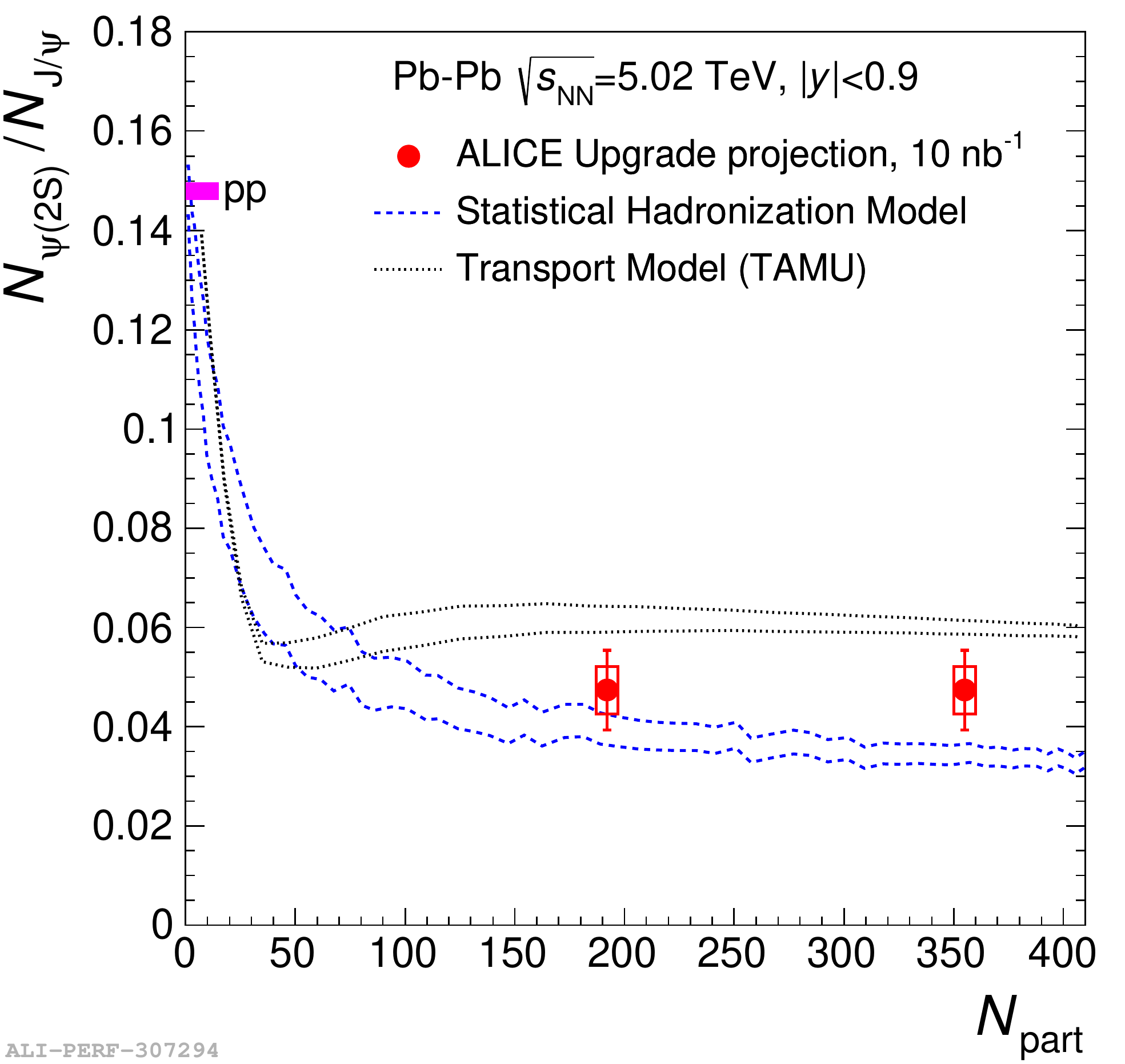}
 \includegraphics[width=0.49\textwidth]{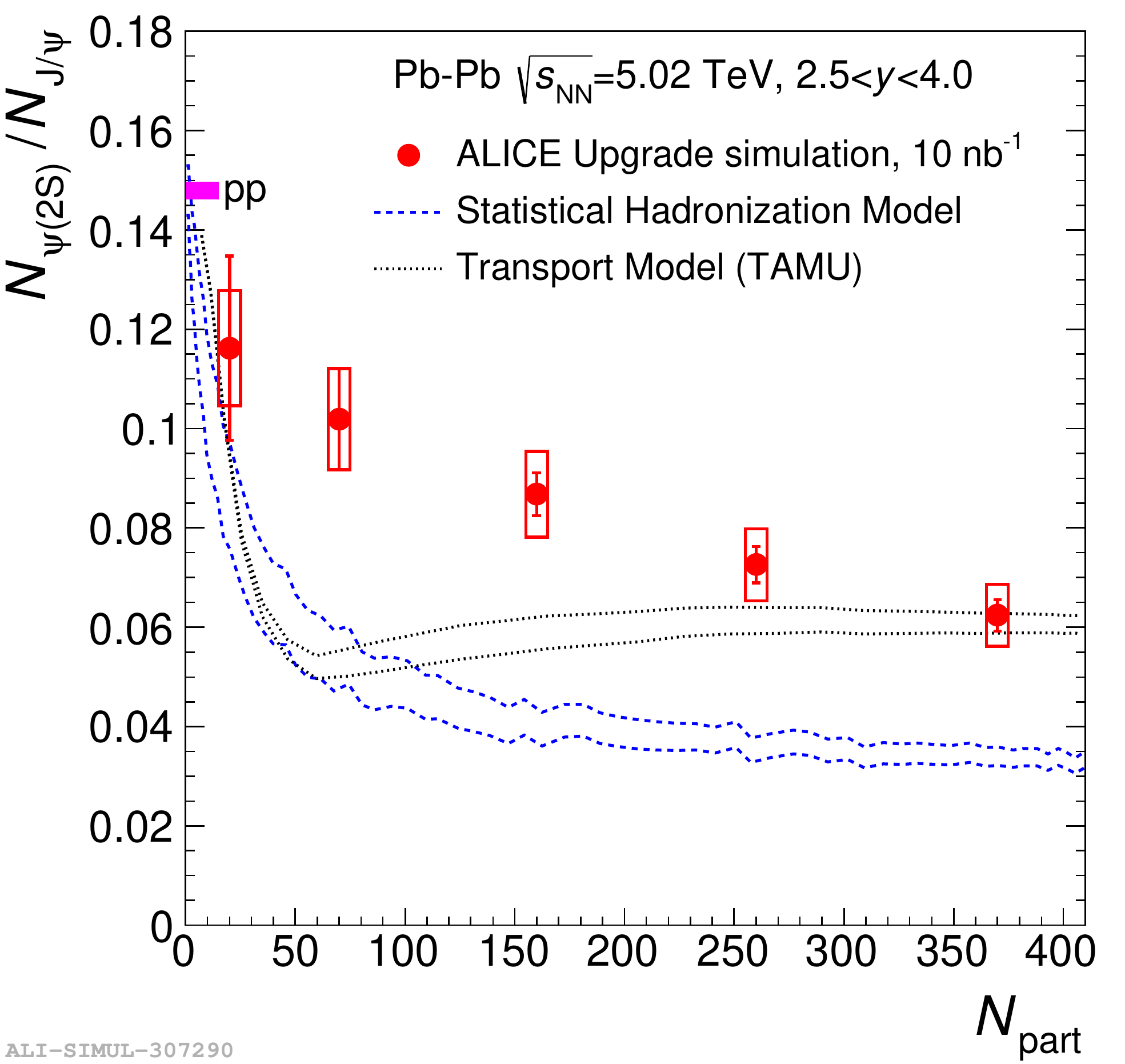}
\end{center}
 \caption{Production ratio $\PGyP{2S}/\PJgy$ vs. $N_{part}$ for $|y|<0.9$ (left) and $2.5<y<4$ (right) \cite{Abelevetal:2014cna,CERN-LHCC-2013-014}. Model predictions in the transport approach \cite{Du:2015wha} and from statistical hadronisation \cite{Andronic:2017pug} are included. The values of the ratio used for the projections are quasi-arbitrary. Figures from Ref.~\cite{ALICE-PUBLIC-2019-001}.}
\label{FigQ:psi2SPbPb}
\end{figure}

The measurement of \PGyP{2S} mesons is more difficult than that of \PJgy, because of a much smaller production cross section times branching ratio and even larger suppression in \PbPb, yielding a very low signal to background ratio.
The projections for the measurement of the \PGyP{2S} state in ALICE are shown in Fig.~\ref{FigQ:psi2SPbPb} as a function of centrality and compared to model predictions in the transport approach \cite{Du:2015wha}  and from the statistical hadronisation model \cite{Andronic:2017pug}. This (\pT-integrated) measurement will significantly contribute to make a distinction between the two models. Projections are also available from the CMS experiment~\cite{CMS-PAS-FTR-17-002}.
Other states, for instance \PGcc, may be measured too, albeit the measurement down to $\pT=0$ will remain challenging. \PBc\ mesons can also be measured, either in the $\PJgy\PGp$ or $\PJgy\PGm\PGnGm$ channel, to further study recombination in the beauty sector.

\subsection{Bottomonia in \PbPb collisions} %

The study of bottomonia with \PbPb data from the \RunsThreeFour of the LHC will bring further information on the physics aspects described above.
Although their production is a priori sensitive to the same effects as charmonia, in practice the two quarkonium families feature some fundamental differences.
Binding energies differ, which is reflected in the different dissociation temperatures.
Experimentally, compared to charmonia, the absence of contribution from B meson decays and the more similar cross section times branching ratio between the ground and excited states make bottomonia measurements easier. 
At the same time, in \pp collisions, up to 30--50\% of the measured 
\PGUP{1S} and \PGUP{2S} yields actually result from the feed-down from other states~\cite{Andronic:2015wma,Aaij:2014caa}:
a large portion of measured \PGUP{1S} suppression can be due to the stronger suppression of the feed-down states -- \PGUP{2S} and \PGUP{3S} mesons also receive a significant contribution from feed-down.
The impact of (re)generation from uncorrelated $\text{b}\bar{\text{b}}$ is also expected to be much smaller than for charmonia, because of the much smaller number of $\text{b}\bar{\text{b}}$
pairs per \PbPb event compared to that of $\text{c}\bar{\text{c}}$ pairs. The importance of regeneration for bottomonia, from correlated or uncorrelated pairs~\cite{Petreczky:2016etz}, is however still very model dependent, and no unambiguous experimental signal
for it has been found yet. Possible ways of constraining this contribution will be discussed in this section.

Experimentally, the higher mass of bottomonia compared to charmonia implies higher $\pt$ decay leptons, allowing the ATLAS and CMS experiments to measure the production down to zero transverse momentum,
as is possible for ALICE for both charmonia and bottomonia~\cite{Abelev:2014nua,Acharya:2018mni}. The proximity in mass between the different mass states, especially between the \PGUP{2S} and \PGUP{3S} states, also 
means that good muon (or electron) momentum resolution is essential to their measurement, especially for excited states. 

It is useful to remind quickly the status in 2018, based on results from Run~1 and early Run~2 LHC data as well as RHIC data. \PGU production is found to be suppressed in \PbPb compared to expectations from a scaling of incoherent \pp collisions, in all rapidity, $\pt$ and centrality ranges measured~\cite{Abelev:2014nua,Acharya:2018mni,Sirunyan:2018nsz,Khachatryan:2016xxp}. Suppression is stronger in central events, as expected from the hotter and longer-lived medium in such events.
The results from the most central collisions suggest that a certain amount of suppression of the directly produced \PGUP{1S} might be needed
  to explain the data in addition to cold nuclear matter effects and melting of the excited \PGU and \PGcb states.
The excited states \PGUP{2S} and \PGUP{3S} show higher suppression wrt the ground state, with \raa values which respect
  the hierarchy expected based on their binding energies. The \PGUP{3S} is still unobserved in \PbPb collisions ($\raa(\PGUP{3S})<0.094$ at 95\% confidence
level, for $\sqrtsnn = \unit[5.02]{TeV}$~\cite{Sirunyan:2017lzi,Sirunyan:2018nsz}). 
No significant dependence of the suppression of \PGU states is found at the LHC on collision energy or rapidity.

\begin{figure}
\begin{tabular}{cc} \begin{minipage}{.49\textwidth}
\centerline{
\includegraphics[width=0.94\textwidth]{\main/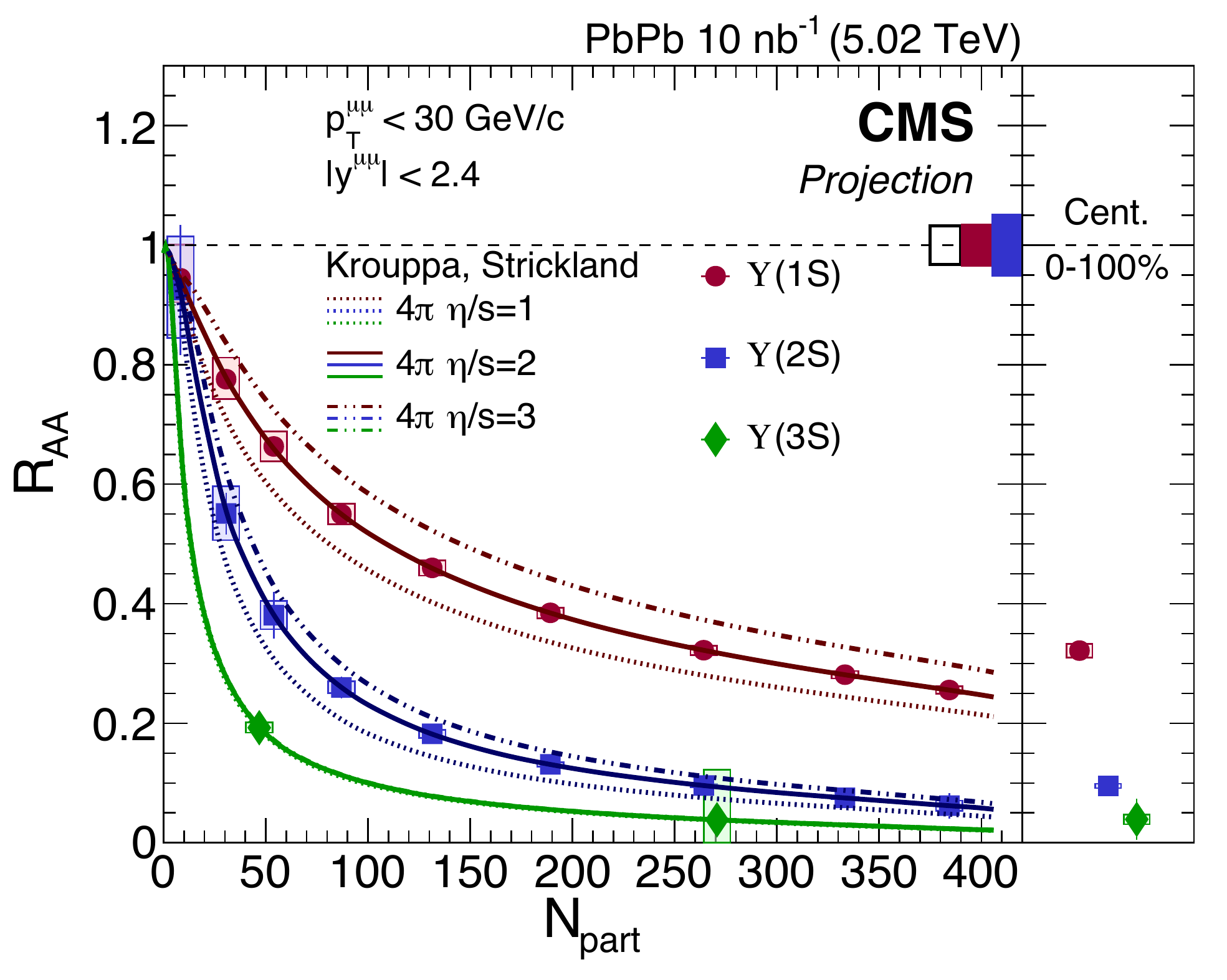}}
  \end{minipage} & \begin{minipage}{.49\textwidth}

    \vspace{-.6cm}
    \centerline{
\includegraphics[width=.94\textwidth]{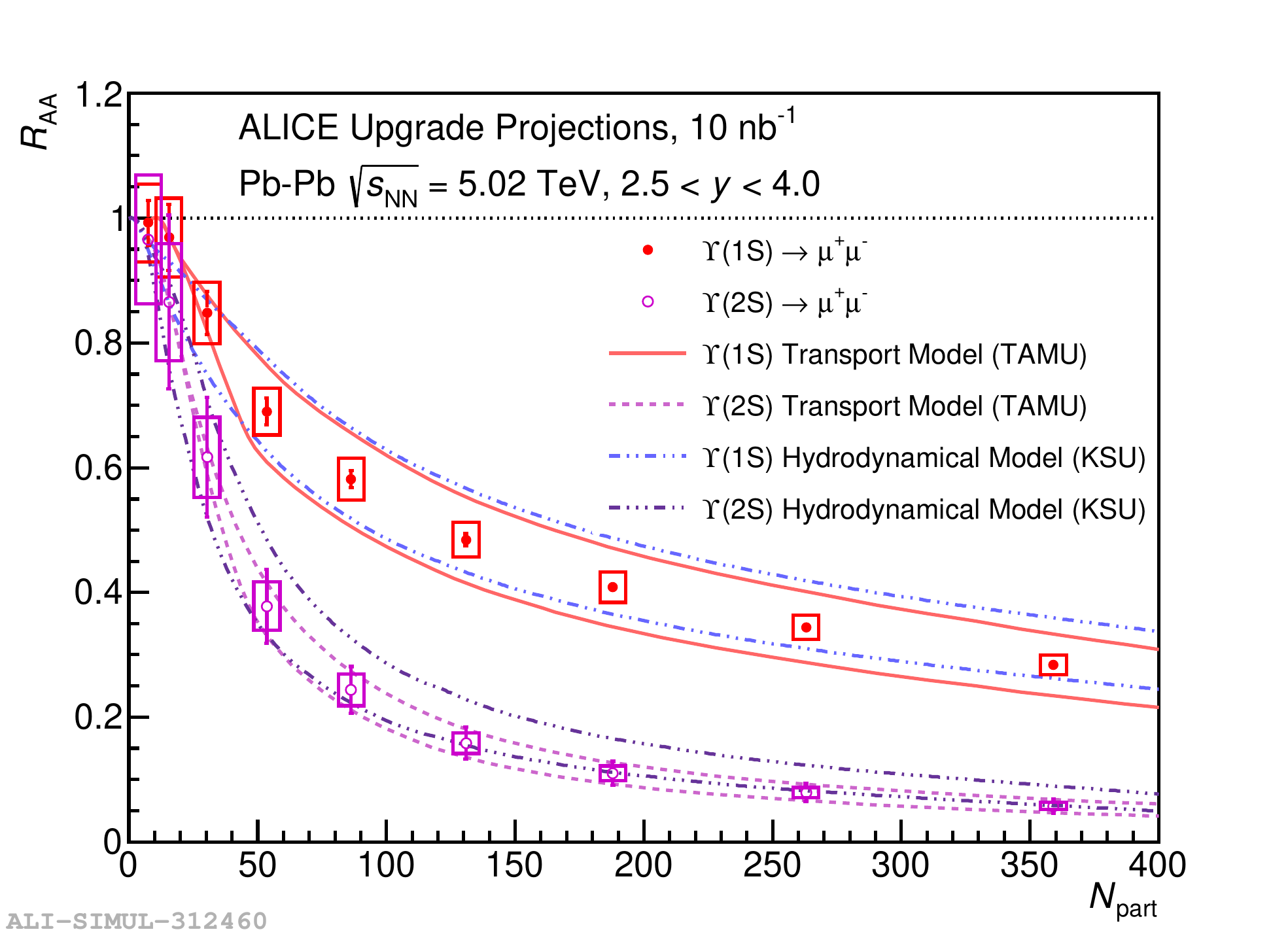}}
\end{minipage}\end{tabular}

 \caption{Centrality dependence of \PGUP{1S}, \PGUP{2S} and \PGUP{3S} \raa, as projected by the CMS\cite{CMS-PAS-FTR-17-002,Krouppa:2016jcl} (left) and ALICE~\cite{ALICE-PUBLIC-2019-001} (right) experiments, and from a transport model\cite{Du:2017qkv}}
 \label{fig:upsi_raa_npart}
\end{figure}

Differences exist between models in the theoretical treatment of the suppression of the bottomonia in the medium, as summarised earlier in Section.~\ref{sec_intro}.
Different assumptions are used regarding the production mechanism, the heavy quark potential, or the evolution of the quarkonia with the medium. 
The understanding of hot medium effects will be also improved thanks to higher precision measurements in \pp collisions of the feed-down fractions
   and to stronger constrains of the cold nuclear matter and initial state effects (including nPDF or coherent energy loss effects~\cite{Arleo:2014oha}) from \pPb collision measurements.
Figure~\ref{fig:upsi_raa_npart} shows that the projected uncertainty on the \raa of \PGU with \unit[10]{nb}$^{-1}$ will be much smaller than the current model uncertainties.
Bottomonia may bring information
complementary to other probes, using the sensitivity of the suppression to the medium shear viscosity or to the initial temperature of the fireball.

\begin{figure}
\begin{center}
 \includegraphics[width=0.38\textwidth]{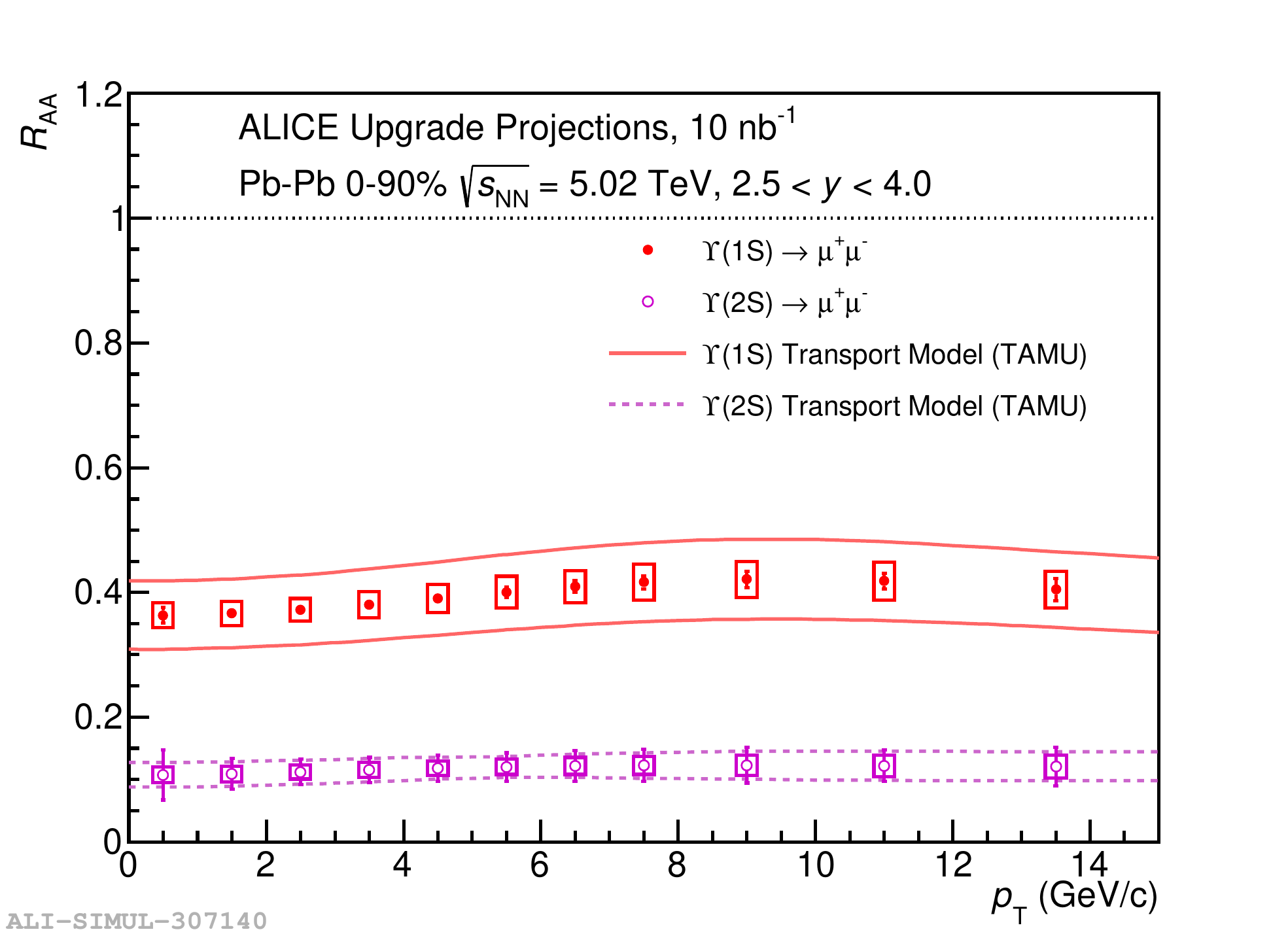}
 \includegraphics[width=0.3\textwidth]{\main/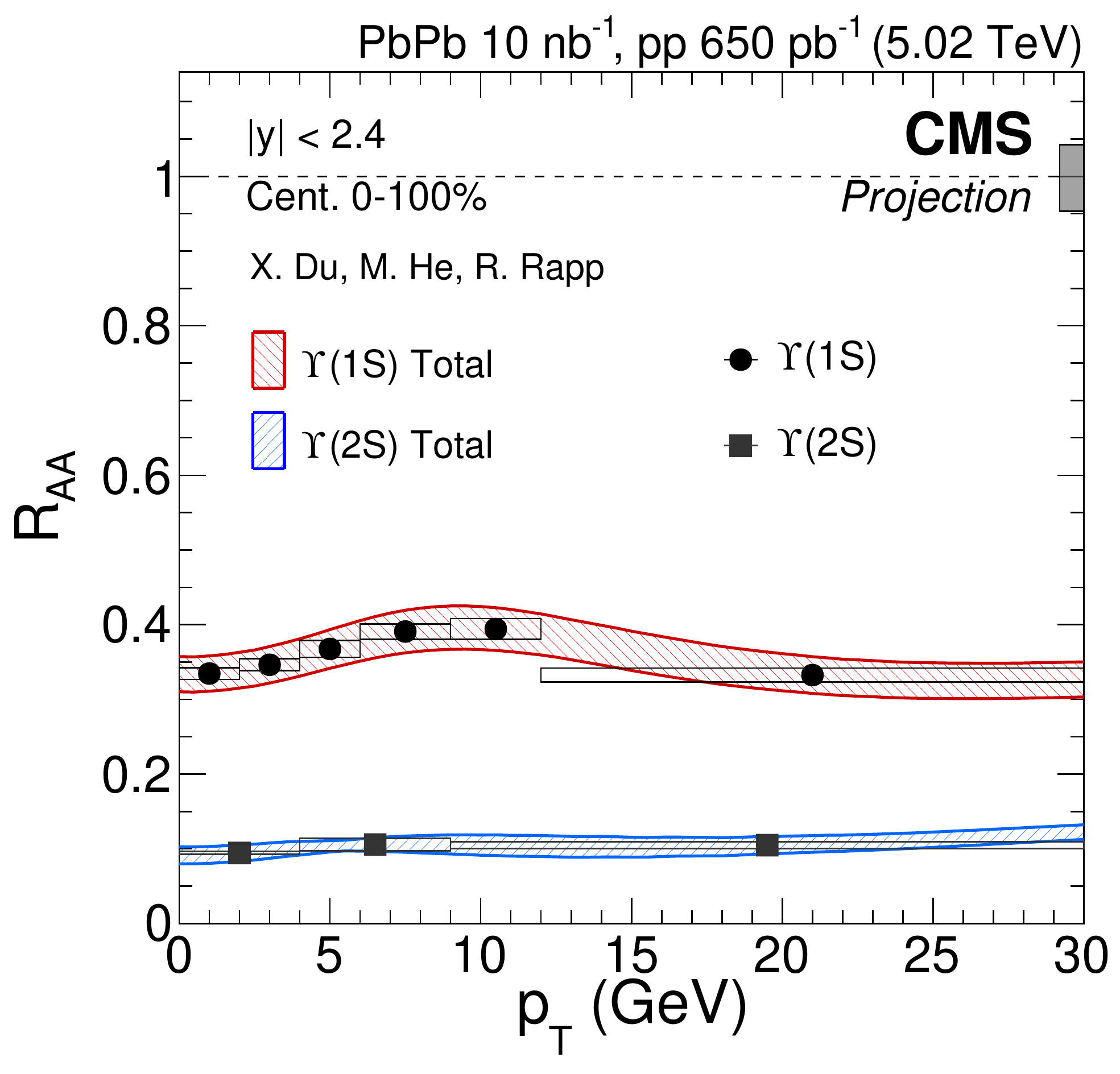}
 \includegraphics[width=0.3\textwidth]{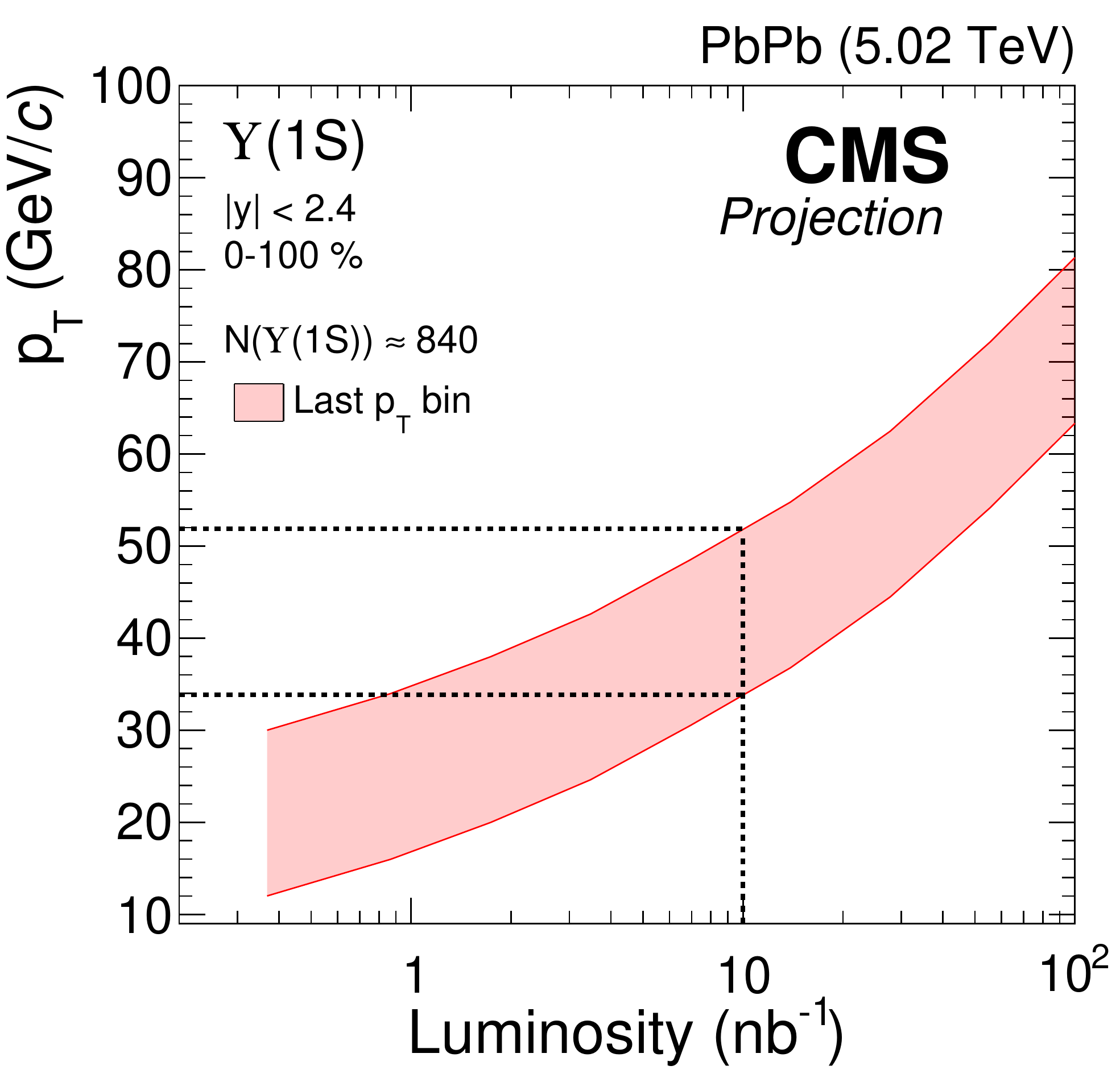}
\end{center}

\caption{Projected \raa for \PGUP{1S} and \PGUP{2S} expected from the ALICE~\cite{ALICE-PUBLIC-2019-001} (left) and CMS~\cite{CMS-PAS-FTR-18-024} (center) experiments, as a function of \pt, with \unit[10]{nb}$^{-1}$ of \PbPb data. The expected \pt reach for \PGUP{1S} from the CMS experiment is
 also shown~\cite{CMS-PAS-FTR-18-024}, as the position of the last \pt bin of the measurement, with constant number of observed \PGUP{1S}, as a function of integrated luminosity.
 }
 \label{fig:upsi_raa_pt}
\end{figure}

\begin{figure}
  \begin{tabular}{cc} \begin{minipage}{.49\textwidth}
      \vspace{-.6cm}
 \centerline{
\includegraphics[width=0.96\textwidth]{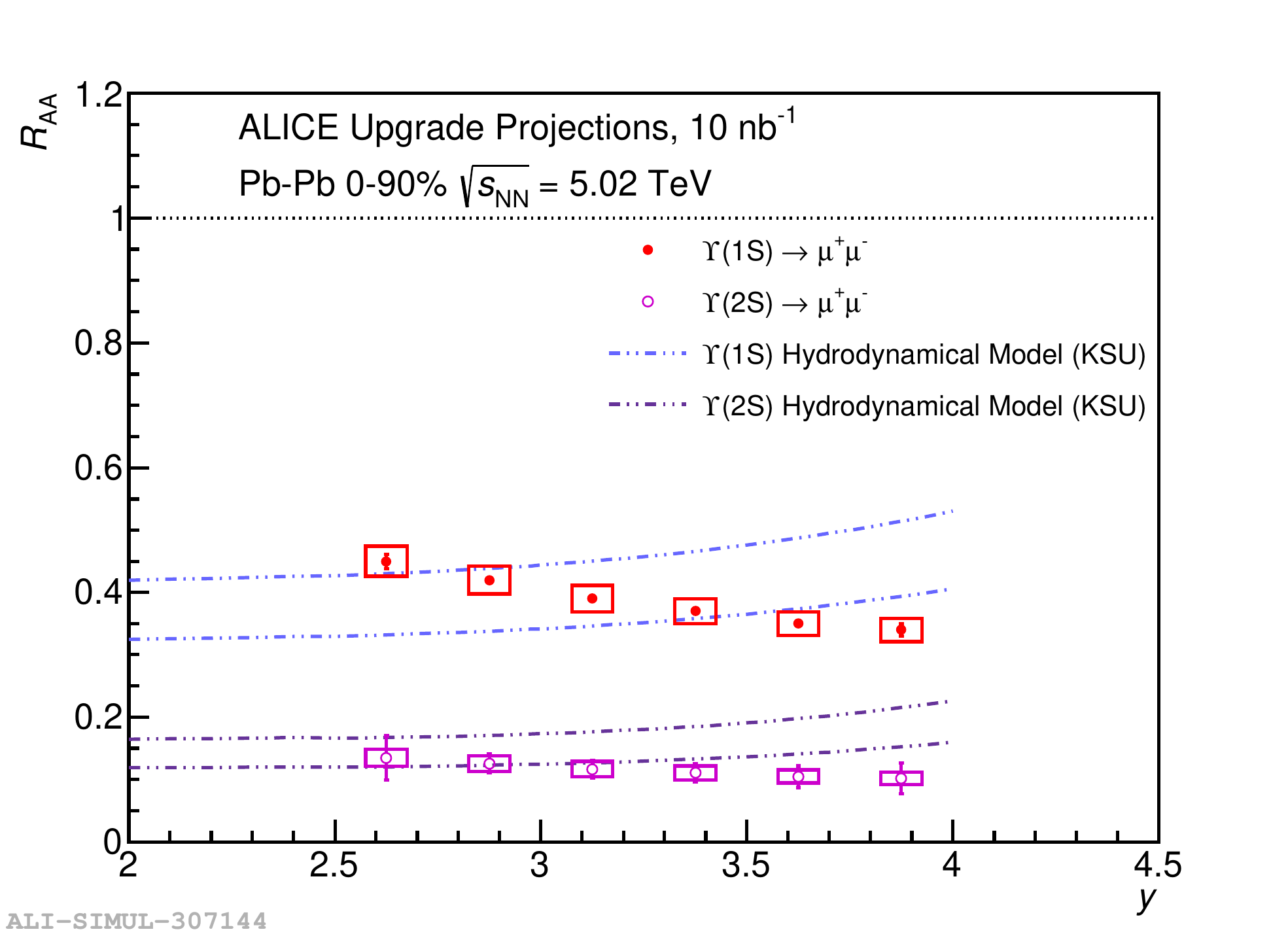}}
  \end{minipage} & \begin{minipage}{.49\textwidth}
 \centerline{
\includegraphics[width=0.8\textwidth]{\main/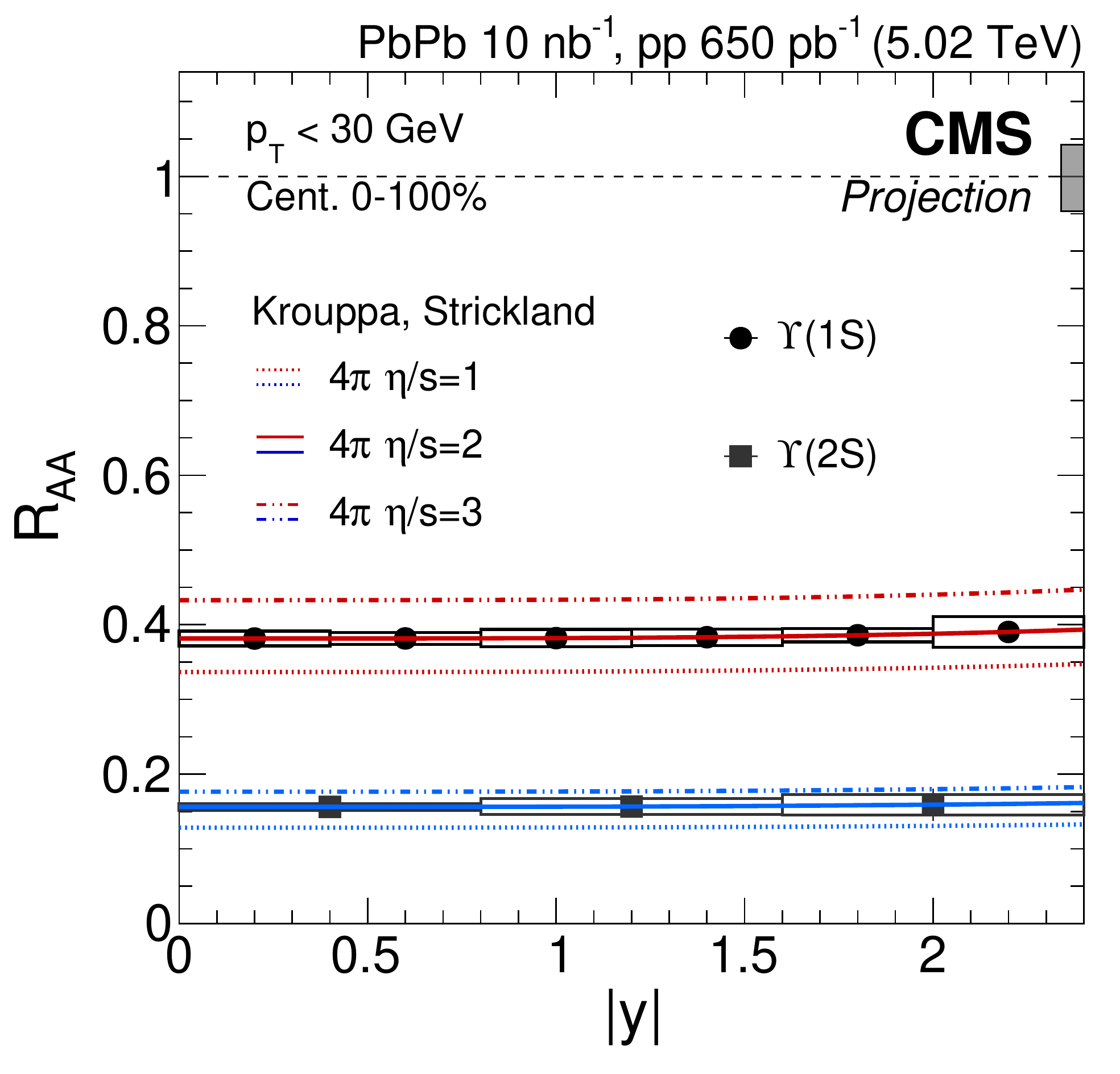}}
\end{minipage}\end{tabular}
 \caption{Projected \raa for \PGUP{1S} and \PGUP{2S} expected from the ALICE~\cite{ALICE-PUBLIC-2019-001} (left) and CMS~\cite{CMS-PAS-FTR-18-024} (right) experiments, as a function of rapidity, with \unit[10]{nb}$^{-1}$ of \PbPb data.}
 \label{fig:upsi_raa_y}
\end{figure}

A precise measurement of the \pt dependence of the \PGUP{1S} \raa will be possible using LHC data from \RunsThreeFour. At low and medium $\pt$, the measurement is
sensitive to the possible regeneration component in \PGU meson production~\cite{Du:2017qkv}. Projections for the expected precision of \PGU measurements from the ALICE and CMS detectors
using an integrated luminosity of \unit[10]{nb}$^{-1}$ after the \RunsThreeFour are shown as a function of $\pt$ in Fig.~\ref{fig:upsi_raa_pt} and $y$ in Fig.~\ref{fig:upsi_raa_y}, and compared to the expectations from two 
models~\cite{Krouppa:2017jlg,Du:2017qkv}. In the Kent state model calculations~\cite{Krouppa:2017jlg} (not shown), where \PGU mesons are originating only from the primordial
production, with no regeneration component, the \raa is rather flat in the low and medium \pt range.
Only at higher \pt (above 10--15\UGeVc) is a small rise predicted, which can be looked for in \RunsThreeFour data: as can be seen in Fig.~\ref{fig:upsi_raa_pt}, it is expected that a measurement up to a $\pt$ of about 50\UGeVc\ can be performed with the ATLAS and CMS detectors with
\unit[10]{nb}$^{-1}$ of data. 
In the TAMU model~\cite{Du:2017qkv} however, a regeneration
component is considered, and several assumptions are explored, especially on the degree of thermalisation of the bottom quarks. 
It predicts a maximum in the \PGUP{1S} \raa at a $\pt$ around 10\UGeVc. The current data is not precise enough to confirm or disfavour such a local maximum in the \raa, but \RunsThreeFour
data will allow to conclude.

Almost no rapidity dependence is expected at the LHC for the nuclear modification factor of \PGU mesons within the acceptance of ATLAS and CMS ($|\eta|\lesssim 2.5-3$), which can be better tested using \RunsThreeFour data. 
This will be further made significant considering the ALICE acceptance ($2.5<y<4$), allowing to confirm or disprove the prediction of the hydrodynamic model, see Fig.~\ref{fig:upsi_raa_y}.

\begin{figure}
\begin{center}
 \includegraphics[width=0.3\textwidth]{\main/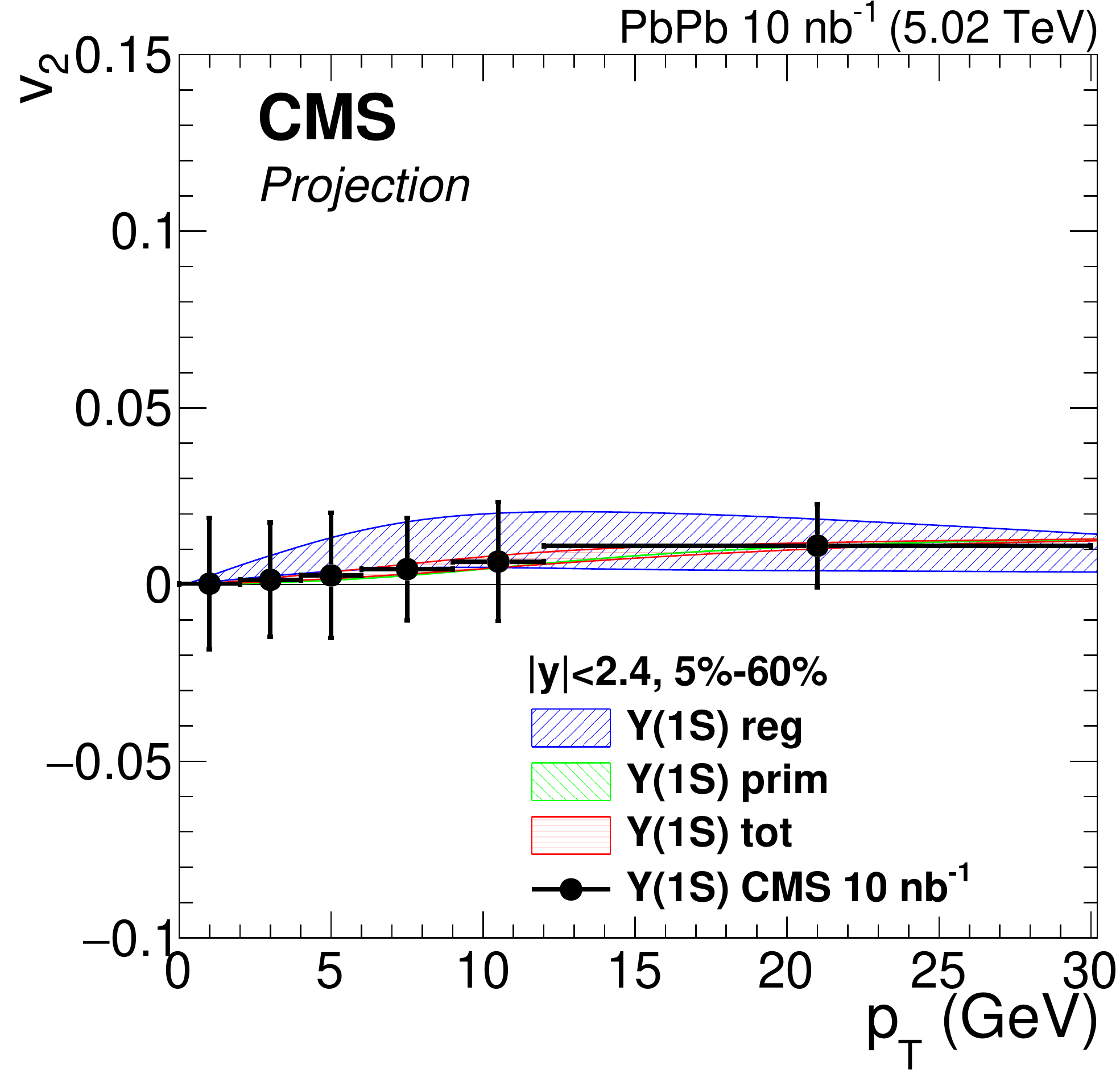}
 \includegraphics[width=0.3\textwidth]{\main/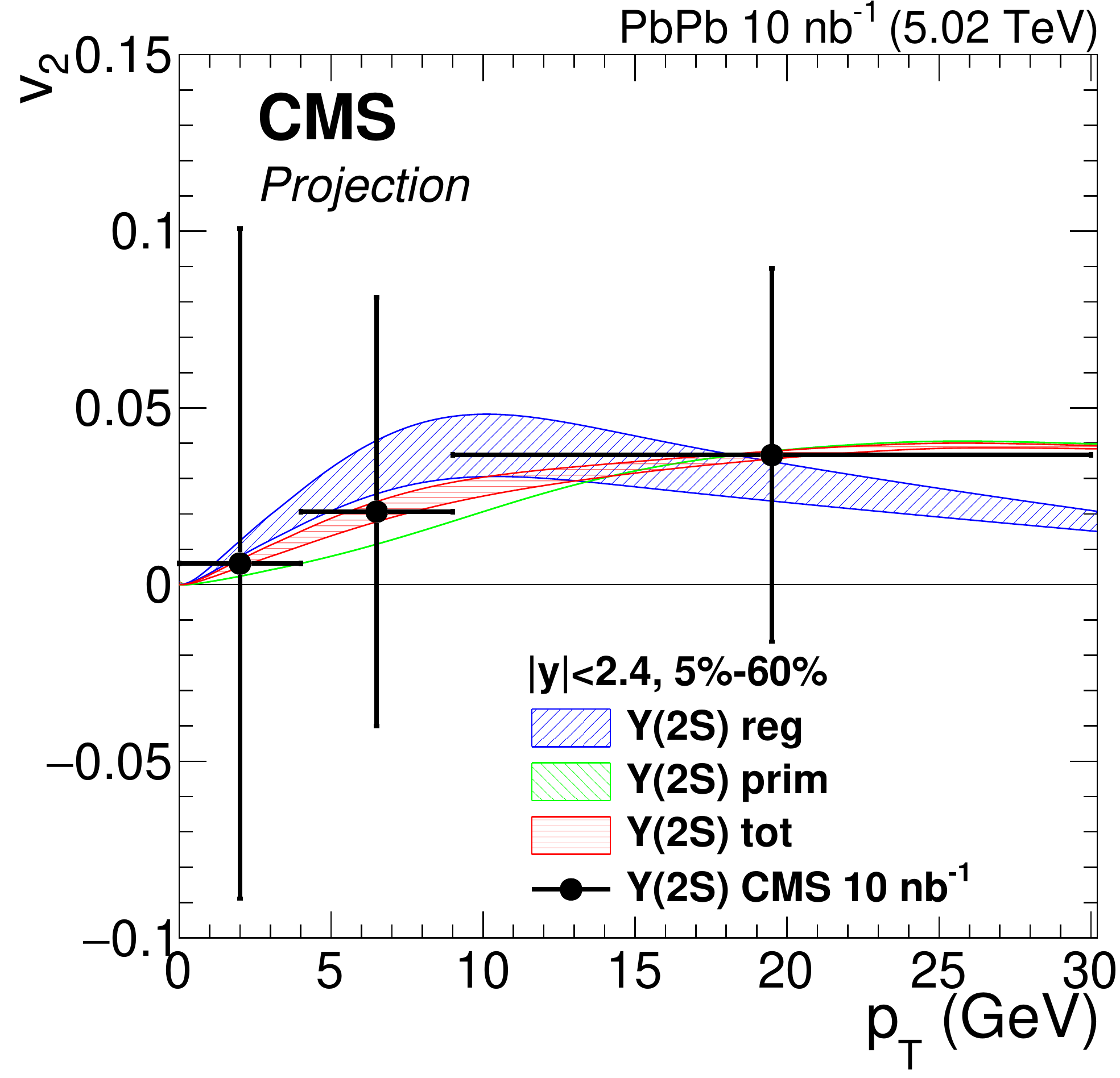}
 \includegraphics[width=0.38\textwidth]{\main/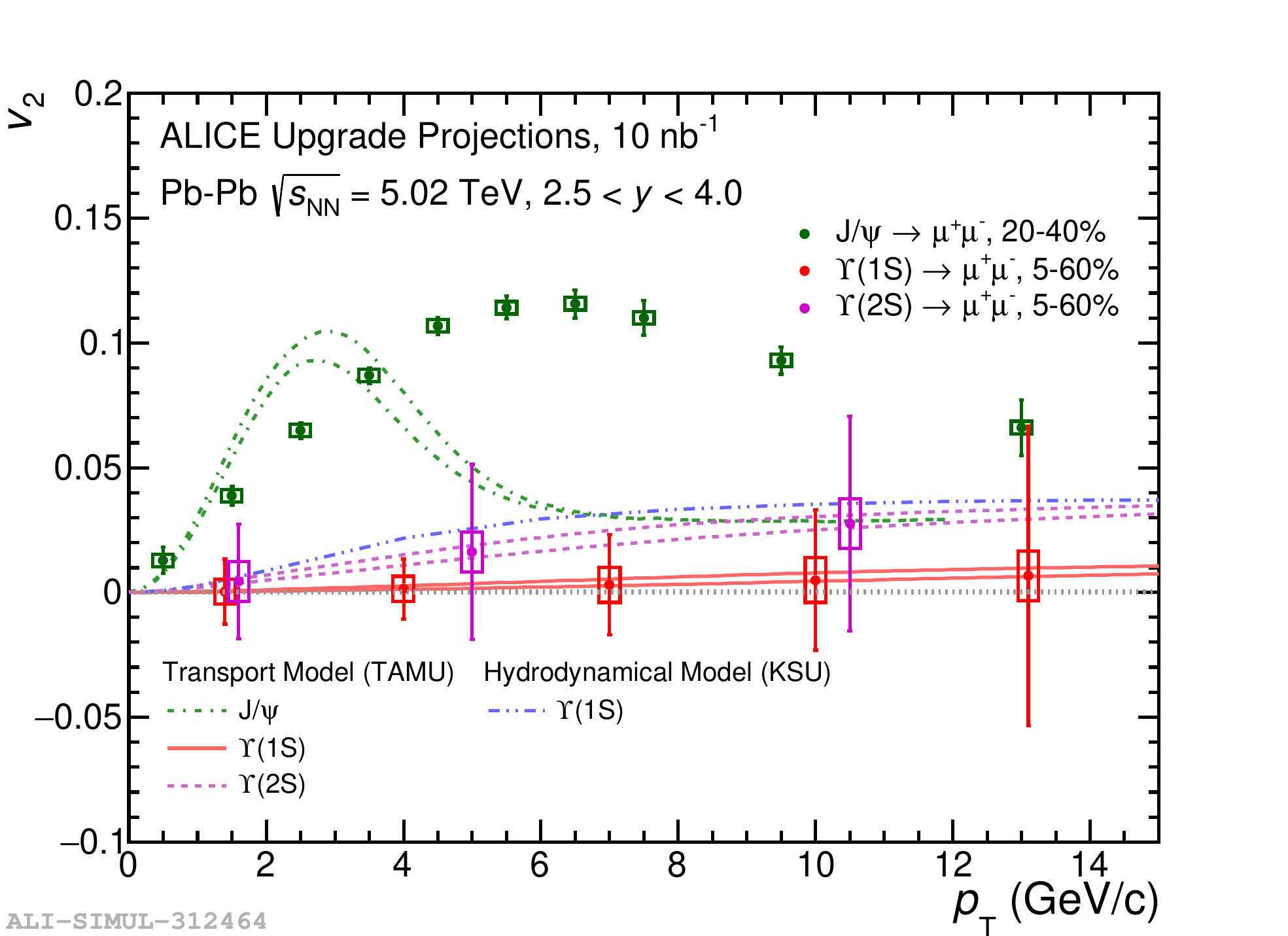}
\end{center}

 \caption{$v_2$ projections for the CMS~\cite{CMS-PAS-FTR-18-024} (left and centre) and ALICE~\cite{ALICE-PUBLIC-2019-001} (right) experiments for the \PGUP{1S} and \PGUP{2} mesons, assuming the predictions from a transport model~\cite{Du:2017qkv}.}
 \label{fig:upsi_v2}
\end{figure}

Coming back to the matter of regeneration, much can be learnt about it by a measurement of the elliptic flow of \PGUP{1S} mesons~\cite{Das:2018xel}, 
unmeasured to date in any collision system. 
A parallel can be
drawn with that of \PJgy, which is still not properly described by models. This observable requires a more detailed implementation of the dynamics of the interactions between the 
quarkonium and the medium: thermalisation of the heavy quarks, time dependence of regeneration, path length dependence of energy loss, as well as initial geometry fluctuations and 
elastic rescattering of the quarkonia in the medium. Thus, collective flow brings complementary information to the \raa, and its measurement can help disentangle some
effects. In the case of \PGUP{1S} mesons, a small $v_2$ (order of 1--2\%) is expected~\cite{Du:2017qkv,Yao:2018zrg,Bhaduri:2018iwr}, as can be seen in Fig.~\ref{fig:upsi_v2}.
The elliptic flow of \PGUP{2S} could be significantly higher~\cite{Du:2017qkv,Bhaduri:2018iwr}, both from the regenerated and primordial
components. For both states, projections show that experimental precision may not be enough for a significant $v_2$ measurement, assuming $v_2$ values as in Ref.~\cite{Du:2017qkv}. For this reason,
combining results between the different LHC experiments would be beneficial to reach a better sensitivity.

While we have focused on the \raa and $v_2$ in this section, bottomonium production can be studied using other observables. For instance, fully corrected yields or cross sections
in \PbPb can be studied, without making the ratio to a \pp measurement in a \raa. Such a measurement, already reported in some of the available experimental results~\cite{Sirunyan:2018nsz},
can directly be compared to production models.

\subsection{Quarkonia in \pPb and \pp collisions} 

\subsubsection{\pPb collisions} %
Quarkonium-production studies in high-energy \pPb collisions are usually carried out to measure how much specific nuclear effects, those which do {\it not} result from the creation of a deconfined state of matter, can alter the quarkonium yields. They should indeed be accounted for in the interpretation of \PbPb  results. They are also interesting on their own as they provide means to probe the modification of the gluon densities in the nuclei, the interaction between such pure heavy-quark bound states and light hadrons, or phenomena such as the coherent medium-induced energy loss of these quark-antiquark pairs.
The measurements as a function of event activity brought several surprises, hotly discussed presently.
 
Usually, a separation into initial-state and final-state effects is done (coherent energy loss effects \cite{Arleo:2010rb} can be seen as an interplay between the two types of effects). Yet, it is probably more instructive to separate out the effects which are believed to impact {\it all} the states of the charmonium or the bottomonium family with the {\it same} magnitude
 from those which are expected to impact differently the ground and the excited states. In principle, initial-state effects (in particular gluon shadowing \cite{Kusina:2017gkz}) are of the first kind as the nature of the to-be-produced quarkonium state is not yet fixed when the effects are at work.  On the contrary, final-state effects (like regeneration \cite{Du:2018wsj}) do depend on the properties of the produced quarkonium state and are thus be of the second kind.
 
However, in \pPb collisions at LHC energies, final-state interactions between the heavy-quark pair and the nuclear matter likely occur {\it before} the pair hadronises. This is due to the large boost between the nucleus and the pair -- and thus the quarkonium. At rest, a $c \bar c$ or $b\bar b$ pair takes 0.3--0.4\,fm/$c$ to hadronise; seen from the nucleus, at, for instance $y^{\rm lab}_{\rm pair}-y_{\rm beam} \sim 7$ , it takes $\gamma=\cosh(7) \simeq 500$ times longer. As such, final-state interactions with the compounds nucleus likely do not discriminate ground and excited quarkonium states, unless rescattering in the nucleus affects the Q$\bar{\text{Q}}$ wave function, overlapping with the quarkonium wave function at large distance~\cite{Benesh:1994du}. Such an argument based on the existence of a large boost is nevertheless not applicable if one considers effects arising from the interactions between the pair and other particles {\it produced} by the \pPb collisions, not those contained in the Pb nucleus. The former are indeed not moving at the Pb projectile rapidity. In fact, some of these particles can have similar rapidities as the quarkonium and can thus be considered as comoving with it~\cite{Gavin:1996yd,Brodsky:1988xz,Capella:2000zp}.
 
The simultaneous study of open-heavy flavoured hadrons along with both ground and excited quarkonium states can shed light on all these phenomena. Along the lines exposed above, one expects forward-quarkonium production in \pPb collisions (namely when the quarkonia flies in the direction of the proton) to be sensitive to low-$x$ phenomena (like the gluon shadowing or saturation in the lead ion) and to the coherent energy loss. On the contrary, the backward production should be sensitive to the gluon antishadowing and to fully coherent energy loss. Moreover, the scatterings of quarkonia with comoving particles occur more often backward than forward, due to the rapidity-asymmetric particle multiplicities, and more often as well with the larger and less tightly bound excited states. 
   
With a wide rapidity coverage spanning from about $-5$ to 5, the LHC data from the 4 experiments are unique as they allow one to probe much smaller $x$ values than at RHIC and with a larger reach in $\pT$. The higher c.m.s. energy, the competitive luminosities and the resolution of the detectors also allow for more extensive studies of the bottomonium family. In fact, an important observation made with Run~1 data was that of a relative suppression in \pPb collisions of the excited \PGUP{2S},\PGUP{3S} states compared to that of the \PGUP{1S} observed by CMS~\cite{Chatrchyan:2013nza} as a function of the event activity (recently confirmed by ATLAS~\cite{Aaboud:2017cif}, but also observed in \pp collisions by CMS~\cite{CMS-PAS-BPH-14-009}). Not only was it unexpected, but it constitutes a challenge to the conventional interpretation of suppression observed in \PbPb collisions~\cite{Chatrchyan:2012lxa,Sirunyan:2017lzi,Sirunyan:2018nsz}, which is of a significantly larger magnitude, but of a similar pattern. Such a relative suppression was also observed in the charmonium sector~\cite{Abelev:2014zpa}, where it is as well remarkable.

As far as the suppression of the \PGUP{1S} and \PJgy is concerned, they seem to follow the expectations based on the RHIC results with a strong forward suppression described by shadowing -- of a compatible magnitude to that observed with HF data~\cite{Kusina:2017gkz}, or with the coherent energy loss mechanism~\cite{Arleo:2010rb}. More data, including that on \PGUP{nS} and Drell-Yan production, are clearly needed to disentangle both effects~\cite{Arleo:2015qiv} (see also Section~\ref{subsubsec:coherent}). More precision for \PGUP{nS} and non-prompt \PJgy is in general critically needed as the typical experimental uncertainties are still on the order of the expected effects. As a case in point, backward $y$ data are not yet precise enough to quantify the magnitude of the gluon antishadowing, see Section~\ref{sec:smallx} for the possible relevance of quarkonium \pPb LHC data on nuclear PDF fits. Direct inclusion of this data in nPDF fits is however not yet possible, pending unambiguous clarification of the different effects impacting quarkonium production in \pPb collisions.

\begin{figure}[h]
 \begin{center}
  \includegraphics[width=0.45\textwidth]{\main/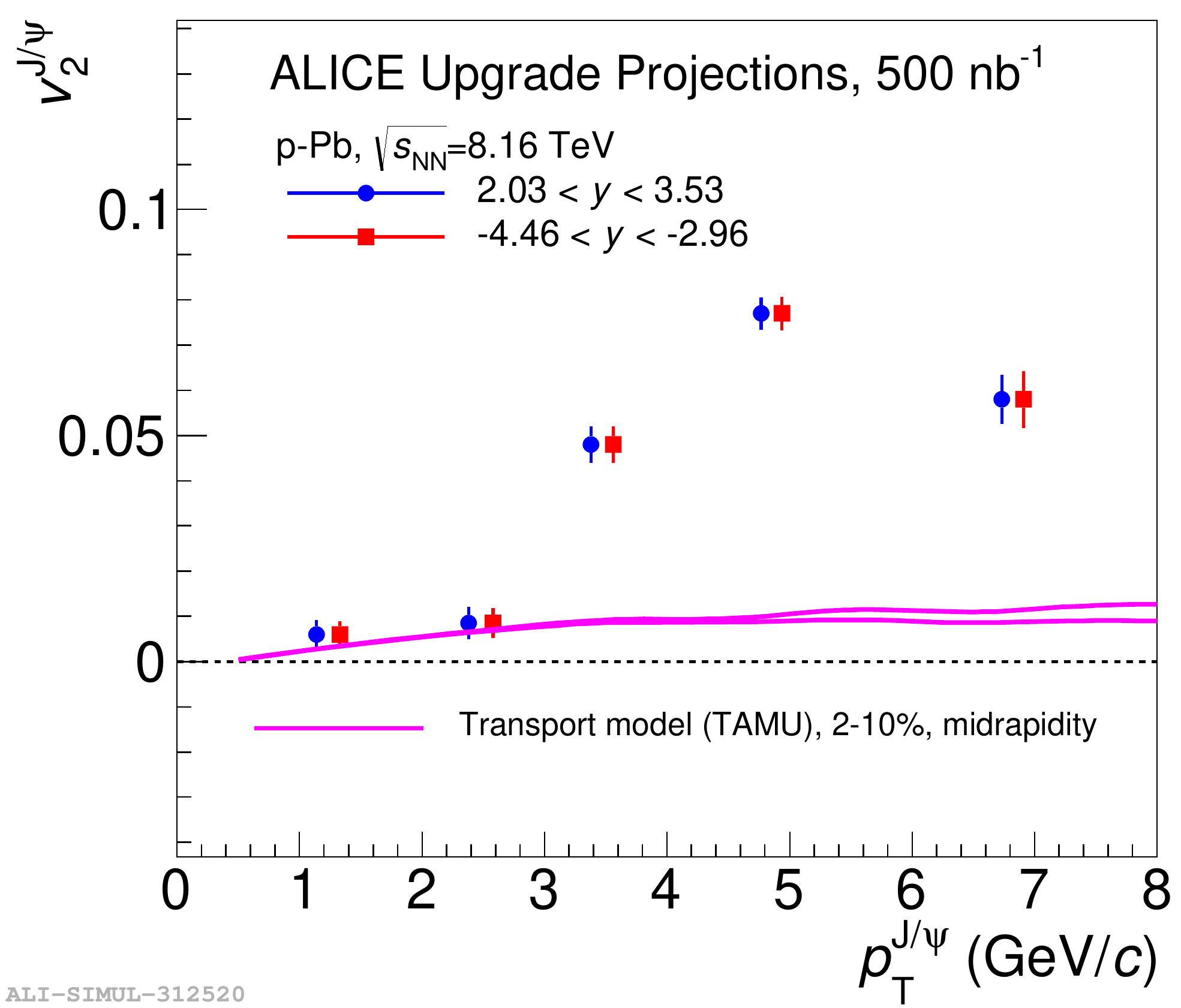}
 \end{center}
 \caption{The \pT dependence of the $v_2$ coefficient of \PJgy mesons in p--Pb collisions, for \unit[500]{nb}$^{-1}$ (ALICE). The projections are based on current ALICE data for 0--20\% centrality \cite{Acharya:2017tfn}  and are shown separately for negative and positive $y_{CM}$), assuming the same magnitude and are compared with transport model (TAMU) calculations \cite{Du:2018wsj} for midrapidity. Figure from Ref.~\cite{ALICE-PUBLIC-2019-001}.}
\label{FigQ:v2pTpPb}
\end{figure}

Recently, the measurement of $v_2$ of \PJgy in \pPb collisions became available~\cite{Acharya:2017tfn,Sirunyan:2018kiz}, indicating a large azimuthal anisotropy, $v_2 \lesssim 0.1$ up to $\pt \lesssim 8\UGeVc$.
Recent transport model calculations \cite{Du:2018wsj}, which are successful in describing the features of the data, including the transverse momentum and centrality dependence of  \PJgy  and \PGyP{2S} production in \pPb, cannot reach the high value of the $v_2$ coefficient seen in data \cite{Acharya:2017tfn,Sirunyan:2018kiz} (see Fig.~\ref{FigQ:v2pTpPb}),
suggesting that the observed $v_2$ in \pPb collisions might also originate from initial state effects.
A precision measurement in \RunsThreeFour for a broad rapidity range will clarify this.

In addition to conventional LHC collider data, one should not overlook the discriminating power of data which can be collected in the fixed-target mode~\cite{Brodsky:2012vg,Lansberg:2012kf}. Not only they correspond to completely different energy and (c.m.s.) rapidity ranges, but extremely competitive luminosities, up to a few fb$^{-1}$, are easily reachable, beyond what can be reached in the collider mode during \RunsThreeFour. The LHCb collaboration has paved the way for a full fixed-target program at the LHC with their SMOG luminosity monitor~\cite{FerroLuzzi:2005em} used as an internal (He, Ne, Ar) gas target \cite{Aaij:2018ogq} (see also Section~\ref{sec:fixedtarget}). It is now clear that corresponding studies to those suggested above are possible~\cite{Hadjidakis:2018ifr} with the LHCb and ALICE detectors with minor technical adjustments. They would drastically expand the scope of current proton-nucleus quarkonium studies.

\subsubsection{High-multiplicity \pp collisions} %

Systematic studies of the quarkonium production in high-multiplicity \pp events can play an important role in understanding hadronisation.
In particular, the correlation of the quarkonium yields with the charged-particle multiplicity can provide new insights into the interplay between hard and soft processes in particle production.
Hidden and open heavy-flavour production measurements as a function of the event activity were carried out  at the LHC during Run~1~\cite{Abelev:2012rz,Chatrchyan:2013nza}.
The striking feature of the data is that the production yields of quarkonia in high multiplicity events are significantly enhanced relative to minimum bias events, like for D mesons~\cite{Adam:2015ota}.
Specifically, the measurements of the self-normalised yields (the yield divided by the mean yield in minimum bias collisions) as a function of the self-normalised charged-particle multiplicity show an increase which is stronger than linear at the highest multiplicities.
The similarity between the D-meson and \PJgy results~\cite{Abelev:2012rz,Adam:2015ota} suggests that this behaviour is most likely related to the production processes, and that hadronisation may only play a secondary role.
When comparing \PJgy preliminary results at $\sqrt{s}=13$\,TeV~\cite{Weber:2017hhm} to the ones previously obtained at $\sqrt{s} = 7$\,TeV~\cite{Abelev:2012rz}, no significant energy dependence is observed, {\it i.e.} the relative \PJgy~yields for events with identical relative multiplicities give similar results. In addition, a dependence of the excited-to-ground-state ratio with charged particle multiplicity is observed in the bottomonium sector in \pp collisions~\cite{Chatrchyan:2013nza,CMS-PAS-BPH-14-009}.

The data are described both by initial-state models as well as by a model assuming hydrodynamic evolution \cite{Werner:2013tya}, considering that the energy density reached in \pp collisions at LHC is high enough to apply such evolution.
Initial-state (saturation) effects are considered within
i) the Color-Glass-Condensate (CGC) framework~\cite{Ma:2018bax}; ii) the percolation approach~\cite{Ferreiro:2012fb,Ferreiro:2015gea}; iii) a model with higher Fock states~\cite{Kopeliovich:2013yfa}, based on parameters derived from \pPb collisions.
The energy dependence of the cross sections is controlled by the saturation momentum $Q_s(x)$ in the case of the CGC or density of colour ropes $\rho_s(y,p_T)$ in the percolation model, which also governs the charged-hadron multiplicity; events at different energies with the same $Q_s$ or $\rho_s$ are therefore identical.
For a given event multiplicity, they predict the relative yields to be almost energy independent.
It seems that, in any case, multiple interactions at the partonic level need to be taken into account in order to reproduce the data~\cite{Sjostrand:2014zea,Skands:2014pea,Sjostrand:2017cdm}.

\RunsThreeFour data, reaching unprecedented high multiplicities because of larger data samples, and allowing for differential studies in \pT, will certainly help discriminate models.
For instance, in the percolation model, where colour interactions produce a reduction of the charged-particle multiplicities, the deviation from the linear behaviour is expected to be steeper for high-\pT quarkonia (and D mesons).
Moreover, measurements of \PJgy  yields relative to those of D mesons with the same transverse mass could help elucidate the relative contribution of hadronisation and initial-state effects.

Studies of double differential ratios of excited-to-ground quarkonium states versus relative multiplicity could help clarify the presence of final-state effects, either QGP-like or the ones proposed by the comover model~\cite{Ferreiro:2014bia,Ferreiro:2018wbd}.
Also, within the CGC+NRQCD framework~\cite{Ma:2018bax}, the relative contributions of the 4 leading \PJgy Fock states have been calculated as a function of the event activity, showing a different dependence for different Fock states.

\clearpage

\section{Electromagnetic radiation}
\label{chapter:electromagnetic_radiation}

{ \small
\noindent \textbf{Coordinator}: Michael Weber (Stefan Meyer Institute Vienna, Austrian Academy of Sciences) 

\noindent \textbf{Contributors}: 
R.~Bailhache (Goethe-University Frankfurt), 
R.~Chatterjee (VECC Calcutta),
T.~Dahms (Excellence Cluster Universe, Technical University Munich), 
T.~Gunji (Center for Nuclear Study, Graduate School of Science, the University of Tokyo), 
M.~He (Nanjing University of Science and Technology),
S.R.~Klein (Lawrence Berkeley National Laboratory), 
A.~Marin (GSI Helmholtzzentrum f{\"u}r Schwerionenforschung GmbH), 
D.~Peresunko (National Research Centre Kurchatov Institute, Moscow),  
R.~Rapp (Texas A\&M University, College Station), 
K.~Reygers (Heidelberg University), 
T.~Song (University of Gie{\ss}en), 
A.~Uras (Universit{\'e} de Lyon, CNRS/IN2P3, IPN-Lyon),
G.~Vujanovic (Ohio State University and Wayne State University)
}

The strongly interacting system formed in ultrarelativistic heavy-ion collisions emits electromagnetic radiation that can be detected using different probes: real {\it direct} photons or virtual photons measurable via dilepton pairs. 
Direct photons can be split into {\it prompt} photons, emitted by the partons of colliding nuclei during their inter-penetration, and {\it thermal} photons, emitted by the almost thermalized hot system. 
For dileptons these contributions are called Drell-Yan and thermal, respectively.
In contrast to real photons, dileptons carry a mass and thus can be used to study the decay of massive particles, such as the in-medium modified spectral shape of vector mesons, the \PGr meson being the most prominent one, and the search for particles beyond the Standard Model, \eg dark photons. In this section, we outline the measurement of photons via calorimetry and the so-called photon conversion method, as well as dielectron (\Pepem), and dimuon (\PGmpGmm) pairs in \AOnA collisions in the ALICE detector at the LHC. Moreover, the photoproduction of dilepton pairs in peripheral collisions and the expected sensitivity for the search of dark photons are discussed in subsections \ref{sec:dileptons:peripheral} and \ref{sec:dileptons:darkphotons}, respectively. We begin with a short review of previous experimental results together with a summary of the basic theoretical models employed to describe these data.

\subsection{Thermal radiation and in-medium spectral function}
\label{sec:thermalradiation}

Electromagnetic radiation from the hot and dense system formed in ultrarelativistic heavy-ion collisions in the form of real photons was measured for the first time at the SPS by WA98~\cite{Aggarwal:2000th}. The direct photon spectrum measured in \PbPb{} collisions at $\sqrtsNN = \unit[17.3]{\UGeV}$ showed an excess above the extrapolated prompt photon signal based on measurements in proton induced reactions. 
The excess is described by a large variety of hydrodynamic and cascade models (see \cite{Peitzmann:2001mz} for review), most of which assume the formation of a QGP phase.
Also at the SPS, a modification of low-mass dilepton pairs in S--Au and Pb--Au collisions relative to the expectation of in-vacuum hadron decays was observed by CERES~\cite{Agakishiev:1995xb,Agakishiev:1997au,Adamova:2002kf,Agakichiev:2005ai,Adamova:2006nu} and studied with high precision by NA60 in In--In collisions~\cite{Arnaldi:2006jq,Arnaldi:2007ru,Arnaldi:2008fw,Specht:2010xu}. 
The data are consistent with an in-medium \PGr spectral function that, driven by the coupling to baryons, melts and approaches the one from $\PQq\PAQq$ annihilation in the vicinity of the phase transition~\cite{Rapp:1995zy,Rapp:1999us,Rapp:2009yu}, which is compatible with chiral symmetry restoration~\cite{Bazavov:2011nk,Hohler:2013eba}. On the other hand, the data cannot be described with a dropping mass scenario, in which the \PGr mass drops to zero as chiral symmetry is restored~\cite{Brown:1991kk}. Beyond the issue of chiral symmetry restoration, NA60 measured an excess of prompt dimuons in the intermediate mass region between the \PGf and the \PJGy masses~\cite{Arnaldi:2007ru,Arnaldi:2008fw}. Contrary to transverse-mass spectra of the dimuon continuum at lower masses, this excess shows no increase of the exponential inverse slope with mass, \ie blue shift, that is typical for radial flow. This suggests that the source of this enhancement is from the earliest phase of the collision, before significant radial flow has built up. 
This supports the idea that the inverse slope of the invariant mass spectrum is insensitive to the expansion of the medium and therefore a true measure of the average temperature. NA60 measured a value of $T=\unit[205\pm12]{\UMeV}$~\cite{Specht:2010xu}, which significantly exceeds the temperature of $\unit[154\pm9]{\UMeV}$, above which the formation of a QGP has been predicted~\cite{Borsanyi:2010bp,Bazavov:2014pvz}.

At RHIC energies, PHENIX and STAR have measured an enhancement of \Pepem pairs in the low mass region in \AuAu{} collisions~\cite{Adare:2009qk,Adare:2015ila,Adamczyk:2013caa,Adamczyk:2015mmx} that can be described with the same model of collisional broadening as used at the SPS. STAR measured that the enhancement above the hadron decay background does not change with collision energy between $\sqrtsNN=\unit[19]{}$ and \unit[200]{\UGeV}~\cite{Adamczyk:2015mmx}. Despite a marked decrease of the net-baryon chemical potential in this energy range, the total baryon plus anti-baryon density does not change much, providing further evidence that the $\PGr$ coupling to baryons and antibaryons is responsible for the enhancement.
Real direct photon production in \AuAu{} collisions was measured by PHENIX~\cite{Adare:2008ab,Adare:2014fwh,Adare:2018wgc}. An excess was observed compared to binary scaled direct photon production in \pp{} collisions. The signal was measured via quasi-real virtual photons, \ie \Pepem pairs  with small invariant mass, as well as real photons converting in detector material. The excess yield at low \pT{} appears to have a universal multiplicity dependence, scaling with the charged-particle multiplicity at midrapidity to the power of about 1.25, independent of collision energy between $\sqrtsNN = \unit[39]{}$ and \unit[200]{\UGeV}~\cite{Adare:2018wgc}. The transverse momentum spectrum of the excess yield has an exponential inverse slope of $T=\unit[221\pm19~\mathrm{(stat.)}\pm19~\mathrm{(syst.)}]{\UMeV}$ for central collisions and values close to that for other centralities. The spectrum, however, is strongly blue shifted by radial flow in the later stages of the fireball radiation, which is further supported by a sizeable elliptic flow ($v_2$) of the direct photon signal~\cite{Adare:2011zr}. Therefore, the inverse slope cannot directly be interpreted as an average temperature, which highlights the importance of thermal dilepton measurements as a function of invariant mass. However, the modelling of the space-time evolution offers the possibility of extracting temperature information from the photon data \cite{Shen:2013vja}. The direct photon $v_2$ is indeed comparable to the $v_2$ of pions, which suggests late emission of direct photons dominated by the hadronic phase \cite{vanHees:2011vb}. A simultaneous description of the elliptic flow effect, as well as the large direct photon excess, which implies early production, poses a significant challenge to theoretical models.

\begin{figure}[tb]
\centering
\includegraphics[width=0.6\textwidth]{\main/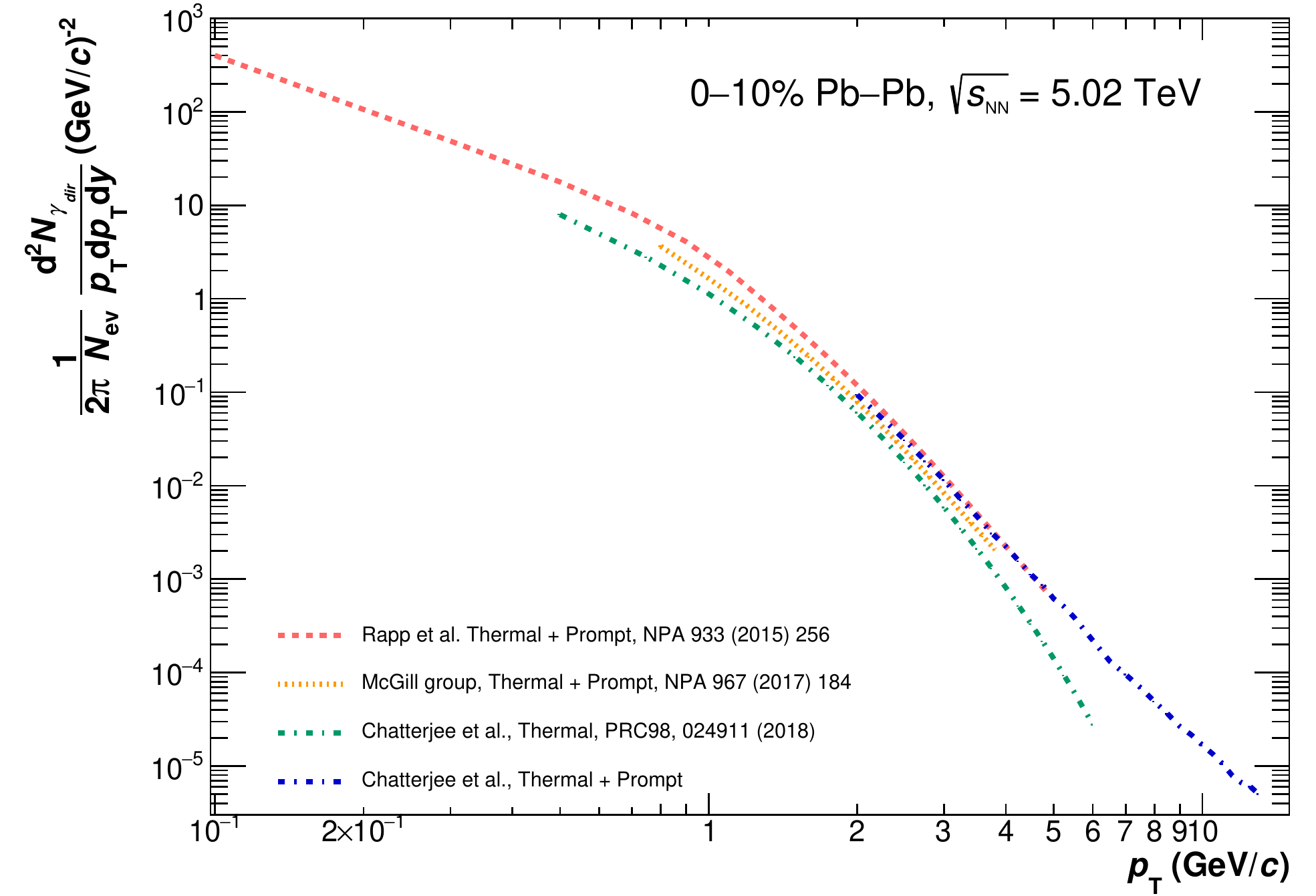}
\caption{Direct photon differential invariant yield for central 0--10\% \PbPb{} collisions at $\sqrtsNN = \unit[5.02]{\UTeV}$ as predicted by several models \cite{Paquet:2015lta,Paquet:2016ime,Paquet:2017wji,Dasgupta:2018pjm,vanHees:2014ida}.}
\label{fig:LHCExpectations_RealPhotons}
\end{figure}

The first measurement of direct photon production in \PbPb collisions with the ALICE detector~\cite{Aamodt:2008zz} at the LHC also show an excess of thermal production at low $\pT < \unit[3]{\UGeVc}$ with respect to the prompt direct photon expectation~\cite{Adam:2015lda}. The extracted effective temperatures $T=\unit[297\pm12~\mathrm{(stat.)}\pm41~\mathrm{(syst.)}]{\UMeV}$ in central collisions and $T=\unit[410\pm84~\mathrm{(stat.)}\pm140~\mathrm{(syst.)}]{\UMeV}$ in semi-central collisions are higher than those at RHIC energies, as expected.
The direct photon elliptic flow was also extracted in central and semi-central collisions~\cite{Acharya:2018bdy}. The measured flow is close to the one at RHIC energy and at low $\pT < \unit[4]{\UGeVc}$ to the one of final hadrons. However, this measurement does not cause the same challenges to models, since the experimental uncertainties are still large at this point.
The reduction of systematic uncertainties of the direct photon measurement is the main objective for Run 3 to improve its significance. Moreover, a low magnetic field run will allow one to access the $\pT < \unit[1]{\UGeVc}$ region where the thermal photon production increases rapidly. Theoretical calculations of thermal and prompt photon productions are available at $\sqrtsNN = \unit[5.02]{\UTeV}$ (Fig.~\ref{fig:LHCExpectations_RealPhotons}) \cite{Paquet:2015lta,Shen:2016zpp,Dasgupta:2018pjm,vanHees:2014ida}. The thermal contribution dominant at low \pT is given by the QGP photon emission rates and the hadronic photon production rates, integrated over the space-time evolution of the system. In \cite{Paquet:2015lta} a (2+1)D hydrodynamic evolution with IP-Glasma initial conditions with initial flow and finite shear and bulk viscosity is followed by a hadronic phase modelled using UrQMD.
A longitudinal boost invariant (2+1)D ideal hydrodynamics is used in \cite{Dasgupta:2018pjm}, while in  \cite{vanHees:2014ida} a (2+1)D ideal hydrodynamic model including non-vanishing initial flow is employed. The prompt photon component, dominant at high \pT, is very similar in all models. It is obtained from NLO pQCD calculations using the BFG-II photon fragmentation function and the CTEQ6.6~\cite{Dasgupta:2018pjm} or nCTEQ15 parton distribution function \cite{Shen:2016zpp}. 
An increase by a factor $\sim$ 1.5 at about $\pT \approx \unit[1]{\UGeVc}$ and by a factor 1.5 to 2 for the prompt photons is predicted compared to yields at $\sqrtsNN = \unit[2.76]{\UTeV}$. The predicted thermal photon elliptic flow parameters for central collisions are close to each other at the two LHC energies  and are very small. Differences become larger as one goes towards peripheral collisions. Simultaneous measurements of photon yields and photon flow with high accuracy and lower \pT{} reach will provide constraints to theoretical models.

Dilepton measurements by ALICE at the LHC are not yet sensitive to possible low-mass enhancement and thermal signals~\cite{Acharya:2018nxm}. A precise measurement of the low-mass dielectron continuum will be one of the main objectives of the ALICE physics programme during the LHC Run 3 and 4. In the meanwhile, the dominant background of dielectrons from correlated semileptonic open heavy-flavour decays is utilised to learn more about open heavy-flavour production in \pp{} collisions at LHC energies~\cite{Acharya:2018ohw,Acharya:2018kkj}.

\begin{figure}[tb]
\centering
\includegraphics[width=0.45\textwidth]{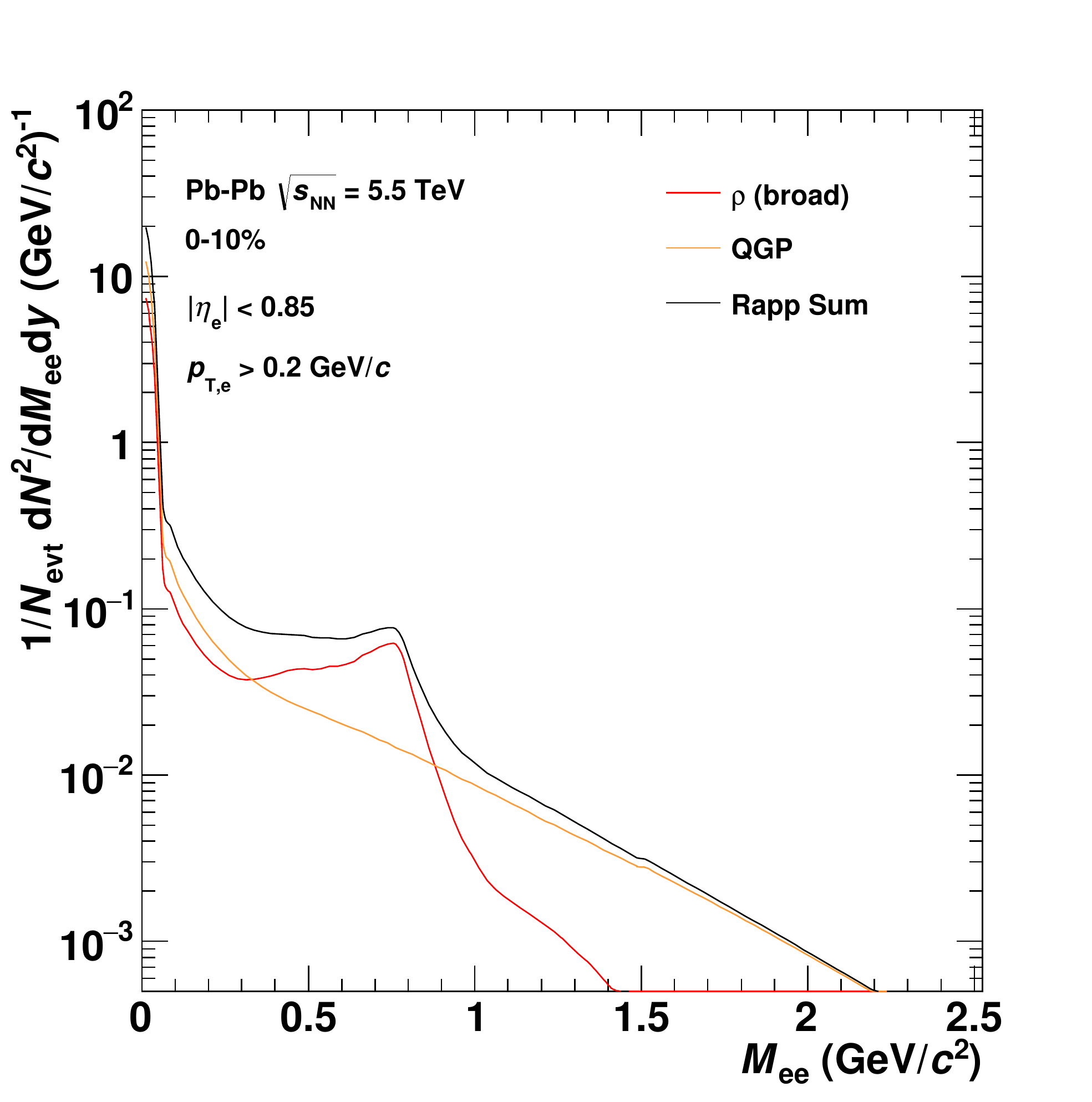}
\includegraphics[width=0.45\textwidth]{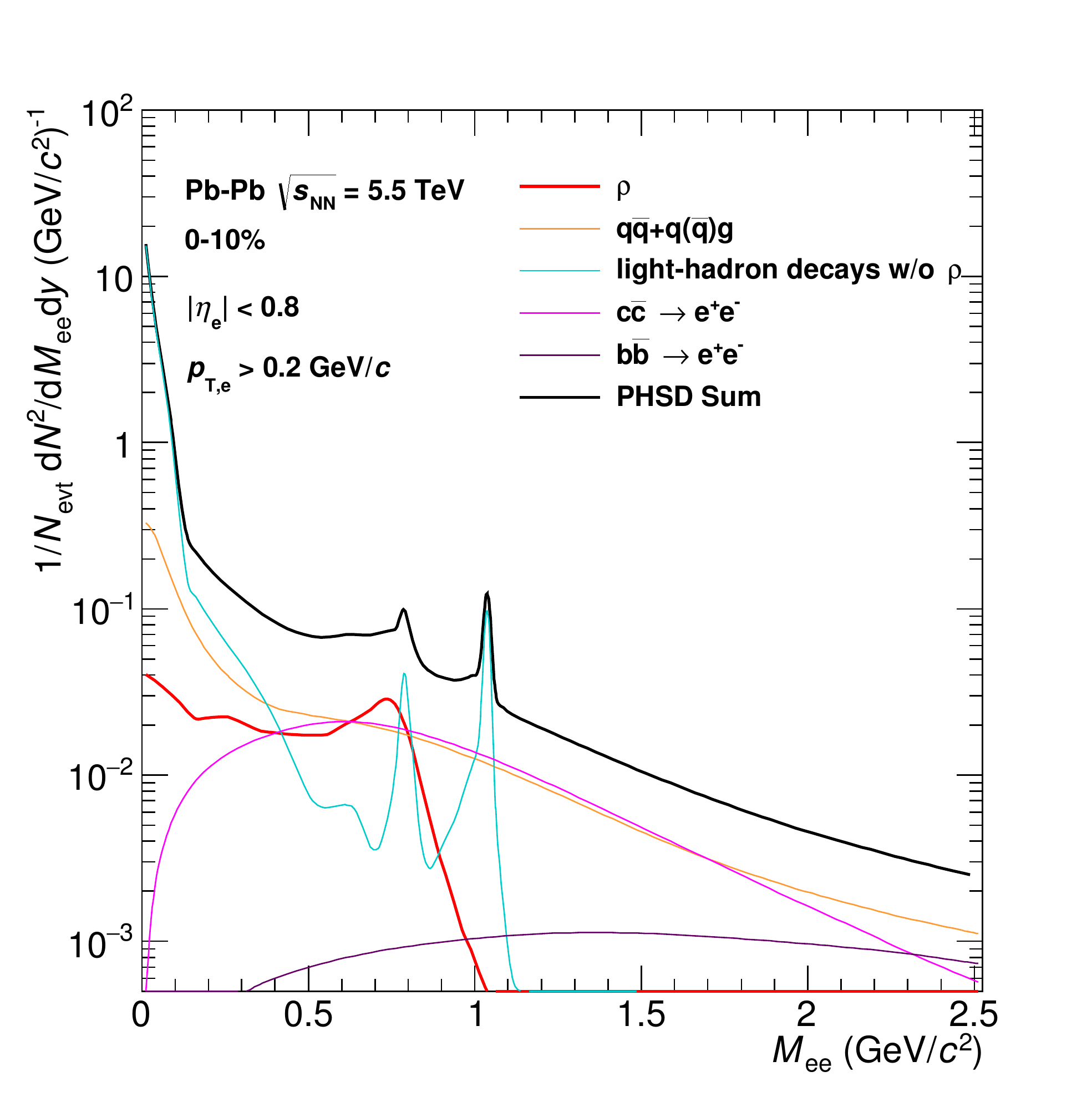}
\caption{Model predictions for the invariant mass spectrum of \Pepem pairs in central (0--10\%) \PbPb{} collisions at $\sqrtsNN = \unit[5.5]{\UTeV}$. Left panel: in-medium radiation plus decays of the \PGr meson at the end of the system evolution by R.~Rapp \etal. Right panel: Expectations from the PHSD model including the in-medium \PGr meson, $\PQq\PAQq\rightarrow\Pepem$, $\PQq\PAQq\rightarrow\Pepem\Pg$, and $\PQq(\PAQq)\Pg\rightarrow\Pepem\PQq(\PAQq)$, hadronic sources, and semileptonic decays of $c\bar{c}$ and $b\bar{b}$.}
\label{fig:LHCExpectations_Rapp_pHSD}
\end{figure}

The model by R.~Rapp \etal, an approach that has been proven to provide a quantitative description of the existing dilepton results \cite{Rapp:2011is}, is based on two ingredients that are put into a realistic space-time evolution \cite{Rapp:2000pe}. The thermal dilepton radiation is modelled by emission rates from the hadronic phase and the Quark--Gluon Plasma \cite{vanHees:2007th,Rapp:2009yu}. A hadronic many-body approach \cite{Rapp:1999us} is used for the medium-modified spectral functions of $\PGr$ and $\PGo$ mesons. In addition, the equation of state is updated to a cross-over transition around $T_{\rm c}=\unit[170]{\UMeV}$ extracted from with recent lattice QCD computations, and hadro-chemical freezeout at $T_{\rm chem}=\unit[160]{\UMeV}$ \cite{He:2011zx}. Figure \ref{fig:LHCExpectations_Rapp_pHSD} (left) shows the calculations performed for central \PbPb{} collisions at $\sqrtsNN = \unit[5.5]{\UTeV}$ for in-medium radiation plus decays of the \PGr meson at the end of the system evolution. The pair-yield is estimated for the rapidity range $|y_{\rm e}|<0.85$ and transverse momentum of single electrons $\pT^{\rm e}>\unit[0.2]{\UGeVc}$ and is normalized to the number of events $N_{\rm evt}$. 
 
A complementary approach to study dilepton spectra and thermal radiation is provided by the parton-hadron-string dynamics (PHSD) transport approach, which also successfully describes the existing experimental data \cite{Linnyk:2015rco,Cassing:2009vt}. The in-medium modification of the \PGr meson is incorporated in PHSD by an off-shell transport of vector mesons with a dynamically changing set of spectral functions \cite{Bratkovskaya:2007jk} evolving towards the vacuum spectral function at the end of the collision history. The electromagnetic radiation of the QGP is modelled by $\PQq\PAQq\rightarrow\Pepem$, $\PQq\PAQq\rightarrow\Pepem\Pg$, and $\PQq(\PAQq)\Pg\rightarrow\Pepem\PQq(\PAQq)$ using effective propagators for quarks and gluons from a dynamical quasi-particle model \cite{Linnyk:2010vb}. Figure \ref{fig:LHCExpectations_Rapp_pHSD} (right) shows both contributions to the dielectron spectrum in central \PbPb{} collisions at $\sqrtsNN = \unit[5.5]{\UTeV}$ calculated from PHSD together with other sources of dielectrons: decays of long-lived light mesons into \Pepem (the so-called hadronic cocktail) and the semileptonic decay of hadrons containing heavy quarks, such as \PD and \PB mesons. 

Important input for models aiming to describe the dilepton yield at LHC energies are the in-medium spectral functions for the vector mesons, most importantly the \PGr meson, as well as the photon and dilepton rates from the QGP. For the latter, Lattice QCD calculations, which are currently limited to the quenched approximation, will hopefully be extended (\eg larger lattices, especially in the time direction, or facilitating extrapolations to the continuum limit) and be available at higher accuracy for realistic systems including light dynamical degrees of freedom in the future. Recent updates on calculations of the photon rate \cite{Ghiglieri:2016tvj}, the electrical conductivity \cite{Aarts:2014nba}, and dilepton rates \cite{Ding:2016hua} are promising.
The photons and dilepton rates from Lattice calculations should in the future be combined with dynamical models like those in Fig.~\ref{fig:LHCExpectations_Rapp_pHSD}, thus improving their results. In addition, the in-medium spectral functions could also use direct input from Lattice QCD \cite{Aarts:2005hg,Brandt:2015aqk} or from a functional renormalization group approach \cite{Jung:2016yxl}. These models can further be refined by including the effects of dissipation, and in that case the electrical conductivity will become of interest to both the dynamical evolution of the medium as well as the electromagnetic rates. In order for that to be achieved self-consistently, the evolution of the medium and the electromagnetic rates need to be modified to account for dissipative effects, which is a currently ongoing effort \cite{Paquet:2015lta,Vujanovic:2017wtw,Vujanovic:2017psb}.

More differential information can be used to study the equation of state of the system throughout the full collision history. The measurement of the elliptic flow coefficient $v_2$ of thermal photons and dileptons, especially if combined with results from hadronic channels, should put tighter constraints on fundamental properties of the medium (\eg transport coefficients), as well as its "initial conditions" or "pre-equilibrium" dynamics \cite{Vujanovic:2016anq}. For example, owing to the penetrating nature of dileptons, the invariant mass dependence of the dilepton $v_2$ is sensitive to the temperature dependence of both shear \cite{Vujanovic:2017psb} and bulk viscosity \cite{Vujanovic:2017wtw} in a way that is difficult to access using hadronic observables alone.

\subsubsection{Real photons}

\begin{figure}[t]
\centering
\includegraphics[width=0.46\textwidth]{\main/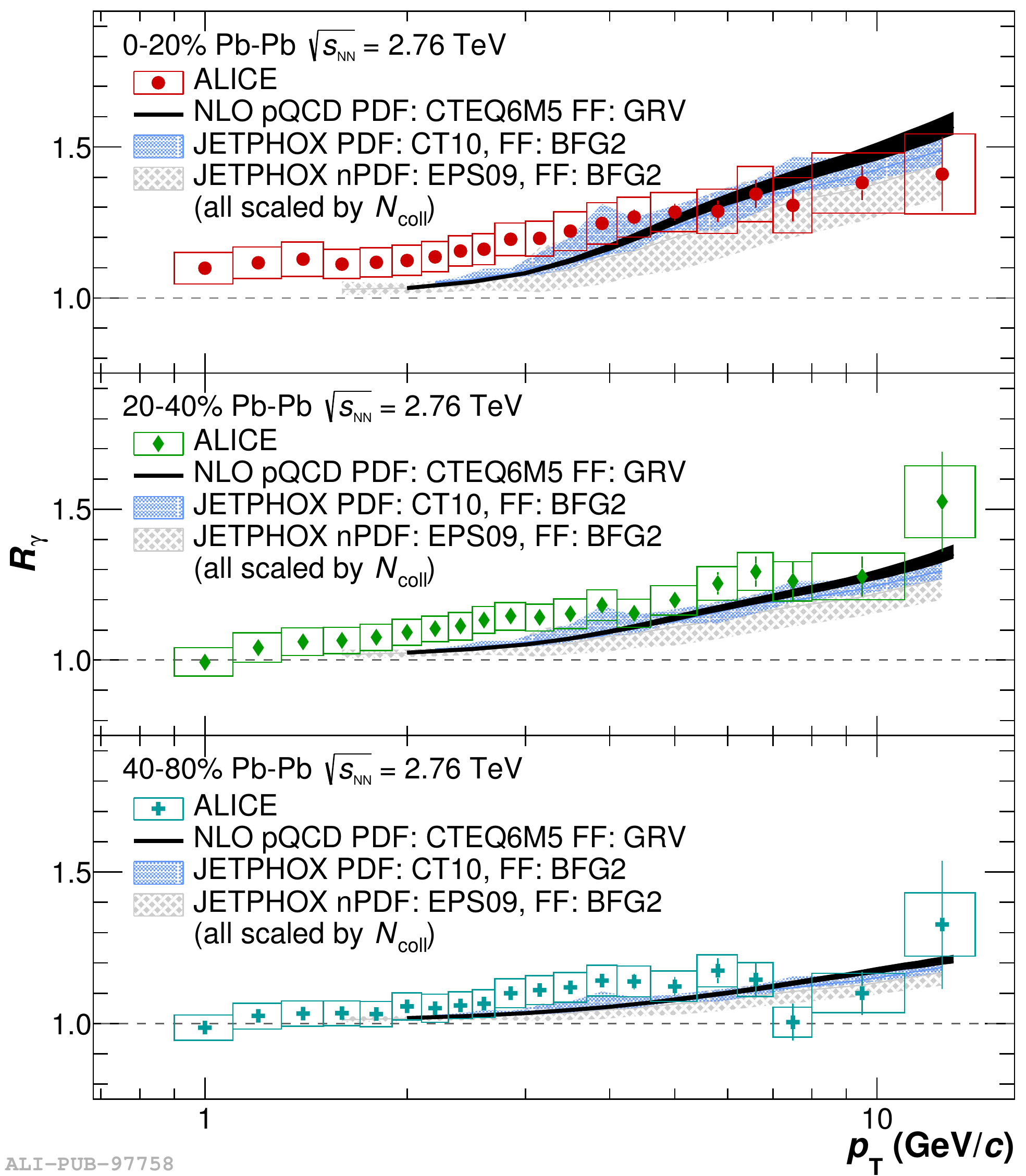}
\hfill
\includegraphics[width=0.425\textwidth]{\main/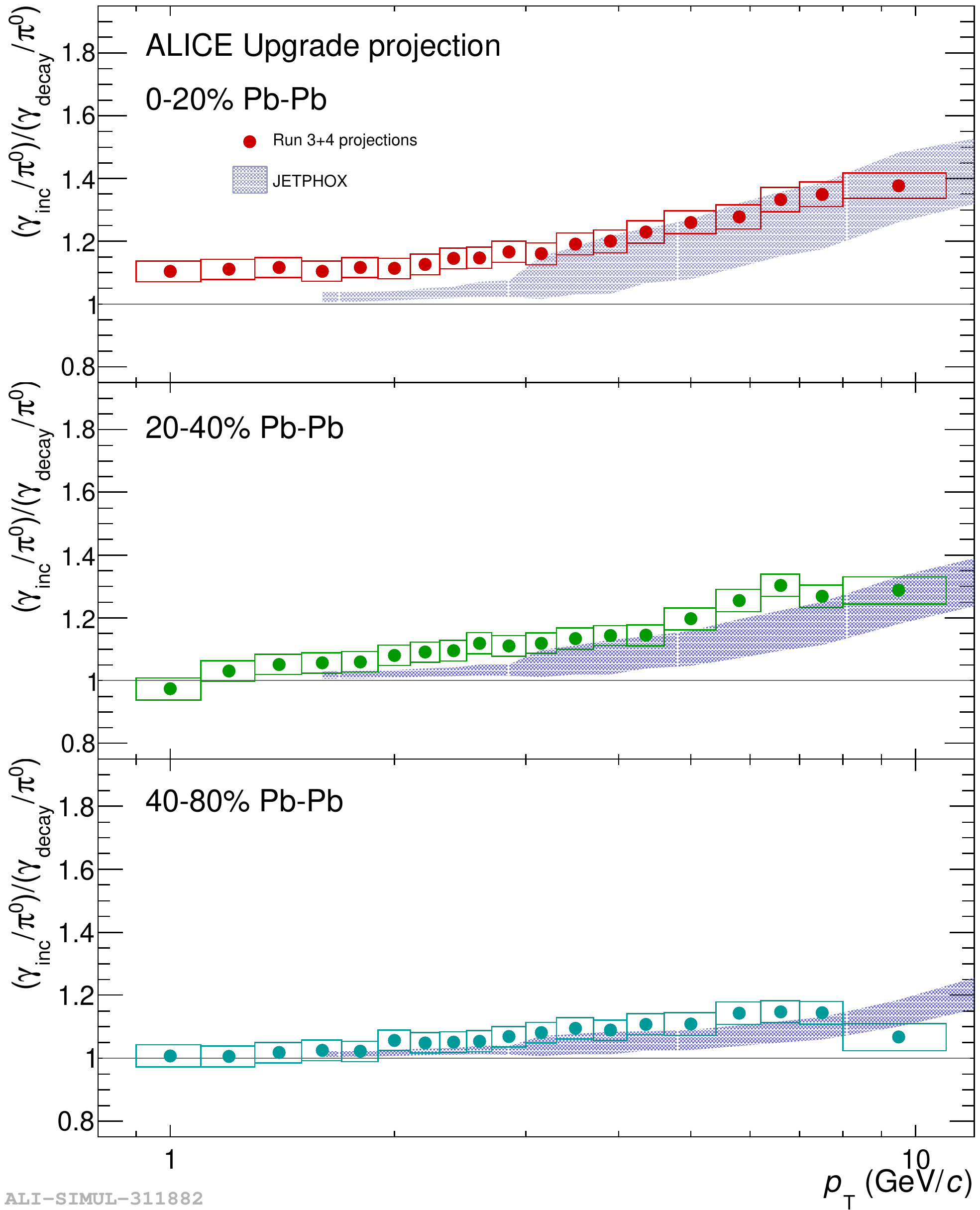}
\caption{$R_{\PGg}$ measured \cite{Adam:2015lda} (left) and $R_{\PGg}$ projected~\cite{ALICE-PUBLIC-2019-001} keeping the measured values of $R_{\PGg}$ and recalculating the uncertainties as explained in the text.}
\label{fig:RealPhotonsRg}
\end{figure}

ALICE has measured direct photon spectra in three centrality classes in \PbPb collisions at $\sqrtsNN=\unit[2.76]{\UTeV}$ \cite{Adam:2015lda}. 
An excess of direct photons was quantified by the $\pT$ dependent double ratio
\begin{equation}
  \label{eq:doubleratio}
  R_{\PGg}  \equiv \left . \frac{\PGg_{\mathrm{incl}}}{\PGpz_{\mathrm{param}}} \right / \frac{\PGg_{\mathrm{decay}}}{\PGpz_{\mathrm{param}}}
 = \frac{\PGg_{\mathrm{incl}}}{\PGg_{\mathrm{decay}}}, 
\end{equation}
where $\PGg_{\mathrm{incl}}$ is the measured inclusive photon spectrum, $\PGpz_{\mathrm{param}}$ a parametrization of the measured $\PGpz$ spectrum, and $\PGg_{\mathrm{decay}}$ the calculated decay photon spectrum. The double ratio has the advantage that some of the largest systematic uncertainties cancel partially or completely. 
The measurement combines results of the Photon Conversion Method (PCM) and of the Photon Spectrometer (PHOS), see Fig.~\ref{fig:RealPhotonsRg}, left. In central collisions at low $\pT  < \unit[4]{\UGeVc}$ an excess with respect to prompt photon predictions is observed that is attributed to thermal photon emission from the QGP. In the 20\% most central collisions the low \pT{} excess is of the order of 10--15\%, while the total uncertainty of the order of 6\%. A signal of direct photons is found in central collisions, but on the level of $\sim 2\sigma$, while in mid-central and especially in peripheral the significance is even smaller. On the other hand, peripheral collisions are important since there one can estimate and restrict the contribution from prompt direct photons.  

For Run 3 the PCM measurement will be influenced by the ALICE Inner Tracking System (ITS) and Time Projection Chamber (TPC) upgrades, while PHOS and the Electromagnetic Calorimeter (EMCal) will be kept unchanged. 
The new ITS shows an improved low \pT{} tracking efficiency, that will partially compensate the efficiency loss due to the $\sim30\%$ reduction of its material thickness. Two \unit[1]{\Umm} tungsten wires with well known thickness will be installed parallel to the beam direction for precise calibration of the material thickness as described later. 
The TPC continuous readout mode together with large pile-up may prevent the use of photon conversions beyond a radius of \unit[35]{\Ucm}. This restriction will translate into a $\sim35\%$ lower photon efficiency. On the other hand, the PCM measurement will also profit from the dedicated heavy-ion run with reduced magnetic field of the ALICE solenoid, which will considerably increase the low \pT{} reconstruction efficiency.
To estimate how one can improve the accuracy of the measurement, the uncertainties are split into three classes: those which can be improved with increase of statistics (statistical uncertainties, uncertainties related to \PGpz spectrum extraction, $\PGh/\PGpz$ ratio); uncertainties which can be reduced using new techniques and some special methods (material budget estimate - with calibrated material analysis, energy scale in calorimeters with new hybrid \PGpz methods); and uncertainties related to the properties of the detector which can not be improved (hadron contamination in calorimeters, electron identification in conversion method \etc). 
To estimate the improvement of the uncertainties it is assumed that the integrated luminosity will reach $\Lint=\unit[3.1]{\Unb^{-1}}$ per \PbPb run and in total $L_{\rm int}=\unit[13]{\Unb^{-1}}$ at the end of Run 4, which is more than a factor of 100 larger the than integrated luminosity $L_{\rm int}=\unit[10]{\Uub^{-1}}$ used in the published analysis \cite{Adam:2015lda}. 
The major improvement foreseen for Run 3 is the use of calibrated tungsten wires
inserted into the ITS to determine the product of the photon flux times the \PGg reconstruction efficiency. This product would then be used to precisely determine the material thickness in the rest of the ITS (assuming $\varphi$-independent photon flux and taking the radial dependence of the reconstruction efficiency from simulation).
The proposed calibration method is based on weights calculated 
as the double ratio: 
\begin{equation}
\omega_i= 
{\left(\frac{N_{\PGg}^{\rm rec}(r_i)}{N_{\PGg}^{\rm rec}(r_{\rm wire})}\right)^{\rm data}} /
{\left(\frac{N_{\PGg}^{\rm rec}(r_i)}{N_{\PGg}^{\rm rec}(r_{\rm wire})}\right)^{\rm MC}} 
\end{equation}
where $N_{\PGg}^{\rm rec}(r_i)$ and $N_{\PGg}^{\rm rec}(r_{\rm wire})$ are the number of reconstructed \PGg 's in data or in MC simulations in a given radial bin and the calibrated wire, respectively. 
For the Run 3 projections a systematic uncertainty of 1\% on the ITS thickness is taken. The uncorrelated systematic uncertainties on the \PGpz and \PGh measurements will be reduced by a factor 10 due to the increased luminosity as they are defined mostly by the raw yield extraction uncertainties which scale proportional to statistical uncertainties. 
The systematic uncertainties on the photon selection and particle identification are expected to be reduced by 50\%.
Figure~\ref{fig:RealPhotonsRg} (right) shows the projection of the $R_{\PGg}$ measurement 
for Run 3 calculated with these assumptions: the measured values of $R_{\PGg}$ are kept but the uncertainties are recalculated. The total errors are reduced by $\sim50$\%. In addition to the reduction of the uncertainties, the large data set foreseen for Run 3 will allow exploration of the 0--1\% centrality range. 

\begin{figure}[t]
\centering
\includegraphics[width=0.45\textwidth]{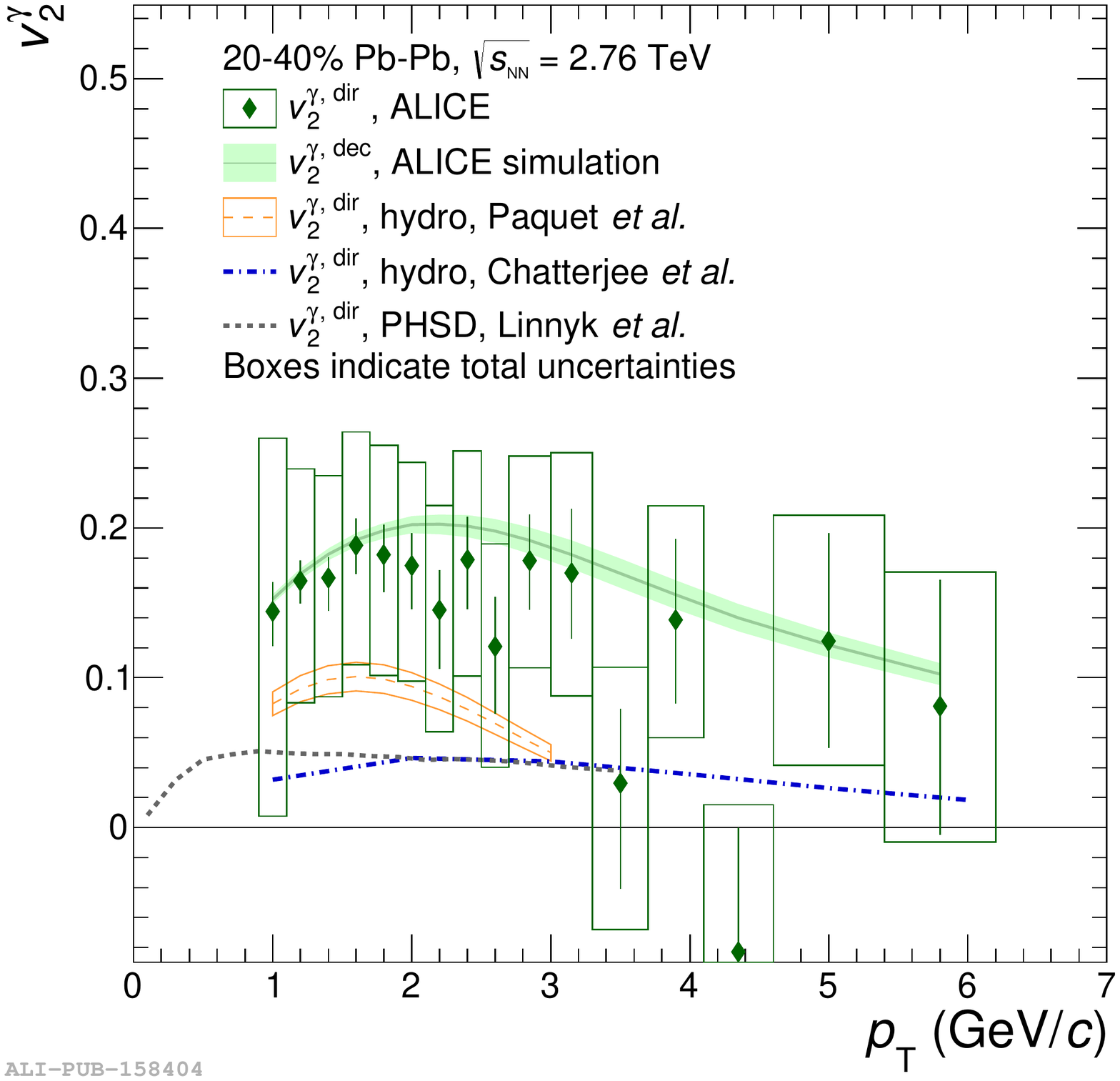}
\includegraphics[width=0.42\textwidth]{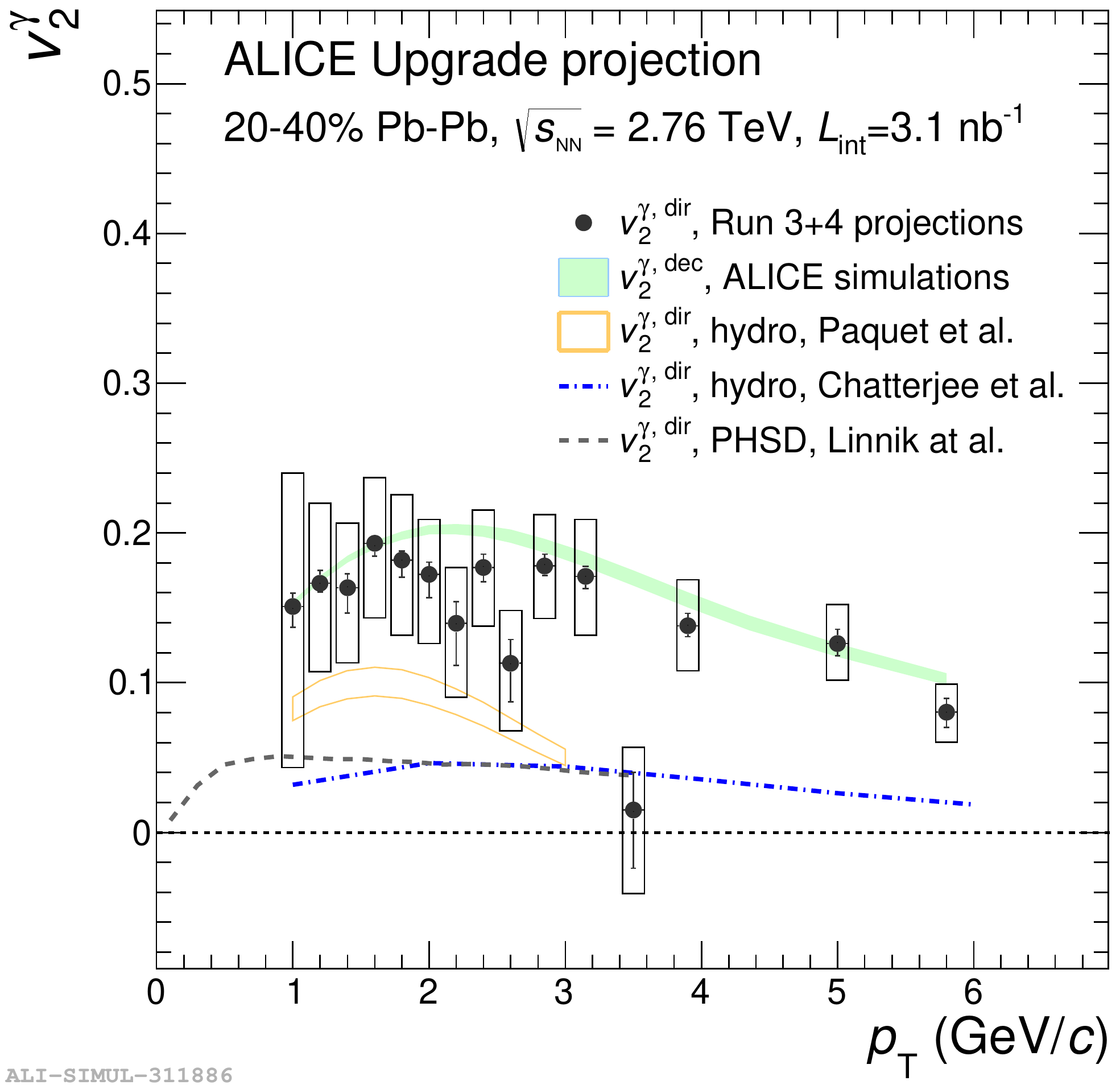}
\caption{Direct photon flow in mid-central collisions. Left: direct photon collective flow measured in \PbPb{} collisions compared to decay photon flow and several theoretical predictions. Right: expected accuracy in Run 3 keeping the  measured values of $R_{\PGg}$ and $v_{2}^{\PGg}$ and recalculating the uncertainties as explained in the text. Figure from Ref.~\cite{ALICE-PUBLIC-2019-001}.}
\label{fig:RealPhotonsV2dir}
\end{figure}

The ALICE Collaboration carried out measurements of the direct photon elliptic flow \cite{Acharya:2018bdy} in \PbPb{} collisions at $\sqrtsNN=\unit[2.76]{\UTeV}$ for the two centrality classes, 0--20\% and 20--40\%, see Fig.~\ref{fig:RealPhotonsV2dir}, left plot for 20--40\% centrality.
The measured direct photon elliptic flow $v_2^{\PGg,\rm dir}$ is compared to the estimated decay photon elliptic flow $v_2^{\PGg,\rm dec}$, marked as cocktail, and to the predictions of several theoretical models. Similar to RHIC measurements, the direct and decay photon elliptic flow are very close and systematically higher than theoretical predictions of hydrodynamic  \cite{Gale:2014dfa,Chatterjee:2017akg} and the transport  \cite{Linnyk:2015tha} models. However, because of the large uncertainties one can not presently exclude either of theoretical calculations. Using the same assumption concerning photon and neutral measurements in Run 3 as for $R_{\PGg}$, the expected accuracy of $v_2^{\PGg,\rm dir}$ measurements in Run 3 is estimated, see Fig.~\ref{fig:RealPhotonsV2dir} (right). The mean values are kept the same but the uncertainties are reduced as expected. Similar to $R_{\PGg}$ with the current assumptions the total errors will be reduced by factor $\sim$ 2 and one will be able to exclude or confirm available theoretical calculations.

\subsubsection{Dileptons}
\label{sec:thermalradiation:dileptons}

The sensitivity to the expected signal of thermal radiation and an in-medium modification of the \PGr spectral function in the dielectron and dimuon channels with the ALICE detector \cite{Aamodt:2008zz,Abelev:2014ffa} was studied already in preparation for ITS upgrade in 2019/20 \cite{Abelevetal:2014cna,Abelevetal:2014dna,ALICE:2014qrd,CERN-LHCC-2013-014}. The measurement of low-mass dileptons after this upgrade will profit from  
\begin{itemize}
\item an improved vertex resolution, which leads to a better separation of electrons from prompt sources, like thermal radiation, and electrons from the decays of heavy-flavour hadrons, for which $c\tau$ is about \unit[150]{\Uum} (open-charm hadrons) or \unit[400]{\Uum} (open-beauty hadrons), 
\item a reduced material budget and improved tracking efficiency at low transverse momentum \pT, which leads to a smaller background of electrons and positrons from photon conversion in the detector material,
\item the installation of the muon forward tracker, that will lead to an improved mass resolution and reduced background in the dimuon channel,
\item and a higher rate capability (\unit[50]{\UkHz} in \PbPb{} collisions) that will increase the expected number of events in the central barrel detector by a factor of 100. The increased rate capability also enables the possibility to record in a single \PbPb{} run a large data sample with reduced magnetic field value in the ALICE central barrel ($B=\unit[0.2]{\UT}$ instead of \unit[0.5]{\UT}), which increases the phase-space acceptance and the reconstruction efficiency of low momentum electrons and positrons.
\end{itemize}
The expected measured spectra discussed in this section closely follow the strategy that is discussed in more detail in \cite{Abelevetal:2014cna,Abelevetal:2014dna,ALICE:2014qrd,CERN-LHCC-2013-014}.

\begin{figure}[tb]
\centering
\includegraphics[width=0.45\textwidth]{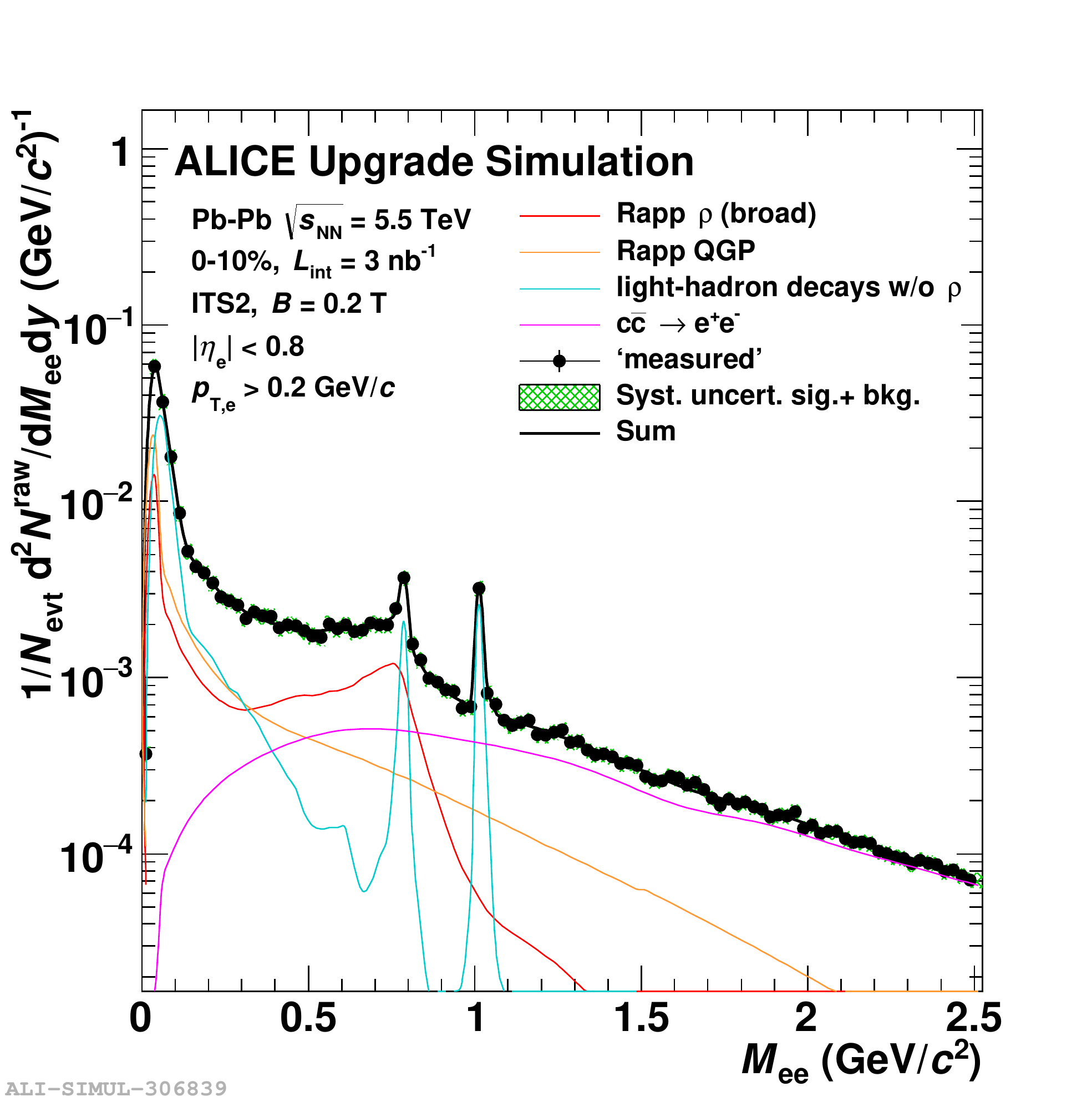}
\includegraphics[width=0.45\textwidth]{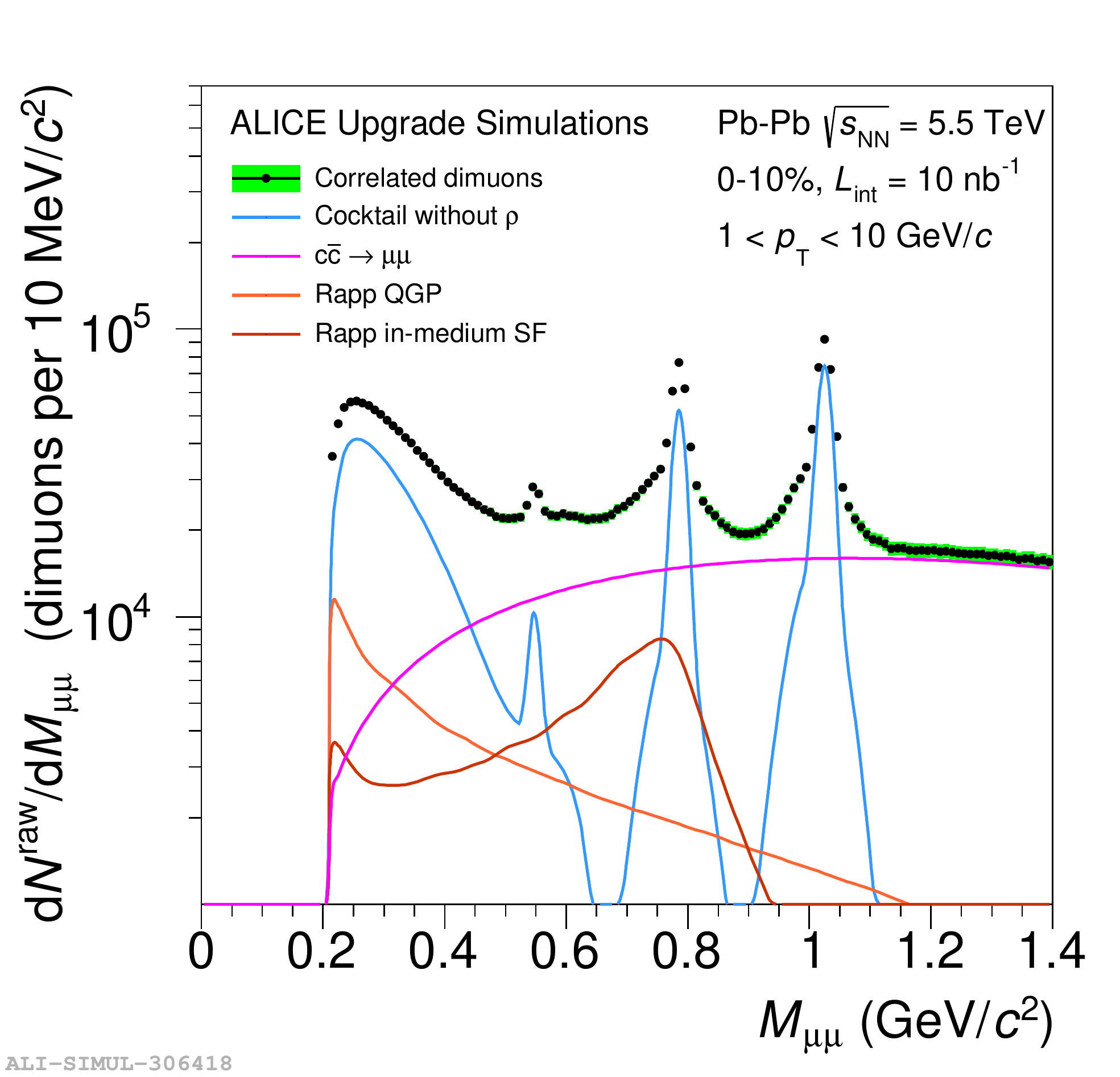}
\caption{Inclusive \Pepem (left) and \PGmpGmm (right) invariant mass spectrum for 0--10\% most central \PbPb{} collisions at $\sqrtsNN = \unit[5.5]{\UTeV}$. The green boxes show the systematic uncertainties from the combinatorial background subtraction. Figures from Ref.~\cite{ALICE-PUBLIC-2019-001}.}
\label{fig:DileptonsSpectra}
\end{figure}

\begin{figure}[tb]
\centering
\includegraphics[width=0.45\textwidth]{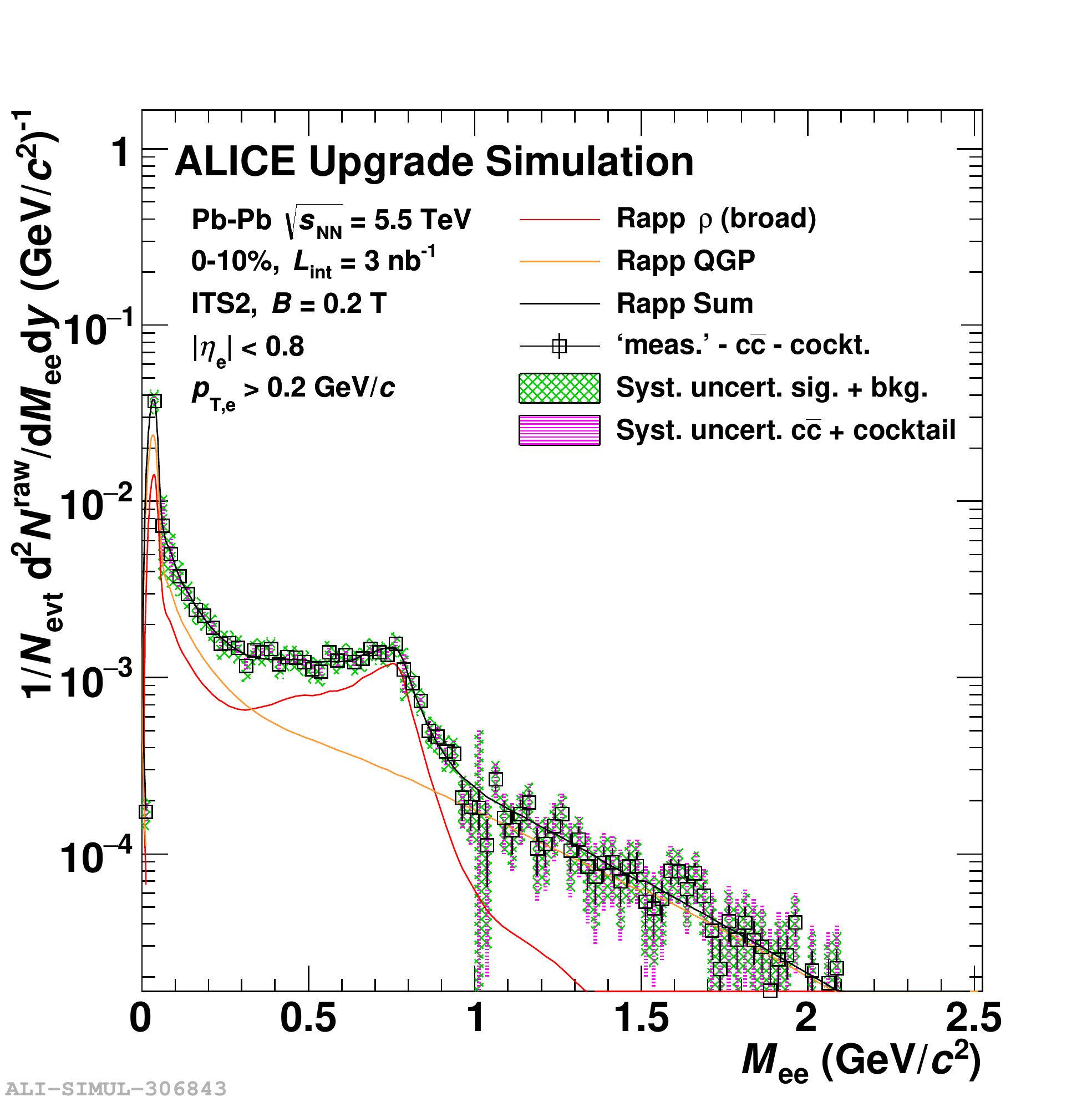}
\includegraphics[width=0.45\textwidth]{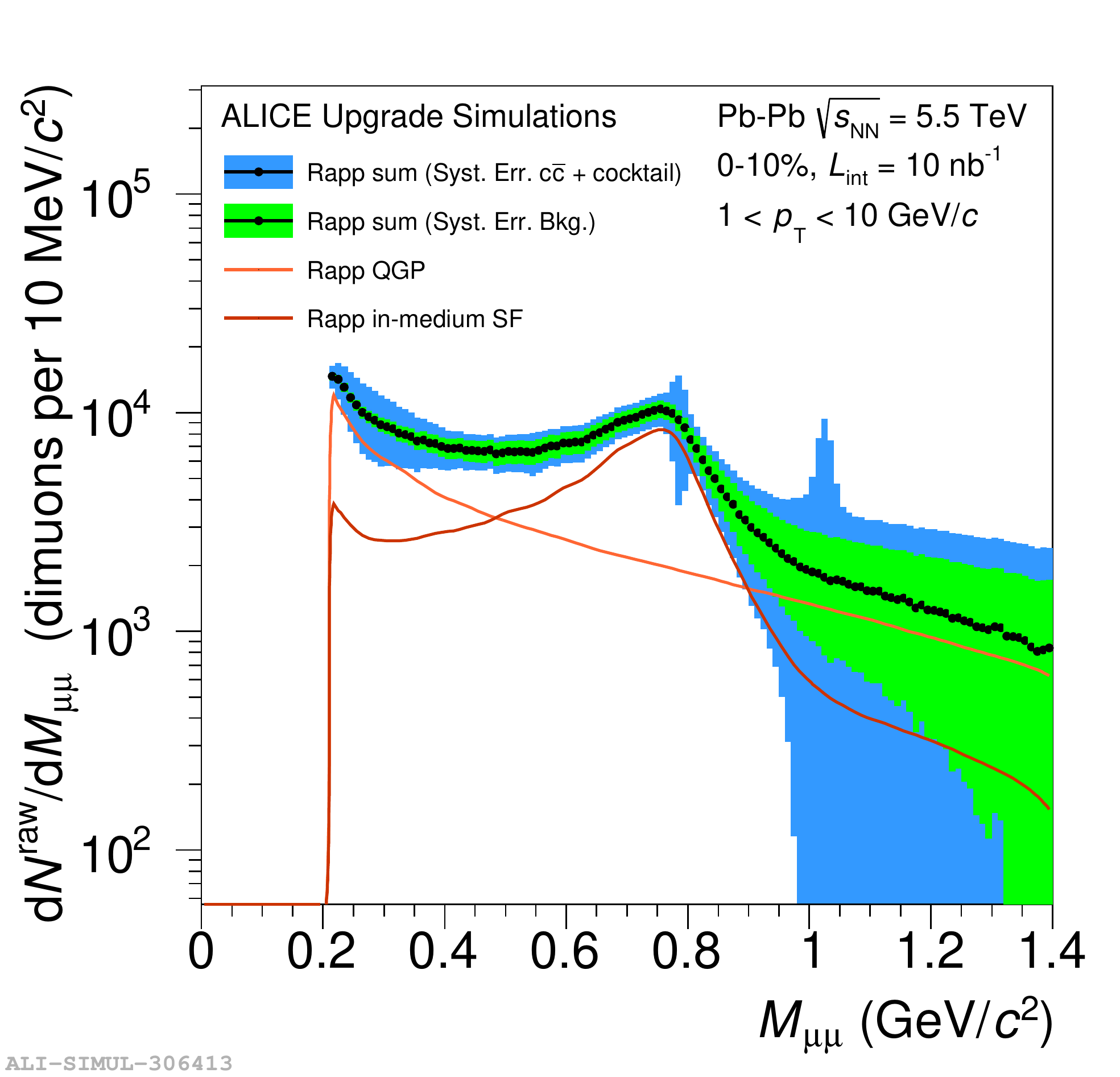}
\caption{Excess (after subtraction of light hadron decays and from correlated charm semileptonic decays) \Pepem (left) and \PGmpGmm (right) invariant mass spectrum for 0--10\% most central \PbPb{} collisions at $\sqrtsNN = \unit[5.5]{\UTeV}$. The green boxes show the systematic uncertainties from the combinatorial background subtraction, the magenta (left) and blue (right) boxes indicate systematic errors related to the subtraction of the cocktail and charm contribution.
Figures from Ref.~\cite{ALICE-PUBLIC-2019-001}.}
\label{fig:DileptonsSpectraSubtracted}
\end{figure}

For the dielectron channel an integrated luminosity $\Lint\approx \unit[3]{\Unb^{-1}}$ is assumed, which should be collected in the dedicated \PbPb{} run at low field. The corresponding number of events in central (0--10\%) collisions is $2.5 \times 10^{9}$.  
The input for the signal is composed of:
\begin{itemize}
\item contributions from the decays of long-lived light pseudoscalar and vector mesons (hadronic cocktail consisting of \PGpz,\PGh,\PGh',\PGo, and \PGf), with particle ratios and spectral shapes extrapolated from existing heavy-ion data at lower energies,
\item correlated semileptonic charm decays based on calculations from the PYTHIA event generator \cite{Sjostrand:2006za},
\item and the radiation of thermal dileptons and a medium-modified spectral function for the \PGr meson in a realistic space-time evolution (see Fig.~\ref{fig:LHCExpectations_Rapp_pHSD}~(left)).
\end{itemize}
With respect to earlier calculations \cite{Abelevetal:2014cna,Abelevetal:2014dna,ALICE:2014qrd} a fast simulation of central \PbPb{} collisions is used here to estimate the combinatorial background and the statistical significance of the signal. The particles are produced with the event generator HIJING \cite{Wang:1991hta} and then propagated through the detector material by GEANT3 \cite{Brun:1994aa}. An updated geometry of the ITS is utilised in the detector description and leads to a more realistic treatment of conversion electrons and the subsequent background. 
Electrons are reconstructed and identified via signals in the ALICE Time Projection Chamber (TPC) and Time-Of-Flight (TOF) detector, a parametrised efficiency from runs at low magnetic field during LHC Run 2 is applied. After pairing electrons and positrons an additional selection on the pair distance of closest approach 
\begin{equation}
DCA_{ee}(\sigma)=\sqrt{(\mathrm{DCA}_{\mathrm{xy},1}/\sigma_{\mathrm{xy},1})^2+(\mathrm{DCA}_{\mathrm{xy},2}/\sigma_{\mathrm{xy},2})^2}
\end{equation}
is applied to reduce the contribution from correlated semileptonic charm decays. The selection is chosen such that 95\% of these pairs are rejected, while having an efficiency for prompt pairs of $\sim17$\%.
The signal distribution $S$, which includes the remaining charm and beauty hadron decays, is obtained by subtraction of the combinatorial background from all \Pepem pairs. The combinatorial background $B$ is estimated from like-sign pairs and a correction factor $R$ that takes into account the different acceptance of the apparatus for unlike- and like-sign pairs \cite{Acharya:2018kkj,Acharya:2018ohw,Acharya:2018nxm}. The significance that is used to project the statistical uncertainty on the measurement is calculated as $S/\sqrt{S+2B}$. The signal $S$ is shown in Fig.~\ref{fig:DileptonsSpectra}~(left) together with all input distributions. In order to extract the QGP component and the in-medium modified \PGr spectral function, the hadronic cocktail and the contribution from correlated semileptonic charm decays is subtracted and shown in Fig.~\ref{fig:DileptonsSpectraSubtracted}~(left). In addition, the systematic uncertainties from the combinatorial background and signal extraction, as well as physical backgrounds after subtraction are shown. The relative systematic uncertainty of the signal, from tracking, track matching, and particle identification, is assumed to be 10\%. For the systematic uncertainty of $B$ a mass independent uncertainty of the $R$ factor of 0.02\% is used. Relative systematic uncertainties from the light-hadron cocktail and the total charm cross section of 10\% and 15\%, respectively, are applied.

In the dimuon channel, the integrated luminosity of \PbPb{} collisions ($\Lint\approx 10 \invnb$) is used. In this channel, the main source of background is represented by the combinatorial pairs of muons coming from uncorrelated semimuonic decays of light-flavoured mesons, mainly pions and kaons, copiously produced in high-energy nuclear collisions. The opposite-sign dimuon mass spectrum obtained after the subtraction of the combinatorial background evaluated by means of an event mixing technique, results from the superposition of several opposite-sign correlated dimuon sources, represented in the right panel of \figurename~\ref{fig:DileptonsSpectra}. 
In order to isolate the thermal dimuon radiation and the in-medium modified line shapes of the $\PGr$ meson,
the known and well-identifiable sources of the hadronic cocktail --- 2-body and Dalitz decays of the $\PGh$, $\PGo$, $\PGf$ mesons, for which no in-medium effect is expected --- are subtracted from the total opposite-sign correlated dimuon mass spectrum. A~10\% systematic uncertainty in the evaluation of the shape and the normalization of these sources has been considered in the performance studies. The same procedure has been also applied for the subtraction of the dimuons from the open charm and open beauty processes; alternatively, these two sources could be separated from the prompt ones by means of an analysis based on the discrimination of the dimuon offset at the primary vertex.

The spectral function of low-mass dielectrons and dimuons in the mass region of the modified \PGr-meson spectral function $M_{\rm ee}\approx\unit[0.5]{\UGeVcc}$ can be extracted with a systematic uncertainty of $\approx 15$\% and $\approx 20$\%, respectively (see Fig.~\ref{fig:DileptonsSpectraSubtracted}). 
The sizeable contribution of thermal dilepton pairs above $M_{\rm ee}>\unit[1.1]{\UGeVcc}$ can be used to extract the temperature of the system. An exponential fit with d$N$ /d$M_{\rm ee} \sim  M_{\rm ee}^{3/2}$ exp( - $M_{\rm ee} /T_{\rm fit}$ ) to the subtracted \Pepem spectra in the invariant mass region $1.1 < M_{\rm ee} < \unit[2.0]{\UGeVcc}$ was performed. Comparing the fit parameter $T_{\rm fit}$ to the real temperature $T_{\rm real}$ from the fit to the thermal contribution, a statistical uncertainty of 5\% and systematic uncertainty of 10\% and 20\% for the background and the charm subtraction, respectively, were estimated. The same kind of measurement is also expected to be possible in the dimuon channel, considering a dedicated set of cuts optimized for the analysis of the intermediate mass region (the cuts considered in the right panel of \figurename~\ref{fig:DileptonsSpectraSubtracted} being optimized for the signal extraction in the mass region below $\sim \unit[1]{\UGeVcc}$).

An alternative method to separate the thermal component from the modified heavy-flavour production in the intermediate mass range, is to fit the measured $DCA_{ee}$ distribution as a function of the dielectron invariant mass and pair transverse momentum with a three component function, including the contributions from prompt dielectron sources, from open-charm hadron decays and from open-beauty hadron decays. Since the shape of the heavy-flavour $DCA_{ee}$ spectra is quasi model independent, the dielectron yield of open heavy-flavour decays in the ALICE acceptance can be determined from the data with small uncertainties, without relying on theoretical calculations. Such fits were performed already with the Run 1 data in \pp{} collisions at $\sqrts = \unit[7]{\UTeV}$ \cite{Acharya:2018ohw}. Nevertheless, the statistics available did not allow for a differential study.    

The measurement of the dielectron elliptic flow coefficient $v_2$ as a function of $M_{\rm ee}$ in peripheral \PbPb{} collisions (40--60\%) was studied already in \cite{Abelevetal:2014cna}. It was shown that an absolute statistical uncertainty on $v_2$ of $\sigma_{v_2} \approx 0.01$ can be achieved. 
Such uncertainties present a very encouraging prospect for dilepton studies since temperature dependent shear viscosity \cite{Vujanovic:2017psb}, bulk viscosity \cite{Vujanovic:2017wtw,Vujanovic:2017fkh}, and early stages of reaction dynamics \cite{Vujanovic:2016anq} have effects on the order of a few up to tens of percent on dilepton $v_2$, and thus future constraints on these properties will greatly benefit from a statistically improved dilepton $v_2$ measurement.
 
\subsection{Two-photon and photonuclear interactions}
\label{sec:dileptons:peripheral}

Heavy ions carry strong electromagnetic fields.  Their electric and magnetic fields are perpendicular, so may be treated as a flux of nearly-real photons \cite{Bertulani:2005ru}.  These photon fields can give rise to photonuclear (photon on nucleus) and two-photon interactions.  Although these interactions are expected to occur in both ultra-peripheral (UPC) and more central collisions, they were not generally expected to be visible in non-UPC collisions.  The few final state particles from the photon-mediated interactions were expected to be swamped by the more copious hadronically produced particles.   That expectation changed recently, when ALICE \cite{Adam:2015gba} and then STAR \cite{Adam:2018tdm,Zha:2018ohg} and ATLAS \cite{Aaboud:2018eph} observed excesses of dileptons produced at very small pair \pT, $\pT < \unit[100]{\UMeVc}$.   These pairs were prominent in \PbPb{} and \AuAu{} collisions, but not in \pp{} interactions; the excess corresponded to $\Raa >5$.  This is inconsistent with all expectations for hadroproduction, but  consistent with photoproduction, where the pair \pT{} scale is set by the nuclear radius $R_{\rm A}$, with $\pT \approx \hbar/R_{\rm A}$.  The kinematics of these pairs may be affected by the medium in which they are produced or propagate, so they may probe the evolving Quark--Gluon Plasma or associated magnetic fields. 

UPC photon-mediated interactions have been studied at both RHIC and the LHC \cite{Baltz:2007kq,Bertulani:2005ru,Klein:2017nqo,Bertulani:1987tz,Baur:2001jj}.  The agreement between data and calculations is quite good.  Photoproduction of \PGr, \PGo, \PGr' , \PJGy, \PJGy', \PGU and direct \PGpp\PGpm pairs has been observed, along with two-photon production of dilepton pairs and light-by-light scattering.  In peripheral collisions, photon-mediated interactions might be used to probe the nuclear medium that they may occur in, including the QGP \cite{Aaboud:2018eph,Adam:2018tdm}. The produced leptons may interact with this medium, leading to alterations in their momentum.    

Peripheral collisions introduce several new considerations for  photon-mediated reactions, particularly evolving coherence conditions for both photon emission and coherent photon-nucleus scattering.  Photon emission in both $\PGg\PGg$ and photonuclear interactions is expected to be completely coherent, governed by the nuclear form factor $F(q)$ \cite{Vidovic:1992ik}.   The photon emission from a nucleus moving with Lorentz boost $\gamma$ should occur before the hadronic interaction (which is taken to occur at $t=0$), at a retarded time, $t-x/c$ \cite{Zha:2017jch}, where $x=|b|/\gamma$; $|b|$ is the transverse distance from the photon emission point to where it interacts. For very small impact parameters, some coherence may be lost, and a more detailed calculation is needed.  For photon-nucleus collisions, the situation is more complicated, and will be discussed below. 

Here, two-photon interactions and coherent photonuclear interactions are discussed. 

\subsubsection{Two-photon interactions}

In two-photon interactions, each nucleus emits a photon, which then interact and form a lepton pair. In UPCs, this process is well described by the Weizs{\"a}cker-Williams approach (where each photon is treated as real), except at very low pair \pT, where a lowest-order QED calculation works better \cite{Adams:2004rz}. UPC calculations can be easily extended to include peripheral collisions \cite{Klein:2018cjh, Zha:2018ywo,Klusek-Gawenda:2018zfz}.  The kinematic distributions are similar to those in UPCs, and the cross-section depends on the range of impact parameters.

\begin{figure}[tb]
\centering
\includegraphics[width=0.85\textwidth]{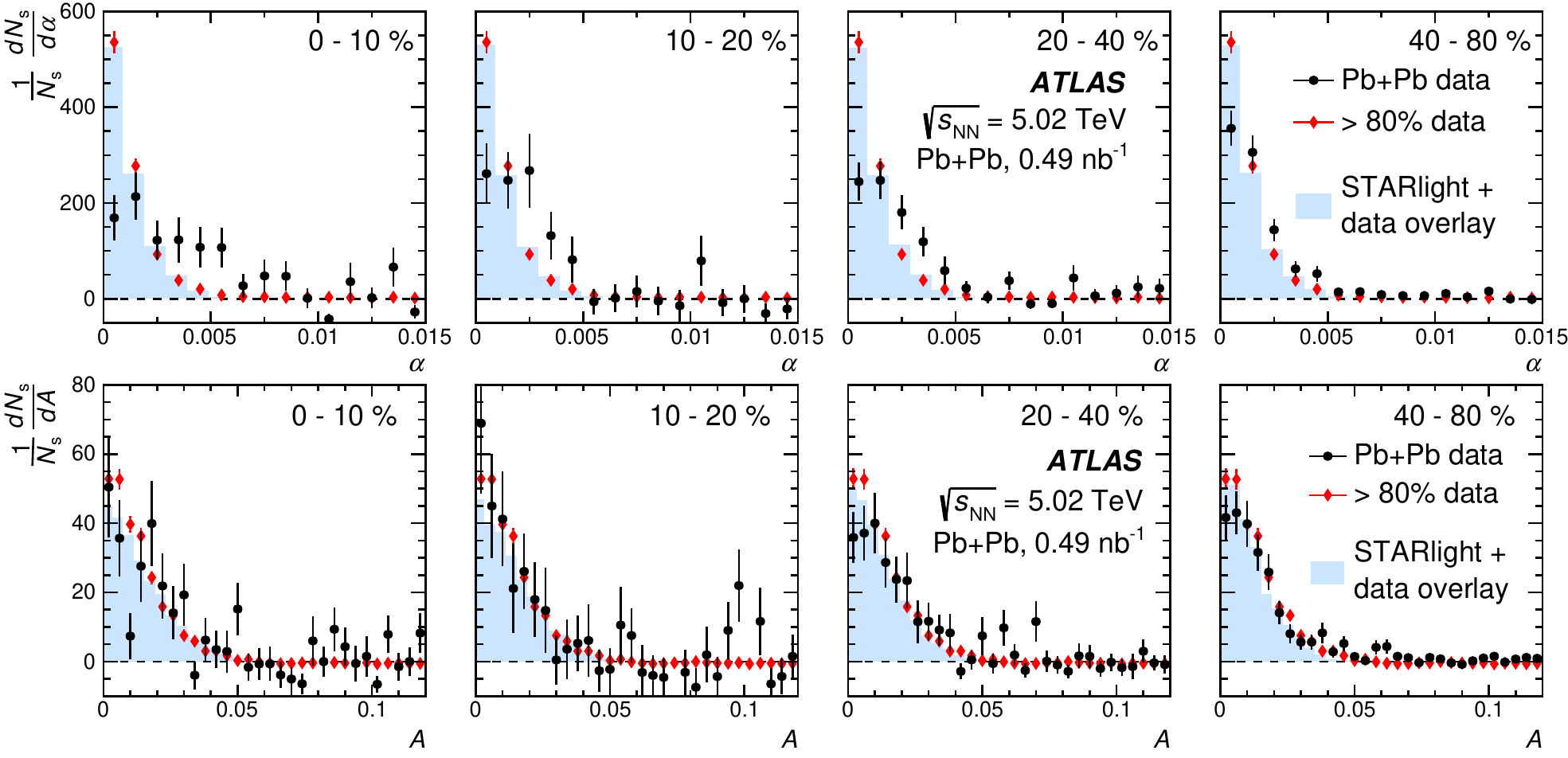}
\caption{Acoplanarity ($\alpha$, top) and lepton energy imbalance ($A$, bottom) as a function of centrality, for dimuon pairs with pair mass above $\unit[10]{\UGeVcc}$, observed in the ATLAS detector.  From Ref. \cite{Aaboud:2018eph}.}
\label{fig:ATLASacoplanarity}
\end{figure}

Recently, the ATLAS collaboration \cite{Aaboud:2018eph} presented results showing a dramatic modification to $\PGg\PGg\rightarrow\PGmpGmm$ in peripheral collisions.  Figure \ref{fig:ATLASacoplanarity} shows the pair acoplanarity $\alpha$, the azimuthal angular deviation from being perfectly back-to-back, and $A$, the energy imbalance between the two leptons.  For UPCs, they found good agreement with the STARlight \cite{Baltz:2009jk,Klein:2016yzr} reference, with the data and calculations peaked at small $\alpha$ and $A$.   More central collisions show dramatic changes with the low-$\alpha$ peak largely disappearing, and the $A$ distributions only minimally changed.  ATLAS described this as "Consistent with order of magnitude estimates from kinetic theory for multiple scattering off electric charges in thermal plasma."  Multiple scattering would remove the peak at low $\alpha$, but leave $A$ largely unaffected.  A recent calculation finds that the magnitude of the change in $\alpha$ is at least roughly consistent with that expected for leptons propagating through a Quark--Gluon Plasma \cite{Klein:2018fmp}.  If multiple scattering is large, though, one might also expect some bremsstrahlung, which should increase $A$.   To evaluate this further requires a calculation of how many of the produced leptons are produced in the medium, and/or traverse it.  An alternate explanation could involve the leptons bending in the magnetic field from the QGP.  If a QGP is electrically conducting, then it may acquire an induced magnetic field from the short-lived magnetic fields carried by the two nuclei \cite{Kharzeev:2012ph}.  The QGP field, however, will be longer lived, and could bend the produced leptons in opposite directions, reducing their coplanarity. Symmetry also predicts that it should disappear for the most central collisions \cite{Klein:2018fmp}, except possibly for event-by-event fluctuations. 

The STAR Collaboration also has studied two-photon \Pepem production in peripheral \AuAu{} collisions; they found a small difference between their pair  \pT{} spectrum and calculations, and suggest that it might be due to medium effects\cite{Adam:2018tdm}. ALICE has not yet seen these pairs \cite{Adam:2015gba}, likely because their pair acceptance requires  lepton $\pT > \unit[1]{\UGeVc}$, eliminating most pairs from $\PGg\PGg$ reactions. 

\begin{figure}[tb]
\centering
\includegraphics[width=0.5\textwidth]{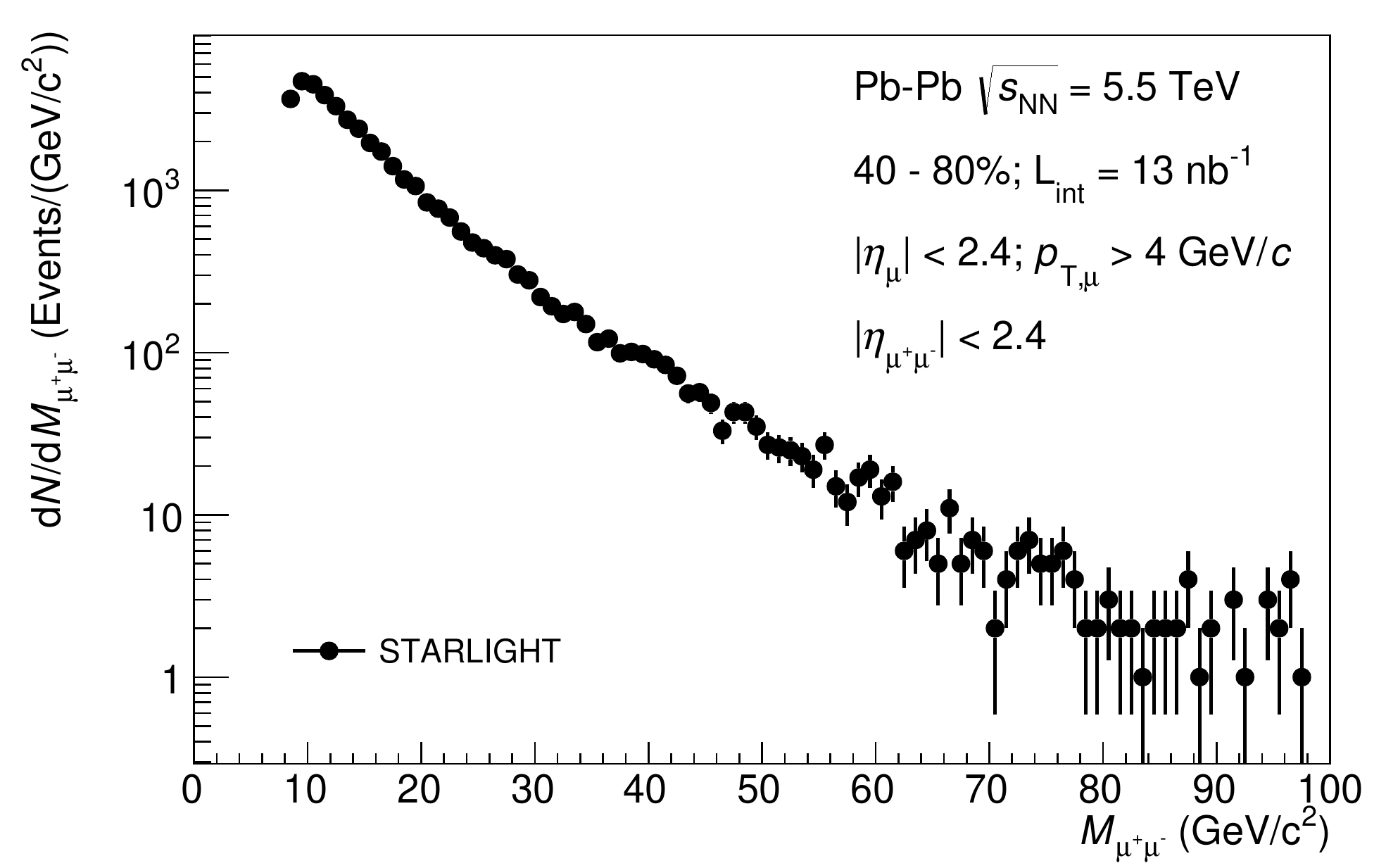}
\caption{Expected dimuon yield in ATLAS acceptance (both muons with $\pT>\unit[4]{\UGeVc}$ and $|\eta|<2.4$), for 40--80\% centrality \PbPb{} collisions and the expected Run 3/4 integrated luminosity of $\unit[13]{\Unb^{-1}}$.  Masses up to \unit[100]{\UGeVcc} are accessible.  The effective \unit[8]{\UGeVcc} minimum mass is because of the nearly back-to-back topology and the \unit[4]{\UGeVc} minimum muon \pT{} cut. This was calculated using STARlight \cite{Klein:2016yzr,Klein:2018cjh}.
}
\label{fig:project} 
\end{figure}

Coupled with better theoretical calculations, the large \PbPb{} integrated luminosity in Run 3 and 4 can confirm and dramatically expand our understanding of this effect.   One important goal is to expand the study to cover a much wider range of masses.  Figure \ref{fig:project} shows the expected mass spectrum obtainable by ATLAS for a $\unit[13]{\Unb^{-1}}$ integrated luminosity run, assuming no changes in the trigger; masses up to \unit[100]{\UGeVcc} should be accessible. These high mass pairs correspond to two-photon interactions in or very near the two nuclei, so should show increased effects due to interactions with the medium or magnetic fields associated with the Quark--Gluon Plasma. 

In contrast, lower masses correspond to larger distances between the dilepton production point and the nuclei, so in-medium effects may be smaller.  These lower masses should be accessible with a softer requirement on the muon momentum.   It would also be interesting to compare \Pepem with \PGmpGmm (and possibly \PGtp\PGtm), since the lighter leptons should interact more.  If the leptons interact with the medium, then the electron $A$ distribution should show more change than that for muons.  

\subsubsection{Photonuclear interactions}

In photonuclear interactions, a photon emitted by one nucleus fluctuates to a quark-antiquark dipole, which then scatters elastically from the other (target) nucleus, emerging as a real vector meson.  The scattering occurs via Pomeron exchange, which preserves the photon quantum numbers.  In perturbative QCD, Pomerons are made up of gluons, so the process is sensitive to the gluon distribution in the target nucleus.  UPC measurements are consistent with moderate gluon shadowing.   In coherent scattering, the typical pair $\pT$ is $\hbar/R_{\rm A}$.  Incoherent scattering is also possible, with a lower cross-section.  There the quark-antiquark dipole scatters elastically from a single nucleon (or, at still higher $\pT$ inelastically from a single nucleon), producing a vector meson with a typical $\pT$ of a few hundred \unit[]{\UMeVc}.

\begin{figure}[tb]
\centering
\includegraphics[width=0.85\textwidth]{\main/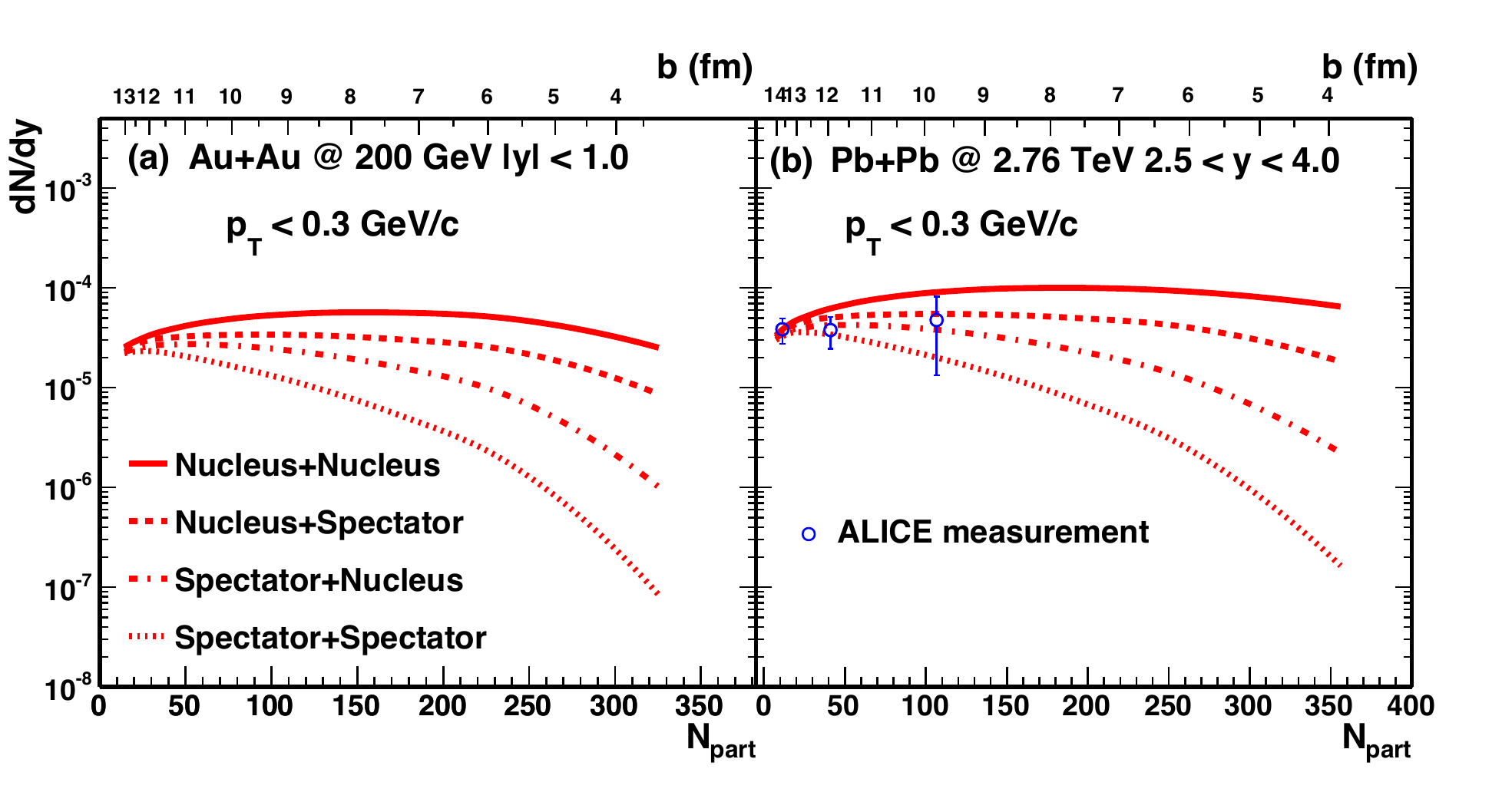}
\caption{$J/\psi$ coherent photoproduction cross-sections in peripheral collisions, as a function of the number of participants (bottom), and impact parameter (top) with \AuAu{} collisions at RHIC (left) and \PbPb at the LHC (right).  The four curves are for different assumptions regarding centrality for the photon emitter (first particle listed) and the target (second particle listed).  From Ref. \cite{Adam:2015gba,Zha:2017jch}.}
\label{fig:jpsiinpcs} 
\end{figure}

Both ALICE \cite{Adam:2015gba}  and STAR \cite{Zha:2018ohg} have observed coherent \PJGy photoproduction in peripheral heavy-ion collisions.  There are a number of parallel theoretical calculations \cite{Klusek-Gawenda:2015hja,Zha:2017jch}.  The photon emission process is similar to the two-photon case, but the dipole-nucleons scattering happens at the same time as the hadronic interaction, introducing several complications to the calculations.  This immediately raises several questions:  What happens to the coherence if a target nucleon is involved in an interaction?   Does the dipole-nucleon interaction occur before or after the nuclear collisions?    If the hadronic interaction occurs first, the target nucleon will have lost energy, so the photon-nucleon cross-section will be smaller.  A detailed calculation should consider both possibilities.   There is also destructive interference between photoproduction from the two possible target nuclei \cite{Abelev:2008ew}; this interference extends to higher $\pT$ for more central collisions, and should reduce the cross-section for the region where nuclear collisions occur. At $b=0$, we expect complete destructive interference.  Ref. \cite{Zha:2017jch} makes predictions for a variety of coherence conditions, and as Fig. \ref{fig:jpsiinpcs} shows, finds that the ALICE and STAR data likely lie below the region where there is complete coherence for both photon emission and scattering, but probably above that where coherence is limited to only the spectator nucleons.  This is not surprising, but there is at least one element missing from this calculation. The lifetime of \PJGy particles is of the order $\unit[10^{-20}]{\Us}$, far shorter than that of the expanding Quark--Gluon Plasma.  Coherently photoproduced \PJGy have $\pT \sim \unit[100]{\UMeVc}$, so, near mid-rapidity,  are moving at a small fraction of the speed of light.  Particularly for more central collisions, one would expect many of them to be engulfed by the expanded QGP, before they have a chance to decay.  

The ALICE error bars are large, and more data, from the current and future runs are needed to pin down the centrality dependence of the cross section.  More data will also allow access to additional observables. A detailed study of the shape of d$\sigma/$d$\pT$ would shed more light on the possible loss of coherence in more central collisions. There are also expected correlations between the reaction plane, which can be determined from the hadronic part of the collision, with the photonuclear interaction.  Because the destructive interference between photoproduction at mid-rapidity on the two nuclei goes as $\sigma \sim |1-\exp{(i{\vec{b}\cdot\vec{\pT})}}|^2$ \cite{Klein:1999gv}, the azimuthal direction of $\vec{\pT}$ provides information about the azimuthal direction of $\vec{b}$, \ie the reaction plane.  Thus, it can be used either as an independent measurement of the reaction plane, or as a test of the loss of correlation.  Also, the \PJGy polarization follows that of the photon that produced it, so it also follows $\vec{b}$, providing another probe of the reaction plane.  With a large data sample, one may also be able to probe incoherent \PJGy photoproduction, at least in very peripheral hadronic collisions, where the signal-to-noise ratio is high.  

It will be very interesting to study \PGy' and \PGU photoproduction in peripheral collisions.  Since these mesons have different sizes from the \PJGy, they should interact with the medium with different strengths.  These studies should be possible at HL-LHC.  

\subsection{Dark photons}
\label{sec:dileptons:darkphotons}

Dark Matter is a hypothetical form of matter that is responsible for accounting for approximately 80\% of the matter in the Universe~\cite{Planck:2013jfk}.
Dark matter cannot be incorporated into the Standard Model, so the introduction of dark matter requires new interactions between dark matter particles 
and the ordinary Standard Model particles via unknown dark-sector forces~\cite{Alexander:2016aln}. 
The dark sector could have a rich structure with a few possible candidates, 
where one of them is regarded as Dark Photon ($A'$) with 
$L_{mix} \propto\frac{g}{2}F^{\mu\nu}X_{\mu\nu}$.
The dark photon is introduced as an extra-$U(1)$ gauge boson and acts as a messenger particle of a dark sector 
with the residual interaction ($g$) to the Standard Model particles.
Understanding of possible interactions of dark photons has been motivated by 
a number of astrophysical anomalies such as antiproton spectrum in 
the cosmic rays measured by AMS Collaboration,
positron excess in the cosmic rays observed earlier by PAMELA~\cite{Adriani:2008zr}
and confirmed by FERMI~\cite{FermiLAT:2011ab} and AMS~\cite{Aguilar:2013qda}, 
and the long standing discrepancy
between the measured and the calculated anomalous magnetic moment of 
the muon $(g-2)_{\mu}$, where
the difference is more than three standard deviations away from zero~\cite{Bennett:2006fi}.

If the dark photon is the lightest state of the dark sector 
and therefore can decay only into the Standard Model particles, dark photons with mass $m_{A'} \le 2m_{\mu}$ decay only into electron-positron pairs. 
For dark photons above 2 muon threshold ($m_{A'} \ge 2m_{\mu}$), 
dark photons can decay into muon pairs. For ($m_{A'} \ge 2m_{\pi}$), 
dark photons can decay into hadrons as well. 
A lot of experimental activities have been seen recently 
and constraints of mixing parameter ($g^2$) as a function 
of dark photon mass ($m_{A'}$) has been done from many experiments. 
They are, for example, 
beam-dump experiments (measurement of lepton pairs from dark photons 
behind a sufficiently long shield. 
Examples are E141~\cite{Riordan:1987aw} and E137~\cite{Bjorken:1988as}
at SLAC, E774~\cite{Bross:1989mp} at Fermilab), 
fixed-target experiments (by scattering the electron beam on a nuclear target, 
the dark photon may be emitted in the initial or final state and coupling to 
electron-positron pairs is studied by looking for a bump in the 
electron-positron invariant mass. Examples are A1~\cite{Merkel:2014avp} 
at MAMI in Mainz, APEX~\cite{Abrahamyan:2011gv} at JLAB, 
DarkLight~\cite{Balewski:2013oza} at JLAB)
and collider experiments (BABAR~\cite{Lees:2014xha}, 
NA48/2~\cite{Batley:2015lha} at SPS, WASA~\cite{Moskal:2014dsa} at COSY, 
HADES~\cite{Agakishiev:2013fwl} at GSI, 
PHENIX~\cite{Adare:2014mgk} at RHIC, LHCb~\cite{Aaij:2017rft} and ALICE at LHC).
Since any process in which a virtual photon couples to lepton pairs or hadrons 
can be used to search for dark photons, 
following processes are used in the collider experiments: measurements of Dalitz decays of the $\PGpz/\PGh/\PGh' \rightarrow \PGg A'$ mesons and rare meson decays such as $\PK\rightarrow \PGp A'$, $\PGf \rightarrow \PGh A'$, 
and $\PDst \rightarrow \PDz A'$,
Bremsstrahlung process 
($\Pem Z \rightarrow \Pem ZA'$ with $A'$ emitted at very forward direction),  
radiative decay of vector resonances and initial state radiation 
(done by BABAR using radiative decays of $\PGUP{3S}$ and 
done by KLOE~\cite{Archilli:2011zc} using $\PGf \rightarrow \Pepem$).
The 79-string IceCube search for dark matter in the Sun public data is 
used to test Secluded Dark Matter models~\cite{Ardid:2017lry}. 
Dark matter particles can be captured by the Sun, annihilate, and 
produce a neutrino flux that can be observed at Earth and that 
depends on the dark matter scattering cross section off nuclei
and on the dark matter annihilation rate and final states.
This analysis constrains a kinetic mixing parameter $g \sim 10^{-9}$
between 0.22 and $\sim\unit[1]{\UGeV}$~\cite{Ardid:2017lry}.

\begin{figure}[tbp]
\begin{center}
\includegraphics[width=120mm]{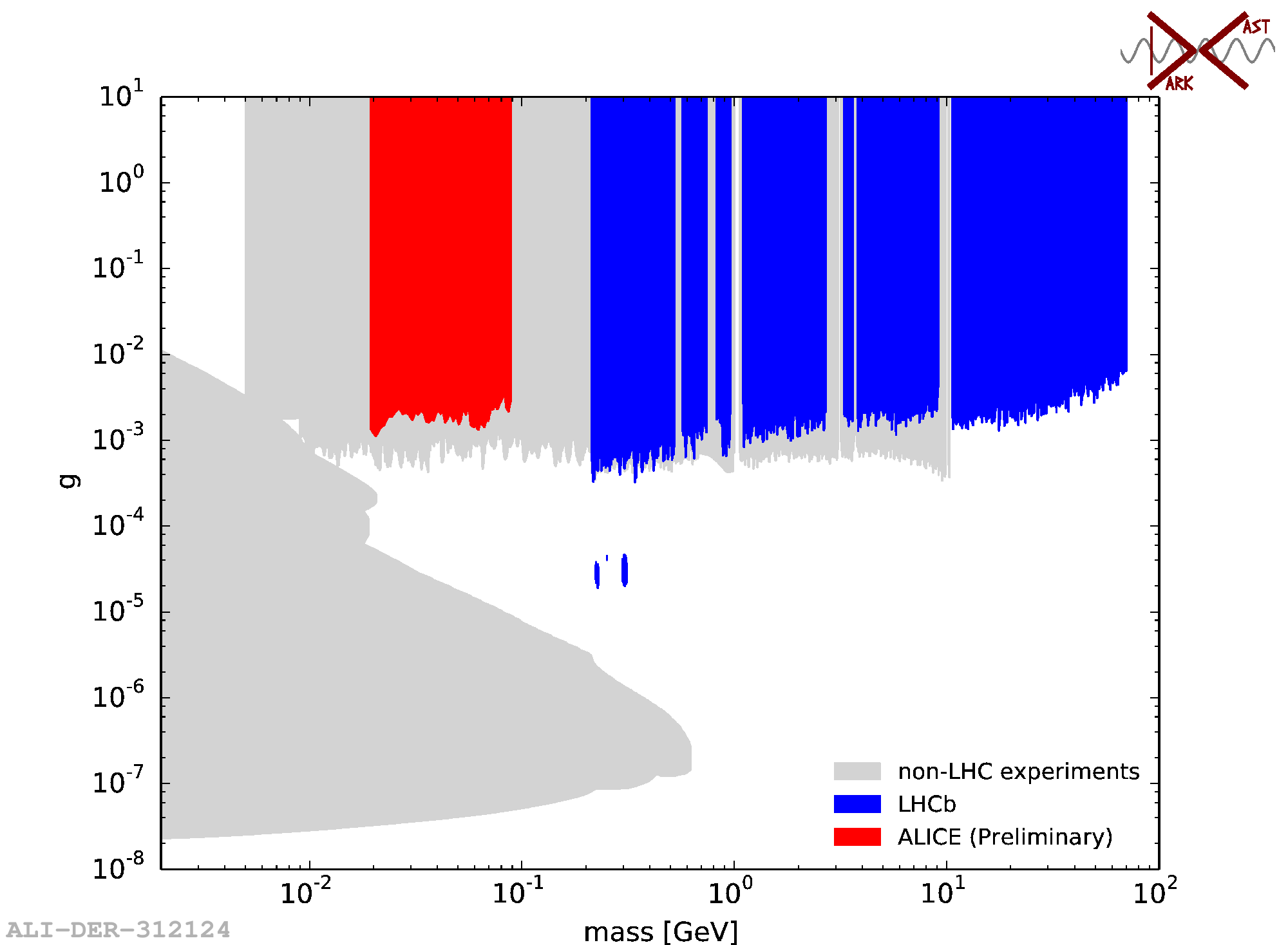}
\end{center}
\caption{90\% of confidence level of mixing parameter as a function of dark photon mass. Figure is adapted from Ref.~\cite{Ilten:2018crw}. 
Red and blue are from ALICE and LHCb~\cite{Aaij:2017rft}. 
Light grey band contains results from BABAR, KLOE, A1, APEX, NA48/2,  E774, E141, E137, KEK, Orsay, BESIII, CHARM, HPS, NA64, NOMAD, NuCAL, and PS191~\cite{Ilten:2018crw}.}
\label{fig:darkphoton_alice_lhcb}
\end{figure}

ALICE has good capabilities for electron identification in the low 
transverse momentum region, that enables the measurement of a large sample of 
the \PGpz Dalitz decays~\cite{Acharya:2018ohw}. 
ALICE searches for possible decays of 
$\PGpz \rightarrow \PGg A', A' \rightarrow \Pepem$ 
by examining the electron-positron invariant mass in a large sample
of \PGpz Dalitz decay for $20 \le M_{\rm ee} \le \unit[90]{\UMeVcc}$ in 
\pp{} collisions at \unit[7]{\UTeV} ($\Lint \sim \unit[4]{\Unb^{-1}}$) and 
\pPb{} collisions at \unit[5.0]{\UTeV} ($\Lint \sim \unit[40]{\Uub^{-1}}$) as shown in 
Fig.~\ref{fig:darkphoton_alice_lhcb}.\\
LHCb has good capabilities to measure muons and hardware and software 
triggers enable the accumulation of a large sample of dimuon pairs. 
LHCb searches for prompt-like and long-lived dark photons 
produced in \pp{} collisions at \unit[13]{\UTeV}, using $A' \rightarrow \PGmpGmm$ decays 
from a large data sample corresponding to $\Lint \sim \unit[1.6]{\Ufb^{-1}}$ 
collected during 2016, where 
the prompt-like $A'$ search is shown in Fig.~\ref{fig:darkphoton_alice_lhcb}~\cite{Aaij:2017rft}.

\begin{figure}[tbp]
\begin{center}
\includegraphics[width=120mm]{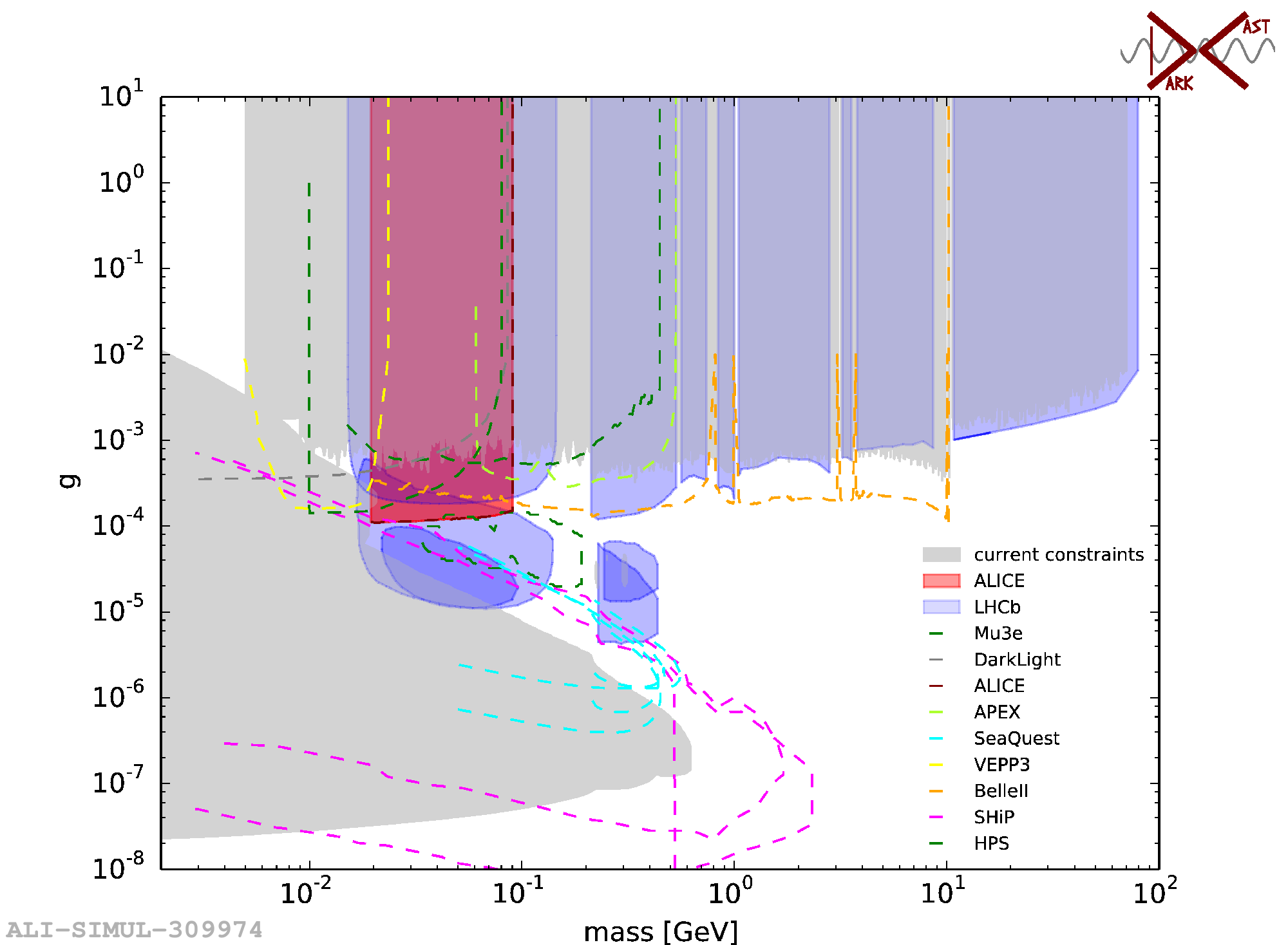}
\end{center}
\caption{90\% of CL constrained by ALICE and LHCb in HL-LHC era.
Constraints by ALICE are based on 
$\unit[6]{\Upb^{-1}}$, $\unit[0.3]{\Upb^{-1}}$, $\unit[10]{\Unb^{-1}}$, $\unit[0.3]{\Upb^{-1}}$, and $\unit[3]{\Unb^{-1}}$
of \pp, \pPb, and \PbPb{} collisions at \unit[0.5]{T}, and \pPb{} and \PbPb{} collisions at \unit[0.2]{T}
and by LHCb are based on $\unit[15]{\Ufb^{-1}}$.
ALICE projection from Ref.~\cite{ALICE-PUBLIC-2019-001}.
The other projections are adopted from Ref.~\cite{Ilten:2018crw}.}
\label{fig:darkphoton_alice_lhcb_hllhc}
\end{figure}

The ALICE upgrade during LS2 will greatly improve 
the efficiency of electron-positron measurements and data taking capability.
Figure~\ref{fig:darkphoton_alice_lhcb_hllhc} shows 
expected constraints that will be achieved by ALICE and LHCb
together with the future experiments. 
After the major ALICE upgrade, ALICE will accumulate 
$\unit[6]{\Upb^{-1}}$, $\unit[0.3]{\Upb^{-1}}$, $\unit[10]{\Unb^{-1}}$, $\unit[0.3]{\Upb^{-1}}$, and $\unit[3]{\Unb^{-1}}$
of \pp, \pPb, and \PbPb{} collisions at \unit[0.5]{T}, and \pPb{} and \PbPb{} collisions at \unit[0.2]{T}, 
respectively. 
LHCb will improve sensitivity of dark photon searches to large regions 
of the unexplored space. These new constraints leverage the improved invariant-mass and vertex resolution, 
as well as the unique capabilities of the particle-identification and real-time 
data-analysis with triggerless readout, 
that enables to accumulate $\Lint \sim \unit[15]{\Ufb^{-1}}$~\cite{Ilten:2016tkc}.

\subsection{Limitations and outlook}

While the statistical precision for the measurement of low mass dielectrons and dimuons as well as real photons will be sufficient in LHC Run 3 and 4 to study their yield as a function transverse momentum and with respect to the event plane (elliptic flow), more differential measurements might still be limited. 
The measurement of the photon polarization via the angular distribution of dileptons can not only provide information on the thermalization of the system, but also on the early stages of the collision \cite{Baym:2017qxy}. Experimentally these distributions have been measured in the NA60 experiment \cite{Arnaldi:2008gp}, where no polarization was found concluding that the observed excess dimuons are in agreement with the thermal emission from a a randomized system. In order to study the angular distributions, for example in the Collins-Soper reference frame \cite{Collins:1977iv,Lam:1978pu,Lam:1980uc} in the polar angle $\theta$ and the azimuthal angle $\varphi$, a large data set is needed (NA60 used $\sim50000$ excess \PGmpGmm pairs).

Another promising direction is measurement of Bose-Einstein (BE) correlations of direct photons. With this probe one can trace space-time dimensions of the hottest part of the fireball and moreover, varying \kT{} of the photon pair, one can select pairs coming mostly from earlier or later stages of the collision and thus look at evolution of the fireball. On the other hand, from the correlation strength parameter one can extract the direct photon spectrum down to very low $\pT \sim \unit[100]{\UMeVc}$. So far there was one successful measurement of direct photon BE correlations by the WA98 Collaboration \cite{Aggarwal:2003zy}, while at RHIC and LHC energies these measurements are still unavailable. The reason is that the expected strength of these correlations $\lambda_{PGg}= 1/2(N_{\PGg}^{\rm dir}/N_{\PGg}^{\rm tot})^2$ is extremely small. Moreover, in contrast to massive particles, averaging of full 3D correlation function $C_2(q_{\rm out},q_{\rm side},q_{\rm long})$ to 1D $C_2(q_{\rm inv})$ results in further dramatic decrease of correlation strength \cite{Aggarwal:2003zy}. This requires very large statistics in addition to understanding the detector response. 

A first step to increase the statistical precision and the available data set for low-mass dileptons could be a further upgrade of the inner barrel of silicon detectors of the ALICE apparatus that is currently under discussion \cite{ALICE:ITS3LoI}. The planned reduction of the material budget would reduce conversion probability. In addition, an improvement of the tracking efficiency especially at low momentum would increase the conversion rejection efficiency even further. First studies \cite{ALICE:ITS3LoI} showed that the statistical uncertainty can be reduced by a factor 1.3, while the systematic uncertainty from the subtraction of the combinatorial background would be reduced by a factor of two. With a better pointing resolution the rejection of charm background is improved and would lead to a reduced systematic uncertainty from the subtraction of the light-hadron and charm decay backgrounds by a factor of two.

\clearpage

\section{Emergence of hot and dense QCD matter in small systems}
\label{chapter:smallsystems}

{ \small
\noindent \textbf{Coordinators}: Jan Fiete Grosse-Oetringhaus (CERN) and Constantin Loizides (Oak Ridge National Laboratory)

\noindent \textbf{Contributors}:
R.~Bi (Massachusetts Institute of Technology),
C.~Bierlich (Lund University and Niels Bohr Institute),
E.~Bruna (University and INFN Torino),
Z.~Chen (Rice University),
C.~Cheshkov (IPN Lyon),
Z.~Citron (Ben-Gurion University of the Negev),
A.F.~Dobrin (CERN),
M.~Dumancic (Weizmann Institute of Science),
M.~Guilbaud (CERN),
P.M.~Jacobs (Lawrence Berkeley National Laboratory),
J.~Jia (Stony Brook University and Brookhaven National Lab),
A.P.~Kalweit (CERN),
F.~Krizek (Academy of Sciences, Prague),
A.~Kurkela (CERN and Stavanger University),
Y.-L.~Lee (Massachusetts Institute of Technology),
N.~Mohammadi (CERN),
D.V.~Perepelitsa (University of Colorado Boulder),
R.~Rapp (Texas A\&M University, College Station),
B.~Schenke (Brookhaven National Lab),
K.~Tatar (Massachusetts Institute of Technology),
M.~Weber (Austrian Academy of Sciences),
H.~Zanoli (Universidade de Sao Paulo),
M.~Zhou (Stony Brook University)
}

\subsection{Introduction}

In the program of proton--proton collisions at the LHC, the main effort is focused on hard processes which are embedded in an underlying event consisting of soft low-\pT particles. The underlying event is described using models, such as Pythia~\cite{Sjostrand:2014zea} or HERWIG~\cite{Bellm:2015jjp}, based on essentially free streaming (i.e. no final-state interactions) of the produced particles, supplemented by a non-perturbative cluster or string fragmentation picture~\cite{Andersson:1983ia,Webber:1983if} to model the non-perturbative soft-particle production. The same models are used to describe minimum-bias events, i.e. events without any signal trigger, primarily consisting of soft QCD interactions. 
In the past years at LHC, during Run 1 and 2, this picture was challenged by several observations that qualitatively differ from the model expectations and cannot be accommodated by tuning of the existing models used to describe minimum-bias collisions and the underlying event~\cite{Fischer:2016zzs}.

The first such observation was the unexpected discovery in 2010 of azimuthal correlations of final-state hadrons in very high multiplicity proton--proton collisions~\cite{Khachatryan:2010gv}, referred to by \emph{the ridge}. These persist at large separation in rapidity on the near side surrounding the jet-like peak. A few years later, a similar observation was made in high multiplicity \pPb collisions~\cite{CMS:2012qk}. By subtracting the jet-like contribution in \pPb collisions, a second long-range rapidity correlation on the away side, back-to-back in azimuth to the first observed correlation, was extracted~\cite{Abelev:2012ola,Aad:2012gla}. Even later, the procedure was adapted to \pp collisions, allowing one to identify two long-range contributions also in high-multiplicity \pp collisions~\cite{Aad:2015gqa,Khachatryan:2015lva}. Under certain assumptions even lower-multiplicity \pp collisions show the same features~\cite{Aad:2015gqa}. With these observations the similarity of small and large collision systems with respect to azimuthal correlations had been clearly demonstrated.

The second observation was that of enhanced production of multi-strange hadrons in high-multiplicity \pp collisions extending the puzzle from final-state particle kinematics to include also hadrochemistry. Already after Run 1, several experiments reported that ratios of strange to non-strange particle yields, in minimum-bias collisions, could not be described using model fits obtained from LEP data~\cite{Khachatryan:2011tm,Aaij:2012ut,Aamodt:2011zza,Abelev:2012jp}. After systematic studies of this discrepancy, it was found that not only does strangeness increase smoothly with particle density at mid-rapidity in \pp collisions, the dependence on this observable continues smoothly to \pPb and \PbPb collisions~\cite{ALICE:2017jyt}.

Initially, these collision systems were thought as a reference for the effects observed in \PbPb collisions. But the discovery of these qualitatively new features has turned the study of small systems into a field on its own, with significant interest in both the heavy-ion and the high-energy physics community.

In ultra-relativistic nucleus--nucleus collisions, ranging from early SPS experiments at CERN through the Relativistic Heavy-Ion Collider (RHIC) at Brookhaven National Laboratory (BNL) to LHC, similar observations have been interpreted as evidence of formation of a droplet of thermalised Quark--Gluon Plasma. The long-range azimuthal correlations, and in particular their lowest harmonic component $\vtwo$, have been used in combination with relativistic fluid-dynamical modeling to constrain the material properties of the plasma.  The striking result from RHIC was that the plasma formed in central nucleus--nucleus collisions flows as a liquid nearly without dissipation such that its specific shear viscosity $\eta/s$ -- quantifying the dissipative properties of the medium -- was found to be smaller than that of any other known substance. The inferred value of the specific shear viscosity $\eta/s\sim 0.07-0.16$ was found to be significantly smaller than the expectation from perturbative QCD and other quasiparticle models, and closer to the expectation of holographic model calculations of strongly coupled (maximally supersymmetric $N=4$) gauge theories in the limit of large number of colors $N_c \rightarrow \infty$. These models can be seen as models of fluids with minimal dissipation allowed by basic principles of quantum mechanics thus giving rise to the paradigm of Quark--Gluon Plasma as a perfect liquid. Perfect liquid models do, by definition, not have any quasi-particle structure. This means that they do not have any degrees of freedom which can free stream for an appreciable amount of time, compared to their de Broglie wavelength. In that context it is notable that the observation of fluid--like signatures in small systems can be described as a small modification of the free streaming evolution, challenging the perfect-fluid paradigm.

There are several theoretical pictures that have been suggested to explain the smooth onset of signals of collectivity in small systems. The \pp event generators have been supplemented on the one hand with elements describing string or cluster fragmentation in a dense medium~\cite{Bierlich:2014xba,Gieseke:2017clv} to address the hadrochemistry, and on the other hand with final-state interactions between the fragmenting strings to account for the final-state kinematical correlations~\cite{Bierlich:2017vhg}. The models underlying \pp event generators can then in turn be extrapolated to cover \pPb and \PbPb collision systems, which is an approach used since the 1980's~\cite{Andersson:1986gw,Wang:1991hta}. Recent theoretical developments~\cite{Bierlich:2018xfw,Bellm:2018sjt} have improved the state of such extrapolations to a degree where also the supplemented hadronisation models can be extrapolated in order to provide a microscopic picture of the QGP even in large systems. The question whether such extrapolations will give even a qualitative description of the observed features is still open.

At the same time, the description of large systems has been employed in regimes initially thought to be not accessible by models implementing the perfect-fluid paradigm. Their application down to proton--proton collisions~\cite{Werner:2007bf,Weller:2017tsr,Aidala:2018mcw}, taken at face value, would imply the formation of a nearly perfect liquid even in the smallest collision systems. Furthermore, pQCD based saturation models can describe the emergence of charged-particle azimuthal anisotropy ($\vn$)~\cite{Schenke:2016lrs}. In these models the final-state azimuthal correlations can arise either from the intrinsic correlations in the nuclear wave function (initial-state correlations) as correlated anisotropic particle production or as a final-state interaction after the initial particle production.
In heavy-ion collisions, statistical models~\cite{BraunMunzinger:2003zd} have been very successful in describing the hadrochemistry of particle yields. Their extension to \pp collisions shows promise, but similarly with points of tension~\cite{Vislavicius:2016rwi}.
Regardless of whether the approach is to extrapolate from \pp to \PbPb collisions or the other way around, it is crucial to establish that any such model can capture the essential features of intermediate systems. Asymmetric collision systems such as p–Pb, provide challenges and opportunities for both approaches. They may be used as a necessary intermediate stepping stone for pp to A–A extrapolations, and they provide possible discrimination between saturation and fluid approaches as a possible point of tension~\cite{Mace:2018vwq,Nagle:2018ybc}. 

A question remains to what extent these different models are describing qualitatively different physical phenomena or to what extent they are different representations of the same underlying physics of final-state interactions. For this reason it is important to develop theoretical tools that encompass both the fluid-dynamic limit and the free-streaming limit to theoretically describe how the microscopic physics that leads to fluid-dynamic behavior in \AOnA collisions should represent itself in small systems. 
An attempt to do so comes from transport theory, which can describe microscopic interactions but in the limit of large number of final-state interactions allows for a coarse-grained effective description in terms of fluid dynamics.  As such, transport theory has the potential to bridge the gap between small systems, where final-state interactions act as small modification to the free-streaming evolution, and central nucleus--nucleus collisions where the final-state interactions bring the matter to the fluid-dynamical limit~\cite{Kurkela:2018qeb}.

The experimental program in the large intermediate region --- spanning from mid-central \PbPb and \XeXe collisions, through \pPb collisions down to minimum-bias \pp collisions --- offers a possibility to bridge the difference between the two limits by providing a setup where the microscopic final-state interactions that lead in central \PbPb collision to the formation of a QGP may be studied in isolation in the limit of small number of final-state interactions.

The suggested theoretical pictures may have implications for high-energy physics analyses, which depend on reliable models of the underlying event. As an example, it has been recently shown that the discussed long-range correlations are also present in the underlying event of $Z$-tagged \pp collisions~\cite{ATLAS:2017nkt}. The direct implication is the necessity for questioning the correct description of the underlying event of MC models. 
As the usual models used to describe the underlying event do not describe such long-range correlations, even qualitatively, the uncertainty introduced by imposing a model dependence, might be larger than expected, as \textit{e.g.} shown for colour reconnection effects on $t\bar{t}$ final states~\cite{Argyropoulos:2014zoa}. 
As such, better descriptions of collective effects in small systems could also probe vital for reducing uncertainties in high-energy physics analyses.

The main experimental task in future years is a detailed examination and characterization of the observed effects in \pp, \pPb and \PbPb collisions, in order to understand whether such effects are different or similar in origin in small and large systems. For such a task to be successful, all three types of collision systems, \pp, \pPb and \PbPb must be utilized, as they each offer unique features not obtainable from the other systems. The central \PbPb collision system is so far the only one where all features of collectivity (including multi--particle correlations, jet quenching, quarkonia supression, thermal photons and hadrochemistry) have been observed. For the study of small collision systems, central \PbPb offers the only viable \emph{true collective} reference. Conversely, \pp is so far the smallest collision system where collective effects have been observed, and the only system where a smooth transition to the \Pepem expectation could be reasonably expected. In the intermediate region \pPb collisions are the only one of the three collision systems which offer, both, a saturation dominated initial state with a well known geometry, and, in a single event the $\mathrm{Pb}$-going and $\mathrm{p}$-going direction allowing the study between both regimes.
The potential study of \OO collisions provides an interesting system with smaller fluctuations in the number of participating nucleons.
Furthermore, the detailed study of asymmetric collision systems provides valuable input both to models extrapolating \pp dynamics to \PbPb collisions, and for providing quantitative distinction between initial-state saturation effects and final-state interactions.

This chapter is structured as follows: Section~\ref{sect:smallsystems_historicoverview} gives an overview presenting the observations that have been made and comparing them between pp, \pPb and \PbPb collisions. Subsequently, Sect.~\ref{sect:smallsystems_openquestions} summarizes the open questions and discusses how these can be addressed at HL-LHC. 
Sect.~\ref{sect:smallsystems_multiplicity} details the multiplicity distribution which needs to be extrapolated to be used for performance studies, the expected energy densities and the data-taking conditions assumed.
A set of performance studies which address the open questions are introduced in Sect.~\ref{sect:smallsystems_global}--\ref{sect:smallsystems_thermalradiation} ranging from correlation measures, over hadrochemistry to signatures of energy loss and thermal radiation. 
Finally, Sect.~\ref{sect:smallsystems_OO} presents the opportunities of a short run of \OO collisions.

\subsection{Overview of experimental results and critical assessment}
\label{sect:smallsystems_historicoverview}

\afterpage{%
\begin{landscape}
\begin{table}[h!]
\caption{\label{table:smallsystems} Summary of bulk observables or effects in \PbPb collisions, as well as in high multiplicity \pPb and \pp collisions at the LHC. References to key measurements for the various observables and systems are given. See text for details. Table adapted from Ref.~\cite{Loizides:2016tew}.}
\begin{center}
  \small
  \begin{tabular}{p{5.5cm}|p{3.4cm}|p{3.4cm}|p{3.4cm}| p{6cm} }
    Observable or effect                             & \PbPb                                             & \pPb (high mult.)                              & \pp (high mult.)                         & Refs.\\
    \hline
    \hline
    Low \pT spectra (``radial flow'')                & yes                                              & yes                                           & yes                            & \cite{Abelev:2012wca,Abelev:2013vea,Chatrchyan:2013eya,Chatrchyan:2012qb,Andrei:2014vaa,Abelev:2013haa,Acharya:2017dmc,Adam:2016bpr,Adam:2015vsf,Adam:2017zbf} \el
    Intermediate \pT (``recombination'')                & yes                                              & yes                                           & yes                            & \cite{Andrei:2014vaa,Abelev:2013xaa,Abelev:2013haa,Abelev:2014uua,Khachatryan:2016yru,Adam:2015jca,Adam:2016dau,Adam:2017zbf} \el
    Particle ratios                                  & GC level                                         & GC level except $\Omega$                      & GC level except $\Omega$       & \cite{ALICE:2017jyt,Adam:2016bpr,Adam:2015vsf,ABELEV:2013zaa} \\
    Statistical model                                & $\gamma^{\rm GC}_s=1$, 10--30\%                  & $\gamma^{\rm GC}_s\approx1$, 20--40\%         & MB: $\gamma^{\rm C}_s<1$, 20--40\%  & \cite{Floris:2014pta,Adam:2015vsf,ALICE:2017jyt} \el
    HBT radii ($R(\kT)$, $R(\sqrt[3]{\nch})$)        & $R_{\rm{out}}/R_{\rm side}\approx1$          & $R_{\rm out}/R_{\rm side}\lsim1$              & $R_{\rm out}/R_{\rm side}\lsim1$       & \cite{Adam:2015vna,Adam:2015vja,Abelev:2014pja,Adam:2015pya,Aamodt:2011kd,CMS:2014mla,Acharya:2017qtq,Aaboud:2017xpw} \el
    Azimuthal anisotropy (\vn)\nl
    (from two particle correlations)                 & \vone--\vseven                                    & \vone--\vfive                                 & \vtwo--\vfour                  & \cite{CMS:2012qk,Abelev:2012ola,Aad:2012gla,Aamodt:2011by,Chatrchyan:2011eka,Chatrchyan:2012wg,ATLAS:2012at,Aad:2014lta,Aad:2015gqa,Khachatryan:2015lva,Khachatryan:2016txc,Acharya:2017ino,Adam:2016ows,Adam:2016nfo,Acharya:2018zuq,Sirunyan:2017uyl,Aaboud:2017acw,ATLAS:2017nkt} \el
    Characteristic mass dependence                   & \vtwo--\vfive                                   & \vtwo, \vthree                                  & \vtwo                         & \cite{Abelev:2014pua,Abelev:2012di,Adam:2016nfo,Khachatryan:2014jra,ABELEV:2013wsa,Sirunyan:2018toe,Khachatryan:2016txc,Acharya:2018zuq} \el
    Directed flow (from spectators)                  & yes                                              & no                                            & no                            & \cite{Abelev:2013cva}\el
    Charge-dependent correlations                    & yes                                              & yes                                           & yes                            & \cite{Abelev:2013csa,Adam:2015vje,Adam:2015gda,Sirunyan:2017tax,Acharya:2017fau,Sirunyan:2017quh,Khachatryan:2016got}\el
    Higher-order cumulants \nl(mainly $v_{\rm 2}\{n\}$, $n\ge4$)                                              & \mbox{``$4\approx6\approx8\approx$ LYZ''} \mbox{+higher harmonics}&\mbox{``$4\approx6\approx8\approx$ LYZ''} \mbox{+higher harmonics}  & \mbox{``$4\approx6$''}  & \cite{Aad:2013fja,Chatrchyan:2013nka,Khachatryan:2016txc,Aamodt:2010pa,ALICE:2011ab,Chatrchyan:2012ta,Abelev:2014mda,Chatrchyan:2013kba,Aad:2014vba,Khachatryan:2015waa,Adam:2016izf,Sirunyan:2017pan,Sirunyan:2017igb,Aaboud:2017acw,Aaboud:2017blb} \el
    Symmetric cumulants                              & up to $\rm{SC}(5,3)$                            & only $\rm{SC}(4,2), \rm{SC}(3,2)$             & only $\rm{SC}(4,2), \rm{SC}(3,2)$                                   & \cite{Aad:2014fla,Aad:2015lwa,ALICE:2016kpq,Sirunyan:2017uyl,Acharya:2017gsw,Aaboud:2018syf}  \el
    Non-linear flow modes                            & up to \vsix                                    & not measured                                  & not measured & \cite{Acharya:2017zfg} \el
    Weak $\eta$ dependence                           & yes                                              & yes                                           & not measured                  & \cite{Adam:2016ows,Aad:2014eoa,ATLAS:2011ah,Khachatryan:2016ibd,Adam:2015bka,Aaij:2015qcq,Aaboud:2016jnr,Sirunyan:2017igb} \el
    Factorization breaking                           & yes ($n=2,3$)                                    & yes ($n=2,3$)                                 & not measured                  & \cite{Aad:2014lta,Khachatryan:2015oea,Sirunyan:2017gyb,Acharya:2017ino,Aaboud:2017tql}\el
    Event-by-event \vn distributions                 & $n=2$--$4$                                       & not measured                                  & not measured                  & \cite{Aad:2013xma,Sirunyan:2017fts,Acharya:2018lmh} \el
    Direct photons at low \pT                        & yes                                              & not measured                                  & not observed                            & \cite{Adam:2015lda,Acharya:2018dqe}\el
    Jet quenching through dijet asymmetry            & yes 						& not observed   				& not observed 					& \cite{Aad:2010bu,Chatrchyan:2011sx,Sirunyan:2018jju,Khachatryan:2016tfj,Sirunyan:2017bsd,Chatrchyan:2014hqa}\el
    Jet quenching through \RAA                       & yes                                              & not observed                                  & not observed                  & \cite{Aamodt:2010jd,CMS:2012aa,Abelev:2012hxa,ALICE:2012ab,Aad:2014bxa,Adam:2015ewa,Aad:2015wga,Adam:2016jfp,Sirunyan:2016fcs,Acharya:2018qsh,Acharya:2018njl,ATLAS:2014cpa,Khachatryan:2015xaa}\el  
    Jet quenching through correlations               & yes ($Z$--jet, $\gamma$--jet, h--jet)               & not observed (h--jet)                          & not measured                  & \cite{Sirunyan:2017jic,Sirunyan:2017qhf,Adam:2016jfp,Adam:2015doa,Acharya:2017okq,Adam:2016xbp,Sirunyan:2018jqr,Sirunyan:2018qec,Aaboud:2017bzv,Aaboud:2017eww}\el
    Heavy flavor anisotropy                          & yes                                              & yes                                           & not measured                  & \cite{ALICE:2013xna,Abelev:2013lca,Abelev:2014ipa,Adam:2015pga,Acharya:2017tfn,Adam:2016ssk,ALICE:2016clc,Acharya:2017qps,Sirunyan:2017plt,Acharya:2017tgv,Khachatryan:2016ypw,Acharya:2018dxy,Sirunyan:2018toe}\el
    Quarkonia production                             & suppressed$^\dagger$         & suppressed                                    & not measured                  & \cite{Abelev:2012rv,Adam:2015rba,Chatrchyan:2012lxa, Chatrchyan:2013nza, Abelev:2014zpa, Adam:2015jsa, Adam:2016ohd, Adam:2015rta, Adam:2015isa, Adam:2015gba, Adam:2016rdg, Adamova:2017uhu, Acharya:2017tfn, Acharya:2017hjh, Sirunyan:2017lzi, Khachatryan:2016ypw, Sirunyan:2017mzd, Sirunyan:2017isk, Khachatryan:2016xxp, Sirunyan:2016znt, Aaboud:2017cif, Aaij:2017cqq} \\
    \hline
    \hline
  \end{tabular}
\begin{minipage}[l]{22.5cm}
\begin{flushleft}
\footnotesize
\vspace{5pt}
$\dagger$ J/$\psi \uparrow$, $\PGUP \downarrow$ w.r.t. RHIC energies.
\end{flushleft}
\end{minipage}
\end{center}
\end{table}
\end{landscape}
}

This section will give an overview where the measurements in \PbPb, \pPb and \pp collisions at LHC provide a consistent picture and where differences emerge. In addition, it is pointed out where measurements are missing and where additional data is needed which forms the basis for the projections given in the subsequent sections of this chapter. Table~\ref{table:smallsystems} lists the different observables and if they have been measured in the \PbPb, \pPb and \pp collision systems. In the following a critical assessment of the findings is performed.

\textbf{Particle spectra}
In all three systems, the \pT spectra of identified particles harden with increasing multiplicity. If this is interpreted by using a combined blast-wave parametrisation in \PbPb collisions\footnote{A combined blast-wave parametrisation model is a blast-wave model that fits charged pions, kaons and (anti-)protons simultaneously. In \cite{Abelev:2012wca}, combined blast-wave parametrisation perfectly describes $\pi^{\pm}$ ($ 0.5 < \pT < 1$ \UGeVc), $K^{\pm}$ ($ 0.2 < \pT < 1.5$ \UGeVc) and ${\rm p}+\overline{\rm{p}}$ ($ 0.3 < \pT < 3$ \UGeVc).} a larger radial flow is observed in \pp and \pPb collisions at the same multiplicity~\cite{Acharya:2018orn} as expected by Ref.~\cite{Shuryak:2013ke}. In the intermediate \pT region ($2 < \pT < 5 $~\UGeVc), enhancement of baryon-to-meson ratios is observed in all three systems. Recombination models suggest that the number of constituent quarks of the hadrons determine this enhancement~\cite{Andrei:2014vaa,Abelev:2013xaa,Abelev:2013haa,Abelev:2014uua,Khachatryan:2016yru,Adam:2015jca,Adam:2016dau,Adam:2017zbf}. Particle ratios and yields are described as in the Grand Canonical ensemble by the statistical model with the strangeness undersaturation factor $\gamma_{S}\approx 1$ with an accuracy of approximately 10--30\% for \PbPb collisions and at 20--40\% level for \pPb collisions (except for the $\Omega$ meson). The statistical model has been so far applied to minimum-bias \pp collisions and when treated as a canonical ensemble, was found to describe the yields with $\gamma^{C}_{S} < 1$ and deviations of only about 20--40\% from the expected yields~\cite{ALICE:2017jyt,Adam:2016bpr,Adam:2015vsf,ABELEV:2013zaa}.

\textbf{Pressure-driven expansion and anisotropies}
Assuming that the pressure gradients build up early in the evolution of the created system, initial spatial anisotropies ($\varepsilon_{\rm{n}}$) translate into final momentum anisotropies, namely anisotropic flow (\vn) in a system with small viscosity. A large number of detailed studies have been done on different coefficients of anisotropic flow. Higher-order flow harmonics are in particular more sensitive to initial-state fluctuations and therefore can constrain the initial conditions of the system. Anisotropic flow has been measured with two-particle correlation techniques up to \vseven in \PbPb, \vfive in \pPb, and \vfour in \pp collisions for charged particles. These $\vn$ coefficients exhibit weaker multiplicity dependence in \pp and \pPb collisions than in \PbPb collisions where this is closely related to the shape of the overlap region~\cite{CMS:2012qk,Abelev:2012ola,Aad:2012gla,Aamodt:2011by,Chatrchyan:2011eka,Chatrchyan:2012wg,ATLAS:2012at,Aad:2014lta,Aad:2015gqa,Khachatryan:2015lva,Khachatryan:2016txc,Acharya:2017ino,Adam:2016ows,Adam:2016nfo,Acharya:2018zuq,Sirunyan:2017uyl,Aaboud:2017acw}.

Higher-order cumulants have been measured using Lee-Yang Zeros (LYZ) method and multi-particle correlation techniques with up to 8 particles in both \PbPb and \pPb collisions and up to 6 particles in \pp collisions~\cite{Aad:2013fja,Chatrchyan:2013nka,Khachatryan:2016txc,Aamodt:2010pa,ALICE:2011ab,Chatrchyan:2012ta,Abelev:2014mda,Chatrchyan:2013kba,Aad:2014vba,Khachatryan:2015waa,Adam:2016izf,Sirunyan:2017pan,Sirunyan:2017igb,Aaboud:2017acw,Aaboud:2017blb}. Interestingly, for each collision system, the measurements of the cumulants at different orders ($n \geq 4$) are similar within 10\%. The presence of non-zero higher-order cumulants with similar magnitude can be interpreted as evidence for a hydrodynamically evolving system. However, some disfavor this interpretation since models that do not incorporate hydrodynamics have also been able to reproduce these results~\cite{Jia:2014pza,Gyulassy:2014cfa,McLerran:2014uka}. The \pT-differential \vn measurements for identified particles show the characteristic mass dependence of anisotropic flow up to \vfive in \PbPb, \vthree in \pPb, and \vtwo in \pp collisions where heavier particles are depleted at low \pT~\cite{Abelev:2014pua,Abelev:2012di,Adam:2016nfo,Khachatryan:2014jra,ABELEV:2013wsa,Sirunyan:2018toe,Khachatryan:2016txc,Acharya:2018zuq}. In \PbPb collisions this is ascribed to the interplay between radial flow and anisotropic flow harmonics at low \pT and recombination at higher \pT. This characteristic mass dependence has been described by hydrodynamic calculations to a good approximation in all three systems. In the intermediate \pT values in all three systems a meson-baryon grouping can be observed which points to a combination of hydrodynamics and quark coalescence (or recombination).

The non-linear hydrodynamic response of the system has been probed using symmetric cumulants which quantify the correlation between different anisotropic flow harmonics. Symmetric cumulants, that are also known as mixed harmonics, have been measured in all three systems up to SC(5,3) in \PbPb and SC(4,2) in \pPb and \pp collisions~\cite{Aad:2014fla,Aad:2015lwa,ALICE:2016kpq,Sirunyan:2017uyl,Acharya:2017gsw,Aaboud:2018syf}. Different order harmonic correlations have different sensitivities to the transport properties of the system and the initial conditions. Based on the hydrodynamic calculations the data favour a small shear viscosity~\cite{Zhu:2016puf}. In addition, the linear and non-linear hydrodynamic response has been investigated in \PbPb collisions up to the sixth order flow harmonic~\cite{Acharya:2017zfg}. These new observables, i.e. linear and non-linear flow modes, are very sensitive to details of the hydrodynamic modelling, i.e. initial conditions and the transport properties of the system. Current data--model comparison show this sensitivity which help to constrain the transport properties of the QGP created in \PbPb collisions~\cite{Acharya:2017zfg}. Linear ($\vn^{\rm{L}}$) and non-linear ($\vn^{\rm{NL}}$) flow modes in \pp and \pPb collisions are not yet measured and can constrain the transport properties as well as initial conditions of these small systems. Furthermore, hydrodynamic calculations capture qualitatively "higher-order" details, such as the breaking of factorization due to event-plane angle decorrelations in \pT and $\eta$ measured in both \PbPb and \pPb collisions~\cite{Khachatryan:2015oea,Sirunyan:2017gyb,Acharya:2017ino}. With the existing data such measurements are not yet possible in \pp collisions. Similarly, event-by-event \vn measurements have only been done in \PbPb collisions~\cite{Aad:2013xma,Sirunyan:2017fts,Acharya:2018lmh} and it would be interesting to study those in both \pPb and \pp collisions. 

Directed flow, for the rapidity-odd as well as the rapidity-even components, of charged particles at mid-rapidity was measured relative to the collision symmetry plane defined by the spectator nucleons, and evidence for dipole-like initial-state density fluctuations in the overlap region was found in \PbPb collisions~\cite{Abelev:2013cva}. In small systems, the concept of directed flow is less clear, especially in \pp collisions. If there is collectivity in \pp collisions, one could also expect a non-zero directed flow measurement. This is technically challenging since the measurement of the spectator plane is not feasible in small systems and, hence, \vone could only be measured using higher-order ($n\geq 4$) cumulants. The width of the balance functions, $\langle\Delta\eta\rangle$ and $\langle\Delta\varphi\rangle$, have been measured for charged particles in \pp, \pPb and \PbPb collisions \cite{Abelev:2013csa,Adam:2015gda}. The balance function probes the charge creation time and the development of collectivity in the produced system. These measurements are consistent with the picture of a system exhibiting larger radial flow with increasing multiplicity but also whose charges are created at the later stages of the collision. The charge-dependent azimuthal correlations are measured in both \PbPb and \pPb collisions~\cite{Abelev:2014pja,Adam:2015pya,Aamodt:2011kd,CMS:2014mla,Aaboud:2017xpw}. These correlations quantify the influence of the chiral magnetic effect (CME) and the chiral magnetic wave (CMW) on the produced particles. These correlators are also sensitive to strong background contributions, for example from local charge conservation and possibly radial and anisotropic flow. 

The freeze-out radii in three orthogonal directions ("out", "side", "long") can be deduced from measurements of quantum-statistic correlations between pairs of same-charge pions and kaons (HBT) at low-momentum transfer. The HBT radii in all collision systems are found to scale with $\sqrt[3]{\nch}$ indicating a constant density at freeze-out, and to decrease with increasing pair momentum \kT{} as expected from hydrodynamics. The size along the emission direction is similar to the geometric size of the system ($R_{\rm out}/R_{\rm side} \approx 1$) in \PbPb collisions~\cite{Adam:2015vna,Adam:2015vja,Abelev:2014pja,CMS:2014mla,Acharya:2017qtq,Acharya:2017qtq} and $R_{\rm out}/R_{\rm side} \leq 1$ for both \pPb and \pp collisions~\cite{Abelev:2014pja,Adam:2015pya,Aamodt:2011kd,CMS:2014mla,Aaboud:2017xpw}.

\textbf{Direct photons}
Direct-photon measurements in the low \pT region are so far performed in \PbPb and \pp collisions. The measurements are reproduced by models assuming the formation of a QGP in \PbPb collisions \cite{Adam:2015lda}. In this measurement, one cannot discriminate between the available models due to the large systematic uncertainties: models incorporating different initial temperatures, i.e. from 385 to 740 \UMeV\ in the most central \PbPb collisions, are able to reproduce the measurements. Nevertheless, the comparison among these models suggests that the initial temperature in central \PbPb collisions must exceed about \unit[400]{\UMeV}~\cite{Adam:2015lda}. No significant direct-photon signal has been extracted in \pp collisions at current available center-of-mass energies~\cite{Acharya:2018dqe}.

\textbf{Energy loss}
The created system in \PbPb collisions is opaque for high-\pT colored probes. Due to radiational and collisional energy loss high-\pT colored probes are strongly suppressed (jet quenching) whereas the system is transparent for photons and other colorless probes~\cite{Aamodt:2010jd,CMS:2012aa,Abelev:2012hxa,ALICE:2012ab,Aad:2014bxa,Adam:2015ewa,Aad:2015wga,Acharya:2018qsh,Acharya:2018njl}. Jet quenching leads to a large asymmetry in back-to-back jet \pT and slightly modified jet fragmentation functions inside small jet cone sizes ($\rm{R} = 0.4$). Most of the radiated energy appears at large angles ($\rm{R} > 0.8$)~\cite{Aad:2010bu,Chatrchyan:2011sx,Sirunyan:2018jju,Khachatryan:2016tfj,Sirunyan:2017bsd}. 

On the contrary, the picture is different in \pPb collisions: measurements of inclusive high-\pT\ hadron and inclusive jet yields in minimum-bias \pPb collisions at the LHC are consistent with $\RpPb = 1$ within the current accuracy of approximately 20\%; i.e. no evidence of medium-induced modification is observed~\cite{ATLAS:2014cpa,Khachatryan:2015xaa,Adam:2016jfp,Sirunyan:2016fcs,Acharya:2018qsh}.
For event classes split by event activity, neither medium-induced modification in inclusive hadron production nor dijet transverse momentum imbalance are observed~\cite{Khachatryan:2015xaa, Acharya:2018qsh,Chatrchyan:2014hqa}; in contrast, for inclusive jet yields \RpPb is strongly suppressed relative to unity in ``central'' \pPb collisions, and strongly enhanced in ``peripheral'' \pPb collisions~\cite{ATLAS:2014cpa}, attributed to selection biases~\cite{Acharya:2017okq}. 

The semi-inclusive yield of jets recoiling from a high-\pT trigger hadron has been used to search for jet quenching in \pPb collisions~\cite{Acharya:2017okq}. This observable is trigger-normalized and semi-inclusive, and it therefore has greater systematic sensitivity to jet quenching effects in small systems than inclusive jet observables. Nevertheless, no significant jet quenching effects within the uncertainties of the measurement have been observed. These uncertainties can be expressed as an upper limit of \unit[400]{\UMeV} (at 90\% CL) on medium-induced energy transport outside a jet cone with $R=0.4$. This value is a factor 20 smaller than the magnitude of out-of-cone energy transport measured by a similar approach in \PbPb collisions~\cite{Adam:2015doa}.

\textbf{Heavy flavour}
Due to interactions and rescattering with the medium, also heavy-flavour particles exhibit finite anisotropies as shown with non-zero \vtwo measurements for heavy flavour particles in both \PbPb and \pPb collisions~\cite{ALICE:2013xna,Abelev:2013lca,Abelev:2014ipa,Adam:2015pga,Acharya:2017tfn,Adam:2016ssk,ALICE:2016clc,Acharya:2017qps,Sirunyan:2017plt,Acharya:2017tgv,Khachatryan:2016ypw,Acharya:2018dxy,Sirunyan:2018toe}, see also Chapter~\ref{sec:HI_HF}. In addition, J/$\psi$ suppression in \PbPb collisions shows an enhancement w.r.t. RHIC energies \cite{Adam:2015isa}. Models incorporating a J/$\psi$ regeneration component from deconfined charm quarks in the medium can reproduce these measurements \cite{Abelev:2012rv,Adam:2015isa}. The limited understanding of cold nuclear matter effects in the open charm cross section determination, however, restricts the ability of these models to fully describe the experimental data on J/$\psi$ production in \PbPb collisions \cite{Aaij:2017cqq}. The size of these effects can be quantified by measurements in \pPb collision. In \pPb collisions, J/$\psi$ is suppressed relative to \pp collisions \cite{Aaij:2017cqq}. The production of the excited charmonium state, $\Psi(2S)$ as well as different bottomonium states ($\PGUP{nS}$) have been measured in both \PbPb and \pPb collisions \cite{Adam:2016ohd,Adam:2015isa,Aaboud:2017cif,Sirunyan:2016znt,Khachatryan:2016xxp} which shows a suppression w.r.t. the ground state. 

\subsection{Open questions and new opportunities at HL-LHC}
\label{sect:smallsystems_openquestions}

The previous section has extensively reviewed the state-of-the-art experimental knowledge of \pp and \pPb collisions. Certain gaps in knowledge became apparent due to either insufficient available data or shortcomings in the present detectors. The HL era of LHC can make a significant step ahead in many areas. The most relevant ones are discussed with dedicated performance projections in the remainder of this chapter.

Run 3 and 4 will allow the study of unprecedentedly high-multiplicity \pp collisions. In order to do estimates in this regime, Sect.~\ref{sect:smallsystems_multiplicity} will establish a firm extrapolation of the multiplicity distribution based on current LHC data together with a review of the data sample to expect. The large multiplicities bring a qualitatively new feature: a wide overlap between \pp and \PbPb collisions up to about 65\% central collisions allowing a unique opportunity to compare observables in a small (\pp) and large (\PbPb) system at the same multiplicity. Studies in \pPb collisions amend the picture. Given that multiplicity is not the only driving variable of a system, comparisons of estimates of the energy density in \pp, \pPb and \PbPb collisions are made below. The uniqueness of these extreme multiplicity \pp collisions warrants that the study of their global-event properties are an interesting subject in itself, see Sect.~\ref{sect:smallsystems_global}.

Subsequently, a set of key observables is presented which require either the large data samples or the upgraded detectors. In particular the measurement of thermal dileptons profits from the new ALICE pixel detector with reduced material budget, and the measurement of higher-order correlations from the extended tracker acceptance in Run 4 in ATLAS and CMS. Correlations at higher orders using the subevent method will provide an essentially non-flow free measurement of $\vn$ coefficients and their inter-correlations measured through symmetric cumulants, see Sect.~\ref{sect:smallsystems_correlations}. These measurements focus on two interesting regimes: at high multiplicity where the overlap with \PbPb collisions will be studied, and at low multiplicity to answer the question on the onset of collective phenomena.
This section also shows that the measurement of the probability distribution of event-by-event $\vtwo$ becomes for the first time feasible in small systems, a quantity presently completely unknown in \pp collisions.

The smooth increase of strange-particle production across system size is one of the key surprising findings from Run 2 \pp physics. A projection of the reach at HL-LHC is given in Sect.~\ref{sect:smallsystems_strangeness} showing that the question if the thermal limit, given by statistical models in \PbPb collisions, is reached also in \pp collisions can be answered.
A puzzling finding is the absence of jet quenching in \pPb collisions with the measurements performed in Run 1 and 2. If final-state interactions are to explain the observed collective phenomena, also energy loss of traversing partons should be measurable. 
Section~\ref{sect:smallsystems_energyloss} discusses how jet quenching can be observed in Run 3 and 4 if present, or alternatively how a stringent limit can be set. Performance studies are presented for hadron--jet, $\gamma$--jet and $Z$--jet correlations, both, in \pPb and \pp collisions.
Finally, the potential to detect thermal radiation and extract a medium temperature in \pPb collisions is presented in Sect.~\ref{sect:smallsystems_thermalradiation}. Such a measurement would constitute a strong indication of the formation of an emitting medium. Finally, the potential of colliding smaller nuclei, in particular oxygen, is assessed in Sect.~\ref{sect:smallsystems_OO}.

\subsection{Proton--proton collisions at extreme multiplicities}
\label{sect:smallsystems_multiplicity}

\subsubsection{Multiplicity distribution}

For the performance estimates at high multiplicity in \pp collisions, a multiplicity-distribution extrapolation has been used which is based on existing ALICE ($|\eta| < 1.5$)~\cite{Adam:2015gka} and ATLAS ($|\eta|< 2.5$)~\cite{Aad:2010ac,Aad:2016xww} data. Data from CMS~\cite{Khachatryan:2010nk} is compatible with the used distributions and is therefore not explicitly included in the extrapolation. A parameterisation with a single\footnote{At LHC energies two NBDs are needed for a good fit to the full distribution, but one is sufficient for the tail of the distribution.} negative binomial distribution is used to characterize the multiplicity distribution~\cite{GrosseOetringhaus:2009kz,ALICE:2017pcy}.

\begin{figure}[t]
\centering
\includegraphics[width=0.49\linewidth]{\main/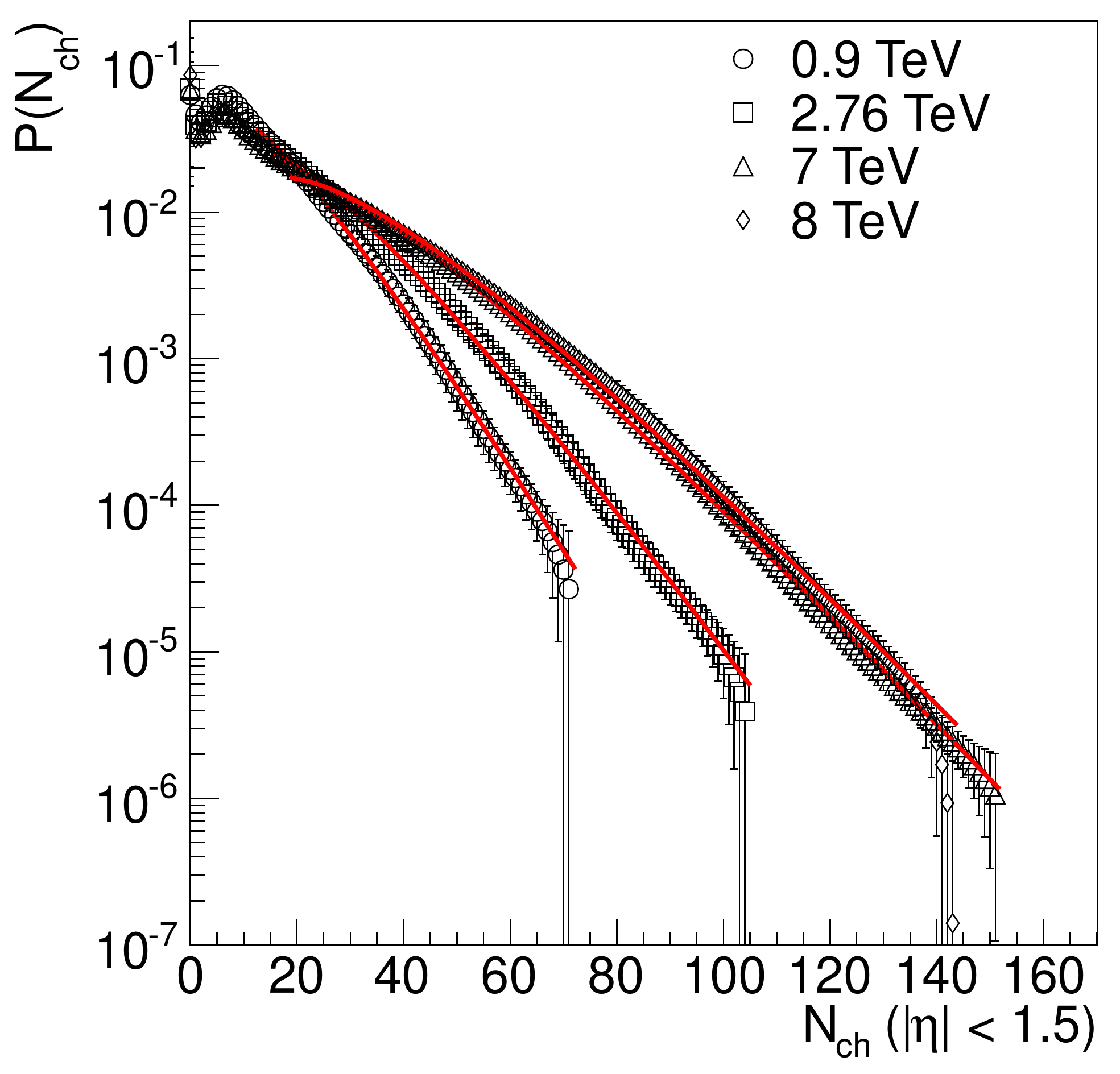}
\hfill
\includegraphics[width=0.49\linewidth]{\main/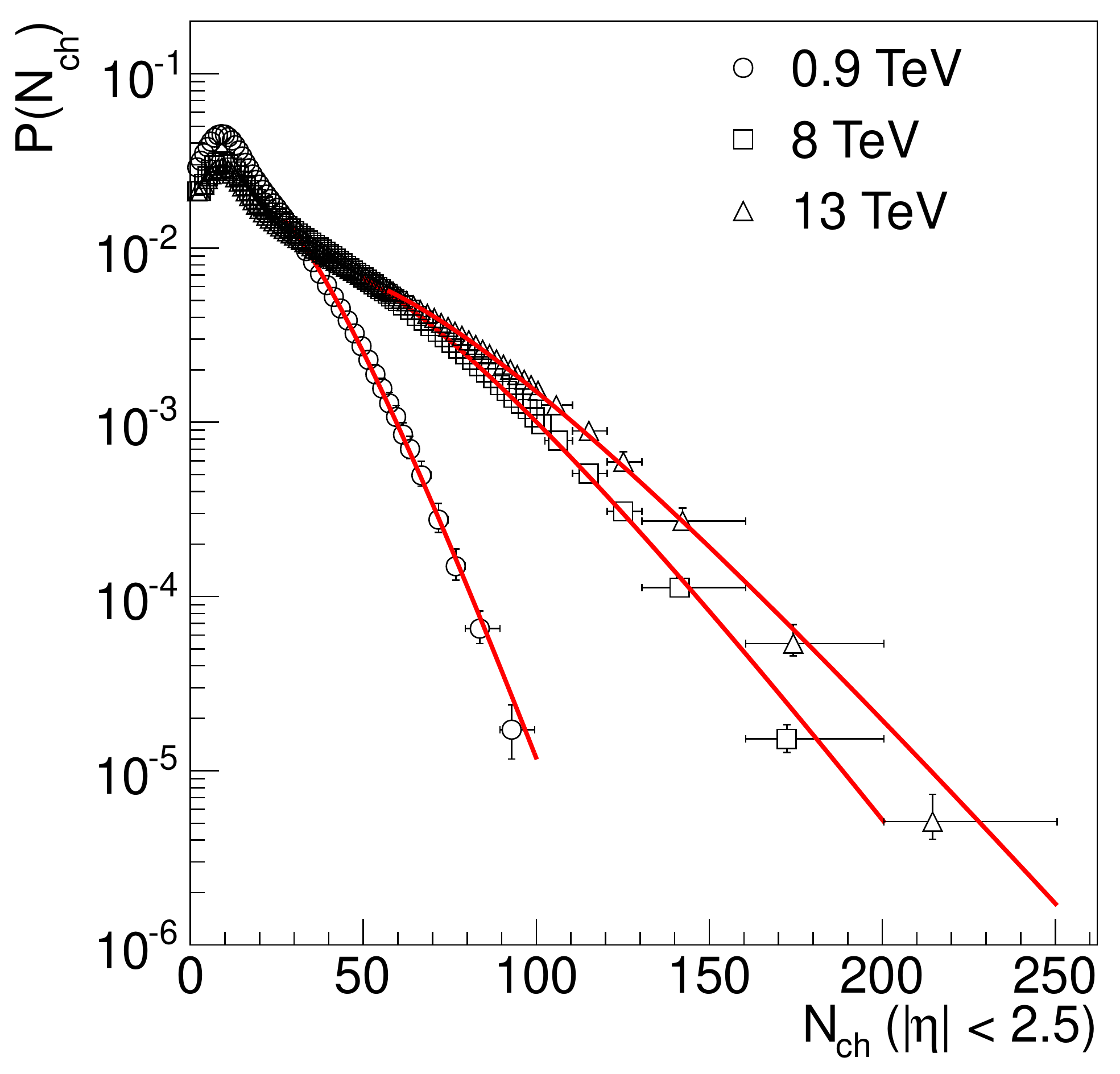}
\caption{Multiplicity distributions measured by ALICE~\cite{Adam:2015gka} (left panel) and ATLAS~\cite{Aad:2010ac,Aad:2016xww} (right panel) overlaid by the fit with a negative binomial distribution. For details see text.}
\label{fig:smallsystems_mult_data}
\end{figure}

The data is shown in Fig.~\ref{fig:smallsystems_mult_data} overlaid with the fit with a single negative binomial distribution of the tail of the distribution (20--40\% of the cross-section). The three parameters of this fit are itself fit with a power law to extrapolate to $\sqrts = \unit[14]{\UTeV}$.

\begin{figure}[t]
\centering
\includegraphics[width=0.32\linewidth]{\main/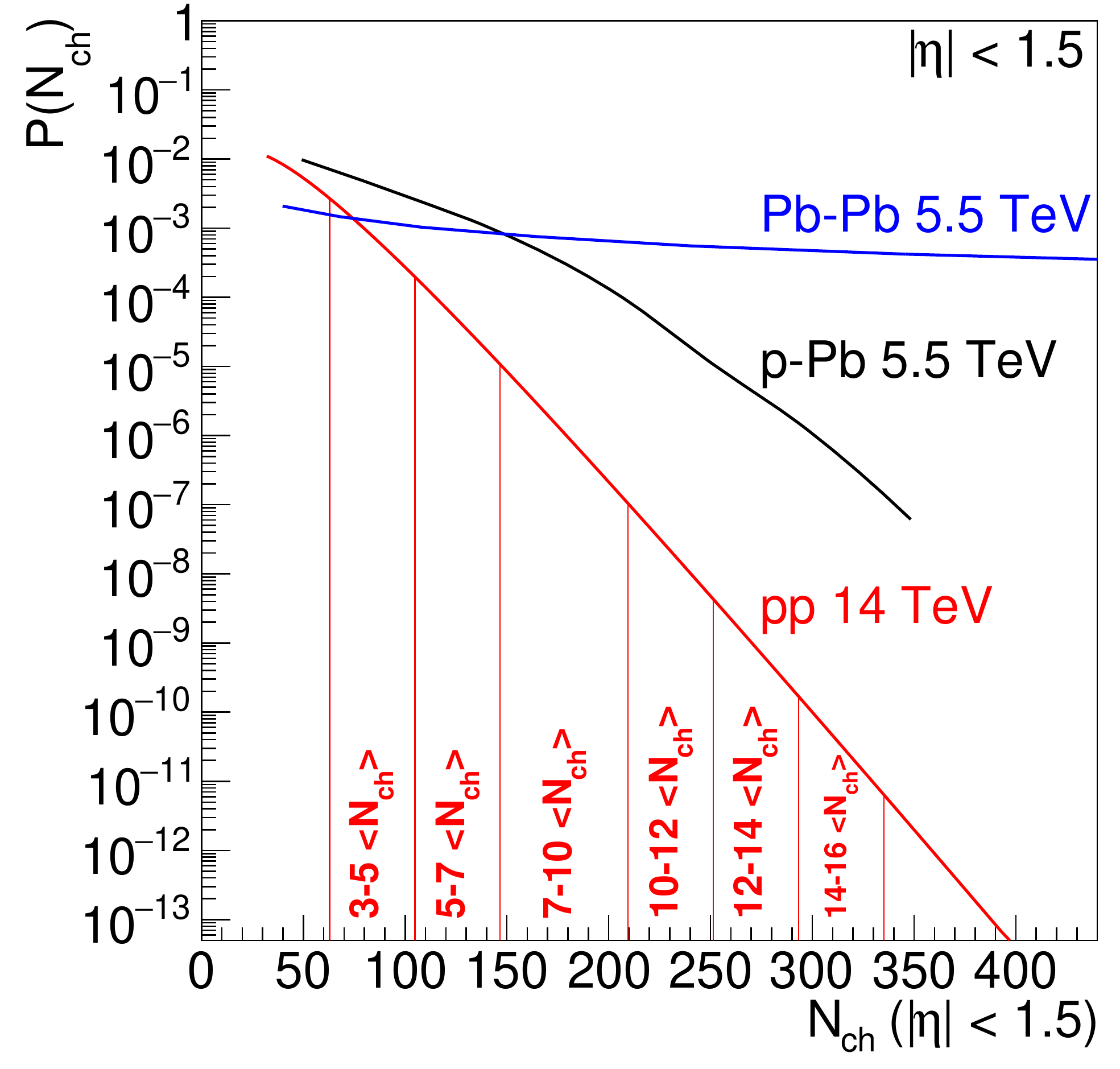}
\includegraphics[width=0.32\linewidth]{\main/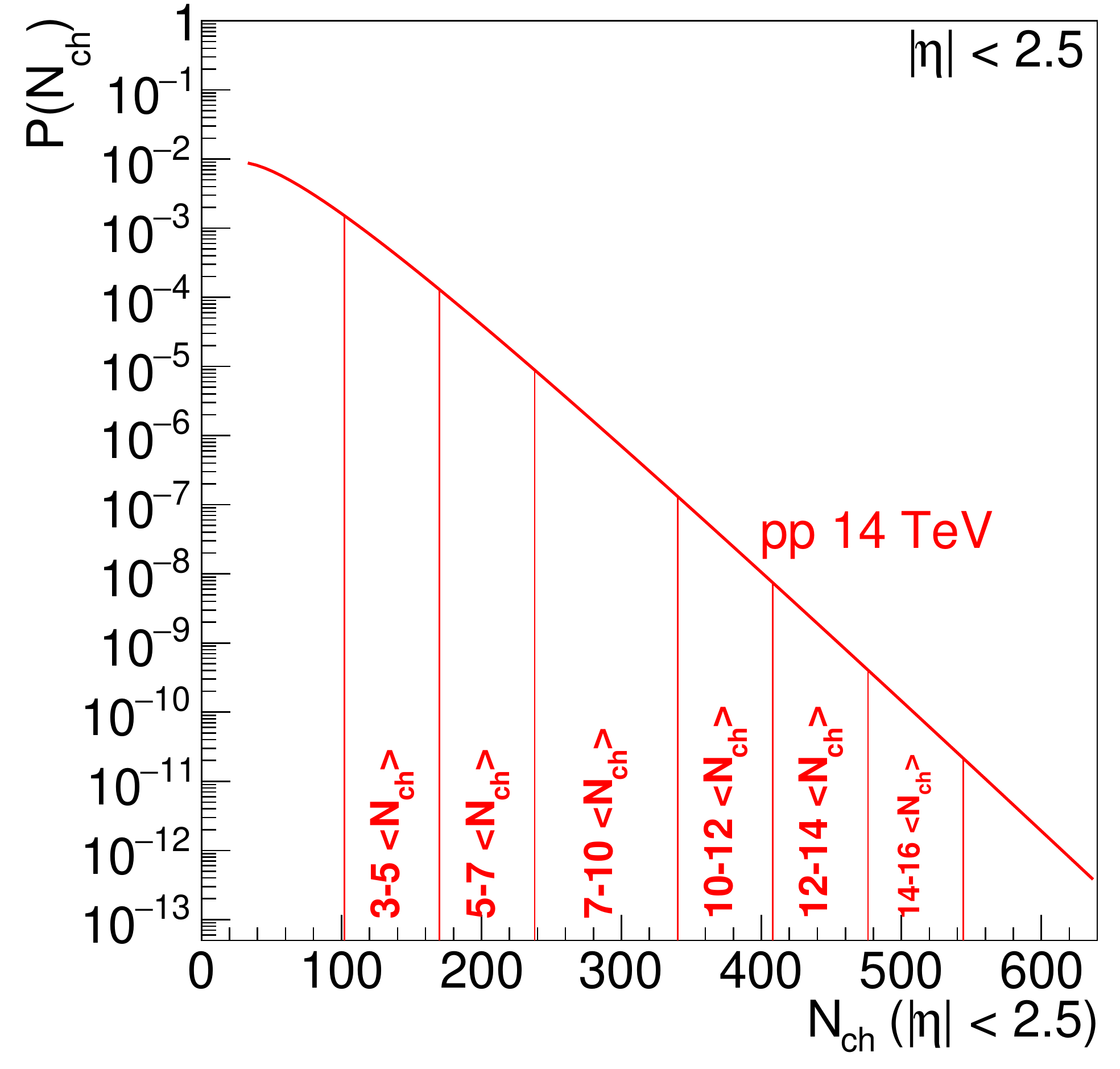}
\includegraphics[width=0.32\linewidth]{\main/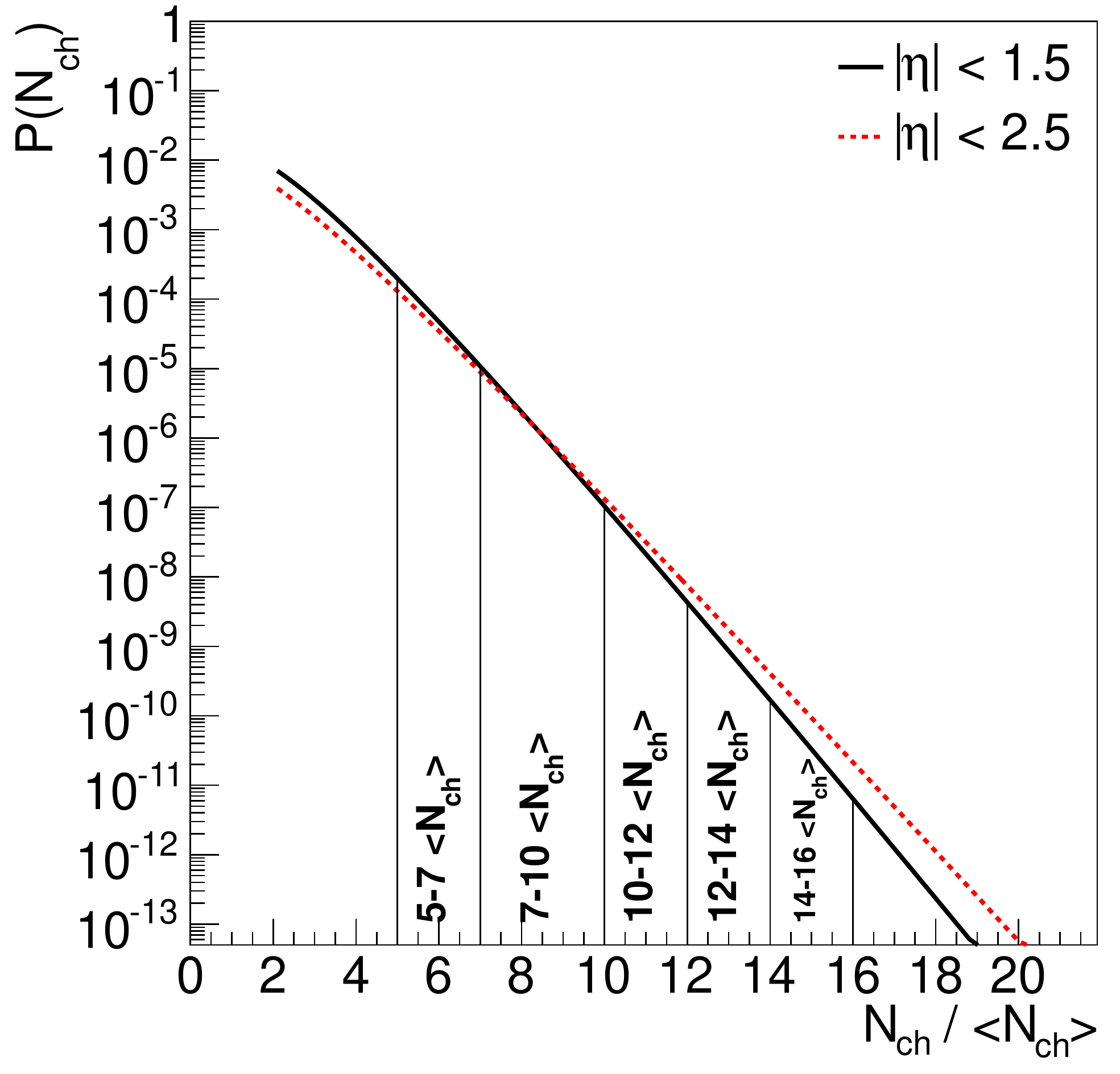}
\caption{Extrapolated multiplicity distributions in \pp collisions within $|\eta| < 1.5$ (left panel) and $|\eta| < 2.5$ (centre panel). The indicated regions are (from left to right) 5--7, 7--10, 10--12, 12--14, 14--16 times the average multiplicity. In the left panel the multiplicity distribution of \PbPb and \pPb collisions is also plotted. The right panel compares these two distributions scaled by the average multiplicity. The extrapolation for $|\eta| < 2.5$ turns out to be a bit wider at large multiplicities; therefore the one based on $|\eta| < 1.5$ is used as baseline.}
\label{fig:smallsystems_mult_extrapolation}
\end{figure}

The resulting extrapolated multiplicity distribution for \unit[14]{\UTeV} is shown in Fig.~\ref{fig:smallsystems_mult_extrapolation} for the ALICE and ATLAS case. In addition, these are compared scaled by their respective average multiplicities. The agreement is rather good, with some discrepancy in the tail of the distribution. The extrapolation based on the smaller phase-space region falls off more quickly with multiplicity, and is therefore used as the conservative estimate for the extrapolations in this chapter.

\begin{figure}[t]
\centering
\includegraphics[width=0.5\linewidth]{\main/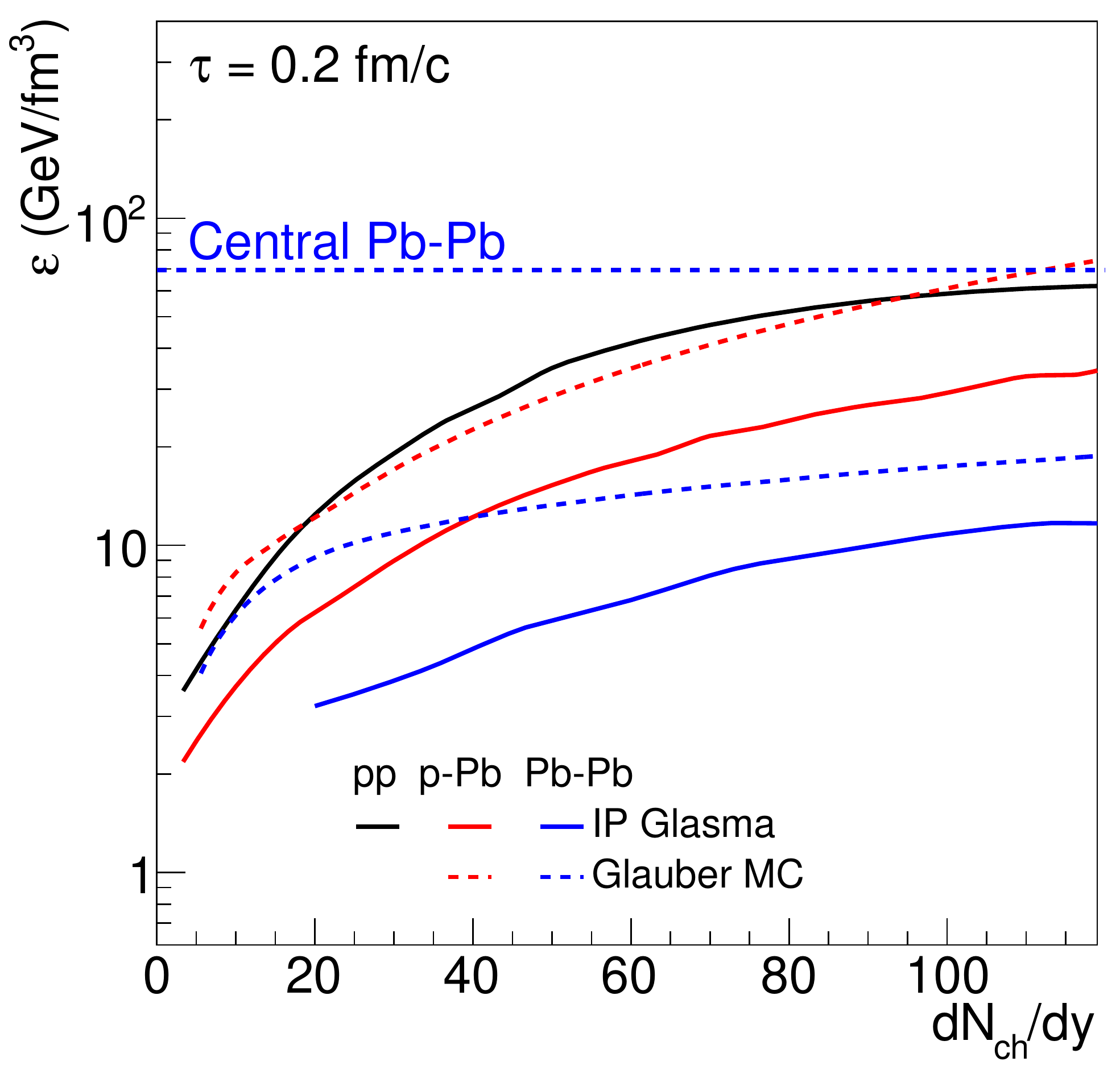}
\caption{Energy density as a function of $\dNdy$ calculated by IP-Glasma (solid lines) and with MC Glauber and the Bjorken formula (dashed lines); for details see text. Compared are \pp ($\sqrts = \unit[7]{\UTeV}$), \pPb ($\sqrtsNN = \unit[5.02]{\UTeV}$) and \PbPb ($\sqrtsNN = \unit[5.02]{\UTeV}$) collisions at $\tau = \unit[0.2]{fm/c}$. The horizontal line indicates the energy density reached in central \PbPb collisions ($\dNdy \approx 2000$).}
\label{fig:small_systems_energy_density}
\end{figure}

\subsubsection{Energy density}
While the multiplicity is a convenient and well-defined observable to compare different collision systems, the underlying dynamics may be driven by other properties. In large collision systems, the energy density $\epsilon$ is often used to characterize the system and the expected effects. Figure~\ref{fig:small_systems_energy_density} shows an estimate of the energy density for \pp, \pPb and \PbPb collisions based on IP-Glasma~\cite{Bzdak:2013zma} as well as on the Bjorken estimate:
\begin{equation}
  \epsilon = \frac{1}{A \tau} \langle E \rangle \frac{3}{2} \frac{dN_{ch}}{dy}.
\end{equation}
For the latter, the input is the multiplicity-dependent $\meanpT$~\cite{Abelev:2013bla,Acharya:2018njl} as well as the multiplicity-dependent transverse overlap from a Glauber MC~\cite{Loizides:2017ack}. The energy density is calculated at fixed $\tau = \unit[0.2]{fm/c}$. It should be noted that these assumptions can be challenged and other ways to calculate $\epsilon$ are available which can lead to largely different value in particular in \pp and \pPb collisions. Here the aim is only to show that the energy density depends on the system at a fixed multiplicity, and can reach large values in \pp and \pPb collisions, of the order of central \PbPb collisions.

\subsubsection{Data-taking conditions and integrated luminosity for \pp collisions}
For the performance studies in this chapter, a high-multiplicity sample of $\Lint = \unit[200]{pb^{-1}}$ is assumed per experiment. In order to assure a clean trigger, collisions at low $\mu \approx 1$ are needed which requires special runs or special conditions at the end of fill for ATLAS and CMS. For LHCb, the comparatively low pileup and good vertex resolution should allow for recording high-multiplicity events during normal running conditions at a pile up of about five visible pp interactions. ALICE generally runs at low $\mu$ and can collect a similar sample over a longer data-taking period. For Run 4, the upgraded tracking and vertex detectors in ATLAS and CMS may allow to isolate high-multiplicity collisions also in a large $\mu$ environment. This option needs to be carefully studied.

Table~\ref{tab:smallsystems_pp} gives the fraction of cross-section and the number of events in five multiplicity classes:  5--7, 7--10, 10--12, 12--14 and 14--16 times the average multiplicity. Table~\ref{tab:smallsystems_pbpb} gives the number of events of bins with equivalent multiplicity than commonly measured multiplicity bins in \pPb and \PbPb collisions. For the calculation of the number of events $\sigma_{\rm inel} = \unit[78.4]{mb}$~\cite{Loizides:2017ack} is used. These tables are the key input for the performance figures presented in this chapter. 
The conversion of the provided $\dNdeta$ to multiplicity ranges with larger pseudorapidity coverage is done for simplicity assuming a flat pseudorapidity distribution within $|\eta| < 2.5$. For the conversion to a phase space with a $\pT$ cut as employed in many current measurements a set of conversation factors is used, listed in Tab.~\ref{tab:smallsystems_conversion}.

\begin{table}
\caption{Number of pp events at $\sqrts = \unit[14]{\UTeV}$ in selected high-multiplicity bins.}
\centering
\begin{tabular}{c|c|c|c|c}
Range & $\dNdeta$ & Fraction & Events per pb$^{-1}$ & Events in \unit[200]{pb$^{-1}$} \\
\hline
5--7 \meannch     & 35--49   & 2.4e-03       & 1.9e+08       & 3.7e+10 \\
7--10 \meannch    & 49--70   & 1.3e-04       & 1.0e+07       & 2.0e+09 \\
10--12 \meannch   & 70--84   & 1.1e-06       & 9.0e+04       & 1.8e+07 \\
12--14 \meannch   & 84--98   & 4.7e-08       & 3.7e+03       & 7.3e+05 \\
14--16 \meannch   & 98--112  & 1.8e-09       & 1.4e+02       & 2.8e+04 \\
\hline
\end{tabular}
\label{tab:smallsystems_pp}
\end{table}

\begin{table}
\caption{Number of events in pp collisions at $\sqrts = \unit[14]{\UTeV}$ sliced in equivalent multiplicity bins as in \pPb and \PbPb collisions.}
\centering
\begin{tabular}{l|c|c|c}
Range & $\dNdeta$ & Events per pb$^{-1}$ & Events in \unit[200]{pb$^{-1}$} \\
\hline
0--5\% \pPb   & 41--56        & 4.9e+07       & 9.8e+09 \\
5--10\% \pPb  & 34--41        & 1.9e+08       & 3.8e+10 \\
10--20\% \pPb & 27--34        & 6.6e+08       & 1.3e+11 \\
\hline
60--65\% \PbPb    & 98--137       & 1.5e+02       & 3.0e+04 \\
65--70\% \PbPb    & 68--98        & 1.6e+05       & 3.1e+07 \\
70--75\% \PbPb    & 45--68        & 2.1e+07       & 4.2e+09 \\
75--80\% \PbPb    & 29--45        & 5.9e+08       & 1.2e+11 \\
\hline
\end{tabular}
\label{tab:smallsystems_pbpb}
\end{table}

\begin{table}
\caption{Conversion factors between $\nch$ with a \pT threshold, and $\nch$ including particles down to $\pT = 0$. The factor shown is $\nch / \nch(\pT > X)$, extracted with Pythia 8, tune CUETP8M1~\cite{Khachatryan:2015pea}. A potential multiplicity dependence of this factor is neglected for the projections in this chapter.}
\centering
\begin{tabular}{c|c|c|c|c|c}
\backslashbox{$|\eta|$}{$\pT$} & $> \unit[0.1]{\UGeVc}$ & $> \unit[0.2]{\UGeVc}$ & $> \unit[0.3]{\UGeVc}$ & $> \unit[0.4]{\UGeVc}$ & $> \unit[0.5]{\UGeVc}$ \\
\hline
$|\eta| < 1.5$ & 1.03 & 1.11 & 1.22 & 1.31 & 1.40 \\
\hline
$|\eta| < 2.4$ & 1.04 & 1.14 & 1.27 & 1.42 & 1.55 \\
\hline
\end{tabular}
\label{tab:smallsystems_conversion}
\end{table}

\subsection{Global-event properties}
\label{sect:smallsystems_global}
  
The measurement of global-event observables in rare high-multiplicity \pp collisions are of interest in itself. The shape of the multiplicity distribution, which has been largely extrapolated in the previous section, is today a challenge for models. 
The dynamics of producing very large multiplicity events is not understood in detail and therefore the shape of the distribution is an important input. 
Furthermore, studies of $\meanpT$ as a function of multiplicity~\cite{Abelev:2013bla} have shown a strong increase with multiplicity. However, those measurements exist only up to $\dNdeta \approx 55$, while the measurements at HL-LHC promise a measurement beyond twice that value. 

The shape of the multiplicity distribution and the growth of $\meanpT$ are closely connected to the physics of multiple parton interactions: high-multiplicity collisions are understood as originating from the collision of multiple partons within the same \pp collisions. It has been shown that the number of (low momentum transfer) parton interactions increases linearly with multiplicity with a possible saturation at large multiplicity~\cite{Abelev:2013sqa}. 
The prospect of showing that adding another parton interaction to an already busy event may be strongly suppressed, is an important ingredient to a revised conceptual understanding of particle production in high-energy \pp collisions. Together with the studies of symmetric cumulants (see the subsequent Section), HL-LHC will determine not only if there is a saturation limit for multiple parton interactions, but also the parton structure within the proton.

\subsection{Particle correlations}
\label{sect:smallsystems_correlations}

\begin{figure}[t]
\centering
\includegraphics[width=.49\linewidth]{\main/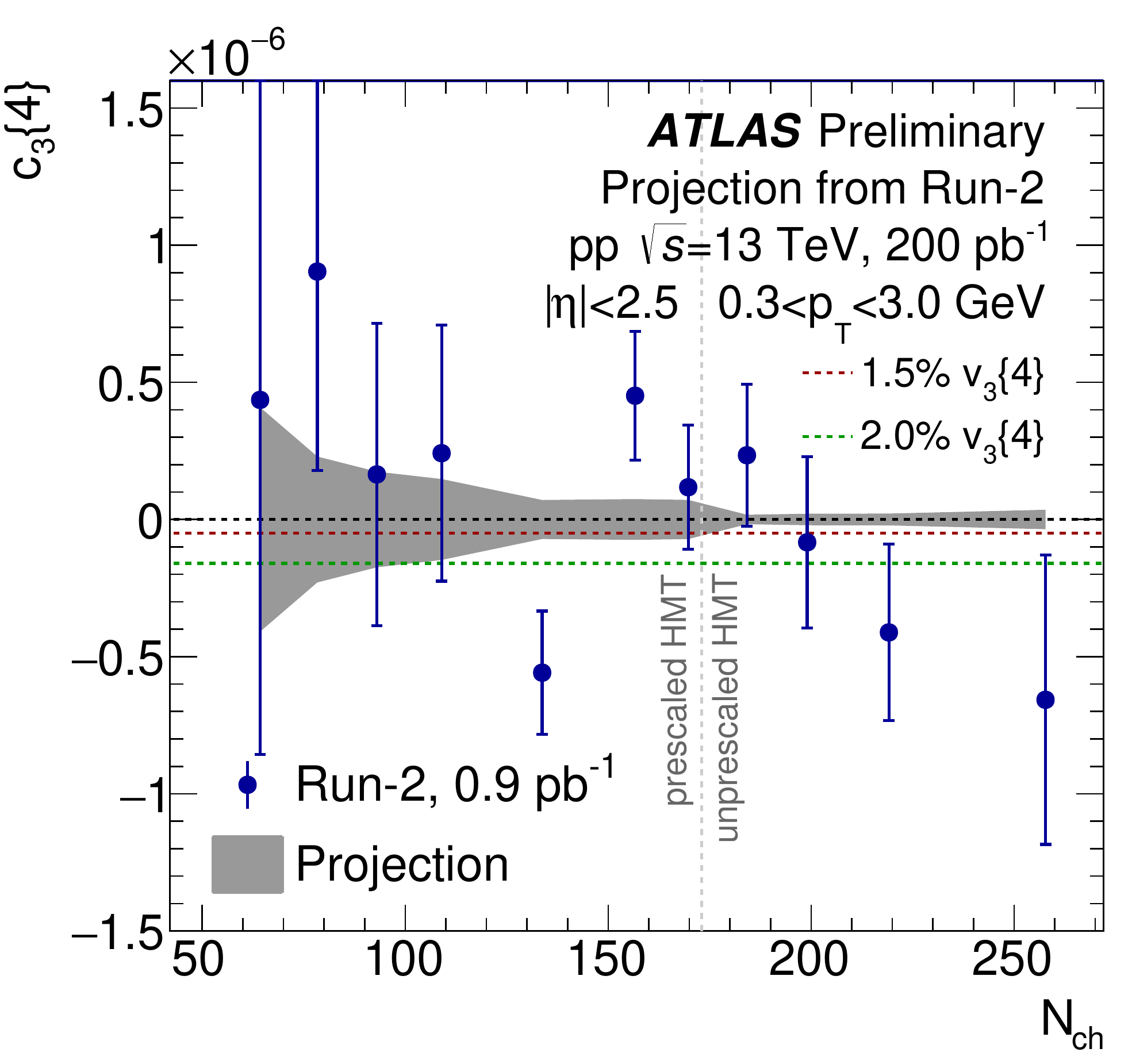}
\hfill
\includegraphics[width=.49\linewidth]{\main/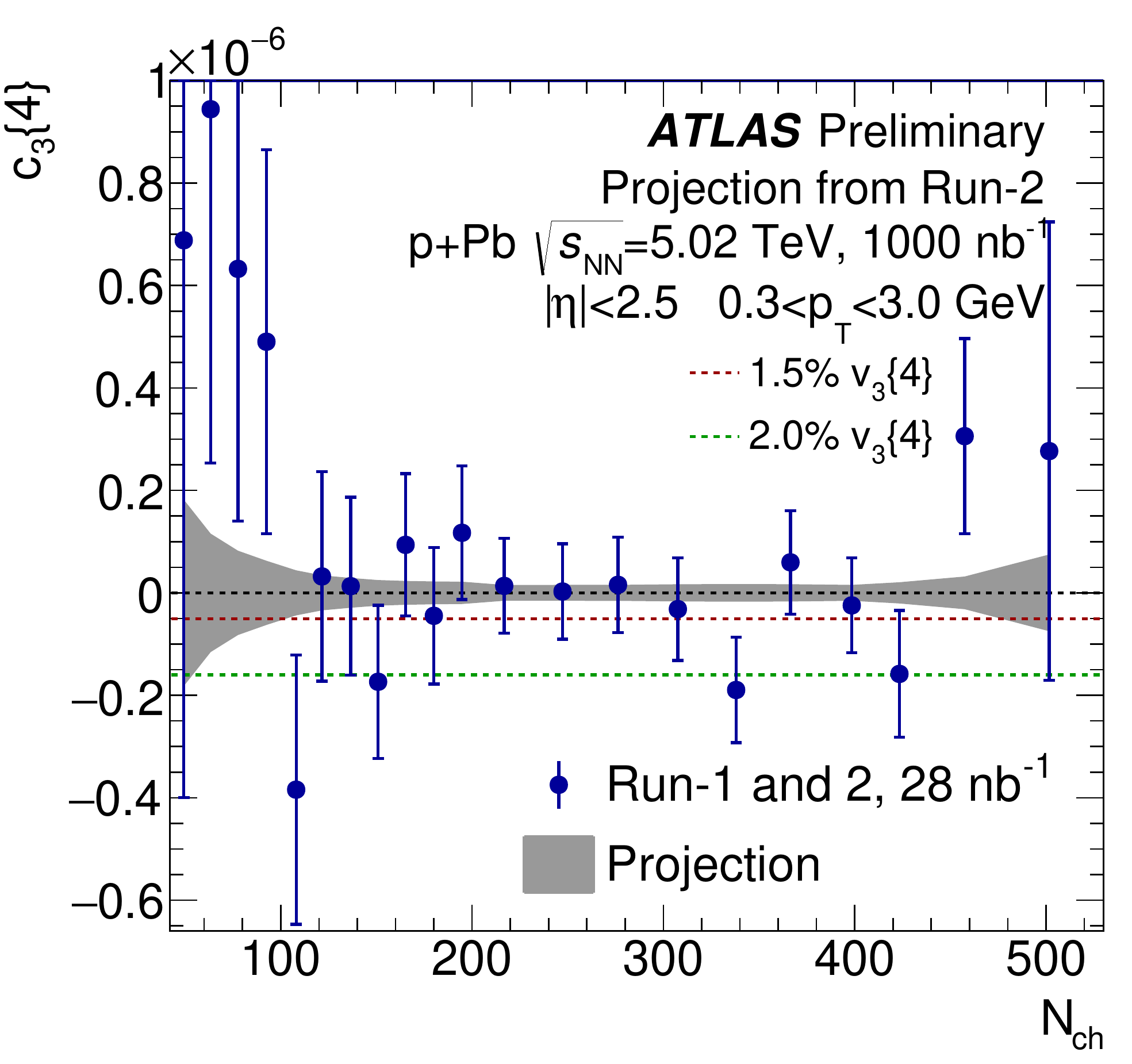}
\caption{4-particle cumulants $c_{\rm 3}\{4\}$ measured with 3-subevent method for \pp (left panel) and \pPb collisions (right panel) as a function of $\nch$ ($|\eta| < 2.5$ and $0.3 < \pT < \unit[3]{\UGeVc}$). Only statistical uncertainties are shown in the figure and the gray band represents the projected statistical uncertainty, with $c_{\rm 3}\{4\}$ assumed to be zero. The red and green dash lines represent $1.5\%$ and $2.0\%$ $\vthree\{4\}$ signal, respectively. The vertical line in the left panel indicates the transition between minimum-bias and high-multiplicity triggered data. Figures from Ref.~\cite{ATL-PHYS-PUB-2018-020}.}
\label{fig:smallsystems_corr_cumulants}
\end{figure}

\begin{figure}[t!]
\centering
\includegraphics[width=0.48\linewidth]{\main/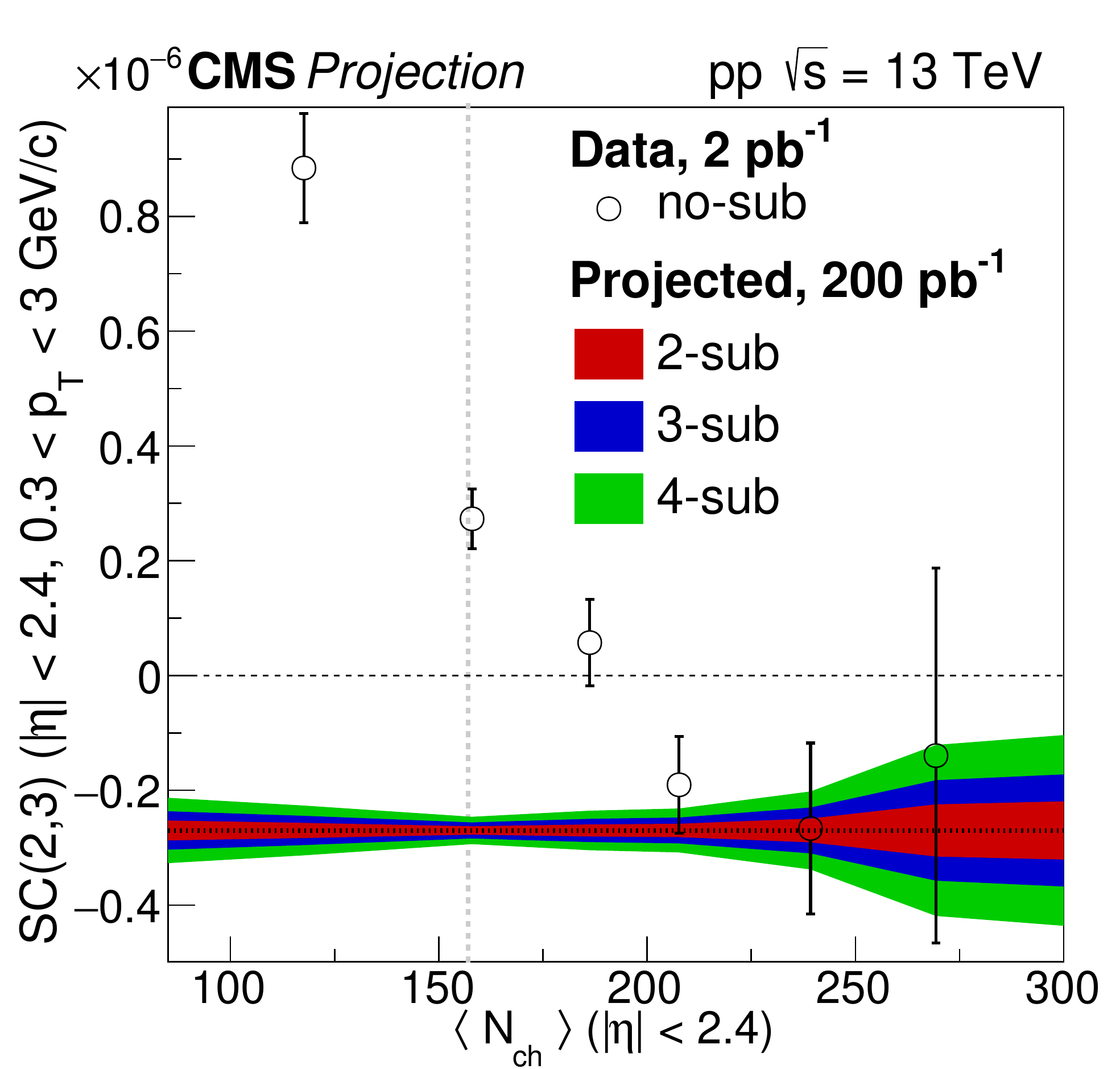}
\includegraphics[width=0.48\linewidth]{\main/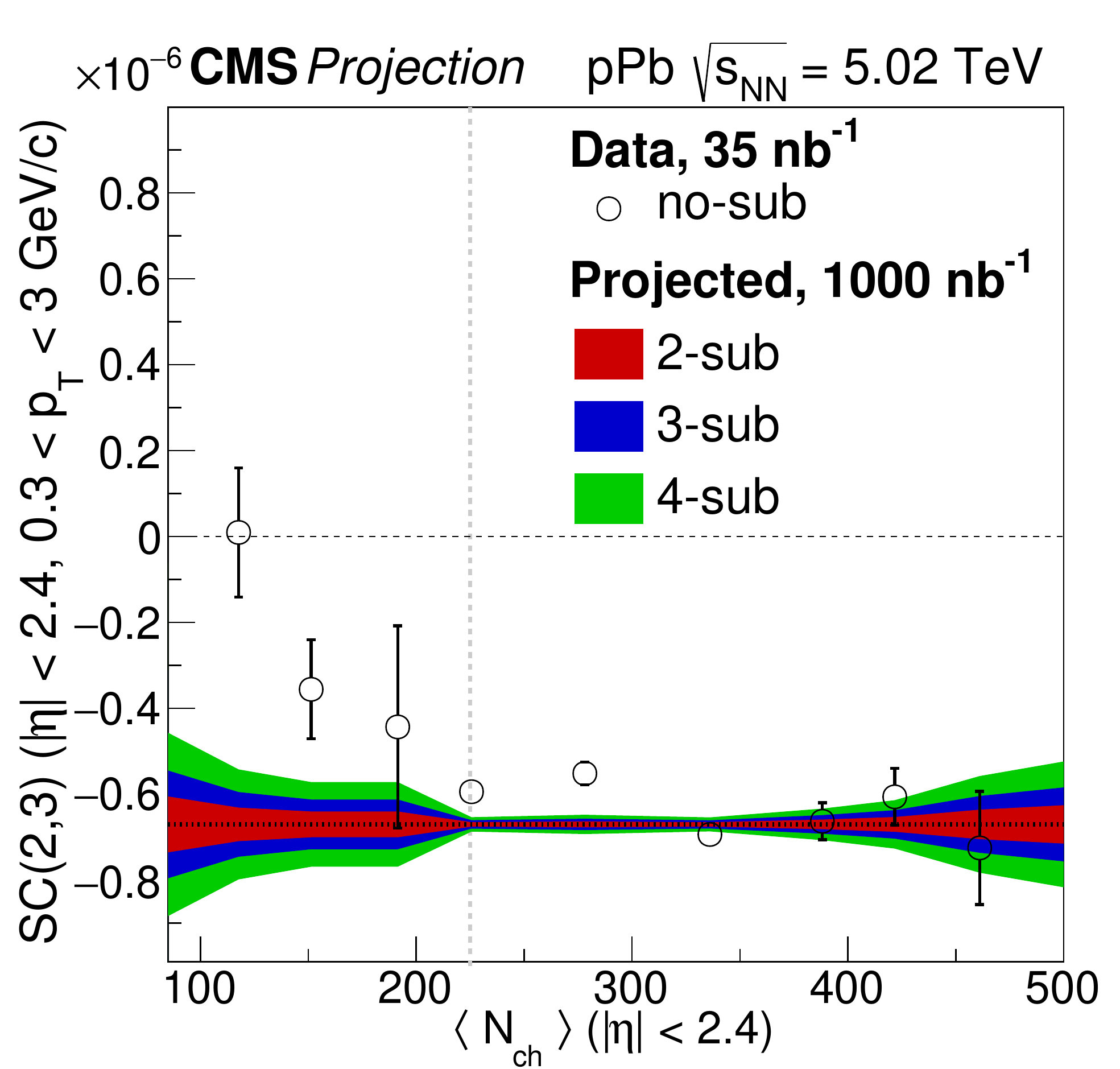}
\caption{Symmetric cumulants extracted with and without applying subevents for \pp (left panel) and \pPb collisions (right panel) as a function of $\nch$ ($|\eta| < 2.4$ and $0.3 < \pT < \unit[3]{\UGeVc}$). The projected reach is shown for the case of 2, 3 and 4 subevents assuming a constant signal as a function of multiplicity indicated by the lower horizontal line. The vertical line indicates the transition between minimum-bias and high-multiplicity triggered data. Figures from Ref.~\cite{CMS-PAS-FTR-18-026}.}
\label{fig:smallsystems_corr_symmetriccumulants}
\end{figure}

The measurements of two-particle correlations and higher-order cumulants have been the initial observations of collective-like effects in small systems. In \pp collisions, two distinct regions are of interest at HL-LHC: the high-multiplicity tail to compare to \pPb and \PbPb collisions and the low-multiplicity region to investigate the onset of these effects. In the following, several performance estimates are given as examples for the rich physics which can be addressed.

\begin{figure}[t!]
\centering
\includegraphics[width=.49\linewidth]{\main/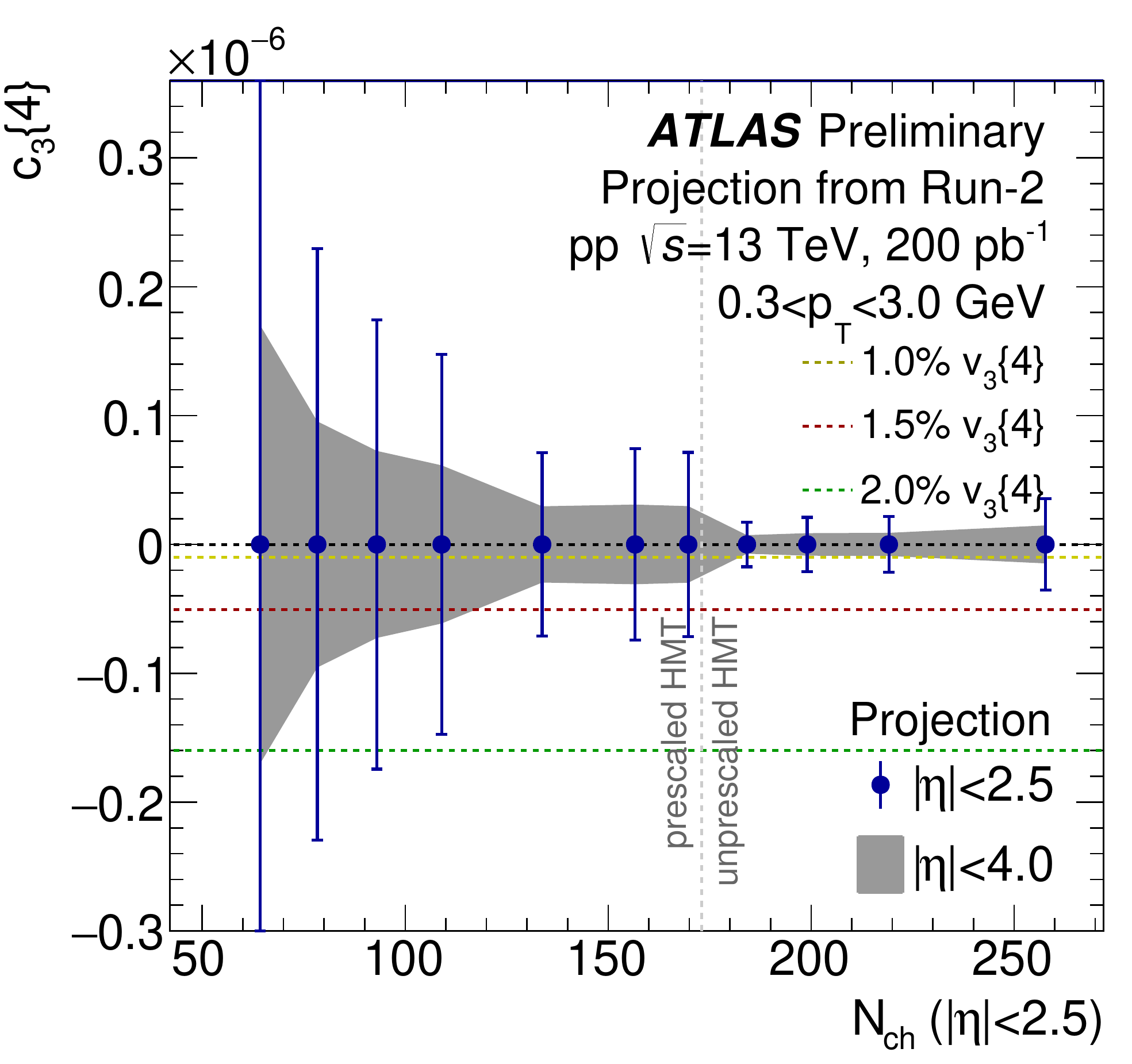}
\includegraphics[width=.49\linewidth]{\main/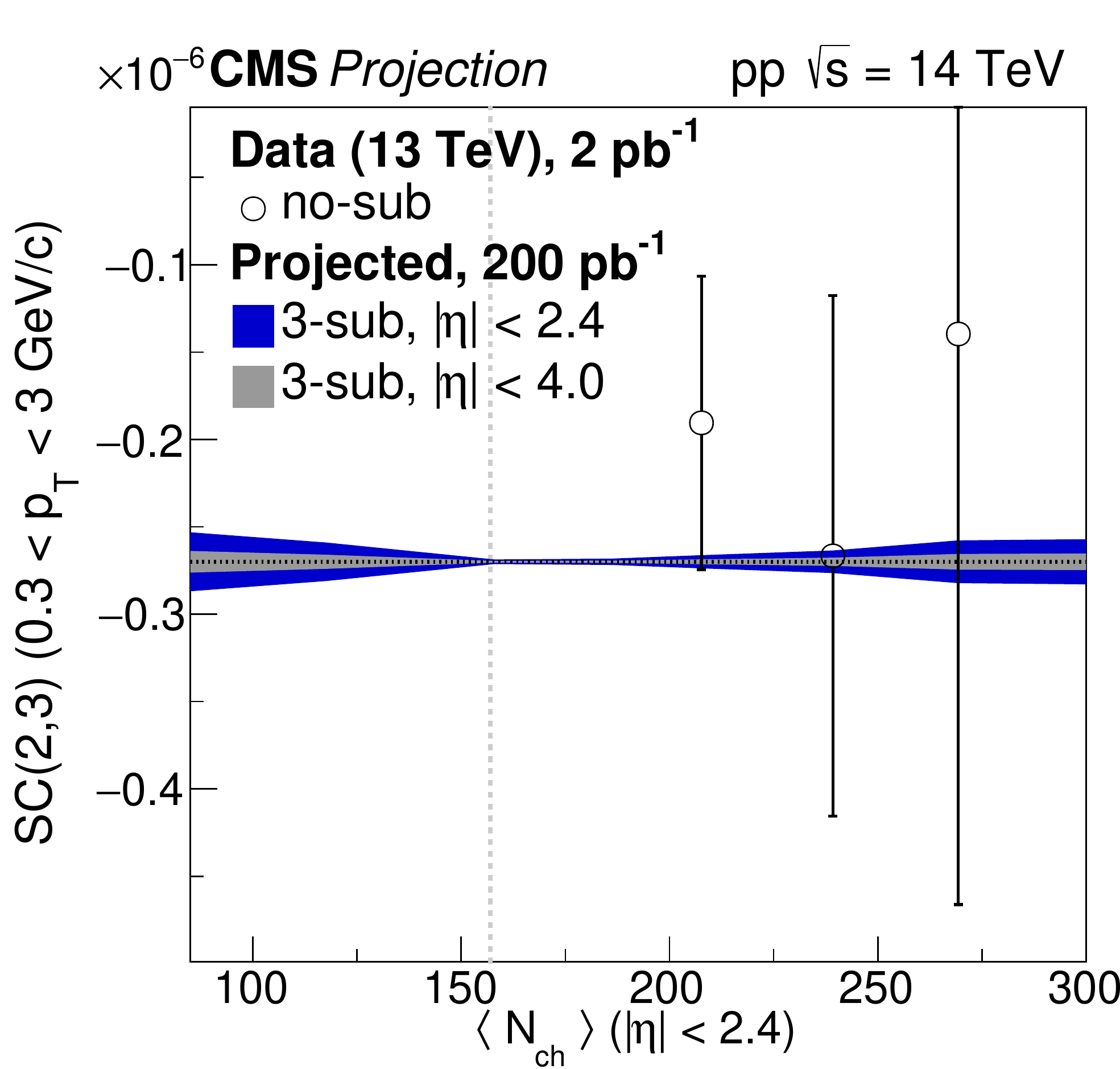}
\caption{Demonstration of the influence of the larger tracking acceptances for ATLAS and CMS available in Run 4. Left panel: 4-particle cumulant $c_{\rm 3}\{4\}$ (as in the left panel of Fig.~\ref{fig:smallsystems_corr_cumulants}) for \pp collisions with $\Lint = \unit[200]{pb^{-1}}$. The data points indicate the reach with the detector in Run 3 ($|\eta| < 2.5$) while the gray band the enlarged acceptance of $|\eta| < 4$ in Run 4. The yellow, red and green dash lines represent $1.0\%$, $1.5\%$ and $2.0\%$ $\vthree\{4\}$ signal, respectively. Figure from Ref.~\cite{ATL-PHYS-PUB-2018-020}. Right panel: Symmetric cumulants with 3 subevents (as in left panel of Fig.~\ref{fig:smallsystems_corr_symmetriccumulants}) for \pp collisions with $\Lint = \unit[200]{pb^{-1}}$. The blue (gray) area indicates the projected uncertainty for Run 3 (4). Figure from Ref.~\cite{CMS-PAS-FTR-18-026}.}
\label{fig:smallsystems_corr_eta4}
\end{figure}

\begin{figure}[h!]
\centering
\includegraphics[width=.49\linewidth]{\main/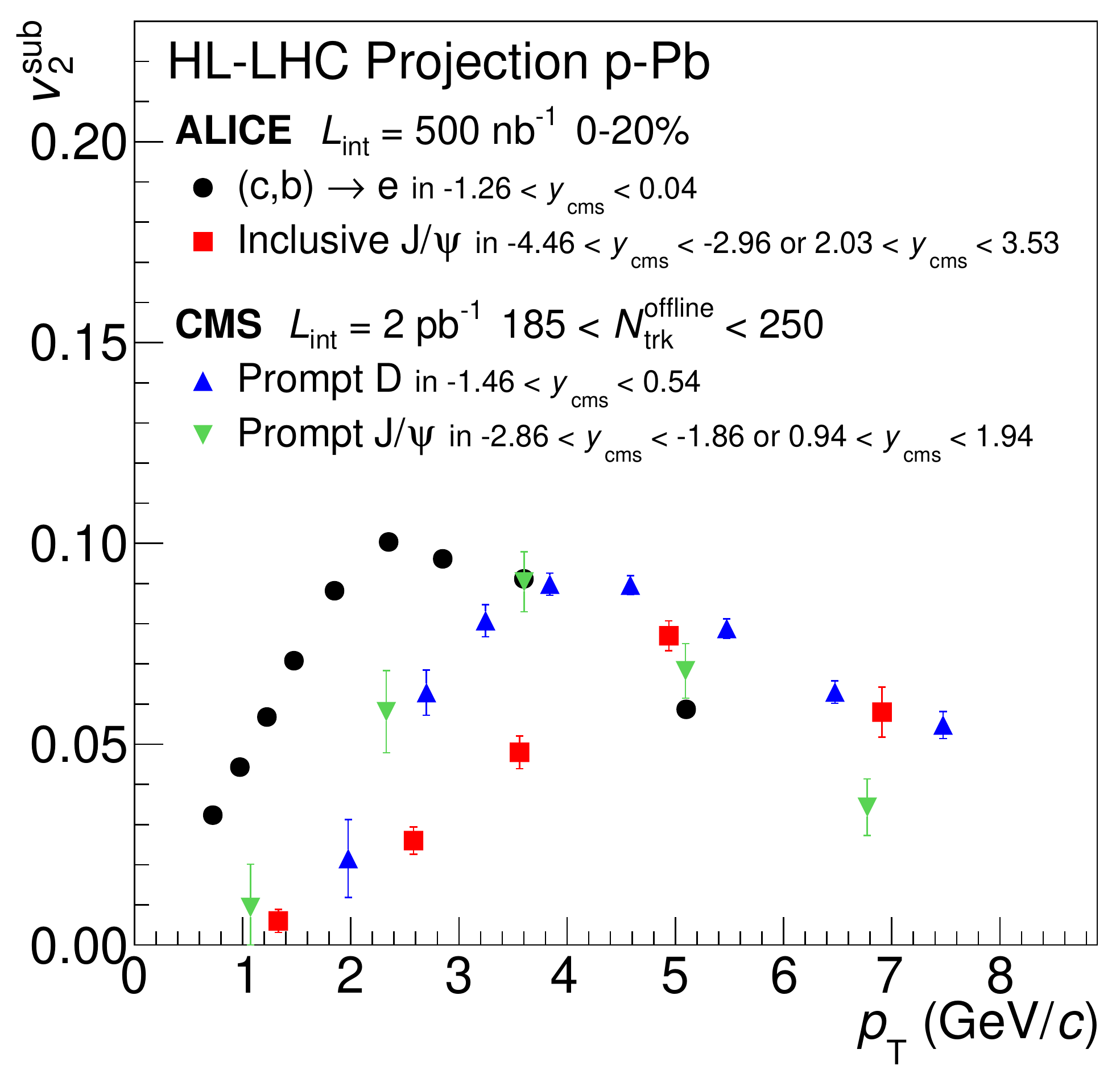}
\hfill
\includegraphics[width=.49\linewidth]{\main/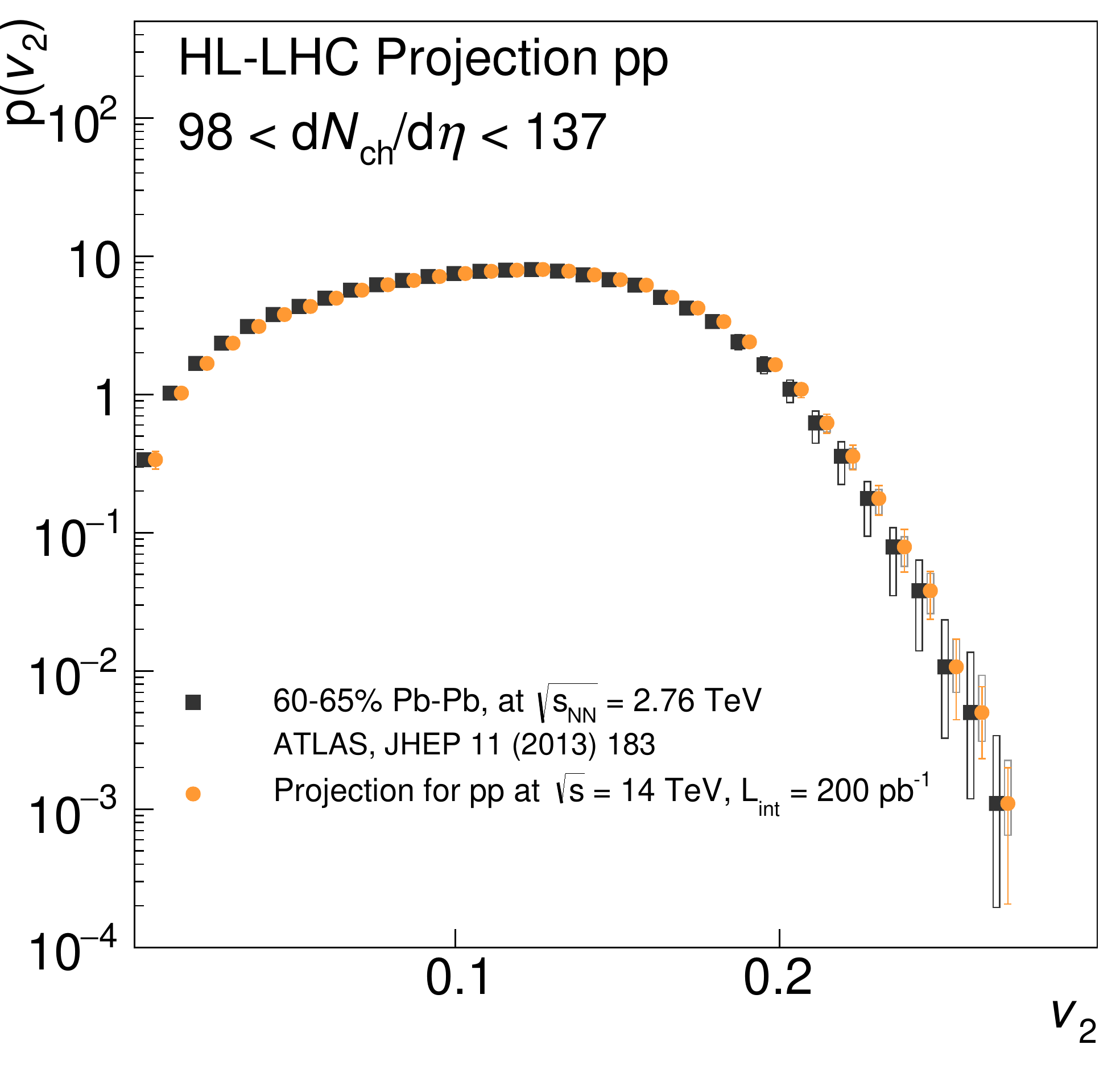}
\caption{Left panel: Particle identified $v_{\rm{2}}$ coefficients for \pPb collisions as a function of $\pT$. Two different cases are shown: the ALICE projections are for the 20\% highest-multiplicity collision ($\Lint = \unit[500]{nb^{-1}}$) demonstrating the negligible statistical uncertainties for heavy-flavour decay electrons and $J/\psi$, while the CMS projection is for a bin with $4-5 \meannch$ ($\Lint = \unit[2]{pb^{-1}}$) demonstrating the wide reach in multiplicity achievable for $D$ mesons and $J/\psi$. 
Right panel: Projection of the measurement in \pp collisions of the probability distribution of \vtwo. To illustrate the reach the same signal as in \PbPb~\cite{Aad:2013xma} is assumed although the mean and the width of the distribution is most likely smaller in \pp collisions. The projection is for the equivalent \pp multiplicity (circles) as in 60--65\% centrality in \PbPb collisions (squares).}
\label{fig:smallsystems_corr_pid_pvn}
\end{figure}

State-of-the-art measured 4-particle cumulants of $\vthree$ ($c_3\{4\}$) in \pp and \pPb collisions are presented in Fig.~\ref{fig:smallsystems_corr_cumulants} overlaid with the projection for HL-LHC.
In order to remove non-flow contributions, the 3-subevent method is applied. In \pp collisions, with the data collected in Run 2, the statistical uncertainties are large and the $c_3\{4\}$ values are consistent with zero in most of the $\nch$ range. On the contrary, in large systems, significant non-zero $c_3\{4\}$ up to $-0.4 \cdot 10^{-6}$ depending on centrality has been measured~\cite{Aad:2014vba}, which reflects the nucleonic fluctuations in the initial state. Whether similar behaviour is observed in small systems still needs to be studied. The increase in luminosity in Run 3 and 4 provides a great opportunity to measure $c_{\rm 3}\{4\}$ in \pp collisions with high precision: the statistics are sufficient to measure a signal as small as $\vthree\{4\} = 1.5\%$ for $\nch \gtrapprox 170$, while 2\% are accessible with large significance over a wide multiplicity range ($\nch \gtrapprox 100$). Similarly, in \pPb collision, the current result shows that $c_{\rm 3}\{4\}$ is consistent with zero, but increased statistics will help to detect a potential non-zero $c_{\rm 3}\{4\}$ smaller than $1.5\%$ for $100 \lessapprox \nch \lessapprox 500$. Similarly, the precision of the already measured non-zero $c_{\rm 2}\{4\}$ will be greatly improved~\cite{ATL-PHYS-PUB-2018-020}.

The correlations of flow harmonics between different orders, called symmetric cumulants, are very sensitive to the initial state and the hydrodynamic evolution in heavy-ion collisions. In \PbPb collisions these are, for instance, used to constrain the shear viscosity over entropy ratio $\eta/s$. In addition, they challenge the description of the observed phenomena within initial-state saturation models.
Their measurement in small systems can provide important insight in the validity of the hydrodynamic description of the observed phenomena. Here, symmetric cumulants probe in particular the proton substructure~\cite{Albacete:2017ajt} which is needed to provide a solid description of the initial state, a necessary ingredient for the hydrodynamic description.
The present uncertainties of such measurement in small systems are too large for a definitive conclusion, in particular in \pp collisions, due to the dominance of non-flow like jets and resonance decays. Figure~\ref{fig:smallsystems_corr_symmetriccumulants} shows the performance projection of ${\rm SC}(2,3) = \langle v_{\rm 2}^2 v_{\rm 3}^2 \rangle - \langle v_{\rm 2}^2 \rangle \langle v_{\rm 3}^2 \rangle$ for HL-LHC for \pp and \pPb collisions. The uncertainties of the measurement without subevents would become practically invisible, however, those stay dominated by non-flow effects. A measurement requiring two, three and even four subevents becomes possible with uncertainties of the order of a few times $10^{-7}$ depending on multiplicity. Such results can give a definitive answer if a similar hydrodynamic footprint is observed in small and large systems.

Figure~\ref{fig:smallsystems_corr_eta4} illustrates the reduction of the statistical uncertainty due to the larger tracker acceptance in Run~4 for ATLAS and CMS. For this 4-particle correlator a reduction of the uncertainties of about 2.5 is expected, and therefore even the measurement of a 1\% $\vthree\{4\}$ signal comes into reach. The influence of the acceptance increase on the uncertainties of 6- and 8-particle cumulants will be larger, factor 4 and 6.5, respectively. Similarly, the uncertainties on the SC measurement reduce significantly at larger \pT.

Figure~\ref{fig:smallsystems_corr_pid_pvn} (left panel) illustrates the reach which can be obtained for the $\vtwo$ measurement of heavy-flavoured objects in \pPb collisions. Shown are projections for heavy-flavour electrons and inclusive $J/\psi$ by ALICE as well as for prompt $D$ and $J/\psi$ by CMS. Minor uncertainties are expected for this observable with the potential to demonstrate for the first time with significance the final-state interaction of charm and beauty in a small collision system.

The \vn fluctuates on an event-by-event basis as no two nuclei have identical parton distribution. The probability density distribution of \vn, ${\rm p}(\vn)$ is closely related to event-by-event fluctuations of the eccentricities, ${\rm p}(\varepsilon_{\rm{n}})$ as $\vtwo \propto k_{\rm{2}}\epsilon_{\rm{2}}$. Therefore its measurement provides crucial information about the initial conditions and the final-state dynamics of the medium. To characterise the initial-state spatial anisotropy these measurements are fitted with Bessel-Gaussian and elliptic power functions. The measurements of probability density distributions for \vtwo at \PbPb collisions are described well by the Bessel-Gaussian function at central collisions and less in peripheral collisions~\cite{Aad:2013xma,Sirunyan:2017fts,Acharya:2018lmh}. This deviation from the Bessel-Gaussian function is expected in peripheral collisions as $k_{\rm{2}}$ increases slightly at large $\epsilon_{\rm{2}}$ values~\cite{Jia:2014jca}. Measurements are well described by the elliptic power functions in all centrality intervals of \PbPb collisions~\cite{Sirunyan:2017fts}. These measurement have not yet been attempted in small systems due to the insufficient available statistics. Figure~\ref{fig:smallsystems_corr_pid_pvn} (right panel) presents a projection for the measurement of ${\rm p}(\vn)$ in \pp collisions. This extrapolation is based on the ${\rm p}(\vtwo)$ measurement in 60--65\% centrality \PbPb collisions at $\sqrtsNN = \unit[2.76]{\UTeV}$~\cite{Aad:2013xma}. In this simple study, the same signal is assumed although the mean and width of the distribution is most likely smaller in \pp collisions. Such a measurement would constitute the first measurement of ${\rm p}(\vn)$ in \pp collisions, and can shed important light on the nature of the observed \vtwo coefficients.

\begin{figure}[t]
\centering
\includegraphics[width=0.49\linewidth]{\main/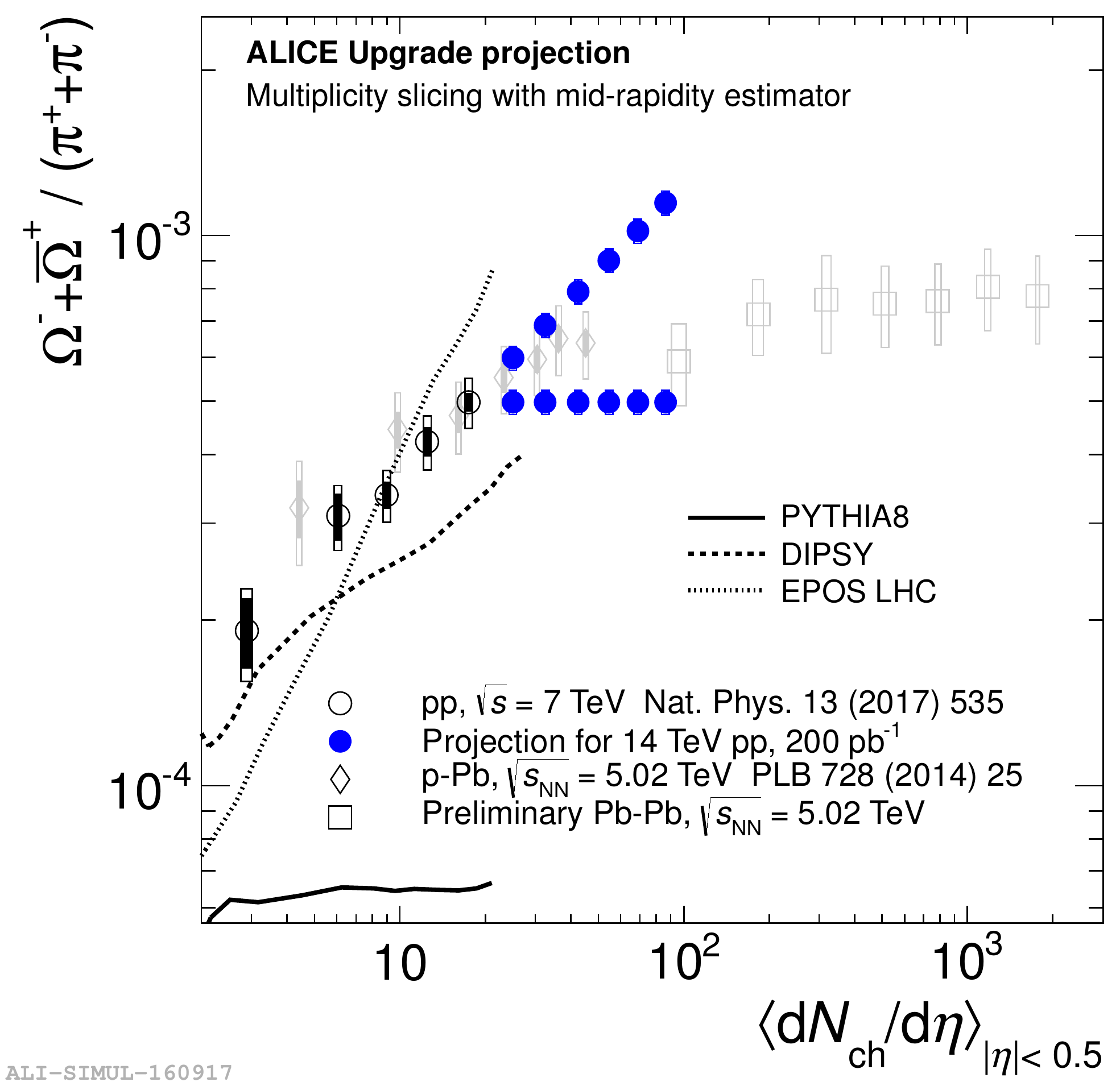}

\caption{$\Omega/\pi$ ratio as a function of $\dNdeta$ for \pp, \pPb, and \PbPb collisions. The existing data (from Ref.~\cite{ALICE:2017jyt}) is shown in open black symbols (\pp), grey diamonds (\pPb) and grey squares (\PbPb), while the extrapolation for \pp collision is shown in blue filled circles. Two scenarios are shown: a) assuming that the ratio continues increasing following the measured trend, and b) assuming that the value stays the same as at the largest measured $\dNdeta$. Figure from Ref.~\cite{ALICE-PUBLIC-2019-001}.}
\label{fig:smallsystems_strangeness_omega_pi}
\end{figure}

\subsection{Strangeness enhancement}
\label{sect:smallsystems_strangeness}

The unexpected increase of the strange-particle yield normalized by the pion yield as a function of $\nch$ is one of the key observations in small systems. In \pp collisions these ratios are measured up to $\dNdeta \approx 17$ with some overlap with \pPb collisions. The most peripheral \PbPb collisions measured have a $\dNdeta \approx 96$, nearly 6 times larger. Figure~\ref{fig:smallsystems_strangeness_omega_pi} presents the expected reach of the $\Omega/\pi$ ratio in \pp collisions which will bridge the present gap between \pp and \PbPb collisions. In particular, if the measured increasing trend would continue, the $\Omega/\pi$ ratio would grow larger than in peripheral \PbPb collisions. 
Assuming that strangeness enhancement scales with the energy density of the system, Fig.~\ref{fig:small_systems_energy_density} suggests that it should indeed be possible to see that the high-multiplicity \pp results exceed the low multiplicity \PbPb results (crossover). Whether the signature will be as striking as the projection in Fig.~\ref{fig:smallsystems_strangeness_omega_pi}, depends on the details of the assumed scaling law. At this point simulations are not precise enough to provide quantitative predictions of such a crossover, and HL-LHC experimental results on strangeness enhancement will as such be driving the theoretical development.
The scenario with a clear crossover will be immediately distinguishable from a scenario where the $\Omega/\pi$ ratio flattens, and connects smoothly with the \PbPb result. Such a result will in itself also be groundbreaking, as it will indicate that the thermal limit reached in \PbPb collisions will already be realized in high-multiplicity \pp collision.

\subsection{Energy loss}
\label{sect:smallsystems_energyloss}

\begin{figure}[t]
\centering
\includegraphics[width=0.49\linewidth]{\main/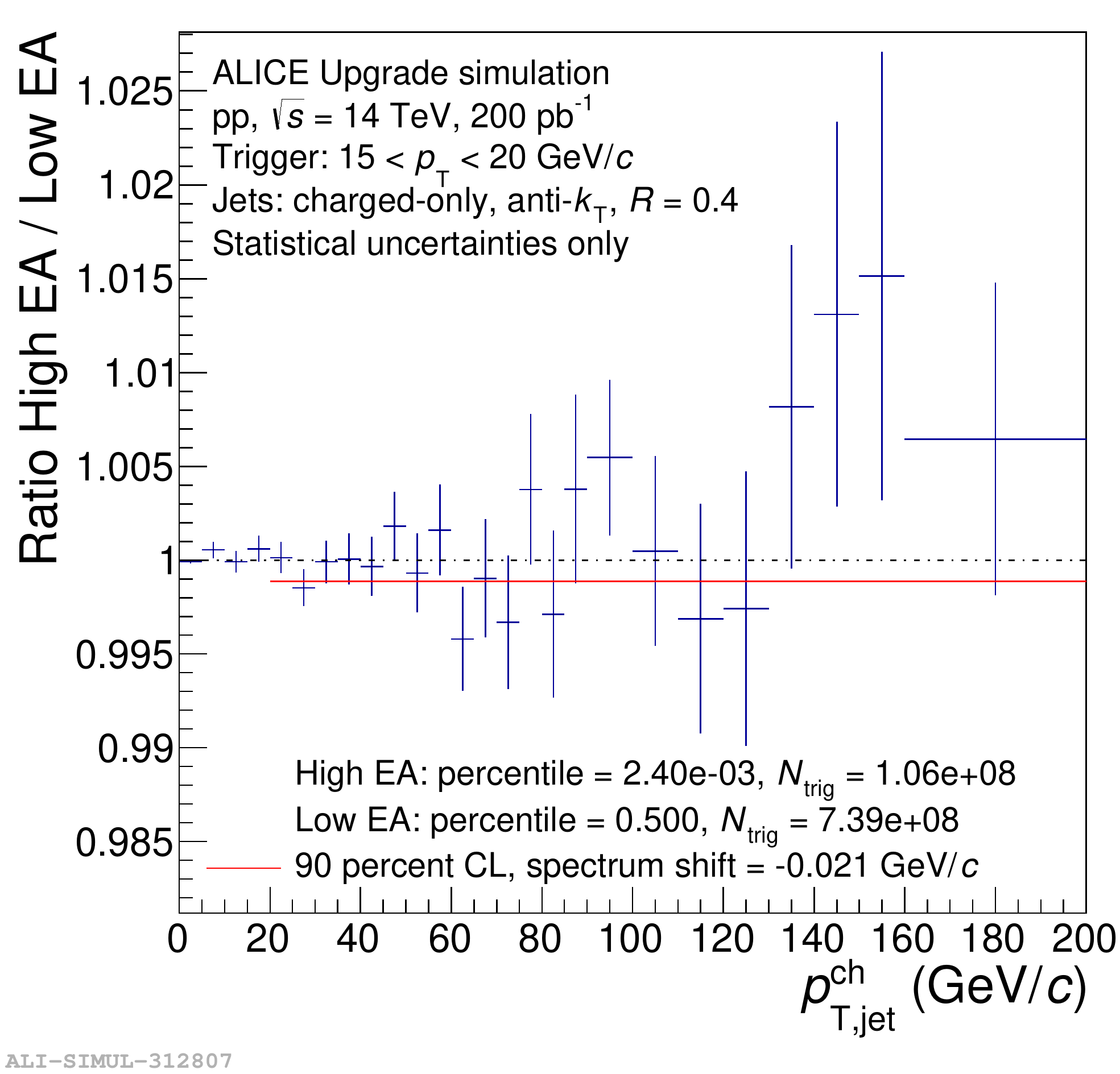}
\hfill
\includegraphics[width=0.49\linewidth]{\main/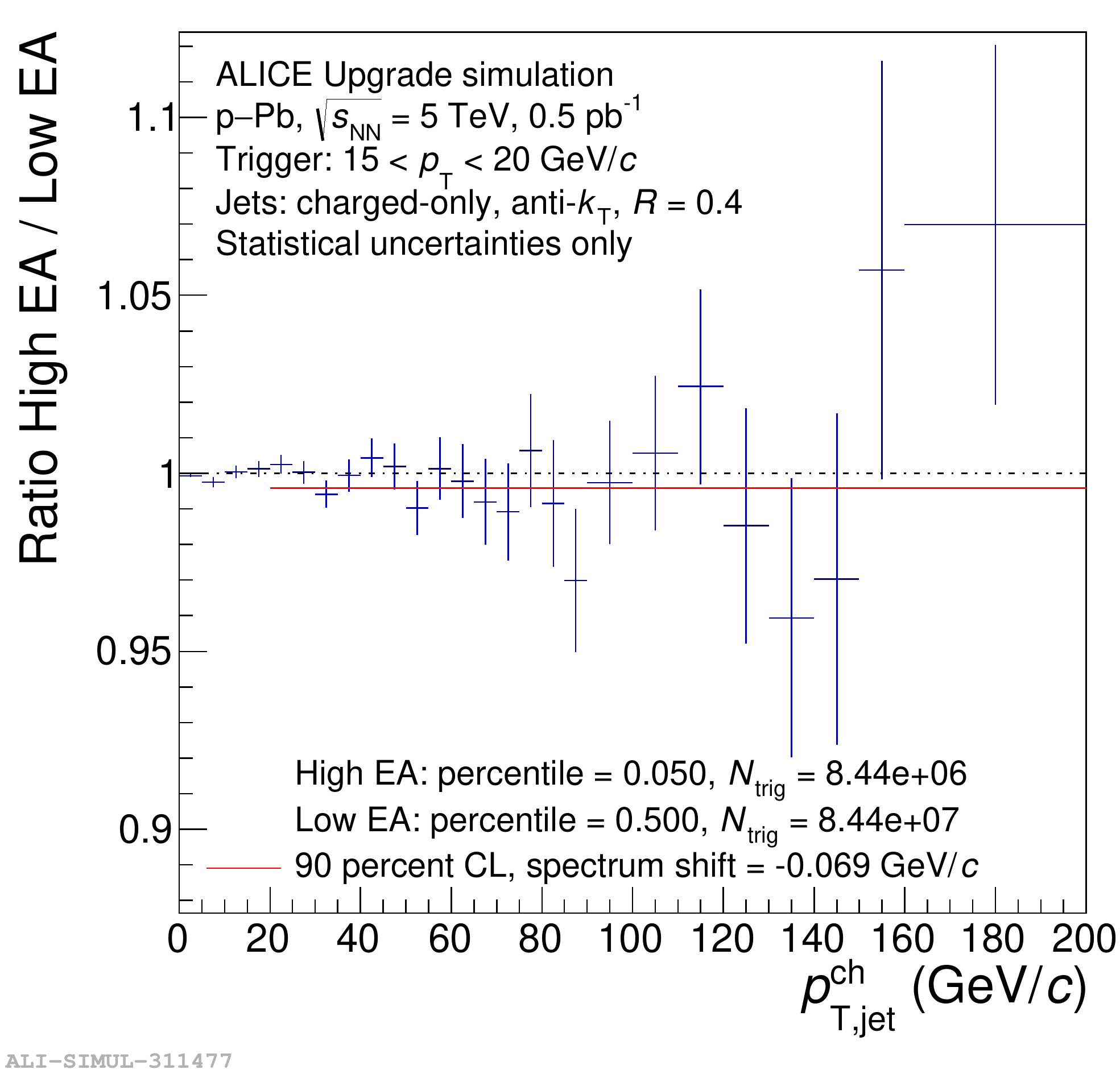}
\caption{Modification of jet-recoil yields extracted from semi-inclusive hadron--jet correlations for \pp collisions (left) and \pPb collisions (right) within the ALICE acceptance. Shown is the ratio  of high event-activity (EA) and low-EA recoil spectra as a function of $p_{\rm T,jet}^{\rm ch}$, with high-EA corresponding to 5--7~$\meannch$ in \pp collisions (left panel), and the 0--5\% bin for \pPb collisions (right panel). Since no EA-dependent shift is imposed, the parent distribution of the ratio has the value of unity at all \pT. The red lines show the 90\% CL limit for a possible EA-dependent spectrum shift. Figures from Ref.~\cite{ALICE-PUBLIC-2019-001}.}
\label{fig:smallsystems_energyloss_hjet}
\end{figure}

\begin{figure}[t]
\centering
\includegraphics[width=0.45\linewidth]{\main/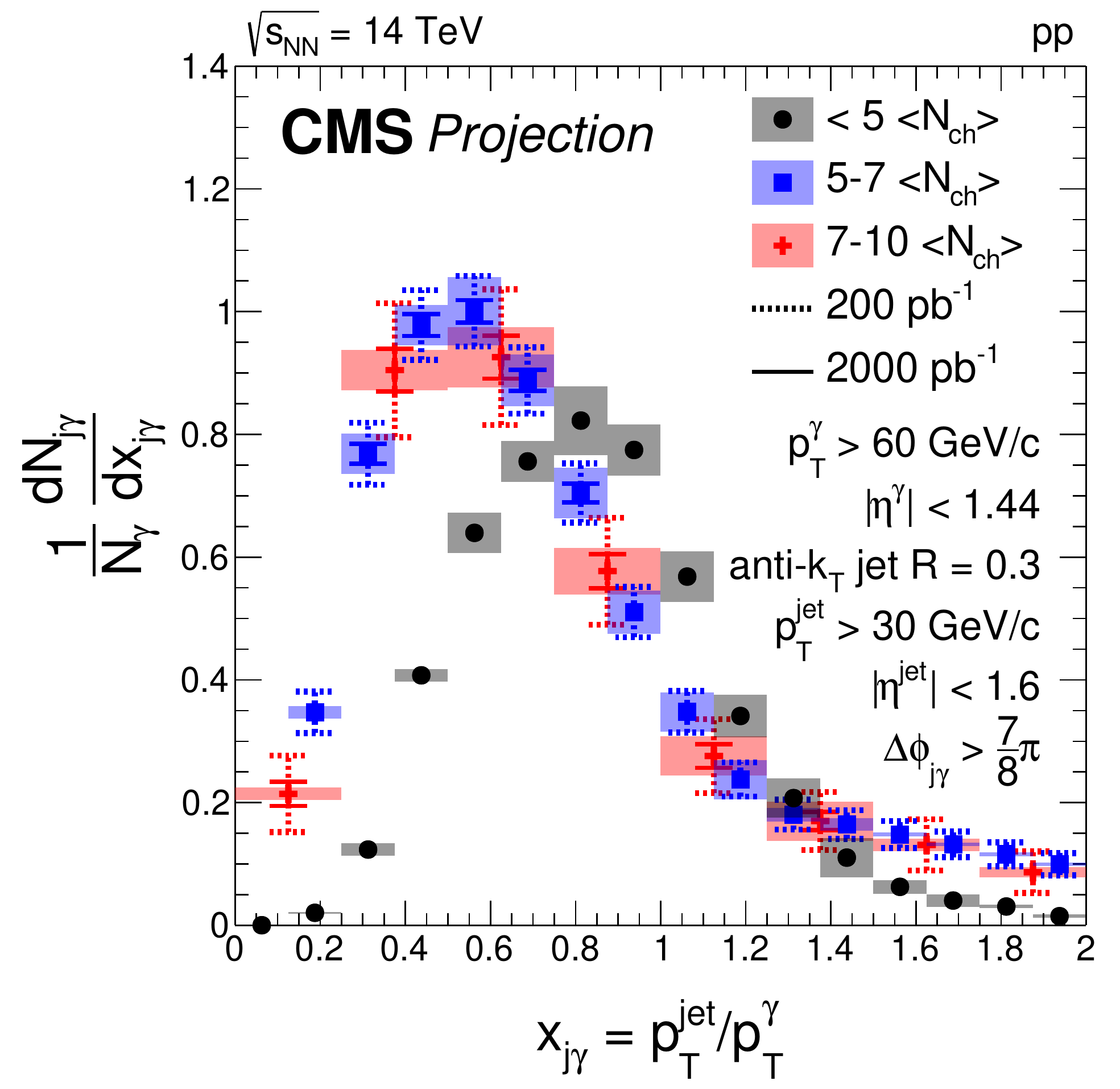}
\hfill
\includegraphics[width=0.53\linewidth]{\main/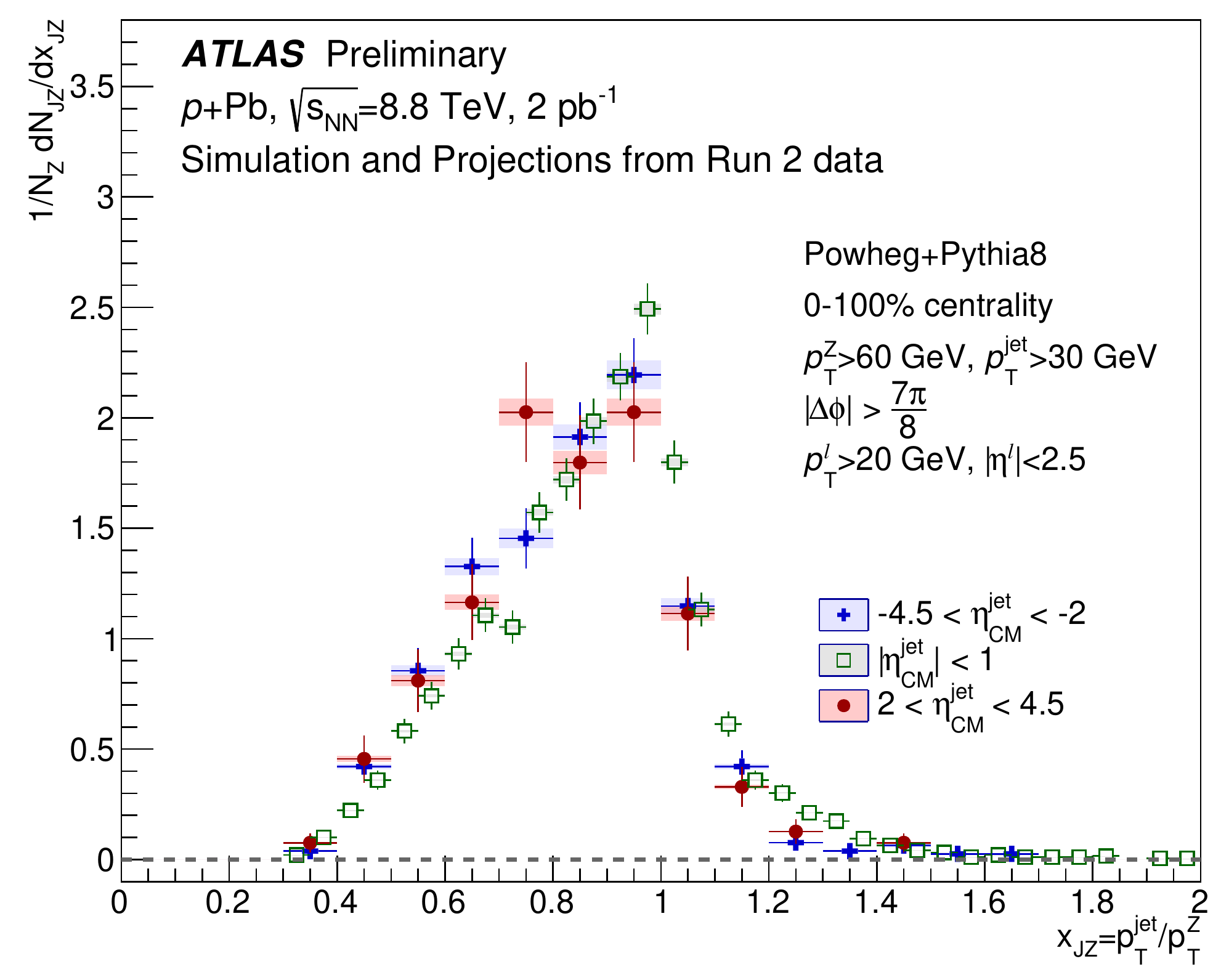}
\caption{Left panel: CMS projection of the measurement of the jet--$\gamma$ momentum fraction $x_{j\gamma}$ in \pp collisions for selected high-multiplicity bins. A jet with $\pT >$~\unit[30]{\UGeVc} is required to be back-to-back ($\Delta\varphi > 7/8\pi$) with a $\gamma$ with $\pT > \unit[60]{\UGeVc}$. The shape is based on Pythia and shifts due to selection biases as a function of multiplicity (see text). Figure from Ref.~\cite{CMS-PAS-FTR-18-025}. Right panel: ATLAS projection of the measurement of the $Z$--jet momentum fraction $x_{jZ}$ in \pPb collisions in different pseudorapidity intervals. The momentum requirements are also \unit[30]{\UGeVc} for the jet and at least \unit[60]{\UGeVc} for the $Z$, with the same back-to-back requirement ($\Delta\varphi > 7/8\pi$). The projection is based on Powheg + Pythia 8 Monte Carlo samples with the CT10 PDF set. Figure from Ref.~\cite{ATL-PHYS-PUB-2018-039}.}
\label{fig:smallsystems_energyloss_xjg_xjz}
\end{figure}

As discussed in Sect.~\ref{sect:smallsystems_historicoverview}, inclusive high-\pT hadron and jet yields show at present no evidence of medium-induced energy loss in \pPb collisions, and suffer from selection biases if measured in event classes. 
Inclusive measurements with the large event set expected at HL-LHC therefore do not help to resolve the question of energy loss in small systems. However, coincidence measurements of jets recoiling against a trigger object are not subject to such biases, and have the potential to identify small energy-loss effects or put stringent upper limits. In this section, projections are given for correlations between high-\pT hadrons and jets, as well as jets and $\gamma$ and $Z$.

Figure~\ref{fig:smallsystems_energyloss_hjet} shows a projection of the measurement of semi-inclusive hadron--jet correlations in LHC Run 3 and 4, for \pp collisions at $\sqrts = \unit[14]{\UTeV}$ and \pPb collisions at $\sqrtsNN = \unit[5.02]{\UTeV}$. The figure shows the ratio of trigger-normalized recoil spectra for events selected on high and low event-activity (EA) classes. 
This projection is based on Pythia simulations for \pp collisions, which gives the expected number of charged-hadron triggers in the interval $15<\ptt<20$~\UGeV\ (scaled by $A$ to model \pPb collisions), and the per-trigger recoil jet spectrum. The measured enhancement in the per-event high-\pT hadron yield for \pp collisions in high-multiplicity collisions~\cite{Acharya:2018orn} has been taken into account.

The projection represents the case where no energy loss occurs for high-EA relative to low-EA collisions, and demonstrates the statistically achievable limit. The 90\% confidence level for a possible EA-dependent spectrum shift due to large-angle energy transport from jet quenching~\cite{Acharya:2017okq} is \unit[70]{\UMeVc} for \pPb (5\% highest EA) and \unit[21]{\UMeVc} for \pp collisions (5--7~$\meannch$). These values are over 100 times smaller than the spectrum shift measured in \PbPb collisions~\cite{Adam:2015doa}. The high statistics of the HL-LHC dataset enables this approach to be applied to yet more stringent EA selections; for 7--10~$\meannch$ (10--12~$\meannch$) the corresponding 90\% CL limit on energy loss is expected to be \unit[69]{\UMeVc} (\unit[590]{\UMeVc}).

\begin{figure}[th!]
\centering
\includegraphics[width=0.65\linewidth]{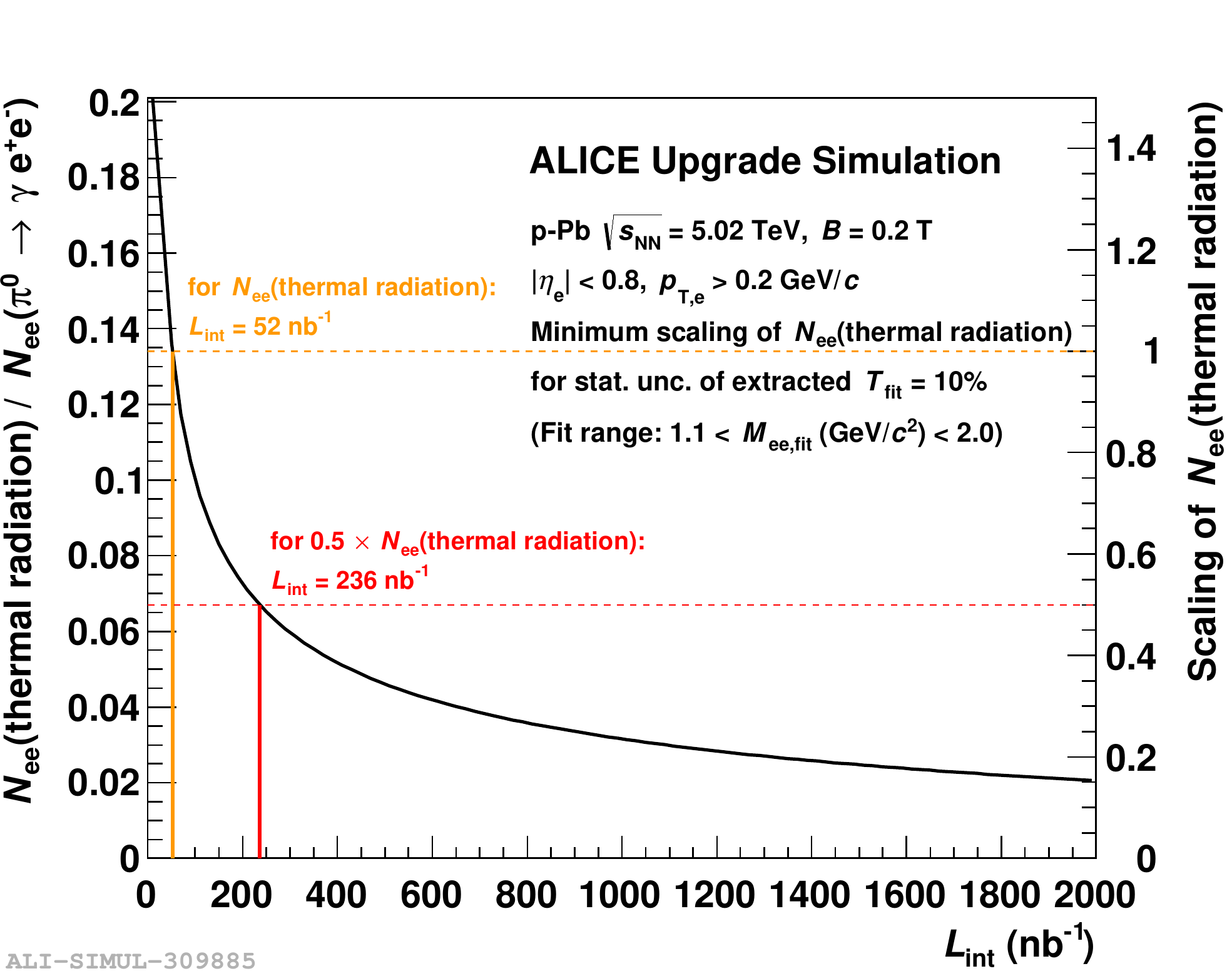}
\caption{Projection of the measurement of the medium temperature extracted from thermal dileptons in \pPb collisions. As the expected signal is uncertain, the figure presents the required integrated luminosity relative to the prediction based on Ref.~\cite{Rapp:2011is} denoted by $N_{\rm ee}$(thermal radiation). It is expressed as the minimum thermal photon $N_{\rm ee}$(thermal radiation) to $\pi^0$ ratio (left axis) and scaling of $N_{\rm ee}$(thermal radiation) (right axis) for achieving a 10\% statistical uncertainty on the extracted temperature. Figure from Ref.~\cite{ALICE-PUBLIC-2019-001}.}
\label{fig:smallsystems_thermal_radition}
\end{figure}

Projections for the correlation of jets and $\gamma$ as well as jets and $Z$ are presented in Fig.~\ref{fig:smallsystems_energyloss_xjg_xjz} for \pp and \pPb collisions. Shown are distribution of the momentum fraction $x_{jX} = \pT^{\rm jet}/\pT^X$ where $X$ is the $\gamma$ or $Z$. Given that the $\gamma$ and $Z$ can be considered unmodified by final-state interactions, a potential energy loss acting on the jet would directly alter the $x_{jX}$ distribution. For \pp collisions, the left panel of Fig.~\ref{fig:smallsystems_energyloss_xjg_xjz} presents the distribution for different classes in multiplicity based on Pythia, demonstrating the reach. 
It can be also seen that the distribution shifts significantly as a function of multiplicity without final-state interactions, but purely due to the presence of an underlying event. 
This shift is caused by selection biases, e.g. the likelihood of multi-jet events is increased by requiring higher event multiplicity. In order to extract a firm conclusion on energy loss, this observable needs to be compared to theoretical calculations or tuned generators which reproduce multiplicity and underlying-event fluctuations in \Pepem and \pp collisions.
The right panel of Fig.~\ref{fig:smallsystems_energyloss_xjg_xjz} presents the projection for \pPb collisions for MB collisions but in different pseudorapidity intervals sensitive to potential differences in the p and Pb hemisphere.

\subsection{Thermal Radiation}
\label{sect:smallsystems_thermalradiation}

The measurement of thermal radiation in \pPb collisions can be considered as a smoking gun for the formation of a system with an energy scale above the phase-transition temperature, see Chapter~\ref{chapter:electromagnetic_radiation}. In order to estimate the sensitivity to the thermal radiation in \pPb collisions, a similar strategy as in Sec.~\ref{sec:thermalradiation:dileptons} was used.
The combinatorial background was scaled from \PbPb collisions to the expected number of pairs in \pPb collisions. The pair efficiency (including the efficiency for rejecting \Pepem pairs from semileptonic charm decays) is assumed to be the same as in \PbPb collisions.
Subsequently, the temperature of the QGP is extracted in the same way as in Sec.~\ref{sec:thermalradiation:dileptons}. The minimum thermal photon to $\pi^{0}$ (both decaying into \Pepem) ratio that is needed for a fit to the invariant mass spectrum with a statistical uncertainty $\sigma_{\rm T,stat} = 10\%$ as a function of \Lint\ up to \unit[2000]{nb$^{-1}$} is shown in Fig.~\ref{fig:smallsystems_thermal_radition}. If the considered prediction is accurate, an integrated luminosity of about $\unit[50]{nb^{-1}}$ is sufficient for the measurement. In case the signal is 50\% smaller about 4--5 times the statistics is needed.

\begin{figure}[t]
\centering
\includegraphics[width=0.8\linewidth]{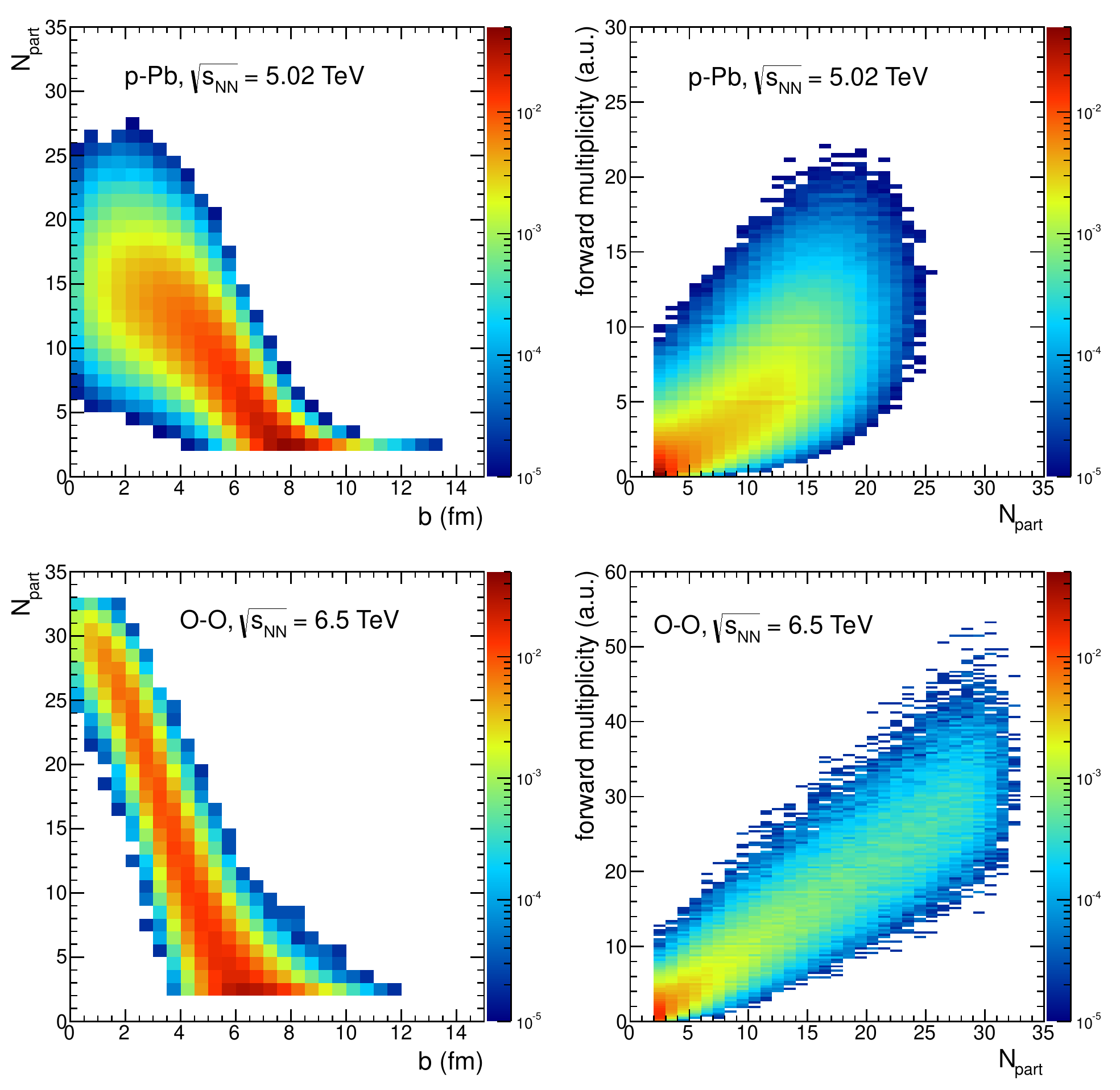}
\caption{Glauber MC calculations are presented for for \pPb (top panels) and \OO collisions (bottom panels). Shown are \Npart as a function of impact parameter (left panels) and forward multiplicity as a function of \Npart (right panels).}
\label{fig:smallsystems_oo_glauber}
\end{figure}

\begin{figure}[t]
\centering
\includegraphics[width=0.5\linewidth]{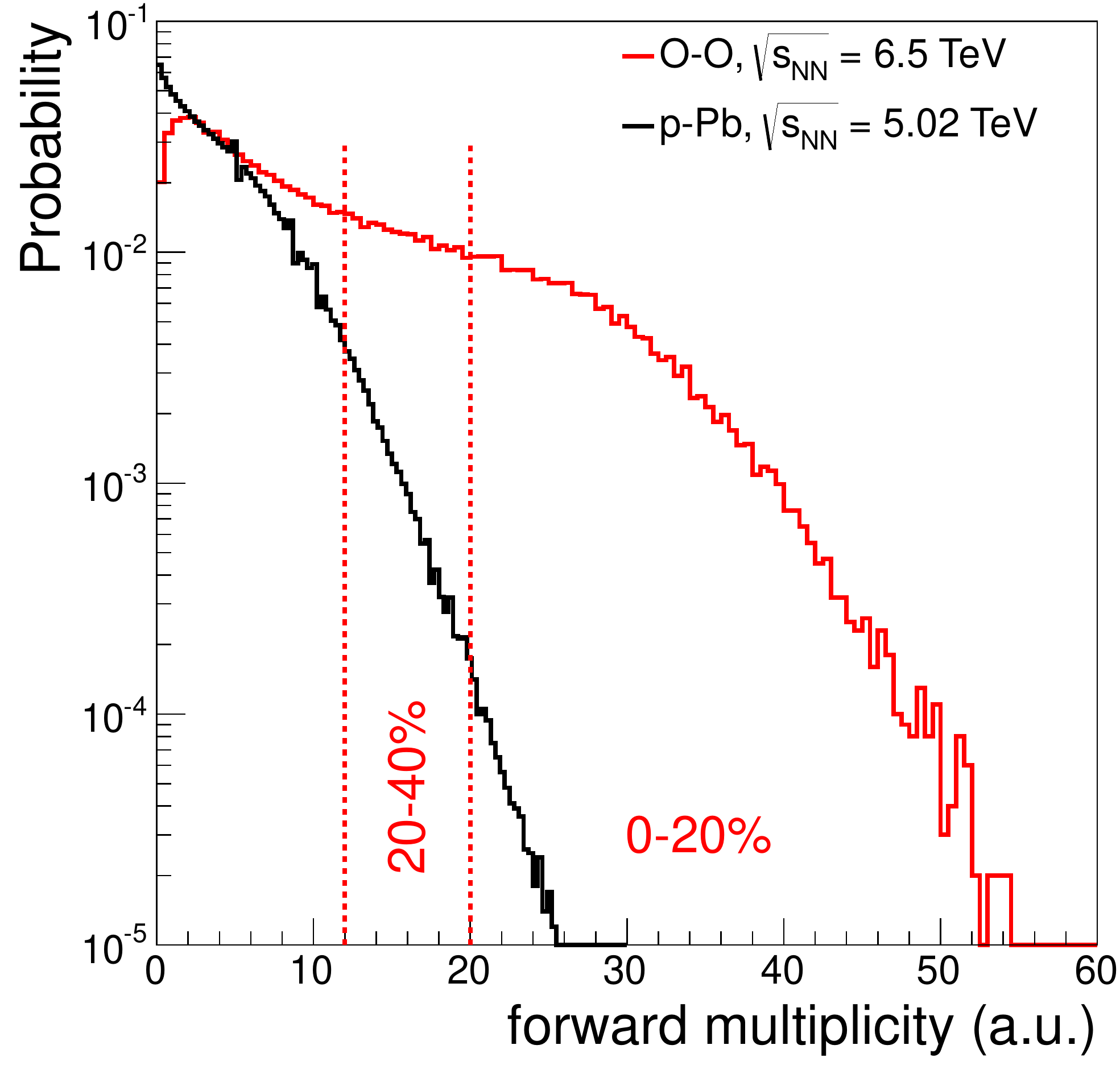}
\caption{Estimated multiplicity distributions in the forward region in \OO and \pPb collisions. The event classes of highest multiplicity in \OO collisions are indicated with 0--20\% and 20--40\%.
The \OO studies have been been performed for the 2018 beam configurations at $\sqrtsNN = \unit[6.5]{\UTeV}$, while the Run 3 configuration would yield $\sqrtsNN = \unit[7]{\UTeV}$, with the advantage that a large \pp reference data set already exists from earlier LHC running (2010--11).
}
\label{fig:smallsystems_oo_mult}
\end{figure}

\subsection{Potential of \OO Collisions}
\label{sect:smallsystems_OO}

A promising opportunity to study the emergence of collective phenomena further as well as the presence of possible parton energy loss in small collision systems, are collisions of smaller nuclei. 
In particular, collisions of oxygen are an efficient way of investigating the properties of low-multiplicity heavy-ion collisions, which in large \AOnA systems only occurs for peripheral geometries.
The achieved multiplicities in \OO collisions are similar to \pPb collisions with the significant advantage that the collision geometry is much better defined.
This is demonstrated in Fig.~\ref{fig:smallsystems_oo_glauber}, which shows the correlations between number of participants, multiplicities and impact parameter in \OO and \pPb collisions. The correlations between \Npart and impact parameter as well as multiplicity and \Npart is much narrower in \OO collisions as compared to \pPb collisions. Consequently, highest multiplicities in \pPb collisions are only accessible in the tail of the distribution while similar multiplicities are already reached in \OO collisions in the plateau region. This is shown in Fig.~\ref{fig:smallsystems_oo_mult} illustrated for the 20--40\% event class.

Scaling the measured nuclear-modification factor in \PbPb collisions at \unit[5.02]{\UTeV}~\cite{Acharya:2018qsh} at similar multiplicities as for central \OO collisions while roughly accounting for the artificial suppression caused by the multiplicity bias present in such peripheral \PbPb collisions~\cite{Acharya:2018njl}, allows to estimate the expected effect on the \RAA in \OO collisions to about 20\%.
Out of this deviation from unity, about half can be attributed to biases due to the multiplicity selection in such small collision systems already in absence of nuclear effects~\cite{Adam:2014qja}. An observable deviations from unity of about 10\% caused by energy loss in the produced medium remains. While this expected suppression may seem small, it should be possible to measure it already with an $\Lint$ of a few $\unit[100]{\mu b^{-1}}$. 
In case such a suppression was absent, the conclusion can be drawn that small collision systems do not exhibit measurable energy loss, while other collective features are present, challenging the role of significant final-state interactions as underlying mechanism.
It should be noted that the absence of suppression can most likely not be taken as a proof against the formation of the QGP, as it may be that the fast partons are emerging without seeing the medium, either because they are emitted from the surface or because their formation time is longer than the time within the medium.
On the contrary, as discussed in Sect.~\ref{sect:smallsystems_energyloss} the first observation of energy loss in small systems would clearly confirm models in which final-state interactions play an important role

\begin{figure}[h!]
\centering
\includegraphics[width=0.48\linewidth]{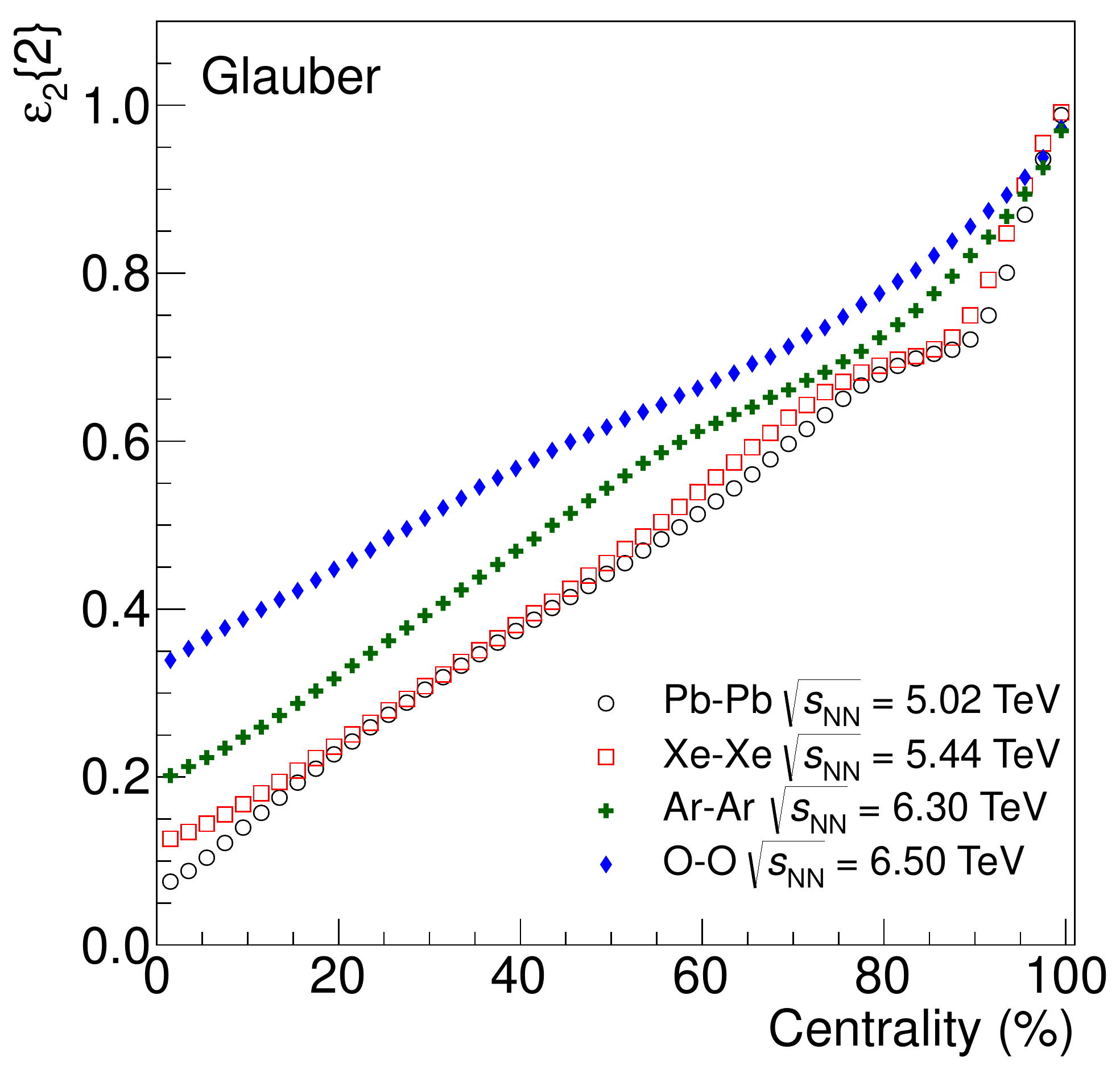}
\hfill
\includegraphics[width=0.48\linewidth]{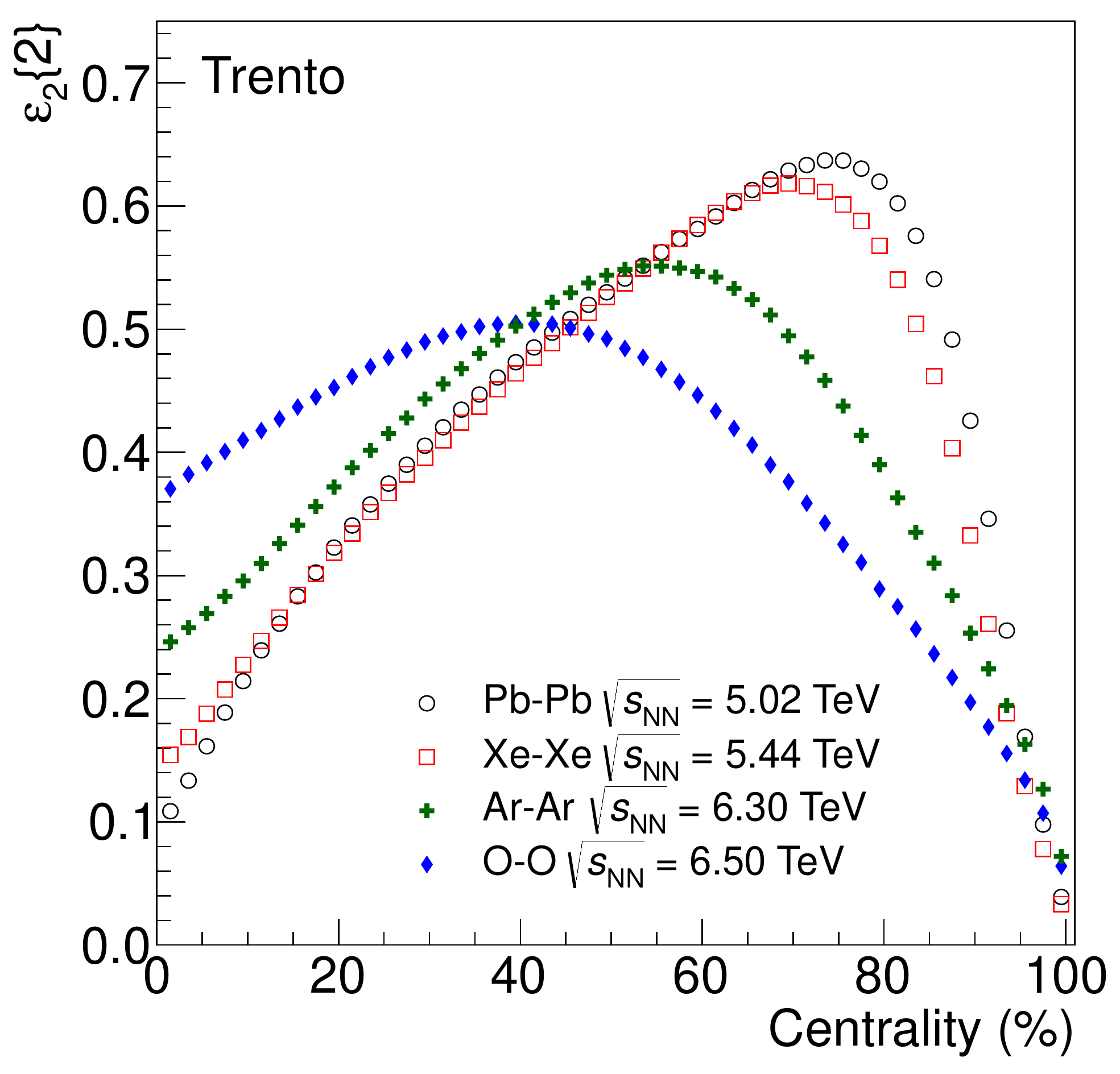}
\caption{Second-order eccentricity coefficient $\epsilon_2$ as a function of centrality for \OO, \ArAr, \XeXe and \PbPb collisions from MC-Glauber (left panel)~\cite{Miller:2007ri,Loizides:2017ack} and Trento (right panel)~\cite{Moreland:2014oya} initial conditions.}
\label{fig:smallsystems_ecc}
\end{figure}

The study of bulk particle production in \OO collisions has twofold interest:
Firstly, the dependence of the pseudorapidity density of charged particles scaled by participant pairs on \sqrts is stronger in \AOnA collisions than in \pp and \pA collisions~\cite{Acharya:2018egz}. The rise for \pPb and \dAu collisions is similar to the one of inelastic \pp collisions indicating that the stronger rise in \AOnA collisions might not be solely related to the multiple collisions undergone by the participants since the proton in \pA collisions also encounters multiple nucleons. High-energy \OO collisions promise to study this \sqrts dependence further in a regime of small number of participants.
Secondly, a strong correlation between initial state geometry (collision eccentricity) and observed flow has been established since RHIC~\cite{Drescher:2007cd}, and pertains at LHC, in \PbPb and \XeXe collisions~\cite{Acharya:2018ihu}. As shown in Fig.~\ref{fig:smallsystems_ecc}, the eccentricity profiles of \XeXe and \PbPb collisions are, however, quite similar, and collision systems exhibiting a different geometry as \OO could therefore provide further insight into the connection between initial-state geometry and multi-particle correlations~\cite{Lim:2018huo}.

\subsection{Summary}

The discoveries made in recent years in small collision systems have challenged two paradigms, the modelling of heavy-ion collisions, as well as the modelling of the underlying event of elementary \pp collisions. The experimental observations of strangeness enhancement and multi-particle correlations suggests that energy loss should also be present, as both are a consequence of significant final-state interactions. But up to this point no hint of energy loss in \pp or \pPb collisions has been seen. The increased luminosity will allow both for precision studies of the established signatures of small system collectivity, and to either establish evidence or place exclusion limits on the latter.

This chapter has presented a HL-LHC small system experimental program promising significant progress shown by the set of performance studies, ranging from largely non-flow suppressed high-order correlations, over measurement of strange-particle yields and thermal radiation, to energy-loss signals. In addition, new theoretical insights and phenomenological developments are also needed for developing a unified picture of parton dynamics and particle production valid for both small and large systems. This working group encourages both investment in the necessary theoretical development, and facilitation of collaboration between theorists and experimentalists. 

The physics community can look forward to a deepened understanding of hot and dense QCD and a universal description of small to large collision systems.

\clearpage

\section{High energy QCD with proton-nucleus collisions and ultra-peripheral collisions}
{ \small
\noindent \textbf{Coordinator}: Michael Winn (LAL and IRFU/DPhN, CEA Saclay) 

\noindent \textbf{Contributors}:
A.~Angerami (LLNL Livermore), 
F.~Arleo (LLR/{\'E}cole Polytechnique), 
N.~Armesto (Instituto Galego de Fisica de Altas Enerxias (IGFAE) Universidade de Santiago de Compostela), 
R.~Bi (Massachusetts Institute of Technology), 
{\'E}.~Chapon (CERN), 
Z.~Citron (Ben-Gurion University of the Negev), 
J.~G.~Contreras~Nuno (Czech Technical University in Prague),  
M.~Dumancic (Weizmann Institute of Science),
D.~d'Enterria (CERN), 
G.~Giacalone (IPhT, CNRS/CEA), 
A.~Giammanco (Universit\'{e}~Louvain), 
V.~Guzey (Helsinki University, University of Jyvaskyla, Kurchatov institute, Gatchina),
I.~Helenius (University of Jyvaskyla and Universit\"{a}t T\"{u}bingen), 
P.~Janus (AGH University of Science and Technology),
H.~Kim (Chonnam National University), 
M.~Klasen (M\"{u}nster University), 
S.R.~Klein (Lawrence Berkeley National Laboratory), 
J.~Kremer (AGH University of Science and Technology),
G.~Krintiras (Universit\'{e}~Louvain),
E.~Kryshen (Kurchatov institute, Gatchina), 
A.~Kusina (IFJ PAN, PL-31342 Krak\'ow, Poland), 
Y.-J.~Lee (Massachusetts Institute of Technology), 
M.~van~Leeuwen (Utrecht University/Nikhef), 
C.~Marquet (CPHT/{\'E}cole Polytechnique), 
C.~Mayer (IFJ PAN, Krak\'ow, Poland), 
M.~Mulders (CERN),
M.~Murray (University of Kansas), 
F.~Olness (SMU Dallas), 
P.~Paakkinen (University of Jyvaskyla), 
H.~Paukkunen (University of Jyvaskyla, Helsinki Institute of Physics), 
I.~Schienbein (LPSC/Universit{\'e} Grenoble Alpes),
P.~Silva (CERN), 
D.~Stocco (Subatech Nantes), 
M.~Strikman (Pennsylvania State University), 
D.~Tapia~Takaki (University of Kansas). 
}

\label{sec:smallx}


\subsection{Introduction}
\label{sec:smallxIntro}

Proton-nucleus~\cite{Salgado:2011wc} and ultraperipheral (UPC)~\cite{Baltz:2007kq} collisions offer the opportunity to study the behaviour of QCD at high energies and large partonic densities~\cite{Kovchegov:2012mbw}. 
As the theory of the strong interaction, QCD is analytically well understood only in a perturbative regime of small coupling constant and where radiation of gluons and quarks is a linear process that can 
be described with linear evolution equations of the non-perturbative 
parton densities, i.e.,\,hadrons and nuclei are considered as dilute partonic objects. 
However, non-linear effects are unavoidable in QCD, and they should in principle dominate at large densities reached at high collision energies and for large nuclei.  
It was proposed long ago~\cite{Gribov:1984tu,Mueller:1985wy} that at such large densities a resummation of powers of density scaled by the strong coupling constant is possible, resulting in a non-perturbative but weak coupling regime where parton densities saturate, and whose effective field theory incarnation is the Color Glass Condensate~(CGC)~\cite{Gelis:2010nm}.
Particle production in \pA collisions in the forward rapidity region is dominated by small$-x$ partons in the nucleus. 
Therefore, saturation effects are expected to be largest there. Furthermore, UPCs as a source of large fluxes of quasi-real photons, provide the opportunity to study the partonic structure of protons (in \pp and \pA) and nuclei (in \pA and \AOnA).

The structure of nucleons and nuclei, and the mechanism of particle production at small~$x$, are also key ingredients for a detailed description of heavy-ion collisions and for the characterisation of the produced hot and dense medium. 
For hard probes, the nuclear modification of standard collinear parton densities is one of the uncertainties in the extraction of medium characteristics~\cite{Rapp:2018qla,Andrews:2018jcm}. 
For soft and semihard observables, both the parton densities and the detailed transverse spatial structure of nucleons and nuclei are crucial ingredients for their description, e.g.\,for the initial conditions for hydrodynamical evolution~\cite{Liu:2015nwa,Weller:2017tsr,Mantysaari:2017cni}.

\begin{figure}[tb]
\centering
\includegraphics[width=0.45\textwidth]{\main/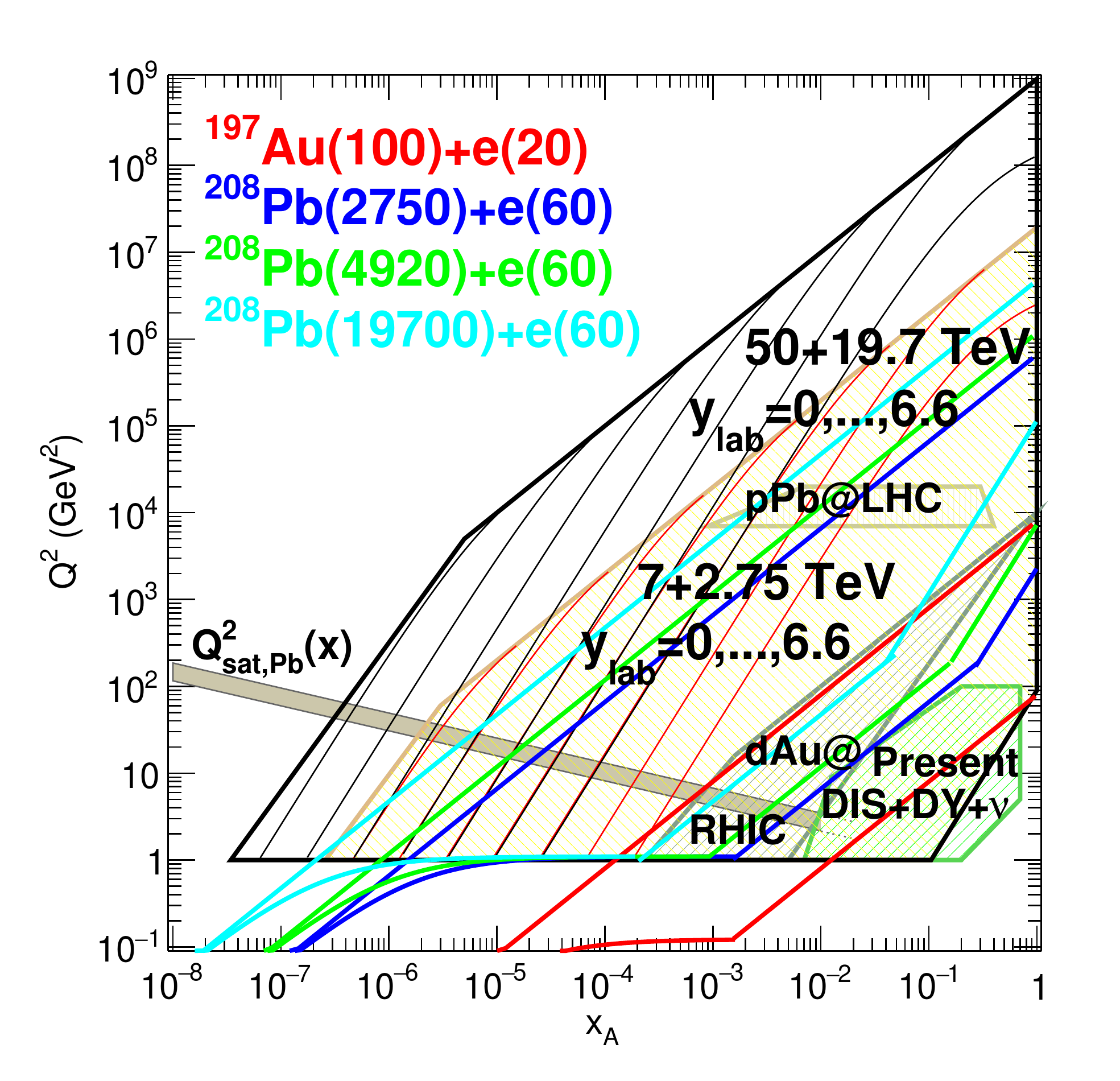} 
\hfill 
\includegraphics[width=0.48\textwidth]{\main/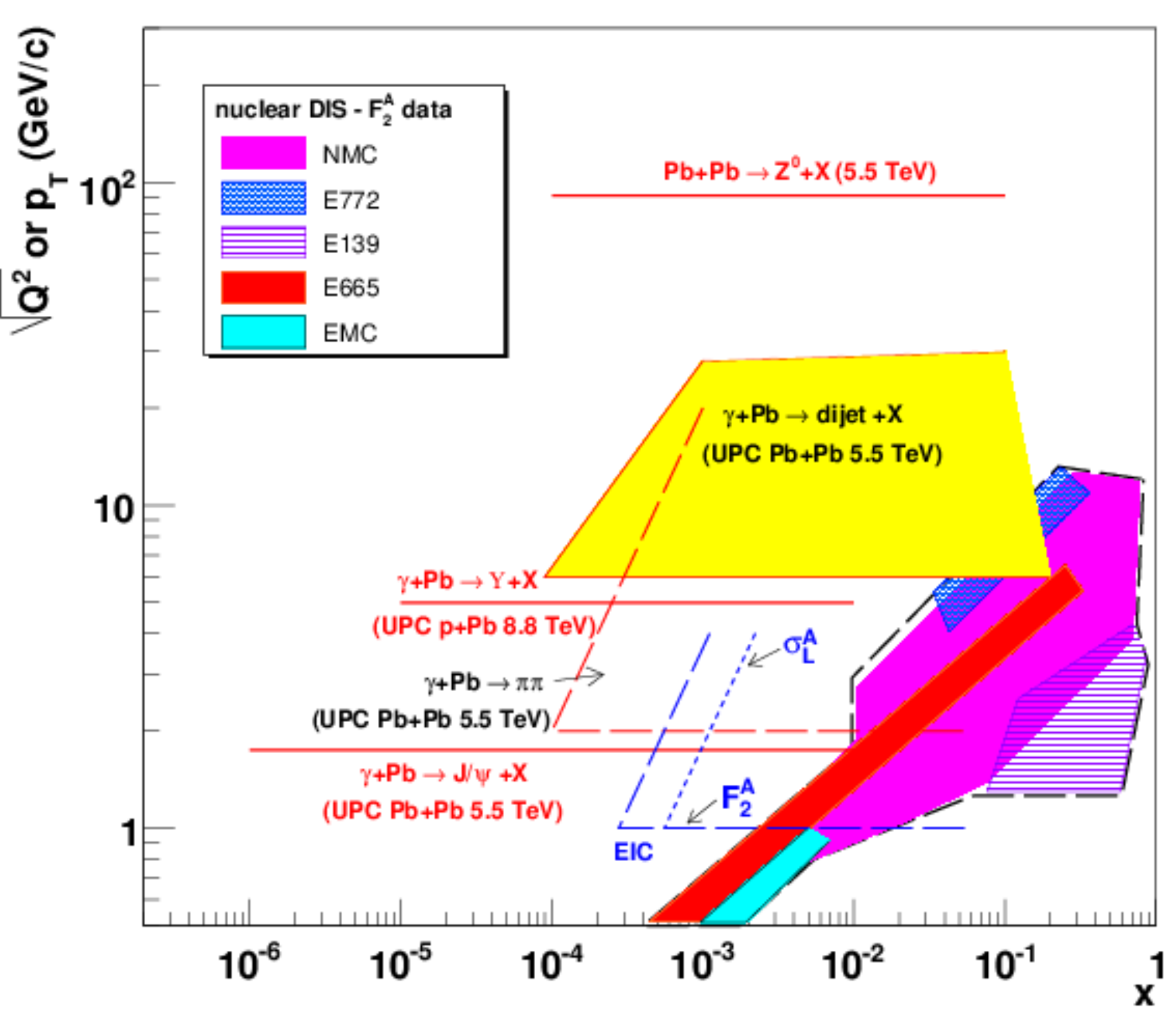}
\caption{Left: $x$-$Q^2$ plane to be explored in proton-nucleus at the LHC and the FCC, and in  proposed electron-nucleus colliders, compared with the regions where the experimental data presently used in the EPPS16 analysis \cite{Eskola:2016oht} lie. Right: $x$-$Q^2$ plane to be explored in UPCs, taken from \cite{Baltz:2007kq}.}
\label{fig:smallx1}
\end{figure}

In the collinear framework, parton densities inside nuclei (nPDFs)~\cite{Paukkunen:2017bbm,armestodis2018} are poorly constrained due to two primary factors. 
On the one hand, data come from a large variety of nuclei and the number of data points for any of them is very small compared to the proton analysis. In particular, for the Pb~nucleus, there are less than 50 points coming from fixed target Deep Inelastic Scattering~(DIS) and Drell-Yan experiments and from \pPb collisions at the LHC. 
The fit for a single nucleus is therefore impossible and the modeling of the $A$-dependence of the parameters in the initial conditions becomes mandatory~\cite{Eskola:2016oht,Kovarik:2015cma}. On the other hand, the kinematic coverage in $Q^2$ and $x$ with existing data is very small compared to the requirements of present hadronic colliders, see Fig.~\ref{fig:smallx1}. 
In spite of the uncertainties in the applicability of collinear factorisation, scale choices and other theoretical caveats for nPDFs extraction in hadronic collisions and UPCs, these are the only experimental collision systems where the nPDFs can be constrained before electron-ion colliders become available. 
The most up to date analyses include between 1000 and 2000 data points for 14 nuclei and are performed at next-to-leading accuracy~\cite{deFlorian:2011fp,Kovarik:2015cma,Eskola:2016oht},  there even exists a first attempt at next-to-next-to-leading~\cite{Khanpour:2016pph}.
Differences between them mainly arise from the different sets of data included in the analysis and from the different functional forms employed for the initial conditions. 
All in all, all parton species are very weakly constrained at small~$x<10^{-2}$, gluons at large~$x>0.2$, and the flavour decomposition is largely unknown - a natural fact for $u$ and $d$ due to the approximate isospin symmetry in nuclei. 
The impact of presently available LHC data, studied in~\cite{Eskola:2016oht}, is quite modest with some constrains on the gluon in the region $0.01<x<0.3$. 
On the other hand,  theoretical predictions for nuclear shadowing of quark and gluon PDFs  based on $s$-channel unitarity and diffractive nucleon PDFs are available down to $x \sim 10^{-4}$ --$10^{-5}$~\cite{Frankfurt:2011cs,Armesto:2003fi}. 

In the context of phenomena beyond collinear factorisation and PDF evolution in ln($Q^2$), there have been recent claims~\cite{Ball:2017otu,Abdolmaleki:2018jln} that resummation of logarithms of $x$ may be required for a better description of DIS data from HERA at small $x$, and searches for long range azimuthal correlations are undergoing~\cite{zeusichep2018}. 
But no conclusive evidence of saturation, i.e.,\,of non-linear dynamics, has been found in hadronic collisions. While the CGC provides a calculational framework for several observables in \pp, \pA and \AOnA, see e.g.\,the reviews~\cite{Albacete:2013tpa,Lappi:2015jka}, like the ridge, back-to-back hadron correlations in the forward region, multiplicities and transverse momentum distributions,$\dots$,  there is no consensus in the field in the interpretation of these results, or they involve non-perturbative modeling, or they are affected by large theoretical uncertainties and, for some of them, higher-order calculations are missing, or the data lie at the border of phase space where extracting clear conclusions is very delicate. 
Therefore, high-energy \pA collisions and UPCs are two promising systems where data can offer clear evidences of non-linear effects.

In Fig.~\ref{fig:smallx1}, the kinematic regions covered by proton-nucleus collisions at the LHC and the FCC~(the left panel, \cite{Dainese:2016gch}) and UPCs at the LHC (right) are shown and compared with the regions where data currently used to constrain nPDFs lie.
A huge enlargement is evident with respect to the presently existing data at the LHC. The HL-LHC offers new improved detectors and larger statistics for some observables like dijets or photon-jet correlations. The HE-LHC would enlarge the kinematic plane in a region intermediate between the LHC and the FCC. On the other hand, electron-nucleus collisions~\cite{AbelleiraFernandez:2012cc,Accardi:2012qut}, if they eventually happen in the 2030's, would be complementary. They offer measurements in a cleaner experimental environment (no pileup, full kinematic reconstruction) and under better theoretical control as first-principle calculations are easier in DIS albeit in a more restricted kinematical region. 
The comparison between the kinematic regions covered by the LHC and Future Circular Collider~(FCC) in \pA mode, and the Electron-Ion Collider~(EIC)~\cite{Accardi:2012qut} (10-20~\UGeV electrons combined with the RHIC nuclear beams or with a new hadron machine at Jefferson Lab) and the LHeC~\cite{AbelleiraFernandez:2012cc} (60 \UGeV electrons colliding with the HL-LHC, HE-LHC or FCC nuclear beams) is shown in Fig.~\ref{fig:smallx1}~(left).

\subsection{The physics of ultra-peripheral collisions}
\label{sec:UPCqcd}
Ultra-peripheral collisions are interactions with impact parameter larger than the radial size of the colliding nuclei or protons. In these collisions, one nucleus acts as a source of quasi-real photons that interact with the crossing nucleus or proton. The electromagnetic field intensity and hence the photon flux scales with the charge number squared.  At the LHC, these collisions occur abundantly and serve as a laboratory for strong and electromagnetic interactions~\cite{Baltz:2007kq}. After an experimental overview discussing the opportunities, the discussion focuses on vector meson production studies to probe nuclear gluons and dijet production in photonuclear or photon-proton interactions in LHC Run 3 and 4 as a direct access of nPDFs. A dedicated section~\ref{sec:upc} addresses the physics of $\gamma\gamma$ interactions as a probe of QED and beyond the standard model physics. The investigation of $\gamma$-proton interactions in pp~collision allowed interesting measurements on vector meson production by LHCb~\cite{Aaij:2013jxj,Aaij:2014iea,Aaij:2015kea,Aaij:2018arx} and will allow for interesting measurements with already recorded and future data. This possibility will not be discussed in detail because the standard running conditions in terms of pile-up and beam optics relevant for forward proton tagging in ATLAS/CMS and the pile-up in LHCb in Run 3 and Run 4 will make comparatively low $Q^2$ measurements primarily discussed here challenging and likely restricted to short special runs whereas in ALICE, the equivalent luminosity $\gamma$-proton luminosity will be smaller or similar than in \pPb collisions but with the ambiguity which beam emitted the photon.
\subsubsection{Experimental overview}
An overview of the detector capabilities in Run 3 and Run 4 is given by the four collaborations ALICE,ATLAS,CMS and LHCb in the following. Subsequently, the expected statistics for vector meson observables is given in Tab.~\ref{tab:PbPbUPCrates},\ref{table:leadshine},\ref{table:protonshine} followed by an estimate of the available precision for coherent heavy vector meson production and the theoretical assessment of vector mesons and dijet production in UPCs.

\begin{itemize}
\item \textbf{ALICE}~\cite{Abelevetal:2014cna} will take data in
both in triggered and in continuous readout mode during Run~3 and Run~4~\cite{Buncic:2015ari}. Using continuous
readout~\cite{Antonioli:1603472} essentially the full delivered luminosity can be integrated
without significant trigger and dead-time inefficiencies. Therefore the total acceptance-efficiency factor for UPC events is significantly larger
than in LHC Runs~1 and 2; it is determined by the tracking 
efficiencies and the geometrical acceptances of the Inner Tracking
System (ITS)~\cite{Abelevetal:2014dna}, of the Time Projection Chamber (TPC)~\cite{ALICE:2014qrd}, and of the Muon Spectrometer~\cite{Uras:2012af}. The geometrical acceptances of these detectors correspond to the narrow central and forward acceptances defined in Tab.~\ref{tab:UPCtoyacceptance}. Vector meson yields in the corresponding acceptance are given in Tab.~\ref{tab:PbPbUPCrates},\ref{table:leadshine},\ref{table:protonshine}.
Final-state neutron emission in UPC events can be detected by the
zero-degree calorimeters (ZDC) which will also take data in continuous
readout mode, and vetoes can be imposed using the fast interaction
trigger detector~(FIT). 
\item  \textbf{ATLAS} UPC measurements in Run~3 and Run~4 will continue to be an important part of the experiments heavy-ion programme and will benefit from several detector upgrades in that period.  Prior to Run~3, the Zero Degree Calorimeter (ZDC), which is key for UPC event identification by measuring forward neutrons, will be upgraded.  At mid-rapidity, the Inner Tracker (ITk) upgrade \cite{ATL-PHYS-PUB-2016-025} will increase the acceptance for charged particles to $|\eta|<4$ for Run 4.   The High Granularity Timing Detector (HGTD)  \cite{Collaboration:2623663} will complement the spatial information of the ITk with timing information in the region 2.4 $<\eta<$4.0, with a resolution of 30 \,\unit{ps}.  In addition, upgrades to the ATLAS trigger and data-acquisition system will take place prior to both Run~3 and Run~4 \cite{Collaboration:2285584}, and will enable advanced triggering capabilities.
Together these will aid the study of low-mass resonances and the continuum as well as jets in UPC events.   
\item \textbf{CMS} UPC studies will benefit from the upgraded inner tracker for Run 4 which will provide a large acceptance for charged particles up to $|\eta|<4$~\cite{CMS-TDR-15-02}. The improved CMS level-1 trigger and data acquisition rate (up to 60\,\unit{GB/s}) will provide opportunities for more innovative and sophisticated triggers to capture a wide variety of processes. 
In addition, the proposed MIP Timing Detector~\cite{Collaboration:2296612}, which is around 1.16\,\unit{m} away from the beam pipe could provide a time resolution around 30\,\unit{ps}. By combining it with other detectors, MIP will provide proton, pion, and kaon separation for $\pT$ values between 0.7 to 2\UGeVc in the midrapidity region ($|\eta|<1.5$). These detector upgrades and the increased trigger performance summarised in~\cite{CMS-FTR-18-012} will facilitate the study low-mass UPC resonant states and UPC heavy flavour studies. In addition, the reconstruction algorithms will be improved with the addition of the four-layer pixel system. 
The CMS-acceptance in Run~3 corresponds to the wide central range defined in Tab.~\ref{tab:UPCtoyacceptance}. Yields in acceptance for vector mesons are provided in Tab.~\ref{tab:PbPbUPCrates},\ref{table:leadshine},\ref{table:protonshine}.

\item \textbf{LHCb} is well suited for exclusive production studies in ultra-peripheral collisions. In particular, its optimisation for flavour physics within its acceptance 2$<\eta<$5 provides an excellent resolution for typical momenta in quarkonium and heavy-flavour exclusive production as demonstrated in~Ref.\cite{LHCb-CONF-2018-003}. Its particle identification capabilities allow to measure final states with charged muons, pions, kaons and protons.  The upgraded detector will be able to efficiently sample the full delivered luminosity for UPC final states based on a purely software based trigger scheme~\cite{LHCb-TDR-016}. Exclusive diphoton analyses as pioneered in ATLAS~\cite{Aaboud:2017bwk} and CMS~\cite{CMS-PAS-FSQ-16-012} are also conceivable with lower \ET-thresholds discussed in Section~\ref{sec:upc}. The feasibility of inclusive $\gamma$-induced measurements in LHCb will require further studies. In Tab.~\ref{tab:PbPbUPCrates},\ref{table:leadshine},\ref{table:protonshine} conceivable vector meson final states are given for LHCb represented by the  wide forward acceptance defined in Tab.~\ref{tab:UPCtoyacceptance}.
\end{itemize}

\subsubsection{Vector meson production}
\subsubsubsection{Experimental reach}
The number of vector mesons expected in the upcoming runs provides an estimate of the expected physics reach.   Four toy-model experimental acceptances are shown in Tab.~\ref{tab:UPCtoyacceptance}. The effects of cuts on the $\pT$ of the final state daughter particles are not estimated, because this depends on the analysis and trigger conditions.   In general, for two-prong decays, as long as the minimum detectable $\pT$ is less than about 1/4 of the final state mass, the effect on the efficiency is limited.  The acceptance cuts are applied to both the vector meson rapidity and the daughter particle pseudorapidities.  These two selection types have similar effects on the production phase space, and the results would not be very different without the cut on the vector meson rapidity. 

Table~\ref{tab:PbPbUPCrates} shows the expected cross sections and rates for five representative vector meson decays, including the decay branching ratios.  The rates are calculated using STARlight~\cite{Klein:2016yzr}, which has been shown to predict the cross sections for all mesons on proton targets and for $\rho$~photoproduction on heavy targets~\cite{Adam:2015gsa,Abelev:2007nb}. Since the $\rho$ vector resonance is broad, the mass range from $2m_\pi$ up to $M_\rho + 5\Gamma_\rho$ is considered.  Non-resonant $\pi^+\pi^-$~production is not included which would lead to a 8\% increase of the production rate.  
The $\rho'$ state represents $\pi^+\pi^-\pi^+\pi^-$ states with masses in the range from 4$m_{\pi}$ up to 2.5~\UGeVcc. In absence of other guidance and the possibility of more complex resonance structure, the calculation is anchored to the STAR measurement of the $\pi^+\pi^-\pi^+\pi^-$ final state in Au--Au collisions at $\sqrtsnn=200$~Gev~\cite{Abelev:2009aa} and uses STARlight to extrapolate in collision energy and collision system. 

Since STARlight does not include nuclear shadowing, a rapidity-dependent nuclear shadowing correction following Ref. \cite{Guzey:2016piu} for the heavy quarkonium (J$/\psi$, $\psi$(2S) and $\PGUP{1S}$ was applied.  The cross sections and rates  are hence reduced by factors of 0.42, 0.475 and 0.77 for the J/$\psi$, $\psi$(2S) and $\PGUP{1S}$ respectively.  

\begin{table}[h]
\centering
\caption{Table of toy-model acceptance cuts for the different experiments.}
\begin{tabular}{|c|ccccc|}
\hline
Condition & Tot. & Central 1 & Central 2 & Forward 1 &  Forward 2   \\
                 &       & Narrow  & Wide    &      Narrow     &    Wide         \\
\hline
Rapidity & - & $|y|<0.9$ & $|y|<2.4$ & $2.5 < y < 4.0$ &   $2 < y < 5$ \\
$\mathrm{e/\pi/\mu}$ pseudorapidity & - & $|\eta|<0.9$ & $|\eta|<2.4$  & $2.5 < \eta < 4.0$ & $2 < \eta <5 $ \\
\hline
\end{tabular}
\label{tab:UPCtoyacceptance}
\end{table}

\begin{table}[h]
\centering
\caption {Table of cross sections and numbers of events in 13~nb$^{-1}$ integrated luminosity for the different mesons in \PbPb collisions.   B, M and K denote $10^9$, $10^6$ and $10^3$ respectively. 
Both the rates and cross sections include the relevant branching ratios.  The cross sections and toy-model acceptances are determined using STARlight~\cite{Klein:2016yzr}.  For the J$/\psi$, $\psi$(2S) and $\PGUP{1S}$, rapidity-dependent nuclear shadowing cross sections have been applied following the approach in Ref. \cite{Guzey:2016piu}. 
}
\begin{tabular}{|c|rrrrrr|}
\hline
\multicolumn{7}{|c|} {PbPb} \\
\hline
            &     $\sigma$              &  All          & Central 1   & Central 2   & Forward 1 & Forward 2 \\
Meson &                                 & Total        & Total          & Total          & Total         l &  Total  \\
\hline
$\rho\rightarrow \pi^+\pi^-$        & 5.2b & 68 B  & 5.5 B  & 21B & 4.9 B & 13 B \\
$\rho'\rightarrow \pi^+\pi^-\pi^+\pi^-$ & 730 mb & 9.5 B  & 210 M & 2.5 B & 190 M   & 1.2 B \\
$\phi\rightarrow \mathrm{K}^+\mathrm{K}^-$           & 0.22b & 2.9 B & 82 M & 490 M  & 15 M & 330 M  \\
J$/\psi\rightarrow\mu^+\mu^-$  & 1.0 mb & 14 M & 1.1 M & 5.7 M & 600 K &  1.6 M \\
$\psi\mathrm{(2S)}\rightarrow\mu^+\mu^-$    & 30$\mu$b & 400 K& 35 K & 180 K & 19 K &  47 K \\
$\PGUP{1S}\rightarrow\mu^+\mu^-$ & 2.0 $\mu$b & 26 K & 2.8 K & 14 K  & 880 &  2.0 K \\
\hline
\end{tabular}
\label{tab:PbPbUPCrates}
\end{table}

The rates for light mesons are very large, enough to support billion-event samples of the $\rho$ and $\rho'$, and hundreds of millions of $\phi$, allowing the studies discussed below.  Beyond precise cross section measurements detailed below, the rates for J$/\psi$, $\psi$(2S) and $\PGUP{1S}$ should allow tomographic measurements which can be used to infer information on the nuclear wave function and which is outlined in Section~\ref{sec:nuclimage}.  
In the $\pi^+\pi^-$ channel, the study of pairs with masses above 2~Gev/$c^2$ comes in reach. 

In the hadronic decay $\phi \rightarrow K^+K^-$, the kaons have a momentum of only 135 \UMeVc\ in the $\phi$ rest frame. Since the kaon momentum is dominated by the longitudinal momentum acquired from the $\phi$-meson, the kaons are produced with large pseudorapidity. Therefore, the acceptance for this channel is low at around midrapidity even without a considering a minimal kaon-$\pT$.   Consequently, an observation is very difficult with the potential exception of the far forward region where the kaons are significantly Lorentz boosted.  Alternately, measurements exploiting the leptonic decay channels $\phi \rightarrow \mu^+ \mu^- \mathrm{or} \mathrm{e}^+ \mathrm{e}^-$ despite the small branching ratios. The feasibility can be conservatively estimated by scaling the $\phi\rightarrow \mathrm{K}^+\mathrm{K}^-$ rates down by Br ($\phi \rightarrow \mu^+ \mu^-$)/Br($\phi \rightarrow \mathrm{K}^+\mathrm{K}^-$)$\approx 5.9 \times 10^{-4}$ neglecting the acceptance increase for leptons w.r.t. kaons. 

The exploration of double vector mesons photoproduction by a single ion-ion pair by exchange of two independent photons becomes available.   The expected ratio of $\rho\rho$ photoproduction to single $\rho$ photoproduction is about $1:600$ while the predicted ratio of $\rho \mathrm{J/}\psi$ to $\rho$ is about $1:160$ \cite{Klein:1999qj}.  
These events should display significant quantum correlations.  

In \pPb collisions, the per-nucleon centre-of-mass system is boosted by 0.465 units of rapidity from the lab frame, and there are two possible directions for the beams, protons from the $+z$ direction, or from the $-z$ direction influencing strongly the available kinematics for the forward detectors.  Furthermore, there are two possibilities for the photon emitter, 'lead-shine' ($\gamma$p), the photon from the lead nucleus, or 'proton-shine' ($\gamma$Pb), the photon from the proton. These two vector meson production channels can be in principle separated by their different~\pT scale corresponding to $\hbar/R_{\rm proton}$ for 'lead-shine' ($\gamma$p) and $\hbar/R_{\rm Pb}$ for 'proton-shine' ($\gamma$Pb).     
The rates are calculated for these two possibilities in separate tables.

The total \pPb luminosity of 2000 nb$^{-1}$ assumed to be equally divided between the two possible proton beam directions.  For the asymmetric (around $y=0$) detectors, these two Runs are considered separately, listing them as 'forward' (in the proton-going hemisphere) and 'backward' (in the lead-going hemisphere).  

Tables~\ref{table:leadshine} and~\ref{table:protonshine} show the cross sections and rates for the lead-shine and proton-shine cases respectively.   Lead-shine is dominant, with proton-shine contributing less than 10\% to the total rate.  The extraction of the proton-shine component by fitting to the different $\pT$ spectra for proton-shine and lead-shine will be challenging and the measurement precision depends strongly of the momentum resolution of the detector.  Nuclear shadowing corrections are not applied, the reduction factors are similar for the proton-shine cross sections as for the lead-lead collisions.  No attempt to calculate the rates for the $\rho'$ on proton targets is undertaken due to the large rate uncertainties.  

\begin{table}[h]
\centering
\caption {Table of cross sections and numbers of events for the different mesons in \pPb collisions for 'lead-shine' (a photon from the lead scattering from the proton).  The rates are for the 2000 nb$^{-1}$ integrated luminosity noted above, split evenly between the two possible proton directions.  For the central regions, the net luminosity is 2000 nb$^{-1}$  since both directions contribute, but for the forward (FW) and backward (BW) directions, the net luminosity is only 1000~nb$^{-1}$  each. B, M and K denote $10^9$, $10^6$ and $10^3$ respectively. 
Both the rates and cross sections include the relevant branching ratios.}
\begin{tabular}{|c|rrrrrrrr|}
\hline
\multicolumn{9}{|c|} {pPb - lead shine, $\gamma$p} \\
\hline
            &     $\sigma$              &  All & Ctl. 1 & Ctl. 2 & FW 1 & FW 2 & BW 1 & BW 2 \\
Meson & & Total & Total & Total &Total  & Toal  & Total & Total \\
\hline
$\rho\rightarrow\pi^+\pi^-$ & 35 mb& 70 B & 3.9 B& 15 B& 2.0 B & 5.5 B& 850 M& 2.0 B\\
$\phi\rightarrow \mathrm{K}^+\mathrm{K}^-$ & 870 $\mu$b & 1.7 B & 65 M & 290 M & 22 M &120 M  &9.7 M  & 52 M\\
J$/\psi\rightarrow\mu^+\mu^-$ & 6.2 $\mu$b  & 12 M & 1.0 M &5.2 M & 260 K & 800 K  & 180 K & 430 K\\
$\psi\mathrm{(2S)}\rightarrow\mu^+\mu^-$   &134 nb & 270 K & 22 K & 110 K & 6.0 K &  18 K & 3.2 K & 7.7 K \\
$\PGUP{1S}\rightarrow\mu^+\mu^-$ & 5.74 nb & 11 K &  1.1 K & 5.4 K & 310 & 880  & 41 & 100 \\
\hline
\end{tabular}
\label{table:leadshine}
\end{table}

\begin{table}[h]
\centering
\caption {Table of cross sections and rates for the different mesons in \pPb collisions for 'proton-shine' (a photon from the proton scattering from the lead nucleus).  The rates are for the 2000 nb$^{-1}$ integrated luminosity noted above, split evenly between the two possible proton directions.  For the central regions, the net luminosity is 2000 nb$^{-1}$  since both directions contribute, but for the forward (FW) and backward (BW) directions, the net luminosity is only 1000 nb$^{-1}$ each. B, M and K denote $10^9$, $10^6$ and $10^3$ respectively. 
Both the rates and cross sections include the relevant branching ratios.}
\begin{tabular}{|c|rrrrrrrr|}
\hline
\multicolumn{9}{|c|} {pPb - proton shine, $\gamma$A} \\
\hline
            &     $\sigma$              &  All & Ctl. 1 & Ctl. 2 & FW 1 & FW 2 & BW 1 & BW 2 \\
Meson & & Total & Total & Total & Total & Total  & Total & Total \\
\hline
$\rho\rightarrow\pi^+\pi^-$ & 531$\mu$b & 1.1 B& 83 M  &360 M & 20 M & 44 M & 56 M& 150 M \\
$\phi\rightarrow \mathrm{K}^+\mathrm{K}^-$ &  23 $\mu$b &46 M & 1.3 M & 8.0 M & 120 K & 1.7 M  & 210 K & 3.9 M \\
J$/\psi\rightarrow\mu^+\mu^-$ &  333 nb & 670 K  & 55 K & 290 K & 14K & 36 K  & 15 K & 41 K\\
$\psi\mathrm{(2S)}\rightarrow\mu^+\mu^-$   & 8.9 nb & 18 K & 1.5 K & 7.9 K & 380 & 990  & 380 &  1.0 K  \\
$\PGUP{1S}\rightarrow\mu^+\mu^-$ & 0.43 nb &  860 & 93  & 460 & 14 & 34 & 14 &  30 \\
\hline
\end{tabular}
\label{table:protonshine}
\end{table}

In case of coherent heavy vector meson production in UPC of lead nuclei, the expected experimental uncertainties are evaluated by the ALICE and CMS collaborations.
The vector meson cross section in \PbPb UPC can be expressed as a sum of two terms reflecting the fact that either of the colliding ions can serve as a photon source:
\begin{equation}
\sigma(y) = n(+y) \sigma_{\rm \gamma Pb}(+y) + n(-y) \sigma_{\rm \gamma Pb}(-y)
\end{equation}
The photoproduction cross sections $\sigma_{\rm \gamma Pb}(y)$ and $\sigma_{\rm \gamma Pb}(-y)$ are coupled and cannot be extracted unambiguously from the measured rapidity differential cross section. However, one can decouple them by measuring vector meson production in UPC with and without additional neutron activity in Zero Degree Calorimeters~\cite{Guzey:2013jaa}. Measurement of $\sigma_{\rm 0N0N}(y)$ (no neutrons on both sides) and $\sigma_{\rm 0NXN}(y)$ (at least one neutron on one of the sides) cross sections provides a system of two equations of two unknown photoproduction cross sections $\sigma_{\rm \gamma Pb}(\pm y)$:
\begin{eqnarray}
\sigma_{\rm 0N0N}(y) &=& n_{\rm 0N0N}(+y) \sigma_{\rm \gamma Pb}(+y) + n_{\rm 0N0N}(-y) \sigma_{\rm \gamma Pb}(-y), \\
\sigma_{\rm 0NXN}(y) &=& n_{\rm 0NXN}(+y) \sigma_{\rm \gamma Pb}(+y) + n_{\rm 0NXN}(-y) \sigma_{\rm \gamma Pb}(-y),
\end{eqnarray}
where $n_{\rm 0N0N}(\pm y)$ and $n_{\rm 0NXN}(\pm y)$ are corresponding photon fluxes, calculable with high accuracy.
Solutions of this system of equations can be used to extract photoproduction cross section $\sigma_{\rm \gamma Pb}$.

The expected experimental uncertainties are evaluated in terms of the nuclear suppression factor $R_{\rm Pb}$ which is defined as root square of the ratio of photoproduction cross section $\sigma_{\rm \gamma Pb}$ measured in \PbPb UPC and photoproduction cross section in the Impulse Approximation calculated as a reference photoproduction cross section off proton scaled by the integral over squared Pb form factor~\cite{Guzey:2013xba}:
\begin{equation}
R_{\rm Pb} (x) = \left(\frac{\sigma_{\rm \gamma Pb}(x)}{\sigma_{\rm IA}(x)}\right)^{1/2}, \qquad {\rm where}\qquad x  = \frac{m_{V}}{\sqrtsNN}\exp(-y).
\end{equation}
Under the assumption that the coherent photoproduction cross section is proportional to the squared gluon density at the scale $Q = m_{V}/2$, where $m_V$ is the mass of the produced vector meson, this nuclear suppression factor can be used to constrain nuclear shadowing at different scales $Q$. The theoretical discussion is given in the following Section~\ref{sec:vmtheo}.

\begin{figure}
\centering
\includegraphics[width=0.6\textwidth]{\main/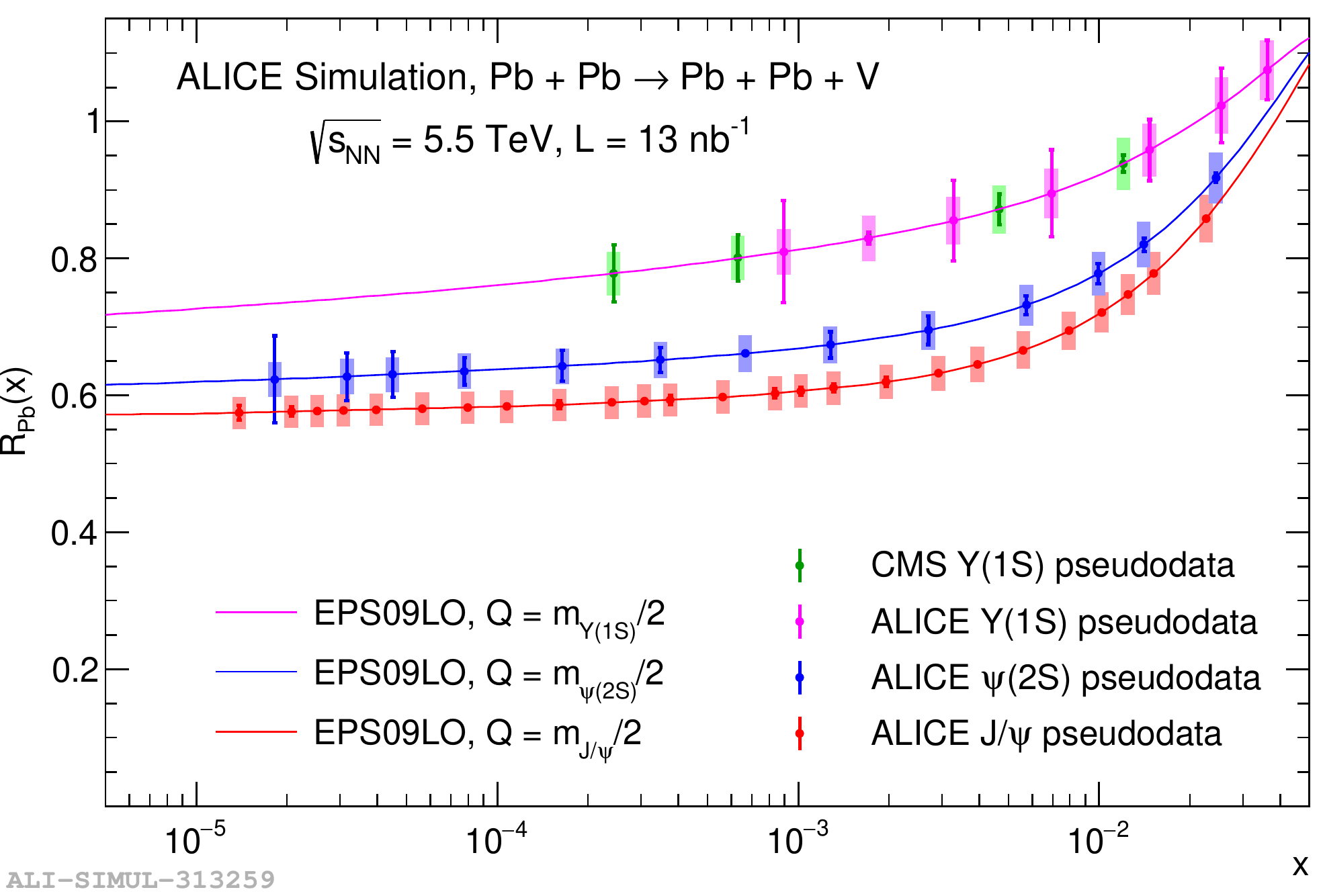}
\caption{Pseudodata projections for the nuclear suppression factor by ALICE~\cite{ALICE-PUBLIC-2019-001} and CMS measured with the photoproduction of three heavy vector mesons in \PbPb ultra-peripheral collisions are shown. The pseudodata points are derived from EPS09-based photoproduction cross section projections following the method described in Ref.~\cite{Guzey:2013xba}.}
\label{fig:r}
\end{figure}

The ALICE and CMS collaborations estimate that the uncertainties on luminosity (4\%), reference cross section (5\%) and photon flux (5\%) result in $\sim8$\% systematic uncertainty on the ratio $\sigma_{\rm \gamma Pb}(x)/\sigma_{\rm IA}(x)$ and $\sim 4$\% uncertainty on the nuclear suppression factor $R_{\rm Pb} (x)$. Detailed information about the uncertainty calculation can be found in existing publications~\cite{Abelev:2012ba,Abbas:2013oua,Adam:2015sia}. The pseudodata projections for the nuclear suppression factor are shown in Fig.~\ref{fig:r} at different scales corresponding to J$/\psi$, $\psi$(2S) and $\PGUP{1S}$ photoproduction measurements demonstrating that precision measurements with a range of different scales become available. 

\subsubsubsection{Coherent vector meson production off nuclei}
\label{sec:vmtheo}

Extensive data were produced in the previous LHC heavy ion Runs on coherent photoproduction  of $\rho$ mesons~\cite{Adam:2015gsa}, 
$\mathrm{J/}\psi$~\cite{Abbas:2013oua,Abelev:2012ba,Khachatryan:2016qhq} and to a lesser extent $\psi$(2S)~\cite{Adam:2015sia}: 
$\gamma+ A \to V +A$. Due to the presence of two photon sources, the $x$-range of these studies is largely  limited
to $x\ge m_V/\sqrt{s}$. For light mesons these measurements provide information on the pattern of interactions of extended pion-size mesons and such phenomena as shadowing and color fluctuations. For small dipoles like J$/\psi$ and $\PGUP{1S}$,
it provided information on the leading twist nuclear shadowing at moderate $Q^2$, which is difficult to obtain using other hard probes. The measured factor of $\sim 3$ reduction of the J$/\psi$ 
cross section compared to the $\gamma p$ case has significantly constrained the gluon distribution in leading twist approaches~\cite{Rebyakova:2011vf,Adeluyi:2012ph,Guzey:2013xba,Guzey:2013qza}. In the dipole model framework, e.g.,\,\cite{Goncalves:2011vf,Lappi:2013am},  have had a tendency to predict less supression than seen in the data (in this case, the shape and normalisation of the rapidity distribution rather 
strongly depends on the form of the dipole cross section and charmonium wave function). The gluon nuclear shadowing in coherent J/$\psi$ photoproduction in UPCs was
also studied in the $k_t$-factorization approach~\cite{Cisek:2012yt} in terms of
the unintegrated nuclear gluon distribution, which determines the initial
condition for the non-linear evolution equation.
In the case of $\rho$ meson production, shadowing is a factor of $\sim 2$ stronger~\cite{Frankfurt:2015cwa} than in the approach based on
the Glauber model and the vector meson dominance model.

The higher LHC luminosity and experimental upgrades will allow us to collect vastly improved samples of UPC events.   In particular, the planned ALICE continuous readout~\cite{Krivda:2017sto}, will eliminate many of the trigger-based constraints that have limited UPC data collection, allowing for high-efficiency collection of large samples of photoproduced light mesons.   The increases in sample sizes should be considerably larger than one would expect from merely scaling the luminosity.  

In order to conclude this section on the opportunities with vector meson production, we want to give a list of not yet exploited measurements that provide further insight into photonuclear interactions with heavy, light and multiple vector meson production:
\begin{itemize}
\item
Extend substantially the $x$ range for coherent J$/\psi$ photoproduction on nuclei
using information on the impact parameter distribution in peripheral and ultra-peripheral collisions provided by forward neutron production~\cite{Guzey:2013jaa}. The impact parameter distribution can be accessed in the context of UPCs by exploiting the properties of additional photon or hadronic interactions in addition to the photon that produces the vector meson. The rates for the combined processed can be found in~\cite{Baltz:2002pp} and the relationship between impact parameter and additional photon interactions is discussed in~\cite{Baur:2003ar}.
 The $x$-range can be also extended by using \pA~collisions to probe the nucleus. In the latter case, one would have to separate coherent J$/\psi$ production in $\gamma A$ and $\gamma p$  using a much more narrow $\pT$ distribution of J$/\psi $ produced in coherent $\gamma A$ scattering
 and very good $\pT$ resolution for the transverse momentum of the pair (LHCb). 
\item
Measure with high enough  statistics coherent $\PGUP{1S}$ production in $\gamma p$ and $\gamma A$
scattering to check the expectation of the 20\% reduction of the coherent cross section, which would allow one to probe gluon shadowing at a factor of $\sim 10 $ higher $Q^2$ than in J$/\psi$ production.
\item
Study coherent production of two pions with masses above 1~\UGeVcc to study an interplay of soft and hard dynamics as a function of $M_{\pi \pi}$ and $\pT(\pi)$.   
\item
Measure the production of heavier $2\pi$ \cite{Klein:2016dtn}, $4\pi$ and other resonances on ion targets, and search for the photoproduction of the observed exotic mesons.  
By using data from both proton targets (at HERA or the LHC) and ion targets, it is possible to separate the photon-meson coupling constant and the meson-nucleon interaction cross sections.  
\item
Study the photoproduction of multiple vector mesons by a single ion pair \cite{Klein:1999qj}.   Double photoproduction introduces many quantum correlations, including the possibility of observing stimulated decays of vector mesons.  Since the two photons share the same linear polarisation, it will be possible to study photoproduction with polarised photons. In the discussed process, the charged particles from the vector meson decays are aligned (as $cos^2(\theta)$) with the plane of the linear polarization.  If the two vector mesons are aligned along the same plane, then the planes formed by the decay particles will be correlated~\cite{Baur:2003ar}.
\end{itemize}

\subsubsubsection{Nuclear imaging with coherent photoproduction}
\label{sec:nuclimage}
In coherent photoproduction, production amplitudes from each individual scattering site add with a phase factor $\exp{i(\vec{x}\cdot\vec{k})}$, where $\vec{x}$ is the location within the nucleus and $\vec{k}$ is the momentum transfer from the nucleus to the vector meson.  So, one can Fourier transform the coherent photoproduction cross section, $\rmd\sigma_{\rm coherent}/\rmd t$ to find the location of the scattering sites within the nucleus.  This can provide information on the spatial dependence of nuclear shadowing, allowing us to compare shadowing in the centre of nuclei vs. shadowing in the periphery through the transverse profile $F(b)$.   Assuming azimuthal symmetry \cite{Diehl,Toll:2012mb}
\begin{equation}
F(b) \propto \frac{1}{2\pi} \int_0^\infty  \pT d\pT J_0(b p_T) \sqrt{\frac{\rmd\sigma_{\rm coherent}}{\rmd t}}.
\end{equation}
One complication is that it is necessary to flip the sign of $\sqrt{\rmd\sigma_{\rm coherent}/\rmd t}$ when crossing each diffractive minimum in the sample.

This calculation is  data-hungry, and is subject to a number of theoretical uncertainties.  It is also necessary to separate the cross section into its coherent and incoherent components.  Nevertheless, the STAR Collaboration applied it to $\rho^0$ photoproduction in UPCs \cite{Adamczyk:2017vfu}, finding a relatively boxy shape, inconsistent with a Woods-Saxon distribution, but seemingly consistent with expectations from nuclear shadowing.   A follow-on study explored the  $Q^2$~evolution of this transverse distribution by dividing the sample into three bins with different dipion mass \cite{Klein:2018grn}.  That study also considered some of the uncertainties inherent in the analysis, including the finite experimental reach in $\pT$ and the effects of the photon $\pT$, and the impact of the vector meson wave function. 

Studies at the LHC could avoid many of these issues, by triggering on exclusive vector mesons (STAR required that the vector mesons be accompanied by neutrons from mutual Coulomb excitation).  This would expand the $\pT$ range where a meaningful $\rmd \sigma_{\rm coherent}/\rmd t$ can be extracted, increasing the accuracy of the transform.  Also, because of the higher beam energies, the photon~$\pT$ is less important than it is at RHIC.    More importantly, LHC Run~3 and 4 could also extend this study to a wider variety of mesons, including the $\rho'$ (generically, $4\pi$ final states with a mass in the 1-2~\UGeVcc range) and the J$/\psi$.  The latter is of particular interest because it is heavy enough to probe gluon shadowing, rather than just nuclear shadowing.  

\subsubsubsection{Incoherent vector meson production off nuclei}

Incoherent diffractive processes provide information on nuclear dynamics,  which is complementary to the 
information one can obtain from 
coherent scattering. In particular, incoherent J$/\psi$ photoproduction  probes quantum fluctuations of the target gluon density~\cite{Frankfurt:2008vi,Mantysaari:2016ykx,Mantysaari:2016jaz,Mantysaari:2017dwh,Cepila:2016uku,Cepila:2017nef}.
The corresponding cross section can be measured in a much larger range of $W_{\gamma N}$ than in the coherent case. This is possible since the activity in the nucleus fragmentation region, for example, neutrons in a ZDC can be used to determine which of two nuclei was a source of photons almost in each event.

One can distinguish two contributions to incoherent diffraction: 
quasielastic, when the nucleus involved in the strong interaction breaks down into nucleons and nuclear fragments, 
and inelastic, when hadrons are produced in the nucleus fragmentation  region.
At small $t$ the second  mechanism gives a $\sim 20\% $ contribution to the incoherent cross section~\cite{Frankfurt:2008vi,Guzey:2018tlk}. However, since the $t$-dependence of the inelastic  mechanism is weaker, it dominates the nuclear incoherent cross section for $|t| \ge 0.5$~\UGeVcc. 

While it is generally understood and accepted that J/$\psi$ photoproduction with target dissociation is sensitive to fluctuations
of the gluon density of the target, practical realisations of this notion require modeling. Notably, 
proton size fluctuations at $t=0$~\cite{Frankfurt:2008vi} should be contrasted with proton shape fluctuations~\cite{Mantysaari:2016ykx,Mantysaari:2016jaz,Mantysaari:2017dwh}, which in turn
can be refined by including fluctuations of number of subnucleonic degrees of freedoms 
representing regions of high-gluon density, so-called hot spots~\cite{Cepila:2016uku,Cepila:2017nef}. The latter two approaches 
are assumed to be valid in an entire range of $|t|$.
Hence, studies of incoherent diffraction
in $\gamma p$ scattering via \pA UPCs and $\gamma A$ scattering via \AOnA UPCs would allow one to map variations of the gluon fluctuations with energy ($x$) including a possible approach to the black disk regime, where the 
fluctuations are strongly suppressed~\cite{Cepila:2016uku}.

For very large $|t| \ge 1 - 2$~\UGeVcc one enters the regime of pQCD, which corresponds to elastic scattering of small-size dipoles
off individual small-$x$ partons of the nuclear target~\cite{Frankfurt:2008et,Frankfurt:2008er}.

Note also that for the proton target, one can use the process  $\gamma + p \to VM (\mathrm{J}/\psi) + gap + Y$  at   
$-t \ge \mbox{few \UGeVcc}$ to study the perturbative Pomeron. 
In the kinematics, where $\Delta y$ is fixed, 
$\rmd \sigma /\rmd y_{VM} \propto (y_{VM}- \Delta y)^{2\alpha_{Pomeron}-2}$. In BFKL one expects
$2\alpha_{Pomeron}-2\sim 0.4$ and, hence, a strong rapidity dependence of the corresponding cross section. A larger acceptance of the ATLAS and CMS detectors should allow for a more effective study of these processes than at HERA.

By studying the $t$-dependence and activity in the nucleus 
fragmentation region it would be possible to separate the two mechanisms of incoherent nuclear scattering as a function of $t$.
For small $|t| \le 0.3 - 0.5$~\UGeV$^2$, one can calculate nuclear shadowing for both mechanisms within the leading 
twist shadowing framework~\cite{Guzey:2013jaa}. The shadowing turns out to be large and sensitive to the details of the leading twist shadowing dynamics.  At large $|t|$, one can study the $A$ dependence of the discussed reaction for different rapidity gap  
intervals to track propagation of a small dipole through the nuclear medium.
 By changing $\Delta y$ it would be possible to vary  strongly the relative role of the initial and final state interaction.

\subsubsection{Inclusive and diffractive dijet production in UPC}
\label{sec:upcdijets}

Ultra-peripheral heavy-ion collisions provide an opportunity to study nuclear modifications of the PDFs in clean photon-nucleus interactions. One possible observable is dijet production as suggested in Ref.~\cite{Strikman:2005yv}. Compared to the dijet production in \pPb collisions the photo-nuclear events have less underlying event activity since multiparton interactions are significantly suppressed. This enables jet reconstruction at lower transverse momenta allowing to study nPDFs at smaller scales $Q^2$ and $x$ where the current PDF uncertainties are more pronounced. As the virtuality of the photons emitted by the nucleus is negligible, there are two components that need to be taken into account: the photons may interact as unresolved particles or the quasi-real photons may fluctuate into a hadronic state described with photon PDFs. The relative contribution of the direct and resolved components depends on the kinematics of the final state jets. Hence, the uncertainty related to weakly-constrained photon PDFs can be reduced by focusing on the region where direct processes dominate the dijet production.

\begin{figure}
\includegraphics[width=0.49\textwidth]{\main/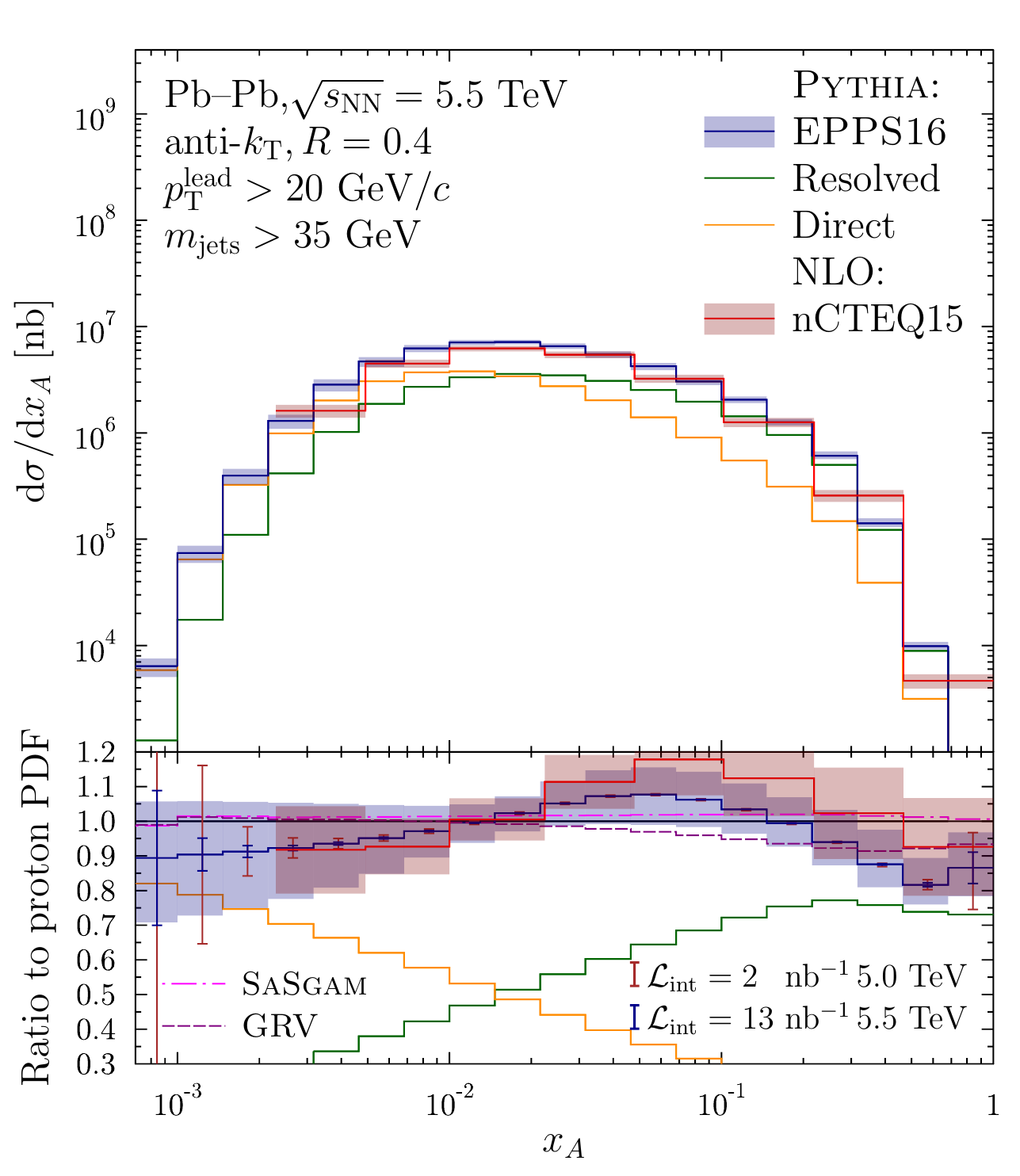}
\includegraphics[width=0.49\textwidth]{\main/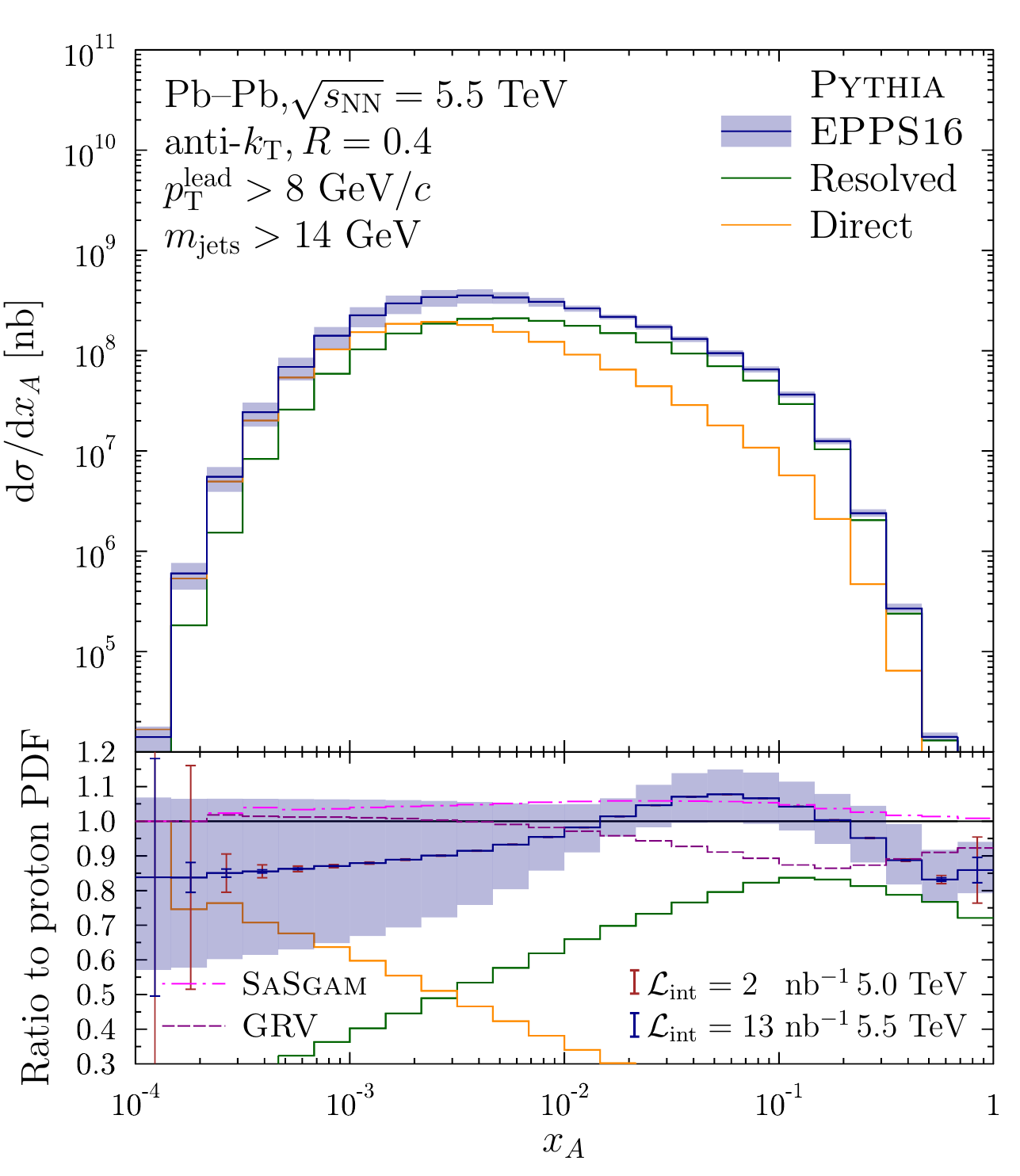}
\caption{Photo-nuclear dijet cross sections in ultra-peripheral \PbPb collisions at $\sqrtsNN = 5.5\UTeV$ with leading jet $\pT$ cut of $20\UGeVc$ (left) and $8\UGeVc$ (right). Results based on \textsc{Pythia} simulations are calculated with EPPS16 nuclear modification (blue) and the contributions from resolved (green) and direct (orange) photons are separately shown. Ratio plots show also results with different photon PDF sets and the expected statistical uncertainties corresponding to the LHC (brown) and the Run~3 and and Run~4 (dark blue) luminosities. Corresponding results based on NLO calculations for \PbPb collisions at $\sqrtsNN = 5.02\UTeV$ with nCTEQ15 nPDFs \cite{Guzey:2018dlm} (red) are shown in case leading jet $\pT$ cut of $20\UGeVc$.}
\label{fig:UPCdijetNPDF}
\end{figure}

Here, the photoproduction framework is applied which has been recently implemented into the \textsc{Pythia 8} Monte-Carlo event generator~\cite{Sjostrand:2014zea} and validated against HERA data~\cite{Helenius:2018bai}, to study the potential of the Run~3 and Run~4 program to constrain nPDFs using photo-nuclear dijets. The relevant part of the photon flux is obtained by integrating the impact-parameter dependent flux from $b_{\mathrm{min}}=2 R_{\mathrm{Pb}} \approx 13.27\Ufm$. Two different jet kinematics are considered, one corresponding to the preliminary ATLAS measurement~\cite{ATLAS:2017kwa} with $\pT^{\mathrm{lead}}>20\UGeVc$ and $m_{\mathrm{jets}} > 35$~\UGeVcc and one similar to HERA dijet photoproduction data~\cite{Derrick:1996zy,Adloff:2000bs} with $\pT^{\mathrm{lead}}>8\UGeVc$ and $m_{\mathrm{jets}} > 14$~\UGeVcc. In both cases the jets were reconstructed from the generated events using the anti-$k_{\mathrm{T}}$ algorithm with $R = 0.4$ implemented in the \textsc{FastJet} package~\cite{Cacciari:2011ma}. The differential cross sections are shown as a function of $x_A$ in Fig.~\ref{fig:UPCdijetNPDF} using \textsc{NNPDF2.3lo} proton PDFs~\cite{Ball:2012cx} with and without EPPS16 nuclear modifications~\cite{Eskola:2016oht}. 
The kinematic variables used in the ATLAS study \cite{ATLAS:2017kwa} are defined as
\begin{eqnarray}
x_{A} & =& \frac{m_{\rm jets}}{\sqrtsNN}e^{-y_{\rm jets}} \,, \quad \quad z_{\gamma}=\frac{m_{\rm jets}}{\sqrtsNN}e^{y_{\rm jets}} \,, \nonumber\\
m_{\rm jets} &=& \left[\left(\sum_i E_i\right)^2-\left|\sum_i \vec{p}_i \right|^2 \right]^{1/2} \,, \quad \quad
y_{\rm jets}=\frac{1}{2} \ln \left(\frac{\sum_i E_i+p_{i,z}}{\sum_i E_i-p_{i,z}} \right) \,.
\label{eq:estimators2}
\end{eqnarray}
where the index $i$ runs over all accepted jets; $E_i$ and $\vec{p}_i$ denote the jet energy and momentum, respectively.
Note that in a leading-order (LO) parton-level calculation, the definitions of $x_A$ and $z_{\gamma}$ would exactly match the momentum fractions probed in the PDFs of the nucleus and the photon. The parton-shower emissions and MPIs, and NLO corrections considered below, smear this connection but $x_A$ and $z_{\gamma}$ do serve as rather precise hadron-level estimators for the momentum fractions \cite{Helenius:2018mhx}.

The uncertainty bands are derived from the EPPS16 error sets and reflect the uncertainties in the current nPDF analyses which are compared to the expected statistical uncertainties of the data in the ratio. Also the contributions from direct and resolved processes are separately plotted. Furthermore, results with the default CJKL photon PDFs~\cite{Cornet:2002iy} are compared to GRV~\cite{Gluck:1991jc} and \textsc{SaSgam}~\cite{Schuler:1995fk} analyses to study the underlying photon PDF uncertainty.

As shown in Figure~\ref{fig:UPCdijetNPDF}(left), the contribution from resolved processes becomes dominant around $x_A > 0.02$ for $\pT^{\mathrm{lead}}>20\UGeVc$. This leads to a more pronounced dependence on the photon PDFs in this region, partly hindering the use of the data from this region in a global nPDF analysis. However, at small-$x$ region, where the nPDF uncertainties are currently large and the dijets in \pPb do not provide additional constraints, the direct processes dominate the dijet production and the dependence on the photon PDFs is negligible. The dijet cross sections fall off rapidly at small-$x_A$ region which increases the expected statistical uncertainty limiting the small-$x_A$ reach of the observable. With an integrated luminosity of $\Lint=2$~nb$^{-1}$ in \PbPb collisions and jet kinematics of the ATLAS preliminary study the expected statistical uncertainties become significant at $x_A \lesssim 2 \cdot 10^{-3}$. The increased luminosity of the LHC Run~3 and 4 increases the potential small-$x_A$ reach only slightly but in the region where nPDF constraints are currently sparse. An effective way to extend the small-$x_A$ reach is to consider jets with lower $\pT$ as demonstrated in Figure~\ref{fig:UPCdijetNPDF}(right).
With a cut of $\pT^{\mathrm{lead}}>8\UGeVc$ and an integrated luminosity $\Lint=\invnb{13}$ in Runs 3 and 4 it is possible to obtain nPDF constraints down to $x_A \approx 10^{-4}$.  Also, the small-$x$ nPDF uncertainties are more pronounced with a lower $\pT^{\mathrm{lead}}$-cut since the nuclei are probed at smaller scales. The theoretical uncertainty related to the limited precision of the photon PDFs could be reduced by performing a similar measurement in \pPb collisions where the photon flux would be dominantly provided by the Pb ion and the jets produced by $\gamma$-p system without any nuclear modifications. This measurement would constrain also the uncertainty related to the impact-parameter rejection that removes the events with hadronic interactions. However, as the minimum allowed impact parameter is smaller in case of \pPb compared to \PbPb and the spectrum of photons is correlated with the impact parameter cut, the kinematics would not be fully comparable preventing a full calibration of the photon flux. As the nPDFs mainly vary the shape of the $x_A$ distributions, part of the theoretical and experimental uncertainties could also be reduced by considering $x_A$ distributions normalized with the integrated cross section.


Inclusive dijet photoproduction in UPCs can also be evaluated to the next-to-leading order (NLO) accuracy of perturbative QCD. The corresponding cross section can be written in the following form~\cite{Klasen:2002xb}:
 \begin{eqnarray}
 &\rmd \sigma(AA \to A+{\rm 2 jets}+X) = \\
& \sum_{a,b} \int \rmd y \int \rmd x_{\gamma} \int \rmd x_b f_{\gamma/A}(y)f_{a/\gamma}(x_{\gamma},\mu^2)f_{b/A}(x_A,\mu^2) \rmd \hat{\sigma}(ab \to {\rm jets}) \,,
\label{eq:cs_jets}
\end{eqnarray}
where $a,b$ are parton flavours; $f_{\gamma/A}(y)$ is the flux of equivalent photons emitted by one ion,
which depends on the photon light-cone momentum fraction $y$;
$f_{a/\gamma}(x_{\gamma},\mu^2)$ is the PDF of the photon, which depends on the momentum fraction $x_{\gamma}$ 
and the factorisation scale $\mu$; $f_{b/A}(x_A,\mu^2)$ is the nPDF with $x_A$ being the corresponding parton momentum fraction;
$\rmd \hat{\sigma}(ab \to {\rm jets})$ is the elementary cross section for production of two- and three-parton final states emerging as jets
 in the interaction of partons $a$ and $b$. The sum over $a$ involves quarks and gluons for the resolved photon contribution 
 and the photon for the direct photon contribution dominating at $x_{\gamma} \approx 1$.
 
 Figure~\ref{fig:UPCdijetNPDF} (left)  presents predictions of Eq.~(\ref{eq:cs_jets}) for the cross section of dijet photoproduction in \PbPb UPCs at $\sqrtsNN=5.02$~TeV in the ATLAS kinematics as a function of $x_A$. The red solid lines and the associated shaded band correspond to the central fit of nCTEQ15 nPDFs and their uncertainty, respectively.
The top panel of this figure demonstrates that NLO pQCD correctly reproduces the shape and, at least 
semi-quantitatively, the normalisation of the preliminary ATLAS data.  The lower panel of Fig.~\ref{fig:UPCdijetNPDF} shows 
the ratio of the curves from the upper panel to the result of the calculation, where nCTEQ15 nPDFs are substituted by 
free proton and neutron PDFs. One can see from this panel that the central value of the ratio of the two cross sections reveals the expected trend of 
nuclear modifications of nPDFs: $\sim 10\%$ shadowing for small $x_A < 0.01$, which is followed by $ \sim 20\%$ antishadowing (enhancement)
around $x=0.1$ and then $\sim 10\%$ suppression for $x_A > 0.3$.
Note that since the uncertainties of nCTEQ15 nPDFs are of the same magnitude as the effect of nuclear modifications, 
inclusion of this dijet data if global QCD fits of nPDFs should in principle reduce the existing uncertainty.

It is also important to study diffractive dijet photoproduction
in UPCs in the reaction $A + A \to A + jet1 +
jet2 + X + A$. NLO pQCD predictions for the cross section of this process in \pp, \pA, and \AOnA UPCs in the LHC kinematics were made in~\cite{Guzey:2016tek}. It was shown that studies of this process on nuclei may shed some light on the mechanism of QCD factorisation breaking in diffractive photoproduction and, for the first time, give access to nuclear diffractive PDFs and test their models. 


\subsubsection{Photoproduction of heavy quarks}
\label{sec:upchqdijets}

Photoproduction of open charm and bottom is a direct probe of the gluon content of the target nucleus \cite{Klein:2002wm,Goncalves:2017zdx}.  The lowest order process, $\gamma + gluon\rightarrow c \overline c$ (or $b \overline b$) dominates over higher order, resolved processes, in which the photon radiates before it interacts with the target gluon~\cite{Klein:2002wm}.  This process type is a subset of flavour untagged dijets, discussed above.  However, open charm and bottom offer some advantages, including very high rates. Based on the leading order calculations in Ref.~\cite{Klein:2002wm}
with the EKS nPDF~\cite{Eskola:1998iy}, a total number of 22 billion (10 million) 
$c \overline c$ ($b \overline b$)  pairs are produced in $\gamma$Pb interactions in a  13~nb$^{-1}$ PbPb data set at $\sqrt{s_{NN}}$=5.5 TeV. In pPb collisions at $\sqrt{s_{NN}}$ =8.8 TeV with a luminosity of 2~pb$^{-1}$, 18 billion (100 million) $c \overline c$ ($b \overline b$)  pairs are produced. The measurement of both of these processes should be well feasible despite the small experimental acceptances and branching ratios. 

Vertex detectors can detect separated vertices from charm production if the charm is moderately relativistic \cite{Bala:2012vg}, so it should be possible to study pairs with an invariant mass of $M_{c\overline c} \approx 4m_c \approx 6$~\UGeVcc.  This will enable to study gluon distributions down to Bjorken$-x$ values of around: $x \approx M_{Q\overline Q}/(4\gamma m_p) \exp(y)$, where the relativistic boost of the nucleus $\gamma$, the rapidity $y$ of the $Q\overline Q$ and the proton mass $m_p$ appear. The corresponding scale is $Q^2 \approx M_{Q\overline Q}$.  Assuming  scales down to $M_{c\overline c}\approx 6$~\UGeVcc, a Bjorken$-x$ lower by a factor 6 can be reached compared to the kinematic selection in the preliminary inclusive dijet ATLAS analysis~\cite{ATLAS:2017kwa} for which the Bjorken-$x$ reach is depicted in Fig.~\ref{fig:UPCdijetNPDF}(left).  By comparing results from proton targets (in \pA collisions) and heavy-ion targets (in \AOnA collisions), it is possible to make a direct  measurement of nuclear shadowing.  

The clear experimental signature for this process consists of a pair of separated vertices each corresponding to a D-meson decay  in an event with one large rapidity gap. The ion moving in the same direction as the rapidity gap should remain intact. The full reconstruction of both D-mesons, the cleanest channel, the  in specific decay channels requires large large event statistics and might be complemented by an analysis with one D-meson decaying semileptonically.  However, the charm production rates are high enough that the low efficiency should be affordable for a precise measurement.

In addition to charm and bottom, it may also be possible to study the photoproduction of top \cite{Klein:2000dk,Goncalves:2013oga}.   The rates are small for lead-lead collisions at the LHC (7 pairs in 13~nb$^{-1}$), but for  pPb collisions at $\sqrt{s_{NN}}$ =8.8 TeV, the calculation of Ref.~\cite{Klein:2000dk} finds that an integrated luminosity of 2~pb$^{-1}$ corresponds to a production of 110 pairs.  These pairs are mostly at fairly central rapidities and the large top semileptonic branching ratio should provide a clear experimental tag.  This process could provide a separate probe of gluon distributions at larger $x$, and very high $Q^2$.  It would also allow a direct measurement of the electric charge of the top quark. 


\subsection{The physics of inelastic p-Pb~collisions}
\label{sec:pPbex}

\subsubsection{Experimental overview} 

\begin{figure}
\centering
\includegraphics[width=0.42\textwidth]{\main/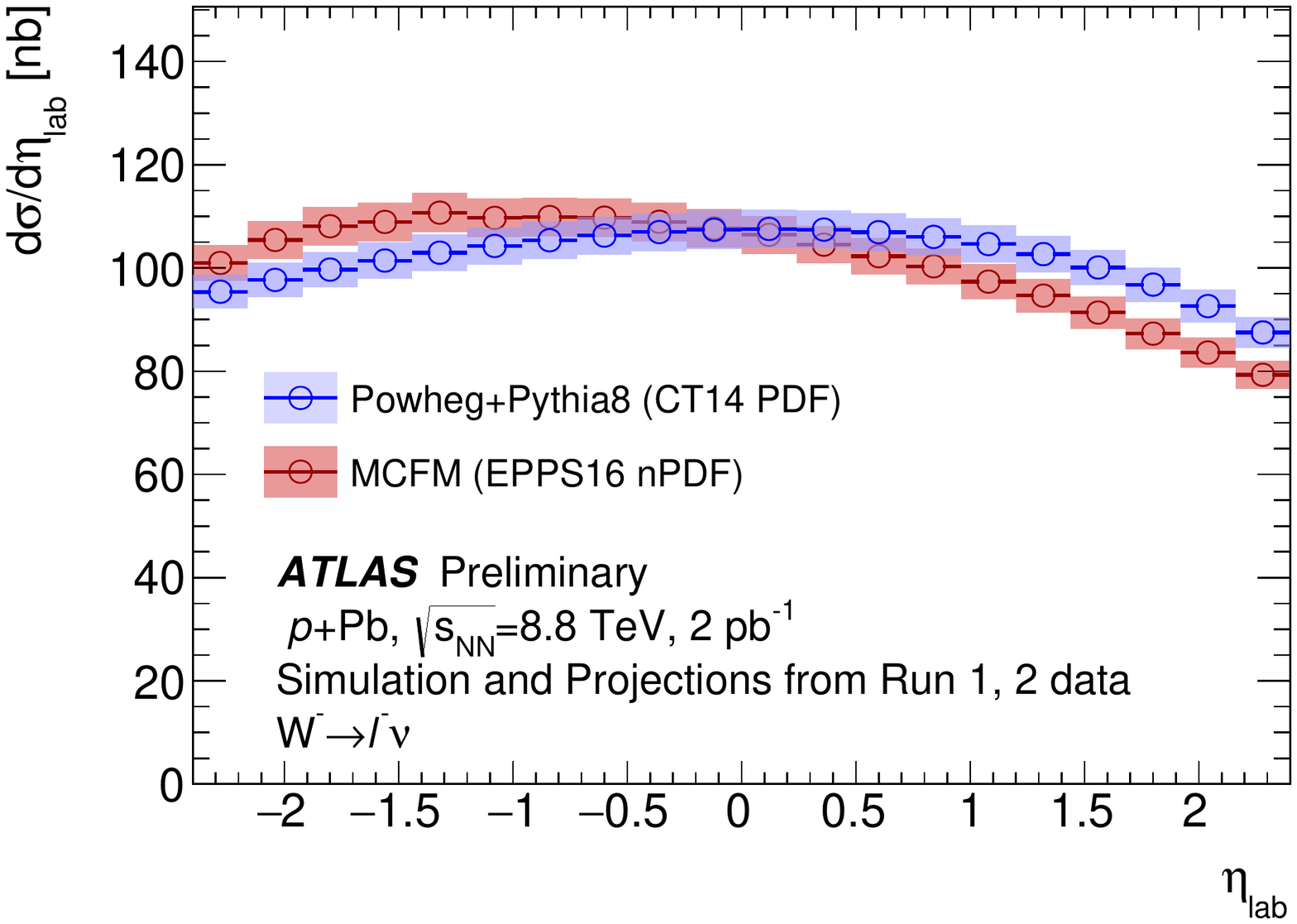}
\includegraphics[width=0.42\textwidth]{\main/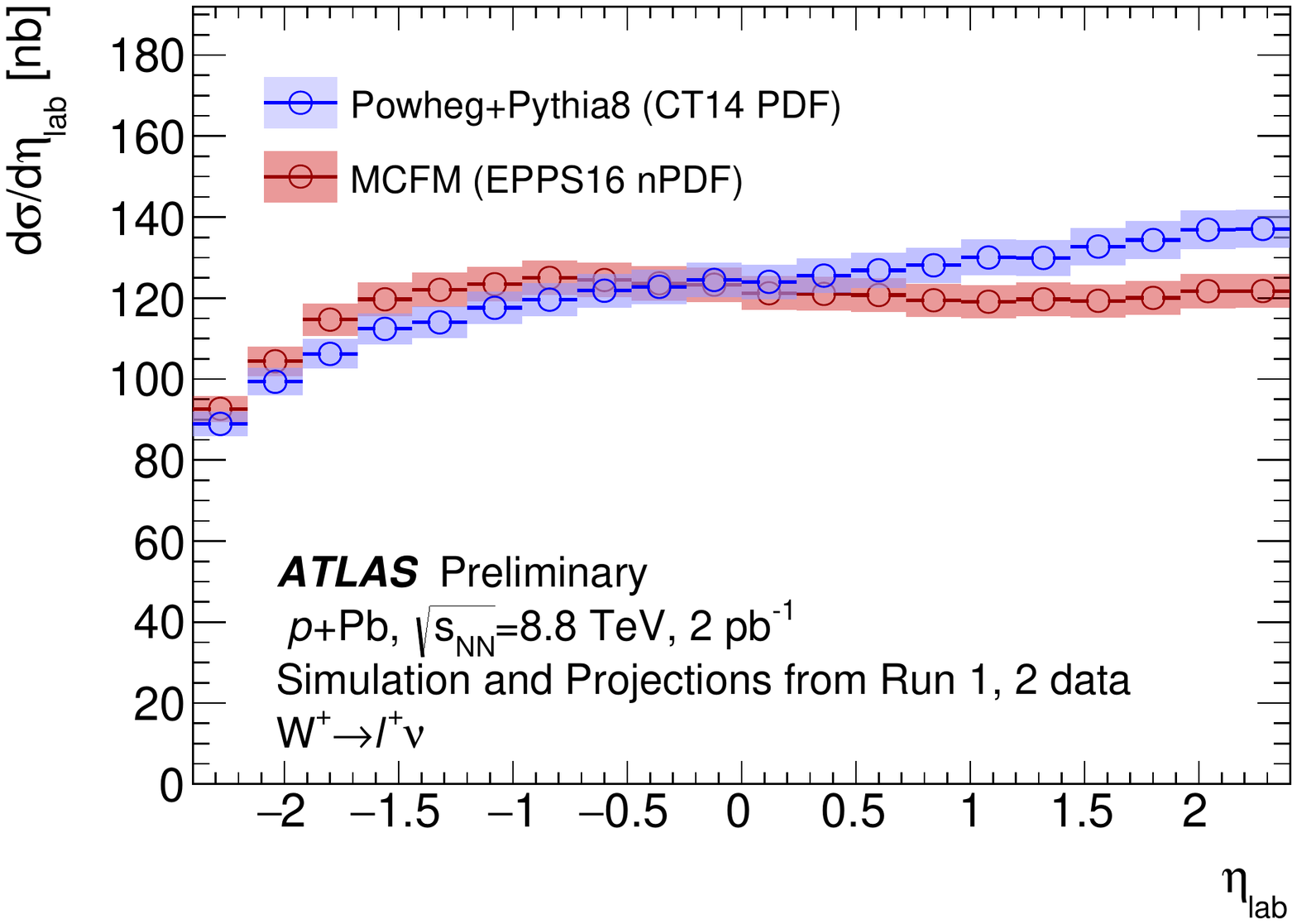}\\
\includegraphics[width=0.42\textwidth]{\main/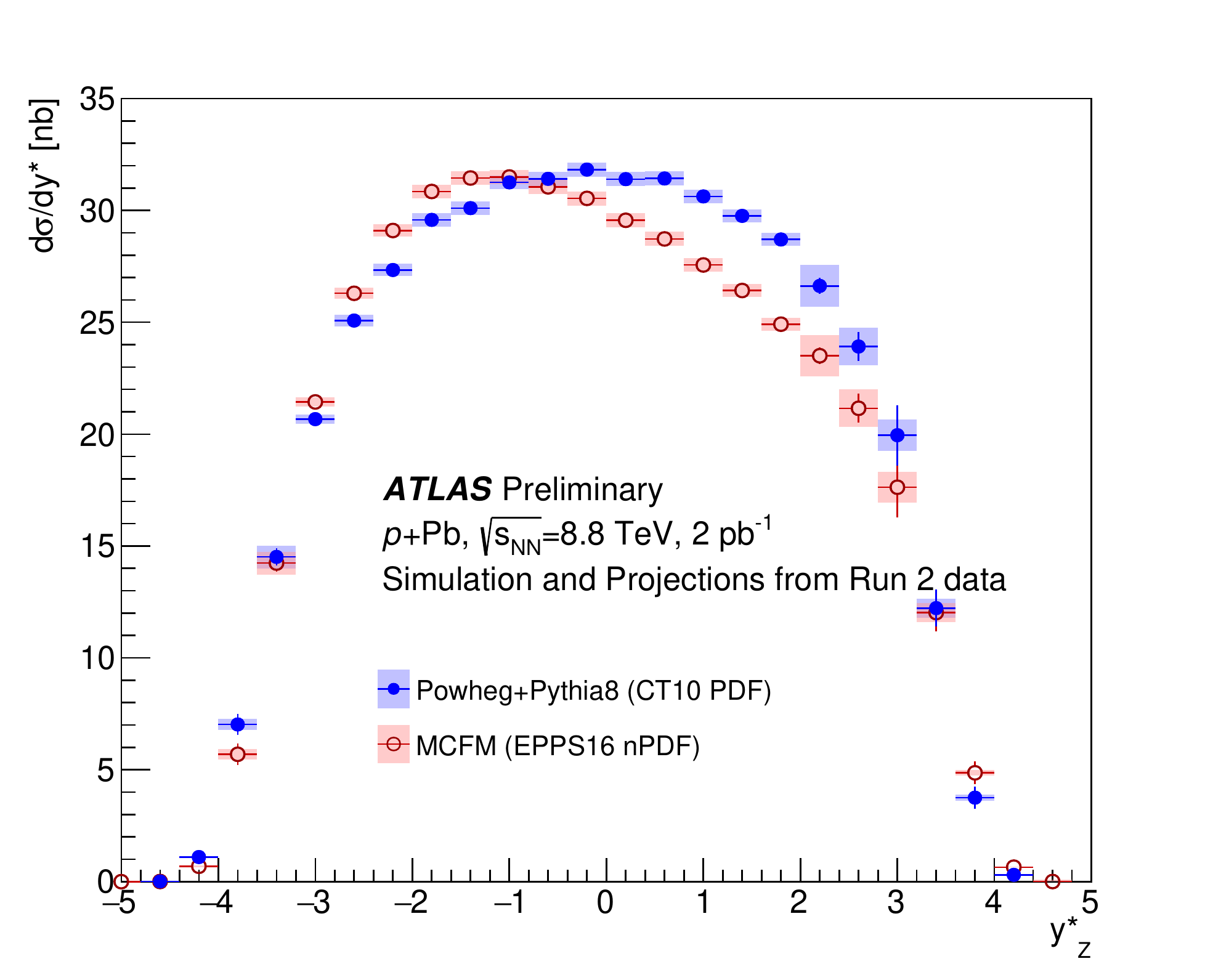}
\includegraphics[width=0.42\textwidth]{\main/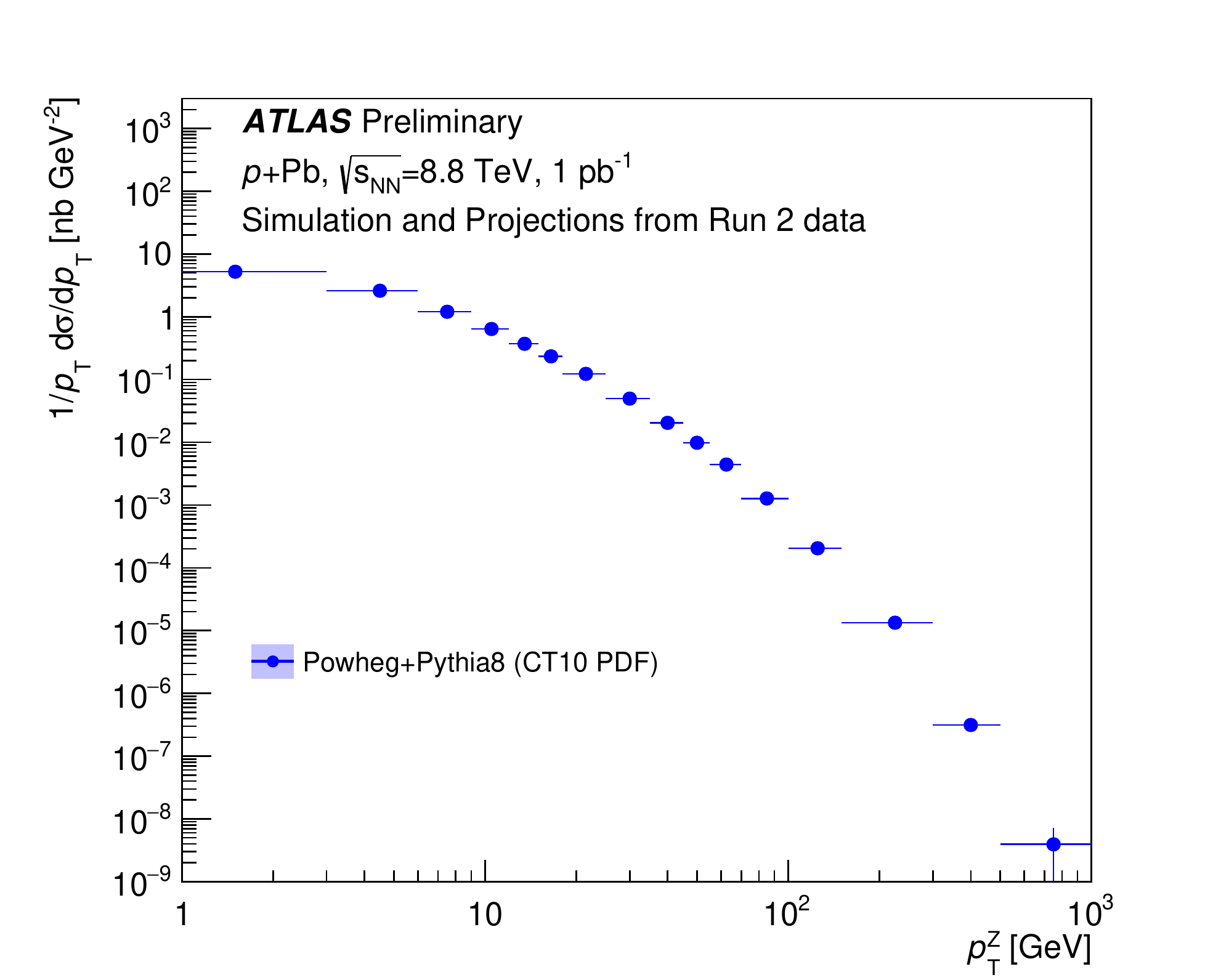}
\caption{Top row: Fiducial cross sections for $W^+$ (left) and $W^-$ (right) boson production in \pPb collisions at $\sqrtsNN=8.8$~TeV differential in the charged lepton pseudorapidity measured in the laboratory frame $\eta_{\mathrm{lab}}$~\cite{ATL-PHYS-PUB-2018-039}. The cross sections are projected with nuclear effects described by the EPPS16 nPDF set and without any nuclear effects. The boxes represent the projected total uncertainties (quadratic sum of statistical and systematic uncertainties), while vertical bars represent statistical uncertainties (smaller than the marker size).
Botton row: $Z$ boson rapidity (left) and transverse momentum (right) differential cross sections~\cite{ATL-PHYS-PUB-2018-039}.}
\label{fig:ATLASWZxsection}
\includegraphics[width=0.42\textwidth]{\main/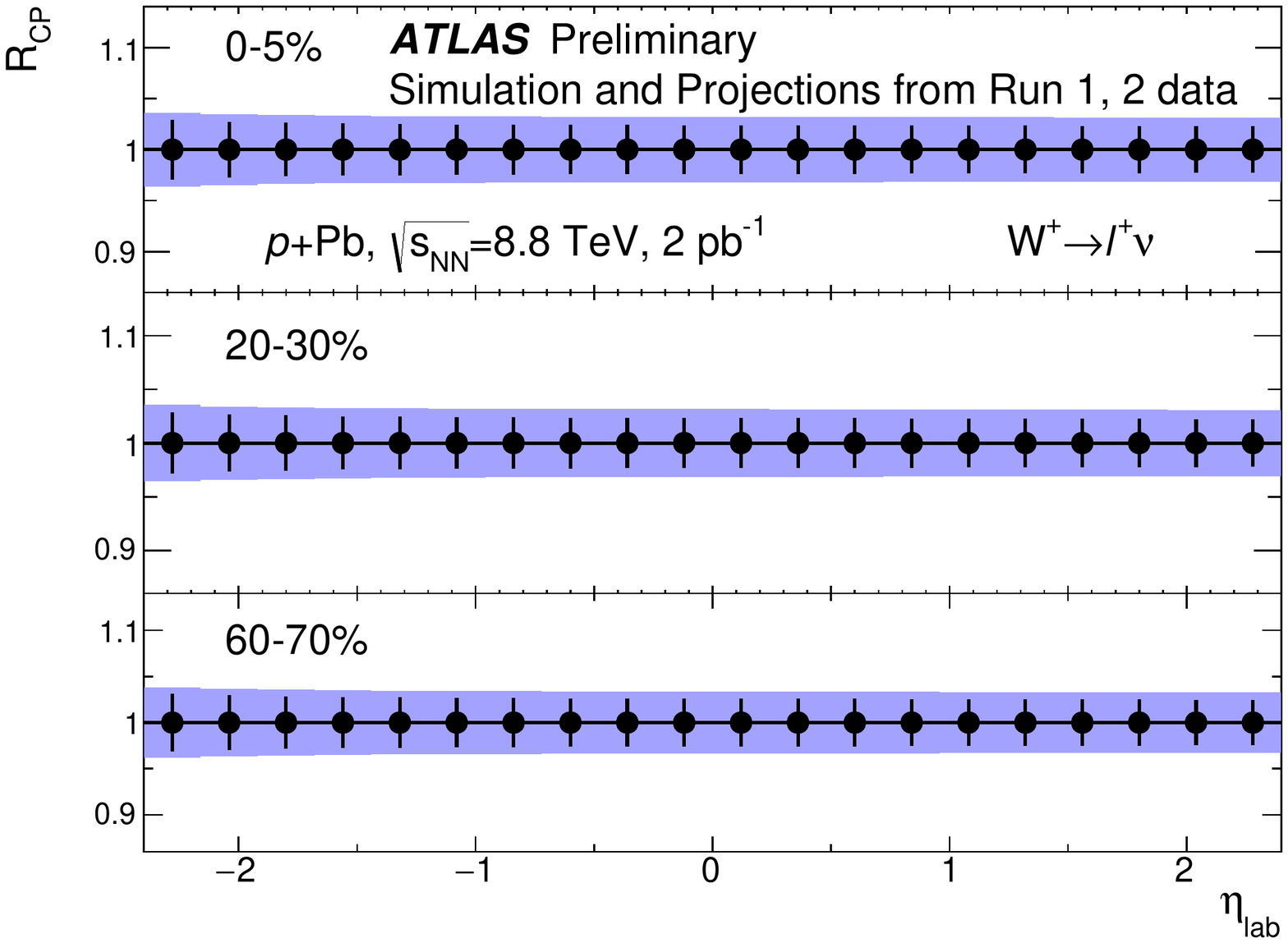}
\includegraphics[width=0.42\textwidth]{\main/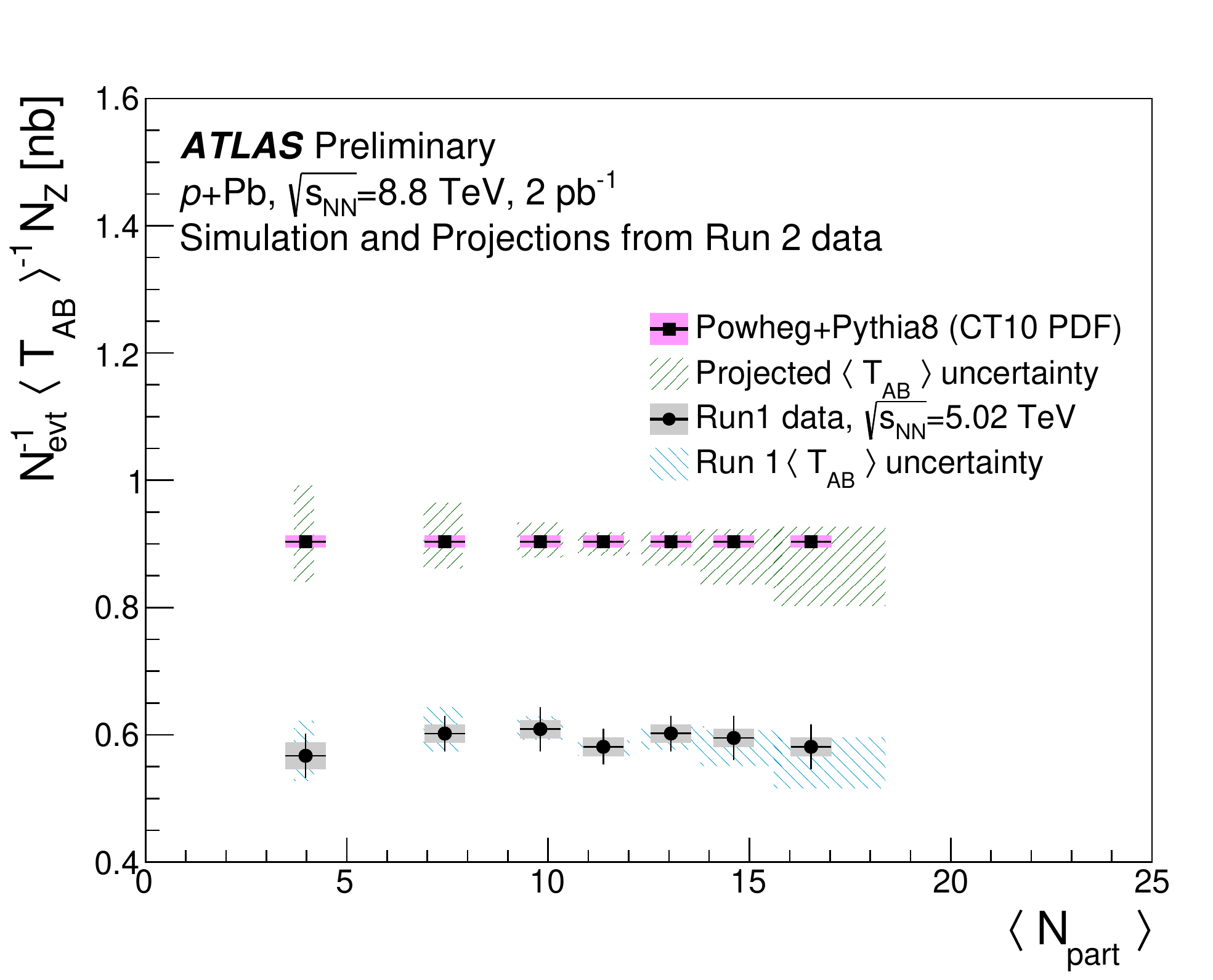}
\caption{Left: Uncertainties on measurments of the $W^+$ (left) and $W^-$ (right) boson yield ratio of central to peripheral collisions $R_{\mathrm{CP}}$.  The boxes represent the projected total uncertainties (quadratic sum of statistical and systematic uncertainties), while vertical bars represent statistical uncertainties~\cite{ATL-PHYS-PUB-2018-039}.  Right: Projected $Z$ boson yields scaled by T$_{\mathrm{AB}}$ as a function of the number of nucleon participants~\cite{ATL-PHYS-PUB-2018-039}.  Statistical uncertainties are everywhere smaller than the marker size, and projected systematic uncertainties (green) are described in the text.  The projections are compared with a previous run 1 analysis~\cite{Aad:2015gta}.
}
\label{fig:ATLASWZcent} 
\end{figure}

\begin{figure}[tb]
\centering
\includegraphics[width=0.44\textwidth]{\main/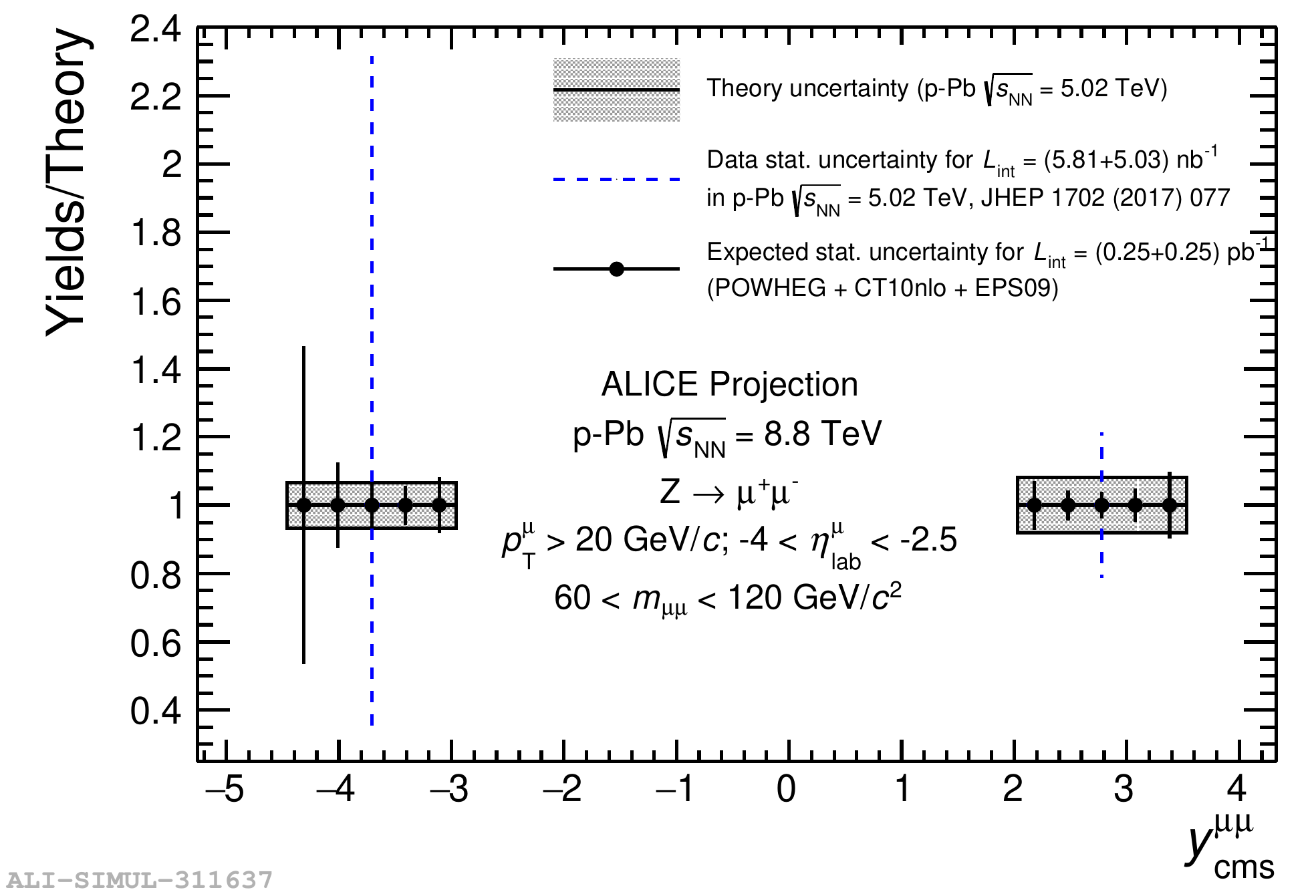}
\includegraphics[width=0.45\textwidth]{\main/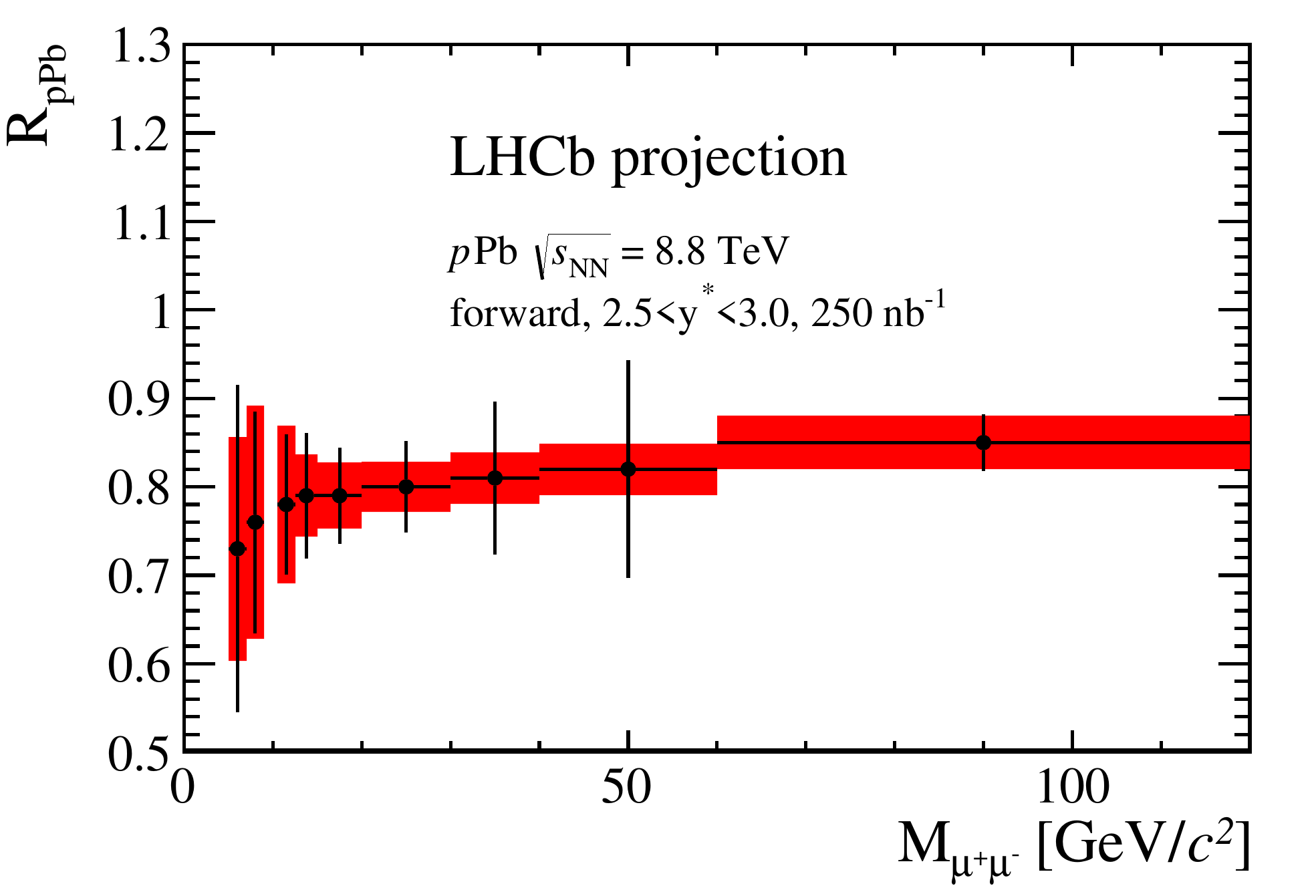}
\caption{Left: Expected statistical significance of the Z boson rapidity distribution for \pPb collisions at $\sqrtsNN = 8.8$~TeV in Run 3-4 assuming an integrated luminosity of 0.25/pb (0.25/pb) obtained with the proton (Pb ion) going toward the ALICE muon spectrometer~\cite{ALICE-PUBLIC-2019-001}. 
The results are compared with the statistical significance of the published data collected in \pPb collisions in 2013. Right: $R_{\rm pPb}$ of inclusive Drell-Yan production in $2.5<y_{cms}<3.0$, one of five rapidity bins accessible with LHCb. The projection is detailed in Ref.~\cite{LHCb-CONF-2018-005}.  
}
\label{fig:plotDY}
\end{figure}

\begin{figure} 
\centering
\includegraphics[width=0.7\textwidth]{\main/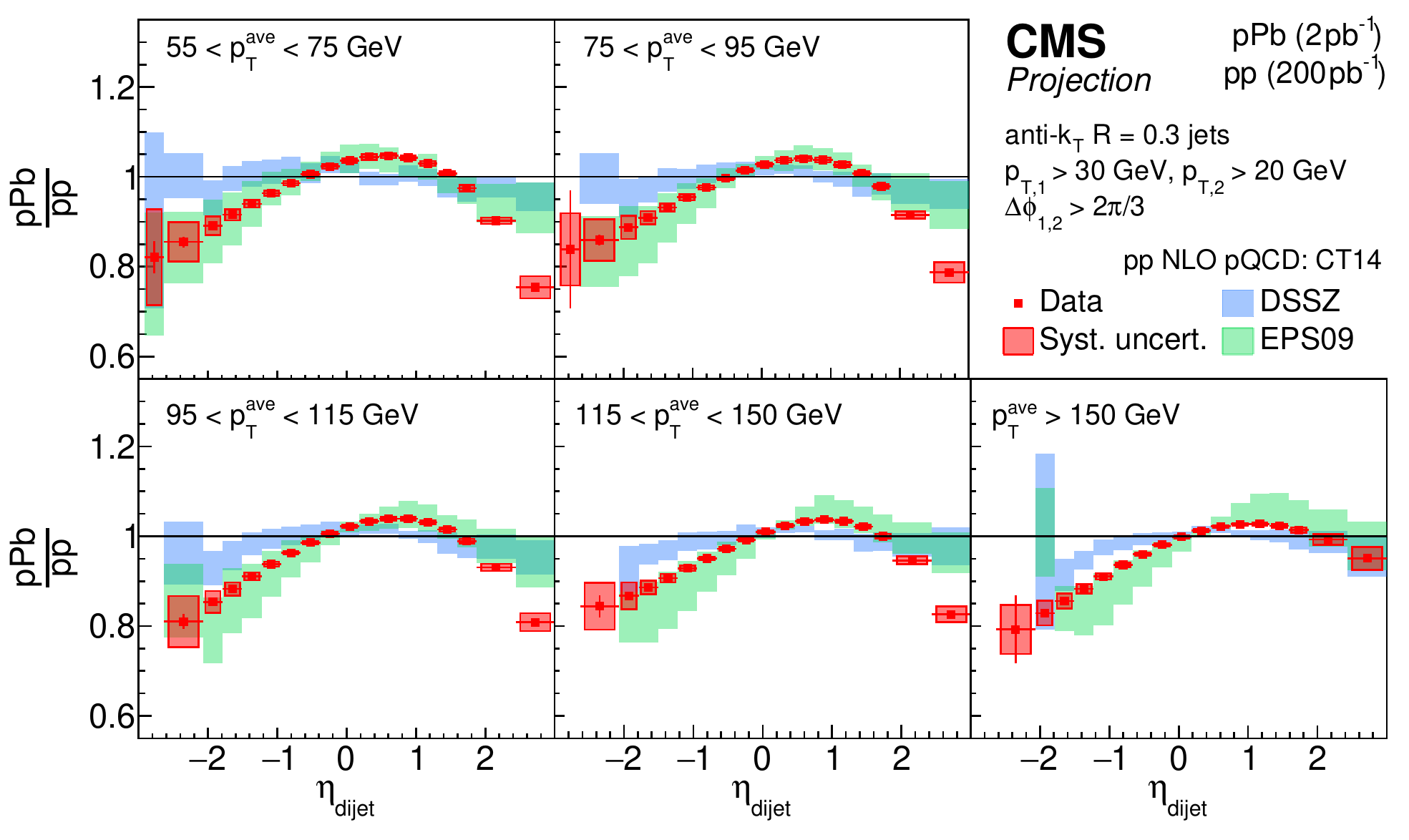}
\caption{
Projected dijet $\eta$ measurements in \pPb collisions with combined Run 3 and Run 4 data~\cite{CMS-PAS-FTR-18-027}.
\label{fig:dijet-eta-projection-cms}
}
\end{figure}

\begin{figure}
    \includegraphics[width=0.3\textwidth]{\main/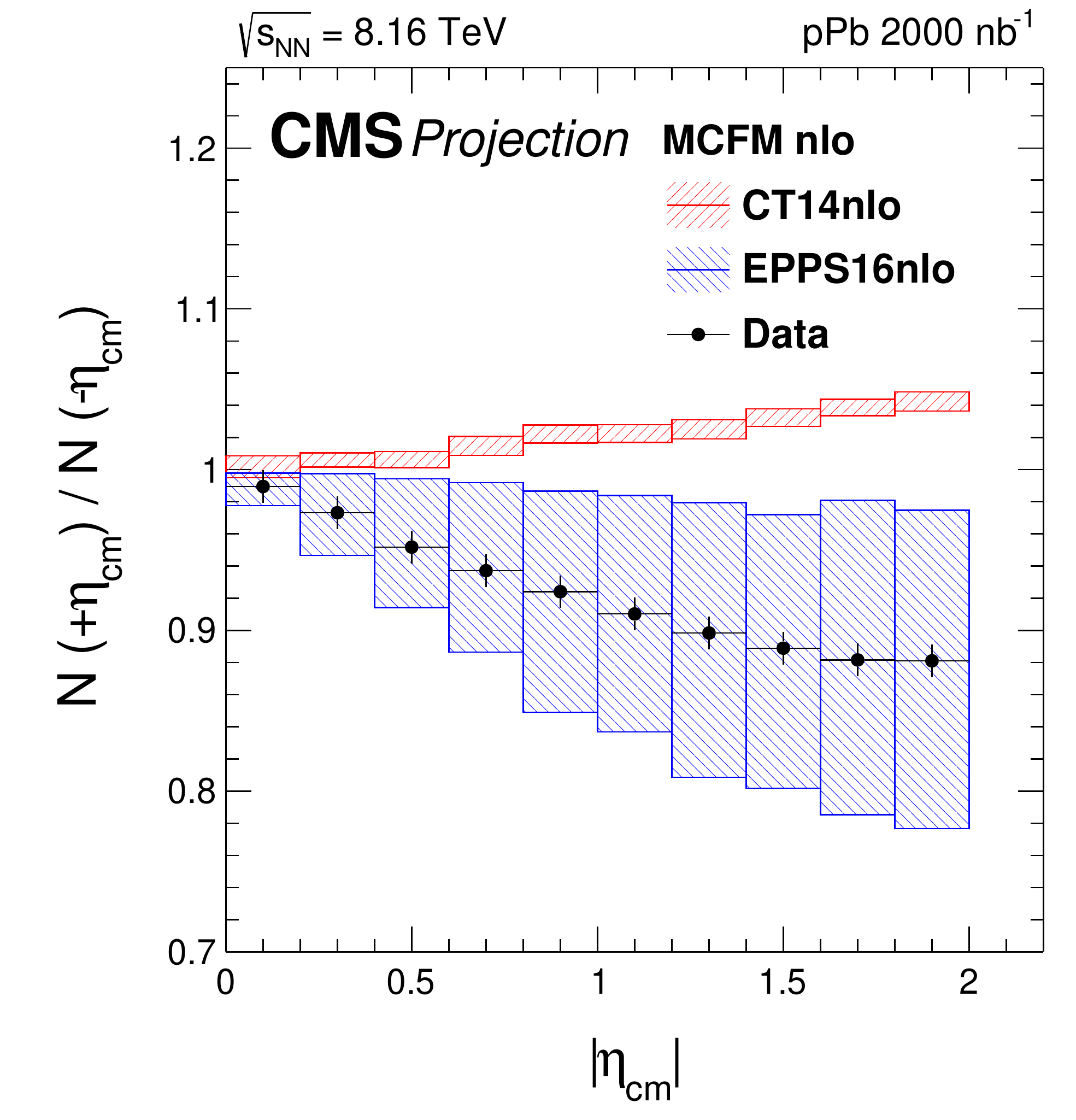}
     \includegraphics[width=0.3\textwidth]{\main/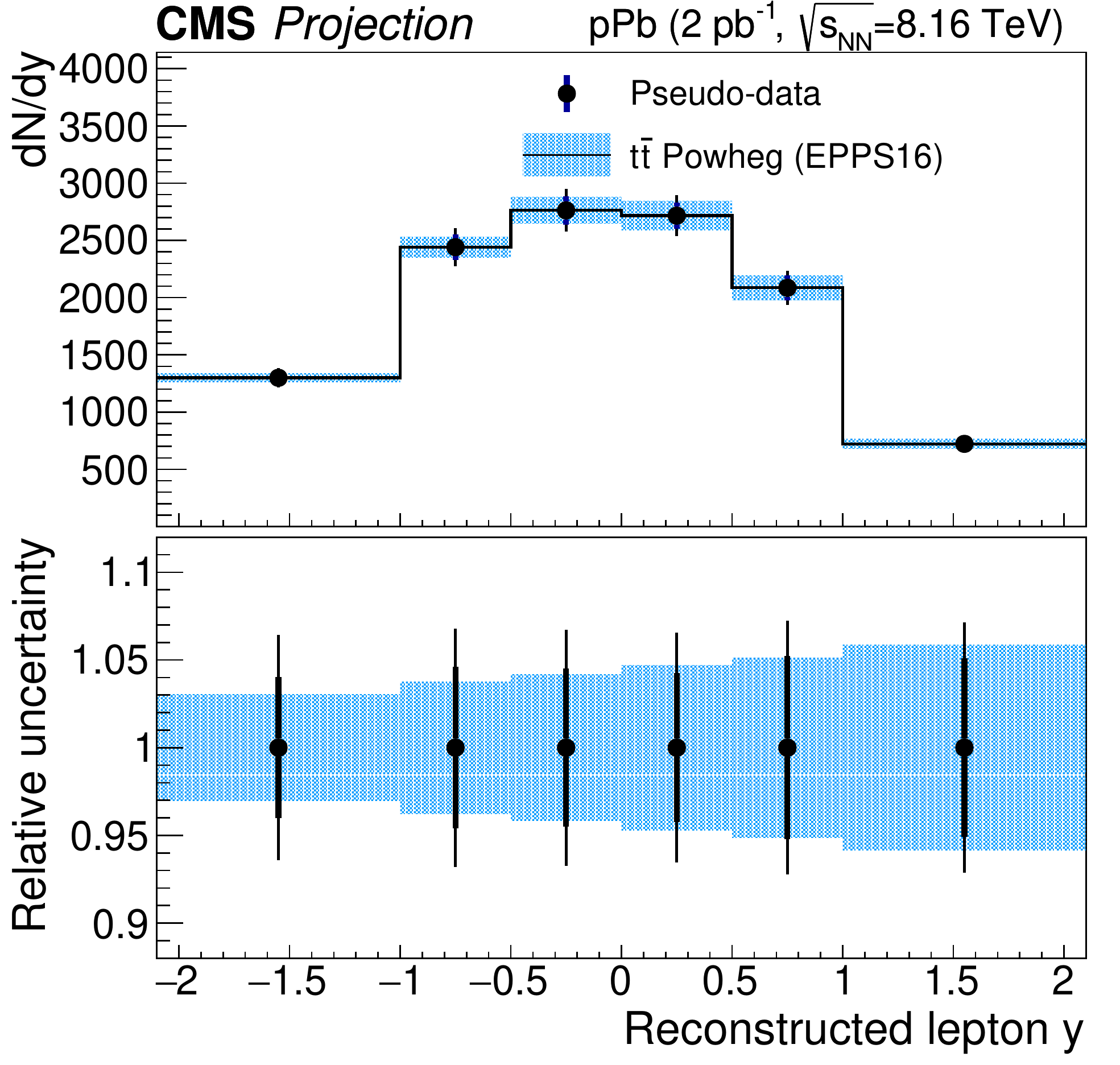}
      \includegraphics[width=0.3\textwidth]{\main/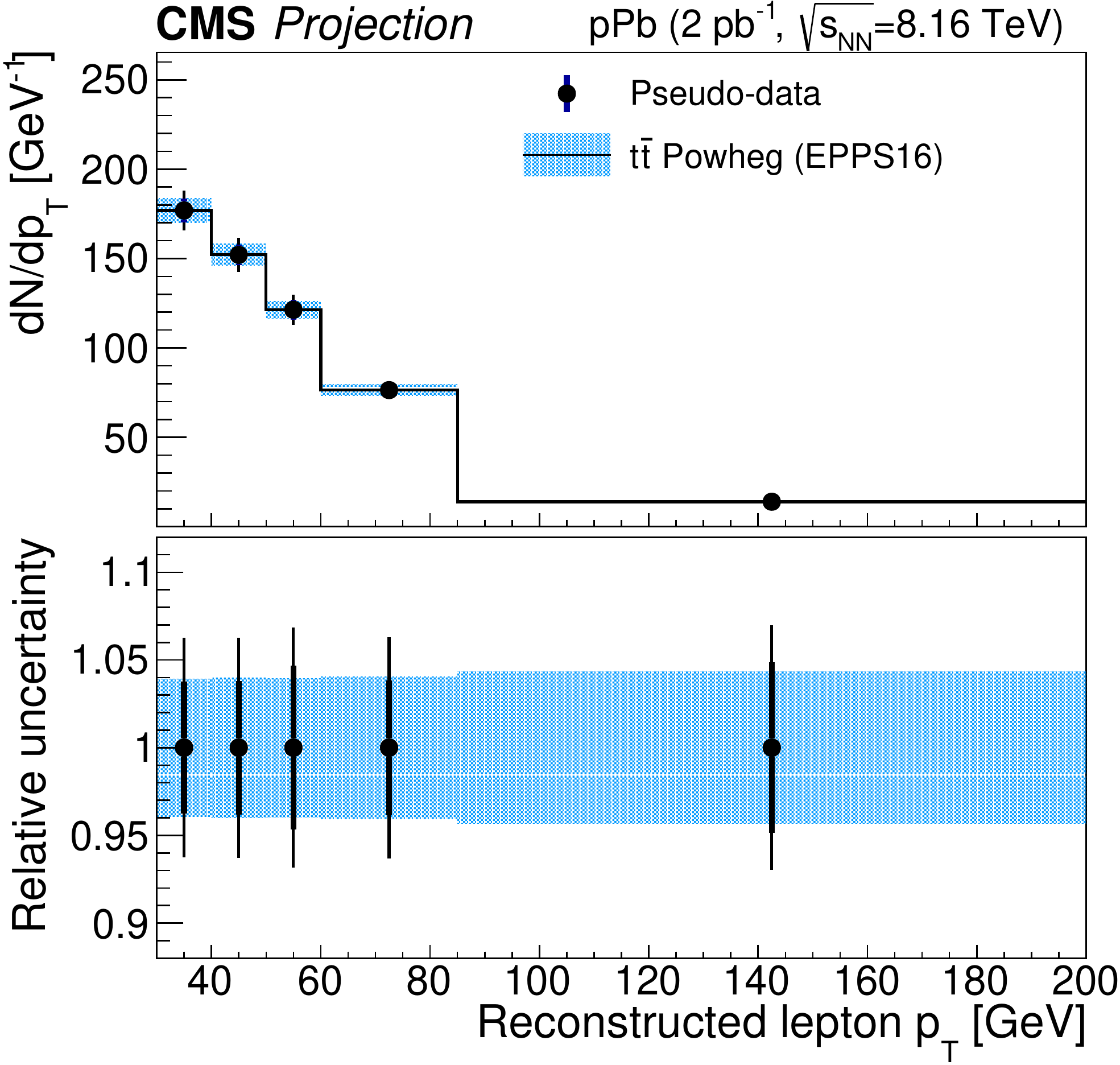}
    \caption{Left: W boson asymmetry measurement with Run 3 and Run 4 with CMS~\cite{CMS-PAS-FTR-17-002}. Middle and right: CMS projected differential \ttbar production cross section as a function of the reconstructed lepton \pT (middle) and rapidity (right)~\cite{CMS-PAS-FTR-18-027}. 
The total and statistical uncertainties are represented by the outer and inner error bars, respectively. The bottom panels represent the relative uncertainties (68 confidence-level error bands) in the projected data	 and theory predictions employing the CT14NLO~\cite{Dulat:2015mca} free proton and EPPS16NLO~\cite{Eskola:2016oht} bound nucleon PDF sets.}. 
    \label{fig:CMShighpt}	
\end{figure}

\begin{figure}
\includegraphics[width=.5\textwidth]{\main/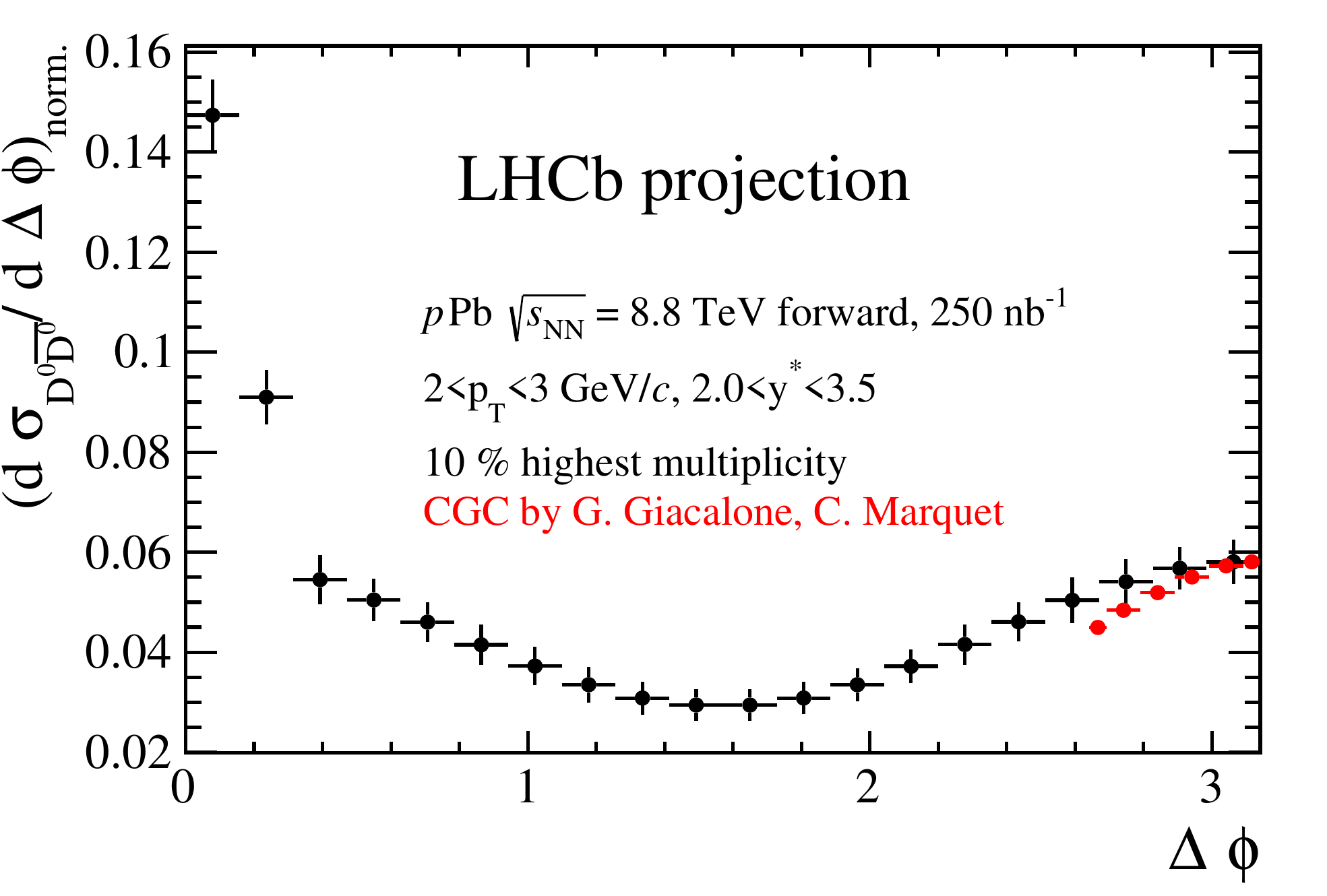}
\includegraphics[width=.5\textwidth]{\main/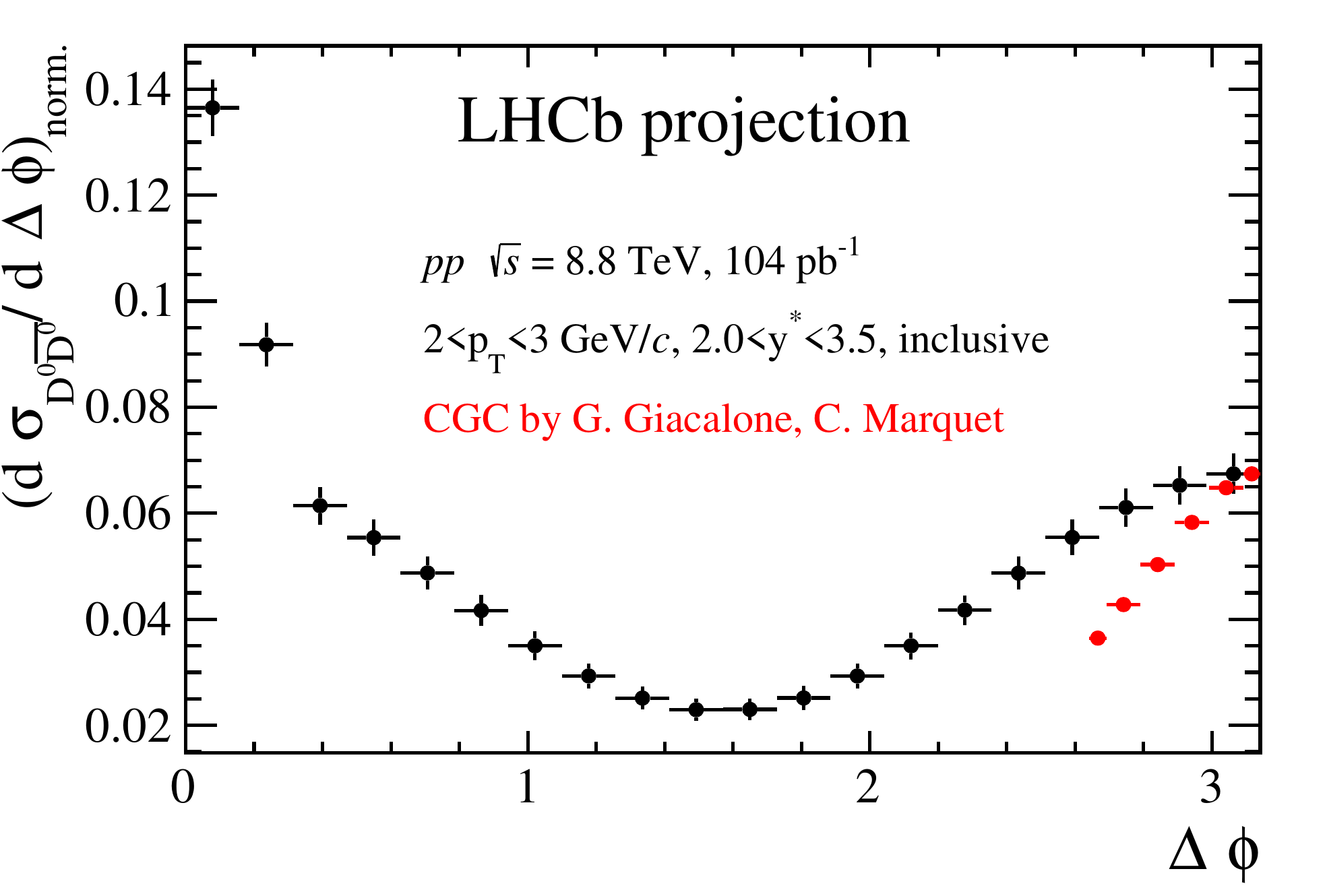}
\caption{LHCb projections~\cite{LHCb-CONF-2018-005} for $D^0 \overline{D^0}$ in LHC Run 3 and Run 4 compared with theoretical calculations decribed in section~\ref{sec:cgctmd} in 10~\% largest multiplicity \pPb~collisions (left panel) and minimum bias, i.e.,\,inclusive, \pp~collisions (right panel). The D-mesons are  both reconstructed in the $\pi K$-decay channel. }
\label{fig:ccbarlhcb}
\end{figure}

Proton-lead  collisions are an integral part of the LHC program since the 2012 pilot run. Within collinear factorisation, constraints on our knowledge of the nuclear wave functions could be extended at high $Q^2$ by dijets and heavy gauge boson available for the first time in nuclear collisions~\cite{Eskola:2016oht}. Insights have been gained at lower~$Q^2$ with heavy-flavour production based on the assumption that their nuclear production modification is dominated by nPDFs~\cite{Kusina:2017gkz}.  In Run 3 and 4, the increased luminosities and detector upgrades  will allow to improve  the statistical precision, to extent the kinematic reach and the available processes. 
The detector capabilities and observables are outlined in view of high-energy QCD studies in \pPb collisions and experimental measurement projections on selected observables are shown. One upgrade dedicated to low-$x$ physics is separately discussed. In all cases, the statistical and point-by-point uncorrelated systematic uncertainties are not taking into account to place the central point.  After this introduction, two theory contributions zoom in on specific aspects of Run 3-4 data. 
\begin{itemize}
\item \textbf{ALICE} will measure heavy flavour and charged particle production at mid-rapidity in \pPb collisions during Run~3 and Run~4. These measurements will constrain the parton densities in the nucleus at moderate $x$. At forward and backward rapidity, measurements of W and Z-boson production, as well as a range of quarkonia (J/$\psi$, $\psi(2S)$, and the $\PGUP{nS}$ family) will be performed. In particular the W and Z boson measurements will constrain the nuclear PDFs, while the quarkonia are also sensitive to final state effects. An example is given for the Z boson in Fig.~\ref{fig:plotDY} on the right. 
\item \textbf{ATLAS} will measure heavy electro-weak bosons, $W$ and $Z$, with the larger \pPb dataset which will be available in Run 3-4.  The much larger luminosity will allow precision in these measurements significantly surpassing the one currently available.  Projected $W$ and $Z$ boson cross sections published in Ref.~\cite{ATL-PHYS-PUB-2018-039} are shown in Figure~\ref{fig:ATLASWZxsection}.  Previous results have suggested that the modification of EW boson cross sections that is described with nPDFs, appears to be stronger in more central collisions \cite{Aad:2015gta,TheATLAScollaboration:2015lnm}.  Figure~\ref{fig:ATLASWZcent} (left) shows the projected boson yield ratio of central to peripheral collisions, $R_{\mathrm{CP}}$, for different centrality bins from Ref.~\cite{ATL-PHYS-PUB-2018-039}. Figure~\ref{fig:ATLASWZcent}(right) shows the yield of $Z$ bosons scaled by the nuclear overlap function, T$_{\mathrm{AB}}$,  as a function of centrality. It indicates that the $Z$ boson yield uncertainties will be considerably smaller than the T$_{\mathrm{AB}}$ uncertainty. 
\item \textbf{CMS}   will exploit the larger \pPb dataset available during Run~3--4 and it will hence further improve the precision on differential measurements of W and Z bosons (constraining quark and antiquark nPDFs)~\cite{Khachatryan:2015hha,Khachatryan:2015pzs}, as well as dijet and top quark pair (\ttbar) production (gluon nPDFs)~\cite{Sirunyan:2018qel,Sirunyan:2017xku}. While first Run~2 results on W boson production already feature experimental uncertainties smaller than the nPDF ones~\cite{CMS:2018ilq}, the larger luminosity by a factor 5 to 10 expected in Runs~3--4 will allow for another large jump in the constraints on nPDFs from data. The CMS dijet capabilities with Run~3--4 data~\cite{CMS-PAS-FTR-18-027} are shown in Fig.~\ref{fig:dijet-eta-projection-cms}. In addition, novel studies will become possible, such as the measurement of differential cross sections for \ttbar~production, for an improved constraining power on nPDFs, as well as the mass dependence of Drell-Yan production down to the $\PGU$ meson mass region. The projection for \ttbar~production~\cite{CMS-PAS-FTR-18-027} is shown together with the precise W boson asymmetry measurement in Run 3--4~\cite{CMS-PAS-FTR-17-002} in Fig.~\ref{fig:CMShighpt}.
Heavy flavour meson cross sections will also be measured, which are sensitive to low-$x$ gluon nPDFs: D mesons ($\pT>0.5$\,\UGeVc), B mesons ($\pT>5$\,\UGeVc), prompt and non-prompt \PJgy ($\pT>3$\,\UGeVc), and $\PGUP{nS}$ down to $\pT=0$. These measurements will benefit from the improved detector and trigger performance in Run~4~\cite{CMS-FTR-18-012}.
\item \textbf{LHCb} will operate during Run 3 and Run~4 with the average charged particle multiplicity in \pPb and \Pbp collisions that are smaller than the nominal conditions in \pp running with an average  pile-up of 5 interactions. All \pPb collisions will be processed in the software trigger. Hence, LHCb can fully profit from the luminosity increase including $\pT=0$ heavy-flavour production.  A natural focus is the study of open and hidden beauty and charm production with improved precision. A novel measurement in the \pPb collision system is discussed in detail in view of saturation physics in section~\ref{sec:cgctmd} and is shown in Fig.~\ref{fig:ccbarlhcb}.  A new focus of LHCb, profitting from the increased ion in one of the considered rapidity range is shown in Fig.~\ref{fig:plotDY} down to 5~\UGeVcc. Direct photon studies in the conversion channel will strongly profit from the increased luminosity. The projections of the LHCb collaboration for \pPb~collisions are discussed in detail in Ref.~\cite{LHCb-CONF-2018-005}.
\end{itemize}

\subsubsection{Forward calorimeter upgrade of ALICE}
\label{Sec:Focal}
The ALICE collaboration is considering to add a high-granularity forward calorimeter (FOCAL) to the experiment to  measure direct photon production in the rapidity range 3.0--5.0 and at low \pT, to probe the gluon density in protons and nuclei at $x\sim 10^{-5}$ where gluon saturation and non-linear effects in the gluon density may become apparent. The FOCAL design is based on the Si-W calorimeter technology, with two or three high-granularity layers with silicon pixel sensors that allow to separate electromagnetic showers with only a few mm distance between them. This unique high granularity makes it possible to reconstruct neutral pions in the forward direction and to reject the decay photon background for the direct photon measurement.

\begin{figure}
\includegraphics[width=0.35\textwidth]{\main/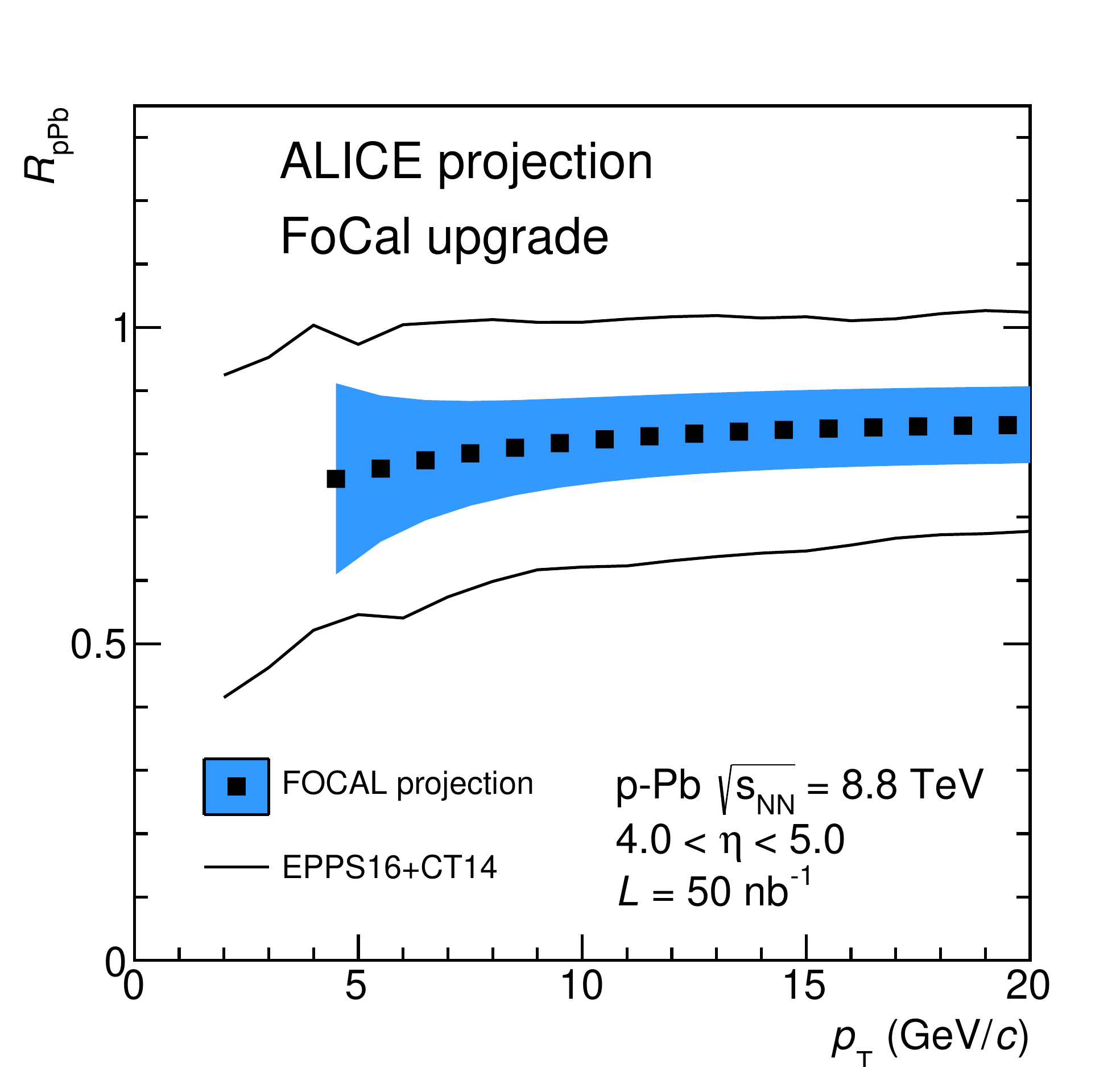}
\includegraphics[width=0.5\textwidth]{\main/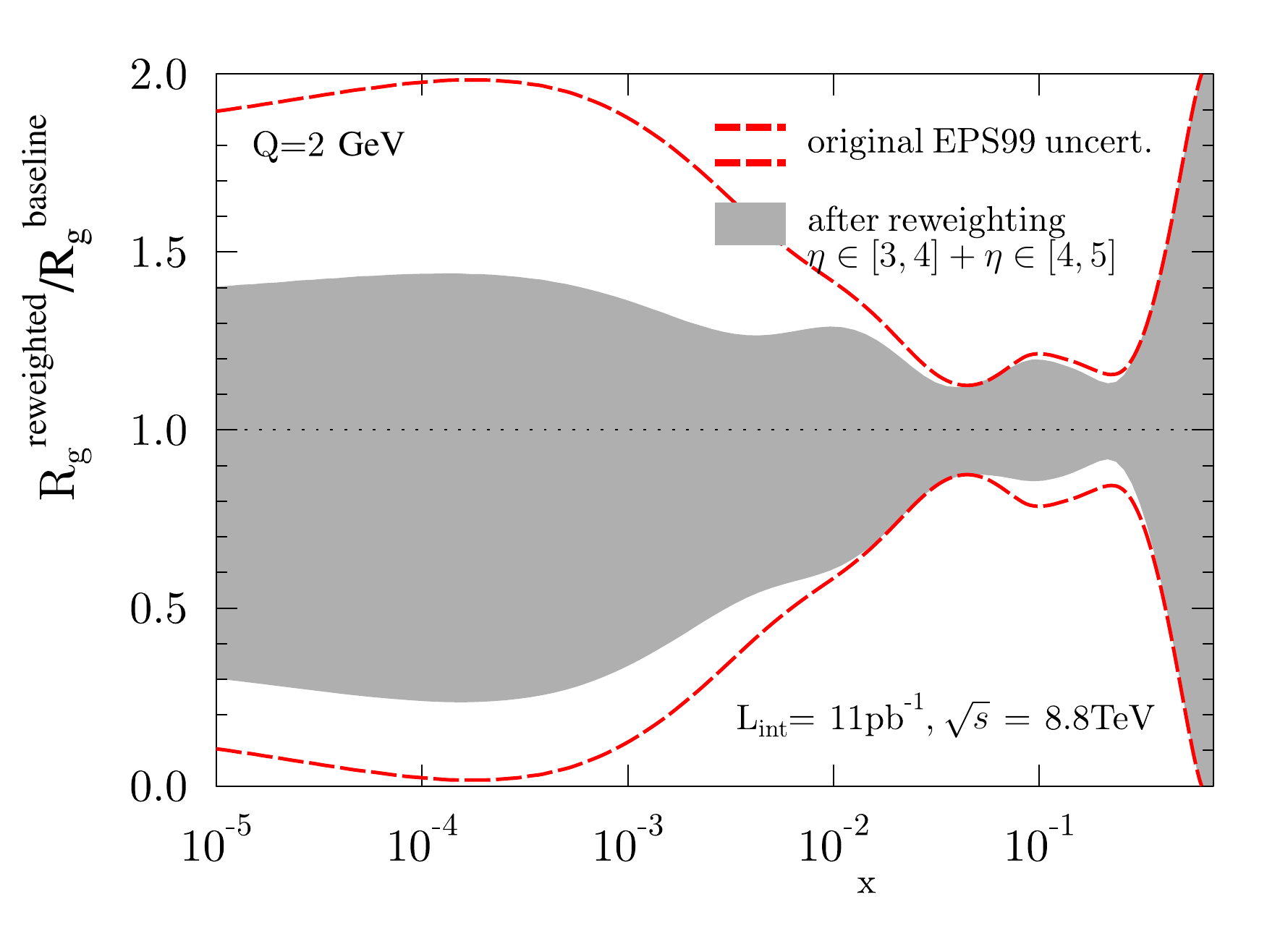}
\caption{\label{fig:focal}Left: Projected uncertainties on the nuclear modification factor for the forward direct photon measurement in \pPb collisions with the ALICE FOCAL upgrade. Right: Uncertainty on the modification ratio of the gluon density in the Pb nucleus from a fit to FOCAL pseudo-data. The red lines show the current uncertainty and the grey band shows the uncertainty after including the FOCAL pseudo-data. Figures from Ref.~\cite{ALICE-PUBLIC-2019-001}.}
\end{figure}

Figure \ref{fig:focal} shows the expected performance of the FOCAL detector for the direct photon measurement in p--Pb collisions. The left panel shows the projected uncertainties, which are 7-8\% at high $p_{T}$ and increase at lower momentum due to the combinatorial background. The right panel shows the impact of the measurement on our knowledge of the nuclear modification of the gluon distribution; the red lines show the current uncertainty, based on the EPS09 nPDFs, but using a broader set of parametrisations for the nuclear PDFs at the initial scale $Q_0$, similar to~\cite{Aschenauer:2017oxs}. The grey band shows the uncertainty after including the FOCAL pseudo-data. The improvement in the uncertainty is about a factor~2, but it should be noted that the direct photon measurements probes the gluon distribution at low $x$ directly, while the most of the existing data only probe $x \gtrsim 10^{-3}$ (see Fig. \ref{fig:smallx1}). At present, the only measurements that probe the very low $x$ region for nuclear PDFs are forward heavy flavour measurements from LHCb \cite{Aaij:2017gcy,Aaij:2017cqq} and ALICE \cite{Acharya:2017hdv}; theoretical developments are under way to use these data to constrain the PDFs~\cite{Kusina:2017gkz,Helenius:2018uul}. The FOCAL program will probe small $x$ in different production channels (quark-gluon Compton scattering vs gluon fusion) and therefore also further test universality and factorisation in this regime \cite{Helenius:2014qla}. Future measurements of Drell-Yan production in LHCb, as shown in Fig.~\ref{fig:DYRpPb}, will also probe this region, but will have a weaker impact for gluons at small $Q^2$ according to current experimental uncertainty estimates. 

In addition to the inclusive direct photon measurement, the FOCAL program will measure forward $\pi^{0}$ production in pp, \pPb and \PbPb collisions, which also provides important constraints for the nuclear PDFs and parton energy loss in \PbPb collisions at large rapidity. Correlation measurements of neutral pions and photons will be used to further probe the gluon density and to search for evidence of multiple-gluon interactions which are expected to be important in the high gluon density of the Color Glass Condensate \cite{Kharzeev:2004bw,vanHameren:2014lna}.



\subsubsection{Transverse momentum dependent and low-$x$ phenomena sensitive observables in $c\bar{c}$ and $b\bar{b}$ production }
\label{sec:cgctmd}

In recent years, many efforts have been devoted to elucidate the properties of transverse-momentum-dependent gluon distributions in the high-energy or small-$x$ limit, and in particular to determine how non-linear saturation effects impact the various distributions. A process which is particularly interesting in this regard is the forward production of a heavy quark-antiquark pair, in high-energy proton-proton or proton-nucleus collisions. For kinematical reasons, in the proton-nucleus case, the proton side of the collision involves large-$x$ partons, while on the nucleus side, small-$x$ gluons participate. Hence, this process can be described in a hybrid approach in which the proton content is described by regular, integrated PDFs, while the small-$x$ dynamics in the nuclear wave function is dealt with using the Color Glass Condensate (CGC) effective theory, from which the gluon transverse momentum dependent distributions~(TMDs)  naturally emerge \cite{Dominguez:2011wm,Marquet:2016cgx}.

In terms of sensitivity to the QCD saturation regime, the same manifestations are expected than with light quarks \cite{Albacete:2018ruq}, although smaller values of x can be reached in the latter case. From the point of view of the TMD content however, heavy quarks are interesting because their production involves linearly-polarized gluons TMDs \cite{Akcakaya:2012si}, in addition to the usual unpolarized gluon TMDs, and because, due to non-linear effects, the linearly-polarized TMDs generally differ from their unpolarized partners \cite{Marquet:2017xwy}. The LHCb detector is well-suited to measure heavy hadrons in the forward rapidity region, and given large-enough statistics, one could potentially try and extract information about the linearly-polarized gluons in protons and lead nuclei, from data obtained with LHC unpolarized beams. Before giving cross section estimates, let us briefly give a bit more details about the process and the physics involved.

Forward quark-antiquark pair production in dilute-dense collisions is characterized by three momentum scales: $P_{t}$, the typical transverse
momentum of a single quark, and always one of the largest scales; $k_{t}$, the total transverse momentum of the pair, which is a measure of the transverse momentum of the small-$x$ gluons coming from the target; and $Q_s$, the saturation scale of the target, which is always one of the softest scales.
The value of $k_{t}$ with respect to $Q_{s}$ and $P_{t}$ governs which factorisation scheme is relevant. Indeed, when $k_{t}\sim Q_{s}\ll P_{t}$ (the quark and the antiquark are almost back-to-back), there are effectively two strongly ordered scales $k_{t}$ and $P_{t}$ in the problem and TMD factorisation applies \cite{Dominguez:2011wm}, implying the involvement of several gluon TMDs that differ significantly from each other, especially in the saturation regime, when $k_t\leq Q_s$ \cite{Marquet:2016cgx}. In the other regime which shall not be discussed here (away from back-to-back production), $k_{t}$ and $P_{t}$ are of the same order and far above the saturation scale, only the linear small-$x$ dynamics is important and the various TMDs differ no more.

In \cite{Marquet:2007vb,Dominguez:2011wm}, the cross section for forward di-jet production in proton-nucleus collisions
was calculated within the CGC. It was then shown that, in the back-to-back limit, a TMD factorisation formula could
be extracted, the result being the same as in a direct TMD approach (i.e.,\,without resorting to the CGC). However, in contrast to the
direct TMD approach, the calculation in the CGC yields explicit expressions for the TMDs in terms of Wilson lines, which can be evolved in rapidity
through the nonlinear Jalilian-Marian-Iancu-McLerran-Weigert-Leonidov-Kovner (JIMWLK) equation. In \cite{Marquet:2017xwy}, those results
were extended to the case of a forward heavy quark-antiquark pair. As already observed earlier (see for instance \cite{Mulders:2000sh,Boer:2009nc,Boer:2010zf,Metz:2011wb,dominguez:2011br,Akcakaya:2012si}), by keeping a non-zero quark mass, the cross section becomes sensitive to additional TMDs, which describe the linearly-polarized gluon content of the unpolarized target.

The three unpolarized gluon TMDs which describe the quark-antiquark pair production are the adjoint-dipole TMD, the Weizs{\"a}cker-Williams TMD, and a third one which is roughly a convolution of two fundamental-dipole TMDs. They are accompanied by three \textquoteleft polarized' partners, which couple through the quark mass and via azimuthal-angle modulation. This is analogous to what happens in the $\gamma^*A\rightarrow q\bar{q}$ process (in that case not only a non-zero quark mass but also a non-zero photon virtuality brings sensitivity to linearly-polarized gluons) described in Section~\ref{sec:upchqdijets},\ref{sec:upcdijets}, although there, only one unpolarized gluon TMD is involved (the Weizs{\"a}cker-Williams distribution), along with its polarized partner \cite{Metz:2011wb,dominguez:2011br}. 

Predictions for D mesons are presented in Figure \ref{fig:ccbarlhcb} as a function of their relative azimuthal angle near $\pi$, along with projections from the LHCb experiment. In the \pPb kinematical range that is investigated here, roughly 2.800~raw counts are expected over the full azimuthal range for the specified decay channel that can be complemented with other channels as in \pp~collisions in Ref.~\cite{Aaij:2012dz}.

\subsubsection{Fully coherent energy loss effects on different final states}
\label{subsubsec:coherent}

The multiple scattering of quarks and gluons traveling in a QCD medium induces the radiation of gluons which carry away some energy of the propagating parton, leading for instance to the jet quenching phenomenon. Therefore, a key ingredient of any parton energy loss calculation is  the medium-induced gluon spectrum radiated by the fast propagating color charge. It is of course of crucial importance to know the correct parametric dependence of the induced spectrum, which in general depends on the parton properties (in particular its energy and mass) and those of the medium, since it has a direct impact on the phenomenology of particle production in \pA and \AOnA collisions. The emission of a gluon radiated by an energetic parton experiencing multiple scattering in a medium takes a typical formation time, $t_f$, which needs to be compared to the length of the medium, $L$.

Over the last few years, it has been realized that in a hard process involving incoming and outgoing energetic color charges (which do not have to be identical) being quasi-collinear in the rest frame of the medium, the associated medium-induced gluon spectrum is dominated by {\it large} gluon formation times, $t_f \gg L$~~\cite{Arleo:2010rb,Peigne:2014rka,Peigne:2014uha}. In this so-called fully coherent (FC) region, the medium-induced radiated energy is similar to the energy loss of an asymptotic charge. In particular it scales as the energy, and thus exceeds at high energy the average parton energy loss in the Landau--Pomeranchuk--Migdal (LPM) regime, $t_f \lesssim L$, for which the energy dependence is at most logarithmic.
This different parametric behavior has important consequences on the phenomenology of hadron production in \pA collisions. In particular, a model based on the fully coherent induced gluon spectrum was shown to describe accurately the quarkonium suppression observed in \pA collisions at all centre-of-mass energies, from the SPS fixed-target experiments ($\sqrt{s}\simeq 20$~\UGeV) to the LHC ($\sqrt{s}=5.02$~\UTeV and $\sqrt{s}=8.16$~\UTeV)~\cite{Arleo:2012hn,Arleo:2012rs,Arleo:2013zua}. It is therefore necessary to investigate further the role of fully coherent energy loss on other processes.
Because the fully coherent induced gluon spectrum arises from the interference between the emission amplitudes off the initial charge and that off the final state, the effects of FC energy loss are process dependent. Let us review, at a qualitative level, the expected nuclear dependence of different processes which could be measured at the LHC with a high luminosity.

In the absence of color charge in the partonic final state, the energy loss in Drell-Yan (DY) production at leading order is expected to be that of a suddenly decelerated parton, that is, independent of its energy (LPM regime). The effects of parton energy loss in nuclei should therefore play almost no role on DY production in high-energy \pA collisions, since $\Delta E / E \to 0$ in the high-energy limit. The inclusive production of DY lepton pairs in \pA collisions at the LHC should therefore be free of any parton energy loss effect. As a result, this process might be used advantageously in order to probe possible nuclear PDF effects at small values of $x$, see \eg Fig.~\ref{fig:plotDY} and Ref.~\cite{Arleo:2015qiv} for more details. Assuming a luminosity of $\Lint=250$~nb$^{-1}$ and taking the conservative value for the cross section in \pPb collisions at $\sqrt{s}=5$~TeV, ${\rm \rmd }\sigma/\rmd y=40$~nb~\cite{Arleo:2015qiv}, the number of forward DY lepton pairs, which could be measured by the LHCb experiment is typically $10^4$ per rapidity unit. Also interesting, and accessible with a high luminosity, would be the production of diphotons in \pA collisions. This process would be free of energy loss effects, for the same reasons as the DY process, while being sensitive to nPDF in the quark sector, $q\bar{q} \to \gamma\gamma$, and in the gluon sector through the `box diagram', $gg \to \gamma\gamma$.

Note that the energy loss scaling as $E$ (FC regime) would come into play only if {\it another} energetic charged particle is produced in the final-state in association with the virtual photon (or the diphoton), such that the final state carries a global color charge. Such a situation typically occurs in DY$+$jet production in \pA collisions. Consider for instance the Compton scattering process, $qg \to q\gamma^\star$. At forward rapidity, an incoming quark from the proton projectile scatters in the medium to produce the final state in a color triplet representation. In this peculiar case of quark to a color triplet final state, one would expect a negative \emph{medium-induced} gluon spectrum (\ie\ with stronger gluon radiation in \pp than in \pA collisions), leading to an energy \emph{gain} (with respect to \pp collisions), $\Delta E \propto (-1/2N_c)\times E$~\cite{Peigne:2014uha}. Such an unusual dependence would manifest by a slight \emph{enhancement} of DY$+$jet production in \pA collisions with respect to \pp collisions~\cite{Arleo:2015qiv}, although a quantitative study would be needed to answer whether this effect could be visible in the experiment. Similarly, should this enhancement be small or negligible, the associate production of a prompt photon with a heavy-quark jet might be sensitive to the nPDF of heavy quarks and gluons, as emphasized in Ref.~\cite{Stavreva:2010mw}. Using $\sigma_{\gamma c} = 1.2\times10^5$~pb in the acceptance of the ALICE calorimeter~\cite{Stavreva:2010mw} leads to $2.4\times10^5$ $\gamma+c$-jet events in \pPb collisions at $\sqrt{s}=8.8$~TeV assuming ${\cal L}=2$~pb$^{-1}$. This observable should be also accessible in ATLAS, CMS and LHCb acceptances as well.

More pronounced effects of fully coherent energy loss are expected when the final state is in a color octet state, or possibly in higher color representations. An example is the production of a jet pair with not too large transverse momenta (ideally only a few times the saturation scale of the target nucleus). In the case of di-gluon production, the final state can be produced in the 27-plet color representation (with Casimir $C_{27}=8$) that would lead to an average coherent energy loss proportional to $(N_c+C_{27})/2=11/2$, that is almost twice larger that expected if the final state is in a color octet state. Such higher color representations could also be probed in the production of $B_c$ mesons (or in the associate production of a D and a B meson), with a complex final state $\rm c\,\bar{c}\,b\,\bar{b}$. From the number of fitted signal candidates ${\cal N}=10^4$ $B_c$ mesons extracted at LHCb in the semileptonic decay channel in \pp collisions at $\sqrt{s}=8$~TeV with ${\cal L}=2$~fb$^{-1}$~\cite{Aaij:2014bva}, the expected $\rm B_c$ rate in the LHCb acceptance using ${\cal L}=0.5$~pb$^{-1}$ in \pPb collisions at the same energy is ${\cal N}=5\times 10^2$ $\rm B_c$ mesons.

\subsection{Constraints on nuclear PDFs}
\label{sec:nPDFs}
\subsubsection{Overview}

As previously discussed in Sec.~\ref{sec:smallxIntro}, 
the nuclear Parton Distribution Functions (nPDFs) 
are poorly constrained as compared to the proton PDFs. 
This is mainly due to the lack of high statistics data 
across the very  large nuclear mass number~(A) range. 
In fact, even the precision of the proton PDFs rely crucially on 
nuclear target data~\cite{Ball:2017nwa,Dulat:2015mca,Alekhin:2017kpj};
for example,  the neutrino-nucleon deep-inelastic-scattering~(DIS) structure functions 
are essential for decomposing the flavour components of the 
proton~\cite{Onengut:2005kv,Goncharov:2001qe,Mason:2006qa,Samoylov:2013xoa,KayisTopaksu:2011mx}.
Consequently, improved determinations of the nuclear PDFs and nuclear correction factors 
could improve the proton PDF precision.
Thus, the future LHC Runs~3 and 4 could provide the opportunity to precisely constrain
the nPDFs for the Pb-nucleus and, in Run~5 and later, one or more lighter nuclei, and thereby
disentangle the nuclear effects from the individual flavour components.

\begin{figure}[tbh] 
\centering{} 
\includegraphics[width=0.68\textwidth]{\main/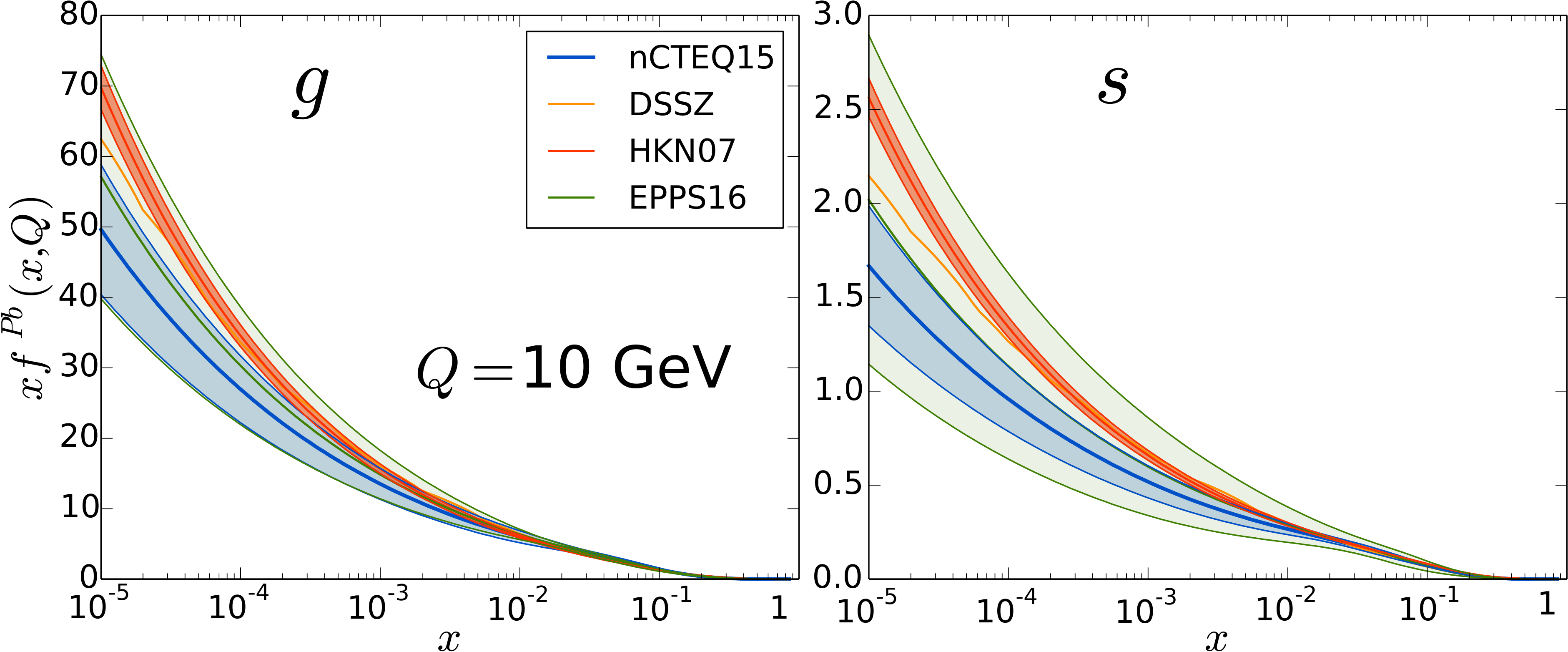}
\caption{
Comparison of the gluon (left) and strange (right) quark nPDF for a lead nucleus as a function of $x$ for 
a selection of nuclear PDF sets:
HKN07~\cite{Hirai:2004wq}, DSSZ~\cite{deFlorian:2011fp}, EPPS16~\cite{Eskola:2016oht}, and nCTEQ15~\cite{Kovarik:2015cma}.
}
\label{fig:npdf}
\end{figure}

In Fig.~\ref{fig:npdf} selected nPDFs from the literature are displayed.

\subsubsection{W and Z boson production}

\begin{figure}[t] 
\centering{} 
\includegraphics[width=0.42\textwidth,height=0.35\textwidth]{\main/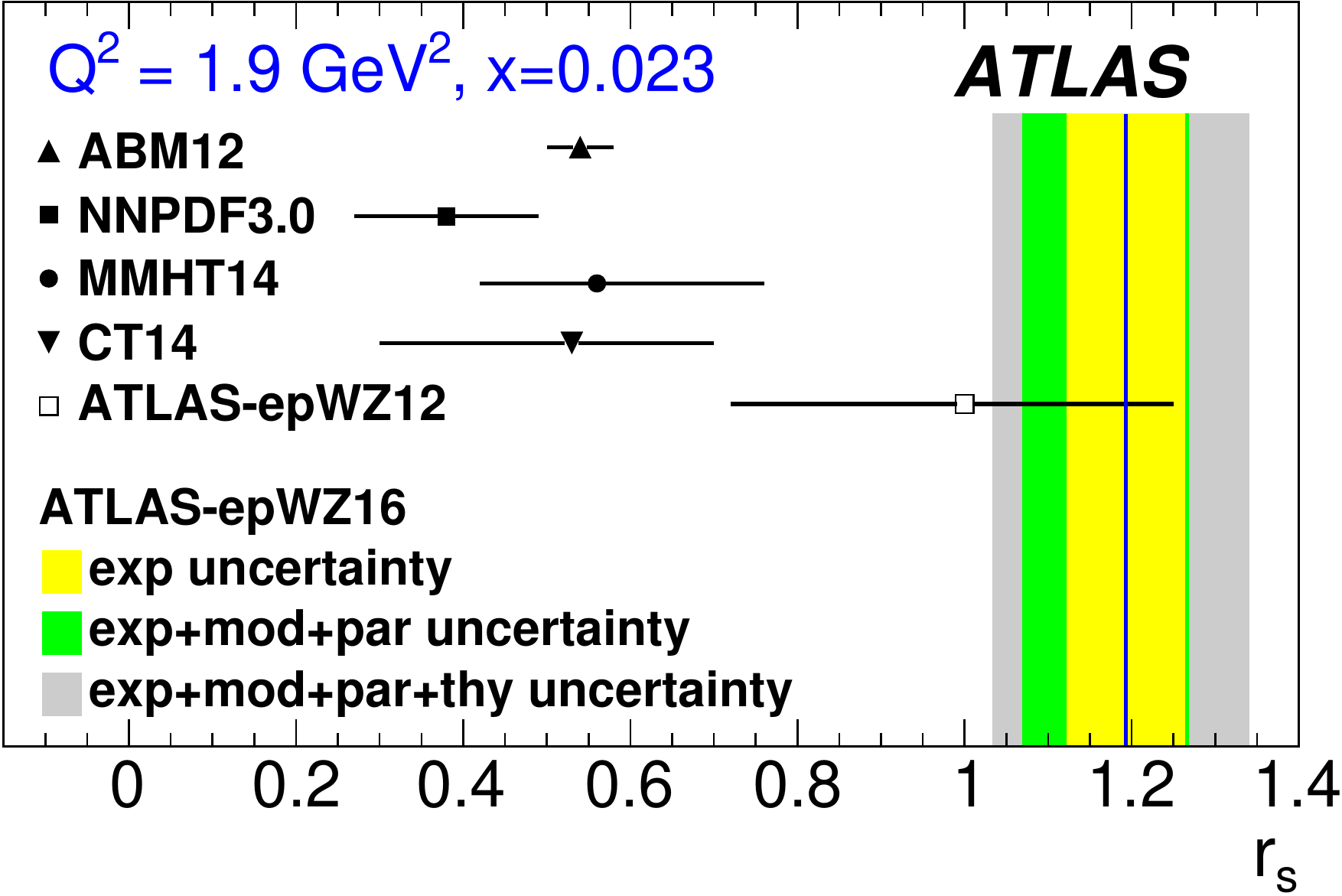}
\hfil
\includegraphics[width=0.48\textwidth,height=0.40\textwidth]{\main/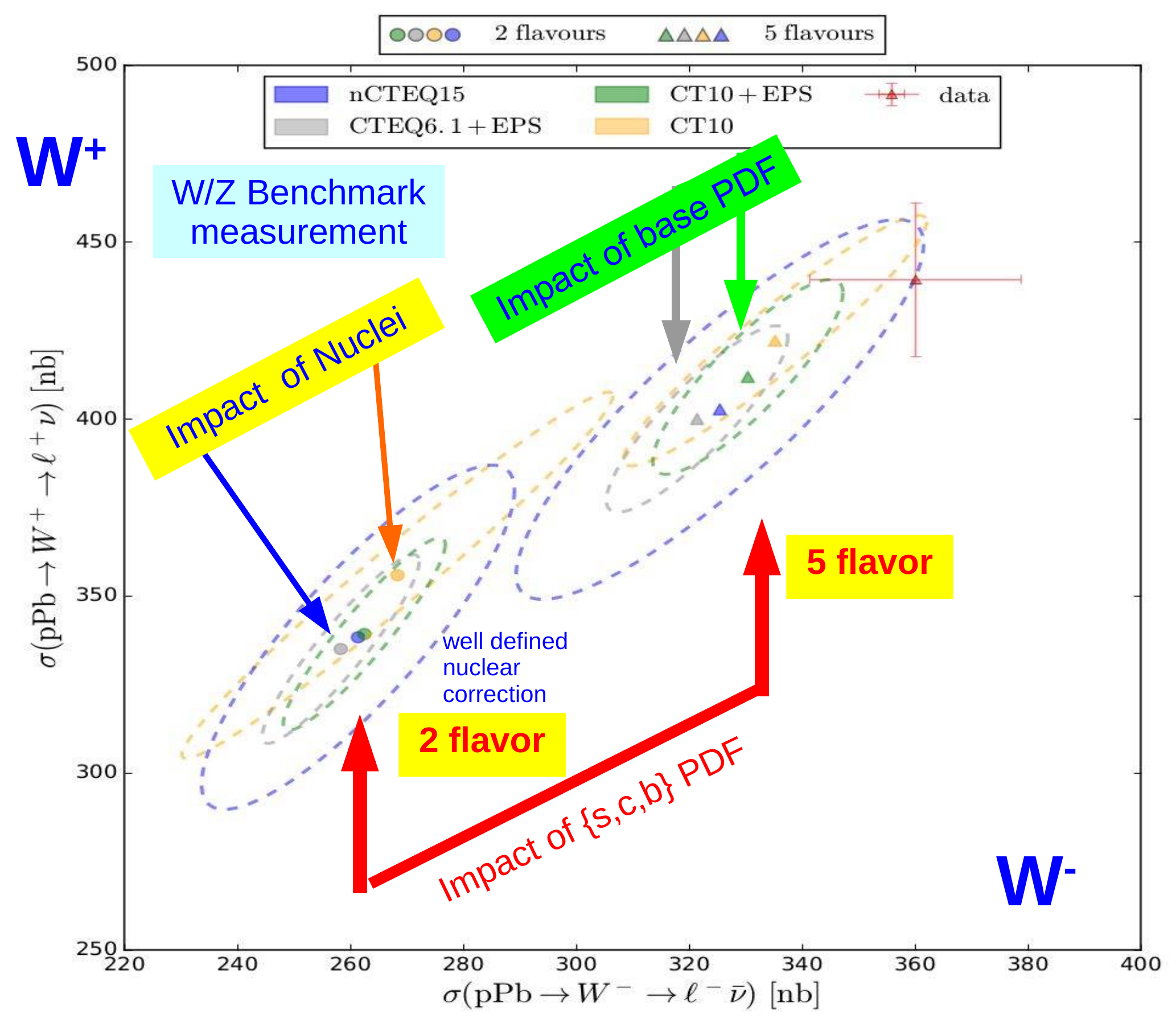}
\caption{
Left: The relative strange-to-down sea quark fractions 
$\rm r_s=0.5(s + \bar{s})/\bar{d}$\  as compared with  predictions from different NNLO PDF sets; figure from Ref.~\cite{Aaboud:2016btc}. 
\quad 
Right: correlations between W$^+$ and W$^-$ \pPb  cross sections calculated with different input PDFs and assumptions   to illustrate the separate impact of the 
i)~nuclear corrections, ii)~heavy flavour components, and iii)~base PDFs~\cite{Kusina:2016fxy,Godat:2018kgp}.}
\label{fig:WZcorr}
\end{figure}

\noindent {\bf  Inclusive W and Z boson production:}
Inclusive production of W and Z bosons in \pp~collisions at the LHC  can provide
new information on the strange, charm and beauty quark PDFs.
Additionally, heavy-ion W and Z production data from \pPb and \PbPb can provide insight on 
the nuclear corrections, 
and this complements other data on nuclear targets as it is at large $A$ (lead) and high energy (and thus, smaller $x$)~\cite{Aad:2015gta,TheATLAScollaboration:2015lnm,Khachatryan:2015pzs,Khachatryan:2015hha,CMS:2018ilq,Aaij:2014pvu,Alice:2016wka}.
For example, ATLAS used  inclusive W/Z production data to extract the strange quark component of the proton 
as displayed in Fig.~\ref{fig:WZcorr} (left). This yielded a larger strange quark PDF than commonly expected~\cite{Aad:2012sb,Aaboud:2016btc}. 
A recent analysis of the ATLAS and CMS inclusive W and Z differential
cross section data at 7 and 8 TeV~\cite{Cooper-Sarkar:2018ufj} and the combined HERA inclusive data indicates that while there is no tension between the
data sets, the  LHC data does support unsuppressed strangeness
in the proton at low $x$ at both low and high scales. 
Certainly this is an area that warrants further study. 

To highlight the sensitivity of the heavy ion W/Z production to both the 
heavy flavour components and the nuclear corrections, 
in Fig.~\ref{fig:WZcorr} (right), the correlations between W$^+$ and W$^-$ 
cross sections for proton-lead interactions calculated with different input PDFs and assumptions~\cite{Kusina:2016fxy} are shown.
By comparing the results with and without the strange, charm, beauty quark flavours, it can be observed that these quarks do have a
large impact on this observable;  hence, this process can provide incisive information about the 
corresponding PDFs. 
To see the effect of the nuclear corrections, the CT10 proton result is compared with the other calculations. 
For the case that only two flavours are considered, the separation of the proton result  (CT10) and the 
nuclear results are quite distinct. In this case, the effect of the specific nuclear correction (nCTEQ15 or EPS09)
or the effect of the underling base PDF (CTEQ6.1 or CT10) is minimal. 
In contrast, when this picture is compared to the five flavour results, the division between the proton and nuclear
result is not as simple as the different nuclear corrections and proton baseline PDFs yield a broader range of results. 
In particular, the strange quark PDFs in the CTEQ6.1 and CT10 proton baseline PDFs are quite different, and this will contribute to the spread of results. 
Thus, proton-lead  production of W/Z  is an ideal ``laboratory'' 
as this process is sensitive to i)~the heavy flavour components,
ii)~the nuclear corrections, and iii)~the underlying ``baseline'' proton PDF. 
Thus, high-statistics heavy-ion run data during Run~3 and Run~4 has the potential 
to reduce  the current uncertainties and improve the nuclear PDF determination as illustrated by the projections from ATLAS in Fig.~\ref{fig:ATLASWZxsection} and CMS~\ref{fig:CMShighpt}.

\begin{figure}[tb!]
\centering
\caption{Left: The CMS projections~\cite{CMS-PAS-FTR-17-002} for the forward-backward asymmetry in W$^\pm$ production (Fig.~\ref{fig:CMShighpt}) compared to the original EPPS16 90\% confidence-level error bands and those after reweighting with these W$^\pm$ data. Right: The change in EPPS16 nuclear PDF modifications for sea quarks and gluons at $Q^2=100\,{\UGeV^2}$ upon reweighting with the data shown in the left-hand panel.}
\includegraphics[width=0.405\textwidth]{\main/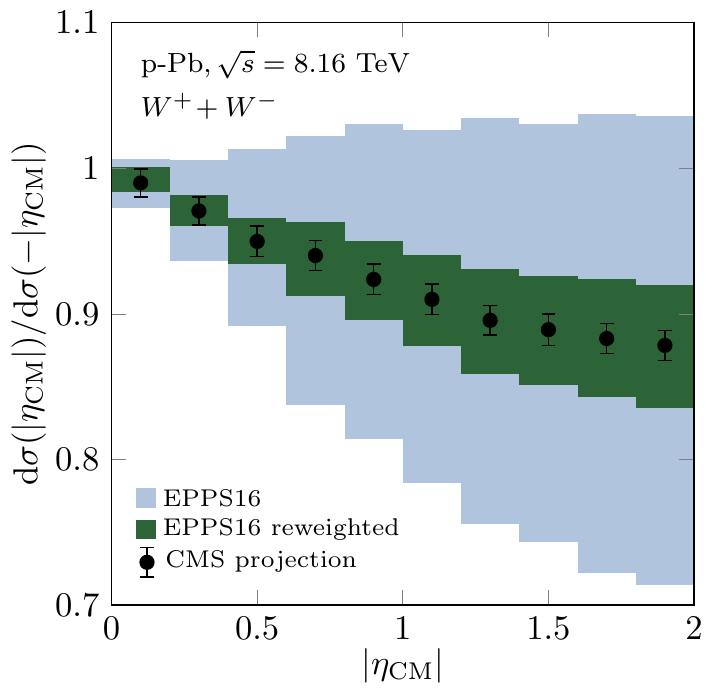}
\includegraphics[width=0.425\textwidth]{\main/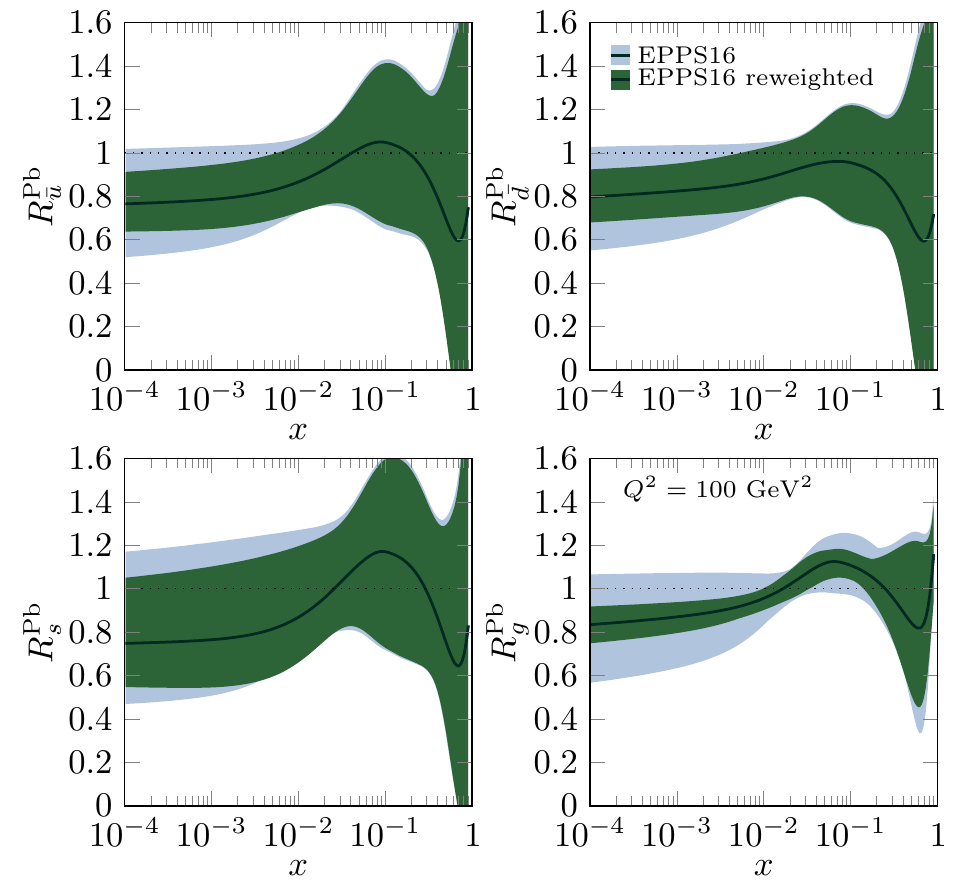}
\label{fig:Wreweight}
\end{figure}

\noindent {\bf W$^\pm$ asymmetries:}
A way to disentangle the nuclear effects from proton PDF and other theory uncertainties like higher-order corrections even in the absence of a \pp~reference, is the forward-to-backward ratio \cite{Paukkunen:2010qg}. To showcase the potential in the case of W$^\pm$ projections (shown in Fig.~\ref{fig:CMShighpt}), Fig.~\ref{fig:Wreweight} presents the effect that PDF reweighting \cite{Paukkunen:2014zia} analysis with these data has on EPPS16 nuclear PDFs and the corresponding theory predictions for the asymmetry. The most notable effect is the dramatic reduction in the uncertainties of the gluon PDF. Indeed, W$^\pm$ production takes place at a high scale $Q^2 \sim M_{\rm W}^2$ and even though it is mostly $q_i\overline{q}_j$ processes that make the W bosons, the $\overline{q}_j$ PDFs probed at $\eta_{\rm CM} \gtrsim 0$ are, in practice, dominated by the evolution effects at small $x$. Thus, it is not that surprising that it is predominantly the gluon component that gets tightly constrained by the W data. The improvement for the light sea quarks (d,u,s) is merely a consequence of better constrained gluons, through QCD dynamics. The large-$x$ ($x \gtrsim 0.1$) part is not really affected by the  W$^\pm$ data.

\noindent Another W$^\pm$ observable that would benefit from a high luminosity is an asymmetry $$[\sigma^{W^+}(y_{\rm CM})-\sigma^{W^+}(-y_{\rm CM})]/[\sigma^{W^-}(y_{\rm CM})-\sigma^{W^-}(-y_{\rm CM})]$$
proposed in Ref.~\cite{Paukkunen:2010qg}. In order to measure such a quantity involving four cross sections and subtractions among them (particularly in the denominator which may fluctuate between positive and negative), an excellent statistical precision --- like that achievable during Run~3 and~4 at the LHC would be advantageous. With the present Run~2 luminosity it appears that the data \cite{CMS:2018ilq} are still not accurate enough to measure this across the full rapidity acceptance.

%
\begin{figure}[tb!]
\centering
\includegraphics[width=0.565\textwidth]{\main/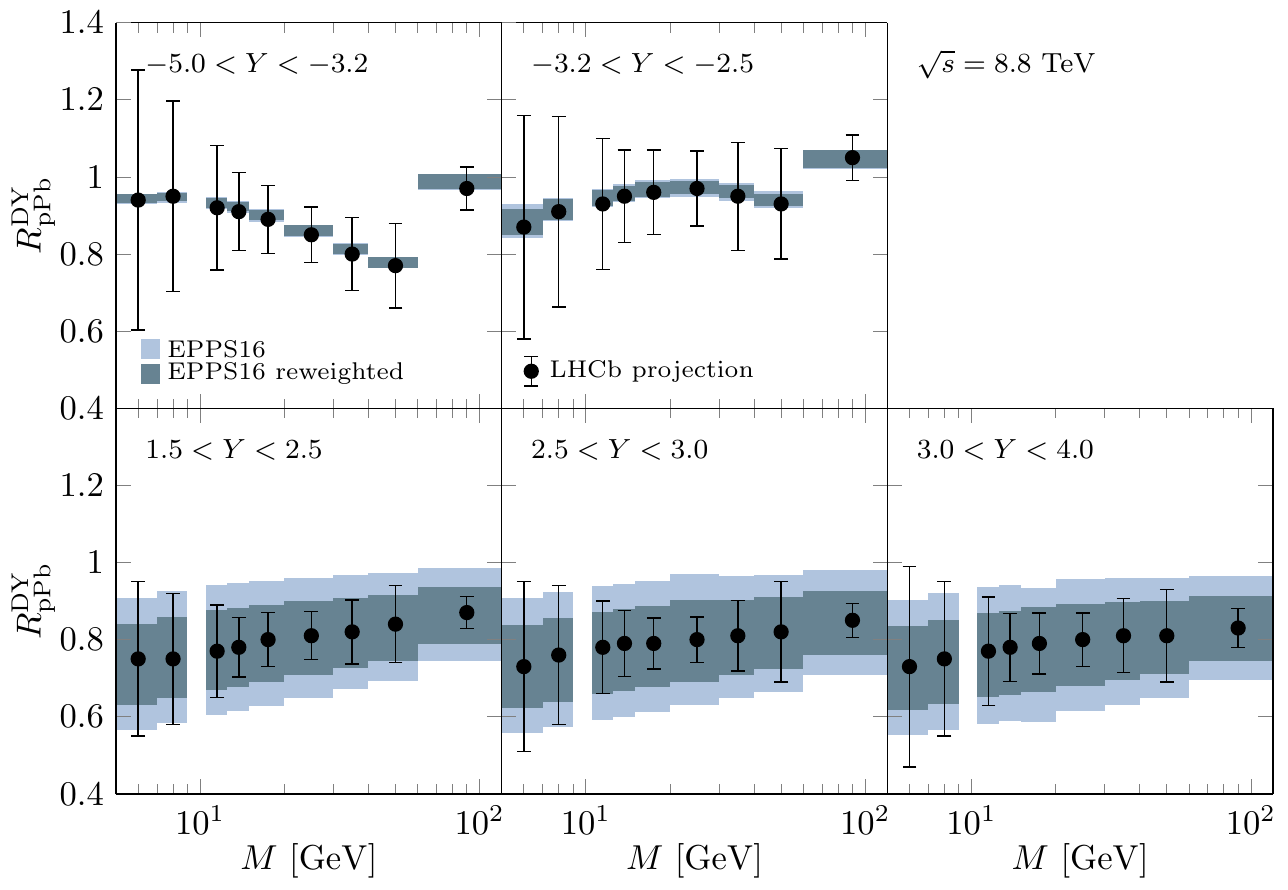}
\includegraphics[width=0.425\textwidth]{\main/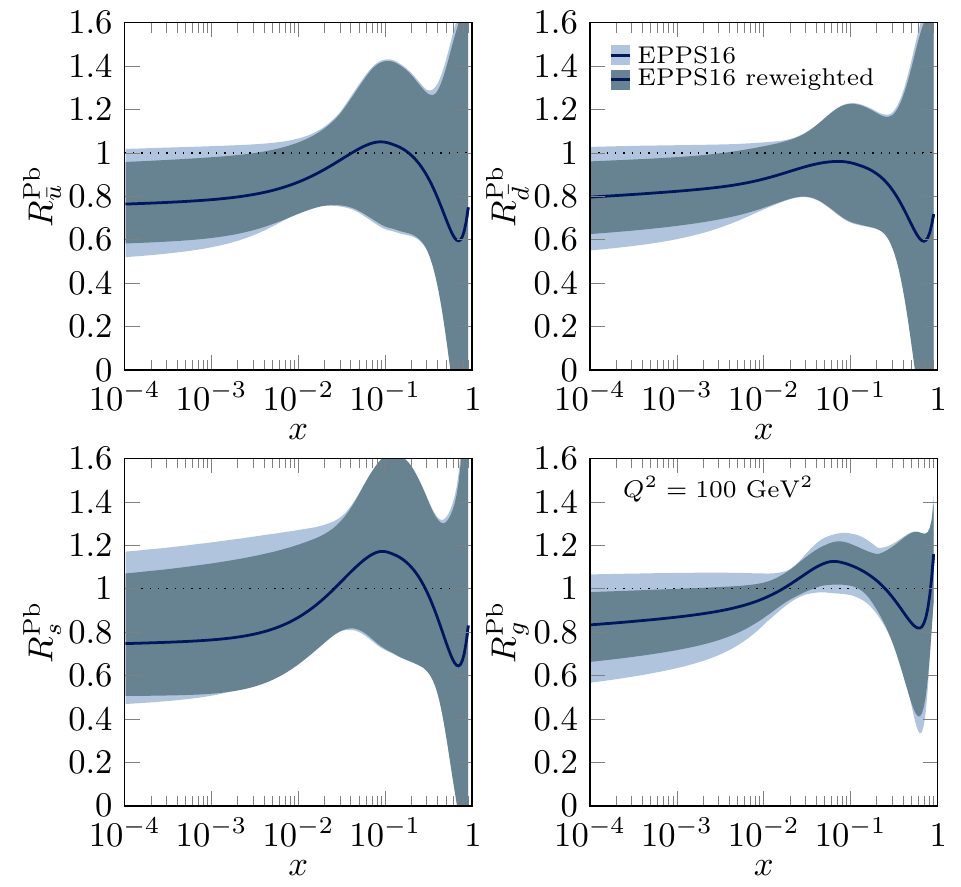}
\caption{Left: The LHCb projections~\cite{LHCb-CONF-2018-005} on $R_{\rm pPb}$ in  Drell-Yan process compared to the original EPPS16 error bands and those after reweighting with these data. Right: The improvement in EPPS16 nuclear PDFs for sea quarks and gluons at $Q^2=100\,{\UGeV^2}$ upon reweighting with the data shown in the left-hand panel.}
\label{fig:DYRpPb}
\end{figure}

\noindent {\bf Low-mass Drell-Yan:}
The Drell-Yan process at low invariant masses is a luminosity-hungry process where LHC Run 3 and Run 4 can make a difference. It would be very much advantageous to reach this low-mass region experimentally as it offers a possibility to constrain the nuclear PDFs at lower factorisation scale where the nuclear effects are larger. The estimated impact is shown in Fig.~\ref{fig:DYRpPb} where the LHCb estimates for $R_{\rm pPb}$ are compared to the reduction of EPPS16 uncertainties upon performing a PDF reweighting~\cite{Paukkunen:2014zia} with these data described in detail in Ref.~\cite{LHCb-CONF-2018-005}. In the calculation, a full decay-lepton phase space has been assumed, but this has a feeble effect on $R_{\rm pPb}$. Although the Drell-Yan process occurs predominantly via $q\overline{q}$ annihilation, the scale-evolution effects are large and these data mostly constrain the gluon PDFs. Here, it should be noted that the probed $x$ values are already so small, that the parametrisation bias which is prominent at small $x$ \cite{Aschenauer:2017oxs, Paukkunen:2017phq} is probably significant and understates the "true" effect of these measurements. This kind of scan over a wide range of invariant masses with the LHC Run~3 and 4 precision would offer a possibility to test the $Q^2$ evolution of nuclear PDFs -- whether there are corrections to standard DGLAP -- much more systematically than only on-shell W$^\pm$- and Z-production measurements do (involving only one scale).

%




\subsubsection{Heavy quark production}

\noindent\textbf{Inclusive Heavy Quark Production:} 
A recent study of  inclusive heavy quark
production in proton--lead collisions at the LHC 
demonstrates this can also help constrain  the gluon distribution in nuclei.
Specifically, Ref.~\cite{Kusina:2018pbp} makes use  of LHC \pPb data on 
$\mathrm{D}^0, \mathrm{J/}\psi, \mathrm{B}\to \mathrm{J/}\psi,   \PGUP{1S}$ meson production~\cite{Abelev:2014hha,Adam:2016ich,Aaij:2017gcy,Adam:2015iga,Aaboud:2017cif,Abelev:2013yxa,Aaij:2013zxa,Aaij:2017cqq,LHCb:2018jln,Abelev:2014oea,Aaij:2014mza,ATLAS:2017dgr,Sirunyan:2018sav,Sirunyan:2018pse,Aaij:2018ogq,Aaij:2018iyy}.
They obtain a consistent description of these data assuming nPDF
modifications are the dominant source of nuclear modifications in \pPb
collisions. Under this assumption, a clear confirmation of gluon
shadowing at small $x$ is found.  Additionally, they demonstrate that
the inclusion of such heavy-flavour data in a global fit can
significantly reduce the uncertainty on the gluon density down to
$x\simeq 7 \times 10^{-6}$ while keeping an agreement with the other
data of the global fits.  A reweighting of the current nPDFs sets with
the LHC heavy-flavour data was performed for the nCTEQ15 and EPPS16
sets.  In a recent analysis of beauty hadrons by the LHCb
collaboration~\cite{Aaij:2013zxa,Aaij:2017cqq,LHCb:2018jln} excellent agreement between the reweighted
predictions and the measured data for the nuclear modification factor
$R_{\rm pPb}$ was found. Furthermore, the precision of these data will
allow to further reduce the uncertainty of the low-$x$ nuclear gluon
distribution. Due to the lower production rates for beauty hadrons,
this kind of observables would clearly profit from a higher
luminosity as shown in Fig.~\ref{fig:POWLANG-small1} in the heavy-flavour chapter.  Note that heavy flavour measurements from LHCb extending
to larger transverse momenta have also the potential to constrain the
intrinsic heavy quark component~\cite{Kniehl:2009ar,Brodsky:2015fna,Lyonnet:2015dca}.
%

\noindent\textbf{Prompt Diphoton Production:} 
As discussed in Sec.~\ref{subsubsec:coherent}, other nuclear effects such
as coherent small angle gluon radiation may explain the heavy-flavour
data without a strong shadowing of the small-$x$ gluon.  A way to
contrast parton shadowing against effects from coherent energy loss discussed in section~\ref{subsubsec:coherent},
will be to study hard processes with color neutral final states in the
context of a global analysis in order to see whether a coherent
description of all the data remains possible. The case of vector boson
production has already been discussed above which is rather sensitive
to the quark distributions.  Interesting processes (with more or less
color neutral final states) which are sensitive to the gluon
distribution are inclusive prompt photon production and diphoton
production. The prompt photon observable has been discussed in Section~\ref{Sec:Focal} in the context of the ALICE upgrade.  The diphoton production is rather clean with an
essentially color neutral final state assuming that the contribution
from the fragmentation of quarks and gluons into photons can be
strongly suppressed by photon isolation criteria~\cite{ATLAS:2017ojy,Acharya:2018dqe}.
Due to the small
diphoton cross section, this measurement requires high luminosities. Based on the available calculations at next-to-leading order~\cite{Catani:2018krb,Boussaha:2018egy}, about 6000~events within the fiducial experimental acceptance used in \pp collisions by ATLAS can be expected with 2~pb$^{-1}$ \pPb~collisions.

\noindent\textbf{Heavy Quark Associated Production:}
The associated production of a heavy quark $Q$ and a vector boson $\gamma/Z/W^\pm$ 
also provides incisive information  about the PDFs.  
For all these processes, the LO contribution   comes from the gluon--heavy-quark (gQ) initiated
subprocess, making this process very sensitive to the gluon and the heavy quark nuclear parton
densities.
For the neutral current processes, a prompt photon $\gamma$ or Z together with a
c or b quark can be considered to obtain information of the c and b PDFs, respectively; 
for the charged current process, the  $Wc$ and $Wb$ final states are sensitive to the strange and charm quark, respectively. 
These channels have been analysed for the LHC \pp data~\cite{Dunford:2013lya,Beauchemin:2013cra,Ciaccio:2013rka,Candelise:2015gxa,Chatrchyan:2013uja,Aad:2014xca}
and the $Wc$ channel is a key input for the ATLAS/CMS comparison of the strange sea quark content of the proton~\cite{Cooper-Sarkar:2018ufj}.
The event statistics in 2~pb$^{-1}$ \pPb~collisions can be expected to be a factor 10 smaller than in the Run~1 CMS~\cite{Chatrchyan:2013uja} and ATLAS~\cite{Aad:2014xca} allowing for a first measurement in \pPb~collisions.

Additionally, the $\gamma/Z$ transverse momentum can be used to gauge the
initial energy of the massive parton propagating through the dense QCD
medium produced in those collisions, making $\gamma$/Z+ Q production a
powerful process in order to probe energy loss dynamics in the
heavy-quark sector.  
Furthermore, the comparison of the photon-jet pair momentum, from pp
to \PbPb collisions, is sensitive to the amount of energy lost by the
heavy-quarks and could therefore be used in order to better understand
parton energy loss processes in the heavy quark sector.

\subsubsection{Top production}
\label{sec:nPDF_top}

\begin{figure}[tb!]
\centering
\includegraphics[width=0.495\textwidth]{\main/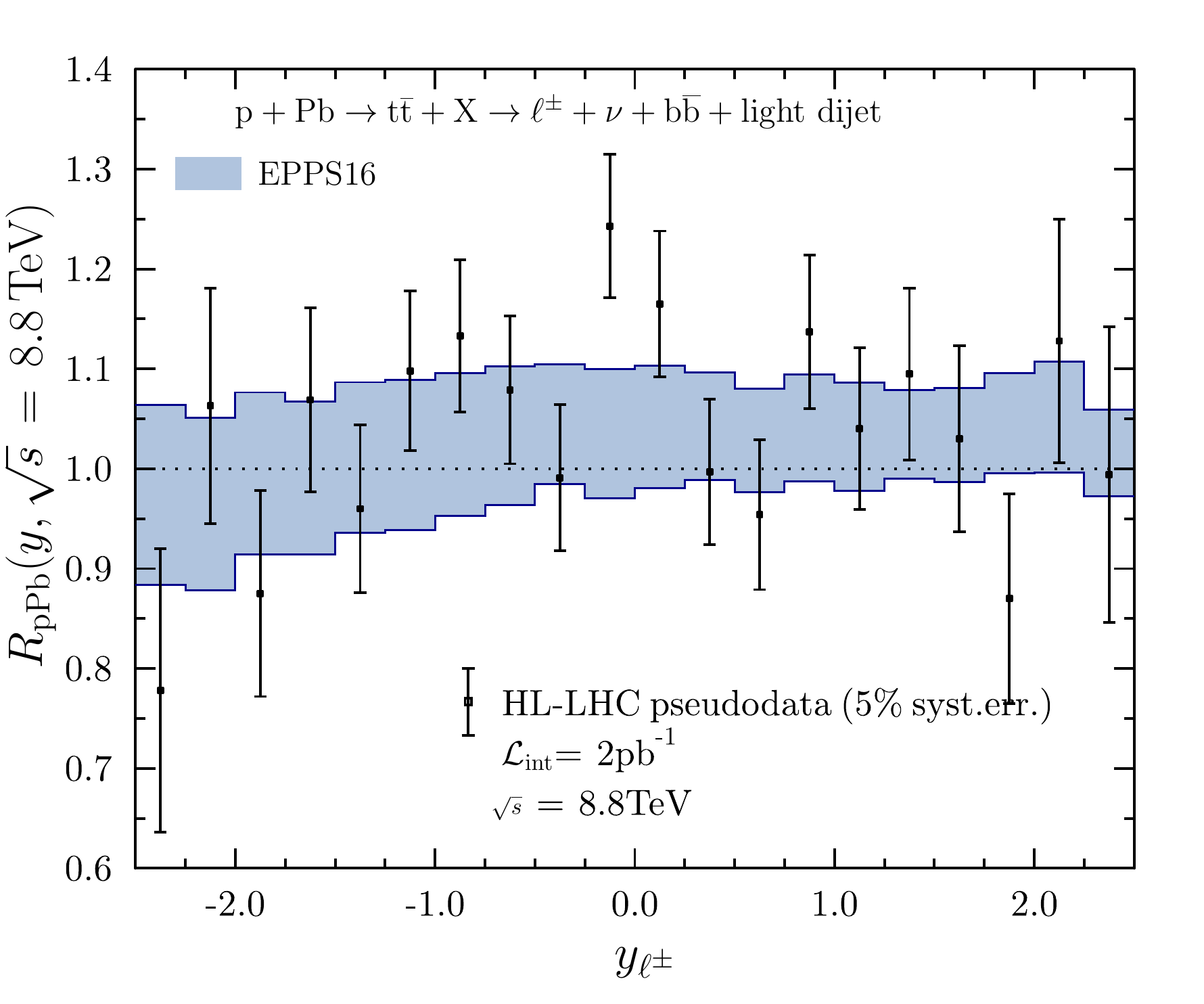}
\includegraphics[width=0.495\textwidth]{\main/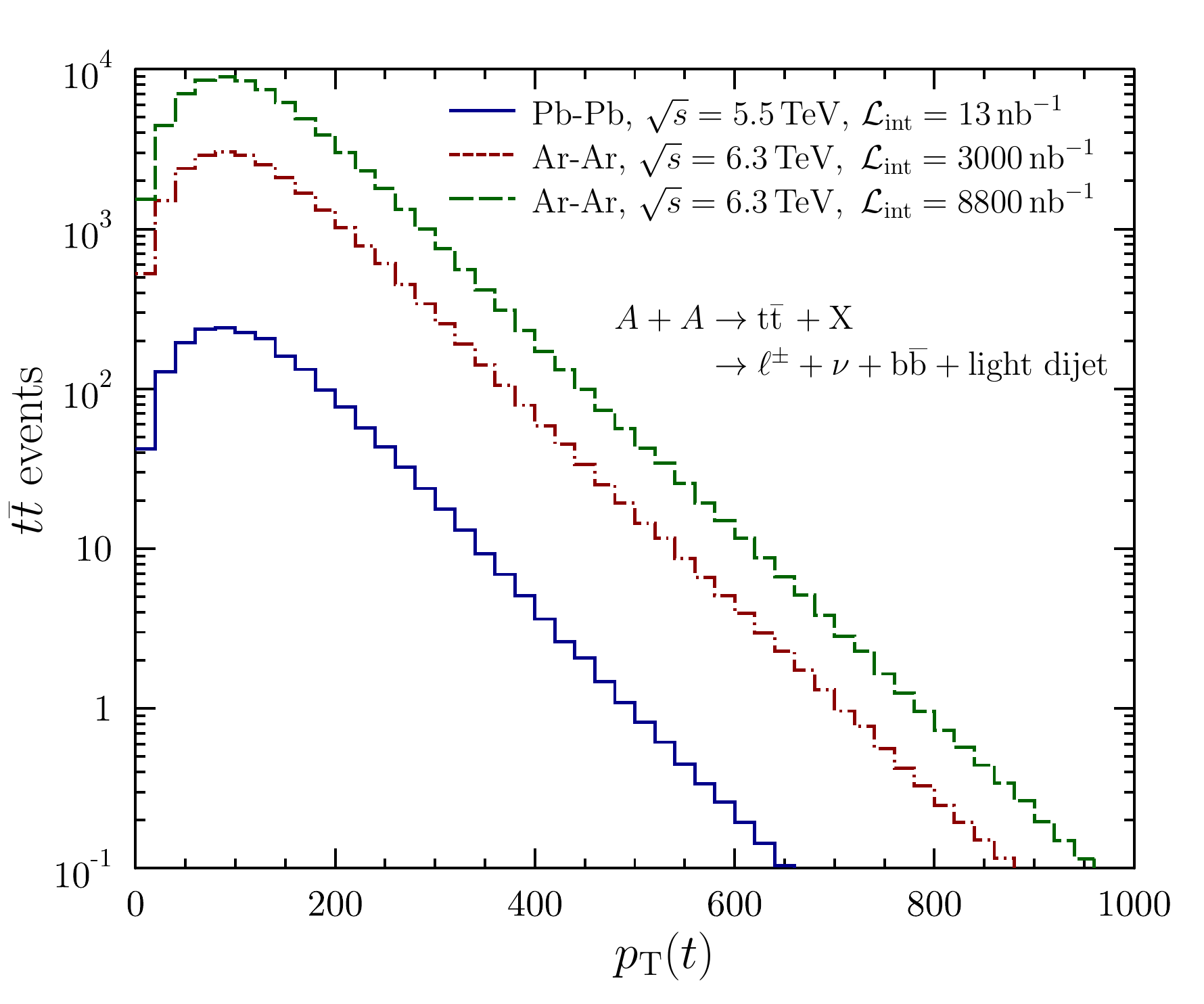}
\caption{Left: Projected data for \ttbar~production during Run~3 and 4 compared with uncertanties of the EPPS16  pdf-set. Right: Number of expected \ttbar events in \ArAr collisions and in \PbPb as a function of the top-quark transverse momentum. the used luminosities are discussed in the text.}
\label{fig:ttbardiffy}
\end{figure}

The $t\overline{t}$ production provides a complementary observable to the dijets for probing high-$x$ gluons \cite{dEnterria:2015mgr,Sirunyan:2017xku}. For the very short t-quark life time, they decay, in practice exclusively, to W~boson and b~quarks. Although, the cleanest channel is the one where both W bosons decay leptonically, it has been recently demonstrated~\cite{Sirunyan:2017xku} that it is possible to get a clear signal when one of the two W~bosons decays leptonically and the other one goes to light-quark jets. This is the preferred observable as the yields are 3 times higher than when both W bosons decay leptonically. The cross sections using the MCFM~code \cite{Campbell:2015qma}, process 146 (where the leptonically decaying W comes from top quark) have been estimated at NLO level. The following fiducial cuts have been applied:
$$ \pT^{\ell^\pm} > 30 \, {\rm \UGeVc}, \ |y^{\ell^\pm}| < 2.5, \ \pT^{\rm missing} > 30 \, {\rm \UGeVc}, $$
$$ \pT^{\rm jets} > 30 \, {\rm \UGeVc}, \ |y^{\rm jets}| < 2.5, R_{\rm isolation}^{\rm jets} = 0.5. $$

In this setup, the NLO calculation with the factorisation and renormalisation scales set to the top-quark mass yields, per-nucleon, $\sigma_{\rm n-n} \approx 4200 \, {\rm fb}$ in \pPb collisions at $\sqrt{s} = 8.8 \, {\rm TeV}$, with CT14NLO~\cite{Dulat:2015mca} proton PDFs and EPPS16~\cite{Eskola:2016oht} nuclear corrections. This is to be multiplied by a factor two to account for the electron and muon final states and by a factor of two to account for the process where the antitop is the one from which the leptonically decaying W originates from. Since this is the nucleon-nucleon cross section, a factor of 208 has to be still multiplied to get the \pPb cross section. This yields $\sigma_{\rm pPb} \approx 3.5 \, {\rm nb}$. Considering the 2000$\, \rm nb^{-1}$ scenario, and efficiency of 0.75~\cite{dEnterria:2015mgr} in b-jet tagging, around 5200 events are expected. Based on this number,  the expected nuclear modification factor $R_{\rm pPb}$ following the steps of Ref.~\cite{dEnterria:2015mgr} has been estimated, assigning each data point a 5\%~uncorrelated systematic uncertainty (in the current \pPb measurement of CMS~\cite{Sirunyan:2017xku} the systematic uncertainty is $\sim$20\%). Dividing the $|y^{\ell^\pm}| < 2.5$ interval to 20 bins, the statistical uncertainty varies from 5\% ($|y^{\ell^\pm}| < 1$) to 10\% (most forward/backward). The resulting $R_{\rm pPb}$ is compared in Fig.~\ref{fig:ttbardiffy} with the EPPS16 uncertainty band. In the considered kinematic configuration the expected $R_{\rm pPb}$ is typically a bit above unity for the gluon antishadowing in EPPS16. As can be seen from the left-hand panel of Fig.~\ref{fig:ttbardiffy}, the expected precision of the measurement is not enough to give significant constraints on nuclear PDFs. In particular, the dijets considered in Section 11.4.3 will probe the same kinematic configuration with a clearly higher precision. However, here much depends on the expected systematic error (taken here to be 5\% for each data point separately) and how large are the bin-to-bin correlations. In the p-Ar mode, the higher c.m. energy of $\sqrt{s} \approx 9.4 \, {\rm TeV}$ increases the yields around 50\% and would also benefit from the higher luminosities, see the next subsection and Sec.~\ref{sec:lightions}.

This, plus a higher achievable nucleon-nucleon luminosity 
would render the above case almost completely systematics dominated. 

Using the same framework and assumptions as above, the fiducial $t\overline{t}$ yields in $A$-$A$ collisions have been estimated. Here, the \PbPb and \ArAr cases have been considered. In \PbPb collisions at $\sqrtsNN = 5.5 \, {\rm TeV}$, the per-nucleon cross section is $\sigma_{\rm n-n} \approx 1200 \, {\rm fb}$ which translates to around 2000 reconstructed events for 13$\, \rm nb^{-1}$ ion-ion luminosity. In the \ArAr option, the c.m. energy is slightly higher, $\sqrt{s} = 6.3 \, {\rm TeV}$, which increases the cross section by some 50\%. In addition, the achievable ion-ion luminosity is much higher, 3000$\,\rm nb^{-1}$-8800$\,\rm nb^{-1}$ within 2.75 months of running, see Tables in Chapter~\ref{sec:lightions}. 
Thus, the estimated amount of $t\overline{t}$ events is clearly larger, around 30000. The right-hand panel of Fig.~\ref{fig:ttbardiffy} shows the event distributions as a function of top-quark transverse momentum $\pT(t)$. This shows that, the HL-LHC may allow to probe the space-time picture of heavy-ion collisions using top quarks \cite{Apolinario:2017sob} up to $p_{\rm T}(t) \approx 400\,{\UGeVc}$ in the \PbPb case, and up to $p_{\rm T}(t) \approx 700\,{\UGeVc}$ in the \ArAr alternative.

\subsection{Perspectives with lighter ions}
\label{sec:nPDF_lightions}

Lighter ions, with the possibility to achieve large integrated luminosities in modest running times, see Sect.~\ref{sec:lightions} in the accelerator chapter, offer several interesting opportunities for the study of the initial stage of ion collisions, for small-$x$ physics and for the determination of nuclear parton densities, see Section~\ref{sec:smallxIntro}.

\begin{figure}[tb!]
\centering
\includegraphics[width=0.485\textwidth]{\main/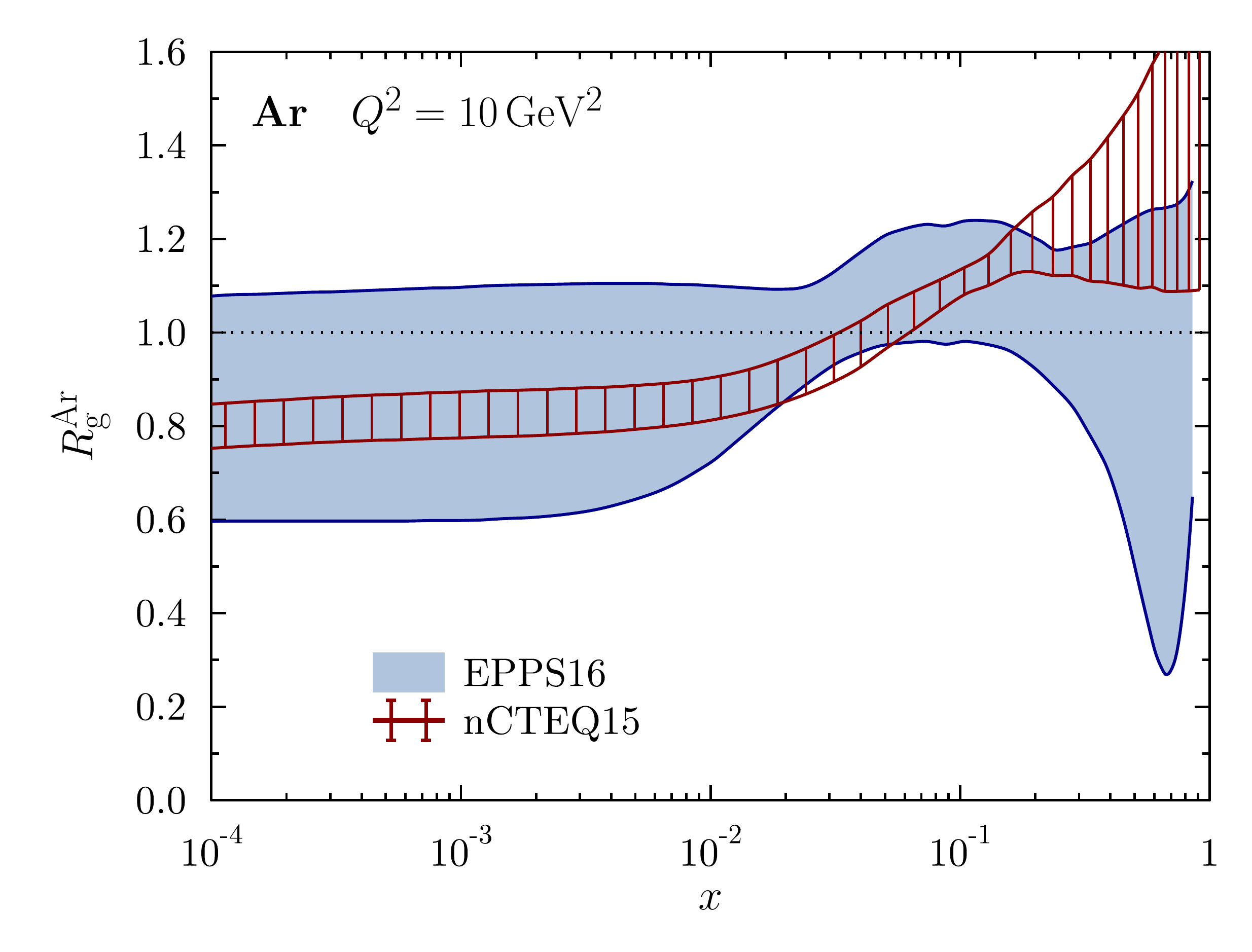}
\includegraphics[width=0.485\textwidth]{\main/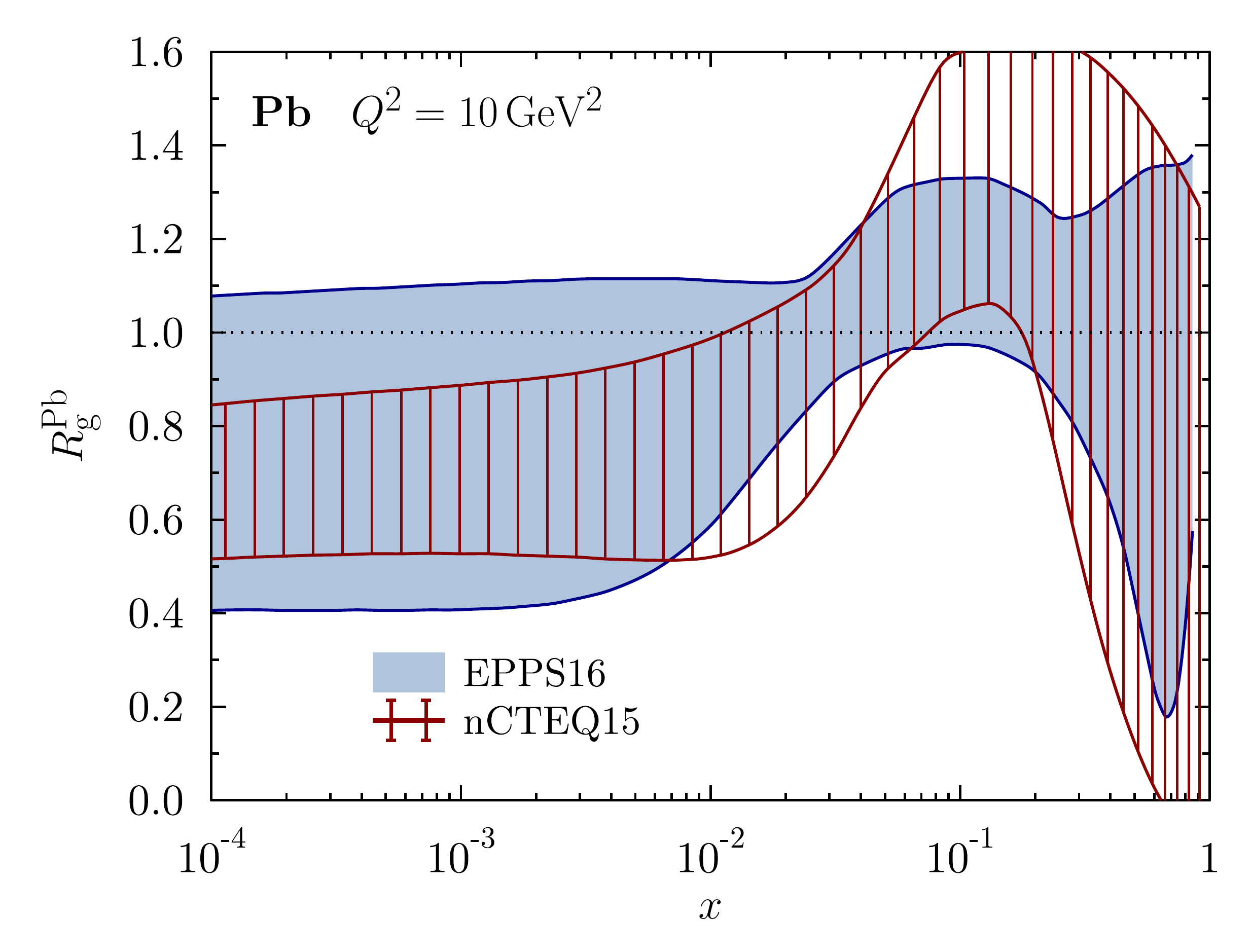}
\caption{The EPPS16 and nCTEQ15 nuclear PDF modifications for gluons at $Q^2=10\,{\rm GeV^2}$. 
Left-hand panel corresponds to the Argon nucleus and the right-hand panel to the Lead nucleus. From Ref.~\cite{Paukkunen:2018kmm}} 
\label{fig:lightnPDF}
\end{figure}

First, concerning nPDFs, it should be noted that due to the scarcity of nuclear data, a PDF fit or a single nucleus is impossible as discussed in Section~\ref{sec:smallxIntro}. The different groups~\cite{deFlorian:2011fp,Kovarik:2015cma,Eskola:2016oht}  have adopted different strategies but, generically, they give the parameters in the initial condition to be fitted a dependence on the nuclear mass number. Such dependence acquires different functional forms and, therefore, it constitutes part of the parametrisation bias in the nPDF set. Data on lighter nuclei may help to constrain such parametrisations, see the discussions for UPCs and \pA~collisions in Section~\ref{sec:nPDFs}. To highlight the current unknowns, Fig.~\ref{fig:lightnPDF} compares the nuclear modifications for Argon and Lead, as given by the EPPS16 and nCTEQ15 global analyses. In particular, the nCTEQ15 prediction varies, even qualitatively, quite significantly between Argon and Lead. This underscores the usefulness of e.g.\,a p-Ar run at the LHC.  

On the other hand, the impact parameter dependence of nPDFs is linked to their dependence on nuclear size. Several models exist (see e.g.\,\cite{Emelyanov:1998phs,Ferreiro:2008wc}), and even less model-dependent approaches like the EPS09s analysis~\cite{Helenius:2012wd} where the dependence on nuclear size was used to constrain the impact parameter dependence. First-principle calculations combining the Gribov theory of inelastic shadowing and factorisation theorems for hard diffraction and DIS relate diffraction in electron-proton collisions with nuclear shadowing. This has been used to predict nuclear shadowing~\cite{Frankfurt:2011cs,Armesto:2003fi}, including its nuclear size and impact parameter dependence. While such relation is exact for the deuteron, its extension to larger nuclei has some degree of model dependence. Lighter ions are the ideal place to test the nuclear size dependence without resorting to centrality selection, whose relation with impact parameter is doomed to be as problematic - at least - as found in \pPb~collisions at the LHC.

Lighter ions also offer large luminosities that are important for several aspects:
\begin{itemize}
\item Data on beauty mesons and bottomonium in \pA~collisions can be used to constrain nPDFs~\cite{Kusina:2017gkz} better than their charm counterparts, because of the larger scale given by the mass and by the opportunity that they are less affected by collective effects. 
But they demand large statistics that can be achieved with lighter ions.
\item Larger luminosities will benefit measurements in \pA collisions for observables with large scales, like high-mass DY or dijets, for precise determination of nPDFs.
\item UPCs and photon-photon studies~\cite{Baltz:2007kq} in \AOnA~collisions will greatly benefit in spite of the $Z^{2(4)}$ factor for the photon luminosities that can be overcompensated by the larger nucleon-nucleon luminosity.
\item Larger luminosities can also benefit small-$x$ forward observables that, like dijets \cite{Dainese:2016gch}, aim to reach quite large transverse momenta in \pA~collisions.
\end{itemize}

Lighter ions offer a bridge between small systems and \PbPb without requiring centrality selection that is problematic both in \pp, \pA and peripheral \AOnA. 
In the framework of saturation models~\cite{Gelis:2010nm} that aim to describe collective effects in small systems without requiring final state interactions (see e.g.\,\cite{Kovner:2016wsq} and references therein), the extension from the proton to the nuclear case in some of the phenomenological realisations  is done by a simple rescaling of the squared saturation momentum $Q_s^2\propto A^{1/3}$.
And the centrality dependence is assumed to be proportional to the nuclear profile, which leads to strong problems in the nuclear periphery where a dilute situation is restored. Lighter ions offer a check of our ideas on the
nuclear size versus energy leading
the density that determines saturation, and the use of minimum bias observables
instead of centrality-sliced ones that
would greatly simplify the phenomenology.

To conclude, lighter ions offer several advantages and disadvantages for initial stage studies. The main disadvantage is the fact that theory calculations usually assume the limit of scattering of a dilute projectile (proton) on a dense target (nucleus). Lighter ion-ion collisions are further from this limit. On the other hand, they offer: (i) a bridge between small and large systems without resorting to
centrality selection that would be useful for constraining the nuclear wave function, the
collision dynamics and the interpretation of collectivity;
(ii) the possibility of larger luminosities for UPCs and forward observables
for nPDFs determination and small-x studies; and (iii)
a more affordable setup for microscopic calculations of nuclear corrections.

\clearpage

\section{Other opportunities with ion and proton beams at the LHC}
\subsection{Physics motivation for collisions of light ions}
\label{sec:smallAsum}
{ \small
\noindent \textbf{Coordinator}: Zvi Citron (Ben-Gurion University of the Negev)

\noindent \textbf{Contributors}:
L.~Apolinario (LIP and IST Lisbon),
A.~Dainese (INFN Padova), 
J.F.~Grosse-Oetringhaus (CERN),
J.M.~Jowett (CERN),
Y.-J.~Lee (Massachusetts Institute of Technology),
C.~Loizides (Oak Ridge National Laboratory),
G.~Milhano (LIP and IST Lisbon, CERN),
A.~Milov (Weizmann Institute of Science),
J.~Pires (CFTP and IST Lisbon),
A.M.~Sickles (U. Illinois, Urbana-Champaign),
U.~Wiedemann (CERN),
M.~Winn (LAL, Orsay and IRFU/DPhN, Saclay)}

The collision of ion species with $A\ll A_\mathrm{Pb}$ is an appealing opportunity to expand the physics programme presented in this document.  The recent \XeXe run of only eight hours has provided valuable input for the physics performance of ion collisions lighter than Pb at the LHC.  Broadly, the advantages of using $A\ll A_\mathrm{Pb}$  collisions are twofold: smaller collision systems sample key physical parameters beyond what can be probed with \PbPb and \pPb collisions, and they allow higher luminosity running to maximize the accumulation of rare events in heavy-ion collisions.  This higher luminosity would enable high-precision measurements for currently rare observables in \PbPb collisions as well as the study of observables totally inaccessible in \PbPb collisions.

A scenario is envisioned in which the programme is extended in two 
directions: a) a short run of \OO ($A=16$) to study system-size dependence and b) longer running of a species of intermediate $A$ to achieve a large luminosity increase. The choice of the intermediate species will be dictated by the competition of increased luminosity with lower $A$ against the goal of studying the properties of an extended QGP system. Optimizing the choice of species will require further study from the accelerator, experimental, and theoretical communities; in this document \ArAr ($A=40$) collisions are considered as a test-case for the choice of intermediate ion. It is understood that any choice of collision species will likely require a pilot run prior to beginning dedicated collisions.

Section \ref{sec:lightions} describes the technical capabilities of the LHC to provide lighter-ion collisions, as well as the expected performance for several ion species.
For example, for \ArAr the expectation for one month of collisions is 1080\,$\invnb$ ($p=1.5$).  In order to compare across different collisions species we consider the nucleon--nucleon integrated luminosity per month of running, which could be larger by a factor 8--25 (for $p=1.5$--1.9) with respect to \PbPb collisions, i.e.\,one month of \ArAr collisions would be equivalent to $\sim 25$--80\,$\invnb$ of \PbPb collisions. This gain would be the same in all centrality classes (defined in terms of percentiles of the hadronic cross section).
Jet quenching estimates for light ion (Ar--Ar) collisions are presented in Sect.~\ref{sec:jetslightions}. 
Section~\ref{sec:flow_sizedep} discusses flow measurements with light ions. A discussion of the role that light ion collisions can play for small-$x$ and nPDF studies is reported in Sect.~\ref{sec:nPDF_lightions}.  Finally, the implications of \ArAr collisions for the study of light-by-light scattering are discussed in Sect.~\ref{sec:upc}. 

Even a short \OO run can help clarify the uncertainty concerning the onset of QGP or QGP-like phenomena in high-multiplicity pA and pp collisions, as discussed in section \ref{sect:smallsystems_OO}.  The search for signals associated with the QGP in \OO collisions should complement the searches in \pp and \pPb collisions.
In particular, the \OO system has well-understood collision geometry as described in detail in Sect.~\ref{sect:smallsystems_OO}, enabling the study of collisions with low values of $\langle N_{\rm part}\rangle$ that are difficult to select and study in \PbPb collisions and that are similar to those associated to high-multiplicity \pPb events. Colliding \OO at the LHC naturally dovetails with \pO collisions whose significance for cosmic-ray physics is detailed 
in Sect.~\ref{sec:pOcosmic}.

\begin{figure}
\centering
\includegraphics[width=0.68\linewidth]{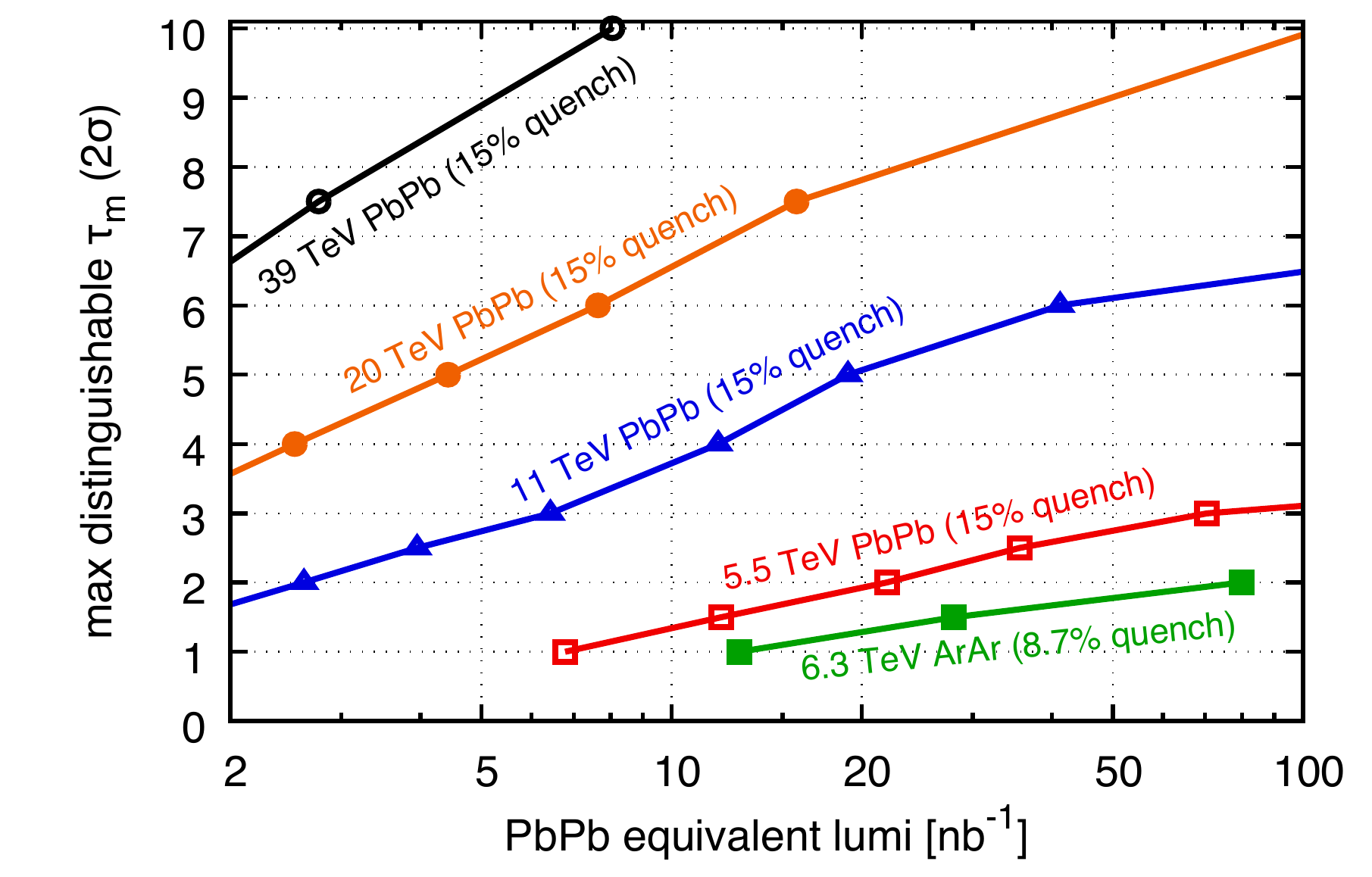}
\caption{Maximum medium lifetime that can be distinguished from a full quenching baseline with a statistical significance of two standard deviations (2\,$\sigma$), as a function of luminosity (shown in terms of equivalent Pb--Pb luminosity) for different species and collider energies. A single \ArAr run is expected to provide $\sim 25$--80 $\invnb$ of \PbPb equivalent luminosity. Adapted from~\cite{Apolinario:2017sob}.}
\label{fig:boosted_tops}
\end{figure}

Complementing \PbPb collisions, 
the possibility of high-luminosity extended LHC runs with intermediate-$A$ nuclei (e.g.\,\ArAr or \KrKr)
is an appealing long-term option.
The chief promise of these collisions is the possibility of reaching much higher luminosity than \PbPb collisions while still producing a QGP over an extended volume of $\sim 1000$~fm$^3$ in central events. Based on a Glauber Monte Carlo simulation \cite{Loizides:2017ack}, the mean number of participants for \ArAr ranges from  $\langle N_{\rm part}\rangle \sim 7$ for 60--80\% centrality to $\sim 70$ for 0--5\% centrality collisions.  
QGP effects are observed in \PbPb collisions with similar number of participants (see e.g.\,\cite{Sirunyan:2018eqi, ATLAS-CONF-2018-007}).  In addition, the much lower underlying event multiplicity in \ArAr relative to central \PbPb collisions is expected to lead to reduced  systematic uncertainties for several observables, from reconstructed jets to all signal affected by large combinatorial backgrounds.  These features, combined with the possibility to increase the nucleon--nucleon luminosity by more than one order of magnitude, make \ArAr collisions an extremely attractive option for hard-probe measurements that in \PbPb collisions are limited or impossible, such as boosted top-quark decay chains for QGP studies~\cite{Apolinario:2017sob}.  Figure \ref{fig:boosted_tops} extends the analysis to lighter nuclei and shows that one month of \ArAr collisions, with nucleon--nucleon luminosity equivalent to 25--80~$\invnb$ for \PbPb, allows a similar physics reach as the entire \PbPb future programme ($13~\invnb$), namely to probe the QGP density evolution up to a time of about 1.5--2~fm/$c$.  Top quark studies in \ArAr collisions in the context of constraints on nPDFs are discussed in Sect.~\ref{sec:nPDF_top}.

\begin{figure}
\centering
\includegraphics[width=0.68\linewidth]{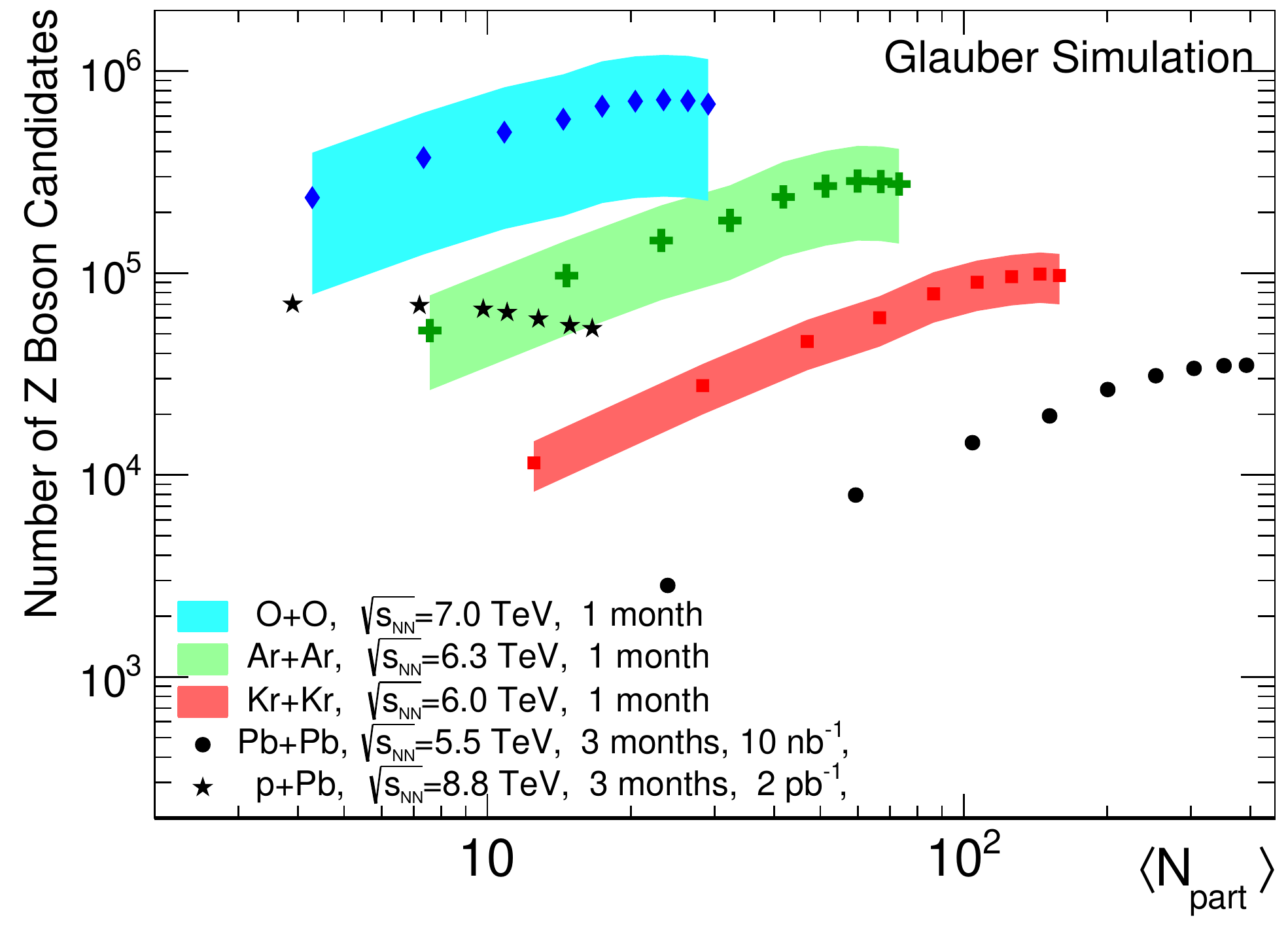}
\caption{
The number of Z bosons as a function of $\langle N_{\rm part}\rangle$ expected for one month of \OO,\ArAr, and \KrKr collisions at the LHC and for the full expected \PbPb and \pPb programmes.  The Z bosons are reconstructed via the di-lepton decay channel with leptonic $\pT>20$~GeV$/c$ and $|\eta|<2.5$, and a mass selection of $66<M_{\ell\ell}<116$~GeV.  
The bands shown indicate the range of the expected luminosity ranging from $p=1.5$ to $p=1.9$, as discussed in section \ref{sec:lightions}.
\label{fig:Zreach}}
\end{figure}

\begin{figure}
\centering
\includegraphics[width=0.68\linewidth]{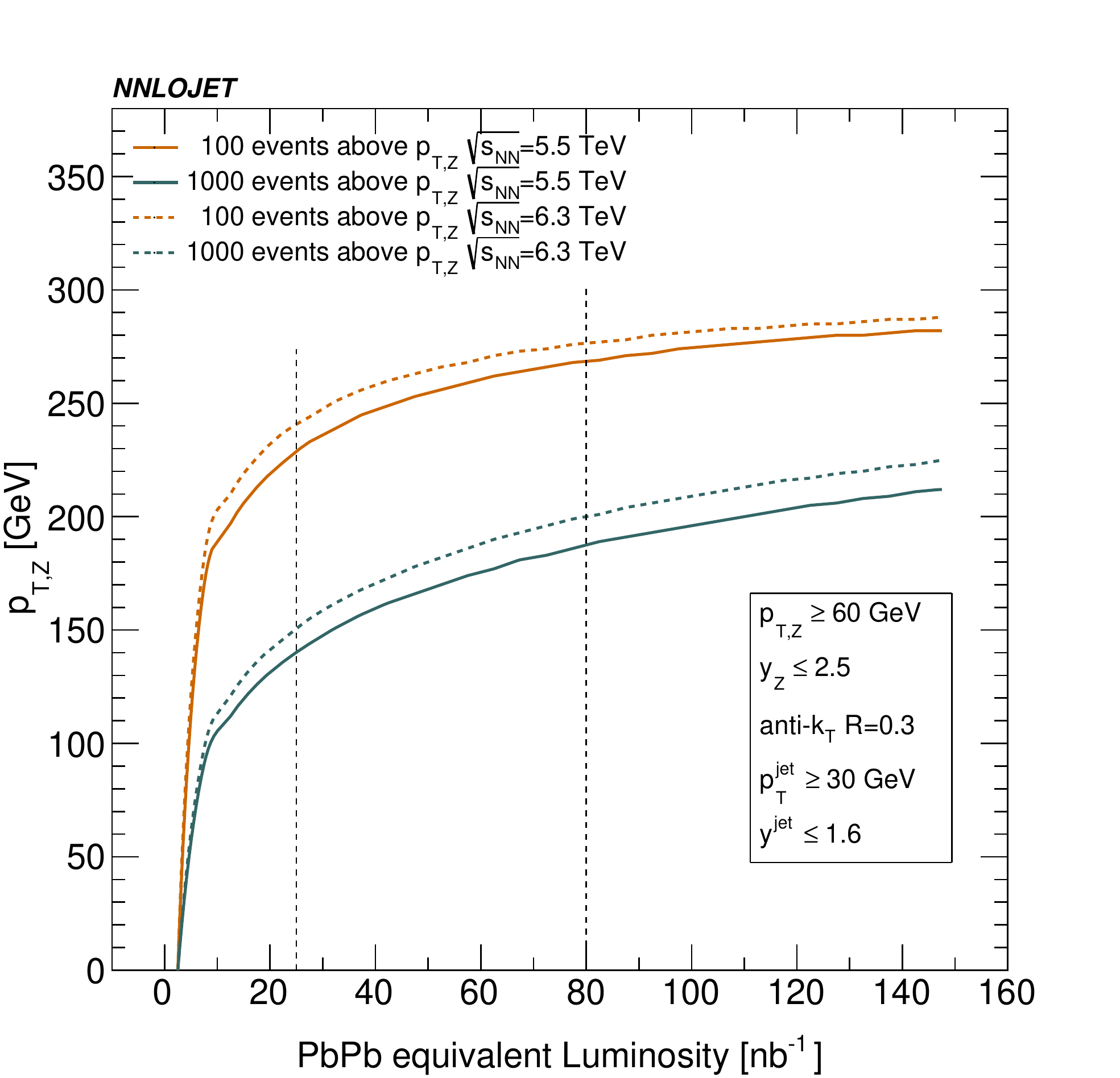}
\caption{
The kinematic reach of Z+jet events as a function of \PbPb equivalent luminosity.  The curves indicate the maximum \pt of the Z boson for which 100 (or 1000) events are expected from 0--10\% centrality collisions for a given \PbPb equivalent luminosity.  25 and 80\,$\invnb$, corresponding to the upper and lower expectations for one month of \ArAr collisions, are indicated by the vertical lines.
\label{fig:Zjetreach}}
\end{figure}

Studies with Z bosons are representative examples of the types of measurements that may be undertaken in a lighter ion rare-probes programme.  In Fig.~\ref{fig:Zreach} the expected number of Z boson candidates (assuming a selection similar to that used by ATLAS and CMS in previous studies) for one month of heavy-ion running  as a function of $\langle N_{\rm part}\rangle$ is shown for several colliding ion species, and compared with the expectation for the full \PbPb and \pPb programmes of Runs 3 and 4. The figure demonstrates that the overall yield of Z bosons would be considerably higher for one \ArAr run than for several years of \PbPb running including both a sufficient number of candidates to study low $\langle N_{\rm part}\rangle$ collisions unreachable with \PbPb collisions as well as moderate $\langle\Npart\rangle$ values in which QGP formation is expected.  
Z bosons are a powerful tool to probe the properties of the QGP in particular in Z+jet events. In these studies the energy of the Z is a direct measurement of the energy of parton that initiated the recoil jet. Therefore, the coverage of a broad range in Z momentum gives access to a jet-energy differential study of jet quenching. the  The expected number of Z+jet events from the 0--10\% centrality class above a given $\pt$ of the Z boson is calculated at NLO (and without including jet suppression) and shown in Fig.~\ref{fig:Zjetreach} as a function of \PbPb equivalent luminosity. The Pb--Pb programme in Runs 3 and 4 (13\,nb$^{-1}$) gives 1000 events with $Z$ $\pt>120$~GeV/$c$, while a single Ar--Ar one-month run extends the coverage with the same number of events to 140--180~GeV/$c$. A three-months Ar--Ar programme extends well above 200~GeV/$c$.

\begin{figure}
\centering
\includegraphics[width=0.58\linewidth]{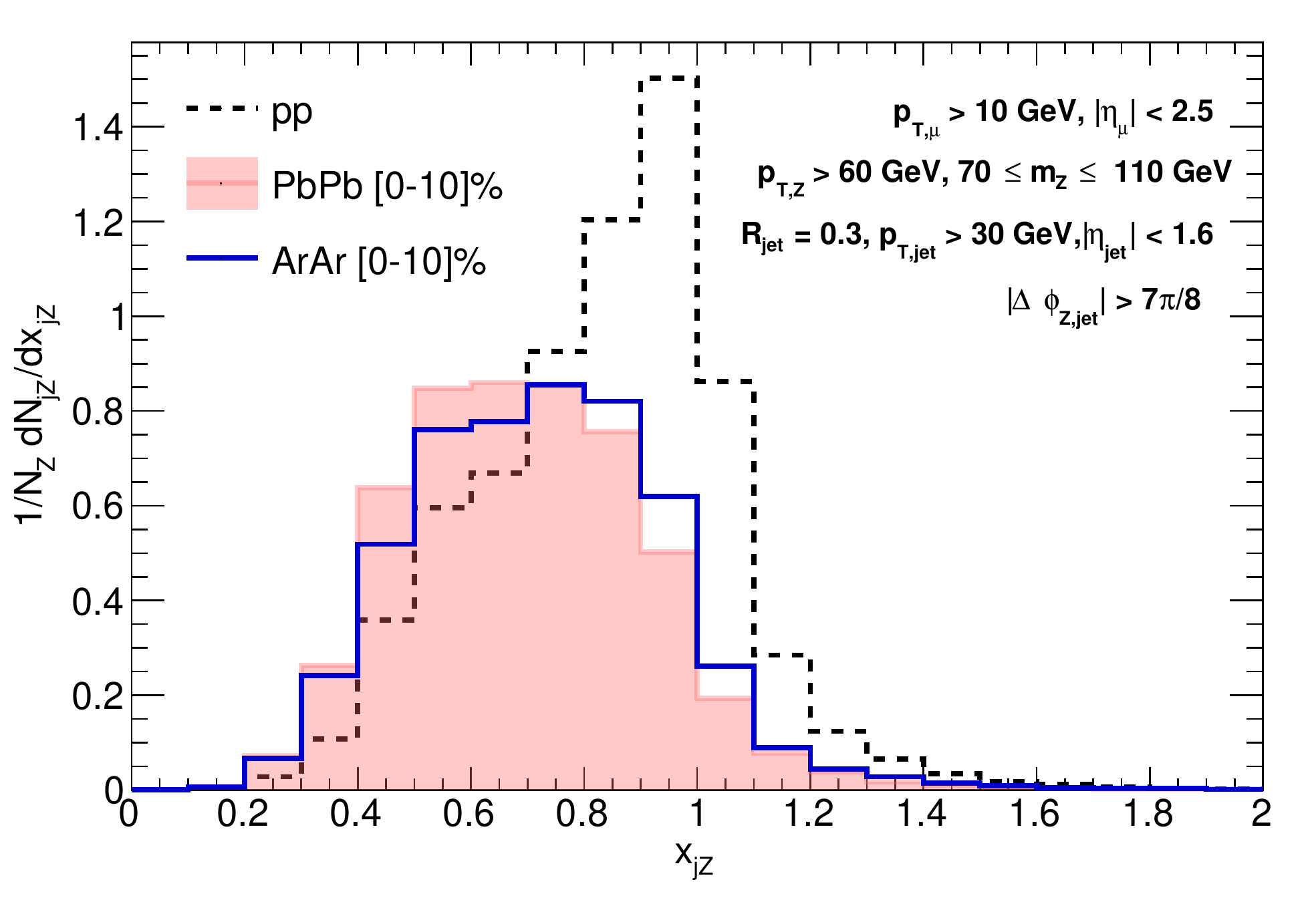}
\caption{
The $x_{\rm jZ}$ distribution for pp, \PbPb, and \ArAr collisions calculated with J\textsc{ewel}.  The 10\% most central events are shown for \PbPb and \ArAr.  \ArAr collisions are calculated as $N_{\rm part} = 60$, $T=318$\,MeV at thermalization time $\tau = 0.63$ fm/$c$ for $\sqrt{s}=6.3$ TeV.  The Z boson must have $\pt > 60$~GeV/$c$ and be back-to-back ($|\Delta\varphi| > 7/8\pi$) to a jet with $\pt > 30$~GeV/$c$.    
\label{fig:ArAr_xjz}}
\end{figure}

For further study, Z+jet events were simulated in the 10\% most central events in \ArAr collisions using Jewel\ \cite{Zapp:2009ud} to estimate the expected jet-quenching effects.  Details of the use of and limitations of J\textsc{ewel} for this purpose are discussed in Sect.~\ref{sec:jetslightions}. Figure \ref{fig:ArAr_xjz} shows the distribution of $x_{\rm jZ}$, the ratio of the jet transverse momentum to that of the Z boson for 0--10\% centrality \ArAr events, as well as \pp collisions and 0--10\% centrality \PbPb events ($N_{\rm part} = 356$, $T=260$\,MeV at thermalization time $\tau = 0.6$~fm/$c$ for $\sqrt{s}=5.02$~TeV).  The Z boson must have $\pt > 60$~GeV/$c$ and be back-to-back ($|\Delta\varphi| > 7/8\pi$) to a jet with $\pt > 30$~GeV/$c$.  The Figure clearly shows that for this observable the jet-quenching phenomena observed in \PbPb collisions as modelled by J\textsc{ewel} are present also in \ArAr collisions.  More studies are needed to refine modelling of more dilute systems and optimize the choice of colliding species, but taken together the available information suggests 
the potential of light ion collision systems like \ArAr for a heavy-ion rare-probes programme.

\clearpage

\subsection{Physics of $\mathrm{\gamma \gamma}$ interactions in heavy-ion collisions}
\label{sec:upc}

{ \small
\noindent \textbf{Coordinators}: Iwona Grabowska-Bold (AGH University of Science and\ Technology)

\noindent \textbf{Contributors}: 
M.~Dyndal (DESY), 
S.~Hassani (Universit\'e Paris-Saclay), 
M.~Klusek-Gawenda (IFJ PAN, PL-31342 Krak\'ow, Poland), 
L.~Schoeffel (Universit\'e Paris-Saclay), Peter Steinberg (BNL)
}

Heavy-ion beams are composed of nuclei which carry electric charge $Ze$~($e$ is the electron charge and $Z$ is the atomic number). They are accelerated to nearly the speed of light, thus they generate large electromagnetic~(EM) fields. 
The EM fields generated by the relativistic ion can interact with the other nucleus or its EM fields.
Therefore, besides nuclear hadronic interactions, EM
interactions also occur in ultra-relativistic heavy-ion collisions.
These EM interactions can be studied in so-called ultra-peripheral
collisions~(UPC) which occur when the distance between two nuclei in the transverse plane is larger than two times the nuclear radius, and hadronic interactions are thus suppressed~\cite{Bertulani:2005ru}.

A broad range of processes can be studied with $\mathrm{\gamma\gamma}$ interactions in UPC. In the following, a few examples of photon-induced processes are considered at the HL-LHC: exclusive production of $\mathrm{\mu^+\mu^-}$ or p$\mathrm{\bar{p}}$ pairs, a rare process of light-by-light~(LbyL) scattering and a potential of searches for axion-like particles~(ALP).

\begin{figure}[!hbt]
\centering
\includegraphics[width=0.50\linewidth]{\main/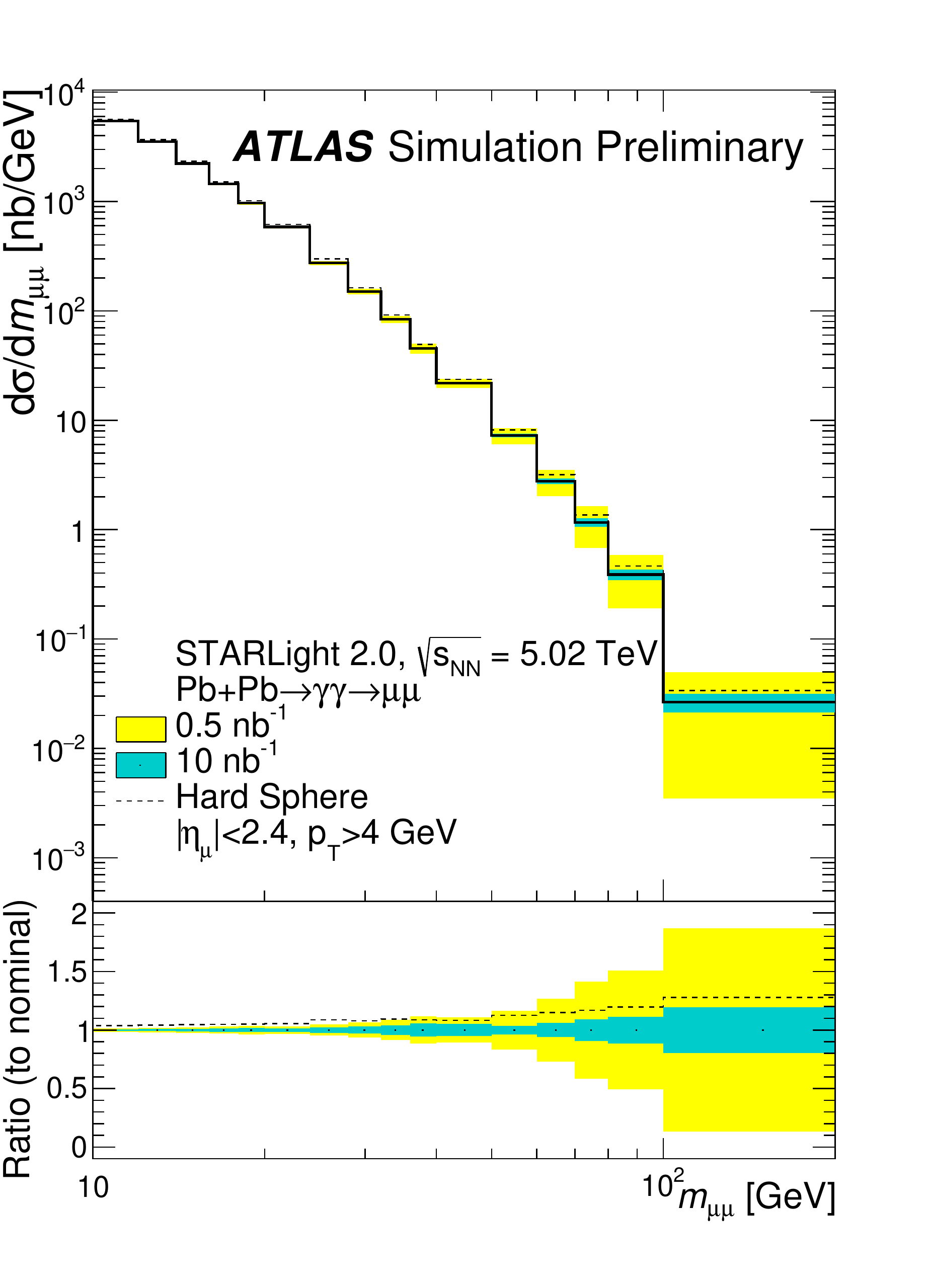}
\caption{
(Upper)~Differential cross section for exclusive production of the di-muon pairs as a function of the di-muon mass for
$10<m_{\mathrm{\mu\mu}}<200$~GeV extracted from STARLight. Two
scenarios are considered for the nuclear geometry: a realistic skin
depth of the nucleus~(solid line) or a hard sphere~(dashed
line). (Bottom)~Ratio to nominal as a function of the di-muon mass,
where "nominal'' stands for the realistic skin depth of the nucleus.
 Shaded bands represent expected statistical uncertainties associated with a number of signal events in each bin for integrated luminosity of $0.5~\mathrm{nb}^{-1}$~(yellow), and $10~\mathrm{nb}^{-1}$~(cyan).}
\label{fig:mumu}
\end{figure}

Exclusive production of di-muon pairs~($\mathrm{\gamma\gamma\rightarrow \mu^+\mu^-}$) in UPC can offer a precision measurement of photon fluxes associated with ion beams, and as such can be used to constrain predictions for the other processes covered in this section. The cross section at high pair mass is also sensitive to the nuclear geometry assumed in the calculations. Figure~\ref{fig:mumu} presents a differential cross section as a function of the invariant mass of the di-muon system in the range of 10--200~GeV with expected statistical uncertainties represented by two bands corresponding to integrated luminosities of $0.5~\mathrm{nb}^{-1}$ and
$10~\mathrm{nb}^{-1}$. Two scenarios are considered for the nuclear
geometry: a realistic skin depth of the nucleus or a hard sphere~\cite{Barrett:1977}.
For the $10~\mathrm{nb}^{-1}$ scenario, a significant reduction of the statistical uncertainty is expected. 
This will help in reducing uncertainties from the modelling of the nuclear geometry.
The expected upgrades of the ATLAS Zero Degree Calorimeters~(ZDC) in the LHC Run 3 will also be important for isolating the contributions to the cross section stemming from dissociative processes.

\begin{figure}[t]
        \includegraphics[scale=0.375]{\main/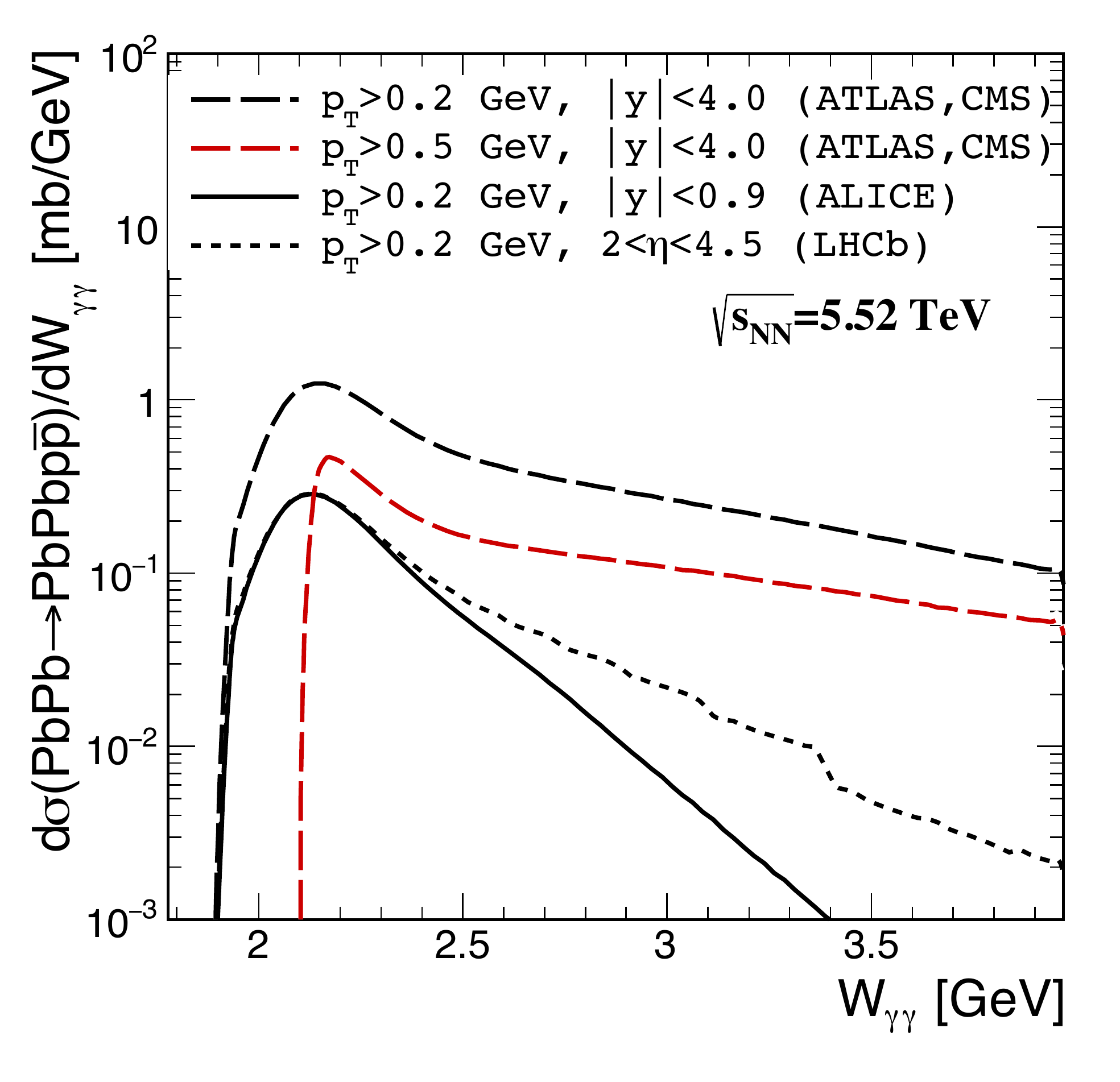}
        \includegraphics[scale=0.375]{\main/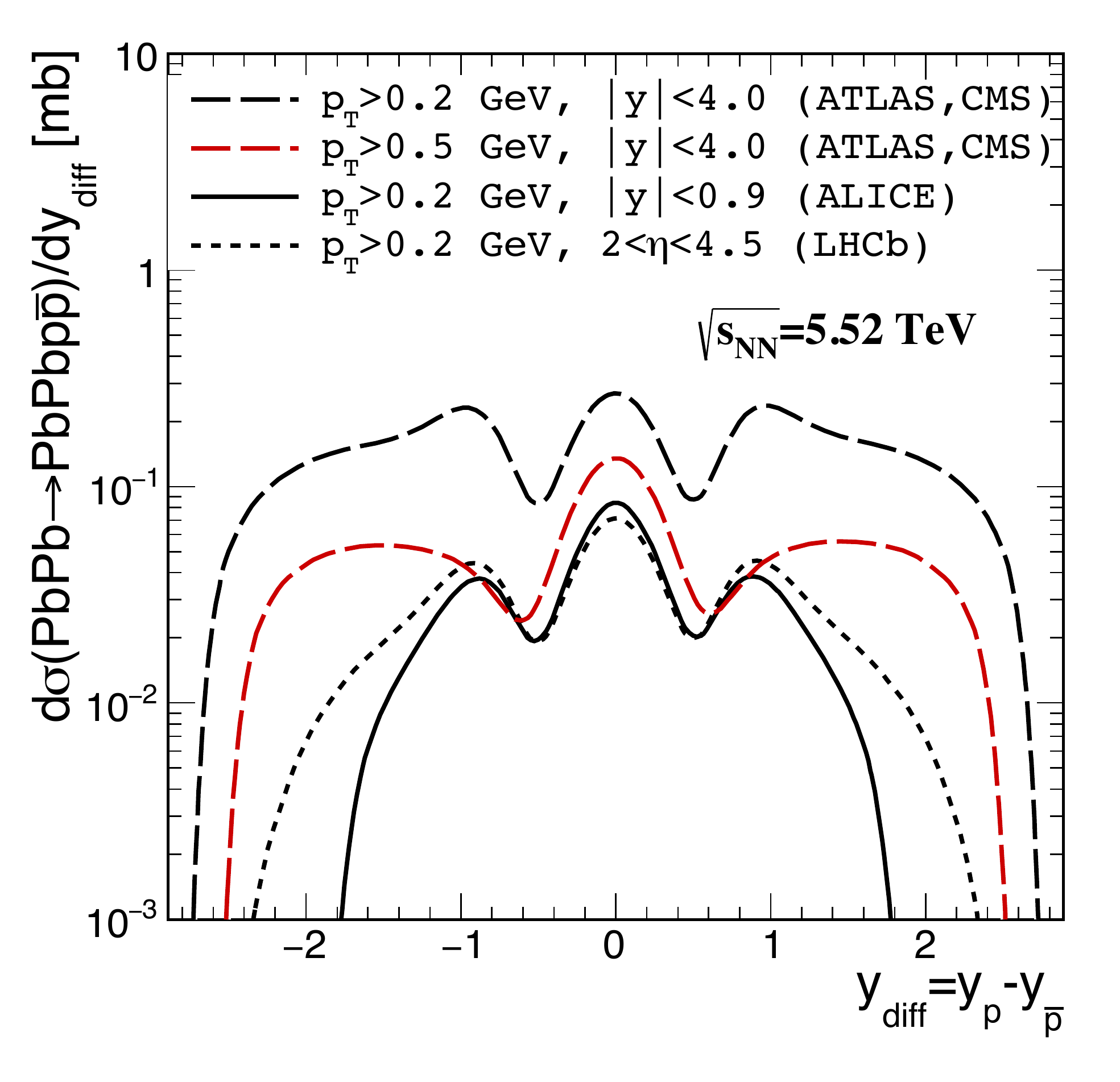}
        \caption{
                Differential cross sections as a function of $\mathrm{p\bar{p}}$
                invariant mass~(left) and rapidity
                distance between proton and anti-proton~(right)
                in Pb--Pb collisions at $\sqrtsNN=5.52$\,TeV
                for four experimental acceptance requirements. 
                For ATLAS and CMS experiments two requirements for proton $\pt>0.2$~GeV/$c$ or $\pt>0.5$~GeV/$c$ are considered.
        }
        \label{fig:ppbar}
\end{figure}

Exclusive production of p$\mathrm{\bar{p}}$ pairs~($\mathrm{\gamma\gamma\rightarrow p\bar{p}}$) in heavy-ion collisions is considered as a process which can help verify the existing theoretical approaches. It has been demonstrated that the  $\mathrm{\gamma\gamma\rightarrow p\bar{p}}$ experimental data~\cite{Kuo:2005nr} from the Belle Collaboration can be successfully described by implementation of several components~\cite{Klusek-Gawenda:2017lgt}: the non-resonant proton exchange, $s$-channel tensor meson exchange and the hand-bag model~\cite{Diehl:2002yh}. Figure~\ref{fig:ppbar} shows the calculated distributions of invariant mass of the p$\mathrm{\bar{p}}$ system, $W_{\mathrm{\gamma\gamma}} = M_{\mathrm{p\bar{p}}}$~(left panel) and of the difference of rapidities for protons and anti-protons, $y_{\mathrm{diff}} = y_\mathrm{p} - y_{\mathrm{\bar{p}}}$~(right panel).
The ALICE Collaboration can measure p$\mathrm{\bar{p}}$ pairs in Pb--Pb collisions at mid-rapidity~($|y|$ < 0.9). 
The LHCb Collaboration could also provide a complementary measurement of p$\mathrm{\bar{p}}$ production in the forward region~($2 <\eta < 4.5$).
The upgraded charged particle tracking capabilities of ATLAS and CMS experiments for Run~4 will measure in $|y|<4.0$.
Corresponding kinematic requirements on transverse momenta and rapidity or pseudorapidity specific for each experiment are presented in the figure legend. The calculations are made for Pb--Pb collisions with $\sqrtsNN = 5.52$~TeV.
The total cross section predicted for the ATLAS and CMS acceptances for $\pt>0.2$~GeV/$c$~($\pt>0.5$~GeV/$c$) is $\sigma = 793~\mu$b~($248~\mu$b), while LHCb and ALICE requirements lead to $\sigma = 125$ and 105~$\mu$b, respectively.

From the left panel of Fig.~\ref{fig:ppbar} one can deduce that the dependence on invariant mass of the $\mathrm{p\bar{p}}$ pair is sensitive to the rapidity/pseudorapidity of the outgoing particle. The cut-off at the minimal value of $W_{\mathrm{\gamma\gamma}}$ is determined by the minimum $\pt$ requirement.
The $y_{\mathrm{diff}}$ distribution shown in the right panel of Fig.~\ref{fig:ppbar} is of particular interest. 
The broad maximum at $y_{\mathrm{diff}}=0$ corresponds to the region with $|\cos \theta| < 0.6$, where $\theta$ denotes the angle of the outgoing nucleon relative to the beam direction in the centre-of-mass frame.  
An observation of peaks at $y_{\mathrm{diff}} = \pm 1$ could be a good test to constraint the theoretical models which predict the elementary cross section. The proposed model has a few parameters~(i.e. vertex form factors for the proton exchange, tensor meson s-channel exchanges and a form factor in the hand-bag contribution) which could be constrained with the help of the $y_{\mathrm{diff}}$ distributions.
The fiducial acceptance requirements imposed on $\pt$ do not distort the maxima.
If the structures in the $y_{\mathrm{diff}}$ distributions indeed exist, the study of $\mathrm{p\bar{p}}$ production in UPC can provide an important complimentary information  to the existing $\mathrm{\gamma\gamma\rightarrow \ell^+\ell^-}$ and J$/\psi\rightarrow\ell^+\ell^-$ data with $\ell=e, \mu$~\cite{Kryshen:2017jfz,Abbas:2013oua}.

\begin{figure}[!hbt]
\centering
\includegraphics[width=0.49\linewidth]{\main/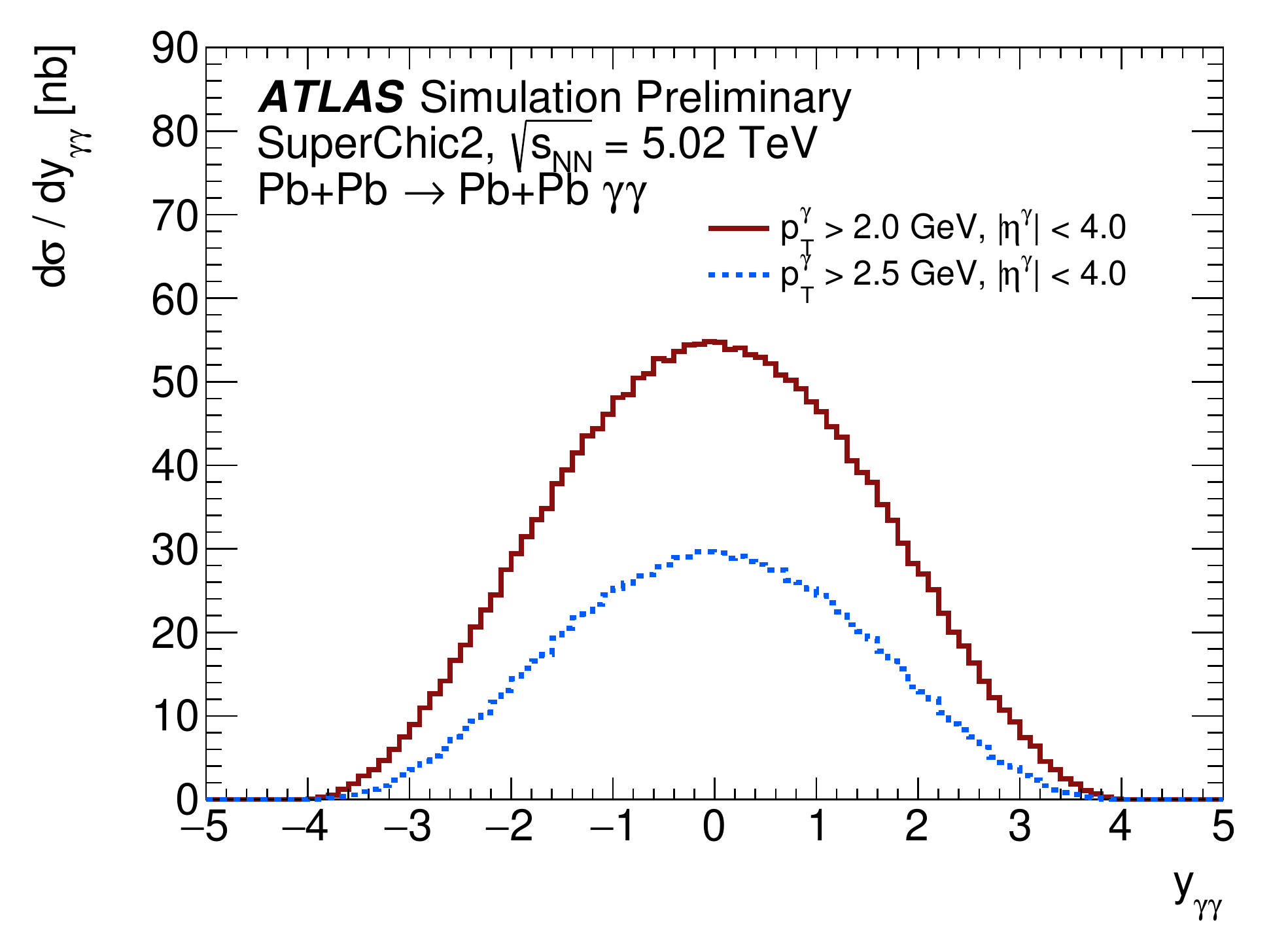}
\includegraphics[width=0.425\linewidth]{\main/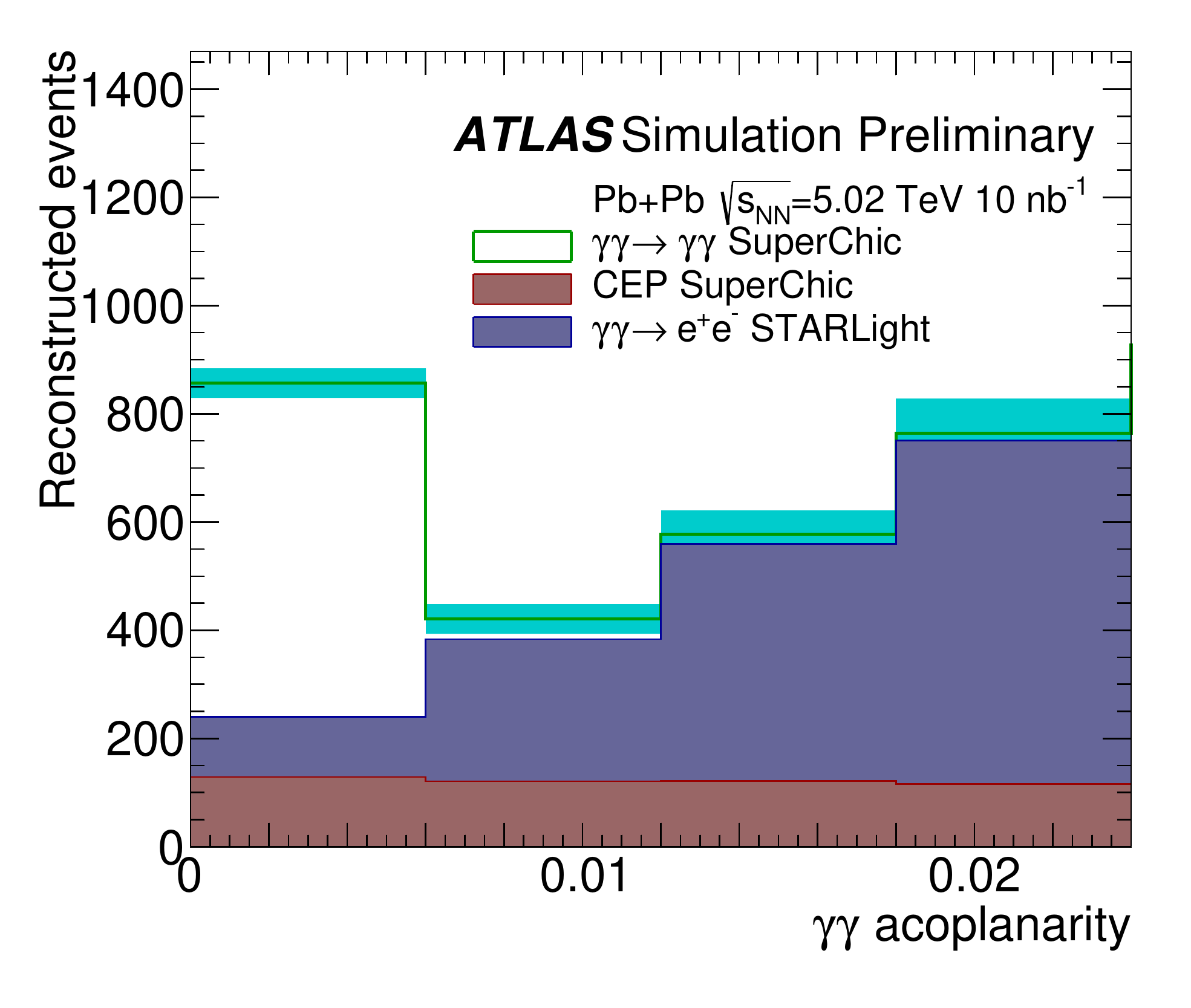}
\caption{
(Left)~Predicted differential cross section as a function of the di-photon rapidity for LbyL scattering for photons with
$\pt^{\mathrm{\gamma}}>2.5$~GeV/$c$~(dashed) or $\pt^{\mathrm{\gamma}}>2.0$~GeV/$c$~(solid), and
$|\eta^{\mathrm{\gamma}}|<4$ extracted from SuperChic~\cite{Harland-Lang:2018iur}.
(Right) Detector-level acoplanarity distribution of the di-photon system for photons from the LbyL signal and background processes in
  5.02~TeV Pb--Pb collisions with an integrated luminosity of
  $10~\mathrm{nb}^{-1}$. The shaded band in cyan represents expected statistical uncertainties.
}
\label{fig:lbyl}
\end{figure}

Evidence of the rare process of LbyL scattering has been established by the ATLAS and CMS Collaborations using Pb--Pb data obtained in 2015~\cite{Aaboud:2017bwk,Sirunyan:2018fhl} with an integrated luminosity of about $0.4\;\mathrm{nb}^{-1}$. That process can be studied with higher precision using heavy-ion data collected at the HL-LHC. The left panel of Fig.~\ref{fig:lbyl} presents a differential cross section as a function of the di-photon rapidity for LbyL scattering for photons with $|\eta^{\mathrm{\gamma}}|<4$ with two photon $\pt^{\mathrm{\gamma}}$ thresholds: 2.0 and 2.5~GeV/$c$. The LbyL scattering occurs in the central region: 91\% of the integrated cross section resides within $|\eta^{\mathrm{\gamma}}|<2.37$. A strong dependence on the $\pt^{\mathrm{\gamma}}$ requirement is observed. The cross section increases by a factor of two when the single photon $\pt^{\mathrm{\gamma}}$ threshold is lowered by half a GeV/$c$ from 2.5 to 2.0~GeV/$c$. The corresponding integrated cross sections in the fiducial region are 112~nb for $\pt^{\mathrm{\gamma}}>2.5$~GeV/$c$ and 221~nb for $\pt^{\mathrm{\gamma}}>2.0$~GeV/$c$. 

The right panel of Fig.~\ref{fig:lbyl} shows a detector-level acoplanarity~(=$1-|\varphi^{\mathrm{\gamma}}_1-\varphi^{\mathrm{\gamma}}_2|/\pi$) distribution for the di-photon system from LbyL signal and two background processes originating from exclusive production of di-electron pairs~($\mathrm{\gamma\gamma\rightarrow\,e^+e^-}$) and di-photons produced in central exclusive production~($\mathrm{gg\rightarrow \gamma\gamma}$). The distributions depict simulated events which passed a full simulation of the ATLAS detector with the extended acceptance in pseudorapidity. About 640 LbyL events pass the selection requirements for acoplanarity below 0.01 and $\pt^{\mathrm{\gamma}}>2.5$~GeV/$c$ in 5.02~TeV Pb--Pb~collisions with an integrated luminosity of 10\,nb$^{-1}$, in comparison to about 13 events observed in the 2015 data set with the $\pt^{\mathrm{\gamma}}>3.0$~GeV/$c$ requirement. 
The signal events are peaked at acoplanarities close to zero, while the background processes are distributed either uniformly (di-photons from central exclusive production) or even grow with acoplanarity~($\mathrm{e^+e^-}$ pairs from exclusive di-electron production). The latter originates from $\mathrm{e^+e^-}$ pairs which trajectories have been bent in the magnetic field before emitting hard-bremsstrahlung photons.
A limitation of the current analysis is lack of simulation of the trigger response. Based on experience from the analyses of 2015 Pb--Pb data, triggering on photons with $\pt^{\mathrm{\gamma}}<3.0$~GeV/$c$ is challenging, and therefore a dedicated trigger strategy needs to be developed for LbyL event candidates exploiting new features of the upgraded trigger system~\cite{Aad:2013tqj,Tapper:1556311}.

The LbyL process can also be studied at lower di-photon masses. The differential cross sections as a function of the di-photon mass can be evaluated taking into account acceptance of the ALICE experiment, i.e.\,pseudorapidity limited to $|\eta^{\mathrm{\gamma}}|<0.9$ or in the forward region defined by $2<\eta^{\mathrm{\gamma}}<4.5$ in the LHCb experiment, and relatively low energies of outgoing photons~\cite{Abelev:2014ypa}. At lower energies~($W_{\mathrm{\gamma\gamma}} < 4$~GeV) meson resonances~\cite{Klusek-Gawenda:2018ijg} may play an important role in addition to the Standard Model box diagrams~\cite{dEnterria:2013zqi,Klusek-Gawenda:2016euz} or double photon fluctuations into light vector mesons~\cite{Klusek-Gawenda:2016euz} or two-gluon exchanges~\cite{Klusek-Gawenda:2016nuo}.
\begin{figure}[t]
        \includegraphics[scale=0.375]{\main/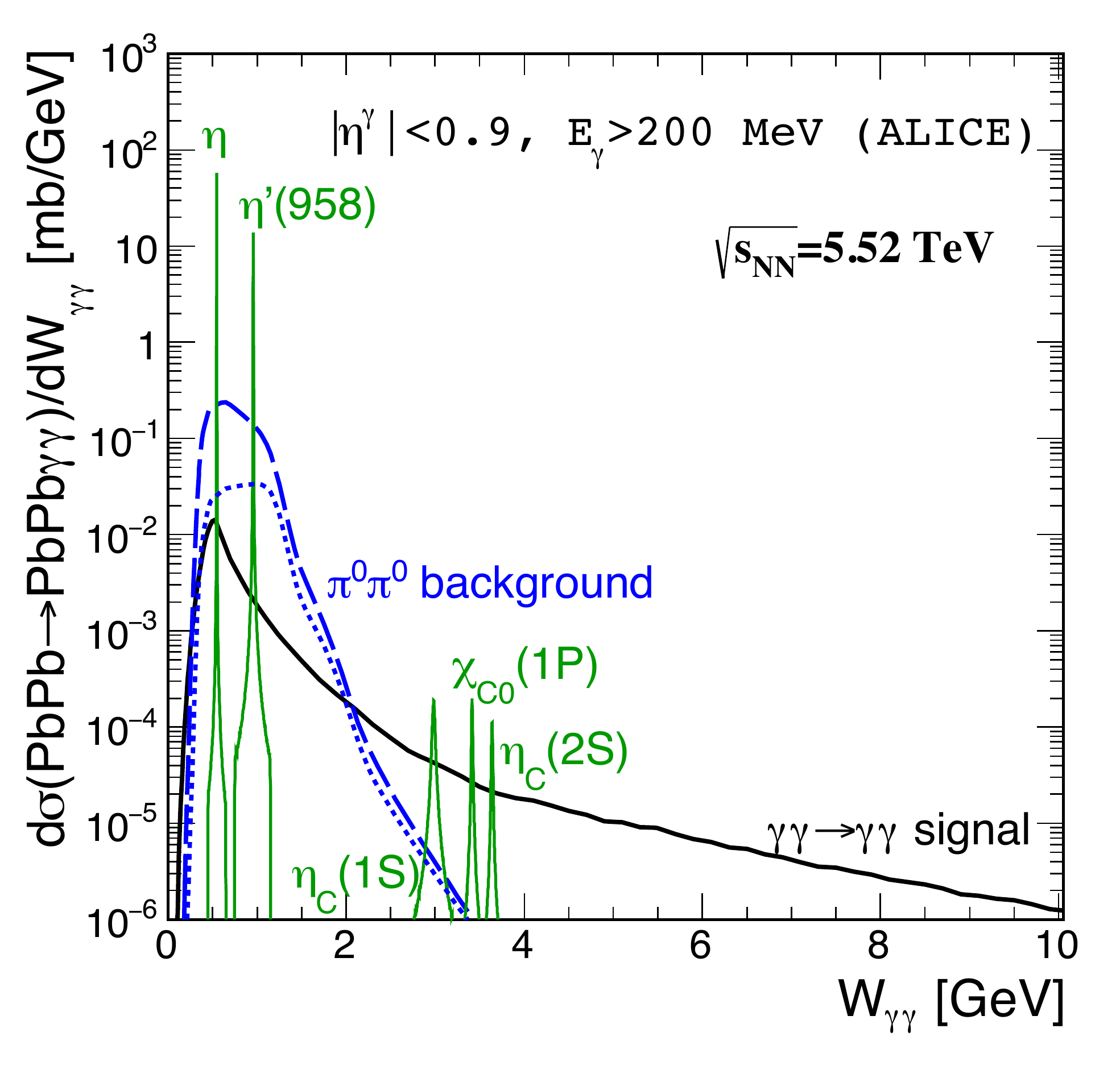}%
        \includegraphics[scale=0.375]{\main/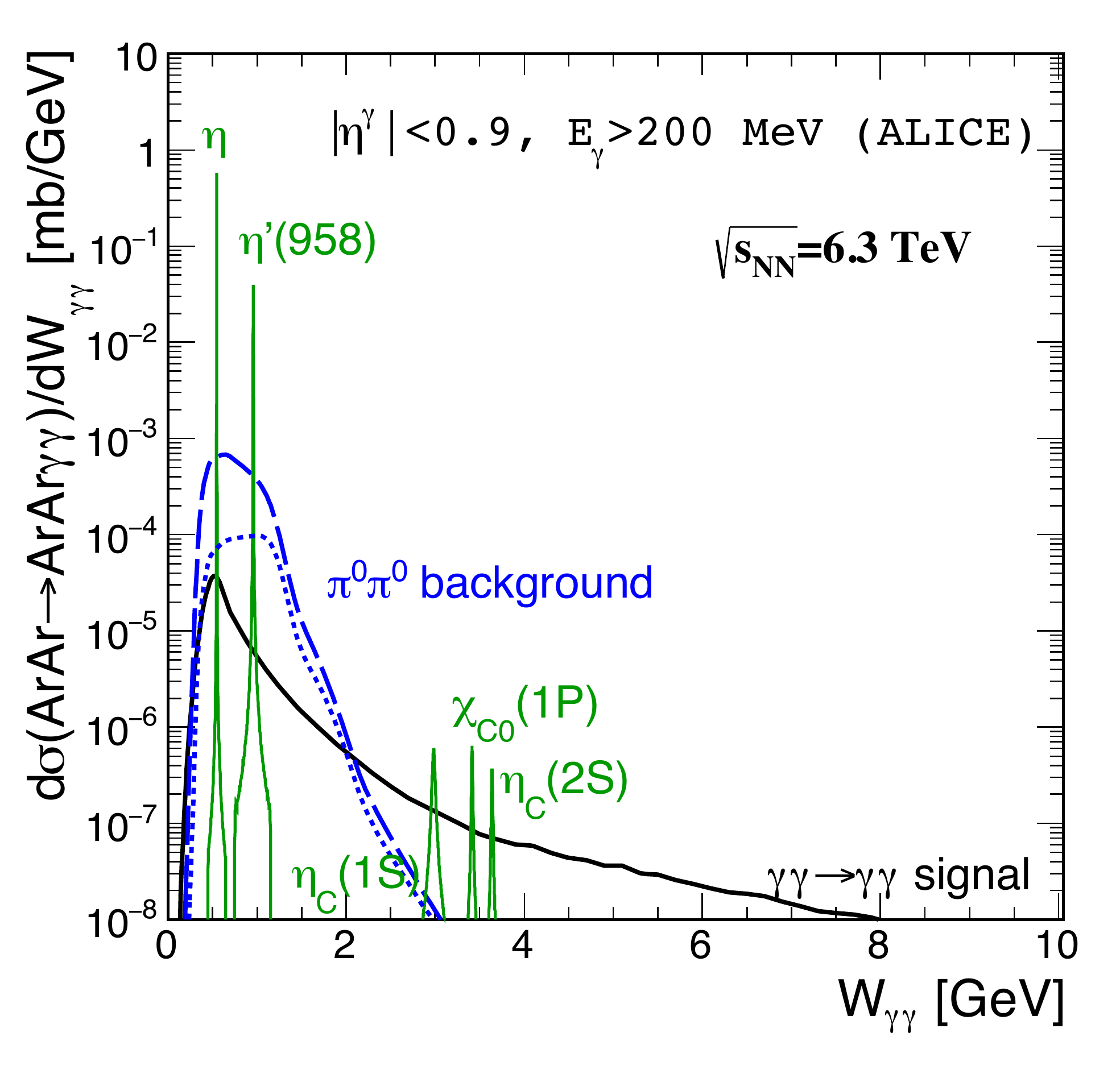}\\
        \includegraphics[scale=0.375]{\main/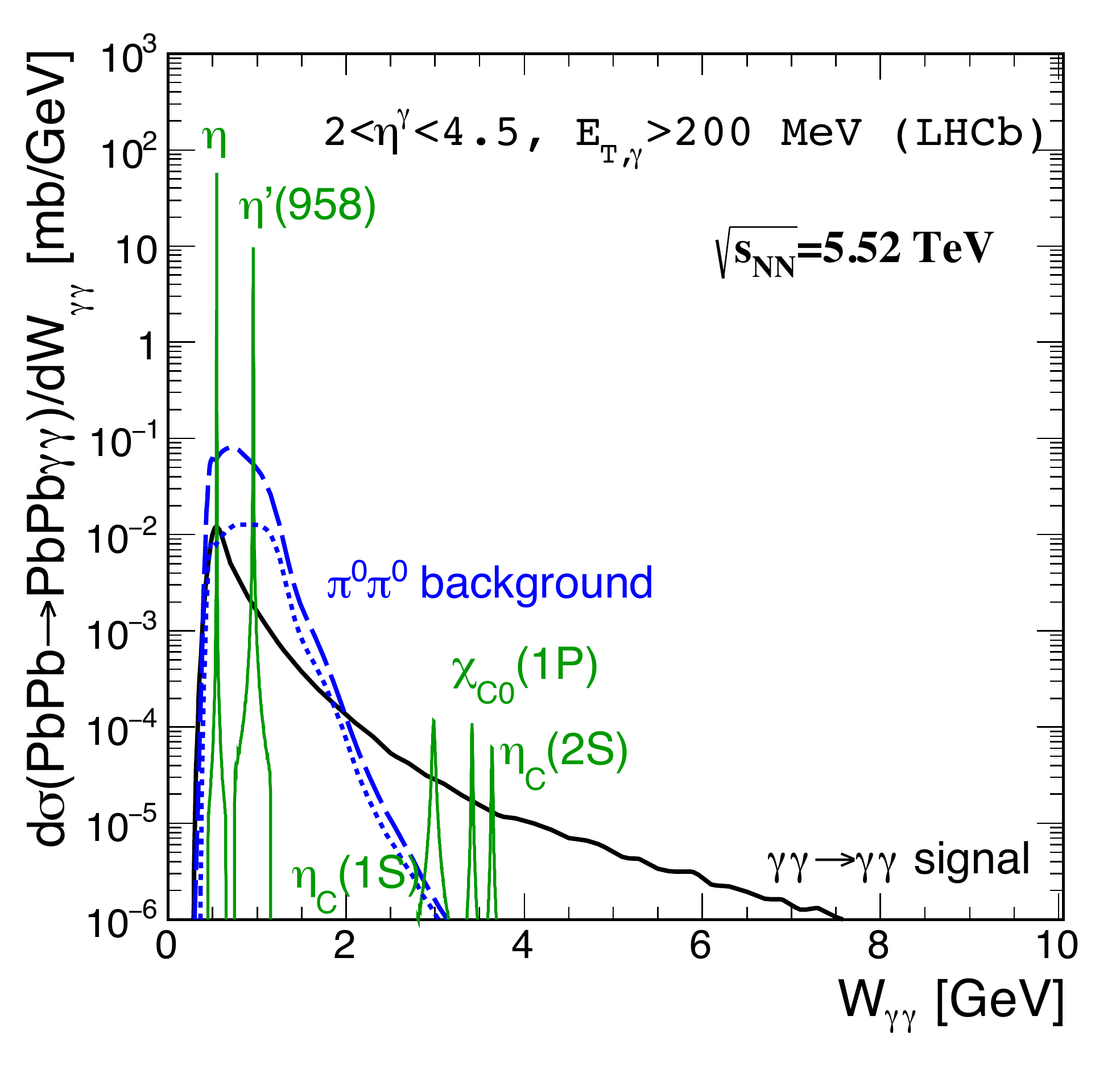}%
        \includegraphics[scale=0.375]{\main/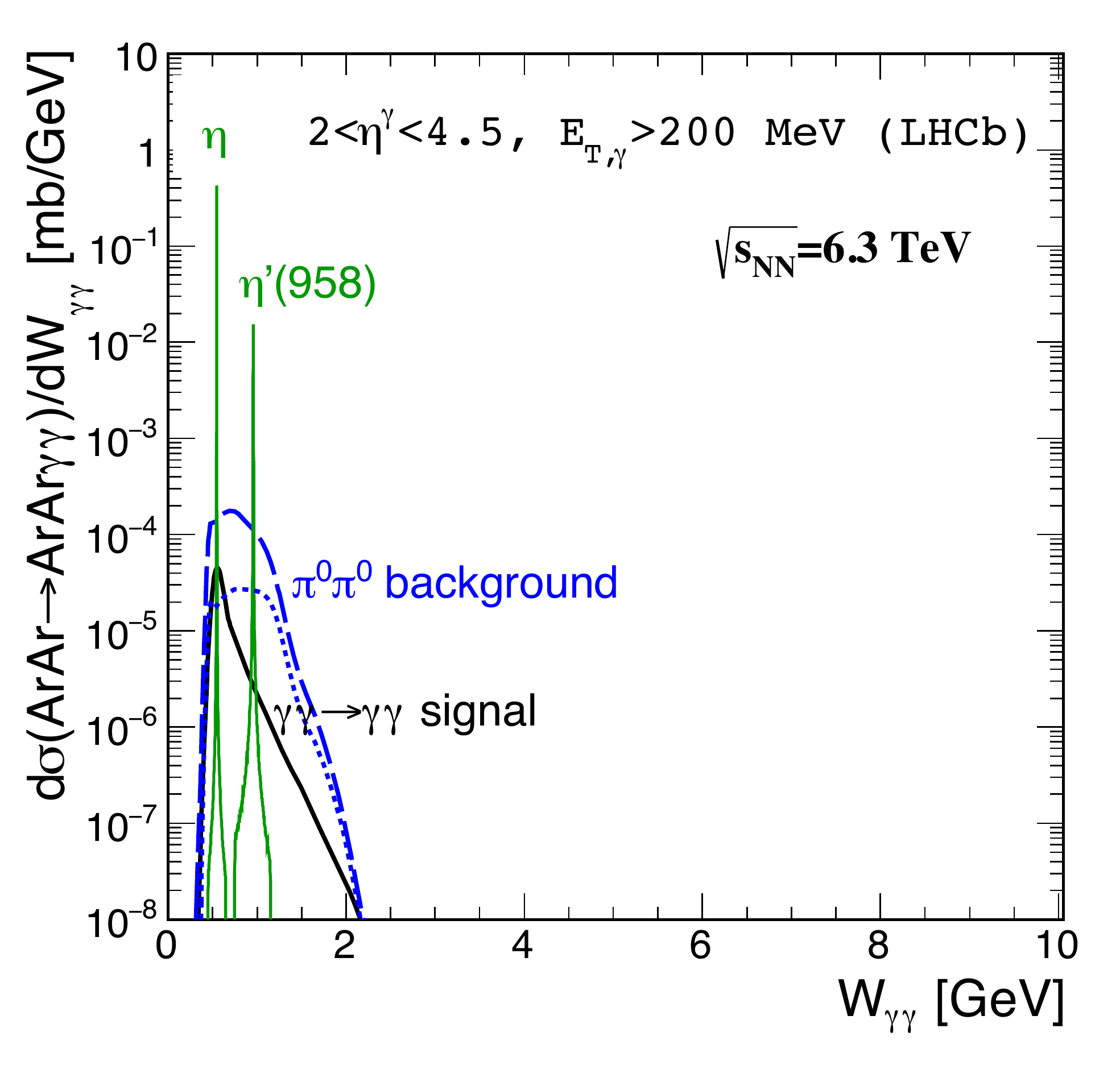}
        \caption{
                Di-photon invariant mass distributions for Pb--Pb collisions at $\sqrtsNN=5.52$\,TeV~(left) and Ar--Ar collisions at $\sqrtsNN=6.3$\,TeV~(right) for ALICE at mid-rapidity~(top) and LHCb at forward pseudorapidity~(bottom). The $\mathrm{\pi^0\pi^0}$ background is shown with the acoplanarity requirement of 0.01~(dotted line) and also without it~(dashed line).}
        \label{fig:lbyl_alice}
\end{figure}
Figure~\ref{fig:lbyl_alice} shows predictions for LbyL and background processes in the ALICE and LHCb experiments with photon acceptance in $|\eta^{\mathrm{\gamma}}|<0.9$ and $E_{\mathrm{\gamma}}$ > 200~MeV~(top panel) or $2<\eta^{\mathrm{\gamma}}<4.5$ and $E_{\mathrm{T,\gamma}}$ > 200~MeV~(bottom panel), respectively, for two systems: Pb--Pb collisions at 5.52~TeV~(left panel) and Ar--Ar collisions at 6.3~TeV~(right panel). Presented results include the effect of the experimental energy resolution~\cite{Acharya:2018yhg,Govorkova:2015vqa}. The black-solid lines depict the LO QED fermionic box mechanism with leptons and quarks. Presented results for the $\mathrm{\gamma\gamma\to\gamma\gamma}$ process are in agreement with calculations from Refs.~\cite{Jikia:1993tc,Bern:2001dg,Bardin:2009gq}. The green-solid lines show results for the $s$-channel $\mathrm{\gamma\gamma}\to$pseudoscalar/scalar/tensor resonances that contribute to the LbyL process.
In the present analysis, $\eta$, $\eta'(958)$, $\eta_c(1S)$, $\eta_c(2S)$, $\chi_{c0}(1P)$ mesons are considered.
Their masses, total widths and branching ratios are taken from the PDG~\cite{Patrignani:2016xqp}.
The contributions of pseudoscalar mesons from radiative decays of a coherently-produced vector meson could be sizeable~\cite{Klein:2018ypk} and should be quantified in future studies.
The dominant background from the $\mathrm{\gamma\gamma\to\pi^0\pi^0}$
process is shown by the blue lines. It becomes non-negligible only when one photon from each $\mathrm{\pi^0\to\gamma\gamma}$ decay is reconstructed in the detector. Two scenarios with and without the acoplanarity requirement of 0.01 are considered. The acoplanarity requirement reduces this background contribution by a factor of 5 in the full $W_{\mathrm{\gamma\gamma}}$ region.
The experimental data for the $\mathrm{\gamma\gamma\to\pi\pi}$ elementary cross section were very well described in Ref.~\cite{Klusek-Gawenda:2013rtu}.
There simultaneously the total cross section and angular distributions for both charged and neutral pions are shown.
Following Ref.~\cite{Klusek-Gawenda:2013rtu}, here nine resonances,
$\mathrm{\gamma\gamma\to\rho^\pm\to\pi^0\pi^0}$ continuum, Brodsky-Lepage and hand-bag mechanism are included. Figure~\ref{fig:lbyl_alice} shows that pionic background dominates at low invariant di-photon mass~(below 2~GeV).
In the same energy region, one can observe a very clear dominance of
$\mathrm{\eta}$, $\mathrm{\eta'(958)}$ mesons over other processes.
The inclusion of energy resolution introduces mainly smearing of the contribution from
$\mathrm{\gamma\gamma\to\eta,\eta'\to\gamma\gamma}$ resonance scattering.
This contribution is supposed to be measured with good precision.
These results suggest that both ALICE and LHCb Collaborations could measure LbyL scattering for $W_{\mathrm{\gamma\gamma}}> 2$~GeV in Pb--Pb collisions. 

In the case of Ar--Ar collisions, the cross sections are about two orders of magnitude lower, because of the  smaller electric charge of Ar nulcei with respect to Pb nuclei. Assuming integrated luminosities of $3.0-8.8$~$\mathrm{pb^{-1}}$ in a dedicated Ar--Ar run, the LbyL production cross section leads to 1460--4280 signal events for ALICE and 11--34 events for LHCb in a range of $W_{\mathrm{\gamma\gamma}}>2$~GeV. A background contribution from $\mathrm{\gamma\gamma\to\pi^0\pi^0}$ is at the level of 20\% for ALICE and 134\% for LHCb in this region.

Axions and axion-like particles~(ALP) are fundamental components of extensions of the Standard Model, occurring in most solutions of the strong CP problem~\cite{Peccei:1977hh,PhysRevLett.38.1440}. Recently an increasing interest has been paid to ALP masses
above 1~GeV~\cite{Bauer:2017ris,Mimasu:2014nea,Jaeckel:2015jla,Knapen:2016moh,Brivio:2017ije}. In particular the Higgs discovery has set spin zero particles in the spotlight of searches for new physics, with scalar and pseudo-scalar particles (elementary or not) as heralds of new phenomena. An interesting feature is that ALP (generically labelled as $\mathrm{a}$ in the following) in this mass range would induce an anomalous contribution to the LbyL, via the reaction:
$\mathrm{\gamma \gamma \rightarrow a \rightarrow \gamma \gamma}$,
under the condition that the magnitudes of the EM fields associated
with the incident photon are large enough, typically $|\vec{E}| >10^{18}$~V/m.
This has triggered  the study presented in Ref.~\cite{Knapen:2016moh},
and then in Ref.~\cite{Knapen:2017ebd} using the recent observation of LbyL scattering published by the ATLAS experiment in Pb--Pb collisions~\cite{Aaboud:2017bwk}, where the electric field produced by the ultra-relativistic Pb is of the order of $10^{25}$~V/m (thus satisfying the above condition).

\begin{figure}[!htbp]
\centering
  \includegraphics[scale=0.45]{\main/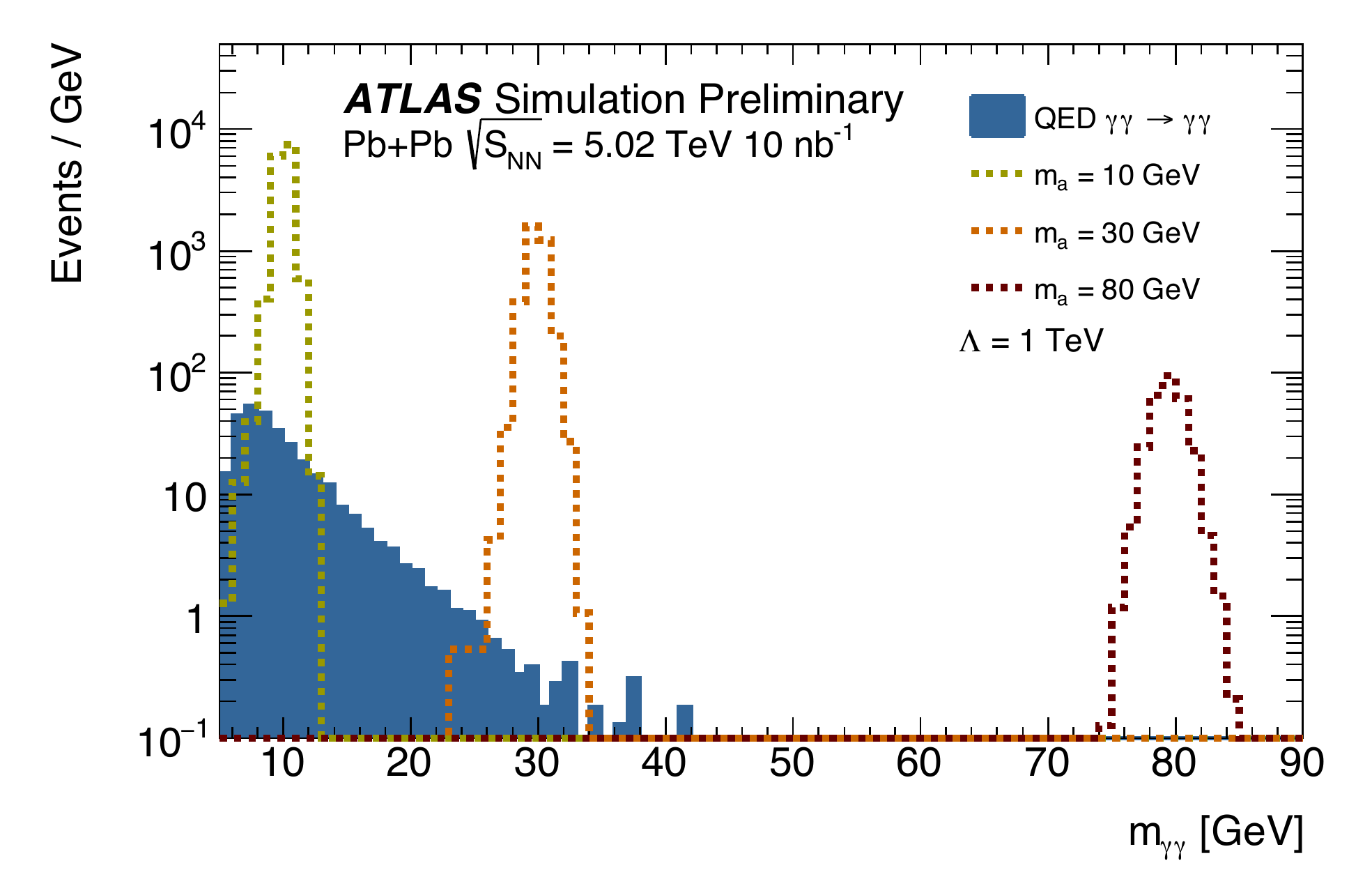}
  \includegraphics[scale=0.34]{\main/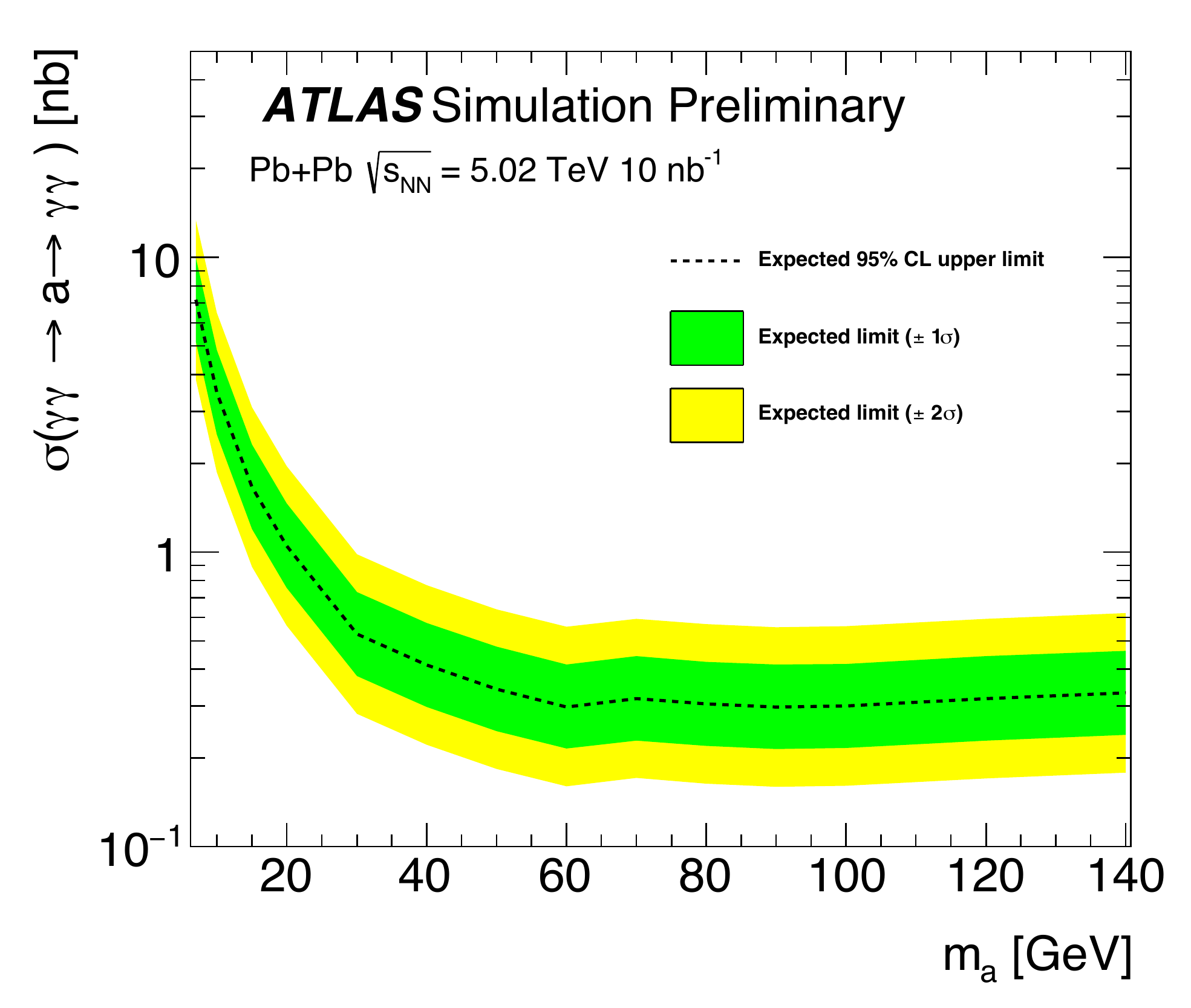}
  \caption{(Left)~Mass distribution for the ALP signal
  shown for three values of the ALP mass: $m_\mathrm{a}=10, 30$ and
  $80$~GeV~(in red). Also shown (in blue) the LbyL background~(see
  text). All ALP mass points are generated with $\Lambda = 1$~TeV~($1/\Lambda$ is the coupling of the interaction) which follows a convention defined in Ref.~\cite{Knapen:2016moh}.
  (Right)~Expected $95$\% CLs upper limits on $\sigma_{\mathrm{a\rightarrow \gamma \gamma}}$.}
  \label{fig:alp}
\end{figure}

The potential of ALP searches in UPC Pb--Pb collisions is studied using detector-level quantities after the LbyL selection requirements are imposed. The overall selection efficiency (times acceptance) relative to generated events increases from about $40$\% to $65$\% for ALP masses ranging from $7$~GeV to $80$~GeV. Also, the mass resolution varies from $0.5$~GeV at low masses (below $15$~GeV) up to $1$~GeV for larger masses. In the left panel of Fig.~\ref{fig:alp} the expected mass distributions for three ALP signal mass values, and the main background from LbyL normalised to integrated luminosity of 10~$\mathrm{nb}^{-1}$ are shown. In this study, other sources of backgrounds are neglected, since they have been found to be small in the LbyL measurement~\cite{Aaboud:2017bwk}. The invariant mass distribution is used as the discriminating variable, with bin widths comparable to the expected resolution of a narrow resonant signal.
Upper limits are set on the product of the production cross section of
new resonances and their decay branching ratio into $\mathrm{\gamma\gamma}$. Exclusion intervals are derived using the CLs method~\cite{Read:2002hq} in the asymptotic approximation. The limit set on the signal strength $\mu$ is then translated into a limit on the signal cross section times branching ratio as presented in the right panel of Fig.~\ref{fig:alp}.

\begin{figure}[!htbp]
\centering
  \includegraphics[scale=0.55]{\main/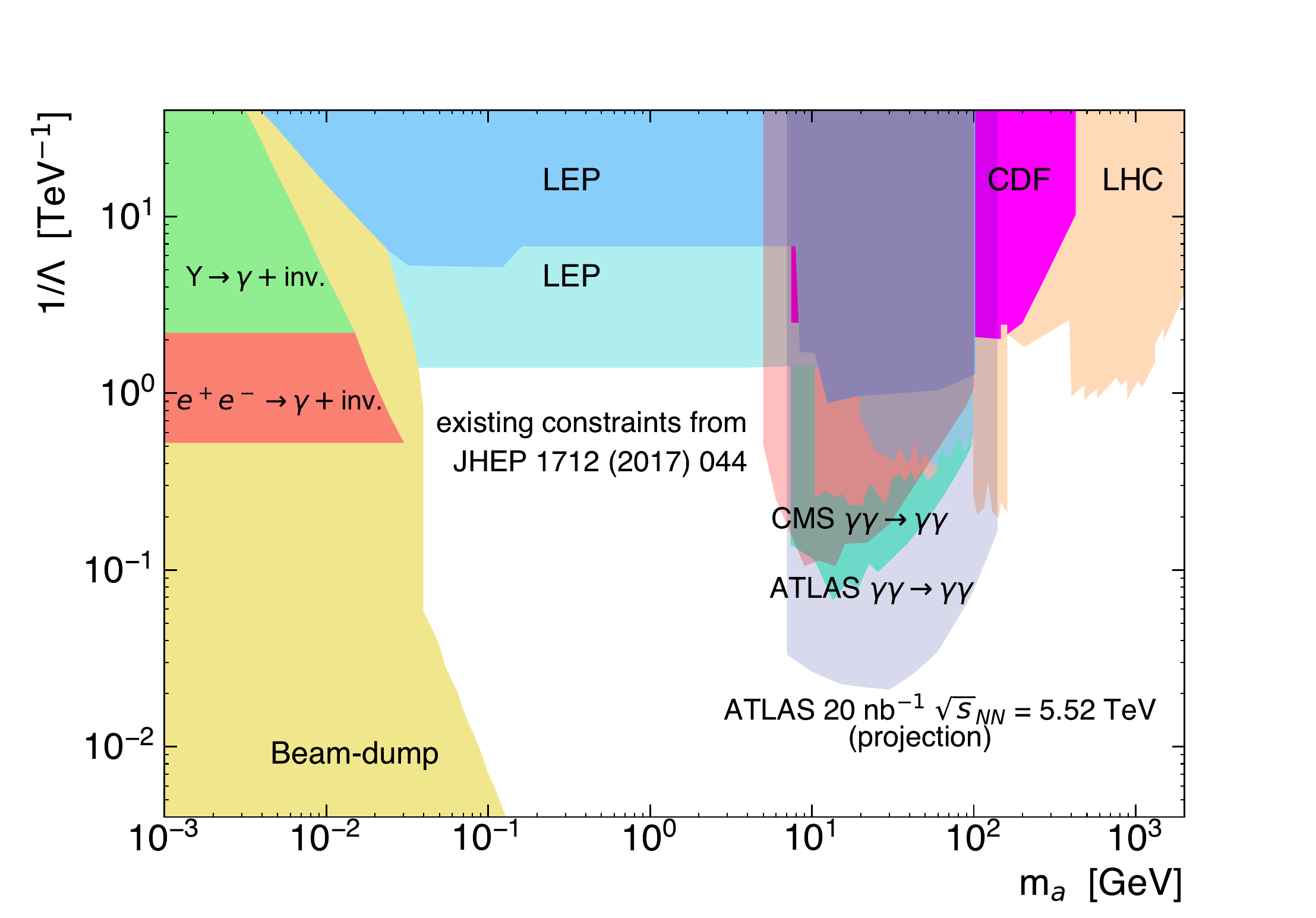}
  \caption{Compilation of exclusion limits obtained by different experiments~(see text). In light grey, the ATLAS $20\;\mathrm{nb^{-1}}$ limit at $\sqrtsNN=5.52$~TeV is presented. The ATLAS $\mathrm{\gamma\gamma\rightarrow\gamma\gamma}$ represents the exclusion limit derived from the LbyL cross section measured in Pb–-Pb collisions by ATLAS~\cite{Aaboud:2017bwk}, while the CMS $\mathrm{\gamma\gamma\rightarrow\gamma\gamma}$ limit comes from the recent analysis described in Ref.~\cite{Sirunyan:2018fhl}. A more complete version of the existing constraints on ALPs masses versus coupling, including the constraints in the sub meV range from astrophysical observations and from dedicated experiments such as CAST can be found in Ref.~\cite{Bauer:2017ris}.
  }
  \label{fig:alp-lambda-limits-cms}
\end{figure}

In Fig.~\ref{fig:alp-lambda-limits-cms} exclusion limits on the coupling, $1/\Lambda$, as a function of $m_\mathrm{a}$ are presented along with the existing results from the compilation discussed in Ref.~\cite{Baldenegro:2018hng}. The ATLAS 20~nb$^{-1}$ limit is derived using Pb--Pb collisions at 5.52~TeV. These results demonstrate that heavy-ion collisions have unique sensitivity to ALP searches in the range of $m_\mathrm{a}=7-140$~GeV, where the previous results based on available Pb--Pb data by ATLAS and CMS~\cite{Knapen:2016moh,Sirunyan:2018fhl} are also shown~(labelled as ATLAS $\mathrm{\gamma\gamma\rightarrow\gamma\gamma}$ and CMS $\mathrm{\gamma\gamma\rightarrow\gamma\gamma}$ in the figure).

\clearpage
\subsection{Proton-oxygen collisions for cosmic ray research}
\label{sec:pOcosmic}
{ \small
\noindent \textbf{Coordinators}: Hans Dembinski (MPI for Nuclear Physics, Heidelberg)

\noindent \textbf{Contributors}:
T.~Pierog (Karlsruhe Institute of Technology),
R.~Ulrich (Karlsruhe Institute of Technology)
}

\begin{figure}
\includegraphics[width=0.5\textwidth,trim=5 -5 20 0]{\main/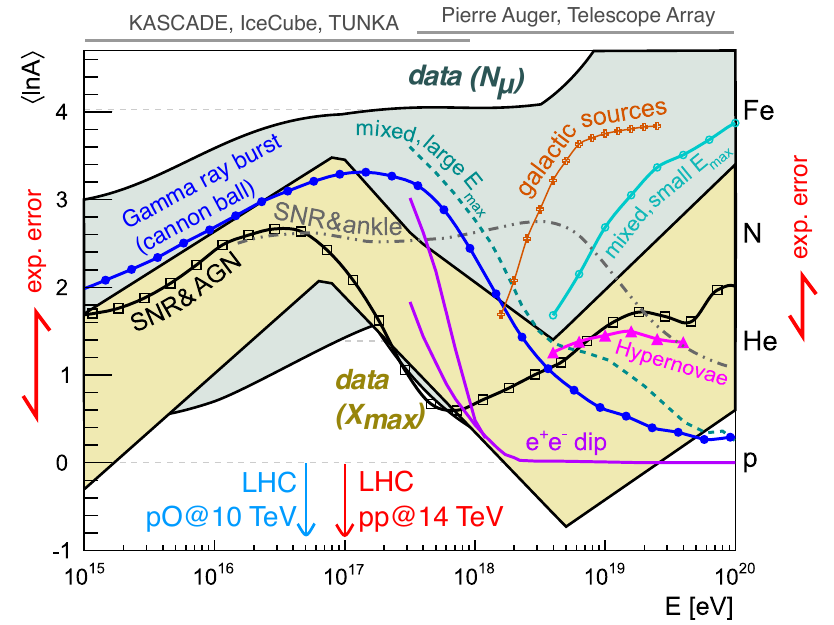}
\includegraphics[width=0.5\textwidth,trim=-20 0 30 0]{\main/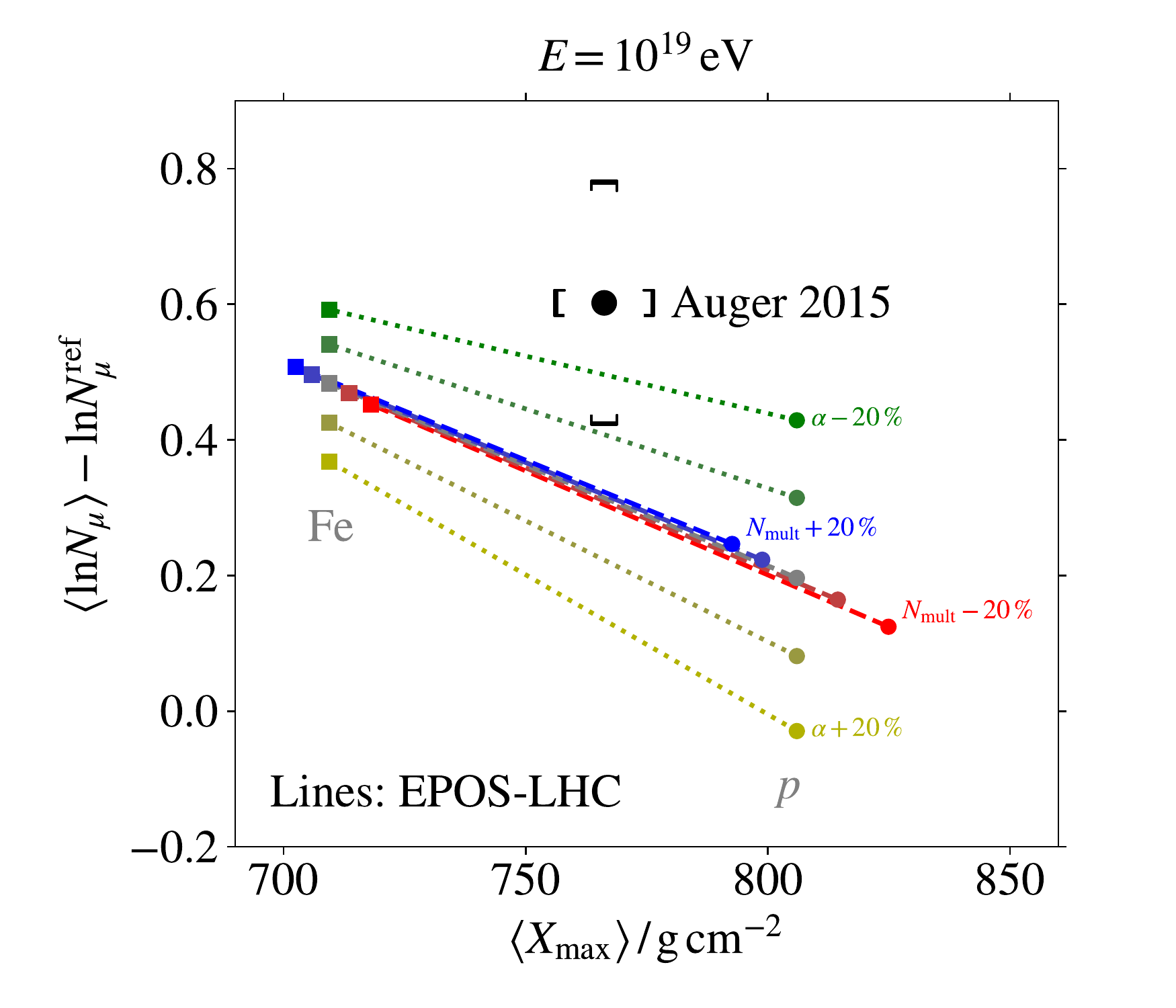}
\caption{Left: Mass composition of cosmic rays quantified by $\mlna$ as a function of cosmic ray energy $E$. See Ref.~\cite{kampert_cr_review} for references to data (bands) and model predictions (markers and lines), and the text for a discussion. Right: Impact of changes of the hadron multiplicity $\nmult$ (dashed lines) and the energy fraction $\alpha$ (dotted lines) which goes into neutral pions in collisions at the LHC energy scale on EPOS-LHC predictions for $\xmax$ and $\lnnmu$ in $10^{19}$\,\si{eV} air showers, compared to Auger data~\cite{Aab:2014pza}. The model lines represent all values that can be obtained for any mixture of cosmic nuclei from proton (bottom right) to iron (top left). The dashed and dotted lines represent modifications of $\nmult$ and $\alpha$ in steps of $\pm$10\% from their nominal values.}
\label{fig:cosmic_rays}
\end{figure}

The recent coincident observations of gamma rays and neutrinos from the flaring blazar TXS 0506+056 confirmed that active galactic nuclei produce high-energy cosmic rays~\cite{IceCube:2018dnn}. This long awaited finding demonstrates that sources of cosmic rays are linked to the most violent places in our universe. Measurements of cosmic rays contribute to the understanding of the high-energy universe. Since cosmic rays are charged and bent by magnetic fields in space onto chaotic paths, their arrival directions at Earth are highly isotropic, but their mass composition carries an imprint of the source physics. Precision measurements of minimum-bias events in proton-oxygen collisions have the unique power to resolve current ambiguities in the mass composition measured with atmospheric air-shower techniques.

Cosmic rays are nuclei from protons to iron (heavier elements are negligible). The energy-dependent mass composition of cosmic rays is characteristic for different source scenarios, as shown in Fig.~\ref{fig:cosmic_rays}, left-hand-side, which displays predictions (lines and markers) of the mean-logarithmic-mass $\mlna$ of cosmic rays. Above particle energies of $10^{15}$\,\si{eV}, $\mlna$ can only be indirectly inferred from extensive air showers, huge secondary particle cascades produced by collisions between cosmic rays and nuclei in the atmosphere. The two leading observables to infer $\mlna$ are the depth $\xmax$ of the shower maximum in the atmosphere (yellow band in Fig.~\ref{fig:cosmic_rays}), and the number $\nmu$ of muons produced in the shower (green band in Fig.~\ref{fig:cosmic_rays}). The width of those bands has two main contributions: the experimental uncertainties, and the hadronic model uncertainties inherent in converting the air shower observables into $\mlna$.

Leading experiments achieve an instrumental accuracy of 10\,\% of the proton-iron difference, which would strongly discriminate between source scenarios, but air shower simulations are required to convert $\nmu$ and $\xmax$ to $\mlna$ and this adds a large model uncertainty. The simulations use the multi-purpose heavy-ion event generator EPOS-LHC~\cite{Werner:2005jf}, or specialized hadronic interaction models such as \mbox{QGSJet-II.04}~\cite{Ostapchenko:2010vb} and SIBYLL-2.3c~\cite{Riehn:2017mfm}. All are designed to describe nucleus-nucleus and soft-QCD interactions by extrapolating combinations of Regge field theory tuned to available data and perturbative QCD. Uncertainties in these models arise from a lack of data on multiparticle production in the very forward phase-space in hadron-nucleus interactions at the TeV scale.

LHC measurements have already reduced the spread of model predictions for $\xmax$ in the latest generation of models. This big improvement was due to high-precision measurements of the inelastic cross-section (see e.g.~\cite{Aaij:2018okq} and references therein). Further measurements now have the potential to make the spread negligible. The model spread for $\nmu$ is still large and predictions are not consistent with $\xmax$ for cosmic rays with the same mass. There is overwhelming evidence from air shower experiments~\cite{Dembinski:2018UHECR,Aab:2014pza,Dembinski:2017zkb,Kokoulin:2009zz,Abbasi:2018fkz} that the muon number $\nmu$ is underestimated in simulations starting at about $10^{16}$\,eV. This corresponds to a cms energy of $4.3$\,TeV, well accessible by the LHC. Shown in Fig.~\ref{fig:cosmic_rays}, right-hand-side, is a representative data point from the Pierre Auger Observatory, which is well above EPOS-LHC predictions -- and EPOS-LHC and SIBYLL-2.3c are already models which produce the highest muon number of all hadronic interaction models models. This is called the \emph{Muon Puzzle}.



Two aspects of multi-particle production with a strong effect on $\nmu$ have been identified\cite{Ulrich:2010rg}, the hadron multiplicity $\nmult$ and the energy fraction $\alpha$ that goes into neutral pions. The impact of changing these variables in EPOS-LHC at $13$\,TeV cms energy and extrapolating upward in energy is also shown in Fig.\ref{fig:cosmic_rays}, right-hand-side. A combined measurement to 5\,\% accuracy of both variables at the LHC would reduce the model uncertainty for the conversion of $\xmax$ to $\mlna$ well below the experimental uncertainty of 10\,\%, and has the clear potential to resolve the discrepancy in the muon number $N_\mu$. To reach the accuracy goal, the following minimum-bias measurements are desired:
\begin{itemize}
\item Double-differential production cross-section for charged pions, kaons, and protons: \\ ALICE $|\eta| < 0.9$, LHCb $2 < \eta < 5$
\item Production cross-section for neutral pions and neutrons: LHCf $\eta > 8.4$.
\item Energy flow over pseudo-rapidity, separated for hadrons and gammas:\\ CMS+CASTOR $-6.6 < \eta < 5.2$, ATLAS $-4.5 < \eta < 4.5$.
\end{itemize}
Energy flow measurements separated by hadronic and electromagnetic energy deposit constrain both $\nmult$ and $\alpha$, and can be done further forward than direct measurements of charged tracks. The particle identification provided by the ALICE and LHCb experiments provides important additional information, needed to tune and test internal parameters of hadronic interaction models. In particular, the number of produced baryons was found to strongly affect the number of muons in air showers at ground, despite their small number compared to pions\cite{Pierog:2006qv}.

\begin{figure}
\includegraphics[width=\textwidth]{\main/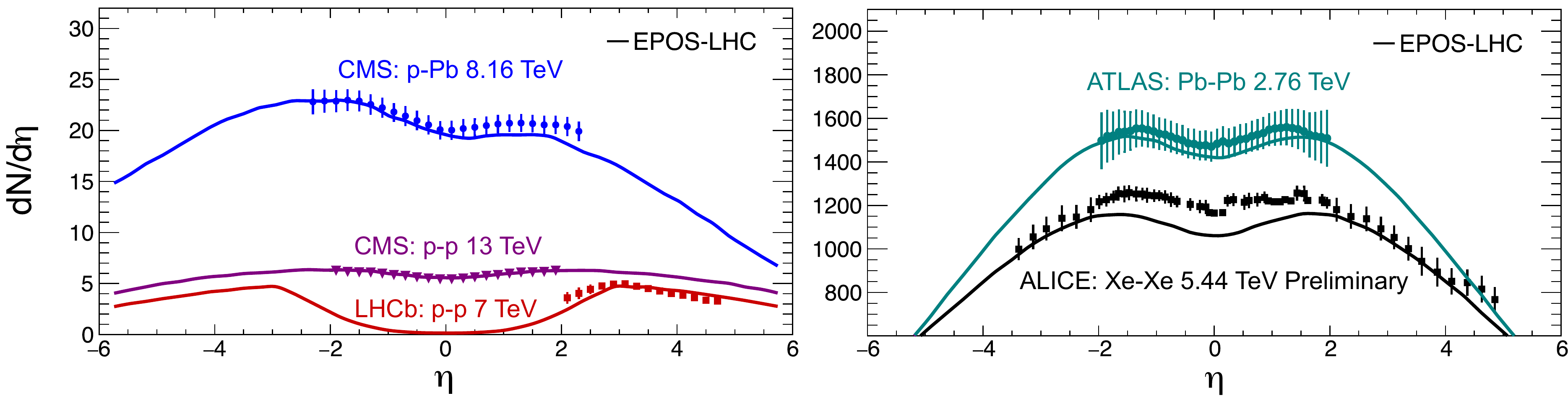}
\caption{Comparison of charged particle multiplicity measurements at different center-of-mass energies and in different colliding systems with the EPOS-LHC model~\cite{Kim:2018ink}. Shown in both plots is $\text{d} N/\text{d}\eta$.}
\label{fig:multiplicity_tuning}
\end{figure}

To meet the accuracy goal, $\nmult$ and $\alpha$ need to be measured in proton-oxygen collisions at the LHC, which directly mimic interactions of cosmic rays with the atmosphere. Constraining $\alpha$ and $\nmult$ to 5\,\% with existing and future \pp and \pPb data is very challenging~\cite{dEnterria:2018kcz}, since forward-produced hadrons experience strong nuclear modification\cite{Aaij:2017cqq,Acharya:2018qsh,Adriani:2015iwv}. A sufficiently accurate theory to predict nuclear modification in the \pO system based on \pp and \pPb data is not yet available, and a simple interpolation is not reliable since both systems are far away in $\ln A$.
The difficulty of predicting hadron production in ion collisions is demonstrated in Fig.~\ref{fig:multiplicity_tuning}. EPOS-LHC predictions for \XeXe collisions significantly underestimate the observed yields in the central region, despite a satisfactory description of \pp, \pPb, and \PbPb collisions. The deviations in \XeXe are much larger than what is expected from a simple interpolation~\cite{Kim:2018ink}. The dominant nuclear effects are expected to be different for light and heavy collision partners. Light nuclei are described by the shell model and nucleon correlations are important. Lead nuclei can be described by a simpler model, essentially a Wood-Saxon potential with reduced nucleon correlations that cannot be probed well in experiments.

Selecting peripheral \pPb collisions to mimic p-air collisions with the same number of binary collisions was considered as an alternative to direct \pO measurements, but this option also increases the uncertainty too much. Centrality in \pPb collisions is extracted from the data using various centrality estimators with different selection biases. These biases would increase the uncertainty of the proposed measurements well beyond the target of 5\,\%~\cite{Adam:2014qja}. However, \pO measurements could provide a sensitive test of centrality estimators since the thickness of the oxygen nucleus and hence the average number of wounded nucleons is about a factor of two smaller. The advantages of estimating centrality in a small ion system are discussed in Sect.~\ref{sect:smallsystems_OO}.

In conclusion, collisions of \pO at the LHC are highly desirable to solve the outlined questions. The corresponding measurements would be a crucial input to cosmic ray physics and have the potential to result in significant advances in the century-long ridde of the origin of cosmic rays. The luminosity requirements to reach the physics goals are moderate. A statistical accuracy better than 5\,\% can be achieved with 100\,M minimum-bias events. Luminosity calculations for light ion systems are given in Sect.\ \ref{sec:lightions}. The setup of \pO collisions would follow the successful rapid set-up procedure previously used in the 2012 \pPb run and the 2017 \XeXe run, as described in Sect.~\ref{sec:pOrun}.

It is worthwhile noting that a period of oxygen acceleration in the SPS would also provide the opportunity to complement cosmic-ray related measurements of nuclear fragmentation at NA61/SHINE\cite{Aduszkiewicz:2287004,Aduszkiewicz:2309890} at beam momenta of $150\,A\,\text{GeV}/c$. These measurements aim at improving our understanding of the cosmic-ray propagation in the Galaxy and to evaluate the cosmic-ray background for signatures of astrophysical dark matter~\cite{Genolini:2018ekk}. Another opportunity is the study of very forward production of hadrons in the \pO system at $\sqrtsnn \sim 100$\,GeV at the LHCb experiment, by colliding the oxygen beam with proton gas provided by an upgraded SMOG system, as described in Sect.~\ref{sec:fixedtarget}.

\clearpage
\subsection{Fixed-target prospects with LHC beams}
\label{sec:fixedtarget}
{ \small
\noindent \textbf{Contributors}: F.~Fleuret (LLLR, Palaiseau), G.~Graziani (INFN, Firenze), C.~Hadjidakis (IPNO, Orsay), E.~Maurice (LLR, Palaiseau), L.~Massacrier (IPNO, Orsay),  P.~Di~Nezza (INFN, Frascati),  L.~Pappalardo (University and INFN Ferrara), P.~Robbe (LAL, Orsay), B.~Trzeciak (Institute for Subatomic Physics, Utrecht)
}


Fixed target experiments present many advantages having the versatility of polarised and nuclear targets and allowing one to reach high luminosities with dense and long targets. The 7~TeV proton and 2.76~A.TeV lead beams allow one to reach a centre-of-mass energy per nucleon pair of $\sqrt{s_{\rm NN}} = 115$ GeV and $\sqrt{s_{\rm NN}} = 72$~GeV with a centre-of-mass rapidity boost of 4.8 and 4.2 units, respectively. These energies correspond to an energy domain between the SPS and nominal RHIC energies. The large rapidity boost implies that the backward rapidity region ($y_{\rm cms} \leq 0$) is easily accessible by using standard experimental techniques or existing LHC experiments such as ALICE or LHCb. 

The physics opportunities offered by a fixed-target programme at the LHC have been developed in several publications of the AFTER@LHC study group \cite{Brodsky:2012vg, Lansberg:Adv2015,Hadjidakis:2018ifr} and can be summarised as follows:
\begin{itemize}
\item{advance our understanding of the large-$x$ gluon, sea quark and heavy-quark content in the nucleon and nucleus,} 
\item{advance our understanding of the dynamics and spin of gluons inside polarised nucleons (if a polarised target were used),}
\item{advance our understanding of the properties of the Quark-Gluon Plasma formed in heavy-ion collisions between SPS and RHIC energies.}
\end{itemize}

\subsubsection{Status and future plans in LHCb}
\label{sec:FTLHCb}

The LHCb experiment has pioneered fixed target physics at the LHC since Run 2, 
using noble-gas targets (helium, neon and argon) obtained by injecting the gas directly in 
the LHC vacuum pipe in the proximity of the LHCb~collision point
through the SMOG device~\cite{smog}.
The nominal target gas pressure of $2 \times 10^{-7}$ mbar corresponds,
for a typical LHC beam of $10^{14}$ protons, to a luminosity of 
$6 \times 10^{29} \text{cm}^{-2}\text{s}^{-1}$ for collisions occurring in one
meter of gas along the beam direction, which is roughly the acceptance of the
LHCb vertex detector.

The forward geometry of the detector is particularly well suited for
this configuration.  It provides three units
of pseudorapidity corresponding to mid and backward rapidities ($-2.8 < y_{cms}< 0.2$
for a beam energy of 6.5 TeV), fully equipped with tracking and particle identification. 
Proton-nucleus and Pb-nucleus collisions using fixed targets of different
nuclear size can be studied at the energy scale of $\sqrtsnn \sim 100$~GeV
with unique coverage of the high-$x$ regime in the target nucleon.

The samples collected during Run 2, corresponding
to integrated luminosities up to about $100~\invnb$, 
allowed to perform 
studies of particle production which are of particular relevance
to cosmic ray physics~\cite{Aaij:2018svt}, and to
collect unprecedented samples of charmed hadrons in
fixed-target collisions at this energy
scale~\cite{Aaij:2018ogq}. 
These data can provide unique inputs to
discriminate cold nuclear matter effects in heavy-flavour production
from the effect of deconfinement, and to study nuclear PDFs 
at large $x$. The physics reach of heavy-flavour studies is presently
limited by the size of these samples. Measurements of absolute
cross-sections are also limited in accuracy by the determination of the
luminosity, since the gas pressure can be controlled only
within $\pm 50\%$ with the SMOG device. For the first fixed-target physics
results, the integrated luminosity has been determined from the rate of
elastically scattered atomic electrons with a precision of
6\%~\cite{Aaij:2018svt}. 

An upgraded gas target device, named SMOG2, is currently being developed, and expected
to be operational already during Run 3. 
In the new setup, the gas is contained in a storage cell, consisting
of a 20-cm-long open-ended tube with a diameter of 1 cm, fed by a capillary.
It allows to increase the gas density in the target
by at least one order of magnitude with respect to SMOG, reaching
luminosities of order $10^{31}~\text{cm}^{-2}\text{s}^{-1}$ with
proton beams.
The target is placed upstream, from $-50$ to $-30$~cm, the nominal  LHCb
collision point and is thus not overlapping the luminous pp region.
This opens the possibility to acquire fixed target events
simultaneously with collision events with negligible impact to the
pp physics program.
The new setup would also allow other gases to be injected, notably hydrogen and deuterium, providing
pp collisions in fixed-target mode as a reference for all pA
collision samples, and extending the physics case to the study
of the three-dimensional structure functions of the nucleon through
spin-independent observables~\cite{3dpdf}. Heavy noble gases as Kr
and Xe would also be usable.
The device will be equipped with a gas feed system, allowing
to know the target gas density at 1\% level.

Assuming that about 10\% of the beam intensity can be exploited for
fixed-target physics, either in synergy with pp data taking or through dedicated runs,
samples corresponding to integrated luminosities of order $0.1~\textrm{fb}^{-1}$
(using proton beams) and $0.1~\textrm{pb}^{-1}$ (using Pb beams) 
can be collected per year, also profiting from the increased beam
intensity provided by the HL-LHC.

Samples of this size would allow copious production of Drell-Yan
and heavy flavour states, including $\textrm{b}\overline{\textrm{b}}$ mesons.
As an example, rough estimates are provided in
Table~\ref{tab:ftyields} for the yields of reconstructed events in an assumed sample of p--Ar collisions corresponding to $0.1~\textrm{fb}^{-1}$.
Substantial advancements in the understanding of parton
distributions for gluons, antiquark and heavy-quarks 
at large $x$, where PDFs are now poorly constrained, are
foreseeable~\cite{Lansberg:Adv2015,Hadjidakis:2018ifr}. The precise determination
of heavy hadron production at large $x$ is expected to clarify
the  extent of the intrinsic heavy quark content
in the nucleon~\cite{Brodsky:1980pb,Brodsky:2015fna}, and to constrain
modifications of the nuclear PDFs due to initial state effects
(anti-shadowing and EMC effect~\cite{Geesaman:1995yd}, saturation
effects~\cite{Albacete:2014fwa}).
Sequential quarkonia suppression is a main signature for
deconfinement~\cite{Matsui:1986dk}, but is also affected by final
state effects as break-up of the heavy quark  pair~\cite{Vogt:2001ky}
and statistical recombination~\cite{BraunMunzinger:2000px}.
The rich samples of  different quarkonia states reconstructed 
in fixed target data will allow to
investigate sequential suppression at an energy scale
between the SPS and RHIC/LHC, for 
collision systems ranging from pp to Pb--Xe.
The study of collisions of Pb beams on heavy nuclei has been 
limited in Run 2  by the detector tracking capabilities and would
greatly profit from the higher detector granularity offered by the 
LHCb upgrade 1 and upgrade 2 detectors.

\begin{table}[tb]
\caption{Expected yields of reconstructed events for some benchmark channels
using the largest fixed-target data sample acquired with SMOG during
the LHC Run~2, and possible with SMOG2, using as example a p--Ar sample of $0.1$~fb$^{-1}$.
}
\label{tab:ftyields}
  \centering
  \begin{tabular}{lcc}
                               & SMOG                   &   SMOG2     \\
                               &  ~~largest sample~~    &    example  \\
                               &  p--Ne@68 GeV            &  ~~p--Ar@115 GeV~~   \\ \hline
Integrated luminosity          & $\sim$ 100~nb$^{-1}$     &    100~pb$^{-1}$  \\
syst. error on J/$\psi$ x-sec.    &     6--7\%            &    2--3  \%     \\
J/$\psi$ yield                    &      15k               &      35M          \\
D$^0$ yield                      &      100k              &      350M         \\
$\Lambda_{\rm c}$ yield                      &       1k               &      3.5M         \\
$\psi$(2S) yield                 &       150              &      400k         \\
\PGUP{1S}~ yield                   &       4               &       15k         \\
Low-mass ($5<M_{\mu\mu}<9~$~GeV/$c^2$) Drell-Yan yield
                               &       5               &       20k         \\
  \end{tabular}
\end{table}

The fixed-target program also presents a very good testbed for the
hydrodynamic description of the QCD medium produced in heavy-ion collisions
down to the energy of $\sqrtsnn \sim 100$~GeV, thanks to the
considerable pseudorapidity coverage, with particle identification capability for pions,
kaons and protons as well as neutral particles $\phi$, $\rm K^0_S$ and $\Lambda^0$.
Measurements of flow observables and correlations can contribute in particular to shed light on the
extension of the hydrodynamic description~\cite{Song:2007ux,Schenke:2010rr} 
successful at top RHIC energy and at the LHC towards lower beam
energies requiring an appropriate 
treatment of baryon density~\cite{Karpenko:2015xea}, and a fully 3 dimensional initial state~\cite{Shen:2017bsr}.
This can provide important inputs to the search for the critical point in
the QCD phase diagram~\cite{Stephanov:2004wx,Luo:2017faz}. 
In addition, the extension of the fluid dynamical paradigm towards smaller collision systems as an
explanation of the experimental findings in proton(deuteron)-nucleus and pp collisions
at the LHC~\cite{Khachatryan:2010gv,CMS:2012qk,Abelev:2012ola,Aad:2012gla,Aaij:2015qcq} and
at RHIC~\cite{Adare:2018toe} can be probed in a novel kinematic regime.

The study of ultra-peripheral collisions in the fixed target
configuration also offers a complementary kinematic regime to
the similar studies in beam-beam collisions discussed in
Section~\ref{sec:smallx}. 
An intriguing possibility would be the observation of $\eta_c$ photoproduction in the PbAr sample, where the cross
section is estimated to be of order $1$~nb~\cite{Goncalves:2015hra} provided that the signal can be cleanly separated from the background due to radiative J/$\psi$ decays. This would constitute 
a confirmation for the existence of the odderon, since the $\eta_{\rm c}$ state cannot be produced in the $\gamma$-pomeron 
process and would be under threshold for the $\gamma$-$\gamma$ process in fixed-target PbA collisions.
Large samples of exclusively produced $\rho^0$ and $\omega$ are also expected~\cite{Goncalves:2018htp}.

Studies of proton collisions on light nuclei provide crucial inputs to the
understanding of cosmic ray propagation in the interstellar medium
(using H and He targets) and in the atmosphere (using N and O
targets). The interpretation of the precise measurements of the
antiproton component in cosmic rays, performed in the last decade by the
PAMELA~\cite{Pamela} and AMS-02~\cite{AMS02} space-based missions, is
presently limited by the knowledge of the antiproton production
in the interstellar medium. The first measurement of antiproton production
in $p$He collisions has been performed by LHCb with Run 2
data~\cite{Aaij:2018svt} and has been used to improve the 
sensitivity to a possible dark matter contribution to the cosmic antiproton
flux~\cite{Reinert:2017aga,Korsmeier:2018gcy}. 
During Run 3, it is planned to extend these studies with a hydrogen
target. The production of antineutrons can also be inferred indirectly
by evaluating isospin violation in antiproton production 
from proton-hydrogen and proton-deuterium collisions.

A better understanding of the bulk of particle production in
high-energy hadronic collisions is also crucial to the modeling of
atmospheric showers induced by ultra-high-energy cosmic rays. 
Despite the moderate energy scale, fixed target data offer the
unique possibility to use a wide range of nuclear targets, including
nitrogen and oxygen, and to study production at large $x$. 
The main background for the observation of the astrophysical
high-energy neutrino flux, recently established by 
the IceCube collaboration~\cite{Aartsen:2013jdh}, originates from
neutrinos of PeV energy produced from decays of charmed hadrons in
atmospheric showers. A relevant uncertainty on this background
is related to the possible intrisic charm contribution to the charm
PDF at large $x$~\cite{Laha:2016dri}.  
Knowledge of light hadron production, notably baryons, at large $x$ will allow 
to improve the modeling of the number of muons produced in the lateral
development of the showers, which is a key observable to infer the
mass composition of ultra-high-energy cosmic rays, as discussed in Sect.~\ref{sec:pOcosmic}.

More ambitious projects for future target upgrades have also been proposed for a possible installation
in LHCb on the time scale of HL-LHC. These are beyond the baseline
LHCb Upgrade 2. A polarised gas target similar to that used in
HERMES~\cite{Airapetian:2004yf}, installed upstream of the LHCb vertex
detector, would make LHCb a key contributor to spin physics. 
With this option, LHCb would have access to single spin asymmetries in an
unique kinematic range for a variety of final states, including unique
measurements with quarkonium and Drell-Yan lepton
pairs~\cite{Kikola:2017hnp}. 
Measurements of the three-dimensional structure of nucleons from hadron collisions would be highly
complementary to the program at the electron-ion collider, which is proposed on a
comparable time scale.

\subsubsection{Opportunities with ALICE}
\label{sec:FTALICE}

The ALICE detector provides many physics opportunities
if running in the fixed-target mode with the LHC proton and lead beams. One of the main strengths of ALICE in the fixed-target mode would be its large rapidity coverage. Assuming a target location at $z=0$, the ALICE muon spectrometer would access the mid- to backward rapidity in the centre-of-mass frame ($-2.3 < y_{\rm cms}\footnote{The rapidity is
calculated assuming massless particles.} < -0.8$) considering an incident proton beam on the target. For an incident lead beam, the muon spectrometer rapidity coverage is $-1.8 < y_{\rm cms} < -0.3$. In addition, the absorber in front of the muon tracking stations is an asset for background rejection and Drell-Yan studies. The ALICE central barrel offers a complementary coverage to the muon arm by accessing the very backward rapidity region ($-5.7< y_{\rm cms} < -3.9$ with the proton beam and $-5.2 < y_{\rm cms} < -3.4$ with the lead beam), reaching the end of the phase space for several probes. Thanks to its excellent particle identification capabilities, particle detection and identification down to low $p_{\rm T}$, unique measurements of soft probes and open heavy flavours can be pursued. Another asset of the ALICE apparatus is the capability to operate with good performance in a high particle density environment. Access to the most central AA collisions at $\sqrtsNN=72$~GeV should be possible if the detector occupancy does not exceed the one expected in Pb--Pb collisions at 50 kHz. In addition, the ALICE Collaboration could potentially devote a significant data taking time to a fixed-target programme (especially with the proton beam), allowing the collection of large integrated luminosities and the investigation of several target species. 

Two main solutions are being investigated to deliver fixed-target collisions to ALICE: an internal gaseous target or an internal solid target (coupled to a bent crystal to deflect the beam halo). On the one hand, a gas-jet or a storage cell with levelled gas pressure would allow to deliver about 45~pb$^{-1}$ of proton-polarised hydrogen collisions to ALICE (260~pb$^{-1}$  in case of unpolarised H$_{2}$ collisions), and 8~nb$^{-1}$ of Pb--Xe collisions per year. For the luminosity calculation, a time duration of $10^7$~s and $10^6$~s, corresponding to one LHC year, is considered for the proton and lead beams, respectively. With a gas system the target can be polarised, but requires large space to be installed, most likely outside the ALICE barrel magnet, 7 metres from the nominal Interaction Point (IP). In that case, additional detectors for vertexing are needed as well as studies of the tracking performances of the TPC in such conditions. A simple unpolarised storage cell might potentially be used closer to the current ALICE IP. On the other hand, the usage of an internal solid target coupled to a bent crystal has the advantage of more portability, allowing one to install the target closer to the IP, from 2.75 to 4.7~m to the IP, and thus benefiting of better performances from the current ALICE apparatus. With such a device 37~pb$^{-1}$ (6~pb$^{-1}$) of p--C (p--W) collisions, and 5~nb$^{-1}$ (3~nb$^{-1}$) of Pb--C (Pb--W) collisions could be registered in ALICE per year. One of these solutions could be
installed during LS3. 

The performance for several key probes has been investigated using fast simulations~\cite{Hadjidakis:2018ifr} and it is summarised in the following.
\begin{itemize}
\item{Quarkonia: Very large yields are expected for charmonia (up to $\sim 10^{6}$ J/$\psi$) in the ALICE muon spectrometer both in pp and Pb-A collisions\footnote{The largest atomic mass number considered here is W for the solid target and Xe for the gas target.} considering one LHC year of data taking. The \PGUP{1S} will also be within reach. Looking at quarkonium suppression as a function of rapidity and the system size would allow one to search for the onset of the QGP formation, and to determine the in-medium modification of the QCD forces for centre-of-mass energies between the ones probed at RHIC and the SPS. In pA collisions, \PGUP{1S} could be used to probe large-$x$ gluons in the target ($0.1<x<1$), in order to constrain the anti-shadowing and EMC effects.}

\item{Drell-Yan: A precise measurement of the Drell-Yan process with the muon spectrometer can probe initial state effects on quarks of momentum fraction $0.05<x<0.8$ at a mass scale of $M_{\mu\mu}> 4$~GeV/$c^2$ from pA to A- collisions. The correlated background from $\rm b\overline{b}$ and $\rm c\overline{c}$ pairs in the dimuon decay channel is largely reduced at the lower centre-of-mass energy of the fixed-target mode, with respect to the TeV energy range.}

\item{Open heavy flavour: At $y_{\rm lab} \sim 1$ about 3000 (up to 100) D$^{0}$ are expected to be produced per 0.1 rapidity unit per year in pp (Pb-A) collisions, respectively. The ALICE central barrel probes the end of the D meson phase-space, in particular if the vertex is located at $z < 4$~m from the IP. This would allow one to reach very large-$x$ gluon close to 1 at low mass scale, where the contribution of the intrinsic charm component in the proton could largely increase the D meson yield. Hence the ALICE central barrel is well suited to study the large-$x$ intrinsic charm component in the proton.}

\item{Identified charged hadrons: In few hours of Pb--A data taking, it would be already possible to collect up to 10$^{6}$ minimum bias events with the ALICE central barrel which would allow one to reach an absolute statistical uncertainty of 0.01 on the elliptic flow coefficient $v_{2}$ for pions and protons, 0.02 for kaons, and 0.05 for antiprotons in semi-central events. Measurements of identified particles up to very large rapidities would complement the limiting fragmentation studies carried out by the BRAHMS and PHOBOS experiments at RHIC. 
In addition, particle yields and flow coefficients measured at large rapidities are powerful tools to constrain the temperature dependence of the medium shear viscosity~\cite{Denicol:2015nhu}. Thermal model calculations also indicate that the temperature and baryonic chemical potential depend on the rapidity~\cite{Becattini:2007qr,Karpenko:2018xam,Begun:2018efg} suggesting that one can perform a rapidity scan of the QGP phase diagram in a complementary approach to the Beam Energy Scan programme of RHIC.
} 

\item{Antiproton production: Measurements of antiproton production cross sections in p--H, p--He, p--C, p--N and p--O collisions are important inputs for theoretical calculations of the secondary cosmic antiproton spectrum \cite{Moskalenko:1997gh, Vladimirov:2010aq, Orlando:2017tde}, where secondary antiprotons originate from the high energy scattering between the interstellar matter and the primary cosmic rays. The measurement of a cosmic antiproton excess with respect to expectations from secondary antiproton production would open new perspectives on the indirect detection of dark matter or unknown astrophysical mechanisms of cosmic ray acceleration. Complementarily to LHCb \cite{Aaij:2018svt}, the ALICE central barrel can measure very slow antiprotons down to few hundred MeV momentum. Measuring slow antiprotons produced with the LHC proton beam on a nuclear target is equivalent to the case where the nuclear target travels at TeV energies, hit an interstellar proton at rest and produces an antiproton with high energy. Thanks to the large antiproton yields expected in p--H$_{2}$ collisions (larger than 10$^{8}$ per 0.1 rapidity unit per year), the ALICE central barrel is well placed to help constraining the uncertainty on the cosmic antiproton spectrum.}

\item{Strangeness: Thanks to the large yields ($\sim 10^{8}$ per 0.1 rapidity unit per year at $y_{\rm lab} \sim 1$) of $\Lambda$ hyperons expected to be produced in the ALICE central barrel by using a longitudinally polarised target with the proton beam, a precise measurement of the longitudinal spin transfer $D_{LL}$ of the $\Lambda$ hyperon could be carried out. So far only limited experimental results exist with poor precision~\cite{Abelev:2009xg,Alekseev:2009ab}. Measurements with ALICE would give a unique opportunity to study the spin-dependent strange quark (antiquark) densities at $0.35<x<0.7$.}

\item{Charmonium and pentaquark photoproduction: Exclusive J/$\psi$ photoproduction is known to be sensitive to gluon Generalised Parton Distributions (GPD) at leading order \cite{Ivanov:2004vd}. ALICE in fixed target mode would have a unique opportunity to study the yet unknown GPD $E_{\rm g}$ thanks to the measurement of single transverse spin asymmetries of photoproduced J/$\psi$ by using a transversally polarised H target with the proton beam~\cite{Massacrier:2017lib,Koempel:2011rc}. About 200 photoproduced J/$\psi$ per year are expected to be produced in the muon spectrometer acceptance. Moreover, the photoproduction of hidden charm pentaquark states \cite{Wang:2015jsa} might be possible in the central barrel acceptance which allows for the access to low photon-proton centre-of-mass energies ($W_{\gamma p} \sim 5$~GeV). About 2 to 20 pentaquarks are expected to be produced in the ALICE central barrel per year in p--H$_{2}$ collisions.}
\end{itemize}

Note that these studies were performed assuming a vertex position at the ALICE IP for the quarkonia, identified hadron production, as well as for charmonium and pentaquark photoproduction, and a vertex position at 4.7~m from the IP for the antiproton and strangeness production. 

Studies are ongoing to address the technical feasibility of the target system integration in the experiment, as well as simulation studies to evaluate the ALICE apparatus tracking performance for target positions displaced by a few metres from the IP. Moreover, investigations are ongoing to extend the ALICE rapidity coverage for several observables thanks to combined measurements of muons detected both in the ALICE central and the muon spectrometer (see as an example the work in \cite{Acharya:2018jua} in collider mode).

\clearpage

\section{Summary of luminosity requirements and proposed run schedule}
\label{sec:schedule}

The physics programme presented in this report requires data-taking campaigns with various colliding systems with centre-of-mass energies and integrated luminosities \Lint\ as outlined in the following. In some cases the requirements are updated or new with respect to the present baseline LHC programme (see Sec.~\ref{sec:LHCpresentbaseline} and Ref.~\cite{Abelevetal:2014cna}). The main variations are: a much larger \Lint\ target for \pPb collisions, motivated by high-precision studies of both initial and final-state effects, following the surprising discoveries of collective-like effects in small collision systems; a large sample of pp collisions at top LHC energy to reach the highest possible multiplicities with the smallest hadronic colliding system; moderate-statistics samples of \OO 
(as mentioned in \label{sec:pOrun}, this would be a limited ``pilot-run'' scenarion that could not achieve the values listed in Tables \ref{tab:species1} and \ref{tab:species2}) 
and \pO collisions, to study the onset of hot-medium effects and to tune cosmic-ray particle production models, respectively. Finally, as discussed in Sec.~\ref{sec:smallAsum}, extended LHC running with colliding intermediate-mass nuclei (as, for example, \ArAr or \KrKr), offers the unique opportunity of a large increase in nucleon--nucleon luminosity to access novel probes of the QGP and to open a precision era for probes which are still rare with the \PbPb system. The working group considers the high-luminosity \PbPb and \pPb programmes to be the priorities that should be pursued in Run 3 and Run 4. High-luminosity runs with intermediate-mass nuclei are regarded as an appealing case for extending the heavy-ion programme at the LHC after LS4.
This case, including the choice of the optimal nuclear species, should be studied further from the theoretical and operational points of view, both of which could be informed with one or two pilot runs with different species. 

\begin{itemize}

\item {\bf Pb--Pb at $\sqrtsNN=5.5$~TeV}, $\Lint=13~{\rm nb}^{-1}$ (ALICE, ATLAS, CMS), $2~{\rm nb}^{-1}$ (LHCb)

\item {\bf pp at $\sqrt s=5.5$~TeV}, $\Lint=600~{\rm pb}^{-1}$ (ATLAS, CMS), $6~{\rm pb}^{-1}$ (ALICE),  $50~{\rm pb}^{-1}$ (LHCb) 

\item {\bf pp at $\sqrt s=14$~TeV}, $\Lint=200~{\rm pb}^{-1}$ with low pileup (ALICE, ATLAS, CMS)

\item {\bf p--Pb at $\sqrtsNN=8.8$~TeV}, $\Lint=1.2~{\rm pb}^{-1}$ (ATLAS, CMS), $0.6~{\rm pb}^{-1}$ (ALICE, LHCb) 

\item {\bf pp at $\sqrt s=8.8$~TeV}, $\Lint=200~{\rm pb}^{-1}$ (ATLAS, CMS, LHCb), $3~{\rm pb}^{-1}$ (ALICE)

\item {\bf O--O at $\sqrtsNN=7$~TeV}, $\Lint=500~{\rm \mu b}^{-1}$ (ALICE, ATLAS, CMS, LHCb)

\item {\bf p--O at $\sqrtsNN=9.9$~TeV}, $\Lint=200~{\rm \mu b}^{-1}$ (ALICE, ATLAS, CMS, LHCb)

\item {\bf Intermediate AA}, e.g.\,$\Lint^{\rm Ar-Ar}=3$--$9~{\rm pb}^{-1}$ (about 3 months) gives NN luminosity equivalent to Pb--Pb with $L_{\rm int} =75$--$250~\invnb$

\end{itemize}

Based on these requirements, the proposed updated running schedule is reported in the following table. It can be seen that the physics programme for Run 3 and Run 4 discussed in this report is achievable by a modest increase of the ``heavy-ion running'' time from 12 to 14 weeks per run.

\begin{center}
\begin{tabular}{llll}
Year & Systems, $\sqrtsNN$ & Time & $\Lint$ \\
\hline
2021 &  Pb--Pb 5.5 TeV & 3 weeks & $2.3~\invnb$ \\
     &  pp 5.5 TeV & 1 week & $3~\invpb$ (ALICE), $300~\invpb$ (ATLAS, CMS), $25~\invpb$ (LHCb)  \\
\hline
2022 &  Pb--Pb 5.5 TeV & 5 weeks & $3.9~\invnb$ \\
     &  O--O, p--O & 1 week & $500~{\rm \mu b}^{-1}$ and $200~{\rm \mu b}^{-1}$ \\
\hline
2023 &  p--Pb 8.8 TeV & 3 weeks & $0.6~\invpb$ (ATLAS, CMS), $0.3~\invpb$ (ALICE, LHCb) \\
     &  pp 8.8 TeV & few days & $1.5~\invpb$ (ALICE), $100~\invpb$ (ATLAS, CMS, LHCb)  \\
\hline
2027 &  Pb--Pb 5.5 TeV & 5 weeks & $3.8~\invnb$ \\
     &  pp 5.5 TeV & 1 week & $3~\invpb$ (ALICE), $300~\invpb$ (ATLAS, CMS), $25~\invpb$ (LHCb)  \\
\hline
2028 &  p--Pb 8.8 TeV & 3 weeks & $0.6~\invpb$ (ATLAS, CMS), $0.3~\invpb$ (ALICE, LHCb) \\
     &  pp 8.8 TeV & few days & $1.5~\invpb$ (ALICE), $100~\invpb$ (ATLAS, CMS, LHCb)  \\
\hline
2029 &  Pb--Pb 5.5 TeV & 4 weeks & $3~\invnb$ \\
\hline
Run-5 & Intermediate AA & 11 weeks & e.g.\,Ar--Ar 3--9~$\invpb$ (optimal species to be defined) \\
     &  pp reference & 1 week & \\
\hline
\end{tabular}
\end{center}

\clearpage

\section{First considerations on a heavy-ion programme at a High Energy LHC (HE-LHC)}
\label{sec:HELHC}

{ \small
\noindent \textbf{Coordinators}: Andrea~Dainese (INFN Padova), David~d'Enterria (CERN) and Carlos~A.~Salgado (Instituto Galego de Fisica de Altas Enerxias (IGFAE) Universidade de Santiago de Compostela)

\noindent \textbf{Contributors}: 
L.~Apolinario (LIP and IST Lisbon), 
N.~Armesto (Instituto Galego de Fisica de Altas Enerxias (IGFAE) Universidade de Santiago de Compostela), 
J.~Jowett (CERN), 
G.~Milhano (LIP and IST Lisbon, CERN), 
U.A.~Wiedemann (CERN)
}

\subsection{Introduction}
\label{sec:HELHC_intro}

In this section the physics opportunities associated with the operation of the HE-LHC with heavy-ion beams are discussed.
These first considerations are based on studies carried out in the scope of the Future Circular Collider (FCC) Study group~\cite{Dainese:2016gch,FCC-CDR}.
For a centre-of-mass energy $\sqrt{s}= 27$~TeV for pp collisions, the relation $\sqrt{s_{\rm NN}}= \sqrt {s} \sqrt{Z_1 Z_2 / A_1 A_2}$ 
gives the energy per nucleon--nucleon collision of $\sqrt{s_{\rm NN}} = 10.6$~TeV for Pb--Pb ($Z=82$, $A=208$) and 17~TeV for p--Pb collisions. 
The present estimate of the integrated luminosity per month of running is larger by a factor two with respect to the current projection for the future 
LHC runs, i.e.\, $L_{\rm int}\approx 6~{\rm nb}^{-1}$ per experiment, see Section~\ref{sec:HLPbPb}. 
The possibility of using nuclei smaller than Pb, like e.g.\,$^{40}$Ar or $^{129}$Xe, to achieve larger instantaneous luminosity is also under consideration. 
For example, the integrated nucleon--nucleon (NN) luminosity per run 
for Xe--Xe collisions at $\sqrt{s_{\rm NN}}= 11.5$~TeV could be 2--3 times larger than for Pb--Pb collisions (see integrated $L_{\rm NN}$ values in Tables~\ref{tab:species1} and \ref{tab:species2}).

The increase in the centre-of-mass energy and integrated luminosity at the FCC with respect to the LHC opens up novel opportunities for physics studies of the Quark-Gluon Plasma (QGP) described in a recent CERN Yellow Report~\cite{Dainese:2016gch}. Most of these opportunities also apply to the HE-LHC scenario, although 
with more moderate reach in terms of available probes and kinematics coverage.
The main scientific motivations for a heavy-ion programme at the HE-LHC can be summarized as follows.

\noindent
	{\bf Novel access to QCD thermodynamics and QCD equilibration processes.}
	Substantially increasing the centre-of-mass energy leads to the creation of initially denser and hotter strongly-interacting systems that expand for a longer duration and over a larger volume, thereby
	developing stronger collective phenomena. 
Extrapolations of LHC measurements indicate that the initial energy density increases by a factor about 1.4 from $\sqrt{s_{\rm NN}} = 5.5$~TeV to 10.6~TeV, up to values of about 22--24~GeV/fm$^3$ (at $\tau=1$~fm/$c$).
These estimates are presented in Section~\ref{sec:HE_qgpglobal}.
The QGP formed at the HE-LHC collision energies reaches closer to a range of temperatures ($T\sim 1$~GeV) where charm quarks start to contribute as active thermal degrees of freedom in the QGP equation of state, thus playing a novel role in QCD equilibration processes. 

\noindent
	 {\bf Characterisation of dense QCD matter through hard-scattering processes.}
As detailed in Section~\ref{sec:HE_hardprobes}, the HE-LHC would provide a much larger abundance of hard-scattering processes than the LHC, as well as novel probes like the top quark and, potentially, the Higgs boson~\cite{Apolinario:2017sob,dEnterria:2015mgr,dEnterria:2017jyt}. 
A notable example is provided by high-momentum (thus, highly boosted) $\rm t \to W \to q\overline q$ decay chains, which are promising probes of the QGP time evolution and of the role of colour coherence~\cite{Apolinario:2017sob}. 
The secondary production of charm quarks in scatterings between quark and gluon constituents of the hot QCD medium could reach a substantial fraction of the initial production in partonic hard scatterings and be observed for the first time. 

\noindent
{\bf Exploration of saturated parton densities in a previously-uncharted, ultra-dense kinematic domain.}
As discussed in Section~\ref{sec:HE_smallx}, the higher centre-of-mass energy of the HE-LHC allows one to explore a wide previously-uncharted kinematic range 
at low $x$ and $Q^2$, where parton saturation is expected to set in.  
Proton--nucleus collisions would have a coverage down  to $x\sim 5\times 10^{-6}$ in the Pb nucleus at a rapidity of $y\approx 5$.

\subsection{Global characteristics of nucleus--nucleus collisions at the HE-LHC}
\label{sec:HE_qgpglobal}

Extrapolating measurements of charged particle multiplicity, transverse energy and femtoscopic correlations 
at lower energies, one can obtain estimates
for the growth of global event characteristics from the LHC to the HE-HC and the FCC. In particular, up to the top LHC energy, the growth of charged-particle 
multiplicity per participant pair per unit rapidity in nucleus--nucleus collisions is consistent with a slowly-rising power-law:
   ${\rm d}N_{\rm ch}/{\rm d}\eta\,(\eta=0) \propto (\sqrtsNN)^{0.3}$ (see e.g.\,\cite{Aamodt:2010pb}).
As shown in Table~\ref{tab:PbPb}, for Pb--Pb this amounts to an increase of a factor $\sim 1.2$ from the LHC to the HE-LHC. 
The multiplicity in central Xe--Xe collisions is expected to be lower by 35\% with respect to Pb--Pb collisions at the HE-LHC, and similar to 
that of Pb--Pb collisions at 2.76~TeV.

\begin{table}[!t]
\caption{Global properties measured in central Pb--Pb collisions (0--5\% centrality class) at $\sqrtsNN=2.76$~TeV and extrapolated to 5.5, 10.6 and 39~TeV. The values for Pb--Pb collisions at the LHC and FCC are from Ref.~\cite{Dainese:2016gch}. The values for Pb--Pb collisions at the HE-LHC are estimated using the same parametrisations as used for the FCC. The values for Xe--Xe collisions at the HE-LHC are all estimated on the basis of the multiplicity extrapolation from the measurement by the ALICE Collaboration~\cite{Acharya:2018hhy} (it is assumed that the transverse energy density scales only with the multiplicity, neglecting possible differences of the average energy per particle between Pb--Pb and Xe--Xe and between the LHC and the HE-LHC).}
\small
\begin{center}
\begin{tabular}{lccccc}
\hline
System, $\sqrtsNN$ (Tev) & Pb--Pb, 2.76 & Pb--Pb, 5.5 & Pb--Pb, 10.6 & Xe--Xe, 11.5 & Pb--Pb, 39.4 \\
\hline
${\rm d}N_{\rm ch}/{\rm d}\eta$ at $\eta=0$ & 1600 & 2000 & 2400 & 1500  & 3600 \\
${\rm d}E_{\rm T}/{\rm d}\eta$ at $\eta=0$ (TeV) & 1.7--2.0 & 2.3--2.6 & 3.1--3.4 & $\approx 1.5$ &  5.2--5.8 \\
Homogeneity volume fm$^3$ & 5000  & 6200 & 7400 & 4500 & 11000 \\
Decoupling time (fm/$c$) & 10 &  11 & 11.5 & 10 & 13 \\
$\varepsilon$ at $\tau=1$~fm/$c$ (GeV/fm$^3$) & 12--13  & 16--17 & 22--24 & $\approx 15$ & 35--40 \\
\hline
\end{tabular}
\end{center}
\label{tab:PbPb}
\end{table}


In general, the global event characteristics listed in Table~\ref{tab:PbPb} determine the spatio-temporal extent QGP system, and they
constrain the thermodynamic conditions that apply after thermalization. The measured transverse energy per unit rapidity ${\rm d}E_{\rm T}/{\rm d}\eta$ (see Table~\ref{tab:PbPb})
is of particular importance since it constrains the initial energy density. 
The energy density is expected to increase by a factor 1.4 from the LHC to the HE-LHC, reaching a value of 22--24~GeV/fm$^3$ at the time of 1~fm/$c$~\cite{Dainese:2016gch}. 
Using the arguments presented in Ref.~\cite{Dainese:2016gch}, an initial temperature as large
as $T_0\approx 600$--$800$~MeV is expected at the time $\mathcal{O}(0.05\, {\rm fm}/c)$ after which both nuclei traverse each other at HE-LHC energies.
In the case of Xe--Xe collisions the energy density is estimated to be significantly lower than that for Pb--Pb and similar to that of Pb--Pb at LHC energies.

\subsection{QGP studies with hard probes}
\label{sec:HE_hardprobes}

\subsubsection{Hard processes in nucleus--nucleus collisions at the HE-LHC}
\label{sec:HE_xsections}

\begin{figure}[!h]
\begin{center}
\includegraphics[width=0.49\textwidth]{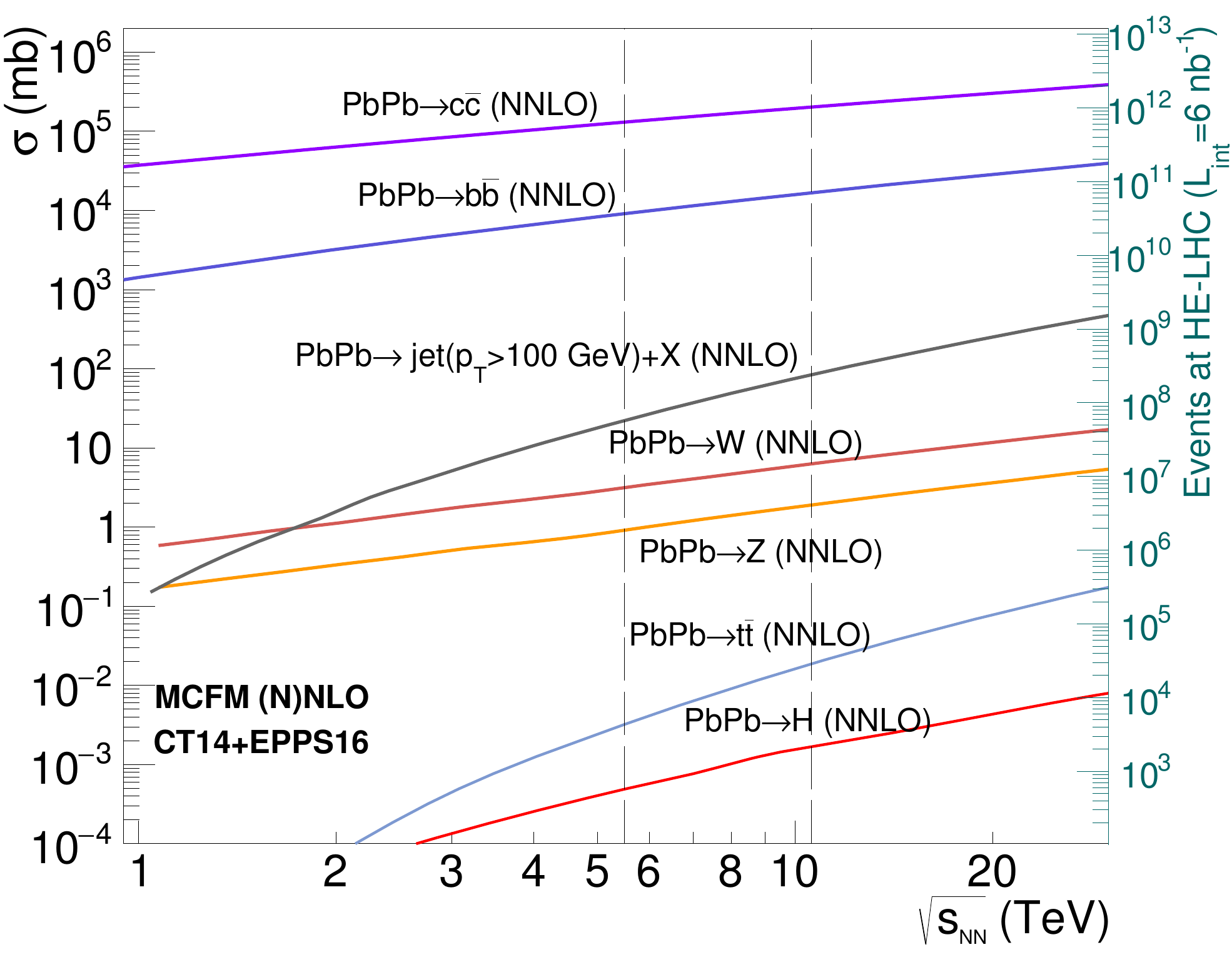}
\hfill
\includegraphics[width=0.49\textwidth]{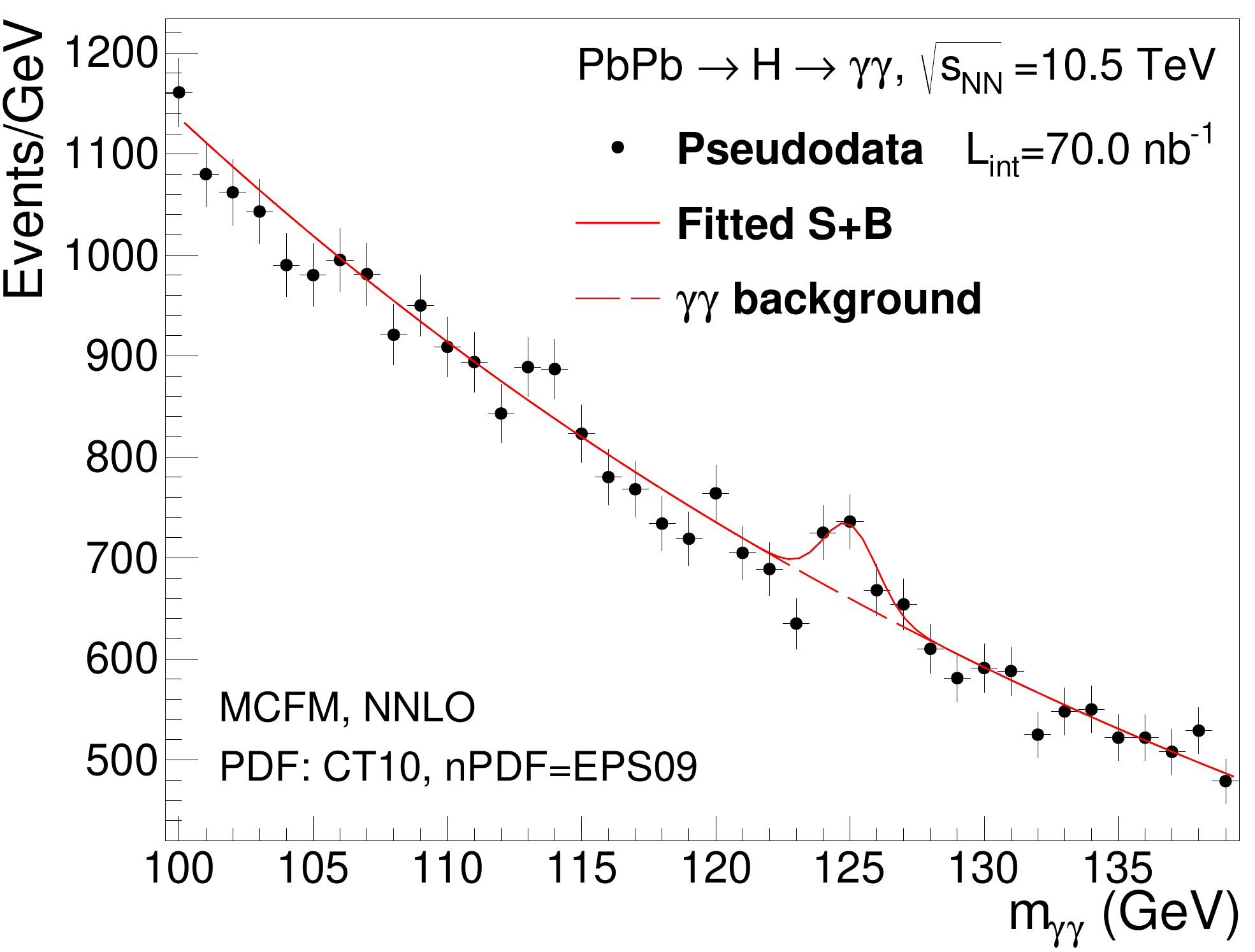}
\caption{Left:$\sqrt s$-dependence of the cross sections for hard processes of
  interest for a heavy-ion programme, calculated with MCFM~\cite{Campbell:2010ff}
at the highest available order. The yields for a one month Pb--Pb run ($\Lint=6~\invnb$) are also shown.
  Right: invariant mass distribution for Higgs boson search in the $\gamma\gamma$ decay channel in Pb--Pb collisions at the HE-LHC with $\Lint=70~\invnb$ using the selections described in Ref.~\cite{dEnterria:2017jyt}.}
\label{fig:hardXsectHIC}
\end{center}
\end{figure}

The increase in energy and luminosity (in the case of Xe--Xe) from the LHC to the HE-LHC provides new tools to study the matter created in the collisions of heavy ions.
In Fig.~\ref{fig:hardXsectHIC} (left), cross sections for different processes
and different energies are computed with  MCFM~\cite{Campbell:2010ff}
at the highest available order. The increases amount to a factor
$\sim 2$ for charm, beauty, W and Z production, $\sim 4$ for jets with $\pT>100$~GeV/$c$ and for Higgs, and $\sim 6$ for top-pair production. 

The motivations for measurements of top quarks in heavy-ion collisions
are multifold. 
In p--Pb collisions the cross sections efficiently probe the nuclear
gluon PDFs in a wide range in momentum fraction
$x$ at high scale $Q\sim m_{\rm t}$~\cite{dEnterria:2015mgr} (see Section~\ref{sec:nPDF_top}). In Pb--Pb collisions, the top-quark observables are sensitive to the energy-loss of heavy quarks~\cite{Baskakov:2015nxa}
and by selecting boosted (very high $p_{\rm T}$) top quarks one could also probe the QGP medium at later times as the decays of boosted top quarks get Lorentz
time dilated (see next section). 
For example, the estimated measurable yields  for $\rm t\bar t\to b\bar b\,\ell\ell\,\nu\nu$ (using the
 per-month luminosities discussed in Section~\ref{sec:HELHC_intro}) with
realistic analysis cuts 
and conservative 50\% efficiency for b-jet tagging are
about $10^4$ in Pb--Pb collisions and $3\times 10^4$ in Xe--Xe collisions (for the case of three-fold increase of NN integrated luminosity with respect to Pb--Pb).

Another potential novel probe of the QGP medium at HE-LHC and FCC energies is the
Higgs boson. The Higgs boson has a lifetime of $\tau\approx 50$~fm/$c$, which is much larger than the 
time extent of the QGP phase~\cite{dEnterria:2018bqi,Berger:2018mtg}. 
In Ref.~\cite{dEnterria:2018bqi}
it has been argued that the Higgs boson
interacts strongly with the quarks and gluons of the QGP and the interactions
induce its decay in the gluon--gluon or quark--antiquark channels, thus depleting the 
branching ratio to the most common ``observation'' channels $\gamma\gamma$ or $\rm ZZ^\star$.
More recent detailed theoretical calculations, including virtual corrections, predict however no visible suppression of the scalar boson~\cite{Ghiglieri:2019lzz}.
The cross section for Higgs boson production in Pb--Pb collisions is expected to
increase by a factor about 4 when going from
$\sqrtsNN=5.5~\tev$ to $\sqrtsNN=10.6~\tev$~\cite{dEnterria:2017jyt}.
A statistically-significant Higgs boson observation in the
$\gamma\gamma$ decay channel in Pb--Pb collisions at the HE-LHC 
requires an integrated luminosity of $70$~nb$^{-1}$ 
(estimated as in Ref.~\cite{dEnterria:2017jyt}), which corresponds to about 12 months with the present machine performance
projections. 
The analysis used similar photon selections as used by ATLAS and CMS in pp collisions: 
$\pt > 30,40$~GeV/$c$, $|\eta| < 4$, $R_{\rm isol} = 0.3$. The backgrounds included the irreducible QCD diphoton continuum plus 30\% of events coming from misidentified 
$\gamma$-jet and jet-jet processes. 
The corresponding invariant mass distribution is shown in the right-hand panel of Fig.~\ref{fig:hardXsectHIC}.
With Xe--Xe collisions the same statistical significance could be reached in 4 months.

\subsubsection{Boosted tops and the time evolution of QGP opacity}
\label{sec:HE_boostedtops}

The HE-LHC would provide large rates of highly-boosted heavy particles, such as tops, Z and W bosons. It is expected that when these particles decay the density profile of the QGP has already evolved. It has been argued that the hadronically-decaying W bosons in events with a $\rm t\bar t$ pair can provide unique insights into the time structure of the QGP~\cite{Apolinario:2017sob}. This is because the time decays of the top and the W bosons are followed by a time-delay in the interaction of the decay products of the W boson with the surrounding medium due to a colour coherence effect. The sum of these three times, several fm/$c$ for boosted tops, would be the time at which the interaction with the QGP begins, providing a unique way to directly measure the time structure of the QGP evolution.
In addition, due to colour coherence effects, energy loss  would be initially absent for the colour-singlet $\rm q\overline q$ decay products of a highly-boosted W boson: the two quarks would start to be quenched only when their distance becomes larger than the colour correlation length of the medium, which depends on the transport coefficient $\hat{q}$ (the average transverse momentum squared that particles exchange with the medium per unit mean-free path)~\cite{CasalderreySolana:2012ef}.
The effect on the reconstructed masses of the top and W was studied with different energy loss scenarios as a proof of concept of the potential of these observables to access completely novel quantities in heavy-ion collisions~\cite{Apolinario:2017sob}.

\begin{figure}[!t]
\begin{center}
\includegraphics[width=0.49\columnwidth]{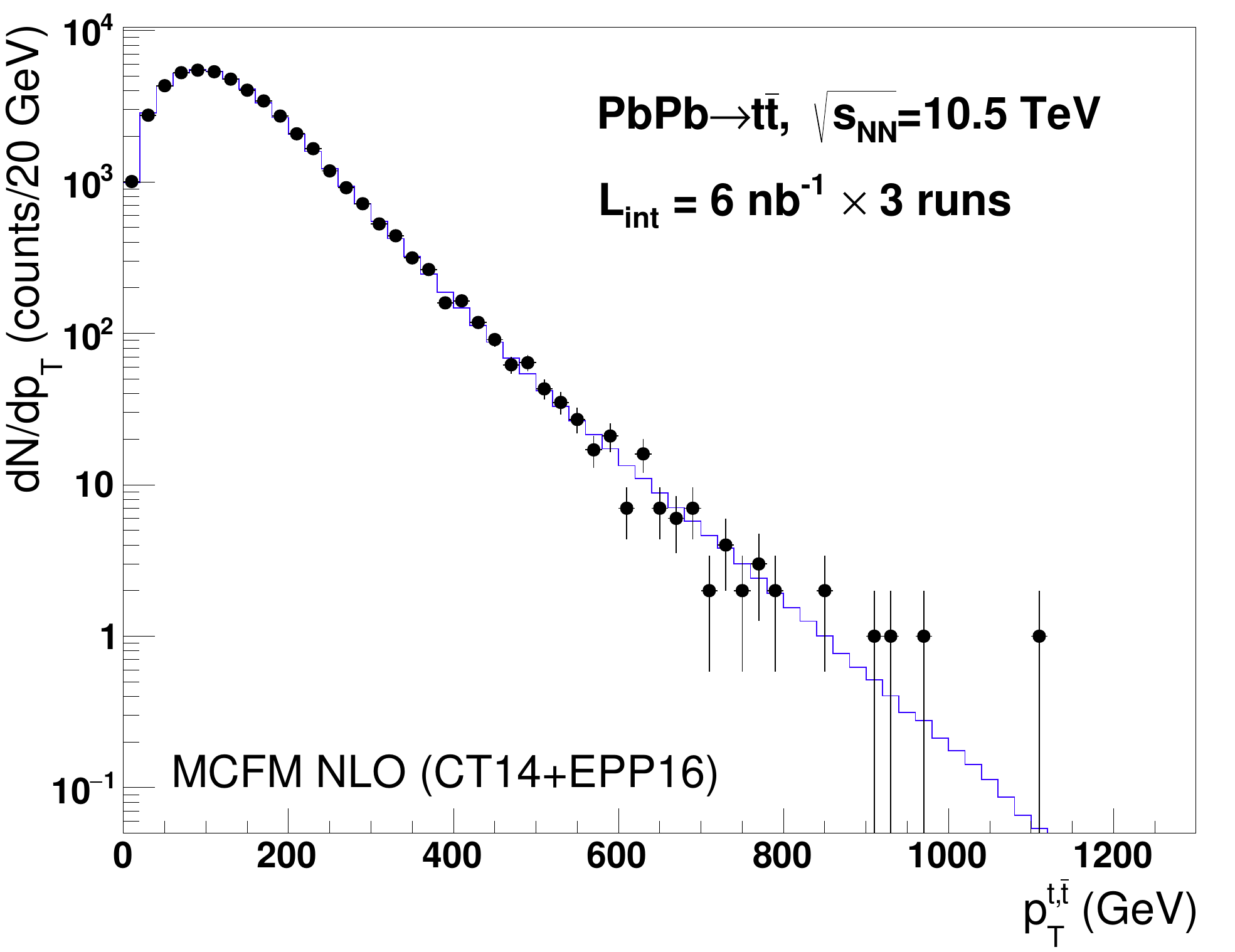}\\
\includegraphics[width=0.46\textwidth]{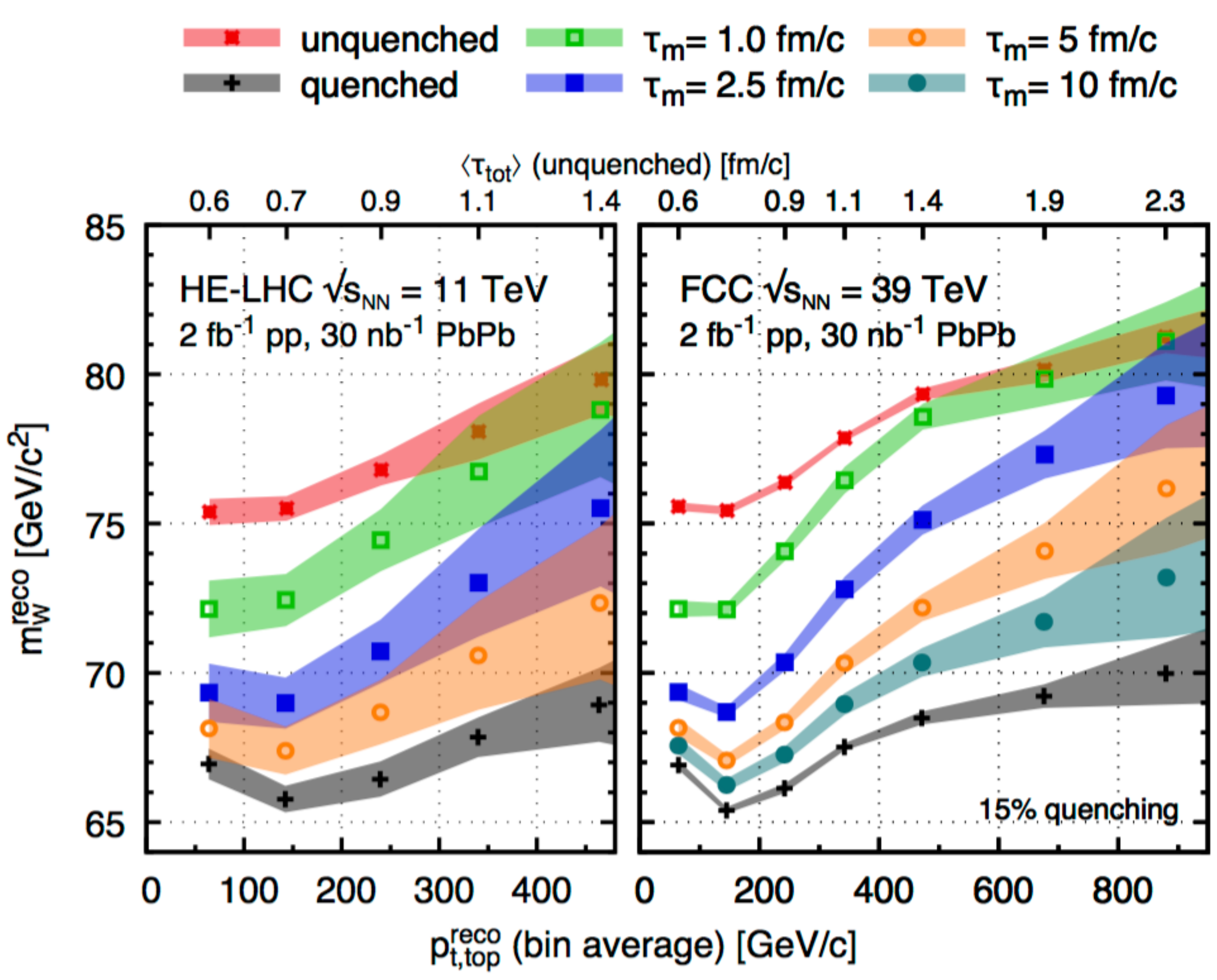}
\hfill
\includegraphics[width=0.5\textwidth]{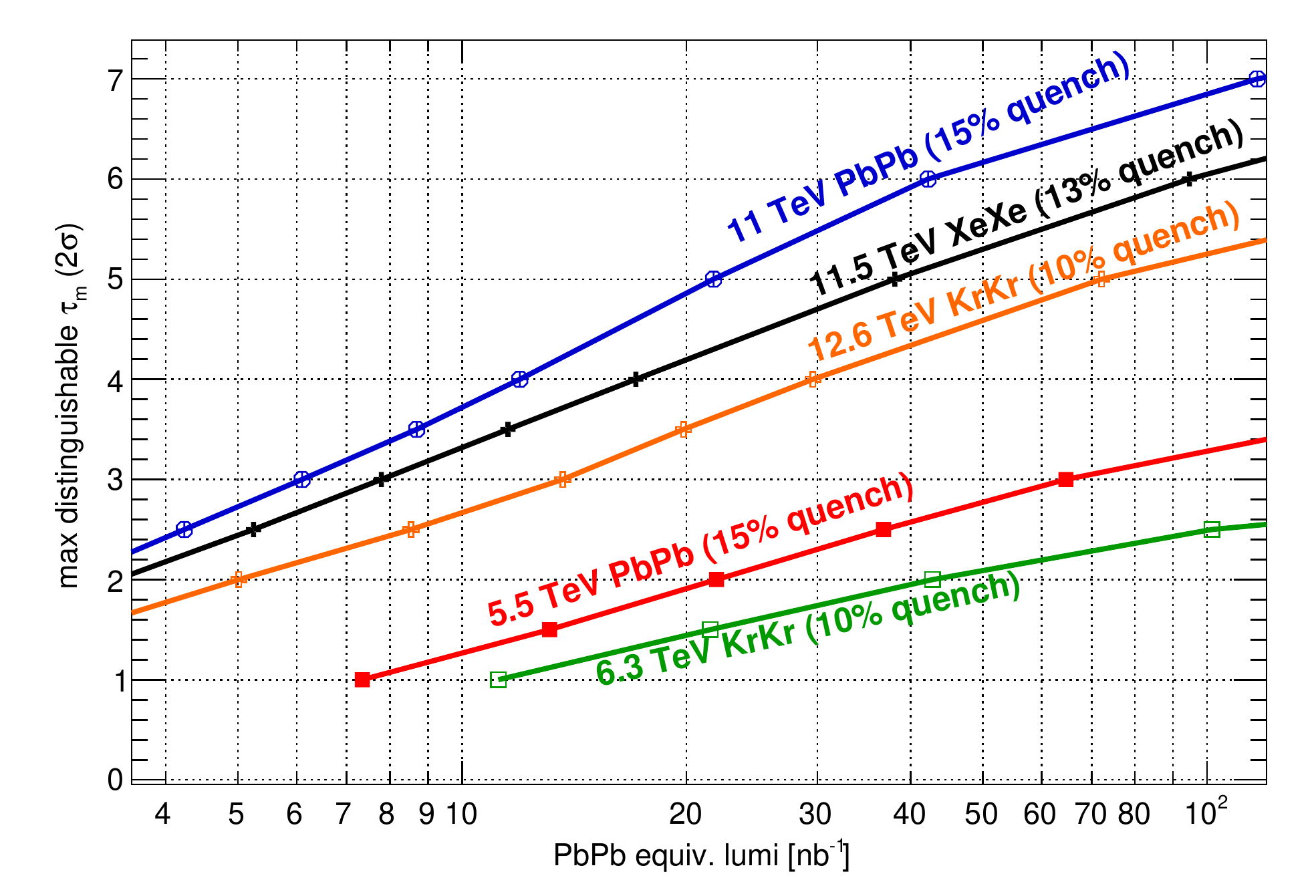}
\caption{Upper panel: expected top-quark $\pt$ distribution in \PbPb\ in the
  decay channels of interest for the boosted top analysis at $\sqrtsnn=10.6$~TeV after acceptance and
  efficiency cuts with the statistical uncertainties
  for $L_{\rm int}=18$~nb$^{-1}$, corresponding to three months of data taking (adapted from~\cite{dEnterria:2015mgr}).
Lower-left panel: reconstructed W boson mass at HE-LHC and FCC
  energies, as a function of the top $\pt$. The upper axis refers to
  the average total time delay of the corresponding top $\pt$
  bin. Lower-right panel: maximum medium quenching
end-time, $\tau_m$, that can be distinguished from full quenching at
two standard deviations, as a function of luminosity for various species at LHC and HE-LHC energies.
The luminosity for nuclei lighter than Pb is scaled to the Pb--Pb equivalent nucleon--nucleon luminosity. 
Figures adapted from Ref.~\cite{Apolinario:2017sob}.}
\label{fig:tops}
\end{center}
\end{figure}

For this study, the $p_{\rm T}$ reach of top quarks in Pb--Pb collisions is of special importance, because it determines the maximum time delay for probing the QGP.
The upper panel of Fig.~\ref{fig:tops} shows the estimated $p_{\rm T}$ distribution
of the reconstructed top yields in Pb--Pb collisions at the HE-LHC with an integrated luminosity of $18~\invnb$, corresponding to three months of data taking. The figure indicates that events with top $\pT$ up to about 500~GeV/$c$ can be studied. 
The reconstructed W-boson mass as a function of top transverse
momentum is shown in Fig.~\ref{fig:tops}
(lower-left panel), together with the FCC case. For details on the simulation and reconstruction procedure
see~\cite{Apolinario:2017sob}. The scale on the upper horizontal axis shows that a $\pT$ reach of 500~GeV/$c$ corresponds to an average total time delay $\langle\tau_{\rm tot}\rangle\sim 1.5$~fm/$c$.
The shaded region corresponds to the statistical uncertainty estimated in central Pb--Pb collisions for
$L_{\rm int} = 30$~nb$^{-1}$ (corresponding to 5 Pb--Pb months or 1.5 Xe--Xe months with the present luminosity estimates)
and $L_{\rm int} = 2$~fb$^{-1}$ for the pp reference.
Energy loss was simulated by assuming that all particles, except the W-boson decay products, lose $15\%$ of their initial momentum. 
Average time delays $\tau_m = 1; 2.5; 5$ and $10~$fm$/c$ were considered as effective QGP time evolution profiles. 
Fig.~\ref{fig:tops} (right) shows the maximum medium quenching
end-time, $\tau_m$, that can be distinguished from full quenching with
two standard deviations, as a function of luminosity for various
collider energies and species. For Pb--Pb with $L_{\rm int}=30$~nb$^{-1}$ (5 months) 
at the HE-LHC a maximum time of 5--6~fm/$c$ can be accessed, which is much
larger than the time up to 1.5~fm/$c$ that can be probed at the LHC with the nominal programme of 
10~nb$^{-1}$. For Xe--Xe collisions, with equivalent NN luminosity larger by a factor 2--3 with respect to Pb-Pb, 
a time range longer by $\sim 1$--2~fm/$c$ can be covered in the same running period.

\subsubsection{Heavy flavour and quarkonia}
\label{sec:HE_hf}

Heavy quarks (charm and bottom) are among the hard probes that have
provided important insights on the formation and the characterics of
the QGP, see Sections~\ref{sec:HI_HF} and~\ref{sec:quarkonia}, and Ref.~\cite{Andronic:2015wma}.
In this section,
a few selected aspects that could
represent novel or particularly remarkable observations at HE-LHC
energy are discussed,
namely: i) large production of thermal charm from
  interactions of light quarks and gluons within the QGP;
ii) observation of an enhancement of charmonium production with
  respect to the binary scaling of the yields in pp collisions, as
  consequence of (re)generation;
iii) observation of a colour screening and (re)generation for the most tightly-bound
  quarkonium state, the $\Upsilon$(1S).

\begin{figure}[!t]
\begin{center}
\includegraphics[width=0.54\textwidth]{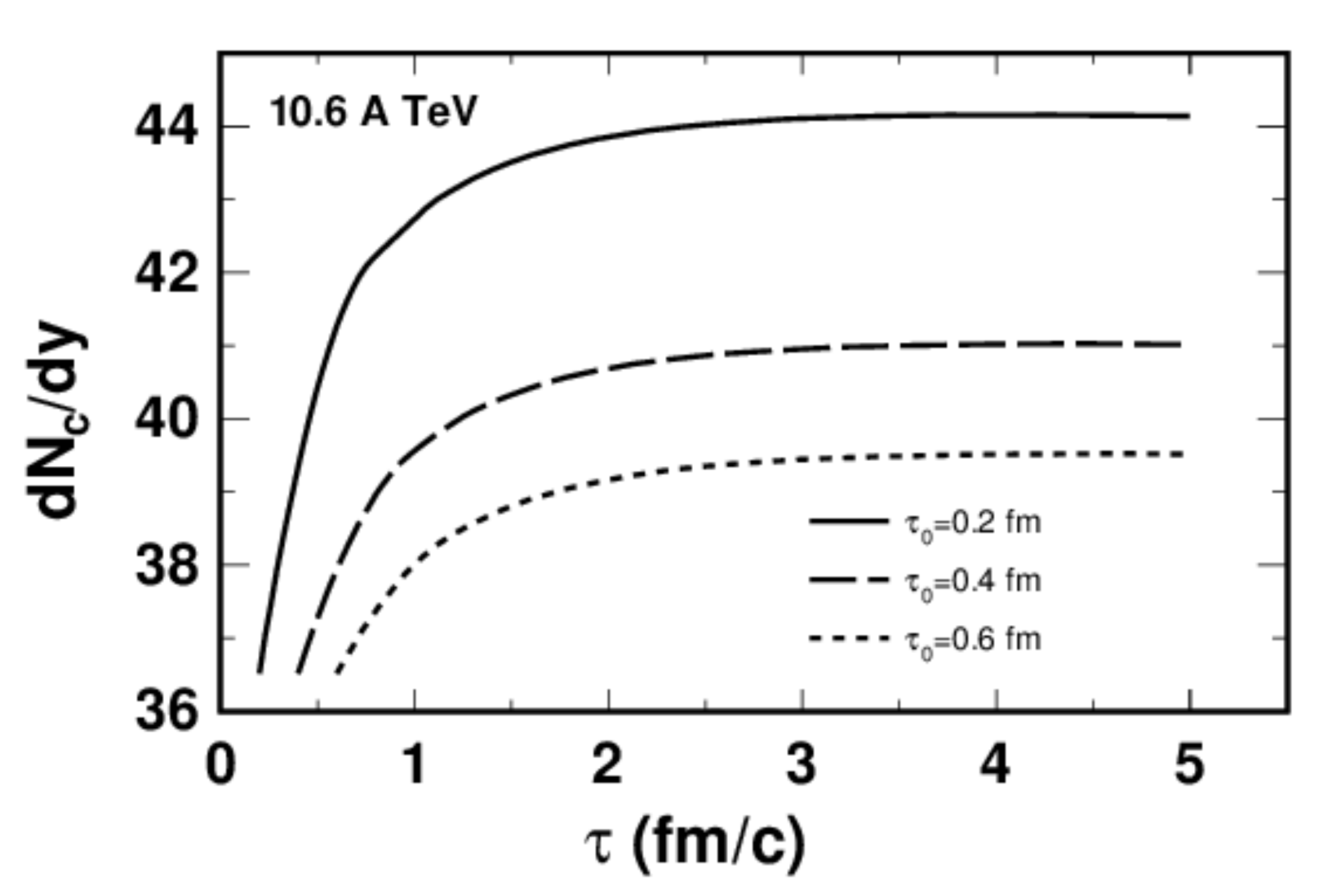}
\caption{Time-evolution of the $\rm c\overline c$ yield
  (per unit of rapidity at midrapidity) for central Pb--Pb collisions
  at $\sqrtsNN=10.6$~TeV, obtained as described in Ref.~\cite{Liu:2016zle}.}
\label{fig:thermalcharm}
\end{center}
\end{figure}

Interactions between gluons or light quarks of the QGP can lead to the
production of $\rm c\overline c$ pairs if the energy in the centre of mass 
of the interaction is of the order of twice the charm quark mass
$\sqrt{\hat s}\sim 2\,m_c\sim 3$~GeV. 
In Section~\ref{sec:HE_qgpglobal} we have estimated 
that an initial temperature $T_0$ of 600--800~MeV could be
reached at the HE-LHC. 
With these QGP temperatures a sizeable fraction of the gluons and
light quarks have energies larger than the charm quark mass 
and $\rm c\overline c$ pairs can be produced in their interactions. 
Figure~\ref{fig:thermalcharm} shows the prediction~\cite{Liu:2016zle} for the time-dependence of the $\rm c\overline c$
rapidity density at mid-rapidity in central Pb--Pb collisons at the HE-LHC. The value at the initial time
$\tau_0$ corresponds to the initial hard-scattering cross section.
Both calculations show a rapid increase
after $\tau_0$ with a final value that is larger by up to 20\% than
the initial production. 
This enhancement could be observed for the first time at the HE-LHC
and provide a handle on the initial
temperature of the QGP. 
The abundance of charm quarks also has an effect on the QGP equation
of the state: 
the inclusion of the charm quark in lattice-QCD calculations results in a sizeable 
increase of $P/T^4 \propto n_{\rm d.o.f.}$ for $T>400$~MeV, as
discussed in the context of the FCC~\cite{Dainese:2016gch}.  

The measurements of the nuclear modification factor of J$/\psi$ at the LHC~\cite{Adam:2015isa,Adam:2015rba,Chatrchyan:2012np} 
are described by models that include dissociation caused by
colour-charge screening and a contribution of recombination
(usually denoted (re)generation) from deconfined c and $\rm \overline c$
quarks in the QGP~\cite{Liu:2009nb,Zhao:2011cv,Andronic:2011yq}. 
The (re)generation contribution the charmonium yield is expected to be proportional to the square of the rapidity density of $\rm c\overline c$ pairs in the QGP. 
Therefore, it is predicted to be much larger
at HE-LHC than LHC energies, as a consequence of the larger hard-scattering
production cross section of $\rm c\overline c$ pairs and the possible 
sizeable thermal production.
This could lead to the observation of an enhancement of J$/\psi$
production with respect to binary scaling of the yield in pp
collisions, i.e.\,$R_{\rm AA}>1$, which would be striking evidence of 
$c\overline c$ recombination from a deconfined QGP.



The measurement of $\PGU$ production would be particularly interesting
at the high
energies and temperatures reached at the HE-LHC.
The LHC data are consistent with a scenario in which the excited
states 2S and 3S are partially or totally suppressed by colour
screening, while the 1S, which is the most tightly bound state, has no
or little direct melting. Its suppression by about 60\% could be explained by the lack of feed-down from the (melted) higher states and the effect of nuclear PDF suppression
(see e.g.\,Ref.~\cite{Andronic:2015wma} for a recent review).
At HE-LHC energies, on the one hand, the temperature could be large
enough to determine a full melting even of the tightly-bound 1S state,
on the other hand the large abundance of $\rm b\overline b$ pairs in the
QGP could induce substantial $\PGU$ (re)generation.
The role of the two effects ---degree of survival of initial bottomonia and contribution of
(re)generation--- could be separated by means of precise measurements of
the $\rm b\overline b$ cross section
and of the B meson and $\PGU$ $R_{\rm AA}$ and elliptic flow $v_2$ 
(the regenerated $\PGU$ states could exhibit a $v_2$ such that $0<v_2^{\PGU}<v_2^{\rm B}$).

\subsection{Nuclear PDF measurements and search for parton saturation}
\label{sec:HE_smallx}

\newcommand{\pizero}{\ensuremath{\pi^{0}}}
\newcommand{\ptjet}{\ensuremath{p_\mathrm{T,jet}}}


Parton saturation~\cite{Gribov:1984tu,Mueller:1985wy} is based on the idea that standard linear parton branching  leads, at small values of momentum fraction $x$, to parton densities so high that non-linear dynamics (like gluon recombination) becomes important and parton densities are tamed to grow from power-like to logarithmically. Non-linear effects are expected to become important when the density of gluons per unit transverse area exceeds a certain limit, the \textit{saturation density}. 

\begin{figure}[h]
\centering
\includegraphics[width=0.45\textwidth]{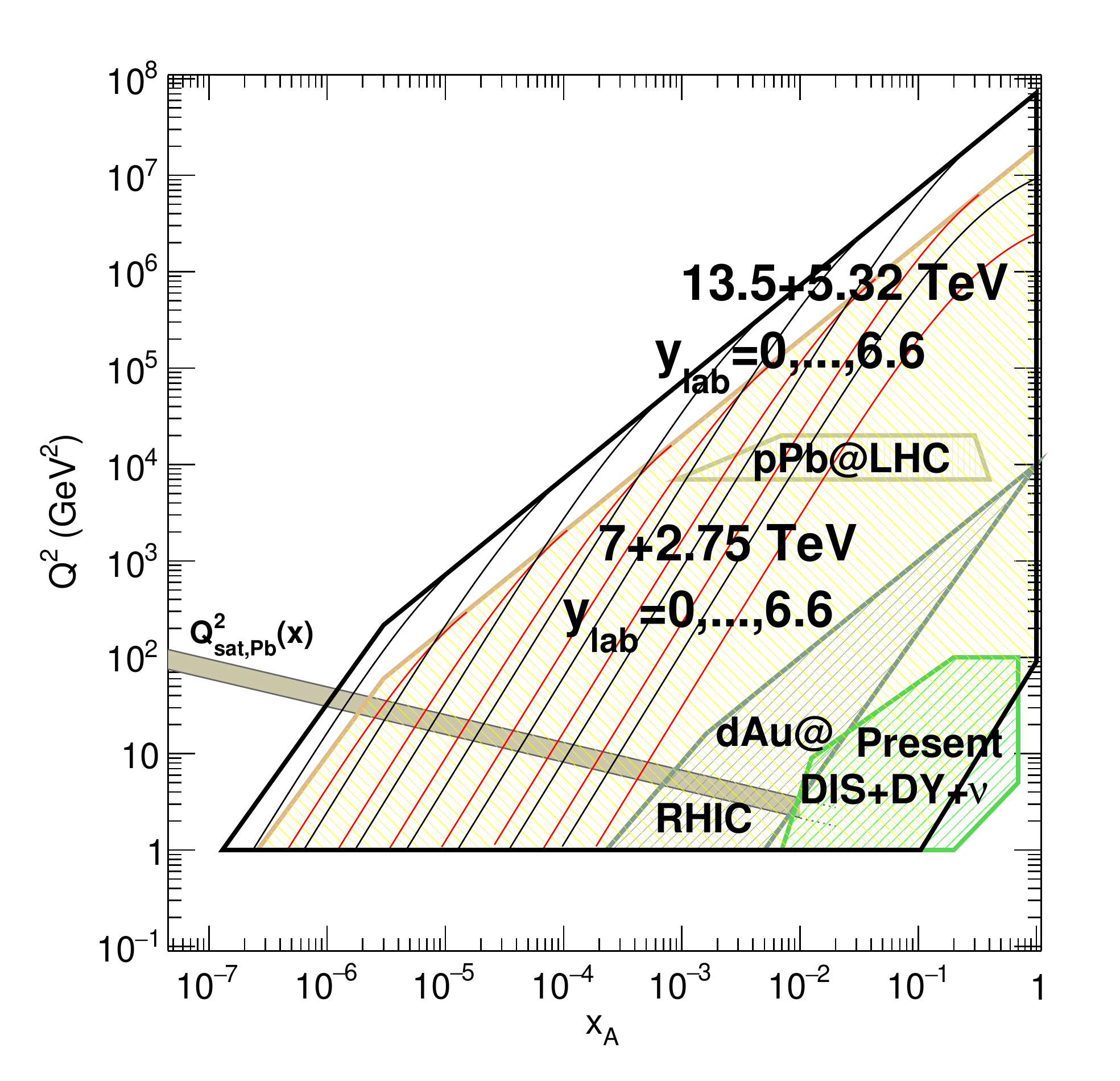}
\includegraphics[width=0.45\textwidth]{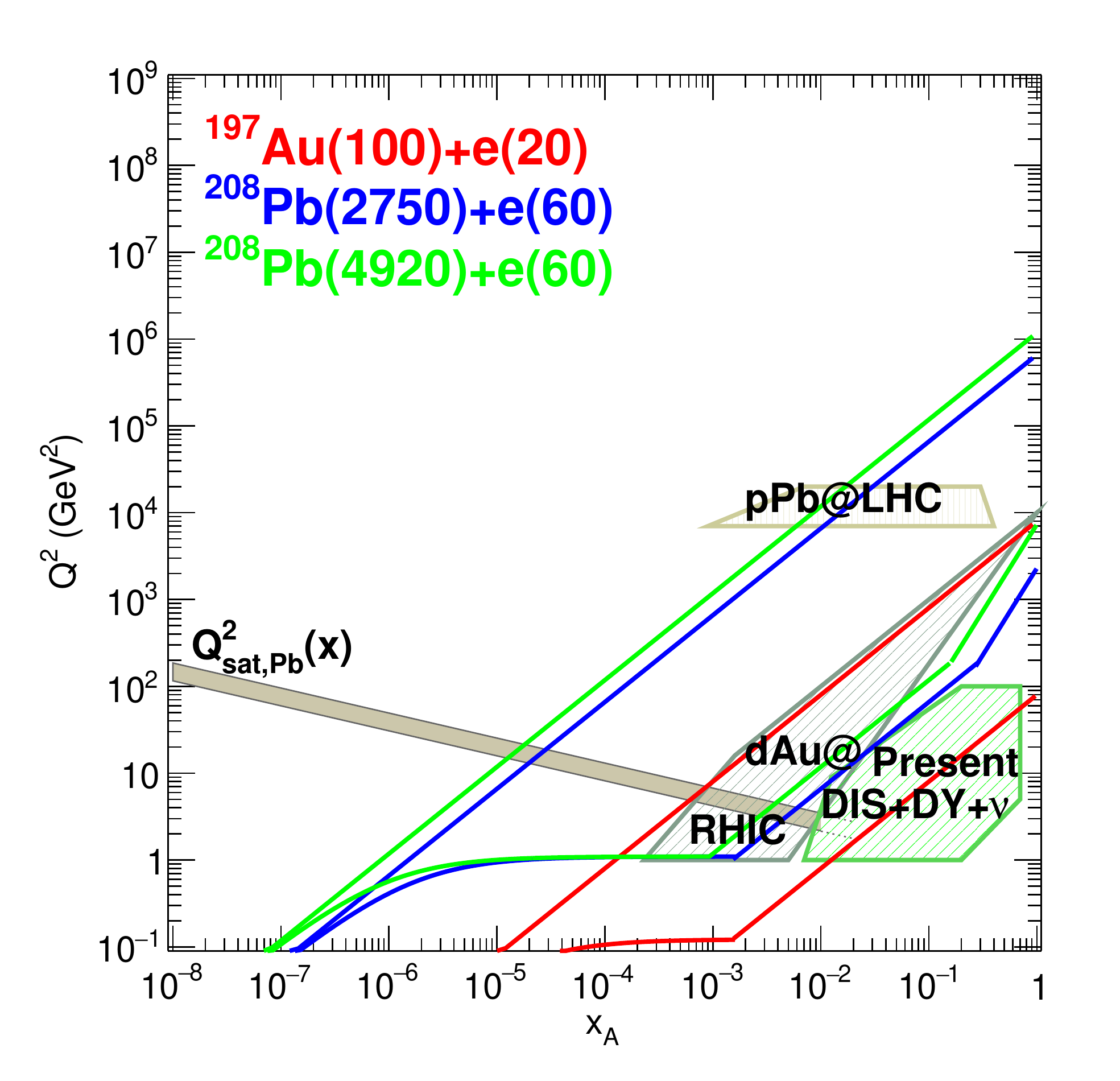}
\caption{Regions of the $x$--$Q^2$ plane covered by the data used for present nuclear PDF fits
(nuclear DIS and Drell-Yan, p--Pb and d--Au at the LHC and RHIC), shown in both panels, by p--Pb collisions at the LHC and the HE-LHC (left) 
and by e--A collisions at the EIC, LHeC and HE-LHC (right). In the left panel, the thin lines correspond to different rapidities in the
laboratory frame $y_{\rm lab} = 0$,1, 2, 3, 4, 5, 6 from right to left, with the left edge defined by $y_{\rm lab} = 6.6$. Values of $Q^2_{\rm S}(x)$ for Pb are shown for illustration in both panels.}
\label{fig:kinplane}
\end{figure}

In the framework of QCD collinear factorization, Parton Distribution Functions of nucleons inside nuclei (nuclear PDFs) can be obtained in standard global fit analysis with usual linear evolution equations. 
The differences with respect to free nuclon PDFs are parametrized in a nuclear modification factor
$R_i^{\rm A}(x,Q^2)$ with $i=\rm g,\,q_{\rm valence},\,q_{\rm sea}$ (see e.g.\,Ref.~\cite{Arneodo:1992wf}).
Collinear factorization 
is expected to break down when the gluon phase-space becomes saturated.
The onset of saturation is usually discussed in terms of the 
saturation momentum $Q_{\rm S}^2$, defined as the 
scale at which the transverse area of the nucleus is completely saturated and gluons start to overlap.
It can be shown that $Q^2_{\rm S} \sim {A}^{1/3} \big(\sqrtsNN\big)^\lambda e^{\lambda y}$,  with $\lambda \approx 0.3$~\cite{Dainese:2016gch}.
Therefore, the regime of high gluon density is best accessed at a high-$\sqrtsNN$ hadron 
collider  with measurements at low $\pt$ and forward rapidity, which probe small $x$ and small $Q^2$.
In order to firmly establish the existence of this new high-energy regime of QCD and clarify the validity of the different approaches to factorisation and evolution, new kinematic regions must be explored using higher collision energies in order to have a large lever arm in $Q^2$ in a region that, while perturbative, lies inside the saturation domain. The HE-LHC extends the small-$x$ coverage by a factor of two with respect to the LHC, as shown in Fig.~\ref{fig:kinplane}.

There is a strong complementarity between the physics programmes at hadron colliders
and at the proposed electron--hadron colliders (Electron-Ion Collider in the USA~\cite{Accardi:2012qut}, Large Hadron Electron Collider LHeC~\cite{AbelleiraFernandez:2012cc}).
With kinematic reach at the TeV scale in the c.m.s.\,(Fig.~\ref{fig:kinplane}, left), 
the electron--nucleus option at the HE-LHC would be well-positioned to reach conclusive evidence for the existence of a new non-linear regime of QCD. It would be clearly complementary with the p--Pb case, providing a precise knowledge on the partonic structure of nucleons and nuclei and on the small-$x$ dynamics. A specific discussion can be found in the electron--nucleus part of the FCC Conceptual Design Report.

\subsection{Photon--photon collisions}
\label{sec:HE_gammagamma}


Photon--photon collisions in UPCs of proton~\cite{dEnterria:2008sh} 
and lead (Pb) beams~\cite{Baltz:2007kq} have been experimentally observed at the
LHC~\cite{Chatrchyan:2012tv,Chatrchyan:2013foa,Abbas:2013oua,Aad:2015bwa}.
The future prospects at the LHC are extensively discussed in Sect.~\ref{sec:upc}. 
Although the $\gamma$ spectrum is harder for smaller charges --which favours proton over nuclear beams in the
production of heavy diphoton systems-- each photon flux scales with the squared charge of the hadron, $Z^2$, and
thus $\gaga$ luminosities are extremely enhanced for ion beams
($Z^4=5\cdot 10^{7}$ in the case of
\PbPb). 
The Pb beam Lorentz factor at HE-LHC ($\gamma = 5\,650$) determines the ``maximum'' quasireal photon energy $\omega_{\rm max} = \gamma/\rm{R_{Pb}} \approx 160$~GeV, leading to photon--photon collisions up to centre-of-mass energies of $\sqrt{s_{\gamma\gamma}} \approx 320$~GeV, twice larger than those reachable at the LHC.

The very rare elastic scattering of two photons in vacuum
$\gaga\to\gaga$  was recently
observed for the first time in UPCs at the LHC~\cite{Aaboud:2017bwk,CMS-PAS-FSQ-16-012}.
At the HE-LHC, due to the higher diphoton masses
reached, this process may be sensitive to physics beyond the SM through new heavy charged particles contributing to the
virtual loop such as, \eg\ from SUSY
particles~\cite{Gounaris:1999gh}. Light-by-light (LbyL) scattering has also been proposed as a tool to
search for monopoles~\cite{Ginzburg:1998vb}, axions~\cite{Bernard:1997kj}, unparticles~\cite{Kikuchi:2008pr}, 
low-scale gravity effects~\cite{Cheung:1999ja}, and non-commutative interactions~\cite{Hewett:2000zp}.

\clearpage
\section*{Acknowledgements}


We would like to thank the LHC experimental Collaborations and the WLCG for their essential support.
We are especially grateful for the efforts by the computing, generator and validation groups who were instrumental for the creation of large simulation samples. We thank the detector upgrade groups as well as the physics and performance groups for their input. Not least, we thank the many colleagues who have provided useful comments on the analyses.

\noindent
Specific authors acknowledge the following forms of support:

\begin{itemize}

\item Nestor Armesto, Elena G. Ferreiro and Carlos Salgado acknowledge the support of the Ministerio de Ciencia e Innovaci\'on of Spain under projects FPA2014-58293-C2-1-P, FPA2017-83814-P and Unidad de Excelencia Mar\'{\i}a de Maetzu under project MDM-2016-0692, by Xunta de Galicia (Conseller\'{\i}a de Educaci\'on) within the Strategic Unit AGRUP2015/11, and by FEDER. 
\item Raphaelle Bailhache acknowledges the support of the German Federal Ministry of Education and Research (BMBF).
\item Zvi Citron acknowledges the support of the ISRAEL SCIENCE FOUNDATION (grant No. 91/6491).
\item Torsten Dahms acknowledges the support of the DFG cluster of excellence ``Origin and Structure of the Universe''.
\item Stefan Floerchinger acknowledges the support of the DFG Collaborative Research Centre SFB 1225 (ISOQUANT).  
\item Iwona Grabowska-Bold acknowledges the support of the Polish National Science Centre grant DEC-2016/23/B/ST2/01409, of the AGH UST statutory tasks No. 11.11.220.01/4 within subsidy of the Ministry of Science and Higher Education, and of the PL-Grid Infrastructure.
\item Ilkka Helenius acknowledges the support of the Carl Zeiss Foundation and the Academy of Finland, Project 308301.
\item Peter Jacobs and Mateusz P\l osko\'n acknowledge the support of the U.S. Department of Energy, Office of Science, Office of Nuclear Physics, under contract DE-AC02-05CH11231.
\item Jiangyong Jia, Peng Huo and Mingliang Zhou acknowledges the support of the U.S. National Science Foundation under grant number PHY-1613294. 
\item Spencer Klein acknowledges the support of the U.S. DOE under contract number DE-AC02-05-CH11231.
\item Filip Krizek acknowledges the support of the Ministry of Education of the Czech Republic (grant InterExcellence LT17018) 
\item Mariola Klusek-Gawenda acknowledges the support of the Polish National Science Center Grant No. DEC-2014/15/B/ST2/02528.
\item Nirbhay Kumar Behera acknowledges the support of National Research Foundation of Korea (NRF), Basic Science Research Program, funded by the Ministry of Education, Science and Technology (Grant No. NRF-2014R1A1A1008246). 
\item Constantin Loizides acknowledges the support of of the U.S. Department of Energy, Office of Science, Office of Nuclear Physics, under contract number DE-AC05-00OR22725.
\item Laure Massacrier acknowledges the support of CNRS under the grant RFBR/CNRS 18-52-15007.
\item Christoph Mayer acknowledges the support of the Polish Ministry of Science and Higher Education and from the Polish National Science Center.
\item Alexander Milov acknowledges the support of  the Israel Science Foundation (grant 1065/15), and the MINERVA Stiftung with the funds from the BMBF of the Federal Republic of Germany.
\item Soumya Mohapatra acknowledges the support of the Division of Nuclear Physics of the US Department of Energy under grant DE-FG02-86ER40281.
\item Petja Paakkinen acknowledges the support of the Magnus Ehrnrooth Foundation.
\item Hannu Paukkunen acknowledges the support by his Academy-of-Finland project 308301.
\item Dmitri Peresunko acknowledges the support of the Russian Science Foundation grant 17-72-20234.
\item Ralf Rapp acknowledges the support of the U.S. National Science Foundation under grant number PHY-1614484, and in part by the ExtreMe Matter Institute EMMI at the GSI Helmholtzzentrum f\"{u}r Schwerionenforschung (Darmstadt,Germany).
\item Kristof Redlich acknowledges the support of the Polish National Science Center NCN under Maestro grant DEC-$\mathrm{2013/10/A/ST2/00106}$.  
\item Mark Strikman acknowledges the support of the U.S. Department of Energy, Office of Science, Office of Nuclear Physics, under Award No. DE-FG02-93ER40771.
\item Adam Trzupek acknowledges the support of the National Science Centre, Poland under the grant no 2016/23/B/ST2/00702.
\item Michael Weber acknowledges the support of the Austrian Academy of Sciences and the Nationalstiftung f\"ur Forschung, Technologie und Entwicklung, Austria.
\item Michael Winn acknowledges the support of the European Research Council~(ERC) through the project EXPLORINGMATTER, funded by the ERC through a ERC-Consolidator-Grant.

\end{itemize}


\footnotesize
\biblio

\end{document}